\documentclass[physics]{puthesis}
\usepackage{amsmath,amssymb,color,pulongtable}
\title{Systematics of identified particle production in \lowercase{pp}, \lowercase{d}A\lowercase{u} and A\lowercase{u}-A\lowercase{u} collisions at RHIC energies}

\author{Levente Moln\'{a}r}{Moln\'{a}r, Levente}

\pudegree{Doctor of Philosophy}{Ph.D.}{May}{2006}

\majorprof{Fuqiang Wang Professor}

\campus{West Lafayette}

\def	\dNdeta	{dN_{\rm ch}/d\eta}
\def	\dNdy	{dN_{\rm ch}/dy}
\def	\dNraw	{dN_{\rm ch}^{\rm raw}/d\eta}

\def	\pt	{p_{\perp}}
\def	\la	{\langle}
\def	\ra	{\rangle}
\def	\meanpt {\langle p_{\perp}\rangle}

\def	\Npart 	{N_{\rm part}}
\def	\Ncoll 	{N_{\rm bin}}

\def	\Tch 	{T_{\rm ch}}
\def	\Tkin 	{T_{\rm kin}}

\def	\pbar	{\overline{p}}

\begin{document}

%
%
%
%
%

\maketitle

\begin{dedication}
Sz\"{u}leimnek - To my parents
\end{dedication}

\begin{acknowledgments}
First, I would like to thank my adviser, Professor Fuqiang Wang. His deeply motivated, diligent attitude toward research provided me great inspiration writing this thesis. I am deeply indebted to him for his help and support.

I also would like to thank Professor Olga Barannikova, who guided me when I joined the Purdue Heavy-Ion group and took my first steps in this field.
 
I would like to thank the members of the Purdue Heavy-Ion group: Professor Rolf Scharenberg, Dr. Brijesh Srivastava, Professor Andrew Hirsch, Professor Norbert Porile, Dr. Blair Stringfellow, for their useful advice and interesting conversation over a cup of coffee. 

I would like to thank Professor L\'{a}szl\'{o} Gutay for his guidance in the first years of the graduate school and his support over the years. I also would like to thank Professor Roberto Colella and Professor Albert Overhauser to be members of my committee.

I also would like to thank the junior members of our group: Terence Tarnowsky, Jason Ulery and Michael Skoby, whom with not only sharing office but friendship through the up and down sides of the graduate school in these years. 

I would like to thank all members of the STAR Collaboration who make possible to complete this work and the members of the Spectra Working Group for their guidance and constant interest on my research.

This work was supported by the DOE grant: DE-FG02-88ER40412 and the Purdue Research Foundation grant: PRF-690 1396-3955.
\end{acknowledgments}


\tableofcontents

\listoftables

\listoffigures





\begin{abstract}

Identified mid-rapidity particle spectra and freeze-out properties are presented for 200 GeV pp, 200 GeV dAu and 62.4 GeV Au-Au collisions, measured in the STAR-TPC. The STAR-TPC is a unique tool to investigate identified bulk particle production from elementary pp to large multiplicity Au-Au collisions. Results are contrasted to previous experiments to provide an overview of bulk properties in heavy-ion collisions.

Evolution of the identified particle spectra ($\pi^{\pm}$, $K^{\pm}$, p and $\overline{p}$) with charged particle multiplicity and event centrality is investigated in detail. Significant hardening of the spectrum of heavy particles (kaons and protons/antiprotons) is found in central Au-Au collisions. The average transverse momentum of kaons and protons/antiprotons in high multiplicity pp and central dAu collisions is larger than in peripheral Au-Au collisions at the same energy. The average transverse momentum in 62.4 GeV and 200 GeV Au-Au collisions seem to only depend on event multiplicity.

Particle production examined through particle-antiparticle ratios ($\pi^{+}/\pi^{-},\\ K^{+}/K^{-},
 \overline{p}/p$) and unlike particle ratios ($K^{-}/\pi^{-},\ \overline{p}/\pi^{-}$) show smooth evolution from pp to dAu to Au-Au collisions. Significant net baryon is present in the central collision zone in 62.4 GeV collisions and 200 GeV collisions.
Strangeness production increases with centrality in peripheral collisions and saturates in medium-central to central collisions in heavy-ion collisions at 62.4 and 200 GeV, in contrast to lower SPS and AGS energies.

Chemical freeze-out properties of the collision systems are obtained from particle ratios and the kinetic freeze-out properties from the shapes of particle spectra. Thermal model fits to the measured particle ratios yield a chemical freeze-out temperature $\sim$ 155 MeV in 200 GeV pp, 200 GeV dAu and 62.4 GeV Au-Au collisions. The extracted chemical freeze-out temperature is close to the critical phase transition temperature predicted by lattice QCD calculations. The kinetic freeze-out temperature extracted from hydrodynamically motivated blast-wave models shows a continuous drop from pp, dAu and peripheral to central Au-Au collisions, while the transverse flow velocity increases from $\sim$ 0.2 in pp to $\sim$ 0.6 in central 200 GeV Au-Au collisions. The kinetic freeze-out parameters in 62.4 GeV and 200 GeV Au-Au collisions seem to be governed only by event multiplicity/centrality.

The kinetic freeze-out results are obtained from blast-wave fit to spectra data treating all particles as primordial ones. However, resonance decays may modify the spectral shapes significantly, and therefore may affect the extrapolated kinetic freeze-out parameters. In order to study this possible effect the data are fitted with the blast-wave model including resonances. It is found that the thus extracted parameters are consistent with those obtained without including resonances. This is because the resonance decays do not modify the spectral shapes significantly in the measured $p_{T}$ region in STAR. 

\end{abstract}

%
%
%

\chapter{Introduction}
Complexity of Nature always fascinated mankind, who tried to interpret its 
environment. Even in the $5^{th}$ century BC, Democritos and Leucippos thought
the world was made of finite set of undividable elements: $atoms$.

Today, we have a more sophisticated and continuously expanding view about
the basic building blocks of Nature. In this thesis we focus on a small 
segment, namely high energy heavy-ion collisions.

The aim of heavy-ion physics is to discover and study the expected new phase of matter, the Quark Gluon Plasma,
which is believed to exist in the early Universe, a few $\mu s$ after the Big Bang, where 
quarks and gluons could roam over large distances.
We hope to recreate the evolution of the early Universe in high energy heavy-ion collisions.
The large number of participating nucleons and the large energy density could
create a suitable environment to study this early phase of matter.
However, this system is far more complex than any elementary collisions. 
Signals to be measured come from a strongly interacting, hot and dense medium, therefore 
proper characterization of this new phase requires combination of them.   

Several experimental facilities have been built since the 1970s, the most recent is the Relativistic
Heavy Ion Collider (RHIC) at Brookhaven National Laboratory. RHIC is capable of
colliding counter rotating Au ion beams at a center of mass energy of 200 GeV per nucleon pair. Hence in 
a central Au-Au collision, in the collision zone, almost 40 TeV of energy is available to create a suitable environment for the search of Quark Gluon Plasma.

Through decades, many observables have been suggested as possible signatures
of the expected new phase. In this thesis we do not attempt to cover every 
aspect of the heavy-ion physics, but only concentrate on the $bulk\ properties$. 
During the 5 years of RHIC running, vast amount of data have been gathered and 
analyzed by the participating experiments. As the result, each experiment has summarized its achievements and addressed the remaining tasks
in the White papers~\cite{Adams:2005dq,Arsene:2004fa,Adcox:2004mh,Back:2004je}. The vast amounts of data has allowed us to characterize the main bulk properties
and the field moves toward more refined and specific measurements. This thesis tries
to provide a summary of the bulk properties of collisions measured at RHIC in 
the STAR detector.

\chapter{QCD and QGP in high energy collisions}

\section{QCD in vacuum}

Elementary particles are divided into two classes: fermions (the building 
blocks: quarks, leptons) and bosons (the glue: gluons). In the search for 
the elementary constituents of matter, the LEP experimental results point 
to the existence of three generations of the basic building blocks, each with 
two quarks (u,d - c,b - t,b) and their antiquarks. There are two leptons 
corresponding to a generation ($e^{\pm}$ - $\mu^{\pm}$ - $\tau^{\pm}$) and their neutrinos ($\nu_{e}$, $\overline{\nu}_{e}$, $\nu_{\mu}$, $\overline{\nu}_{\mu}$, $\nu_{\tau}$, $\overline{\nu}_{\tau}$).
Experiments aimed to search for bare quarks all have failed. Quarks always appear 
bounded in hadrons: in $baryons$ (qqq) or in $mesons$ (q$\overline{q}$). Exploration of 
the baryon spectrum and the prediction of new particles and their experimental 
discovery formed a solid foundation to the constituent quark model~\cite{Gell-Mann:1962xb}. 
The existing hadron spectrum(as known in $\sim$ 1960) could be described by 
conservation laws of the quantum numbers: baryon number, isospin, strangeness
number and hypercharge, electric charge and spin. 

Discovery of $\Delta^{++}$(uuu), $\Delta^{-}$(ddd) and $\Omega^{-}$(sss) 
particles required the introduction of a new quantum number to avoid the 
contradiction to the Pauli Exclusion Principle within the quark model~\cite{Gell-Mann:1962xb}. 
The proposed solution assigns new quantum numbers to the quarks, the $colors$ suggested
by Greenberg~\cite{Greenberg:1964pe} and Gell-Mann~\cite{Fritzsch:1973pi,Han:1965pf}. 
In order to satisfy the Pauli Exclusion Principle, three color states are needed 
(called red, green and blue), but the hadrons remain colorless objects. 

Further development of the quark model based on gauge invariance lead 
to Quantum Chromo Dynamics (QCD), a field theory, which describes the 
strong interactions between $colored$ quarks and the force carriers: 
(eight $colored$) gluons. The color charge is confined to the hadrons, 
according to the $confinement$ $hypothesis$ which can be described by 
the potential obtained from lattice QCD calculations for heavy quarks:
\begin{equation}
V_{\overline{Q}Q}\sim\frac{4}{3}\frac{\alpha_{S}(r)}{r}+\sigma r
\label{eq:Yukawa}
\end{equation}
where $\alpha_{S}(r)$ is the strong coupling constant, $\sigma$ is 
the QCD string tension and $r$ is the distance of the color charges. 
As can be seen from Eq.~\ref{eq:Yukawa}, the potential at small distances 
is Coulomb like, but increases linearly at large distances. The confinement 
hypothesis provides a natural explanation of the observed color neutral 
hadrons and the short range of the strong interaction.

\begin{figure}[!h]
  \begin{center}
  	  \resizebox{.7\textwidth}{!}{\includegraphics{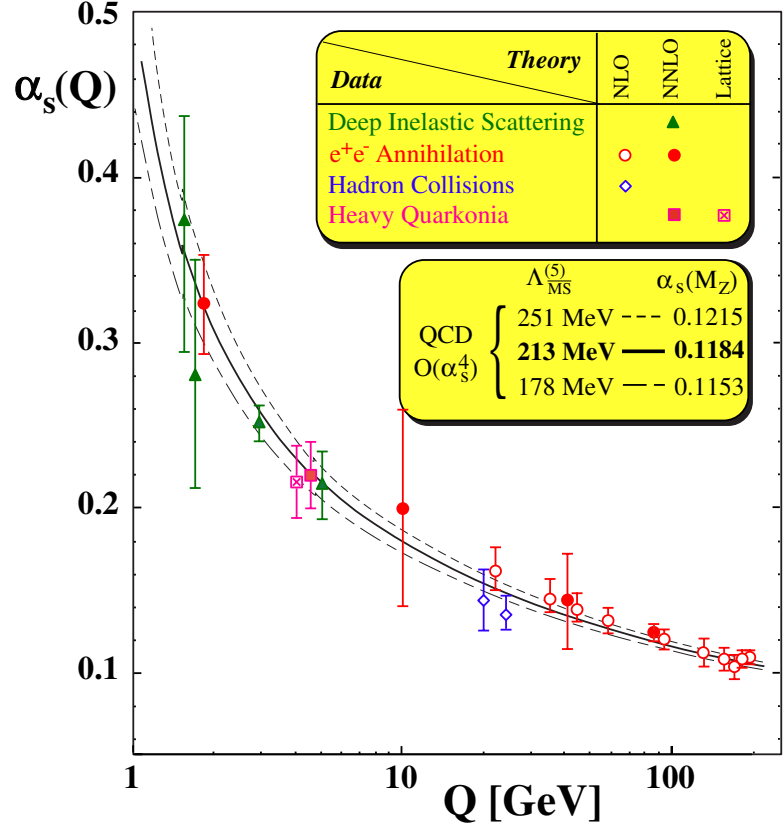}}
    \caption{Summary of the $\alpha_{S}$ measurements. Figure is taken from~\cite{Bethke:2000ai}.}\label{fig:alpha}
  \end{center}
\end{figure}

The gluon fields are non-Abelien, which leads to self interaction between 
the gluons and to the change in the effective coupling constant of the strong 
interaction. Figure~\ref{fig:alpha} shows the change in $\alpha_{S}$ as a function of momentum transfer. 
At large momentum transfer the effective coupling constant becomes small and the 
probed quarks appear to be free objects, as measured in deep inelastic 
scattering experiments. The experimental results 
in this region are well described by perturbative QCD (pQCD). However, in the 
region where the momentum transfer (Q) is small (soft physics region), 
perturbative calculations are not applicable.%
%
%
\begin{figure}[!h]
  \begin{center}
  	  \resizebox{.7\textwidth}{!}{\includegraphics{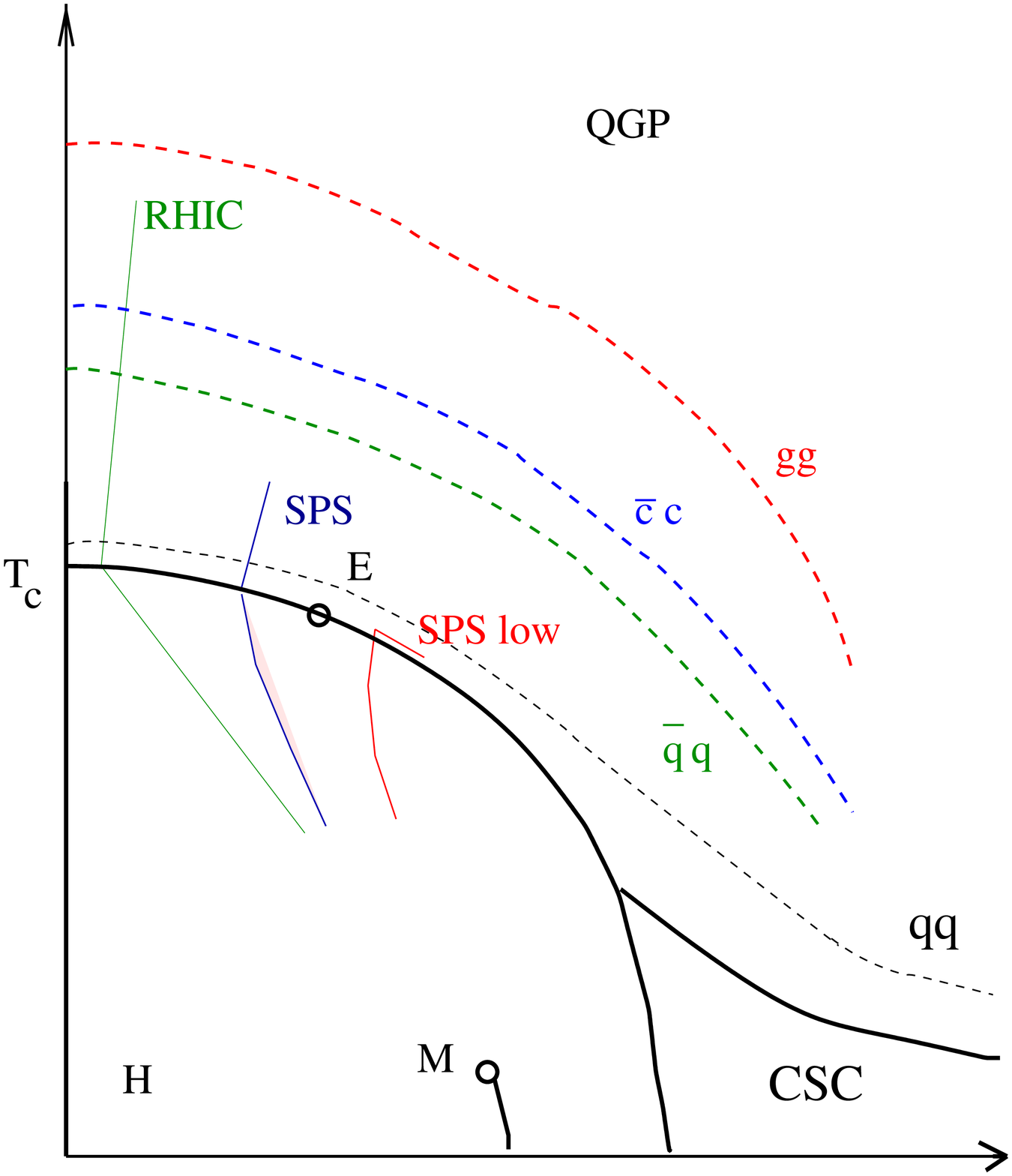}}
    \caption{Schematic view of the QCD phase diagram in the T - $\mu_{B}$ plane. Figure is taken from~\cite{Shuryak:2004cy}.}\label{fig:qcdphasediag}
  \end{center}
\end{figure}
%
%
\begin{figure}[!h]
  \begin{center}
  	  \resizebox{.8\textwidth}{!}{\includegraphics{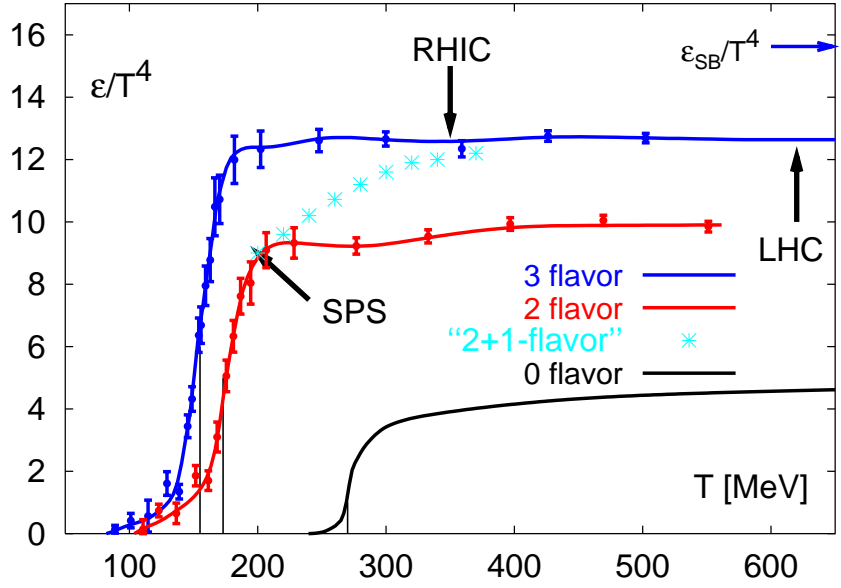}}
    \caption{$\epsilon$/$T^{4}$ as a function of critical temperature. Figure is taken from~\cite{Karsch:2004ti}.}\label{fig:karcsi}
  \end{center}
\end{figure}
\begin{figure}[!h]
  \begin{center}
  	 \resizebox{.8\textwidth}{!}{\includegraphics{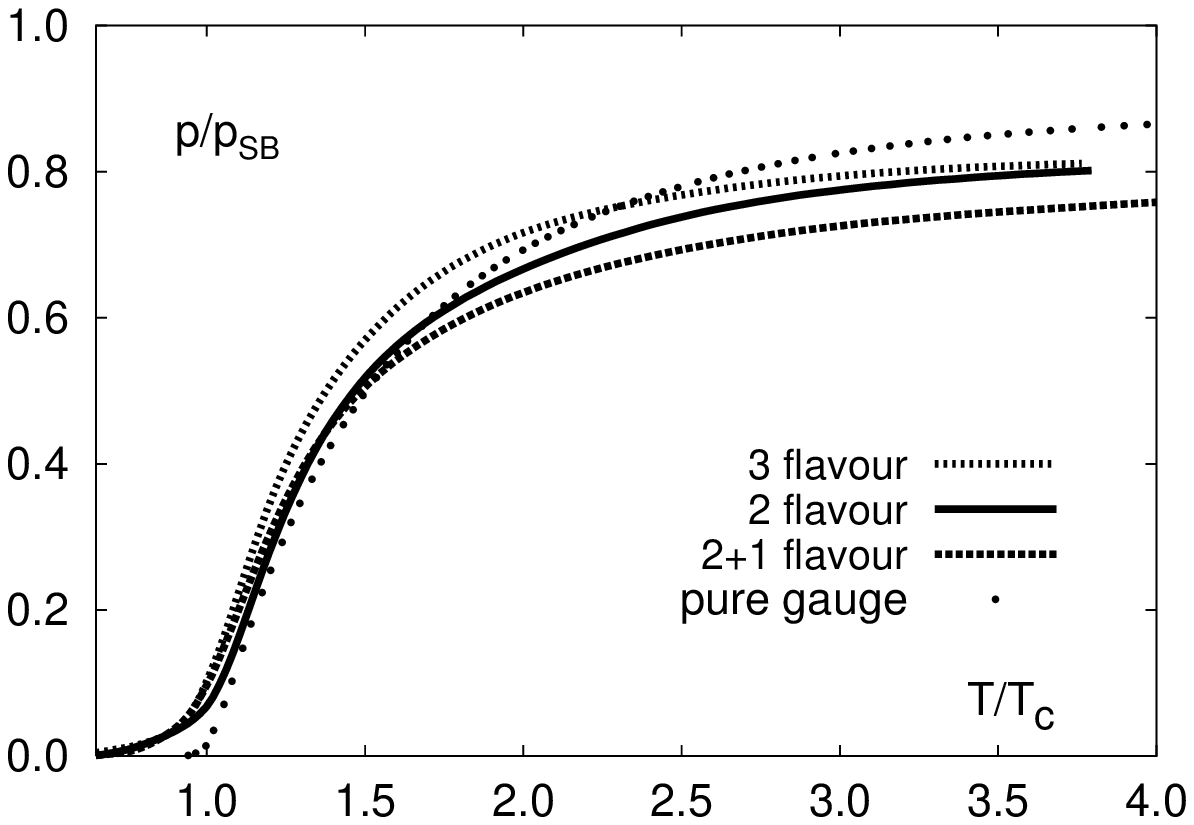}}
  	 \resizebox{.8\textwidth}{!}{\includegraphics{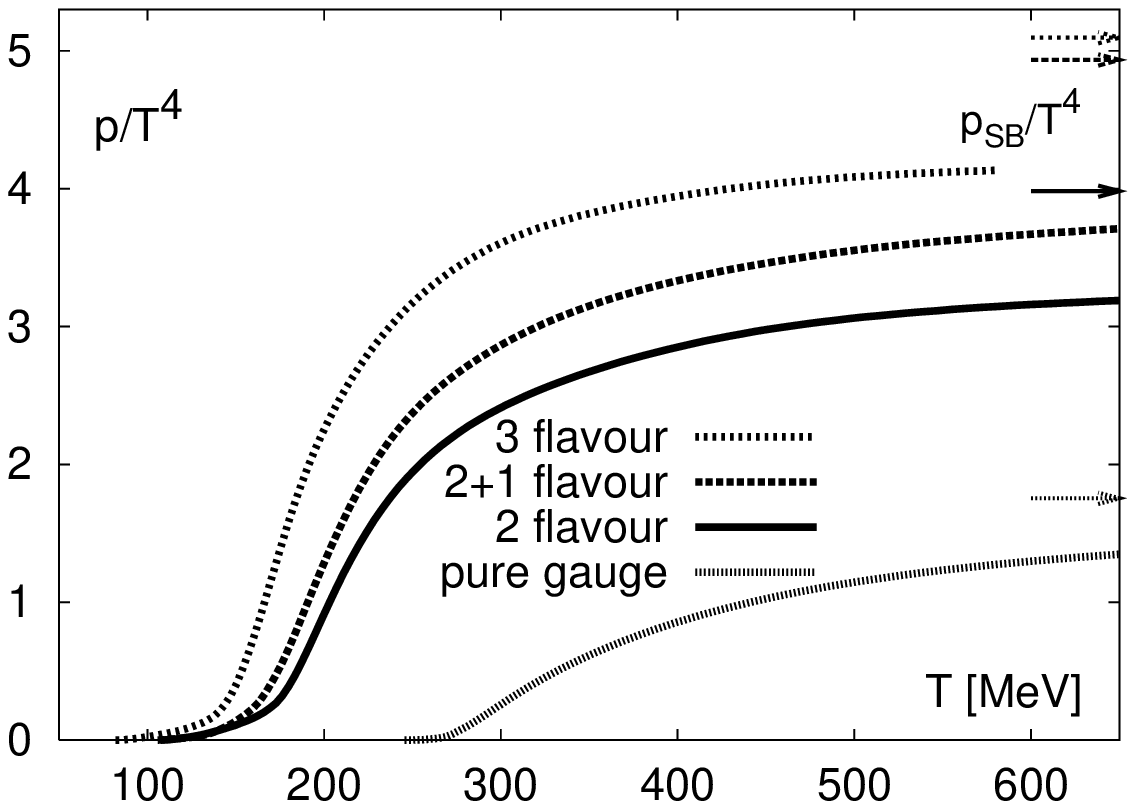}}
    \caption{Lattice QCD calculation of p/$T^{4}$ as a function of critical temperature. Figure is taken from~\cite{Karsch:2001jb}.}\label{fig:karcsiP}
  \end{center}
\end{figure}
\begin{figure}[!h]
  \begin{center}
  	 \resizebox{.8\textwidth}{!}{\includegraphics{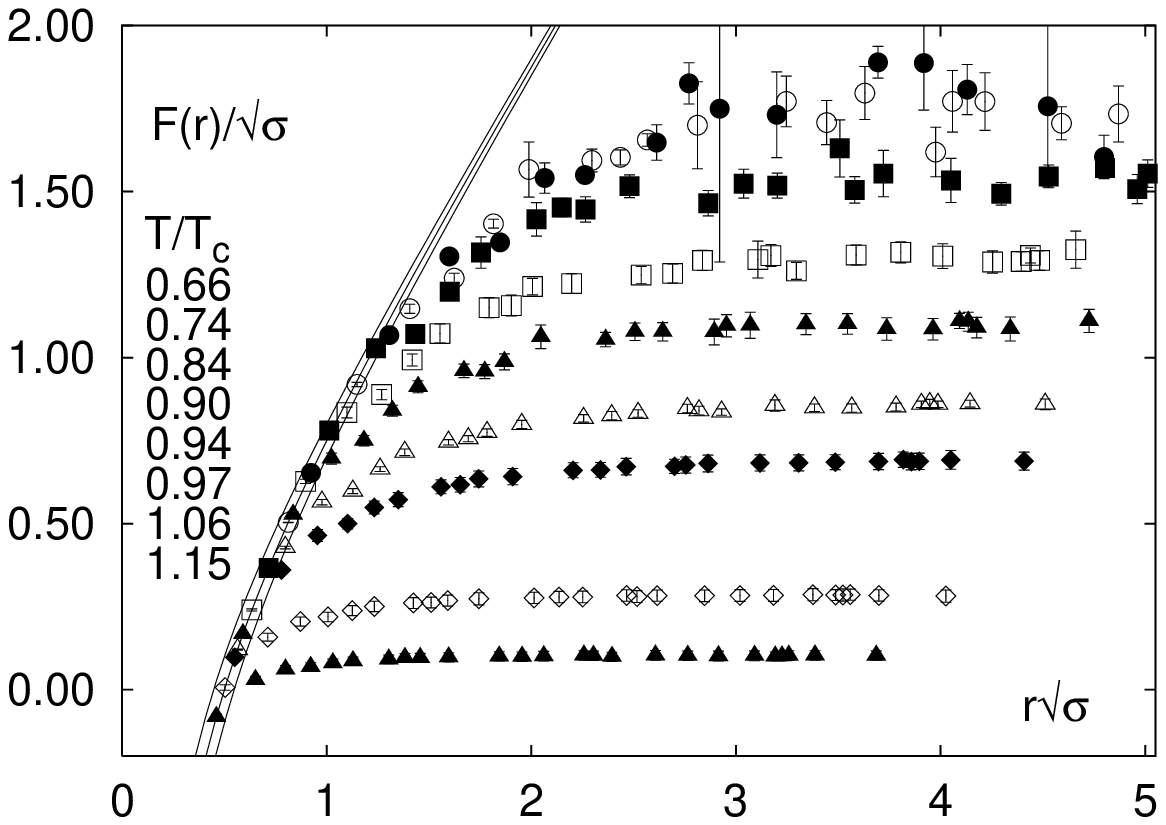}}
  	 \resizebox{.8\textwidth}{!}{\includegraphics{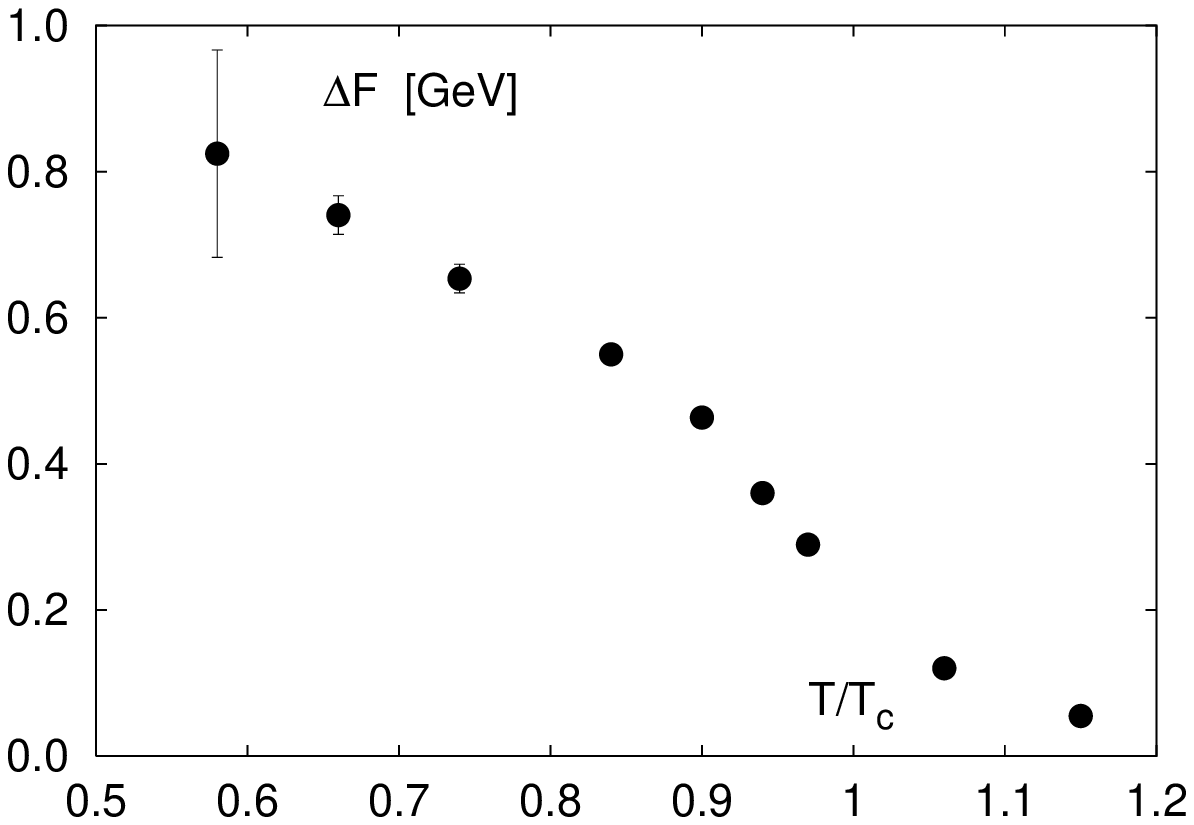}}
    \caption{Left panel: Temperature dependence of the heavy quark free energy in three flavor QCD. Figure is taken from~\cite{Karsch:2001cy}.}\label{fig:karcsiHeavy}
  \end{center}
\end{figure}
The observed behavior of the effective coupling constant can be described by the following expression:
\begin{equation}
\alpha_{S}(Q^{2})=\frac{4\pi}{(11-2N_{F}/3)ln(-Q^{2}/\Lambda_{0})}
\label{eq:alpha}
\end{equation}
where $N_{F}$ represents the number of flavors with mass below $\left| Q^{2}\right|^{\frac{1}{2}}$ and $\Lambda_{0}$ is the scaling parameter $\sim$ 200 MeV. Large momentum transfer corresponds to a small interaction distance; the observed decrease of the effective coupling constant with increasing momentum transfer or decreasing distance is called $asymptotic$ $freedom$.

\section{QCD in colored medium}

In high energy heavy-ion collisions the number of participating nucleons/quarks is large and the behavior of $\alpha_{S}$ is modified compared to the in-vacuum case (as described above). In high energy heavy-ion collisions the average momentum transfer is in the order of $\Lambda_{0}$ limit, therefore the effective coupling constant should be described as a function of the temperature: 
\begin{equation}
\alpha_{S}(Q,T)=\frac{g^{2}}{4\pi(1-\Pi(Q,T)/Q^{2})}
\label{eq:alphatemp}
\end{equation}
where the term: $1/(4\pi(1-\Pi(Q,T)/Q^{2}))$ is the QCD analogue of the Debye screening of a test charge in electrolyte, but includes the effect of the colored medium~\cite{Letessier:2002gp}. 
$\alpha_{S}$ exhibits the same behavior with increasing temperature as with increasing momentum transfer.
Therefore, one can summarize the expectations from QCD of quark confinement: at large momentum transfers (small distances) or at large temperatures the quarks appear to be free, that is, the quarks are deconfined. 

This deconfined phase of quarks and gluons is called Quark Gluon Plasma (QGP). A more precise definition will be given in the next chapter. 

The early theoretical expectations predicted that the phase transition simultaneously occurs with chiral symmetry restoration. Due to the confining nature of vacuum the quark mass is generated dynamically inside the hadrons. The so called quark condensate, which can be regarded as an order parameter has a finite value in vacuum: $\left\langle \overline{\psi}\psi\right\rangle \approx$ -235$(MeV)^{3}$~~\cite{Letessier:2002gp} and is expected to disappear in the QGP phase.  

Figure~\ref{fig:qcdphasediag} shows the phases of QCD matter in the temperature (T) - baryon chemical potential ($\mu_{B}$) plane. Letter $H$ denotes the phase of the normal hadronic matter. Letter $M$ denotes the place of the nucleus in the QCD phase diagram. Black lines represent the phase boundary between hadron gas and the Quark Gluon Plasma at small baryon chemical potential and between hadron gas and Color Super Conductor (CSC) phase at large baryon chemical potential. Letter $E$ denotes the critical end point from Lattice QCD calculations for first order phase transition. Furthermore the accessible regions of the RHIC and SPS experiments are also shown. In the case of RHIC the accessible phase space is well above the phase transition boundary, possibly reaching another newly suggested reign of bound states~\cite{Shuryak:2004tx}. 

\section{Lattice QCD}

Experimental probe of the QCD phase map is limited, and pQCD calculations are limited to interactions involving large momentum transfers. However, the average momentum transfer in high energy heavy-ion collisions is small. Numerical simulation methods of QCD on the lattice are developed to calculate the analytically unaccessible region of QCD. A thorough description of lattice QCD can be found in~\cite{Karsch:2001cy}. 
Lattice QCD calculates Feynman path integrals representing the expectation values of the quantum field theory operators. Integrals are calculated over all gluon and quark fields at all lattice space - time points. After the calculations are performed, the extracted quantities are extrapolated to the continuum limit (lattice spacing $\rightarrow$ 0).

The first calculations are performed with pure gluon fields at vanishing baryon chemical potential. Introduction of the fermion fields on the lattice result a doubling of flavors. On the 4D lattice (3 space, 1 time) each quark specie appears in 16 copies. Different techniques are developed to overcome the doubling problem. The first solution is from Wilson~\cite{Wilson:1974sk}, where the mass of the doublets is inversely proportional to the lattice spacing, hence they disappear at the continuum limit. However, the non zero mass introduces chiral symmetry breaking in the action. To avoid the chiral symmetry breaking, Kogut-Sussking has introduced the $staggered$ fermion action~\cite{Kogut:1974ag,Susskind:1976jm,Bernard:1997an}. A recent development is the $domain$ $wall$ approach~\cite{Furman:1994ky}, where the doubling problem is solved through the introduction of a 5th dimension. Upon interpreting the lattice QCD results the above approximations should be kept in mind to understand the limitation of the calculations/predictions.

Development of the lattice formulation of thermodynamics has lead to several interesting results. Investigation of QCD at non-zero baryon chemical potentials and non-zero temperatures suggest phase transition from the hadronic phase to the Quark Gluon Plasma phase when sufficiently high energy density and temperature is reached, as shown in Fig.~\ref{fig:karcsi}. The $\epsilon$/$T^{4}$ is proportional to the number of degrees of freedom. The arrow indicates the Stefan-Boltzmann limit:
\begin{equation}
\epsilon=g\frac{\pi^{2}}{30}T^{4}
\label{eq:eSB}
\end{equation}
where $g$ is the number of degrees of freedom. For a hadron gas, the basic number of degrees of freedom are given by the three pion states ($\pi^{+}$, $\pi^{-}$, $\pi^{0}$):
\begin{equation}
\epsilon_{HG}=3\frac{\pi^{2}}{30}T^{4}.
\label{eq:eHG}
\end{equation}
In the QGP phase the relative number of degrees of freedom are the quarks and gluons. From the estimate of ideal relativistic boson (gluons) and femion gas (quarks), the following relation can be written for the energy density: 
\begin{equation}
\epsilon_{g+q}=\int{\frac{d^{3}p}{(2\pi)^{3}}\frac{p}{e^{p/T}-1}}+\int{\frac{d^{3}p}{(2\pi)^{3}}\frac{p}{e^{p/T}+1}}
\end{equation}
\begin{equation}
\epsilon_{g+q}=\frac{4\pi T^{4}}{(2\pi)^{3}}\int{\frac{x^{3}dx}{e^{x}-1}}+\frac{4\pi T^{4}}{(2\pi)^{3}}\int{\frac{x^{3}dx}{e^{x}+1}}
\end{equation}
\begin{equation}
\epsilon_{g+q}=\frac{\pi^{2}}{30}T^{4}+\frac{7}{8}\frac{\pi^{2}}{30}T^{4}.
\label{eq:eQGP}
\end{equation}
The energy density can be written as:
\begin{equation}
\epsilon_{QGP}=(2_{spin}\times8_{colors})\frac{\pi^{2}}{30}T^{4}+(2_{quark-antiquark}\times2_{spin}\times3_{colors}\times n_{flavors})\frac{7}{8}\frac{\pi^{2}}{30}T^{4}
\label{eq:eQGP}
\end{equation}
\begin{equation}
\epsilon_{QGP}=(16+\frac{21}{2}n_{flavor})\frac{\pi^{2}}{30}T^{4}.
\label{eq:eQGP}
\end{equation}
From these naive estimates the number of degrees of freedom is significantly increased in the QGP phase with respect to the hadron gas phase. 
Figure~\ref{fig:karcsi} shows significant increase in the number of degrees of freedom around the critical temperature of the phase transition.
The critical temperature depends on the number of flavors and the mass of the quarks. The black curve shows the classical calculation for pure gluon fields. The blue curve shows the expectation for three light quarks, and the red curve shows the two light quarks calculation. In a more realistic calculation shown in light blue, two light quarks (u,d) and a heavy quark (s) are considered. This later case might represent the case at RHIC, where the sharp transition slightly flattens out. Although the critical temperature changes in the above cases (173 $\pm$ 8 MeV for two flavors and 154 $\pm$ 8 MeV for three flavors), the critical energy density is found to be in the range: $\epsilon_{c} \sim$ 0.5 - 1.0 GeV/fm$^{3}$~\cite{Karsch:2004ti}.

In Fig.~\ref{fig:karcsiP} the Stefan-Boltzmann scaled pressure of hadrons is shown as a function of the scaled temperature (left panel) and the pressure of hadrons in units of $T^{4}$ as a function of the temperature (right panel). The evolutions of the pressures (left panel) are similar and they do not reach the Stefan-Boltzmann limit. Lattice QCD calculations show deviation from the ideal Stefan-Boltzmann gas. 

As we mentioned above, QCD predicts the confinement of quarks which can be described by the potential between two heavy quarks as shown in Eq.~\ref{eq:Yukawa}. This effective potential can be calculated on the lattice as well~\cite{Karsch:2001cy}.

Figure~\ref{fig:karcsiHeavy} (left panel) shows the temperature dependence of the heavy quark free energy in three flavor QCD with a quark mass 0.1 GeV. The calculation is performed for static quarks. The three lines represent the Cornell type potential (V(r)/$\sqrt{\sigma}$=-$\alpha$/r$\sqrt{\sigma}$+r$\sqrt{\sigma}$, $\alpha$=0.25 $\pm$ 0.05)~\cite{Karsch:2001cy} in the unit of the square root of the string tension, which coincides with the lattice calculation for temperature T $\leq$ $T_{c}$ and r $\approx$ 1.5/$\sqrt{\sigma}$ $\approx$ 0.3 fm. 
The flat region at T $< T_{c}$ suggests that the two heavy quarks separate into two nearly non-interacting heavy-light mesons, such as D and B. The decrease of the magnitude in the flat region shows that the effective light quark mass decreases due to chiral symmetry restoration. 
Furthermore, Fig.~\ref{fig:karcsiHeavy} also shows $\Delta F$, the difference in the free energy of a quark-antiquark pair at infinite separation and the free energy of the quark-antiquark pair at distance $r_{q}$=$\sqrt{\alpha/\sigma}$. Around the critical temperature the probability of the formation/existence of a heavy bound state is small. The distance and temperature dependence of the free energy will lead to the enhancement of the D meson (bound state of a charm and an up or down quark) with respect to the J/$\psi$ (c$\overline{c}$) which can be tested experimentally.

\chapter{High energy heavy-ion collisions and QGP}

Ultra-relativistic heavy-ion collisions provide possible means to explore the bulk properties of QCD in a volume several times larger than the initial colliding nuclei. These collisions might lead to a high enough energy density for the formation of the Quark Gluon Plasma (QGP). Theoretical expectations and interpretations of the QGP have evolved over two decades. A working definition of QGP is as follows~\cite{Adams:2005dq}:  {\it A (locally) thermally equilibrated state of matter in which quarks and gluons are deconfined from hadrons, so that color degrees of freedom become manifest over nuclear, rather than merely nucleonic, volumes}.

\section{Space-time evolution of high energy heavy-ion collisions}
\begin{figure}[!t]
  \begin{center}
  	  \resizebox{.7\textwidth}{!}{\includegraphics{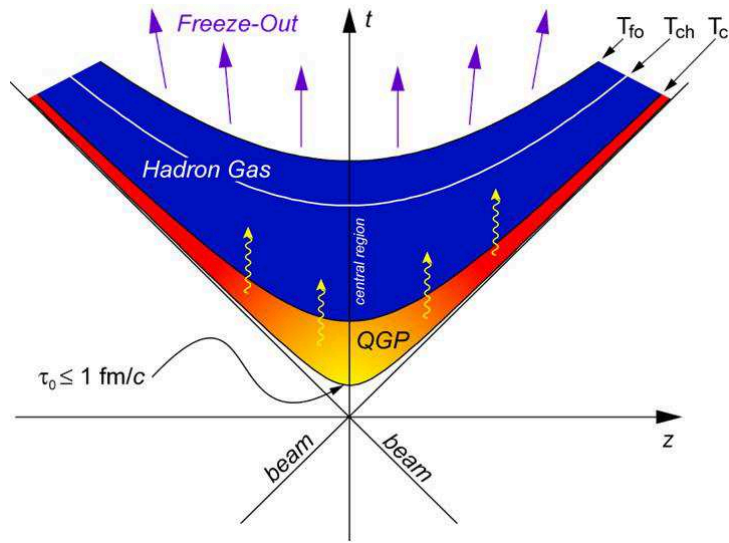}}
  	  \vspace*{-0.1cm}
    \caption{Illustration of the space time evolution of a heavy-ion collision with QGP formation.\label{fig:HIevo}}
  \end{center}
\end{figure}
The space time evolution of the heavy-ion collision is summarized in Fig.~\ref{fig:HIevo}. It is a hard task for theorists (and maybe too much to ask) to provide a coherent picture of all stages of a heavy-ion collision within the same theoretical framework, although the field is rapidly evolving. To build a general picture of heavy-ion collisions, we select narrow topics from the initial collision stage to the final free-streaming of the particles. We try to address gluon saturation, themalization, freeze-out properties and hadron production through experimental measurements.

A general view of heavy-ion collisions is the following.
The two incoming highly Lorentz contracted nuclei approach each other at the interaction point ($z=0$) with speed near the speed of light ($c$). A common understanding of the initial collisions that they are dominated by gluons, hence the number of partons is significantly larger than the constituent quarks of the two nuclei. This initial state is often referred as the Color Glass Condensate. The temperature from the initial stage is increasing and the Color Glass Condensate melts above the critical temperature of the phase transition to the Quark Gluon Plasma. The estimated time from the initial collisions to the formation of the QGP is short, less than: $\sim$ 1 fm/c. This is important, since without rapid thermalization, one cannot treat the system within thermodynamical description. 

\begin{figure}[!t]
  \begin{center}
  	  \resizebox{.95\textwidth}{!}{\includegraphics{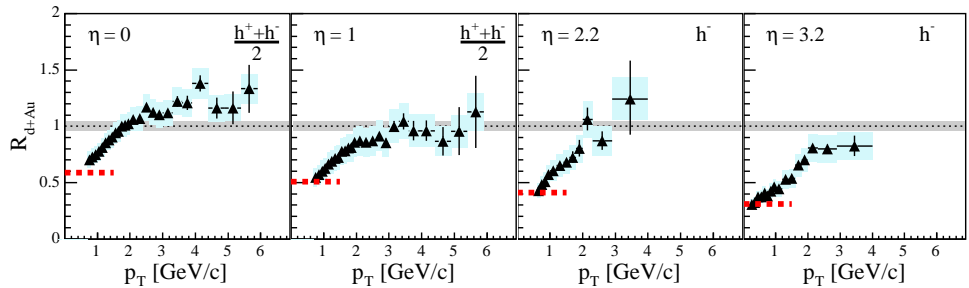}}
  	  \vspace*{-0.1cm}
    \caption{Nuclear modification factor of charged hadrons at different pseudorapidities measured in 200 GeV dAu collisions. Figure is taken from~\cite{Arsene:2004ux}.\label{fig:BRAHMSdAu}}
  \end{center}
\end{figure}
Once the system evolves to the QGP phase, the high energy density and the pressure gradient drive the system to expansion and subsequent cooling. This regime can be described by relativistic hydrodynamics. As the temperature drops and the system becomes dilute, the QGP phase is not longer sustainable, the system freezes out, and the hydrodynamical approach breaks down. When inelastic collisions cease in the system, the chemical composition of the final state will not change. It is referred to as chemical freeze-out and it is characterized by the chemical freeze-out temperature ($T_{ch}$) and the baryon and strangeness chemical potentials ($\mu_{B}$, $\mu_{S}$). Some models introduce an ad-hoc strangeness suppression factor ($\gamma_{S}$) to account for the non-fully equilibrated system.
\begin{figure}[!h]
  \begin{center}
  	  \resizebox{.7\textwidth}{!}{\includegraphics{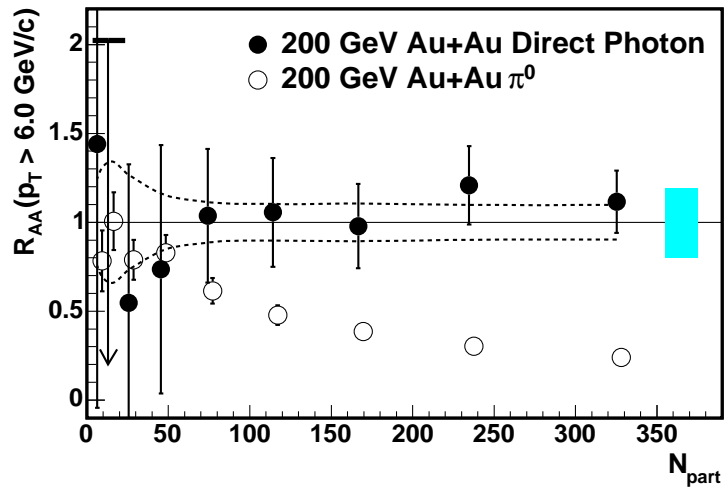}}
  	  \vspace*{-0.1cm}
    \caption{$R_{AA}(p_{T})$ is shown for direct photons (dots) and $\pi^{0}$ (open circles) for central Au-Au collisions at 200 GeV.  Figure is taken from~\cite{Adler:2005ig}.\label{fig:phenixdirphot}}
  \end{center}
\end{figure}
Chemical freeze-out is likely to be a continuous transition rather than a sudden freeze-out. Therefore, between the critical temperature ($T_{c}$) and the chemical freeze-out one would expect the existence of the mixed phase of QGP and hadron gas. Further expansion leads to a more dilute stage; the elastic collisions eventually cease. At this stage the kinetic properties of the system are frozen and it is called kinetic freeze-out. It is characterized by the kinetic freeze-out temperature ($T_{kin}$, denoted by $T_{fo}$ in Fig.~\ref{fig:HIevo}) and the average transverse flow velocity ($\beta$). Hereafter, the particles free stream toward the detector. Below we address recent measurements probing different stages of the collision.

\section{Saturation before collision}

Theoretical models suggest the saturation of gluon densities in the two highly Lorentz contracted ($\gamma \approx$ 100) nuclei receding toward the interaction point. Due to the Lorentz contraction the colored gluon wave functions start to overlap and the collision itself can be pictured as two highly colored rose-windows passing through each other. This hypothetical initial stage of the heavy-ion collision is called the Color Glass Condensate~\cite{Kovner:1995ja,Kovner:1995ts,Krasnitz:1999wc,Krasnitz:1998ns,Krasnitz:2001qu}.

HERA deep inelastic scattering results indicate~\cite{Breitweg:1998dz} a rise in the gluon distribution function ($xg(x,Q^{2}$)) at small momentum fractions, where $x$ is the Bjorken $x$, and $Q^{2}$ is the 4 momentum transfer. 
At momentum transfers of a few GeV scale the gluon distribution function  seems to saturate and gluon fusion ($g + g \rightarrow g$) and gluon splitting ($g \rightarrow g + g$) becomes equally probable.
The latest developments~\cite{Kovner:1995ja,Kovner:1995ts,Krasnitz:1999wc,Krasnitz:1998ns,Krasnitz:2001qu}  indicate that the saturation is achieved at higher $x$ (at a fixed $Q^{2}$) in heavy-ion collisions at RHIC compared to protons at HERA. 
At RHIC the available kinematic region of $x$ is large and the average $Q$ is small, but it still can lead to a rise in the number of low $x$ gluons. The total cross section rises more slowly than the number of gluons per unit area per unit rapidity, hence the areal density of partons involved in the collision may increase above unity~\cite{Kharzeev:2000ph,Kharzeev:2001gp}.
\begin{figure}[!h]
  \begin{center}
  	  \resizebox{.6\textwidth}{!}{\includegraphics{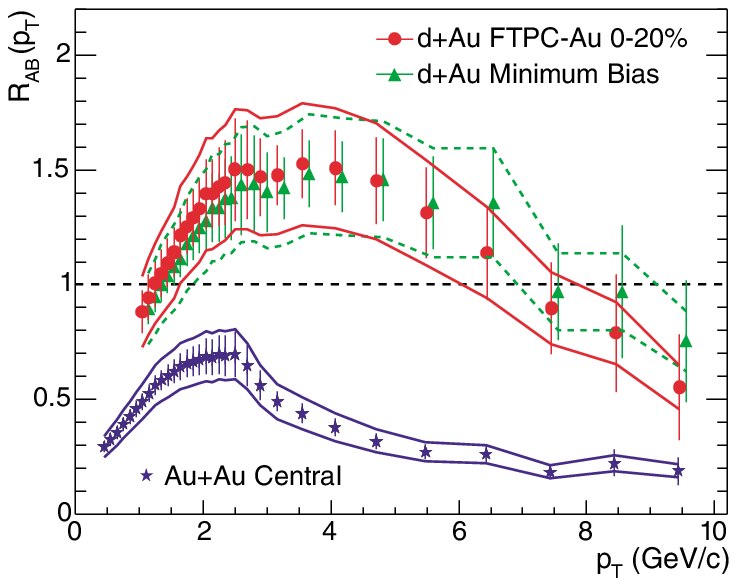}} \\
  	  \resizebox{.6\textwidth}{!}{\includegraphics{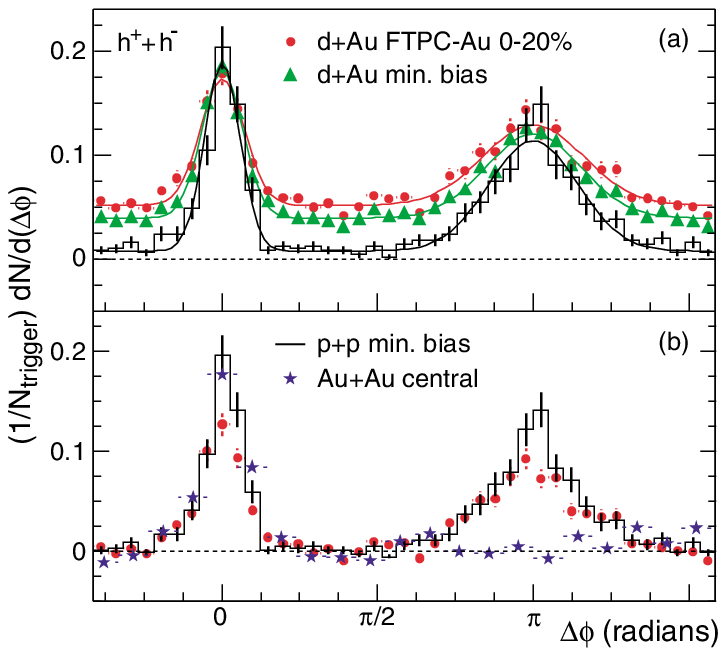}}
  	  \vspace*{-0.1cm}
    \caption{Left panel: $R_{AB}(p_{T})$ for minimum bias and central dAu collisions, and central Au+Au collisions. The minimum bias d+Au data are displaced 100 MeV/c to the right for clarity. Right top panel: Two-particle azimuthal distributions for minimum bias and central d+Au collisions, and for pp collisions. Left lower panel: Comparison of two-particle azimuthal distributions for central dAu collisions to those
seen in pp and central Au+Au collisions. Figure is taken from~\cite{Adams:2003im}.\label{fig:STARjet}}
  \end{center}
\end{figure}

The applicability of the CGC framework is represented by the saturation scale ($Q_{S}^{2}(x,A)$), which depends on the Bjorken $x$ and the mass number ($A$) of the nuclei. The saturation is enhanced by $\sim A^{\frac{1}{3}}$ in the case of low $x$ and moderate $Q^{2}$ in a nucleus with respect to a proton. The proton should be probed at two orders of magnitude lower $x$ to achieve the same enhancement.
Hence, the target nucleus sees an incoming nucleus with a much smaller transverse size compared to the nuclear diameter and longer longitudinal coherence length. The geometrical picture of the collision is similar to the constituent quark based classical geometrical model of heavy-ion collisions, but gluons see a coherent cylinder of gluons in the receding nucleus.  
The BRAHMS experiment has measured the nuclear modification factor in dAu collisions at large rapidities, which is defined as:
\begin{equation}
  R_{d+Au} = \frac{1}{N_{coll}} \frac{dN/d\eta(dAu)}{dN/d\eta(pp)}.
  \label{eq:rdau}
\end{equation}
The observed suppression of the measured nuclear modification factor in dAu collisions has been qualitatively predicted within the CGC framework, including quantum evolution, and is interpreted as the appearance of the Color Glass Condensate. The formation of the QGP may start with the CGC which is not fully understood yet and should be tested in various collisions (ep, eA, pA, AA) at RHIC and LHC energies. 
In the working definition of the QGP we require a thermalized system and the CGC seems to provide a natural explanation for the fast thermalization expected at RHIC~\cite{Kovner:1995ja}.

\section{Early probes of the collision and the medium}

After the initial collisions the QGP is expected to form. The colored medium particle production is expected to be different from vacuum production. Below we discuss some of the observables in the focus of the experimental quark gluon plasma search.

\begin{figure}[!h]
  \begin{center}
  	\resizebox{.8\textwidth}{!}{\includegraphics{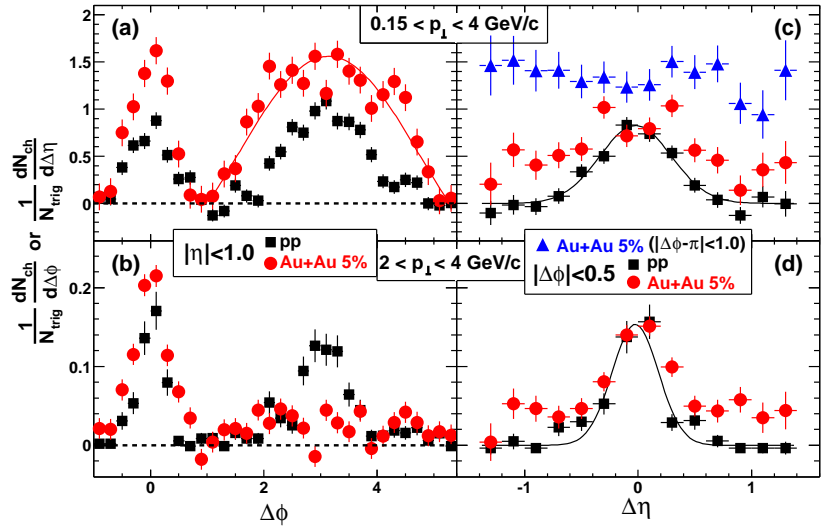}}
   	  
    \caption{Background subtracted $\Delta\phi$ and $\Delta\eta$ distributions. Figure is taken from~\cite{Adams:2005ph}.}
    \label{fig:corrfuncs}
  \end{center}
\end{figure} 

\begin{figure}[!h]
  \begin{center}
  		  	\resizebox{.8\textwidth}{!}{\includegraphics{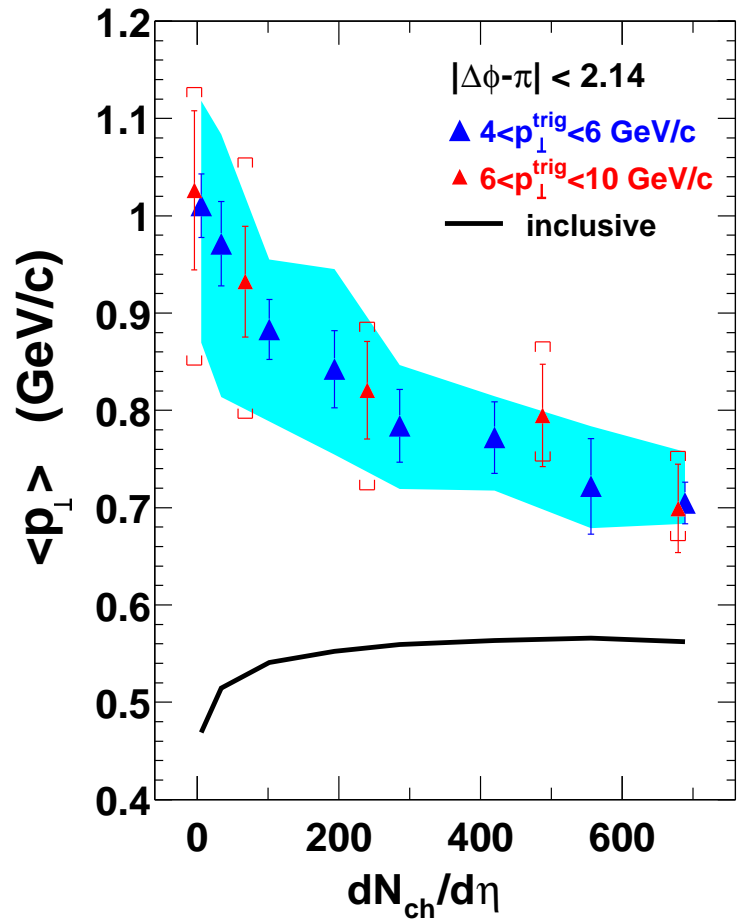}}
  	  
    \caption{Average transverse momenta of the away side associated hadrons and of inclusive particles are shown for 200 GeV Au-Au collisions as a function of centrality. Figure is taken from~\cite{Adams:2005ph}.}
    \label{fig:hardsoftmeanpt}
  \end{center}
\end{figure} 
\begin{figure}[!h]
  \begin{center}
  	  \resizebox{.55\textwidth}{!}{\includegraphics{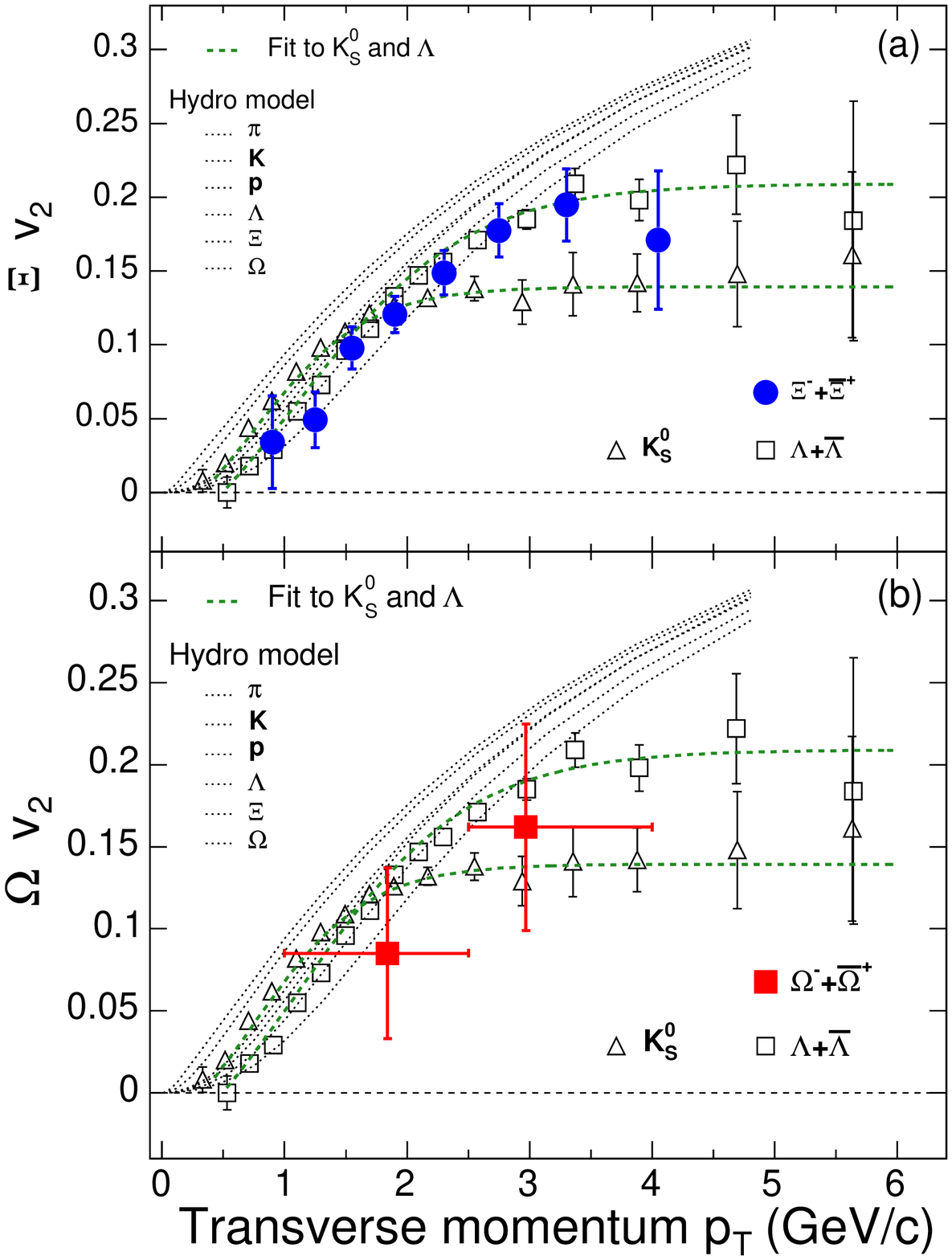}}\\
  	 
  	  \resizebox{.55\textwidth}{!}{\includegraphics{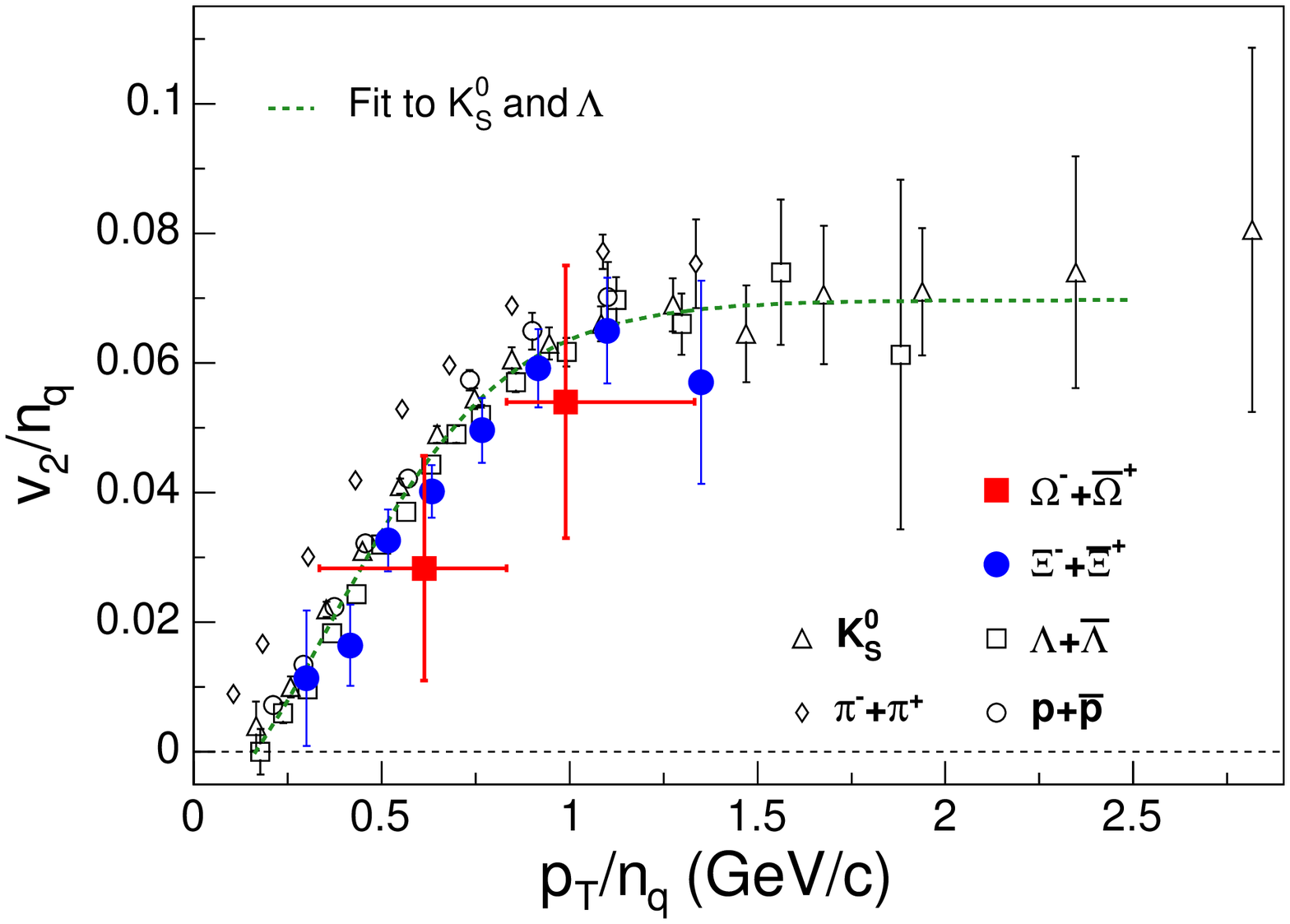}}
  	 
    \caption{Top: $v_{2}$ of multi-strange particles measured at 200 GeV minimum bias Au-Au collisions by STAR~\cite{Adams:2005zg}. 
    Bottom: Number of quark scaled $v_{2}$ as a function of scaled transverse momentum. }
    \label{fig:STARv2MS}
  \end{center}
\end{figure}

\subsection{Direct photons}

Probing the early phase of the collision and the QGP is a difficult task. Direct photons provide a useful tool to measure the very early stage of the collisions. They interact only through the electro-magnetic channel, therefore their mean free path is larger than the expected size of the system. Direct photons are created in the thermally equilibrated quark gluon plasma through gluonic channels: $q + \overline{q} \rightarrow \gamma + g$, $g + q \rightarrow \gamma + q$, $g + \overline{q} \rightarrow \gamma +\overline{q}$.

Besides QGP gluons, there are other significantly contributing photon sources through the evolution of the collision: eg. photons from hard scatterings in the QGP phase or photons from particle decay in the hardonic phase. By filtering out the photons from these background processes, the measured direct photons reflect the thermodynamics of quarks and gluons in the system before hardonization. Therefore the transverse momentum spectrum of direct photons should exhibit enhancement with respect to the photon spectrum measured in the hadronic phase~\cite{Steffen:2001pv}. The PHENIX experiment at RHIC has measured the spectra of direct photons in 200 GeV Au+Au collisions. A clear enhancement is observed above $p_{T} >$ 3 GeV/c, and the data is well described by pQCD calculations~\cite{Adler:2005ig}. As shown in Fig.~\ref{fig:phenixdirphot}, direct photons are not suppressed in central Au+Au collisions at 200 GeV, but $\pi^{0}$s and charged hadrons are (Fig.~\ref{fig:STARjet} (left panel)). These measurements suggest that the suppression of charged hadrons and $\pi^{0}$s is indeed a final state effect due to energy loss and the nuclear modifications of quark and gluon distribution functions are small. 

\subsection{Charmonium suppression}

The evolution of the effective free energy of static heavy quarks with distance and temperature calculated in lattice QCD predicts the enhancement of open charm (eg. D or B meson) with respect to charmonium (J/$\psi$) production. Charmonium suppression is predicted to be a signal of the QGP formation~\cite{Matsui:1986dk}. The cornerstone of the QGP search at CERN was the $J/\Psi$ measurement in central Pb+Pb collisions~\cite{Abreu:1997ji}. The observed $J/\Psi$ cross-section is suppressed by a factor of 2 with respect to the Drell-Yan cross-section from peripheral to central collisions. Theoretical developments revealed that the suppression can be explained by other nuclear effects as well.
\subsection{Jet quenching}

Hard scattering of partons is expected to occur early in nuclear collisions. The energetic partons from hard scatterings, traversing the colored, medium are expected to lose energy via gluon bremsstrahlung, which depends on the color charge density of the medium. Direct observation of jets formed by energetic parton fragmentation is not possible in heavy-ion collisions. But the measurements of spectra and two particle azimuthal correlation of the large transverse momentum particles are possible.  Figure~\ref{fig:STARjet} (left panel) shows the nuclear modification factor (blue symbols) measured in central Au+Au and dAu collisions at 200 GeV. The large transverse momentum spectrum in central Au+Au collisions is suppressed compared to pp collisions. (In the absence of any nuclear effect the ratio of the spectra is situated around 1.) Later, the dAu measurement was performed as a control experiment. On the same panel the central and the minimum bias dAu nuclear modification factors are plotted. Suppression is not observed at large transverse momenta. Therefore, one concludes that partons from hard scattering lose energy in the medium created in central Au-Au collisions, a final state effect (to be distinguished from the expected gluon saturation, which is an initial state effect). Figure~\ref{fig:STARjet} (right panel) also shows the two particle azimuthal correlation. On the top panel the pp and central and minimum bias dAu collisions are shown. As known from high energy physics, particles produced in hard scatterings (appear as jets) are back to back in azimuth. Selecting the largest transverse momentum particle (trigger particle) and calculating the difference in azimuth for each track in the event the two particle azimuthal distribution is expected to peak at $0^{o}$ (in the direction of the trigger particle) and at $180^{o}$. This is clearly shown in pp and dAu collisions. In the bottom panel the correlation from central Au+Au is added. The peak around the trigger particle direction is clear, however the peak expected at $180^{o}$ is missing. Partons from hard scattering traveling through the medium have lost their energy and their memory of the common origin.

\subsection{Soft - Hard correlations}

High transverse momentum particles are expected to lose significant energy traversing through the medium created in high energy heavy-ion collisions. In the previous section, it was shown that the high $p_{T}$ suppression is a final state effect. Energy from the high $p_{T}$ particles should be redistributed in the surrounding medium, which mainly constitutes soft particles ($p_{T} <$ 2 GeV/c). It was also shown that statistical reconstruction of jets is possible in heavy-ion collisions. For technical details and definitions we refer to~\cite{Adams:2005ph}.
Figure~\ref{fig:corrfuncs} shows the $\Delta\phi$ and $\Delta\eta$ distributions. The $\Delta\phi$ distribution clearly shows jet like correlations in pp (top left panel) and Au-Au (bottom left panel) collisions, although it is strongly suppressed in the higher associated particle range. The away side of the $\Delta\phi$ distribution is significantly wider in Au-Au than in pp collisions. 
Transverse momentum distributions of the near and away side $\Delta\phi$ distributions are calculated. Furthermore, the average transverse momenta are extracted as shown in Fig.~\ref{fig:hardsoftmeanpt}.
The black curve represents the average transverse momenta of the inclusive particles. The average transverse momenta from the away side jet and from the inclusive particles converge with increasing centrality. This might indicate the equilibration of the away side jet hadrons and the particles from the medium.

\section{Bulk properties}
\subsection{Elliptic flow}
In a non central heavy-ion collision the collision zone has an anisotropic spatial distribution, which is often referred to as the almond shape. This spatial anisotropy is transformed to momentum space anisotropy by rescattering among the constituents of the system. This can be observed in the final azimuthal distribution of hadrons. The invariant cross section can be decomposed into a Fourier series:
\begin{equation}
E\frac{d^{3}N}{d^{3}p}=\frac{1}{2\pi}\frac{d^{3}N}{p_{T}dp_{T}dy}\left(1+\sum^{\infty}_{n=1}2v_{n}\cos(n\left[\phi-\Psi_{r}\right])   \right)
\end{equation}
where $\Psi_{r}$ is the reaction plane angle. The $reaction$ $plane$ is spanned by the beam direction and the direction of the impact parameter. From the Fourier decomposition the component $v_{2}$ is called elliptic flow and can be expressed as $v_{2}\ =\ \left\langle(p_{x}/p_{T})^{2}\ -\ (p_{y}/p_{T})^{2} \right\rangle$ for particle number distribution.

The elliptic flow is expected to develop early in the collision and survives the hadronization, hence the hadron $v_{2}$ measurements carry information from the partonic and hadronic stage of the collision~\cite{Kolb:2003dz}.
Recently STAR has measured the elliptic flow of multi-strange baryons as shown in Fig.~\ref{fig:STARv2MS}. Measurements of their elliptic flow presumably give information on the early stage of the collision, since they are expected to be less sensitive to hadronic re-scattering due to their small hadronic cross-section~\cite{Adams:2003fy}. 

Each particle follows a distinct trend as a function of transverse momentum. Bulk particles are well described by hydrodynamical calculations, but the pion $v_{2}$ is underestimated. The $v_{2}$ for each particle saturates at transverse momenta ($p_{T} \geq 2.0 - 2.5$ GeV/c). Ideal non viscous hydrodynamical calculations show a monotonically increasing trend even at larger transverse momenta. 

Figure~\ref{fig:STARv2MS} (top panel) also shows results for bulk ($\pi, K, p$) and for singly strange particles ($K_{S}^{0}, \Lambda$), and hydrodynamic model calculations~\cite{Huovinen:2001cy}. The measured $v_{2}$ of multi-strange particles is non zero and similar in magnitude to the singly strange and bulk particles. This implies that the multi-strange particles acquired significant flow in the partonic stage of the collision, hence partonic collectivity is present. Figure~\ref{fig:STARv2MS} (bottom panel) also reveals another important finding, namely the constituent quark scaling. As can be seen in the top panel $v_{2}$ of mesons and baryons clearly follow a distinct trend above $p_{T} >$ 2 GeV/c, but the constituent quarks scaled values for mesons and baryons fall on the same curve within errors. This implies that the partonic degrees of freedom are the constituent quarks. Furthermore, the $s$ quark, which is heavy, flows similarly to $u$ or $d$.

\subsection{Statistical model description}

In relativistic heavy-ion collisions the number of produced particles is $\sim$ 5000. The large number of particles and the large system size compared to the interaction length allows macroscopical treatment of the system created in heavy-ion collisions. Investigation of hadron abundances provides an indirect way to study the degree of thermalization. The formation of the QGP and its subsequent thermalization from a near locally thermal equilibrium leads the constituents of the system to chemical equilibrium~\cite{Hwa:2004yg}. As a consequence of the equilibration the saturation of the strange particles is expected as well. Particle yields measured by identified particle spectra provide the input for the thermal analysis.

We follow the statistical model approach as can be found in~\cite{Xu:2001zj}, based on the grand canonical description of the partition function. 
The system is assumed to be in local chemical and thermal equilibrium. The resulting particle density is given by:
\begin{equation}
n_{i}=\frac{g_{i}}{2\pi^{2}}\gamma_{S}\int^{\infty}_{0}\frac{p^{2}dp}{e^{(E_{i}-\mu_{i})/T}\pm1}
\end{equation}
where $g_{i}$ is the spin degeneracy, $p$ is the total momentum, $E$ is the total energy, and $\mu_{i}$ is the chemical potential which can be written as: 
\begin{equation}
\mu_{i}=\mu_{B}B_{i}-\mu_{S}S_{i}-\mu_{I}I^{3}_{i}
\end{equation}
including the baryon number, the strangeness number, and the third component of the isospin. The $\gamma_{S}$ is the strangeness suppression factor which was introduced ad hoc, to describe strange particle yields~\cite{Hwa:2004yg}. The observed number of strange particles in proton-proton and elementary $e^{+}+e^{-}$ collisions were less than it was expected from the statistical approach. One possible explanation is that the strange quark possesses a mass heavier than the up and down quarks, therefore its production is not energetically favored.

The model contains four free parameters: the temperature (T), the baryo-chemical potential $\mu_{B}$, the strangeness chemical potential $\mu_{S}$ and the strangeness suppression factor: $\gamma_{S}$. The relevant conservation laws are:
\begin{equation}
V\sum\mu_{B}B_{i}=Z+N
\end{equation}
\begin{equation}
V\sum\mu_{S}S_{i}=0
\end{equation}
\begin{equation}
V\sum n_{i}I^{3}_{i}=\frac{Z-N}{2}
\end{equation}
The statistical model has been successfully applied to the available data sets from various heavy-ion collisions. Results from STAR will be presented in the Result section. 

Statistical models are successful to describe heavy-ion data, however, we should mention the caveats as well. Furthermore, statistical models with canonical description are able to reproduce hadron abundances in elementary collisions as well, where thermal equilibrium is not expected due to the small system size and the small number of produced particles.

\section{Collectivity and hydrodynamics}

If the system created in heavy-ion collisions thermalizes rapidly due to the strong initial interactions between its constituents and preserves this thermalized state over a significant period of the evolution time, the system can be treated as a relativistic fluid undergoing collective, hydrodynamical flow~\cite{Kolb:2003dz}. Hydrodynamical models have been applied to heavy-ion collisions at BEVALAC, AGS, SPS and RHIC and they achieved impressive success at RHIC energies. Applicability of these models provide the indirect evidence for local thermal equilibrium. 

Hydrodynamics provide a sensitive tool to study the Equation of State. Since these hydrodynamical models cannot be applied to matter out of local thermal equilibrium, models need initial and final boundary conditions. The initial conditions for hydrodynamical models can be calculated from the CGC approach, or various transport codes such as MPC~\cite{Molnar:2000jh} or AMPT~\cite{Zhang:1999bd}, kinetically treating the period from the initial stage to thermal equilibrium. 

Formalism for ideal relativistic fluid in local kinetic equilibrium is as follows.
The equations of motion for a relativistic fluid element come from the local conservation of energy and conserved charges:
\begin{equation}
\partial_{\mu}T^{\mu\nu} = 0
\end{equation}
and
\begin{equation}
\partial_{\mu}j_{i}^{\mu}=0
\end{equation}
where $T^{\mu\nu}$ is the energy momentum tensor and $j_{i}^{\mu}$ are the currents of the conserved charges ($i=1,2,...,n$).
The energy momentum tensor can be written as:
\begin{equation}
T^{\mu\nu} = (\epsilon+p)u^{\mu}u^{\nu}-pg^{\mu\nu}
\end{equation}
where $\epsilon$ is the energy density, $p$ is the pressure, and $g^{\mu\nu}$ is the metric tensor.
The conserved current can be written as:
\begin{equation}
j_{i}^{\mu}=n_{i}u^{\mu}
\end{equation}
where $n_{i}$ is the number density of charge $i$ and $u^{\mu}$ is the four-velocity of the flow field.

The Equation of State (EoS) is also needed, which connects the energy density, pressure, and number densities: $p=p(\epsilon,n_{1}, n_{2}, ..., n_{n})$. Applying this formalism to measured data, the input EoS can be tested. Although several further assumptions are made to derive a simple applicable model. Calculations including hydrodynamical treatment, usually assume cylindrical symmetry due to the geometry of the collision. Calculation in the  longitudinal direction can be simplified as well, assuming Bjorken scaling~\cite{Bjorken:1982qr}. In the Bjorken picture the longitudinal flow is assumed to scale with the $z$ distance: $v_{z} = z/t$. This assumption leads to boost invariance of the system. This assumption based on the measured particle yields at RHIC, work well in the mid-rapidity region: $|y| <$ 1.5. However the BRHAMS experiment showed that particle yields have nearly Gaussian distribution in a wider rapidity region: $|y| <$ 4. Assumption of the Bjorken scenario limits the sensitivity of the models to the transverse activity of the colliding system.

The high energy density and pressure at the beginning of the hydrodynamical evolution leads to rapid expansion. The average mean free path increases and the density of the system decreases. When the system reaches a dilute state the hydrodynamical evolution stops (the elastic collisions cease) and the kinetic freeze-out happens. The kinetic freeze-out is driven by the expansion rate of the system rather than the size of the system~\cite{Schnedermann:1994gc,Kolb:2003dz}.

The freeze-out is customarily described by the Cooper-Frye formula assuming a sudden break up from the perfect local thermal equilibrium to free streaming particles. If the break up criteria for the given fluid element is satisfied the final spectrum for  particle $i$ can be calculated: 
\begin{equation}
E\frac{dN_{i}}{d^{3}p}=\frac{1}{2\pi}\frac{dN_{i}}{p_{T}dp_{T}}=\frac{g_{i}}{(2\pi)^3}\int_{\Sigma}f_{i}(p\cdot u(x),x)p\cdot d^{3}\sigma(x)
\end{equation}
where $d^{3}\sigma(x)$ is the normal vector of the freeze-out surface $\Sigma(x)$. The phase space distribution ($f$) is calculated just before freeze-out at local equilibrium:
\begin{equation}
f_{i}(E,x)=\frac{1}{e^{\frac{E-\mu_{i(x)}}{T(x)}}\pm1}
\end{equation}
The Cooper-Frye formula has two important aspects, which have to be mentioned. Particles with different momenta freeze-out at the same time from the same fluid element, however high momentum particles require larger number of scatterings to reach thermal equilibrium. Recent developments try to address the whole momentum spectrum of particles at freeze-out. One should note, that some part of the initially produced partons does not participate in the collective hydrodynamical motion of the system.
Without proper treatment of the fluid dynamics, one can assume a freeze-out surface and find a solution for that particular choice. In this case the $d^{3}\sigma(x)$ can be negative, which is the flux of particles entering the hydrodynamical phase from the vacuum. Simplified treatment of this negative contribution has lead to problems with energy-momentum conservation. But, the contribution from this negative current is usually negligible. 

We would like to emphasis once more, that the Cooper-Frye formula describes a sudden break up of the thermalized system, while experimental measurements suggest time-temperature ordered freeze-out for different particles~\cite{Adams:2005dq}.

At mid-rapidity, in near baryon free collisions at RHIC, the surface of constant temperature, energy and particle density is a good approximation.

\subsection{Contribution to heavy-ion physics}

Thermalization has a key role in heavy-ion physics. Since experimental measurements take place after the phase transition in the hadronic stage, direct evidence of thermalization cannot be addressed. However, the excellent agreement between measured data and chemical equilibrium and kinetic freeze-out models serves as a strong hint for thermalization. This thesis completes the currently available systematic measurements of freeze-out properties, because the analysis is carried out with the same STAR detector setup as for 130 and 200 GeV Au-Au collisions. Moreover, the 62.4 GeV measurement is situated between the previously available highest energy at SPS and RHIC data, and provides further constraint for models describing the evolution of bulk quantities and freeze-out properties with collision energy.
From the available low momentum measurements, in light of the AGS and SPS measurements, and the results from RHIC, particle production is statistical and follows the expectations from a source in local thermal and chemical equilibrium. Dynamics of the collision system are governed by hydrodynamical principles. Freeze-out properties evolve smoothly with collision energy in the RHIC regime. Larger collision energy creates a larger system, while particle production is predominantly determined by the net baryon content of the collision zone.

\chapter{The RHIC facility and the STAR detector}

\section{The RHIC facility}

The Relativistic Heavy Ion Collider (RHIC) opened a new era in the exploration
of heavy-ion collisions. The machine is dedicated to the search of the theoretically
predicted Quark Gluon Plasma. The RHIC heavy-ion physics goals were complemented by the installation
of the Siberian Snakes, which allow the study of the spin structure of nucleons in a wide
 range of collision energies. 
\begin{figure}[!h]
  \begin{center}
  	  \resizebox{.7\textwidth}{!}{\includegraphics{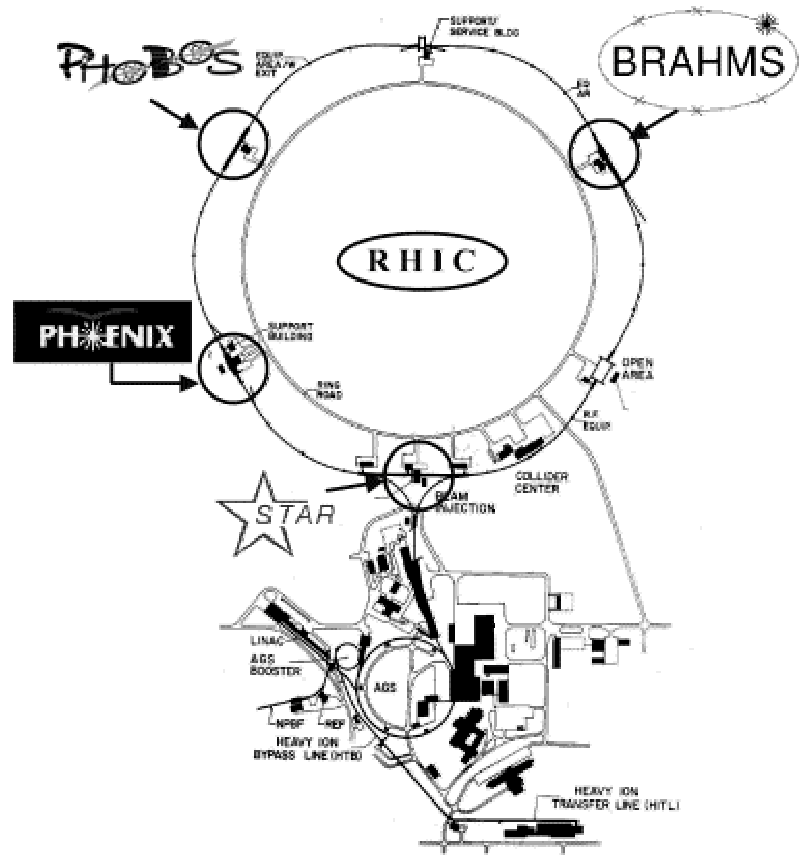}}
  	  \vspace*{-0.1cm}
    \caption{Perspective view of the RHIC complex in BNL. The four experiments are also indicated at the interaction points. Figure is taken from~\cite{Harrison:2003sb}.\label{fig:RHICpersp}}
  \end{center}
\end{figure}
Up to date RHIC provides the best environment to the search of QGP and 
characterization of its properties; colliding two counter rotating
Au ion beams at a center of mass energy per nucleon pair $\sqrt{s_{NN}}=200$ GeV. 
The total kinetic energy available in the collision zone is $\sim$ 40 TeV.
By design RHIC is capable of colliding several ion species from light ions to heavy
ions such as Au, as well as  protons. The magnet system was optimized for Au-Au 
collisions at 100 GeV/u, but the charge to mass ratio allows kinetic energies up 
to $\sim$ 125 GeV/u for lighter ions and $\sim$ 250 GeV for protons. Run 5 gives a 
good example of the versatile utilization of RHIC, when Cu-Cu collisions took place
at 62.4 GeV and 200 GeV and  proton-proton collisions at 400 GeV.
\begin{figure}[!h]
  \begin{center}
  	  \resizebox{.7\textwidth}{!}{\includegraphics{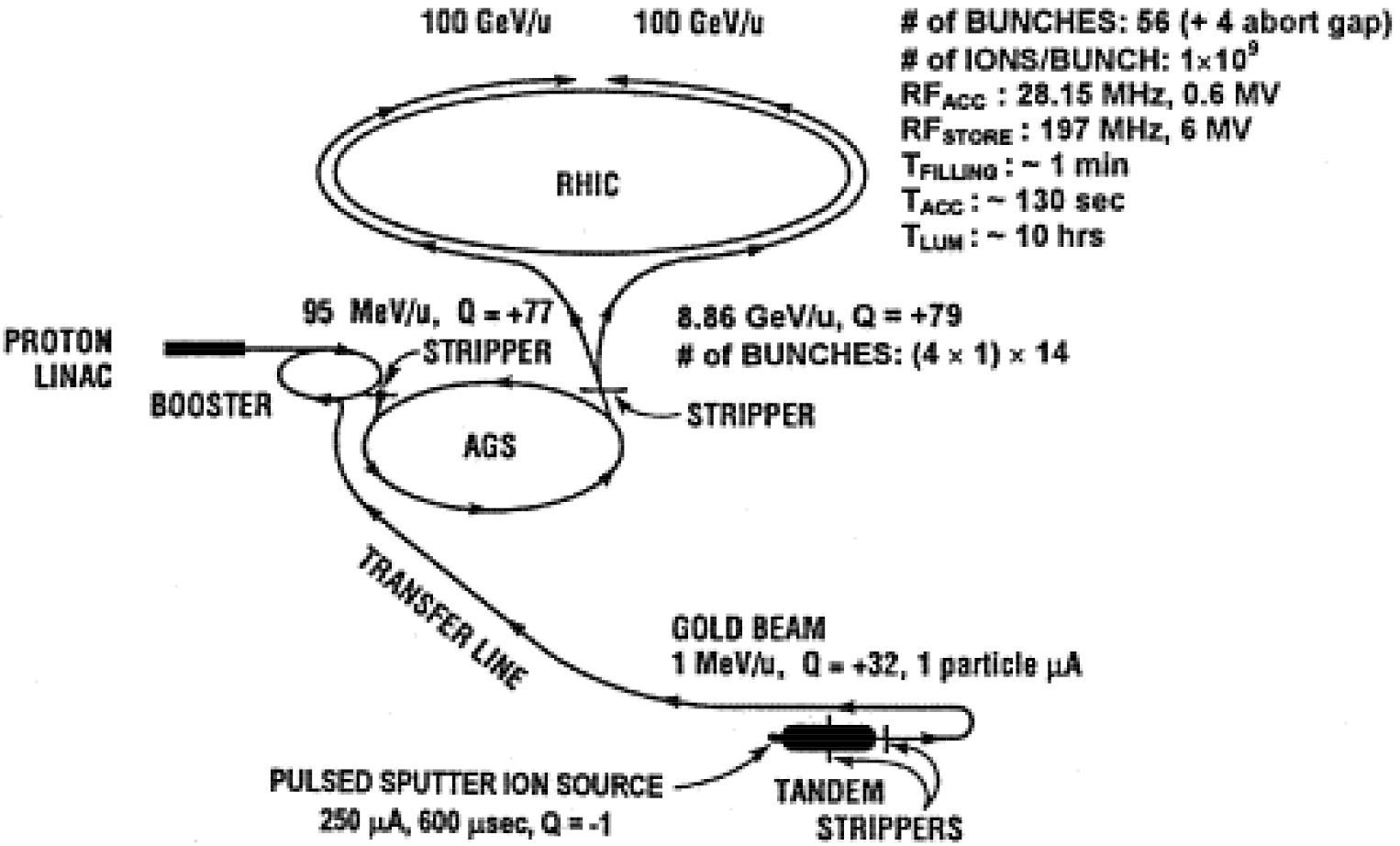}}
  	  \vspace*{-0.1cm}
    \caption{Acceleration scenario for Au ions. Figure is taken from~\cite{Harrison:2003sb}.\label{fig:RHICaccel}}
  \end{center}
\end{figure}

The acceleration of the Au ions takes places in several steps to achieve the 
100 GeV/u. Fig.~\ref{fig:RHICpersp} shows the RHIC complex and the important 
accelerator facilities. The Au ions are initially accelerated in the tandem Van
de Graaff in the charge state -1$\it{e}$ to 15 MeV.  Passing through a stripping
foil in the high voltage terminal, the charge state of the accelerated ions is further increased.
The achievable charge state depends on the ion specie; in case of Au the charge
state becomes +12$\it{e}$. Upon exiting the Van de Graaffs the Au ions are stripped
further to +32$\it{e}$ and injected to the Booster synchrotron which 
accelerates them to 95 MeV/u. In the Booster to AGS transfer line the ions are stripped
to charge state +77$\it{e}$ leaving only the K shell electrons. In the AGS the 
ions are accelerated to full AGS energy of 10.8 GeV/u. The last stripping foil is 
located in the AGS to RHIC transfer line, where the remaining K shell electrons
are removed and the ions are fully stripped to +79$\it{e}$. The Au ions are 
accelerated further in RHIC to the maximum energy to 100 GeV/u. 
The nominal beam life-time is 10 hours. The counter rotating beams can be extracted to collide 
at six interaction points. Currently four of them are utilized, as shown in Fig.~\ref{fig:RHICpersp},~\cite{Harrison:2003sb}.

\section{The STAR detector}

\begin{figure}[!t]
  \begin{center}
  	  \resizebox{5in}{!}{\includegraphics{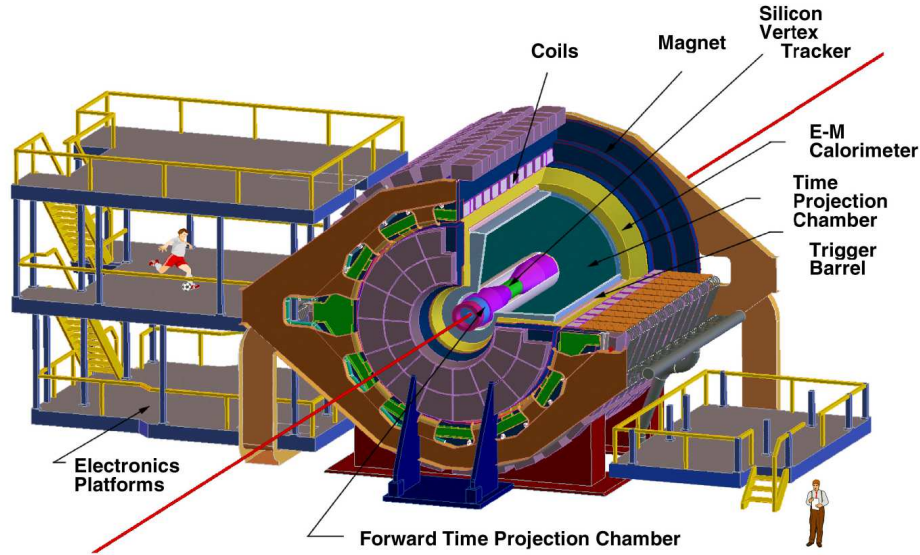}}
    \caption{Perspective view of the STAR detector~\cite{Ackermann:2002ad}.\label{fig:STARbig}}
  \end{center}
\end{figure}

The STAR - Solenoidal Tracker At RHIC - detector is primarily designed to 
measure hadronic observables in heavy-ion collisions but is able to cope 
with the broad physics program of RHIC. STAR is a large solenoidal 
detector system covering 2$\pi$ in azimuth and 3.6 units in pseudo-rapidity
(-1.8 $\leq\eta\leq$ 1.8). Its structure is shown in Fig.~\ref{fig:STARbig} representing the Year 2001 configuration.
The whole detector is enclosed in a solenoidal magnet that provides a uniform 
magnetic field parallel to the beam direction as shown in Fig.~\ref{fig:STARside}. As of 2005, the detector capabilities
are significantly extended compared to the Year 2001 setup. The Ring Imaging Cherenkov (RICH) detector was 
removed and the Barrel Electromagnetic Calorimeter (BEMC)~\cite{Beddo:2002zx} was installed. Next to the pole tips the Endcap EMCs~\cite{Allgower:2002zy} were installed to gain nearly 4$\pi$ coverage of calorimetry together with the BEMC. The Time Of Flight (ToF) detector patch covers -1 $\leq \eta \leq $0 and $\Delta\phi$ = 0.04$\pi$ and will be expanded to match the full TPC coverage. 
The current setup of the STAR detector allows measurements from hadronic to leptonic observables in a broad range.
In the next subsections we describe those subsystems, that are relevant to the analysis presented in this thesis.

\subsection{Trigger Detectors}

STAR has five main trigger detectors relevant to our analysis: the Central Trigger Barrel (CTB), the two Zero Degree Calorimeters (ZDC) and the two Beam Beam Counters (BBC). 

The ZDCs are situated $\pm$ 18 m from the center of the STAR detector and are at zero degrees with respect to the beam direction ($\theta$ $<$ 2 mrad). The ZDCs measure the energy of spectator neutrons, since charged fragments are bent away by the steering dipoles situated between the STAR detector and the ZDCs. Real collisions can be distinguished from background events by selecting events with ZDC coincidence. To ensure comparability of the results all four RHIC experiments have the same ZDC design. Recently, Shower Max Detectors were added to the ZDCs to extend the forward capabilities of STAR and open new analysis possibilities such as strangelet search and directed flow measurements.

The CTB subsystem completely encloses the TPC, and covers $\left|\eta\right|$ $<$ 1.8 and $2\pi$ in azimuth. It comprises 120 trays with 2 scintillator slats each. The photons in the scintillators are collected by photo-multiplier tubes, whose light output is proportional to the measured charged particle multiplicity at mid-rapidity. The flux of charged particles is proportional to collision centrality. In Au-Au collisions trigger selection and centrality are determined from the combined ZDC and CTB signals.

The Beam Beam Counters are hexagonal scintillating tiles mounted outside on the east and west poletips of the STAR magnet.  The inner ring consists 18 small scintillating tiles, and the outer ring consists 18 large scintillating tiles. In pp collisions, the BBC coincidence provides the minimum bias trigger.

\subsection{Forward Time Projection Chamber}
\begin{figure}[!h]
  \begin{center}
  	  \resizebox{5in}{!}{\includegraphics{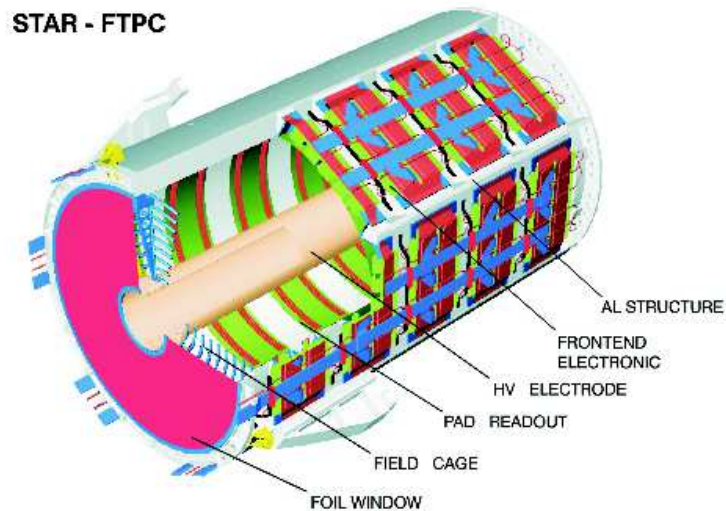}}
    \caption{Schematic view of an FTPC~\cite{Ackermann:2002yx}.\label{fig:FTPC}}
  \end{center}
\end{figure}
%
The Forward Time Projection (FTPC) chambers extend the STAR coverage 
to pseudorapidities between 2.5 $ < |\eta| <$ 4.0. The two FTPCs are located between the beam pipe and the inner field cage of the TPC, at the two sides of the SVT.
Both FTPCs have cylindrical shape with a 75 cm diameter and 120 cm length. Due to the limited space and to achieve good two-track separation close to the beam direction, a radial drift field configuration is implemented with a $Ar-CO_{2}$ gas mixture. The short drift length is not sufficient to provide enough dE/dx information to identify particles, but charged particle momentum can be measured between 2.5 and 4.5 GeV/c in full azimuth and the momentum resolution is estimated to be $12 - 15\%$~\cite{Ackermann:2002yx}. In the Year 3 dAu run the charged particle multiplicity measured in the FTPCs provides the centrality selection for data analysis. 
\begin{figure}[!h]
  \begin{center}
  	  \resizebox{5in}{!}{\includegraphics{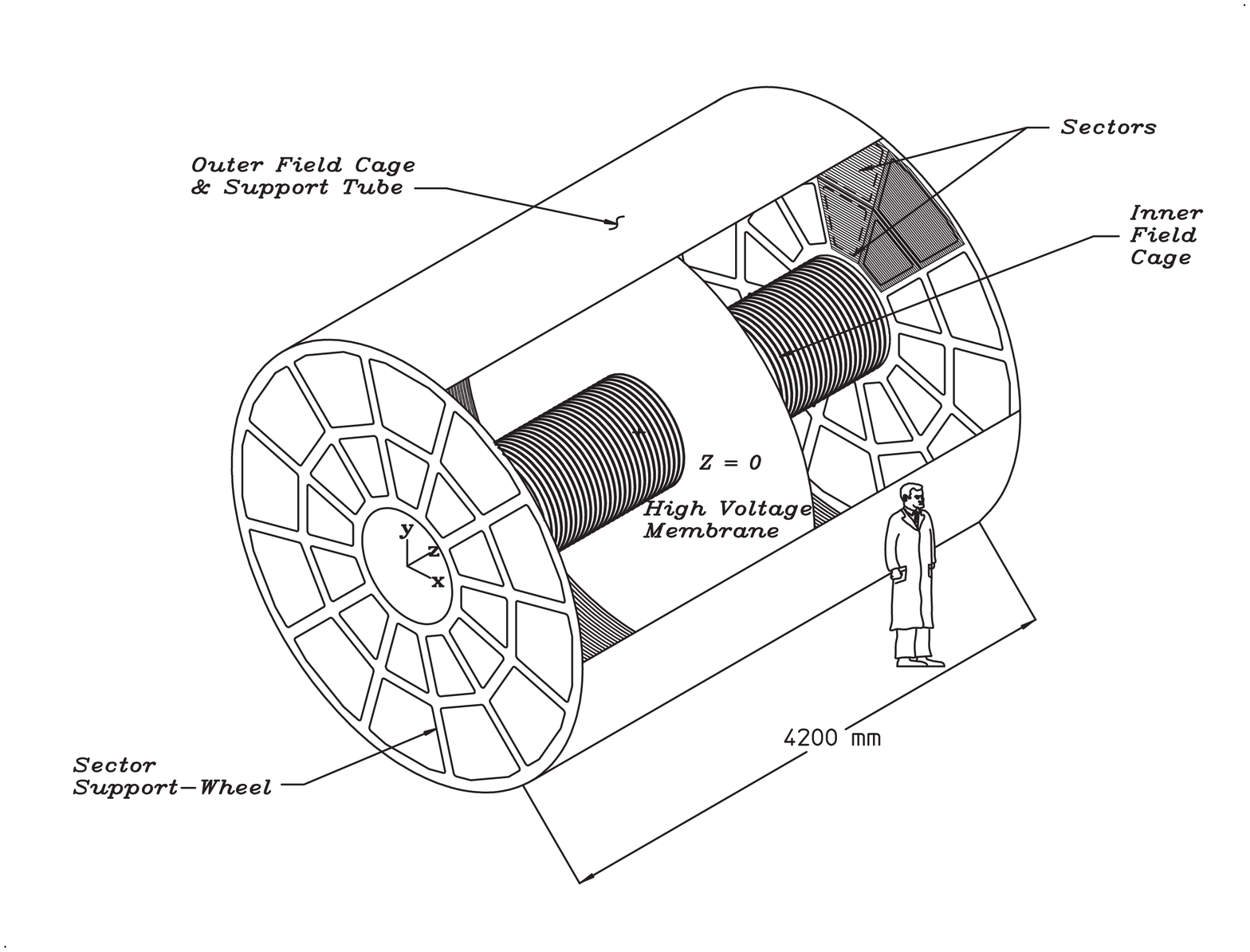}}
    \caption{Schematic view of the STAR TPC~\cite{Anderson:2003ur}.\label{fig:tpcside}}
  \end{center}
\end{figure}
\begin{figure}[!h]
  \begin{center}
  	 \resizebox{.5\textwidth}{!}{\includegraphics{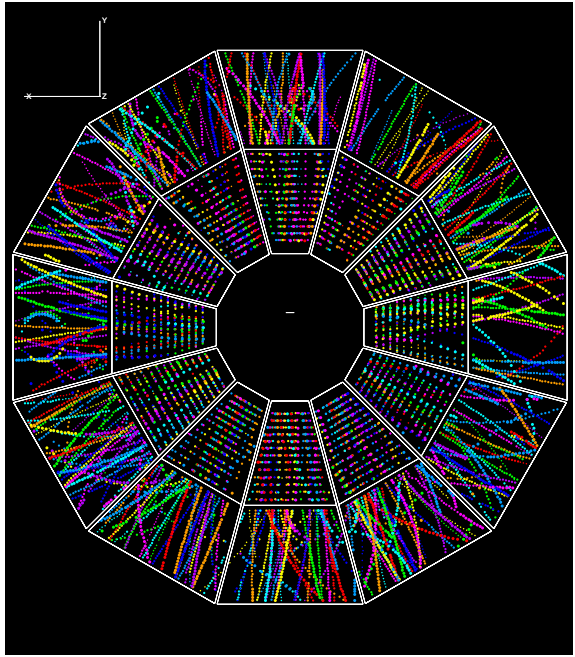}}
    \caption{Bulls eye view of a typical Au-Au collision in the STAR TPC.\label{fig:tpcendevent}}
  \end{center}
\end{figure}

\subsection{The Time Projection Chamber}

The main tracking detector of the STAR experiment is the Time Projection Chamber. It extends $\pm$ 2.1 m from the center of the detector, providing 2$\pi$ azimuthal coverage. The total diameter is 4 m, the inner drift volume starts at 50 cm (radius) and extends to the outer drift volume 200 cm (radius) as shown in Fig.\ref{fig:tpcside} by the inner and outer field cages. Thus the sensitive tracking pseudorapidity interval is $\pm$ 1.2 - 1.5 units. 
\begin{figure}[!h]
  \begin{center}
  	  \resizebox{5in}{!}{\includegraphics{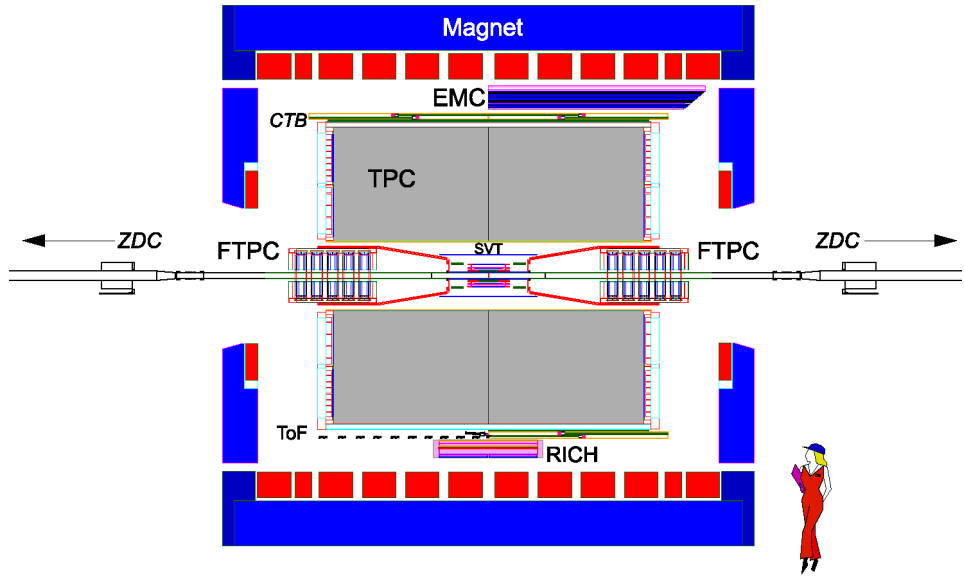}}
    \caption{Side view of the STAR detector~\cite{Ackermann:2002ad}.\label{fig:STARside}}
  \end{center}
\end{figure}

The thin conducting Central Membrane (CM) is situated in the center of the TPC and splits the tracking volume in the beam direction. The CM (cathode) is held at -28 kV while the anodes are at 0 V. The field cage cylinders and the 182 attached rings provide equipotential planes from the CM to the anode planes. The uniform electric field henceforth is created with the careful design of the field cages, the Central Membrane and the anode planes. 
\begin{figure}[!h]
	\resizebox{.5\textwidth}{!}{\includegraphics{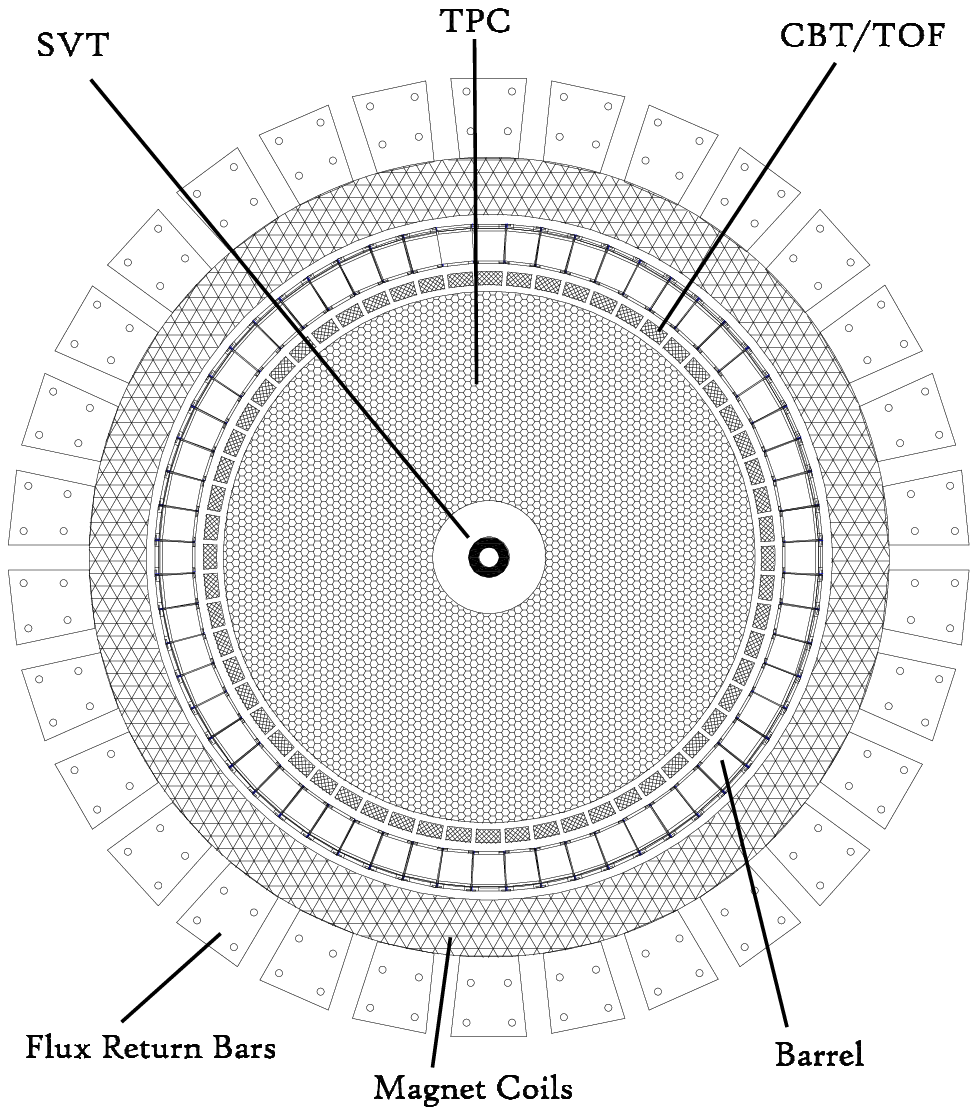}}
	\resizebox{.5\textwidth}{!}{\includegraphics{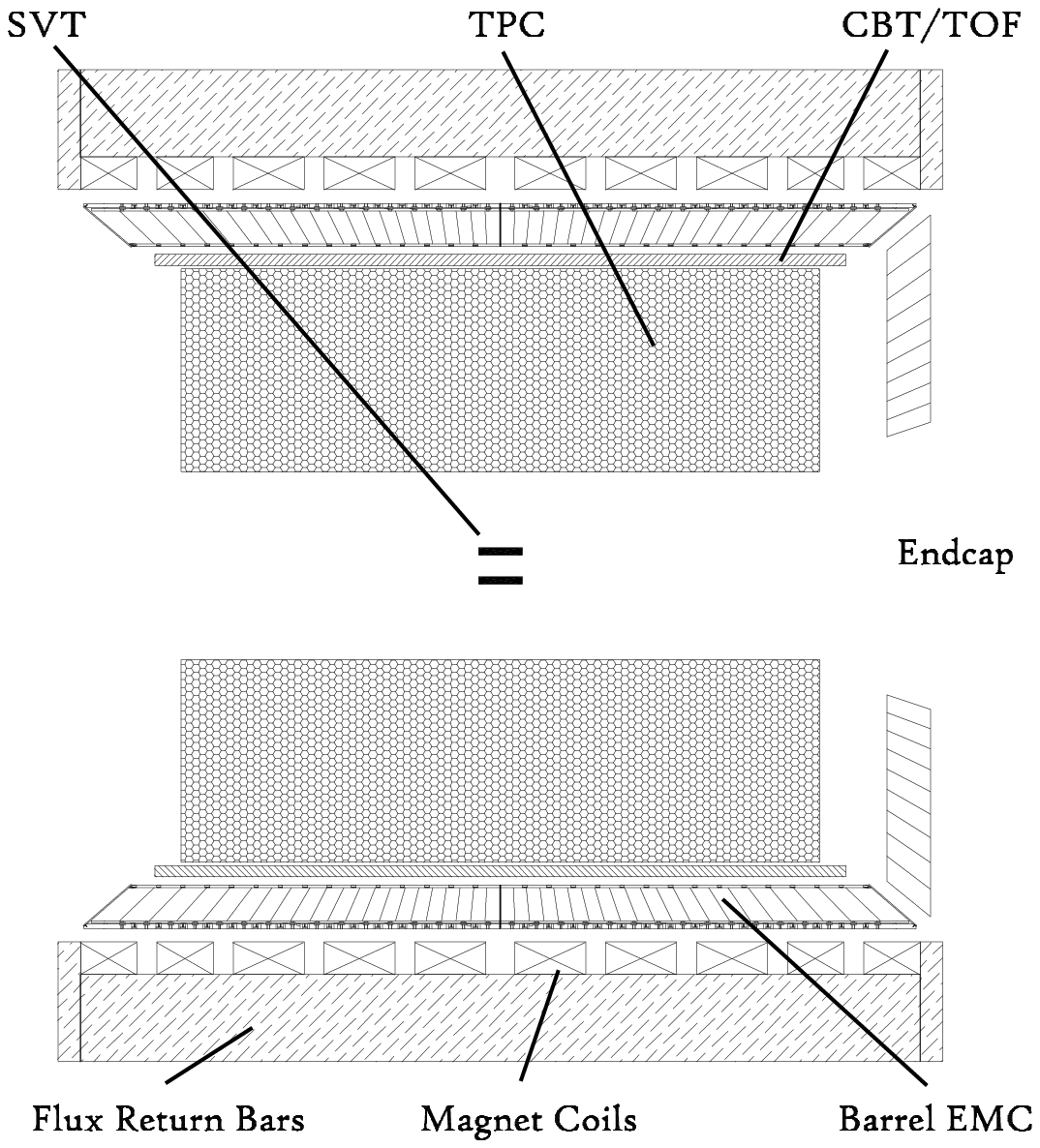}}
    \caption{Bulls eye and cross sectional view of the STAR detector, indicating the position of the subdetectors. Figure is taken from~\cite{Beddo:2002zx}.\label{fig:BEMC_and_subs}}
\end{figure}

Since the directions of the uniform magnetic and electric fields are parallel, the transverse diffusion (with respect to the electric field) is small. The Lorentz force keeps the charged particle on a circular path around the electric field line.

The uniformity of the field is essential since the electrons produced by a charged particle traversing through the TPC volume have to drift $\sim$ 2 m to the read out planes. The gas is required not to attenuate the drifting electrons and provide a pure enough environment to avoid the loss of electrons due capture on oxygen or water molecules.
Hence, TPC is filled with P10 gas that is a mixture of 10$\%$ methane and 90$\%$ argon~\cite{Kochenda:2002zz} and regulated at +2 mbar above atmospheric pressure. The oxygen content is kept below 100 parts per million and the water content is kept below 10 parts per million.

The typical drift velocity in TPC gas is $\sim$ 5.45 cm/$\mu$s and can be monitored in each run with a precision of $\sim$ 0.001 cm/$\mu$s and over several days can change by $\pm$ 0.01 cm/$\mu$s. The large scale of the drift in $z$ direction sets limits on the sampling rate and the resolution. At full magnetic field (0.5 T) the transverse diffusion after 210 cm is about $\sigma_{T}\approx$ 3.3 cm and sets the scale for the read out chambers. The longitudinal diffusion ($\sigma_{L}\approx$ 5.2 mm) limits the time resolution of the clusters traversing the whole TPC volume to $\sim$ 95 ns, or a sampling rate 10 MHz. Here we should note that with the recent DAQ upgrade (DAQ100) the final event collection rate including the TPC and other subdetectors is $\sim$ 60 - 80 MHz.

The geometry of the read out pad planes can be seen in Fig.~\ref{fig:tpcendevent}. The endcaps (Multi Wire Proportional Chambers) of the TPC are divided into 12 sectors. Each sector is divided into two parts: inner and outer sector. The inner and outer pads all together contain 5690 pads which translate to 136,560 channels for the whole TPC. The signals from the pads are amplified, shaped and passed to the ADCs. The combination of X and Y positions and the drift time of electron clusters allow precise measurements.

Besides the position measurement the TPC is capable of momentum determination from 100 MeV/c to 30 GeV/c. Particle identification with the $dE/dx$ method alone is possible in the momentum range of 100 MeV/c to 1.2 GeV/c, but with combined techniques can be extended to very high transverse momentum (above 10 GeV/c)~\cite{Shao:2005iu}. 

\subsection {The STAR magnet}
The STAR magnet provides uniform solenoidal a magnetic field. It is parallel to the beam pipe and encloses most of the detector subsystems.

In Fig.~\ref{fig:STARside}. the magnet is shown in blue and the coils are shown in red. Due to the uniform field the charged particles move on a helical trajectory in the lowest order of the approximation. This enables a fast pattern recognition and track reconstruction. The field strength can be varied between 0 and 0.5 Tesla. Data sets presented in this work are measured at 0.5 T. The magnetic field is reversible, and in each run data are taken at both polarities to account for systematic effects. A thorough mapping of the magnetic field shows that uniformity is achieved on the level of $\pm$ 50 Gauss (25 Gauss) in radial and less than $\pm$ 3 Gauss ($\pm$ 1.5 Gauss) in azimuth for full (half) field setup. Distortion effects on the tracks thus can be calculated to the order of $\sim$ 200 $\mu m$~\cite{Bergsma:2002ac}.

\chapter{Event reconstruction and particle identification}

\section{Event reconstruction}

In this chapter we discuss the event reconstruction and the extraction of raw particle yield via the dE/dx method.
The main detector for the analysis is the Time Projection Chamber (TPC) of STAR.

\subsection{Hit and cluster finding}

Each TPC sector has 45 pad rows as shown in Fig.~\ref{fig:tpcendevent}. 
Hence a charged particle traveling through  the TPC can leave 45 possible hits.
Reading out the pixel information of the padrows allows the location of the hit to be determined.
The hit finder algorithm reads in the time ordered information from each padrow pixel, that are above a trigger threshold and marked as good channels by the DAQ (Data Acquisition System). Gain correction and timing information are also included in the hit finding. In the next step the cluster finder identifies hits in a 2D cluster in the plane of the padrow and the longitudinal direction. In the cluster a single hit as a centroid of the 2D distribution or multiple hits as local maximum in the deconvoluted distribution can be determined. Deconvolution of the hits is very important in particle identification and multiplicity measurements. The deconvoluted hits are converted to real space points in the local coordinate system of the TPC, including the calibration parameters such as drift velocity, trigger timing and geometry. Beside the space information the deposited energy is also stored for each hit. 

\subsection{Track finding} 
The track finder, starting from the outer padrows, assigns hits to a track candidate. In the first iteration many track segments (tracklets) are created as candidates for a track. Then, these track segments are fitted by the algorithm that keeps or rejects the hits, depending on their position with respect to the fitted track. In this stage of fitting the effect of multiple Coulomb scattering and energy loss assuming pion mass are taken into account. At the end of the algorithm, the collection of the tracks is produced with their space coordinates and their 3-momenta.

\subsection{Global and primary tracks}
As the final step in the event reconstruction the $global$ and $primary$ tracks are created. As mentioned in the previous subsection, tracks are reconstructed in the local coordinate system of the TPC. To perform data analysis, global information of the tracks is needed. The global track finder first re-fits the tracks in the TPC, based on a 3D helix model. After the re-fit with the knowledge of the alignment of the different subsystems, the global track finder reconstructs the global tracks from the 'local' tracks in the subdetectors. Based on the information of the global tracks in the event the primary vertex can be found. Those global tracks with a distance to the primary vertex smaller than 3 cm (distance of closest approach, hereafter: $dca$), are re-fitted including the primary vertex as an additional point in the fit. In the analysis these primary tracks are used. Primary tracks are largely the particles produced in the primary interaction. Global tracks 
include a large number of particles from background or pile-up processes.

\begin{figure}[!h]
\begin{center}
	\resizebox{.45\textwidth}{!}{\includegraphics{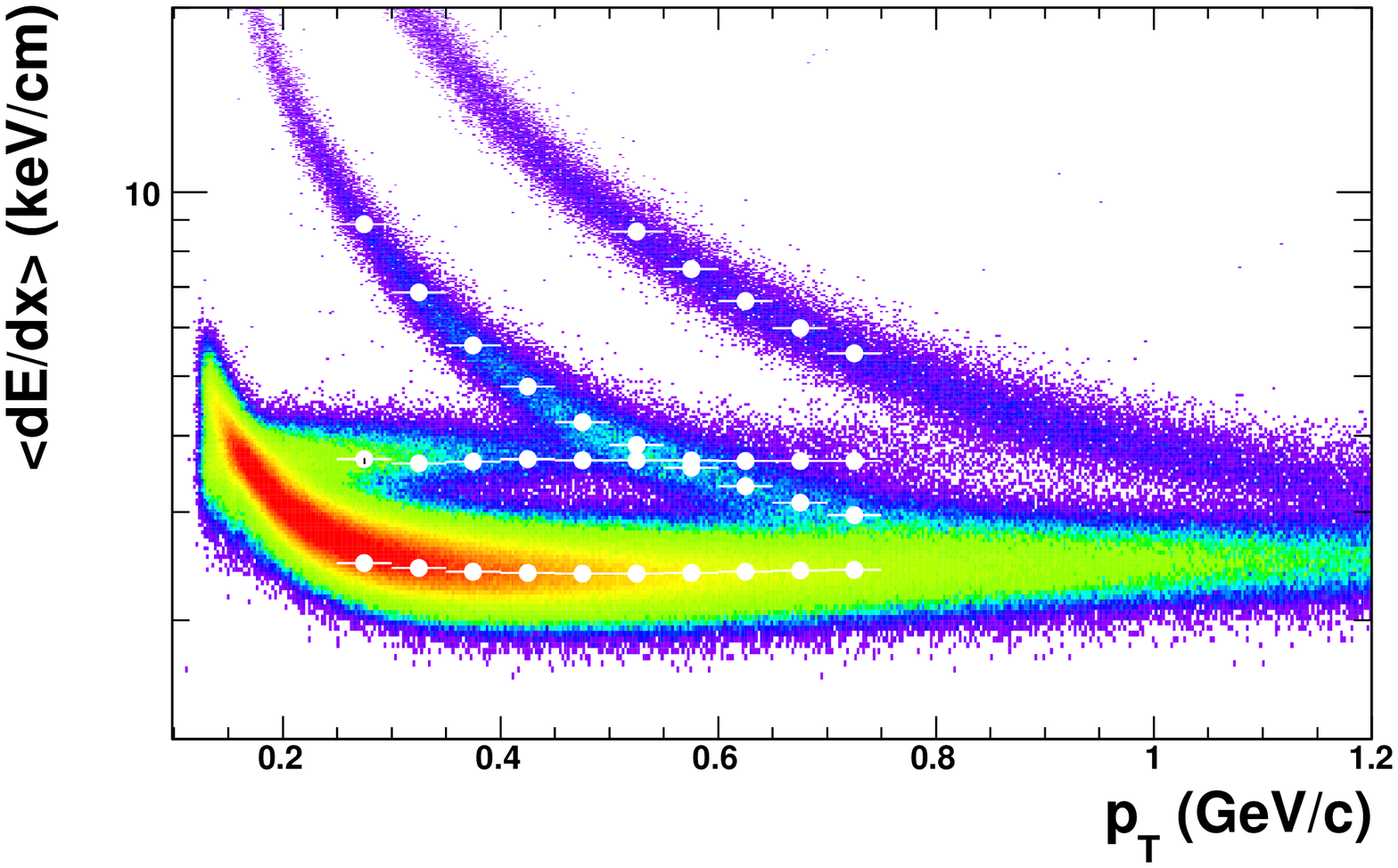}}
	\resizebox{.45\textwidth}{!}{\includegraphics{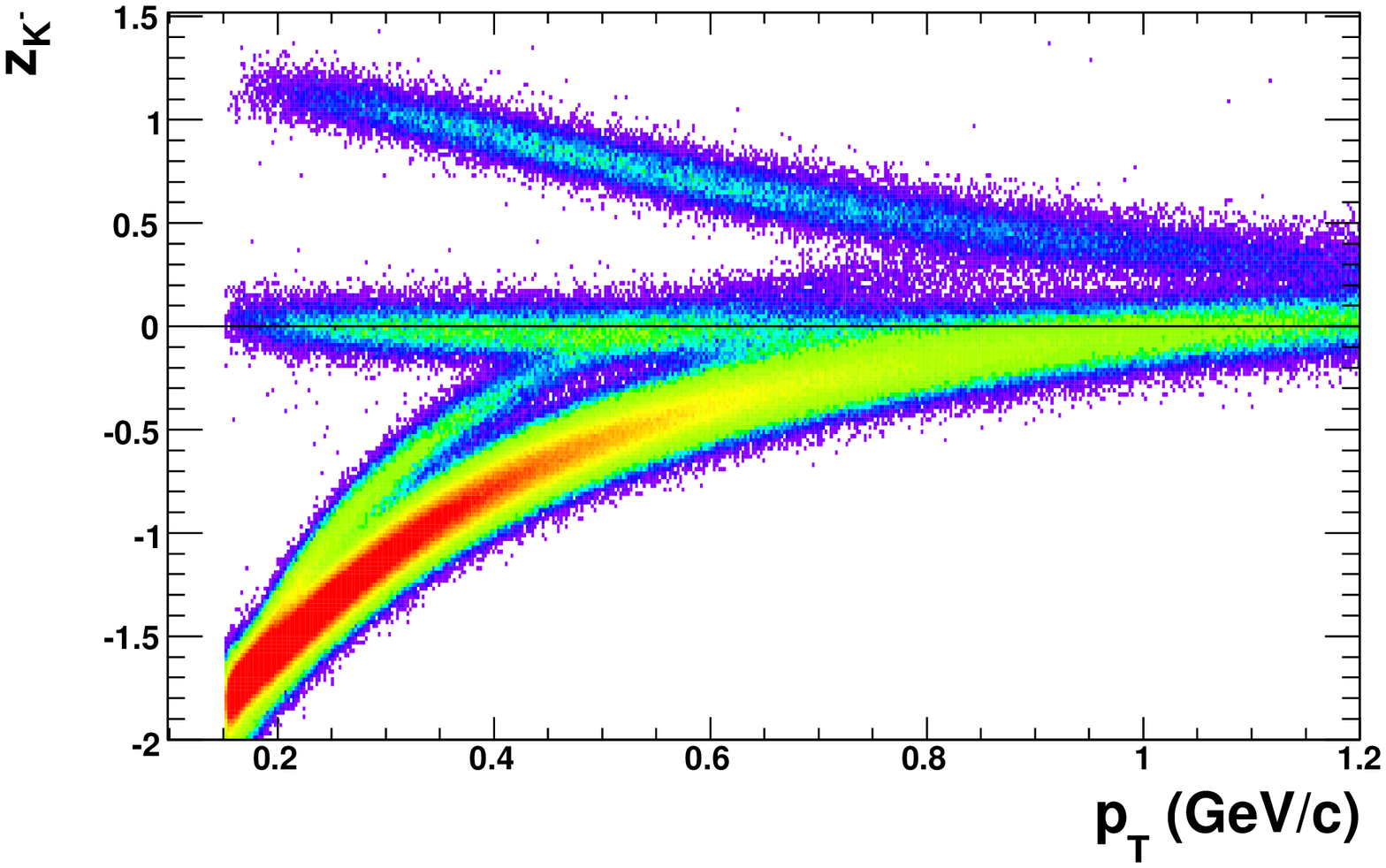}}
	\resizebox{.45\textwidth}{!}{\includegraphics{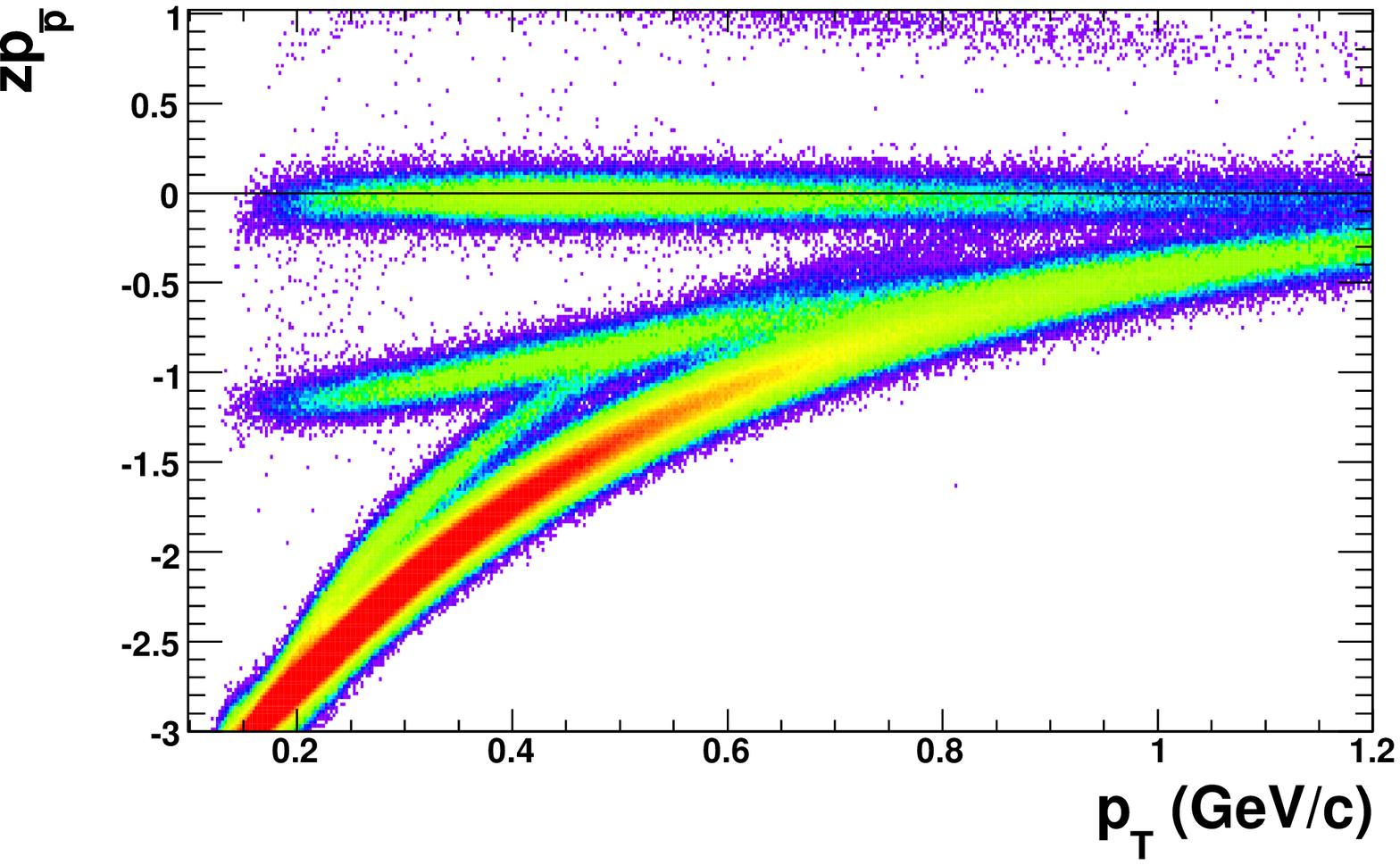}}
	\resizebox{.45\textwidth}{!}{\includegraphics{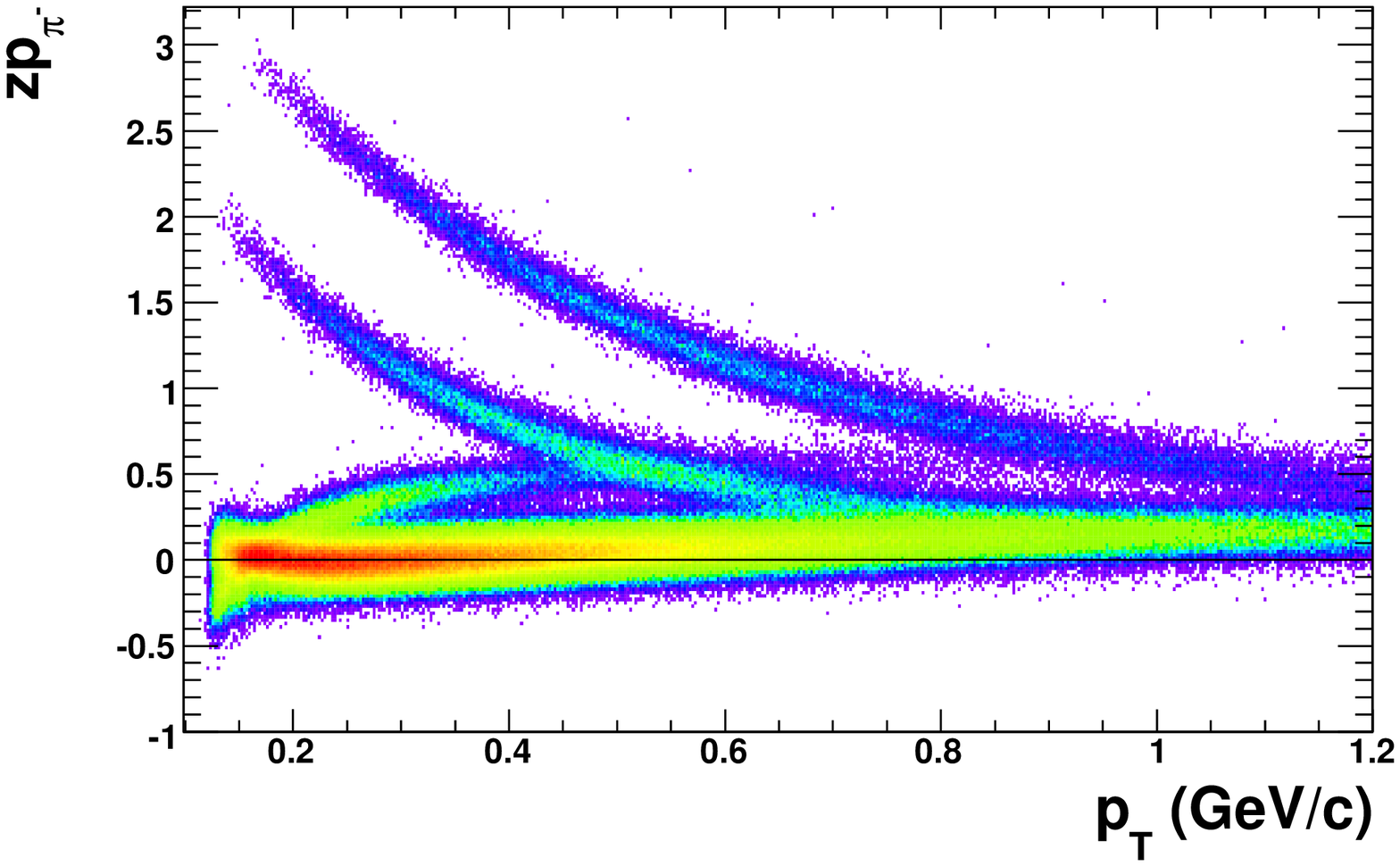}}
   \caption{Top left figure shows the specific ionization energy loss as a function of transverse momentum measured in the TPC. The white dots represent the fitted $\pi^{-}$, $e^{-}$, $K^{-}$ and $\overline{p}$ centroids. Top right figure shows the kaon mass parameterized $z$ distribution.  Left bottom plots shows the proton mass and right bottom plots shows the pion mass parameterized $z$ distributions.\label{fig:dedxandz}}
   \end{center}
\end{figure}
\section{Particle identification}
As the main tracking detector of the STAR, the TPC can identify particles by measuring the mass dependent specific ionization energy loss (dE/dx) at low transverse momentum ($p_{T} < ~$1.5 GeV/c). 
A charged particle traversing the TPC gas volume ionizes the gas atoms. In the electric field these charge clouds drift from their creation point to the two ends of the TPC, where the charges are read out on the padrows. Produced charge in each hit on a padrow is proportional to the energy loss of the particle traversing through  the TPC volume. If a particle travels through  the entire TPC volume, 45 dE/dx points can be measured on the 45 padrows. 

Energy loss of a charged particle for a given track length can be described with a Landau probability distribution.
However, the mean of the distribution is sensitive to the fluctuations in the tail of the distribution. Therefore, the highest 30$\%$ of the measured charge clusters is discarded for each track. The truncated mean is calculated from the remaining 70$\%$ and defines the average ionization energy loss used in the data analysis.

Specific ionization in an isotropic homogeneous medium can be parameterized by the Bethe-Bloch formula:
\begin{equation} 
-\frac{dE}{dx}=4\pi N_{0}r_{e}^{2}m_{e}c^{2}\frac{Z}A\rho \frac{1}{\beta^{2}}z^{2} \left[ ln \left( \frac{2m_{e}c^{2}}I\beta^{2}\gamma^{2}\right) - \beta^{2} - \frac{\delta}2\right]
\end{equation}
where $N_{0}$ is the Avogadro number, $r_{e}$ is the classical electron radius, $Z$ is the atomic number of the medium, $\rho$ is the density of the medium, $z$ is the charge of the particle traveling through  the medium, $I$ is the ionization potential of the medium, and $\delta$ accounts for the density effect of the medium.

As shown in Fig.~\ref{fig:dedxandz} the dE/dx bands for particles with different mass can be separated. 
Figure~\ref{fig:dedxandz} also shows the kinematic range for particle identification: antiprotons can be identified in 0.3 - 1.2 GeV/c, kaons can be measured in 0.2 - 0.7 GeV/c and the pions can be measured in 0.2 - 0.7 GeV/c. To extract the raw yield of the particles one can introduce the so called $z$ variable:  
\begin{equation} 
z = \log\left(\frac{dE/dx_{measured}}{dE/dx_{parameterized}}\right) 
\end{equation}
where $(\frac{dE}{dx})_{measured}$ is the measured energy loss and $(\frac{dE}{dx})_{parameterized}$ is the parameterized form of the energy loss~\cite{Aguilar-Benitez:1991yy}. The $z$ variable has the advantage that each particle has a Gaussian distribution around the expected Bethe-Bloch value. In our analysis a single parameter approximation of the Bethe-Bloch formula is used:
\begin{equation} 
dE/dx_{measured}\ =\ A\cdot\left(1+\frac{m^{2}}{p^{2}}\right) 
\end{equation}
where $A$ is a constant, $m$ is the mass of a given particle and $p$ is the total momentum of the particle.  
Figure~\ref{fig:dedxandz} (top left panel) shows the measured energy loss as a function of transverse momentum in minimum bias pp collisions at 200 GeV. The white dots represent the centroid positions of $\pi^{-}$, $e^{-}$, $K^{-}$ and $\overline{p}$ from the multi Gaussian fits to the energy loss corrected $z_{K^{-}}$ distribution. Figure~\ref{fig:dedxandz} also shows the $z$ distributions of $K^{-}$, $\pi^{-}$ and $\overline{p}$. The energy loss band of the particle of interest is centered around zero and the other bands are well separated.

\section{Extraction of raw particle yields}\label{4Gauss}

\begin{figure}[!h]
\begin{center}
\resizebox{.7\textwidth}{!}{\includegraphics{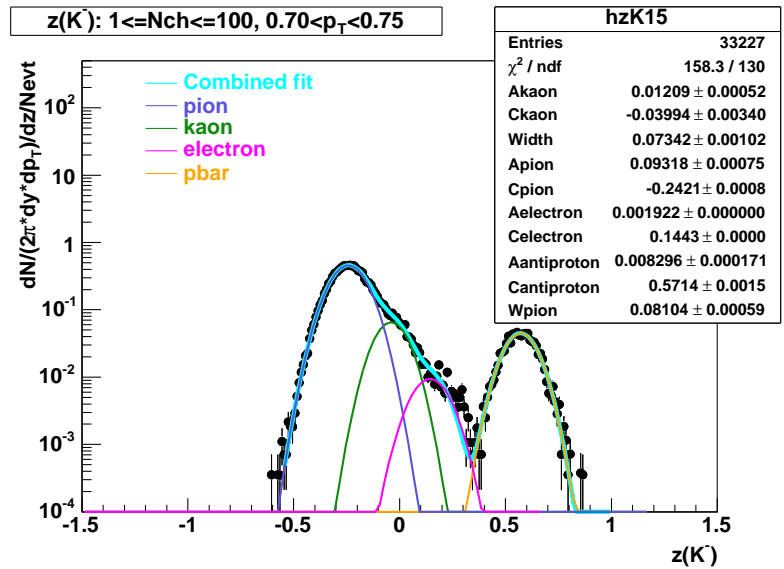}}
	 \caption{Multi-Gaussian fit to the $z_{K^{-}}$ distribution in 200 GeV pp collisions.\label{fig:ppGaussianFitsKaonSample}}
	 \end{center}
\end{figure}

As noted above the specific energy loss is parameterized with the Bethe-Bloch formula for each particle to calculate the $z$ variable which is centered around 0. The parameterization is slightly dependent on the multiplicity/centrality and adjusted for each data production due to calibration. 
To extract the raw particle yields, the $z$ distributions/peaks are simultaneously fitted by multiple Gaussian functions, as demonstrated in Fig.~\ref{fig:ppGaussianFitsKaonSample}.

Charged pions can be separated in the transverse momentum region 200 - 800 MeV/c. Figure~\ref{fig:ppGaussianFitsPion} shows the multi Gaussian fits to the $z_{\pi^{-}}$ distribution in these momentum slices. Black dots represent the measured $z_{\pi^{-}}$ distribution and the colored lines represent the Gaussian fits to $\pi^{-}$, $e^{-}$, $K^{-}$ and the combined fit.
The raw kaon yield can be extracted in the transverse momentum region $\sim$200 - 800 MeV/c. The extraction of the raw yield is more complicated since the electron and kaon peaks start to merge at $\sim$ 450 MeV/c. Therefore, the raw electron yield is extracted in the $p_{T} <$ 450 MeV region and extrapolated in the merged bins to obtain the raw kaon yields. The measured raw electron yield is fitted to an exponential function (inspired by MC studies) in the momentum range 200 - 450 MeV/c. The fit result is fed to the multi-Gaussian fit in the large $p_{T}$ region and the electron yield is either fixed or left to vary within a reasonable range around the fitted value.
The fit results are shown in Figure~\ref{fig:ppGaussianFitsKaon}.

As shown in Fig.~\ref{fig:dedxandz}, electrons are also merged into the $\pi$ band, although to a less degree than the kaon band.
In the merged region the same procedure is applied as for pions. The raw electron yield is estimated from an exponential fit over the momentum range where electrons are well separated.

Protons/antiprotons are well separated from the rest of the particles in the momentum region $\sim$ 300 - 1200 MeV/c, as shown in Fig.~\ref{fig:dedxandz}, and can be fitted to a single Gaussian up to $\sim$ 850 MeV/c, as shown in Fig.~\ref{fig:ppGaussianFitsPbar}. 
In the following bins the electron contamination is estimated the procedure is the same as for kaons. 

Figure~\ref{fig:ppGaussianFitsPion}, Fig.~\ref{fig:ppGaussianFitsKaon} and Fig.~\ref{fig:ppGaussianFitsPbar} show the typical fits to the negative identified particles in 200 GeV pp collisions. Fits to positive particles and to 200 GeV dAu and 62.4 GeV Au-Au data are similar.
The momentum range can vary due to the changing resolution of the dE/dx bands in the different datasets.

\chapter{Analysis method of identified charged particle spectra}

In this section the analysis technique of identified charged particle spectra measurements of $\pi^{\pm}$, $K^{\pm}$, $\overline{p}$ and p are reported for 200 GeV pp, 200 GeV dAu and 62.4 GeV Au-Au collisions. 

\section{General procedure of data analysis}

Before the detailed discussion, a general overview is given to provide a conceptual framework for the data analysis. Our goal is to extract the corrected particle spectra and their properties for identified pions, kaons and protons/antiprotons. Steps of the analysis leading to the fully corrected identified particle spectra are listed below:
\begin{enumerate}
	\item {Good events are selected from data on tape, satisfying trigger and vertex requirements. Event wise variables such as the uncorrected charged particle multiplicity are corrected for vertex inefficiencies upon selecting the good events in pp and in minimum bias and peripheral dAu collisions.}
	\item{Once a good event is identified, good tracks are selected based on the analysis specific quality cuts.
	In the case of kaon or proton/antiproton tracks, each track is corrected for energy loss upon selection.}
	
	\item{At this point selected data includes event and track corrections, which is followed by the extraction of raw yield from the multi-Gaussian fits described in Sec.~\ref{4Gauss}.}
	\item{The extracted raw yield is corrected for tracking efficiency and acceptance depending on particle type, multiplicity and/or centrality.}
\begin{itemize}
	\item{Raw pion yield is further corrected for weak decay and detector background contamination.}
	\item{Raw proton yield is corrected for background contribution from detector material.}
	\item{In the case of minimum bias pp and dAu and peripheral dAu collisions, a fake vertex correction is applied for all particle types.}
	\end{itemize}	
	\item{Finally point-to-point systematic errors are assigned to each spectrum point.}
\end{enumerate}

At the end of this procedure the fully corrected identified particle spectra are obtained and one can proceed to extract the bulk properties of the collisions which will be discussed in the Result section.

\section{Data sets and trigger}

Data presented here are collected in three different RHIC runs: pp collisions at $\sqrt{s_{NN}}$ = 200 GeV in 2002, dAu collisions at $\sqrt{s_{NN}}$ = 200 GeV in 2003, and Au-Au collisions at $\sqrt{s_{NN}}$ = 62.4 GeV in 2004.

Various combinations of the trigger detectors (BBC - CTB, ZDC - CTB) are utilized to measure charged particle and neutral particle multiplicity. 
In pp collisions the minimum bias events are selected by the coincidence of the two BBCs measuring charged particle multiplicity near beam rapidity. In dAu collisions the minimum bias events are obtained from the combination of BBC and ZDC coincidence. In Au-Au collisions the minimum bias events are selected from the CTB-ZDC charged-neutral multiplicity correlation. In each run the magnetic field strength is set at 0.5 Tesla.  

\section{Event selection}

The position of the collision vertices are distributed around the center of the detector. To select events with approximately uniform detector acceptance in pseudorapidity, the primary vertex position has to be limited. Selection on the $z$ component of the primary vertex is specific to the colliding species. Additionally, events have to satisfy the following requirements: $\left|v_{x}\right|< $ 3.5 cm and $\left|v_{y}\right|< $ 3.5 cm.

To experimentally vary the impact parameter/centrality of the collision, cuts on the uncorrected reference multiplicity (or charged particle multiplicity) are applied. The uncorrected reference multiplicity is defined as the number of charged primary tracks in pseudo-rapidity ($\eta)$: -0.5 $<$ $\eta$ $<$ 0.5. 

The specific $z$ vertex and multiplicity/centrality selection is presented below.

\subsection{Proton - Proton collisions}

\begin{figure}[!h]
\begin{center}
	\resizebox{.9\textwidth}{!}{\includegraphics{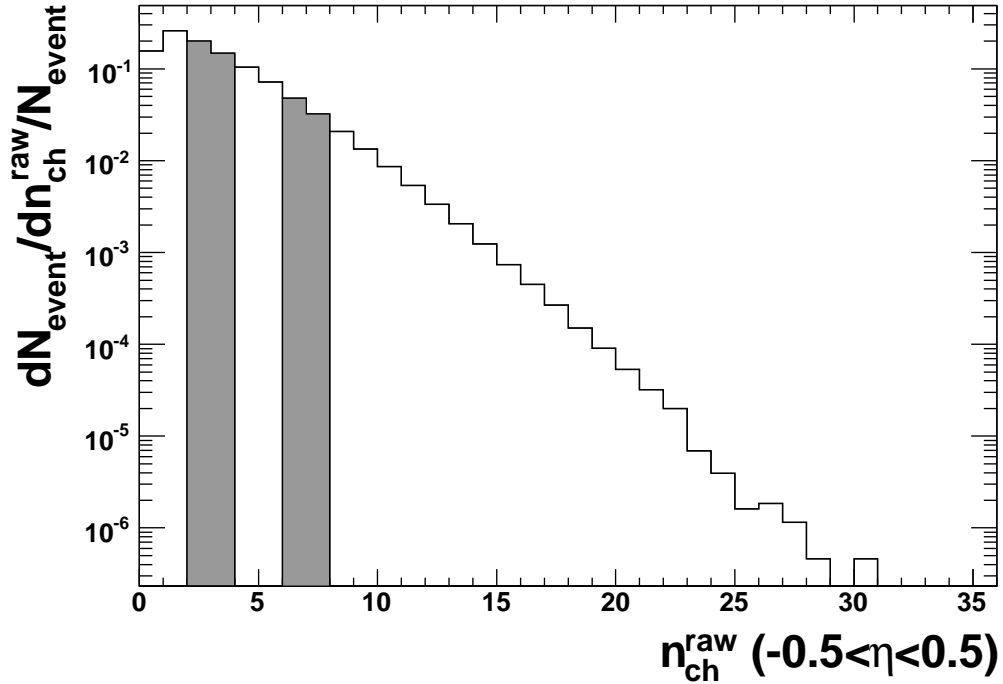}}
    \caption{Uncorrected reference multiplicity distribution in 200 GeV minimum bias pp collisions.   \label{fig:pp_mult}}
    \end{center}
\end{figure}

The $z$ component of the primary vertex in each minimum bias event has to satisfy the following condition: $\left|v_{z}\right|< $ 30.0 cm. With this vertex cut and minimum bias trigger 3.9 M good minimum bias events are selected. Figure~\ref{fig:pp_mult} shows the uncorrected reference multiplicity distribution in minimum bias pp collisions. To gain more insight, we will investigate the bulk properties not only in minimum bias pp collisions but also as a function of charged particle multiplicity. In pp collisions five multiplicity classes are chosen as summarized in Table~\ref{tab:auau_coll_prop}.

\subsection{Deuteron - Gold collisions}
\begin{figure}[!h]
\begin{center}
	\resizebox{.9\textwidth}{!}{\includegraphics{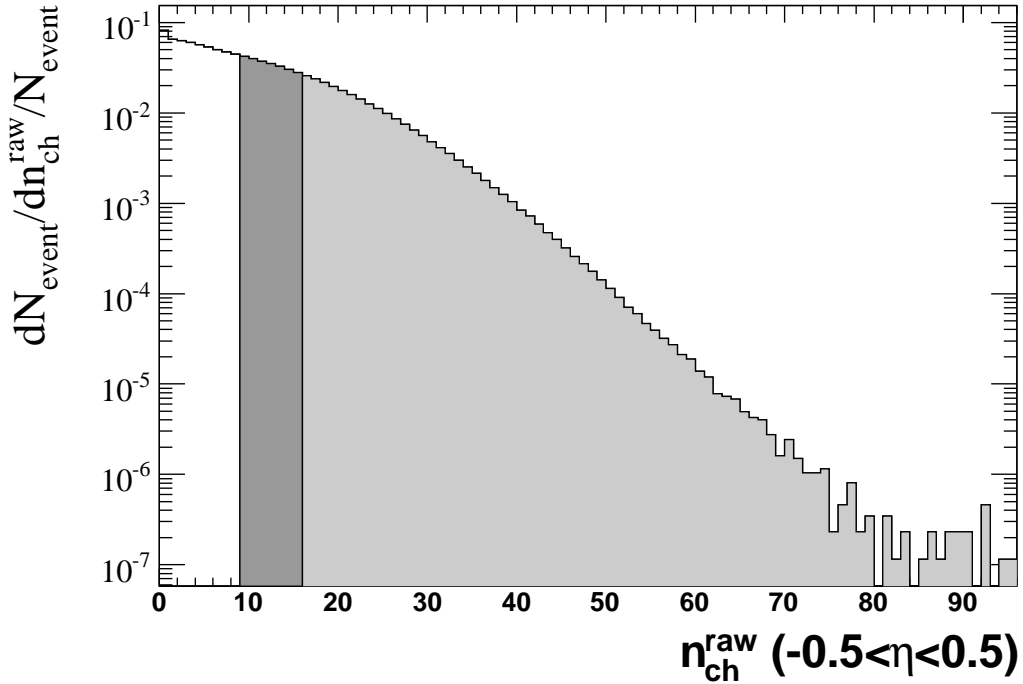}}
\caption{Uncorrected charged particle multiplicity measured in the East FTPC (on the outgoing Au side) in 200 GeV dAu collisions.  \label{fig:dau_mult1}}		   
		   \end{center}
\end{figure}
\begin{figure}[!h]
\begin{center}
	\resizebox{.9\textwidth}{!}{\includegraphics{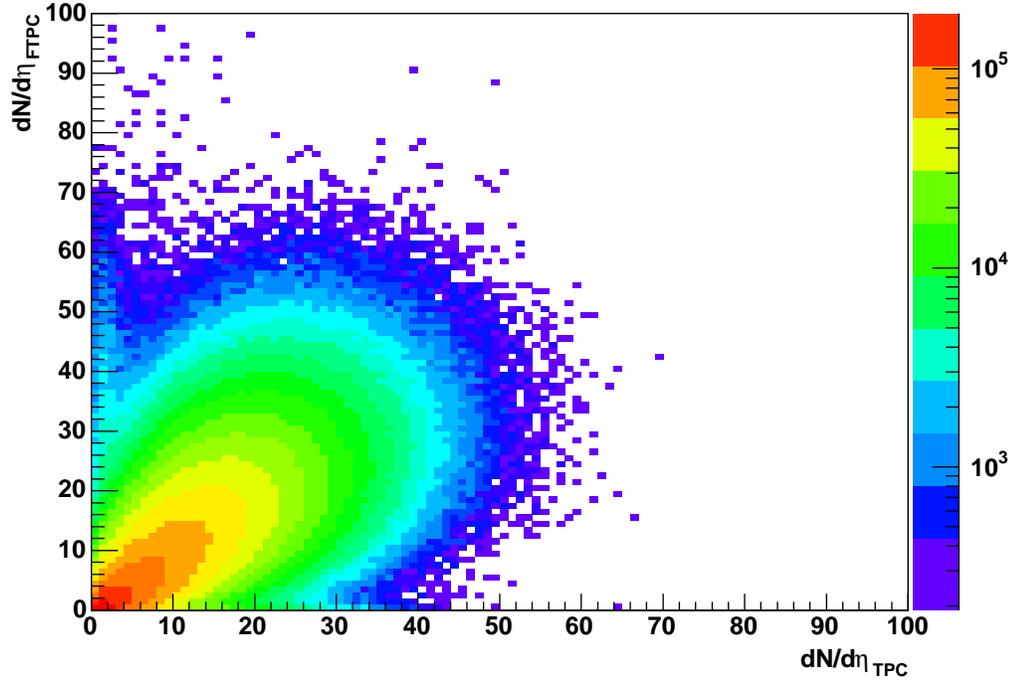}}
		   \caption{Uncorrected charged particle multiplicity measured in the {\bf East FTPC} vs. measured in the {\bf TPC} in 200 GeV dAu collisions.  \label{fig:dau_mult2}}
		   \end{center}
\end{figure}
\begin{figure}[!h]
\begin{center}
	\resizebox{.9\textwidth}{!}{\includegraphics{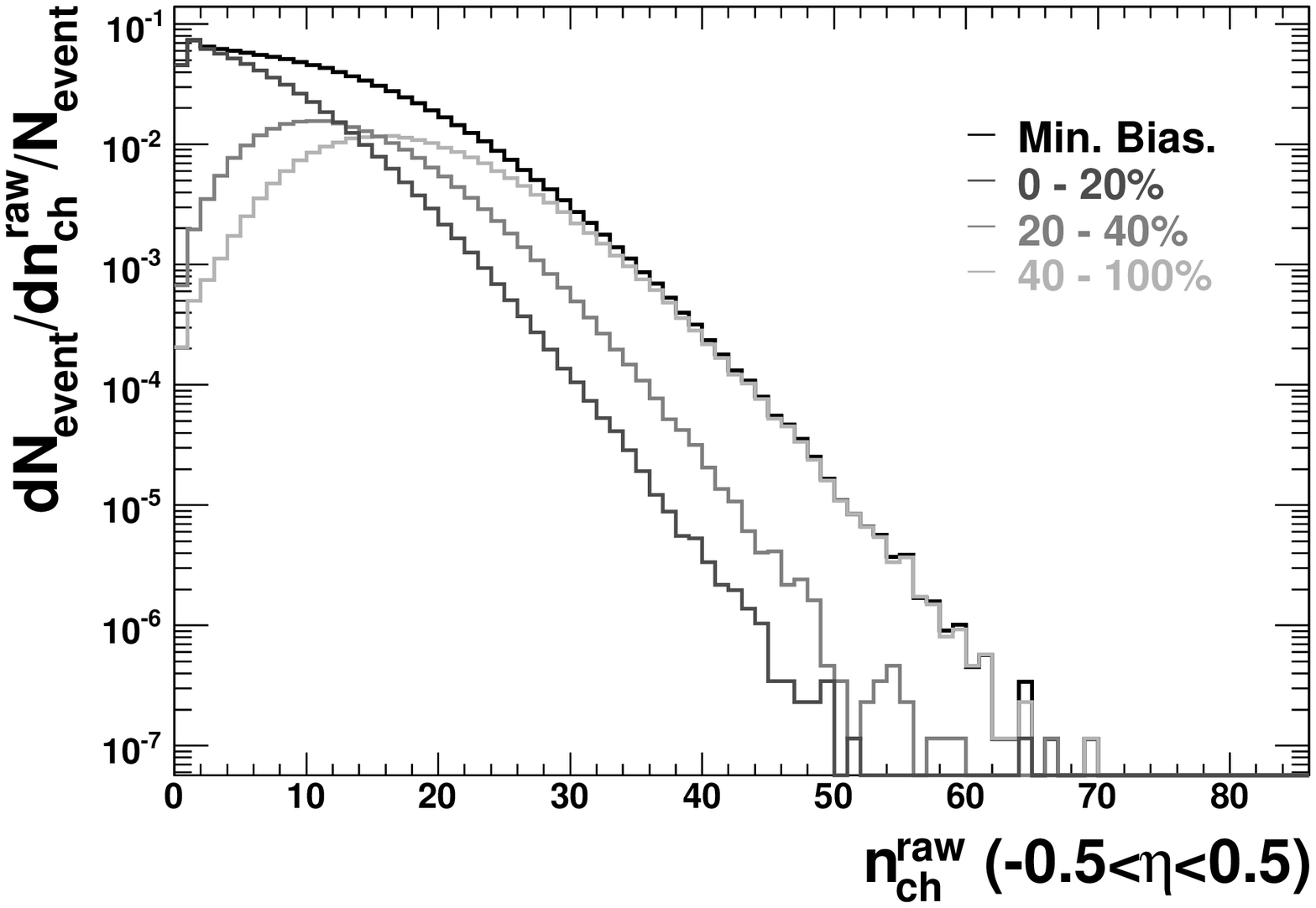}}
		   \caption{Uncorrected charged particle multiplicity measured in the TPC in 200 GeV dAu collisions. Color coding refers to the centrality selection as shown in Fig.~\ref{fig:dau_mult1}. \label{fig:dau_mult3}}
		   \end{center}
\end{figure}
%
%
%
In dAu collisions the z component of the primary vertex in each minimum bias event has to satisfy the following condition: $\left|v_{z}\right|< $ 50.0 cm. With this vertex cut and minimum bias trigger, 8.8 M good events are selected. Broader vertex distribution is used in dAu than in pp collisions because of the asymmetric bunch timing. 
In dAu collisions the uncorrected reference multiplicity is defined in the East FTPC, as shown in Fig.~\ref{fig:dau_mult1}, (situated on the outgoing Au side) as the number of charged primary tracks in the pseudo-rapidity range of: -3.8 $< \eta <$ -2.8. Three centrality classes are selected based on the East FTPC, which represent 0-20$\%$, 20-40$\%$, 40-100$\%$ of the geometrical cross-section. Figure~\ref{fig:dau_mult2} shows the uncorrected charged particle multiplicity measured in the East FTPC as a function the TPC multiplicity. Collisions selected in a FTPC multiplicity window correspond to a broad range of multiplicites in the TPC. Figure~\ref{fig:dau_mult3} shows the multiplicity distributions of the corresponding FTPC centrality selection.

Collision properties for pp and dAu collisions are summarized in Table~\ref{tab:auau_coll_prop}.

\begin{figure}[!h]
\begin{center}
	\resizebox{.9\textwidth}{!}{\includegraphics{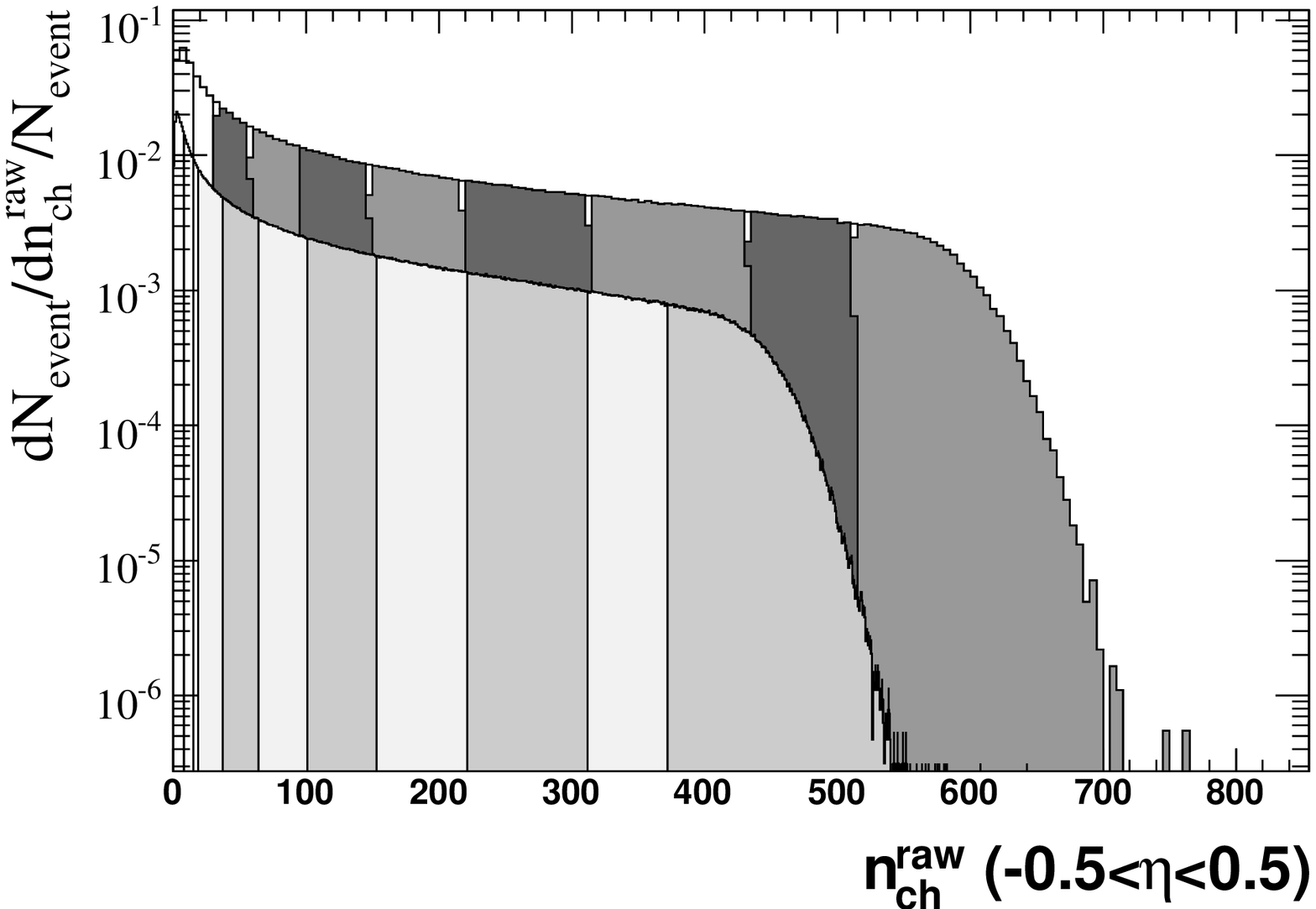}}
		   \caption{Uncorrected charged particle multiplicity measured in the TPC in {\bf 62.4} GeV Au-Au collisions (front) and in {\bf 200} GeV Au-Au collisions (back). \label{fig:auau62_mult}}
		   \end{center}
\end{figure}
\subsection{Gold - Gold collisions}
Events collected in 62.4 GeV Au-Au collisions are required to have a $z$ vertex component in $\left|v_{z}\right|< $ 30 cm. With this vertex cut and minimum bias trigger selection 6.3 M good events are selected. Nine centrality classes are defined based on the charged particle multiplicity measured in -0.5 $<$ $\eta$ $<$ 0.5. Figure~\ref{fig:auau62_mult} shows the centrality selection of the uncorrected charged particle multiplicity in 62.4 GeV collisions and the corresponding multiplicity and centrality selection in 200 GeV collisions. 
The nine centrality bins correspond to the fraction of the total geometrical cross-section: 0 - 5\%, 5 - 10\%, 10 - 20\%, 20 - 30\%, 20 - 30\%, 30 - 40\%, 40 - 50\%, 50 - 60\%, 60 - 70\%, 70 - 80\% as shown in Table~\ref{tab:auau_coll_prop}. The last centrality bin 80-100\% is not used in data analysis due to significant trigger bias.

\section{Track selection}

Tracks selected for spectra analysis are required to satisfy certain quality cuts. The first criterion is the number of fit points cut. Tracks traversing through the TPC volume can leave 45 possible hits. To avoid splitting tracks we require at least 25 fit points on the track. The distance of closest approach (dca) should be less than 3 cm, which ensures that tracks come from the triggered event vertex and not from a secondary collision or interaction. These tracks are called primary tracks. To estimate the systematic errors on track selection three additional variations of quality cuts have been implemented as shown in Table~\ref{Tab:qa_cuts}. Set 1 represents the default quality cuts for the spectra analysis implemented in this work.
\begin{table}[!h]
	\begin{center}
  	\caption{Collection of quality cuts, implemented for systematic studies. }
	 	\begin{tabular}{|c|c|c|c|c|}
	 		\hline 
	 		Cuts & {\bf Set 1} & Set 2& Set 3 & Set 4\\ \hline
	 		$|y|<$ & {\bf 0.1} & 0.1 & 0.1 & 0.3 \\
	 		Number of fit points $\geq$ & {\bf 25} & 35 & 25 & 25\\
	 		dca (cm) $\leq$ & {\bf 3.0} & 3.0 & 1.0 & 3.0 \\ \hline
	 		
	 	 	\end{tabular}
 	\end{center}
			\label{Tab:qa_cuts}	
\end{table}
\section{Short description of Monte Carlo Glauber calculation}\label{sec:glauber}

Sometimes it is desirable to connect measurements to geometrical quantities of the collisions. Typical examples are the number of participants ($N_{part}$) and the number of binary collisions ($N_{bin}$ or $N_{bin}$) or even the impact parameter (b). These parameters cannot be directly measured, but can be calculated in a geometrical model of a nucleus-nucleus collision, namely the Glauber model~\cite{GlModel}. The model is based on individual nucleon-nucleon collisions which are controlled by the elementary nucleon-nucleon cross-section. In the Monte Carlo Glauber calculation nuclei are independently generated, distributing the nucleons according to the Wood-Saxon density profile:
\begin{equation}
    \rho(r) = \frac{\rho_o}{1+e^{\frac{r-r_0}{a}}} .
\end{equation}
Here $r_0 = 6.5 \pm 0.1$~fm and $a = 0.535 \pm 0.027$~fm are experimentally measured in $e$-Au scattering~\cite{Antinori:2000ph} and $\rho_0 = 0.169$~fm$^{-3}$ is fixed by the normalization.
Each nucleon in the nucleus is separated by a distance larger than $d_{min}$ = 0.4 fm. This cut off value is the characteristic length of the repulsive force acting on the nucleons. 

$N_{part}$ is defined as the total number of nucleons that underwent at least one collision. $N_{bin}$ is defined as the total number of interactions in the event. The nuclei generation and the nucleon-nucleon selection is repeated with random impact parameter ($b$) selection, where $b^2$ is a flat distribution. The extracted quantities can be studied as the fraction of the total geometrical cross-section.
The distributions of $d\sigma/db$ (and $d\sigma/dN_{part}$ and $d\sigma/dN_{bin}$) are determined. Each distribution is divided into bins corresponding to the fractions of the measured total cross-section of the used centrality bins and the mean values of $\left\langle N_{part}\right\rangle$ and $\left\langle N_{bin}\right\rangle$ are extracted for each centrality bin. 

Moreover, from the MC Glauber calculation the transverse area ($S_{Glauber}$) of the colliding nuclei can be determined from the spatial distribution of the nucleons. The $S_{Glauber}$ is defined as the average transverse area of the overlapping nucleons in a given centrality bin.
To make a comparison with previously published results, the overlap area ($S$) can also be calculated as:
\begin{equation}
S\ =\ \pi\cdot R^{2} =\ \pi\cdot \left( a_{0}\cdot A^{1/3}\right)^2 =\ \pi\cdot a_{0}^{2}\cdot (0.5\cdot N_{part})^{2/3}
\end{equation}
where $a_{0}$ = 1.12.
For detailed description of the Glauber calculation implemented in STAR, we refer the reader to~\cite{Adams:2003yh}. In the calculations the proton-proton cross-sections are obtained from the Particle Data Group~\cite{Hagiwara:2002fs}. 

The proton-proton cross-section used in the MC Glauber calculation is 36 $\pm$ 3 mb for 62.4 GeV and 41 $\pm$ 3 mb for 200 GeV. Systematic uncertainties are obtained from the variation of the proton-proton cross-section by $\pm$ 3 mb and the variation of the Wood-Saxon parameters.
The calculated MC Glauber quantities are listed in Table~\ref{tab:auau_coll_prop}.

\section{Embedding}

The correction in our analysis relies on good knowledge of the detector and its simulation. The STAR geometry has been implemented in GEANT~\cite{geant,pnevski} with detailed detector material description. Moreover, realistic simulation of the TPC pad response has been implemented~\cite{Gong:00} in the STAR simulation framework. Physical processes such as drift of the electrons in the TPC gas, the amplification of the signal at the sense/read-out wires, the induction on the readout pads, and the response of the readout electronics (ADCs) are encoded in the TPC Response Simulator (TRS).

To obtain realistic corrections, simulated tracks (from GEANT) are embedded into a real event at the raw data (ADC) level. The traces of charged particles in the TPC are simulated, starting with the initial ionization of the TPC gas, then electron transport and multiplication in the drift field, and finally the induced signal on the TPC's read-out pads and the response of read-out electronics (TRS). The obtained raw simulated signal is then embedded into a real event and then passed through the STAR Offline reconstruction chain. The resulting mixed events are of the same format and contain the same information as real raw data delivered by the data acquisition system. This procedure is called $embedding$, providing nearly realistic simulation of the collision environment. 

\subsection{Hit level studies}

\begin{figure}[!h]
\begin{center}
	\resizebox{.9\textwidth}{!}{\includegraphics{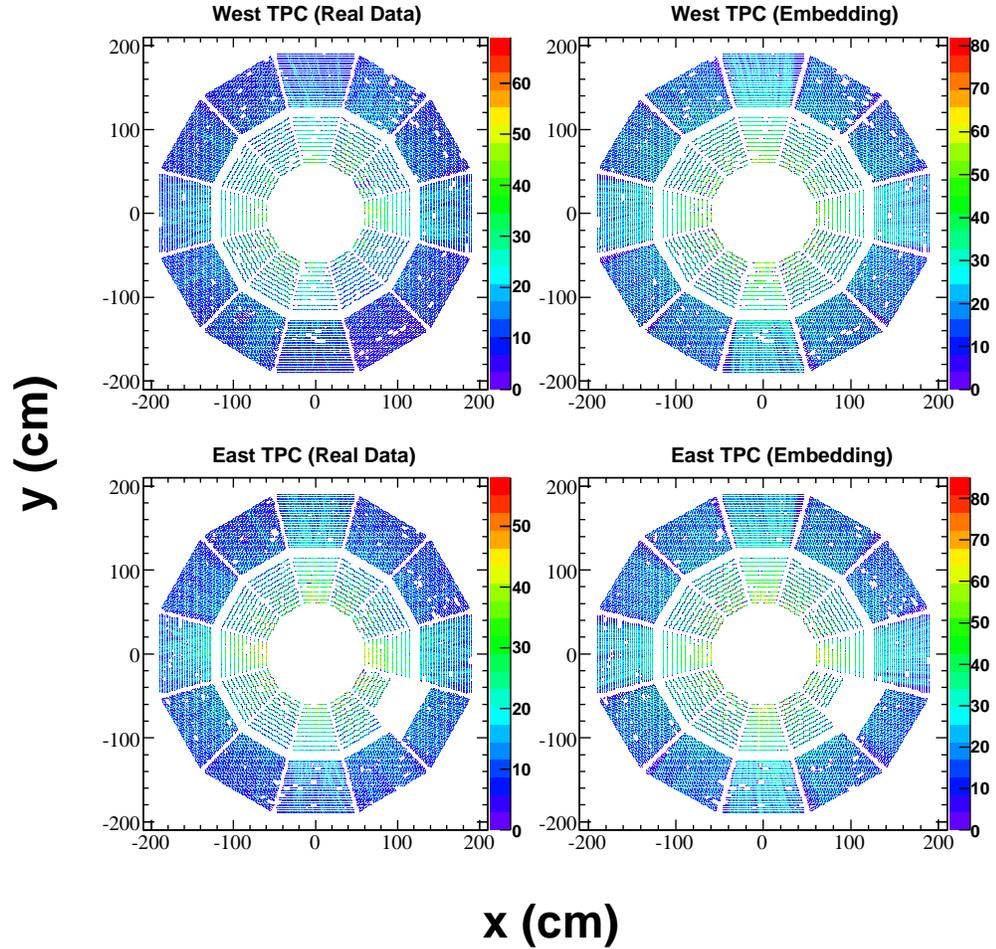}}
		   \caption{Hit distribution in the East and West half of the TPC for real and embedded events in 62.4 GeV Au-Au collisions.}\label{fig:xyhittpc}
		   \end{center}
\end{figure}

To calculate proper efficiencies it is important to check the quality of the embedding process. First, hit level quantities are compared from embedding and real events, such as X-Y hit distributions in the east and west TPC padrows, as shown in Fig.~\ref{fig:xyhittpc}.
Since the embedded MC tracks are reconstructed with real events starting at the raw hit level, the calibration database of the given run has to be propagated into the embedding as well. (Separate off-line event reconstruction chains are used for embedding and real events.) Figure~\ref{fig:xyhittpc} shows the hit distributions from real data (left panels) and embedding (right panel). The hits density is represented in the color coding. The sector structure of the TPC is clearly shown. Empty white spots in the sensitive area of the TPC represent dead sectors and the larger white areas at the 4 o'clock position represents a bad Read Out Board for this particular run. Propagation of the correct hit level calibration information is essential to calculate proper efficiencies.

The amount of embedded tracks is $\sim$ 5$\%$ of the total number of tracks in the real event. To calculate acceptance and tracking efficiency corrections one has to use the reconstructed $associated$ $tracks$. In the reconstruction process hit information of the MC track is kept and can be compared to the hit information of the reconstructed tracks. A MC track is associated to a reconstructed track if they share at least 3 common hit points within 5 mm in $x$, $y$ and $z$ hit coordinates. 
\begin{figure}[!h]
  \begin{center}
  	  \resizebox{.9\textwidth}{!}{\includegraphics{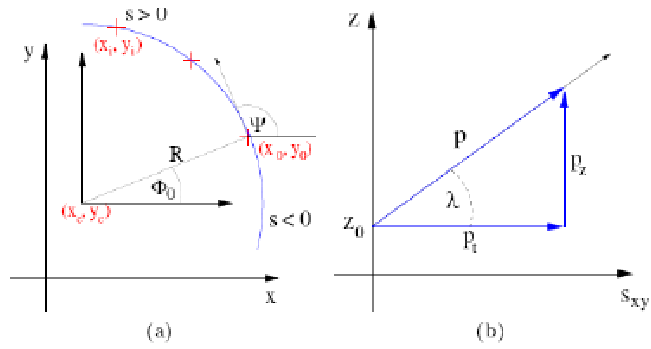}}
    \caption{Schematic view of the crossing angle ($\lambda$) and the dip angle ($\phi$)~\cite{dipcrosstex}.\label{fig:dipandcross}}
  \end{center}
\end{figure}
For embedding calibration purposes, the longitudinal and transverse resolution have to be compared to real data as a function of the longitudinal distance ($z$), the crossing angle and the dip angle. In the local coordinate system of the padrow, a coordinate system can be defined as the $x$ axis is along the padrow direction and the $y$ axis is perpendicular to that, as shown in Fig.~\ref{fig:dipandcross} (left panel). The first points of the track are denoted as $x_{0}$, $y_{0}$, $z_{0}$. 
The crossing angle is the angle enclosed by the momentum of the particle crossing the padrow and the $x$ direction. The dip angle ($\lambda$) is defined as the angle between the momentum of the particle and the momentum component perpendicular to the drift direction, as shown in Fig.~\ref{fig:dipandcross} (right panel). Hit level quantities are propagated to track finding and hence to multiplicity and spectra quantities, therefore embedding has to reproduce data reasonably well. 
\begin{figure}[!h]
\begin{center}
	\resizebox{.45\textwidth}{!}{\includegraphics{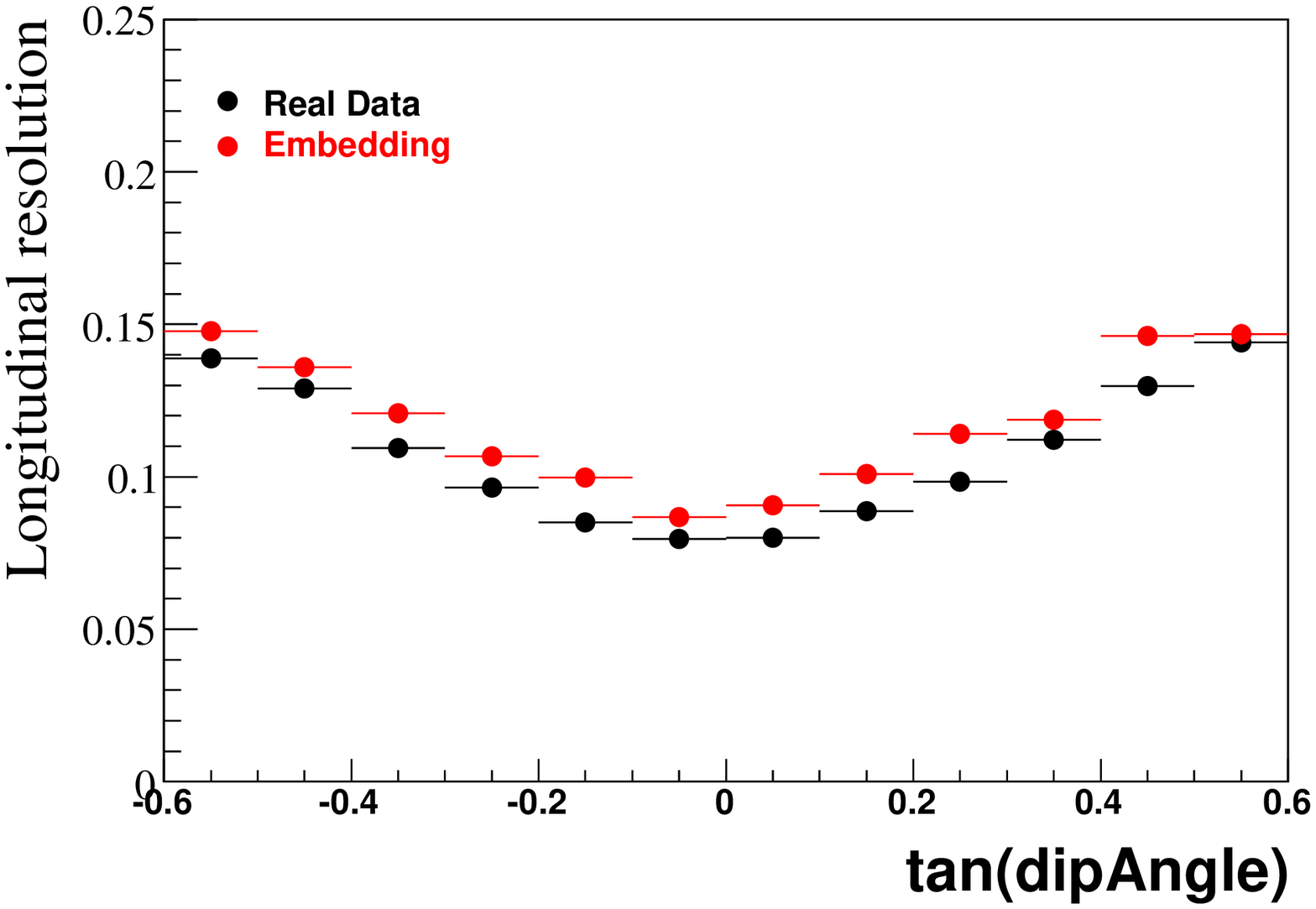}}
	\resizebox{.45\textwidth}{!}{\includegraphics{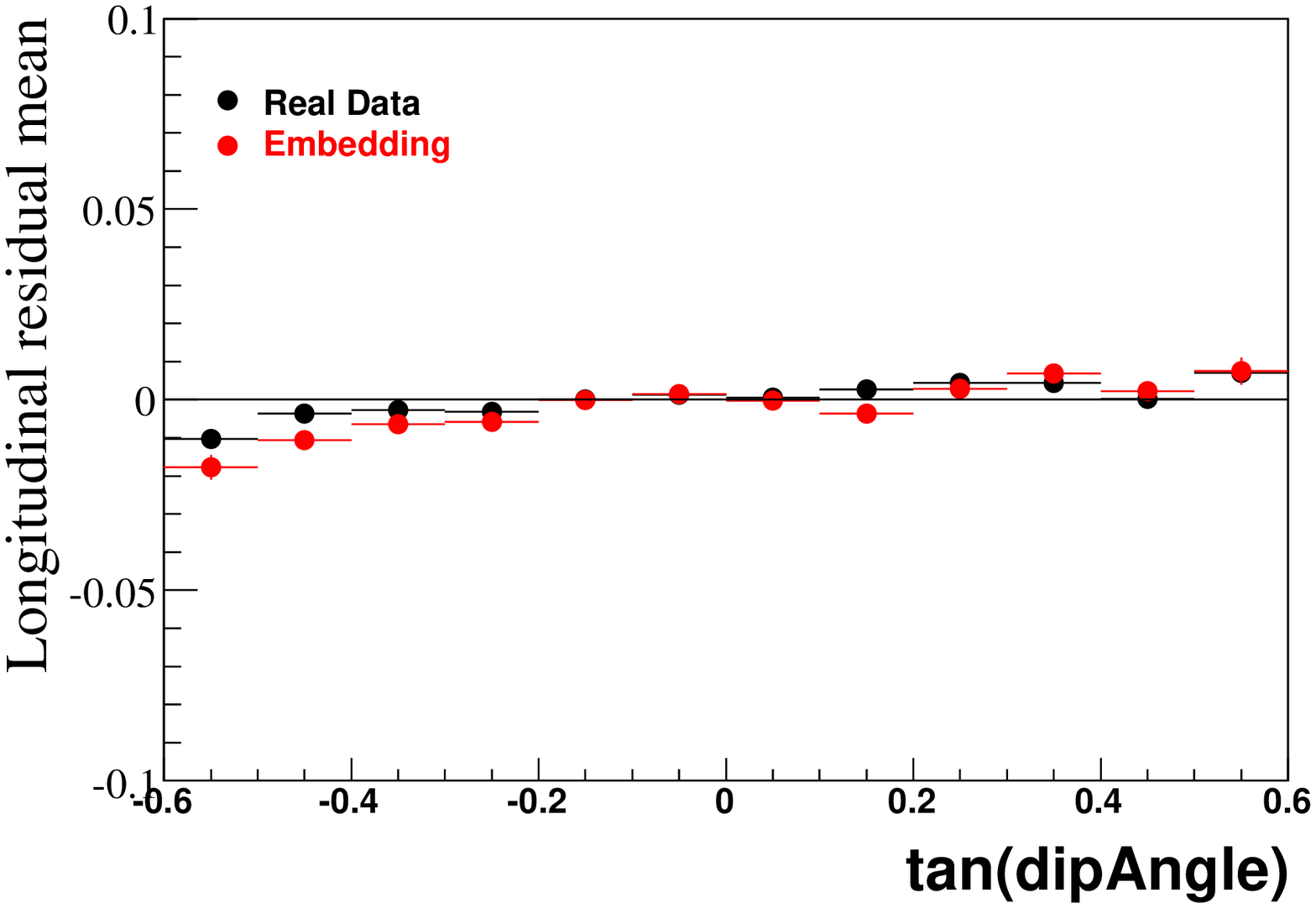}}
		   \caption{Longitudinal hit resolution and mean as a function of the tangent of the dip angle.}\label{fig:longhitdipangle}
		   \end{center}
\end{figure}
\begin{figure}[!h]
\begin{center}
	\resizebox{.45\textwidth}{!}{\includegraphics{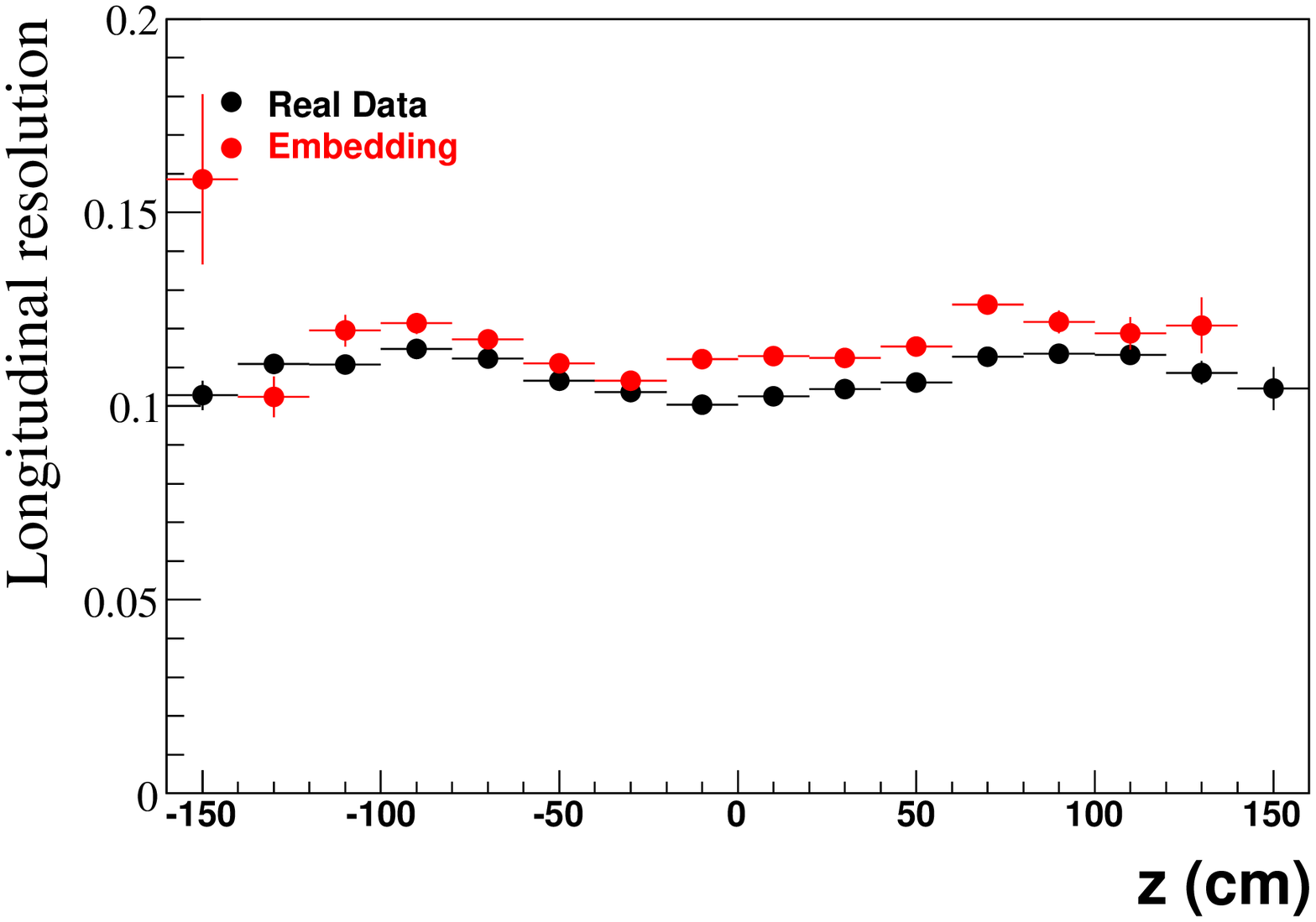}}
	\resizebox{.45\textwidth}{!}{\includegraphics{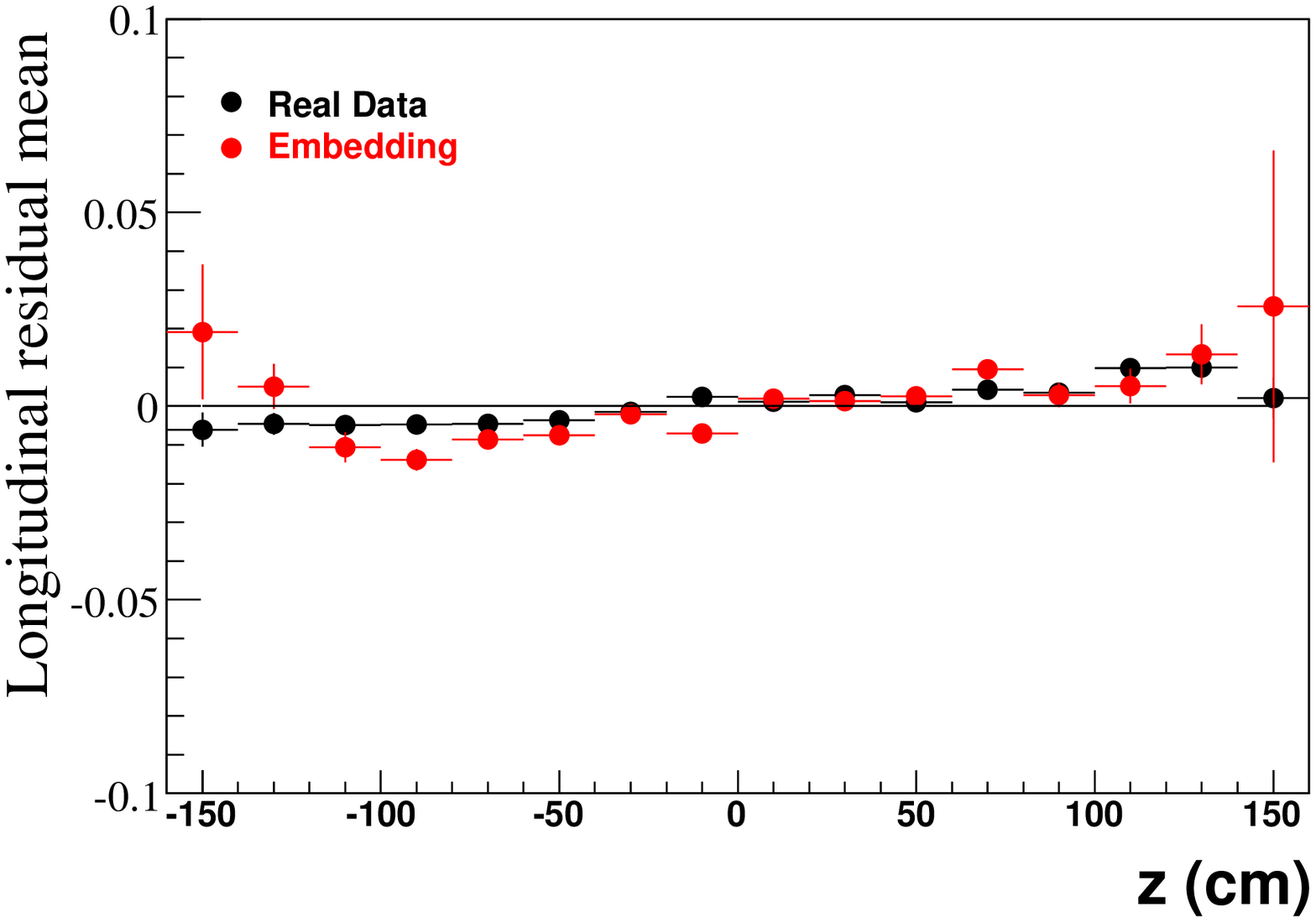}}
		   \caption{Longitudinal hit resolution and mean as a function of the $z$ coordinate.}\label{fig:longhitz}
		   \end{center}
\end{figure}
\begin{figure}[!h]
\begin{center}
	\resizebox{.45\textwidth}{!}{\includegraphics{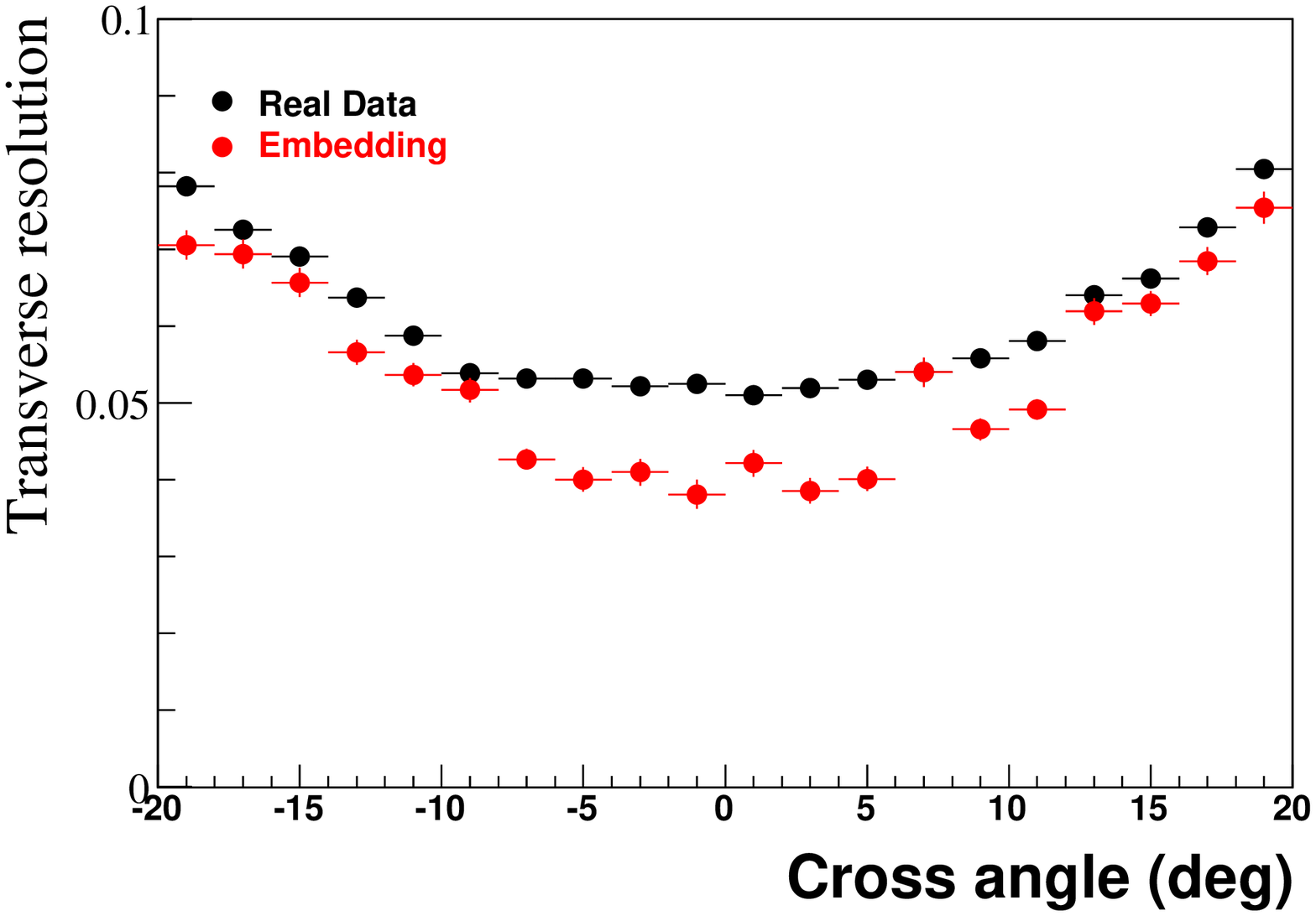}}
	\resizebox{.45\textwidth}{!}{\includegraphics{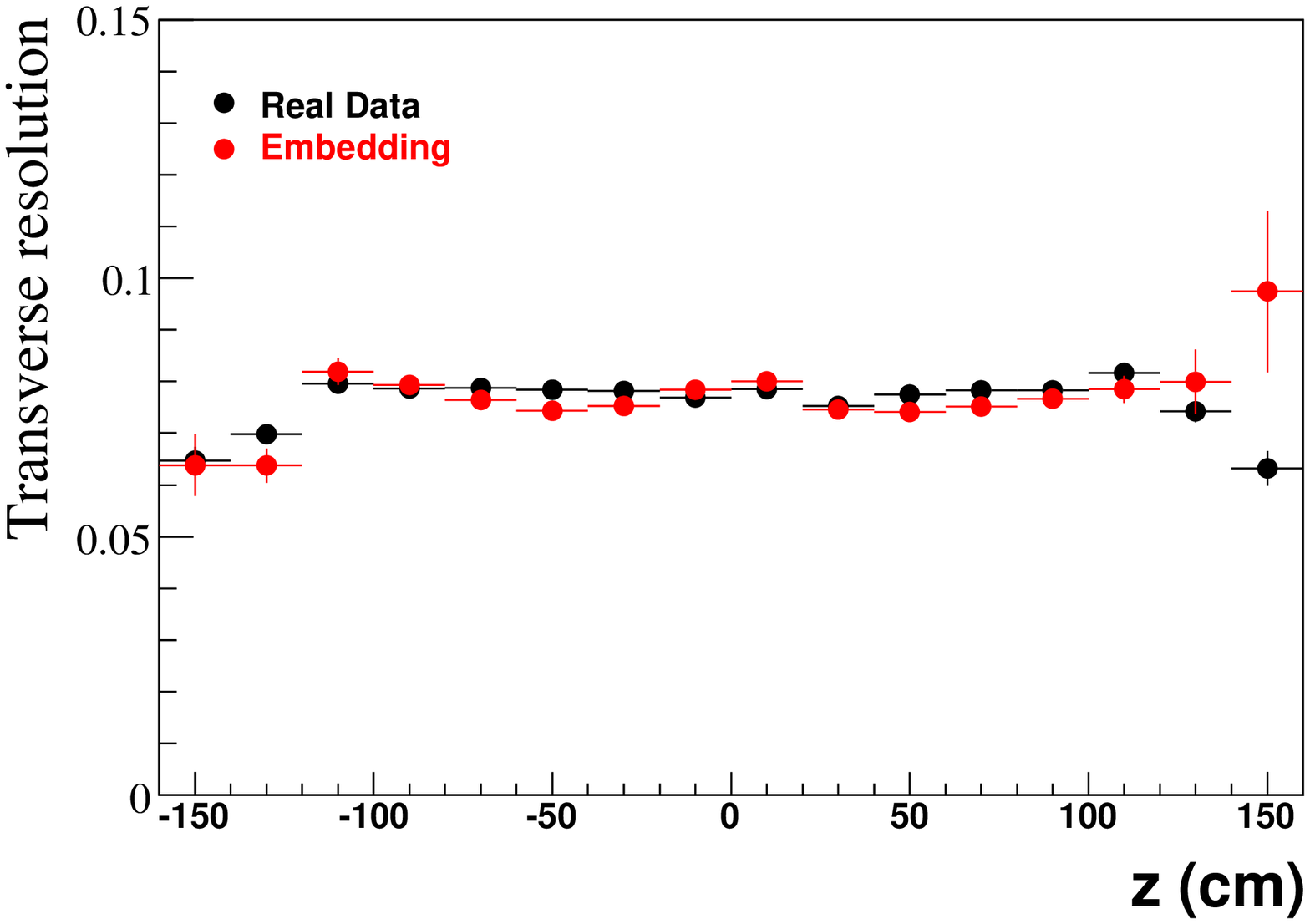}}
		   \caption{Transverse hit resolution as a function of the crossing angle and  the $z$ coordinate.}\label{fig:tranhitcrosz}
		   \end{center}
\end{figure}

As an example of the hit level simulation of the TPC, the comparison of longitudinal and transverse hit resolution between real data and embedding as a function of the dip angle (Fig.~\ref{fig:longhitdipangle}), $z$ vertex coordinate (Fig.~\ref{fig:longhitz}) and the crossing angle (Fig.~\ref{fig:tranhitcrosz}) are shown. Plots are generated from negative kaon embedding and real data produced from 62.4 GeV Au-Au collisions in the transverse momentum range: 400 - 500 MeV/c. The embedding can reproduce real data well, both transverse and longitudinal hit resolution is $\sim$ 10 - 12\%, and deviation form the mean is less than 2 $\%$.  

\subsection{Track level studies}

Since the same analysis cuts are applied on the embedding and on real data, to extract the efficiencies one has to compare the track level distributions (cuts used to select tracks for identified particle spectra): $dca$ and number of fit points ($N_{fit}$). 
\begin{figure}[!h]
\begin{center}	
\resizebox{.45\textwidth}{!}{
\includegraphics{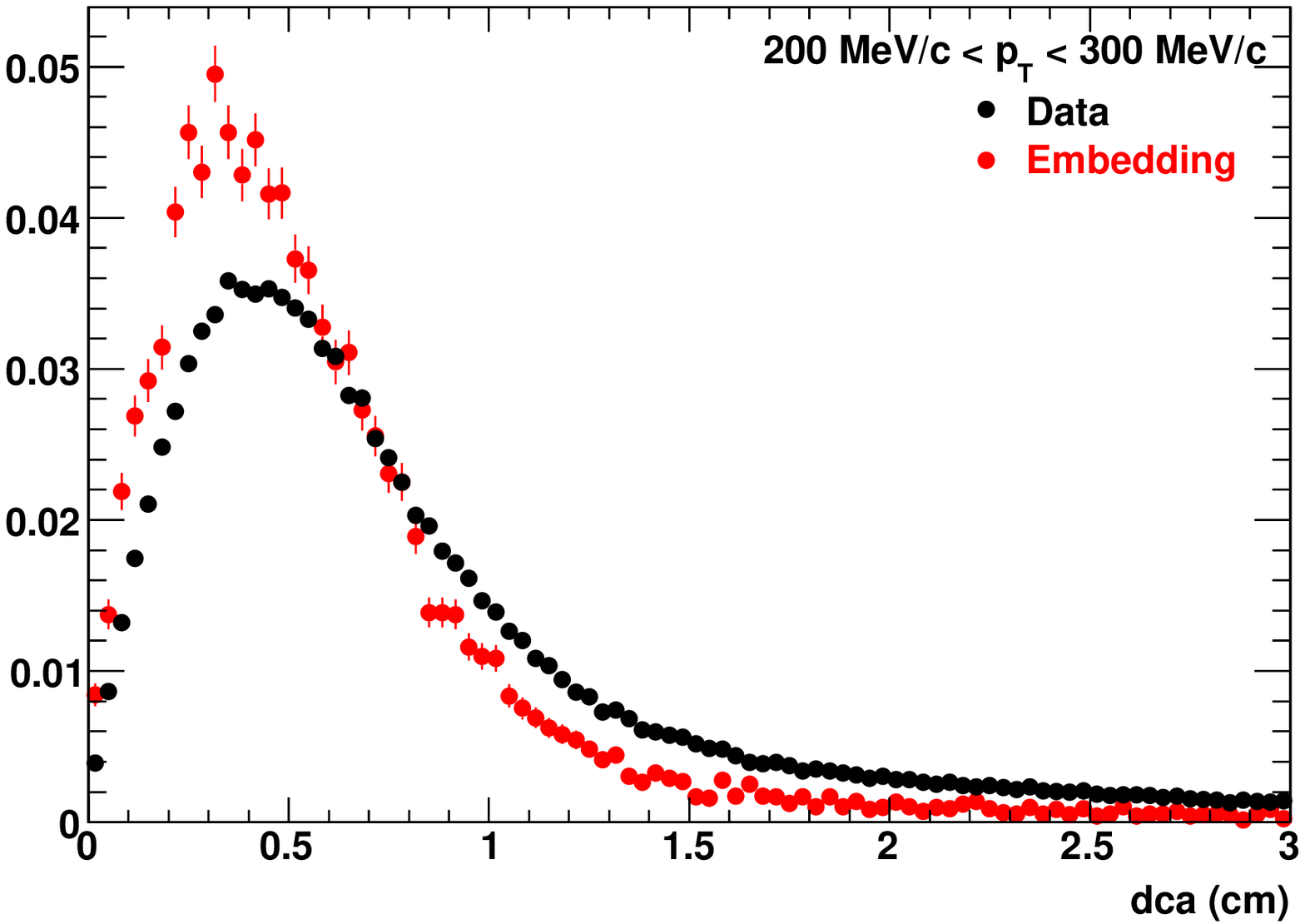}}	
\resizebox{.45\textwidth}{!}{
\includegraphics{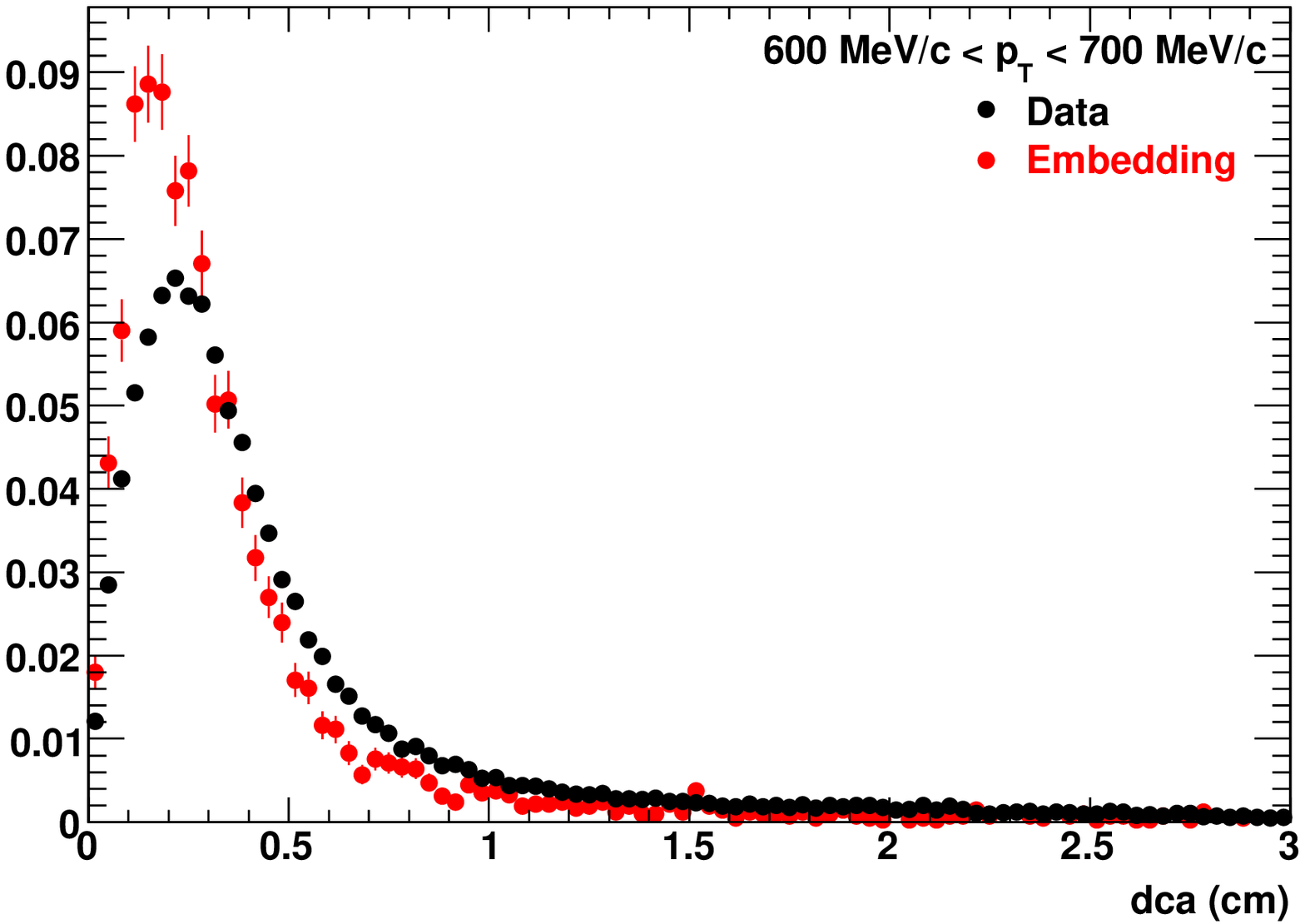}}
		   \caption{Comparison of $\pi^{-}$ $dca$ extracted from data and embedding for 200 GeV pp collisions.}
		   \label{fig:pp_embeddatacomppim1}
		   \end{center}
\end{figure}
\begin{figure}[!h]
\begin{center}
\resizebox{.45\textwidth}{!}
{
\includegraphics{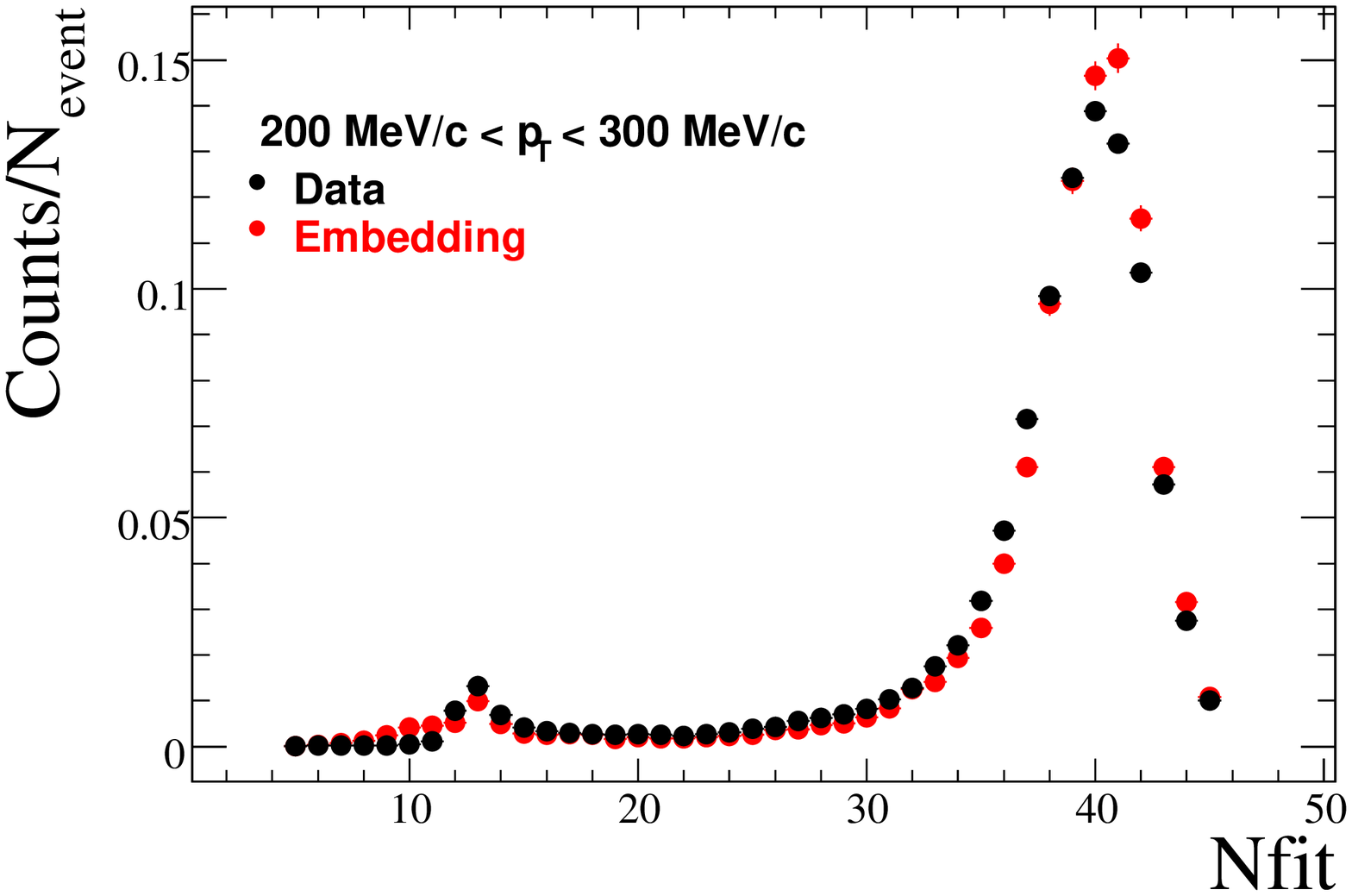}}	
\resizebox{.45\textwidth}{!}{
\includegraphics{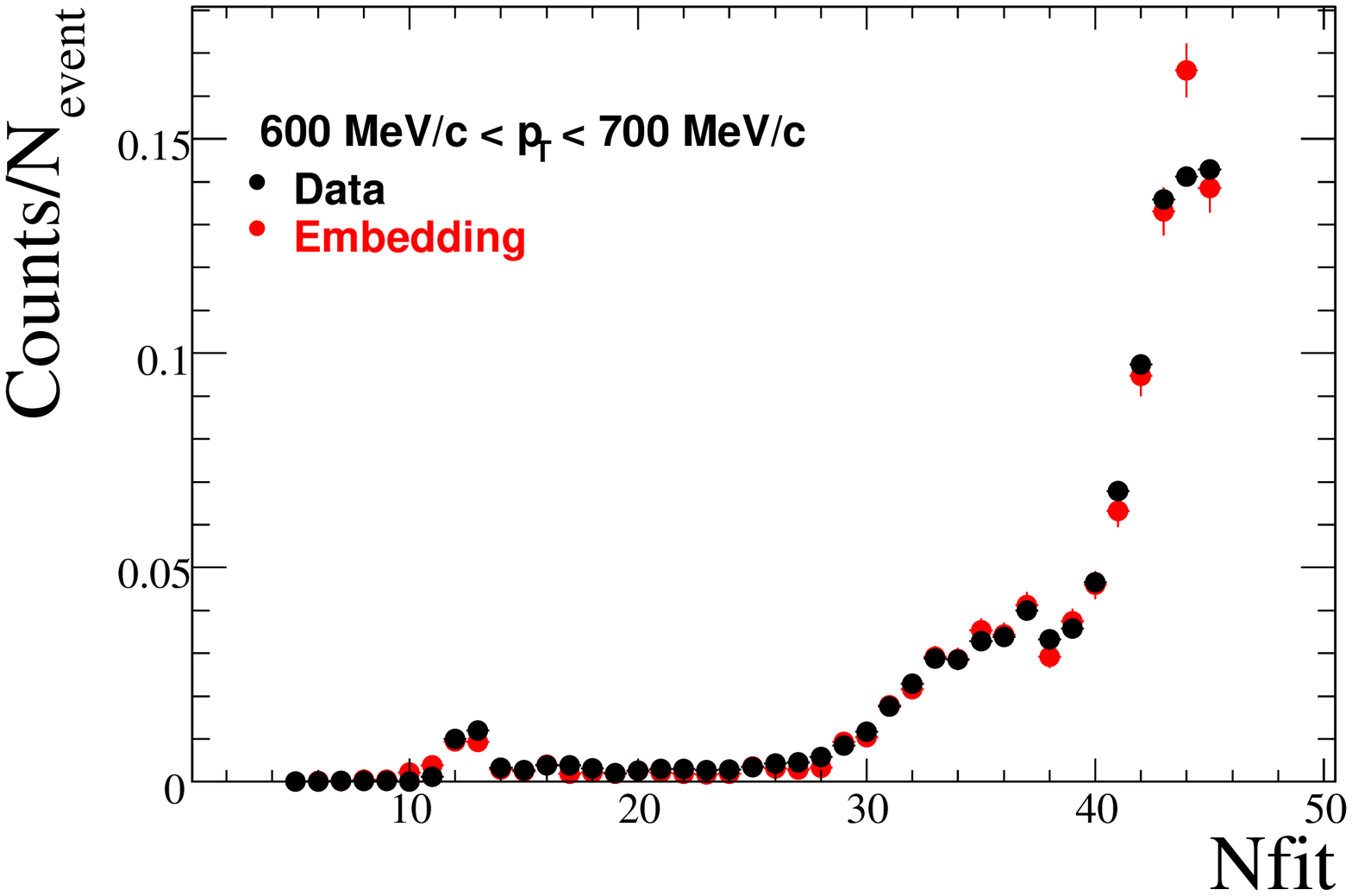}}
		   \caption{Comparison of $\pi^{-}$ fit points, extracted from data and embedding for 200 GeV pp collisions.}
		   \label{fig:pp_embeddatacomppim2}
		   \end{center}
\end{figure}
\begin{figure}[!h]
\begin{center}	
\resizebox{.45\textwidth}{!}{
\includegraphics{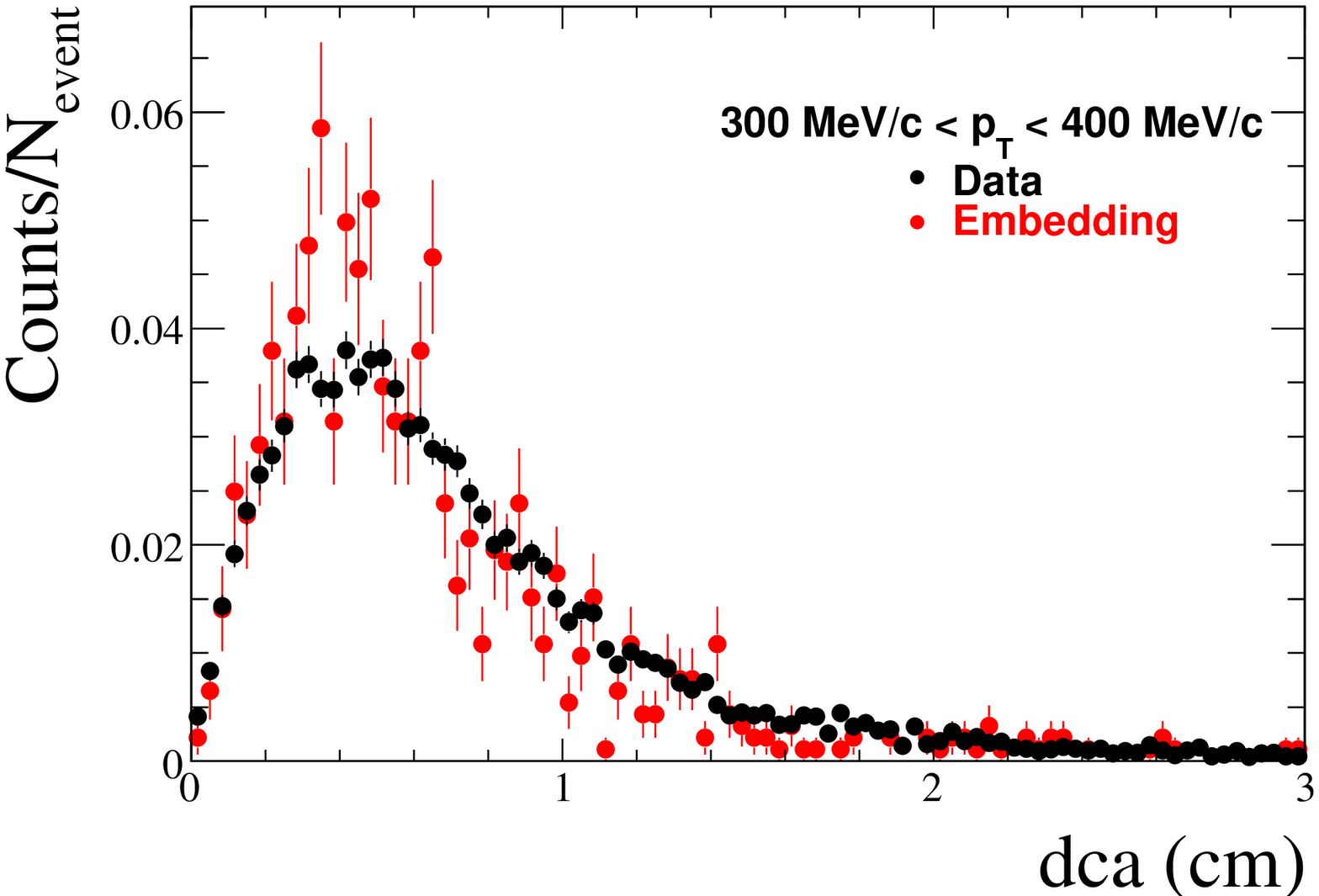}}	
\resizebox{.45\textwidth}{!}{
\includegraphics{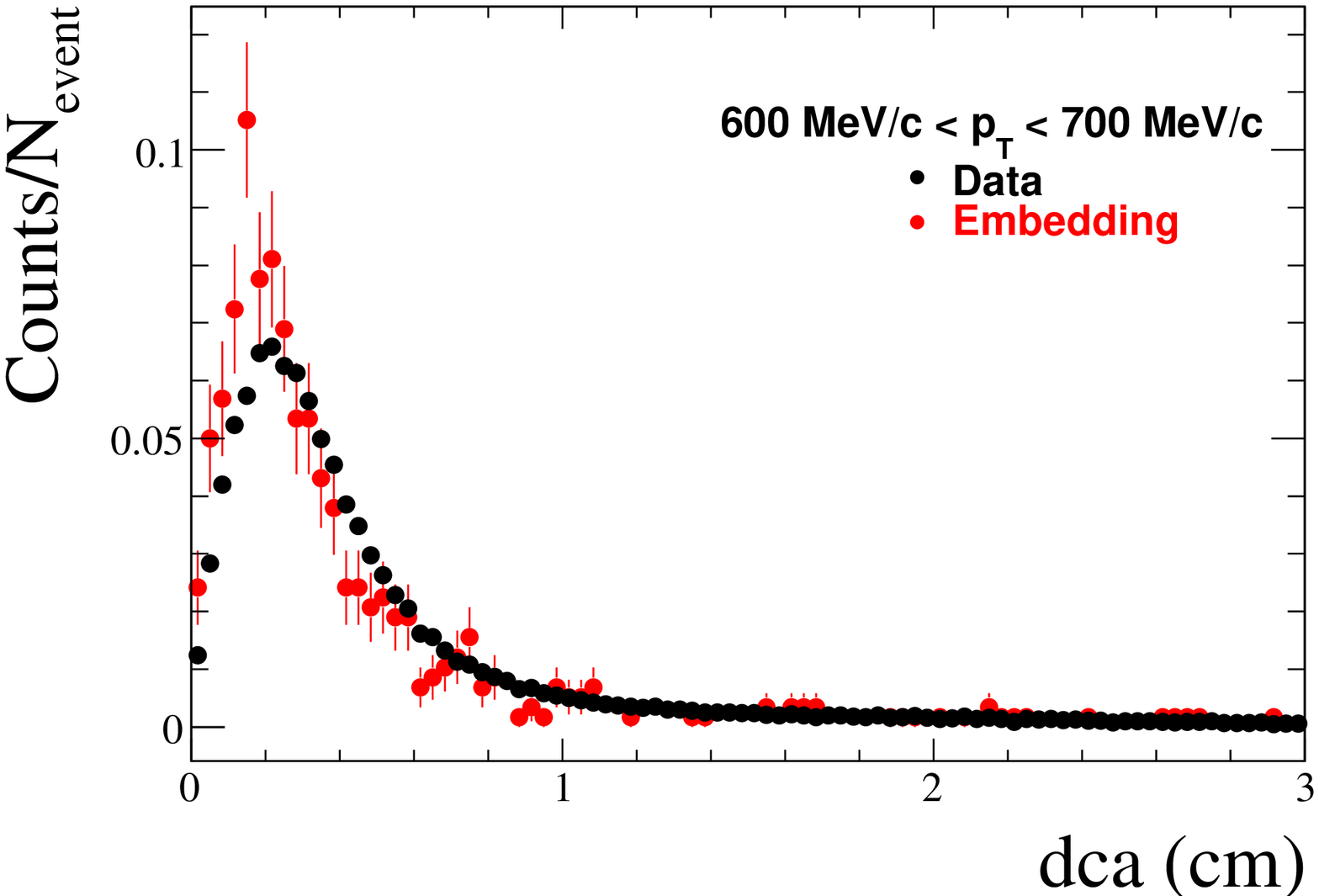}}
		   \caption{Comparison of $K^{-}$ $dca$ extracted from data and embedding for 200 GeV pp collisions.}
		   \label{fig:pp_embeddatacompkm1}
		   \end{center}
\end{figure}
\begin{figure}[!h]
\begin{center}
\resizebox{.45\textwidth}{!}
{
\includegraphics{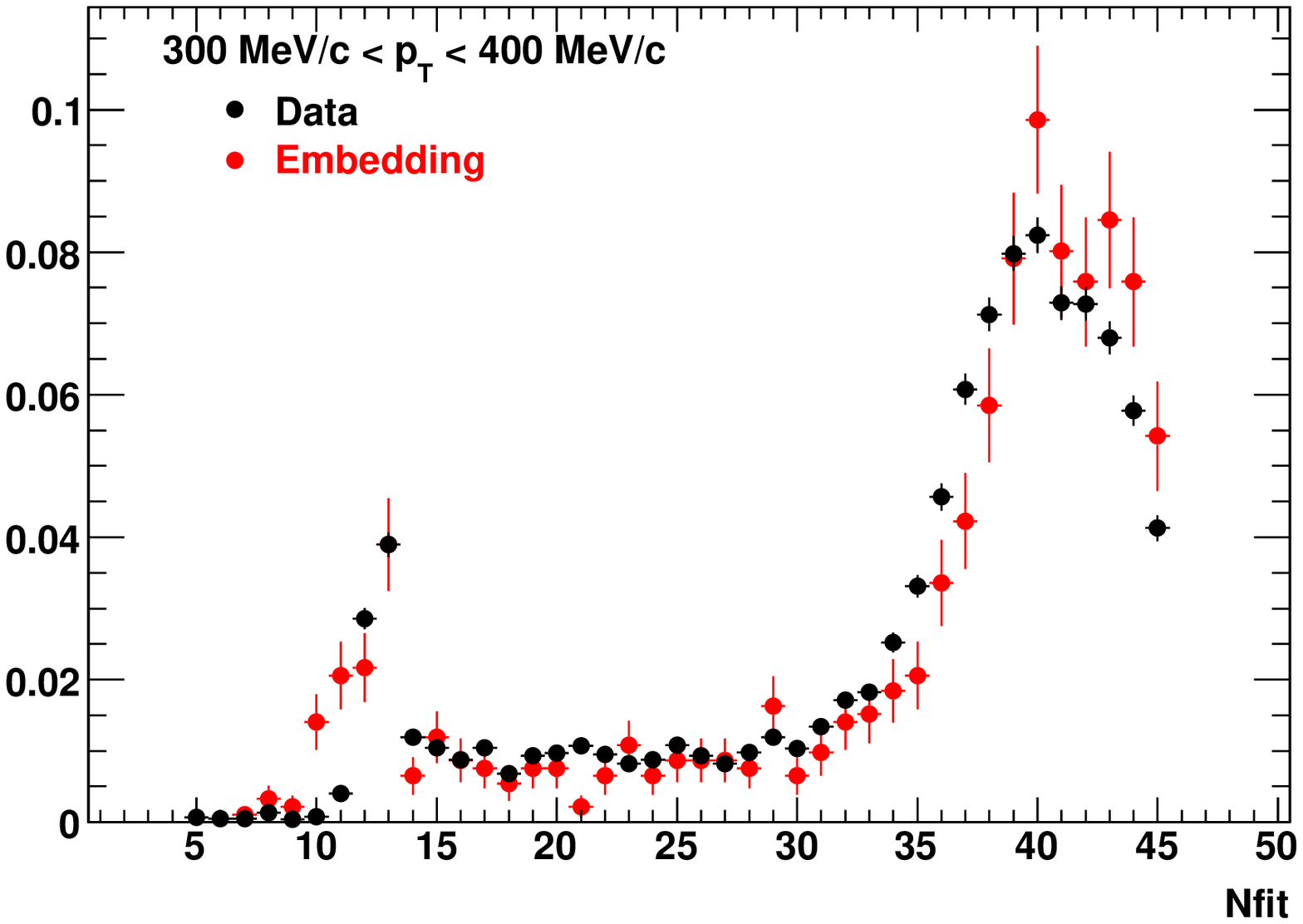}}	
\resizebox{.45\textwidth}{!}{
\includegraphics{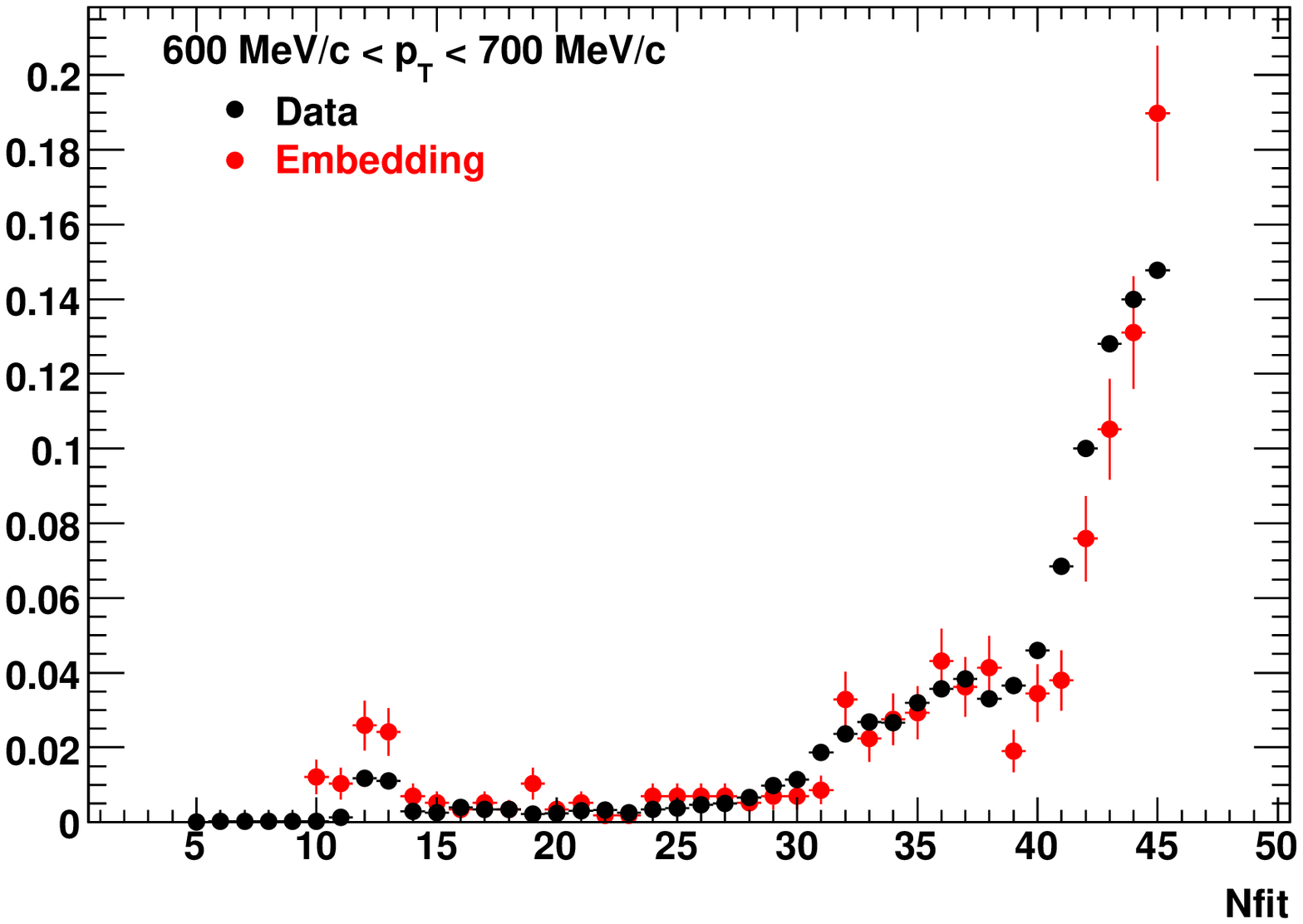}}
		   \caption{Comparison of $K^{-}$ fit points extracted from data and embedding for 200 GeV pp collisions.}
		   \label{fig:pp_embeddatacompkm2}
		   \end{center}
\end{figure}
\begin{figure}[!h]
\begin{center}	
\resizebox{.45\textwidth}{!}{
\includegraphics{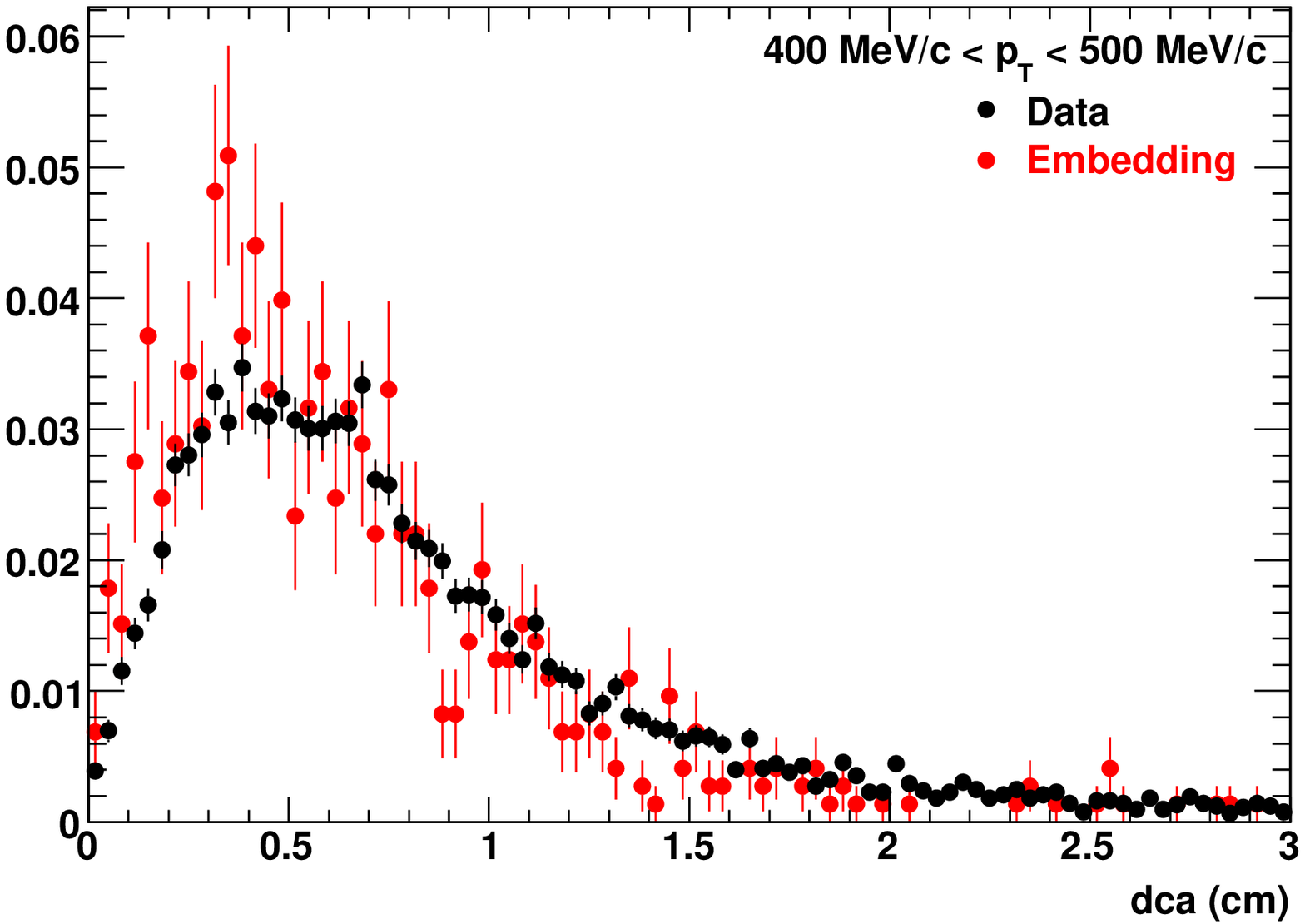}}	
\resizebox{.45\textwidth}{!}{
\includegraphics{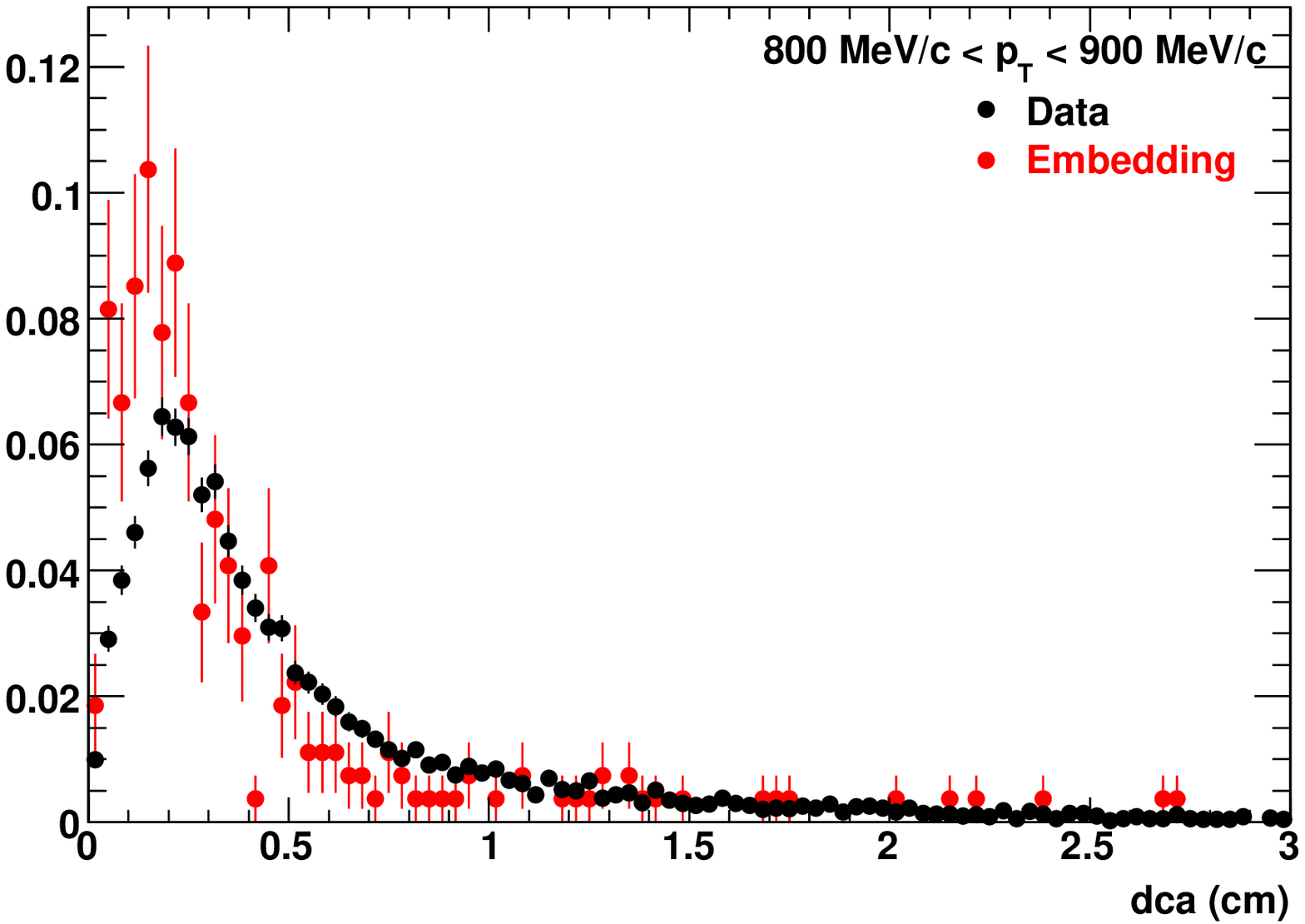}}
		   \caption{Comparison of $\overline{p}$ $dca$ extracted from data and embedding for 200 GeV pp collisions.}
		   \label{fig:pp_embeddatacomppbar1}
		   \end{center}
\end{figure}
\begin{figure}[!h]
\begin{center}
\resizebox{.45\textwidth}{!}
{
\includegraphics{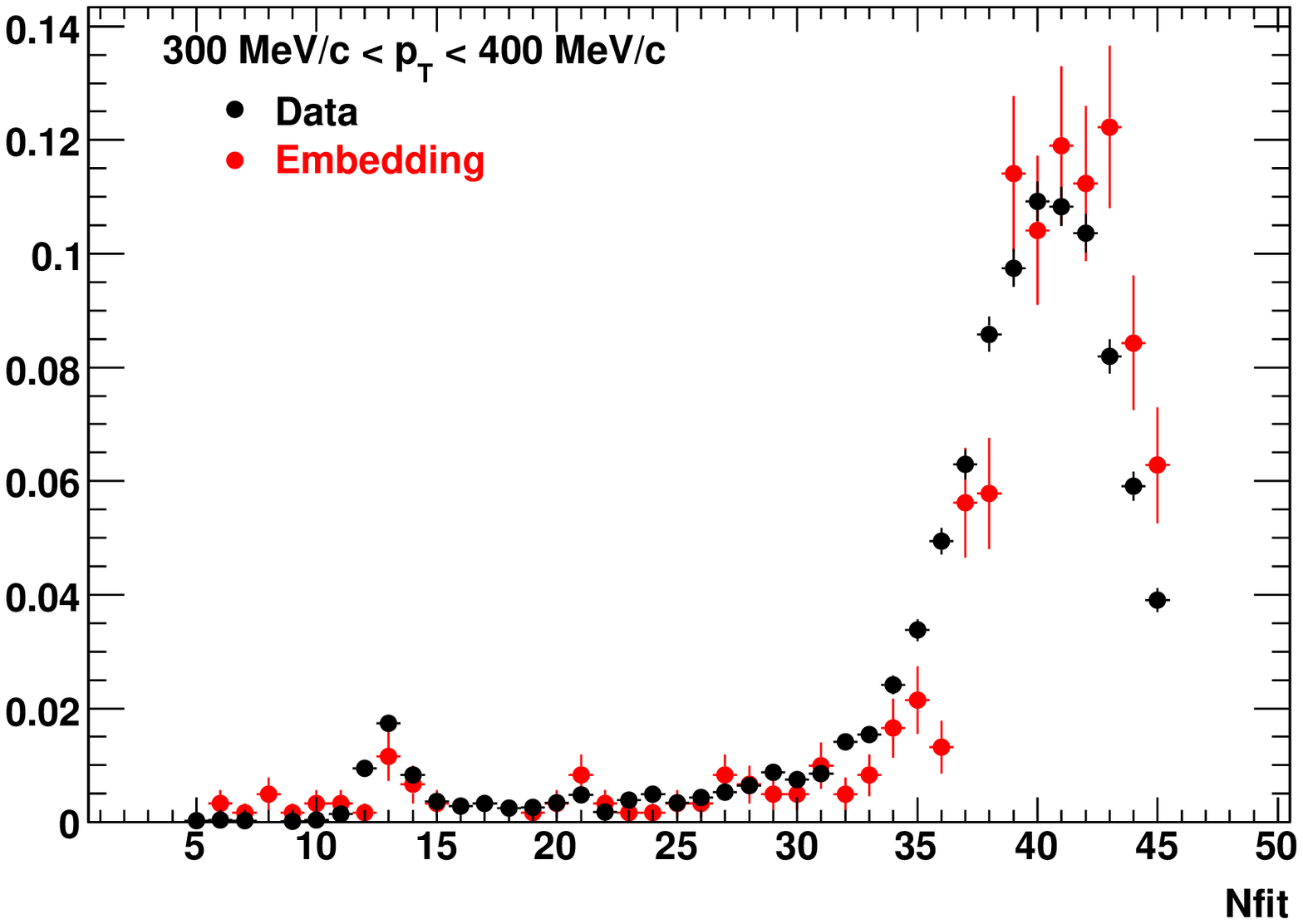}}	
\resizebox{.45\textwidth}{!}{
\includegraphics{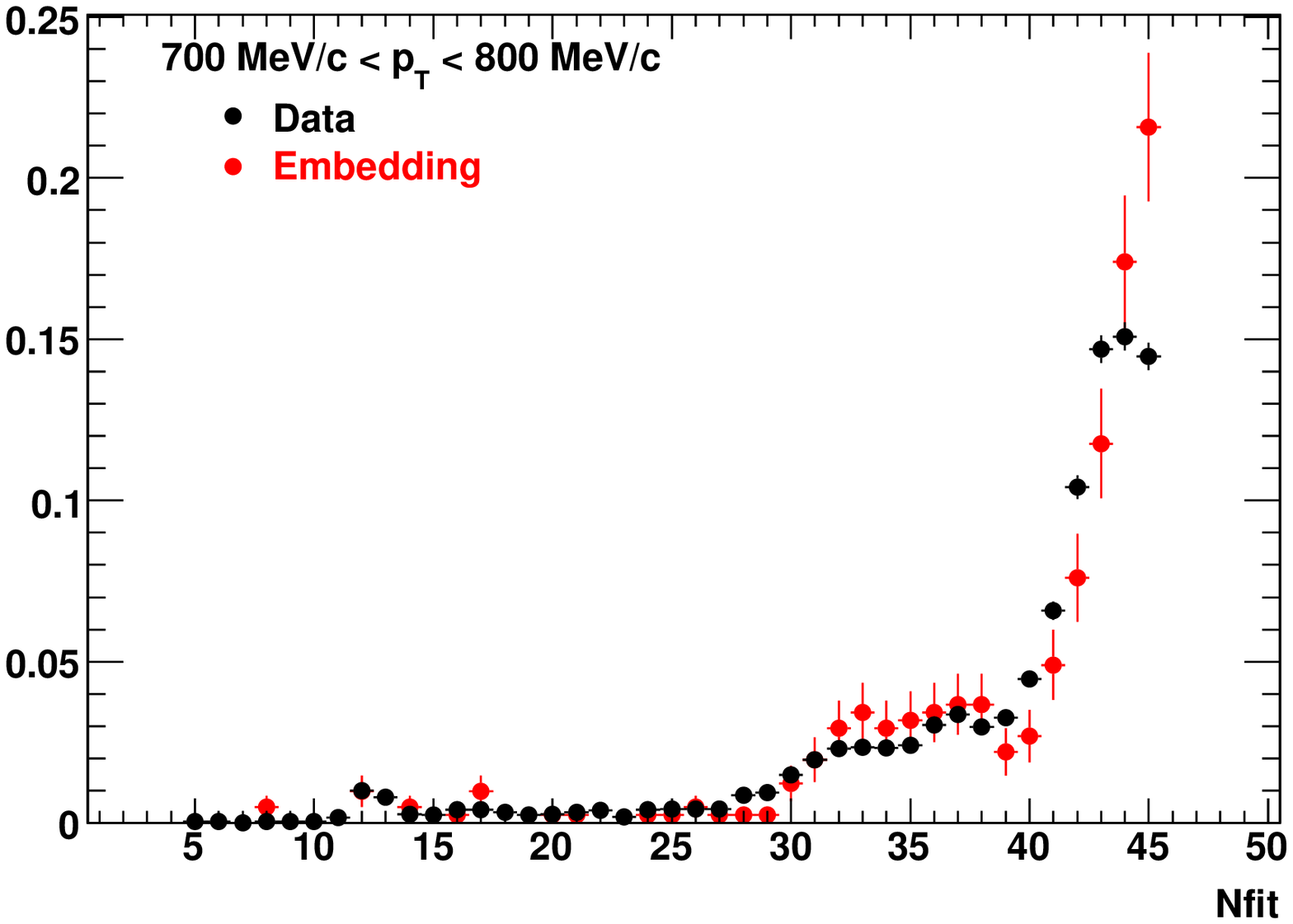}}
		   \caption{Comparison of $\overline{p}$ fit points extracted from data and embedding for 200 GeV pp collisions.}
		   \label{fig:pp_embeddatacomppbar2}
		   \end{center}
\end{figure}
%
%

%
%
%
\begin{figure}[!h]
\begin{center}	
\resizebox{.45\textwidth}{!}{
\includegraphics{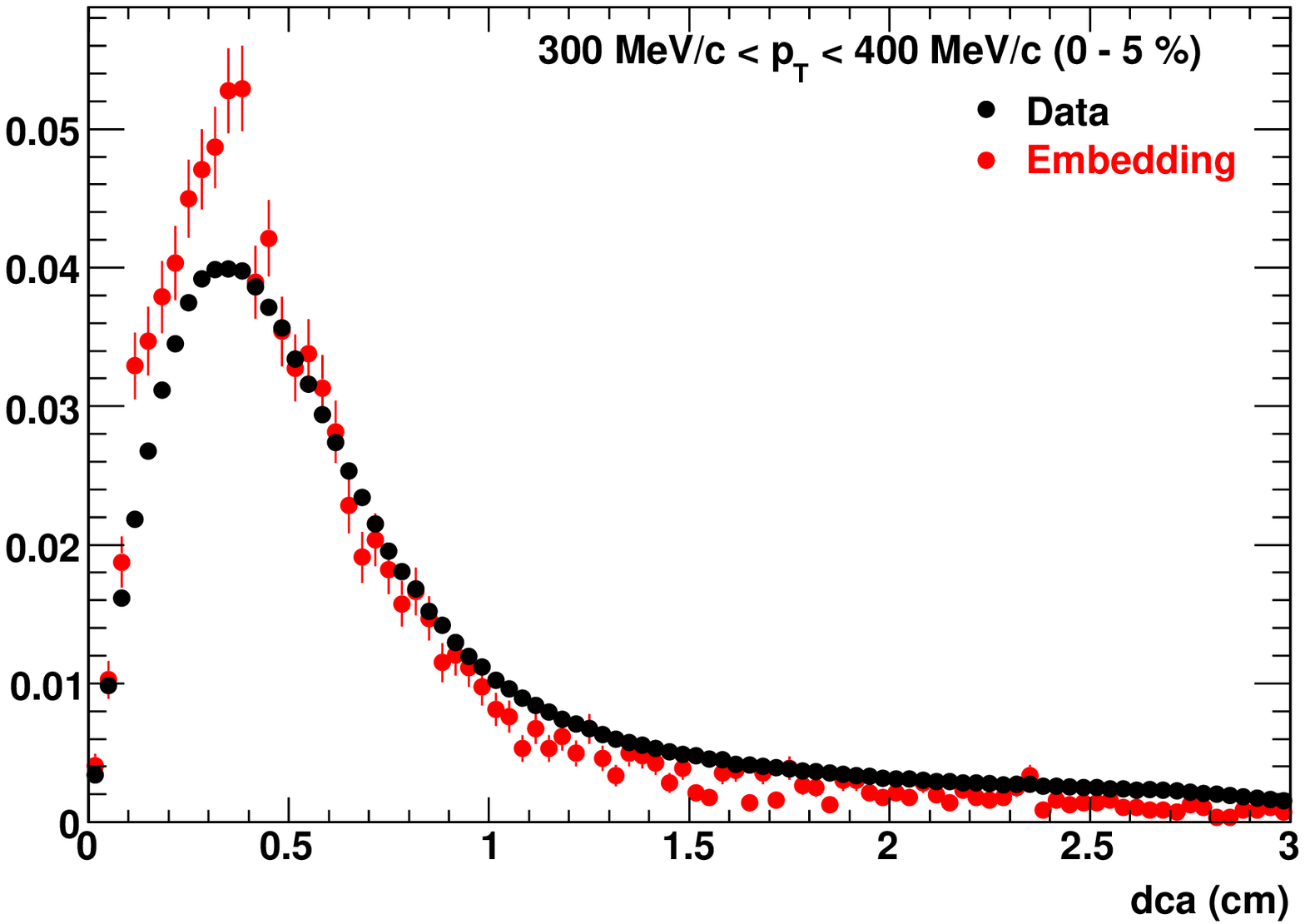}}	
\resizebox{.45\textwidth}{!}{
\includegraphics{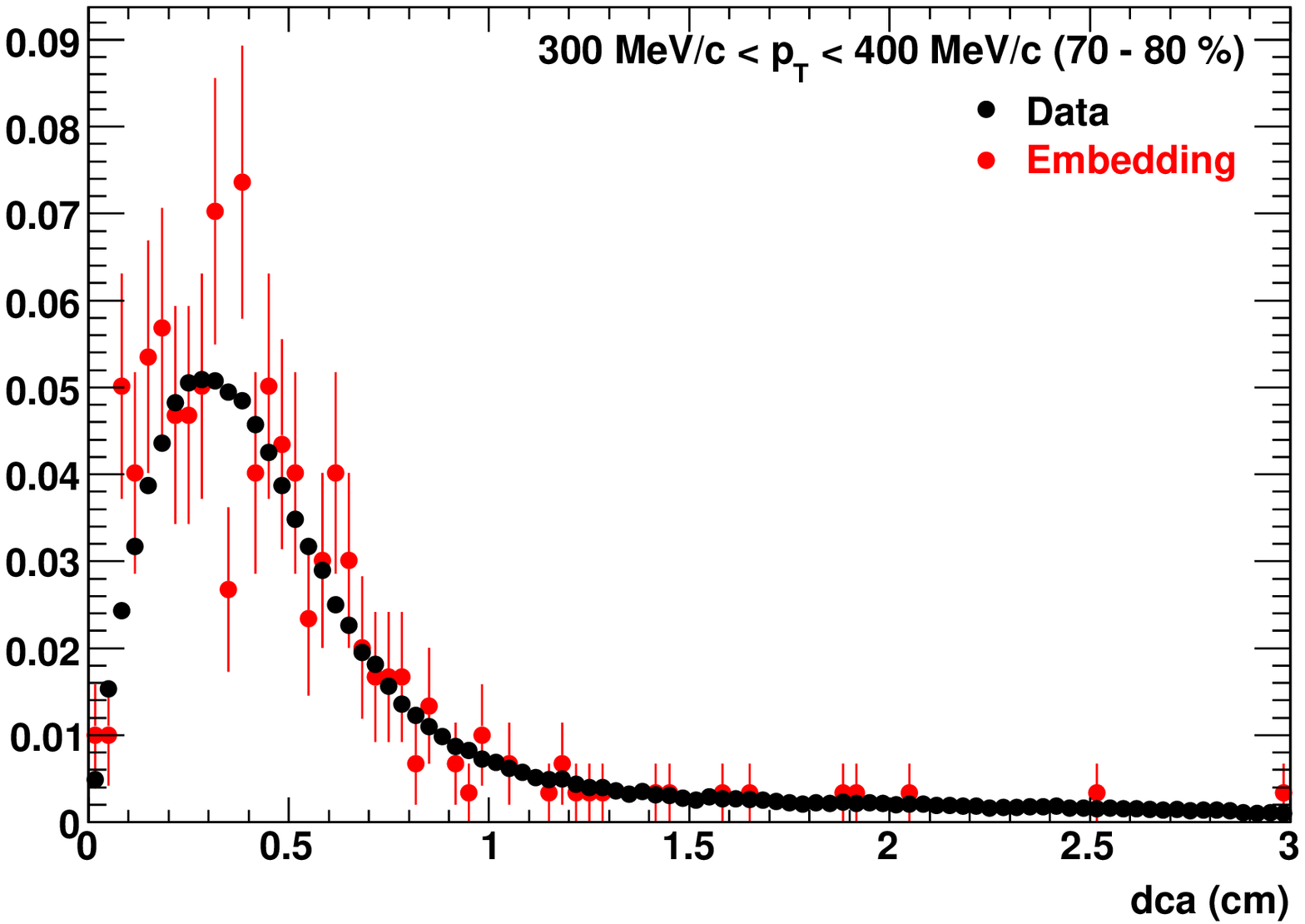}}
		   \caption{Comparison of $\pi^{-}$ $dca$ extracted from data and embedding for 62.4 GeV Au-Au collisions.}
		   \label{fig:auau62_embeddatacomppim1}
		   \end{center}
\end{figure}
\begin{figure}[!h]
\begin{center}	
\resizebox{.45\textwidth}{!}{
\includegraphics{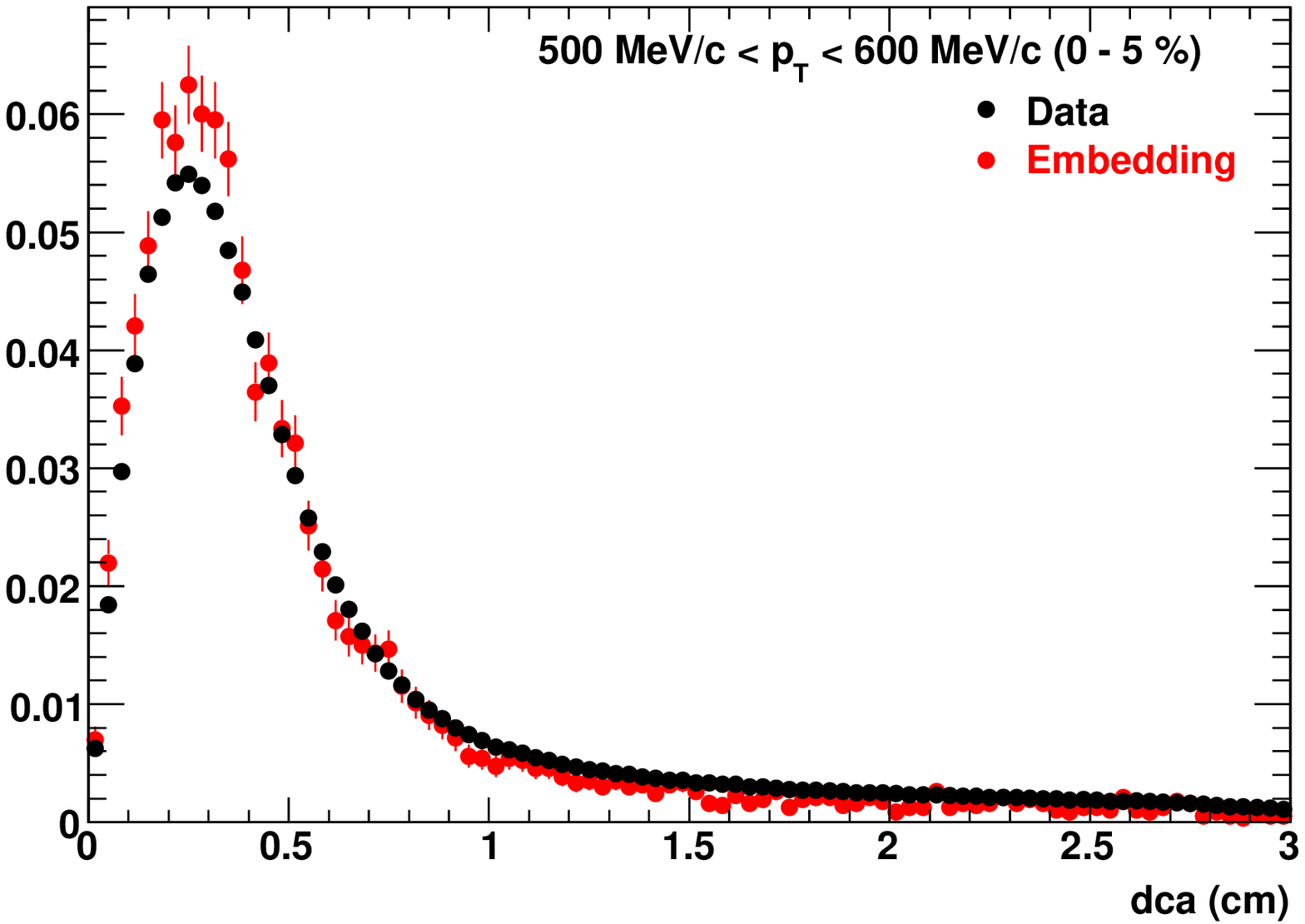}}	
\resizebox{.45\textwidth}{!}{
\includegraphics{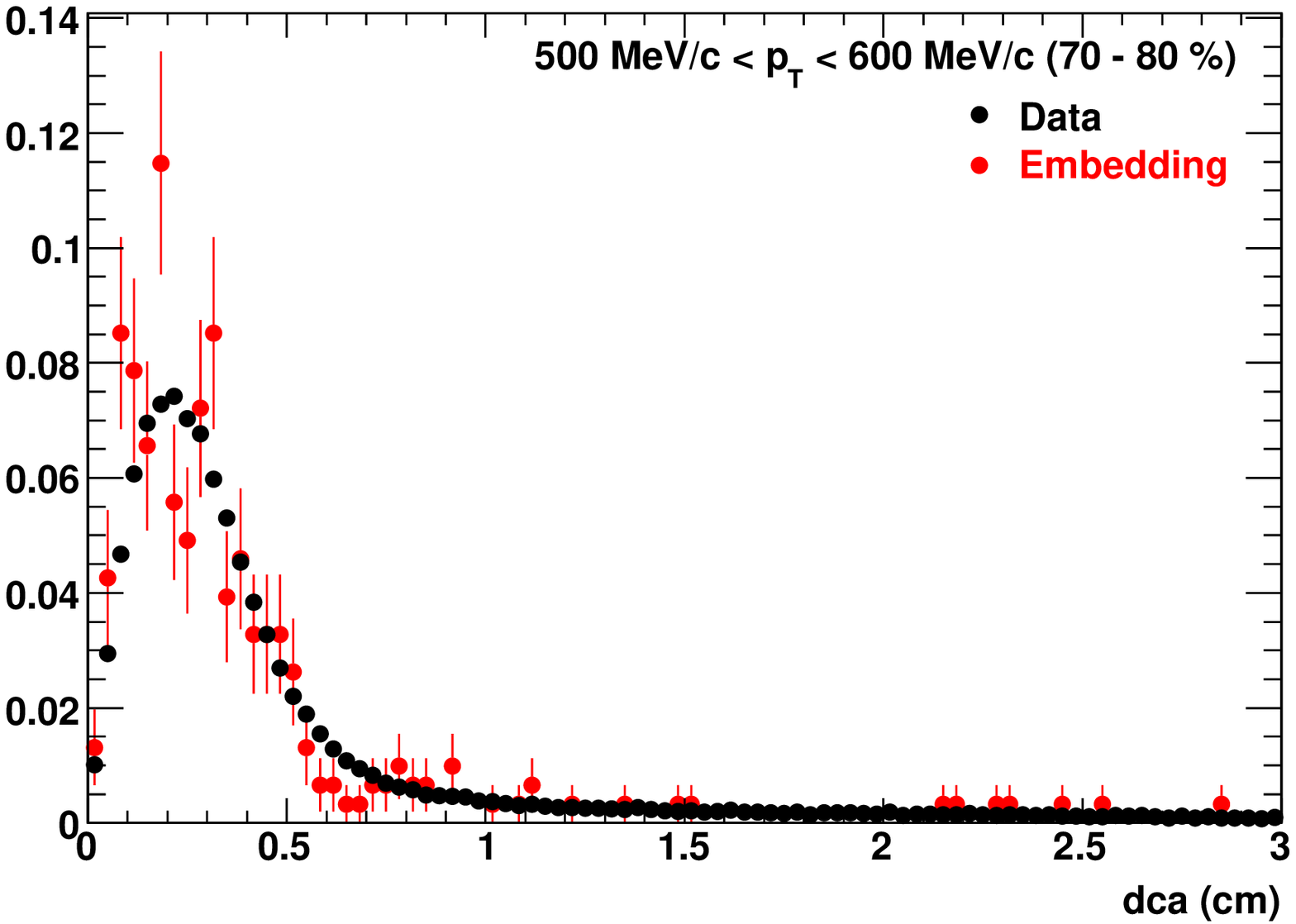}}
		   \caption{Comparison of $\pi^{-}$ $dca$ extracted from data and embedding for 62.4 GeV Au-Au collisions.}
		   \label{fig:auau62_embeddatacomppim2}
		   \end{center}
\end{figure}
\begin{figure}[!h]
\begin{center}	
\resizebox{.45\textwidth}{!}{
\includegraphics{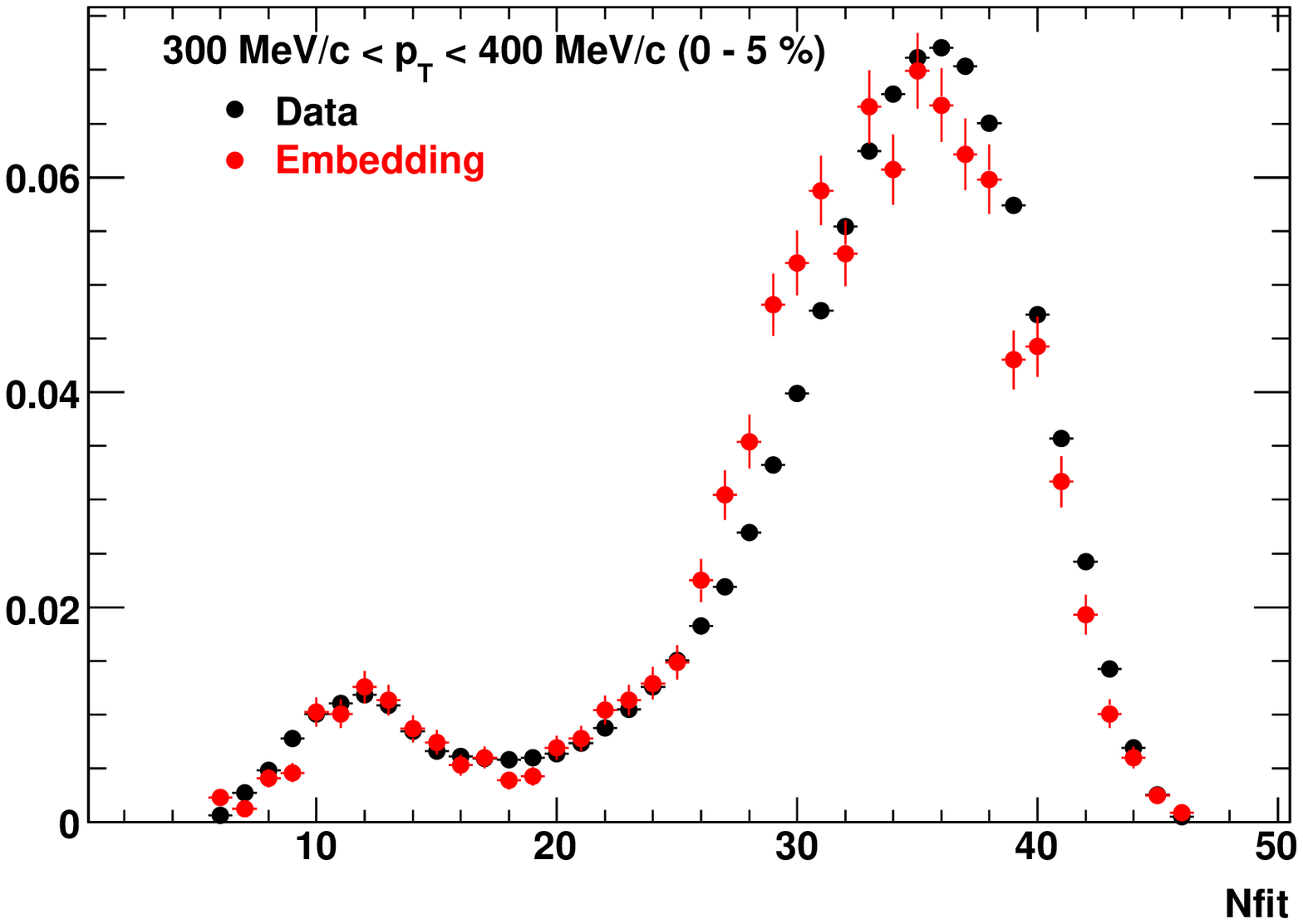}}	
\resizebox{.45\textwidth}{!}{
\includegraphics{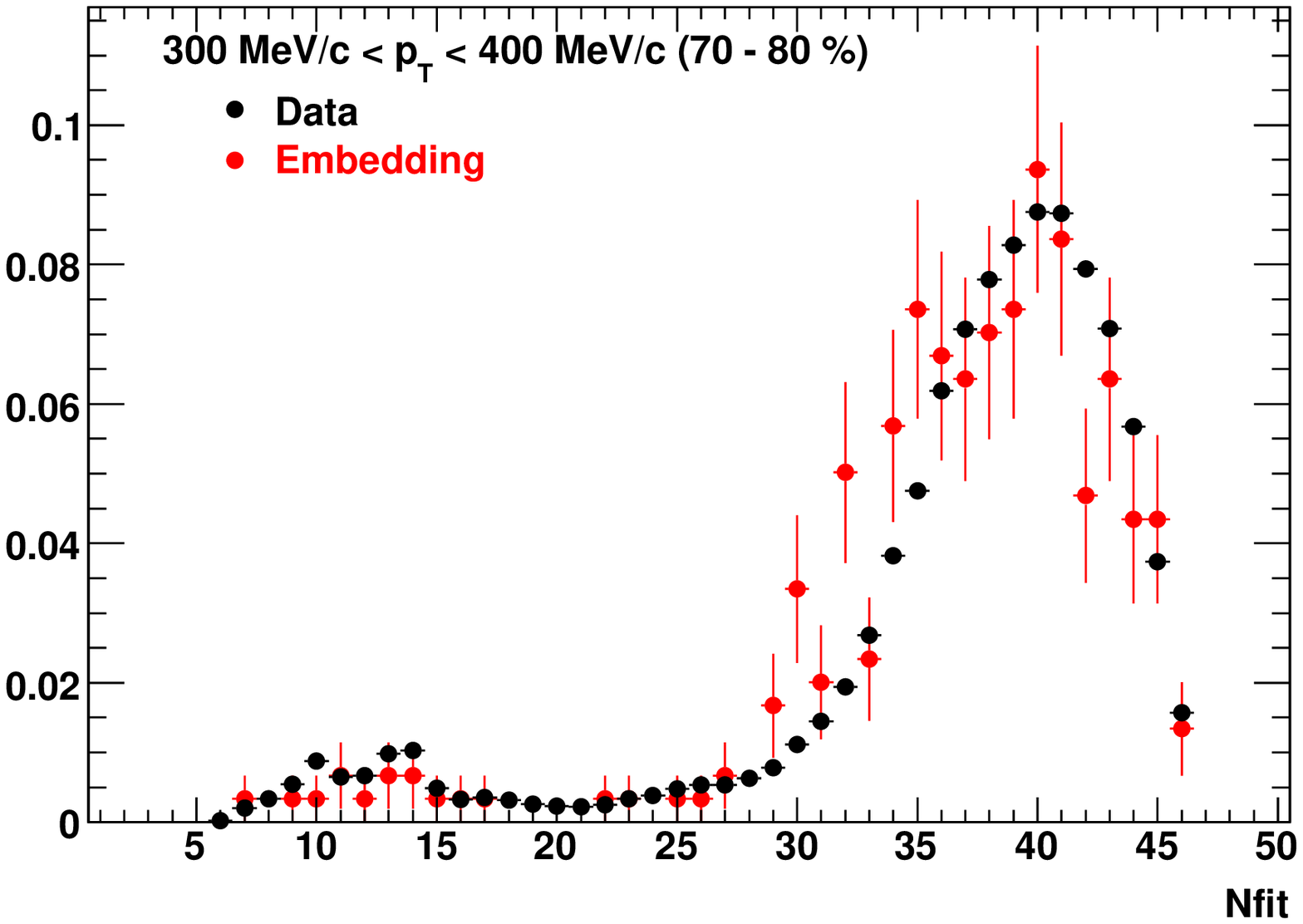}}
		   \caption{Comparison of $\pi^{-}$ fit points extracted from data and embedding for 62.4 GeV Au-Au collisions.}
		   \label{fig:auau62_embeddatacomppim3}
		   \end{center}
\end{figure}
\begin{figure}[!h]
\begin{center}	
\resizebox{.45\textwidth}{!}{
\includegraphics{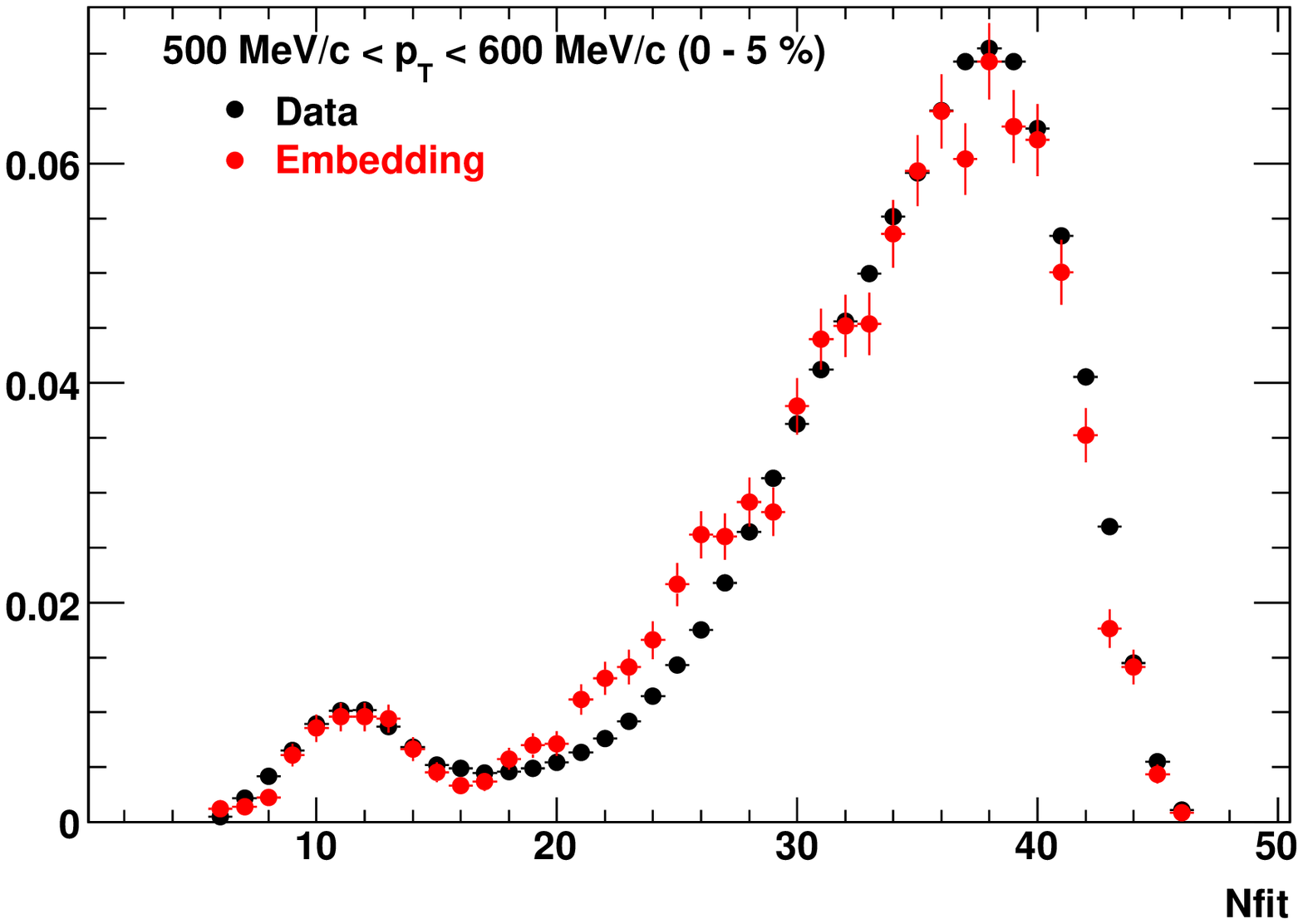}}	
\resizebox{.45\textwidth}{!}{
\includegraphics{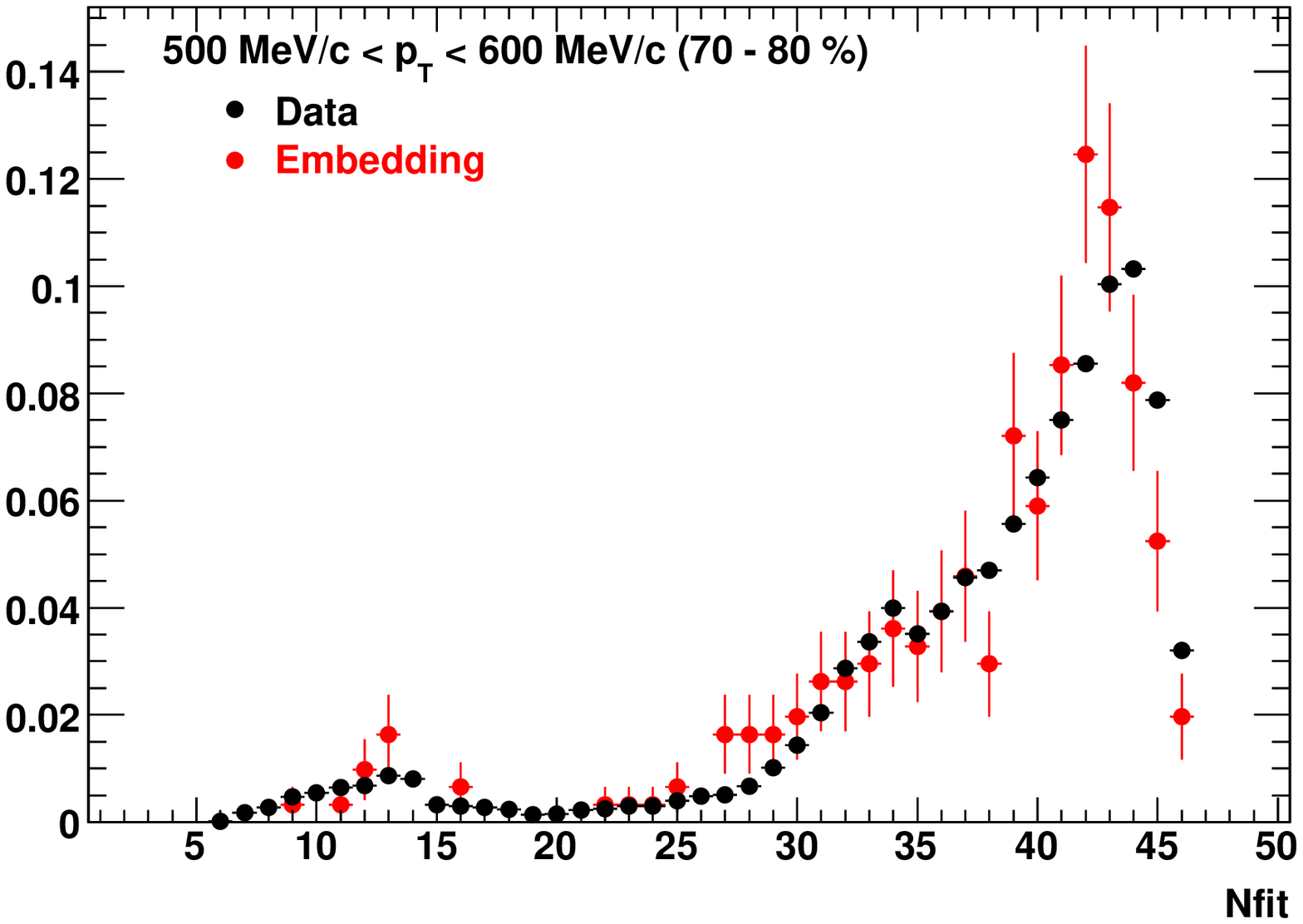}}
		   \caption{Comparison of $\pi^{-}$ fit points extracted from data and embedding for 62.4 GeV Au-Au collisions.}
		   \label{fig:auau62_embeddatacomppim4}
		   \end{center}
\end{figure}

In the presented embedding real data comparisons are repeated for each particle species, collision types and multiplicity/centrality. As one can see in the comparison plots embedding can successfully reproduce real data within 3$\sigma$ particle selection. Figure~\ref{fig:pp_embeddatacomppim1} shows the $dca_{\pi^{-}}$ distribution and Fig~\ref{fig:pp_embeddatacomppim2} shows the Nfit$_{\pi^{-}}$ distribution in 200 GeV pp collisions. Figure~\ref{fig:pp_embeddatacompkm2} and Fig.~\ref{fig:pp_embeddatacompkm2} show the same distribution for negatively charged kaons. Antiproton distributions are shown in Fig.~\ref{fig:pp_embeddatacomppbar1} and Fig.~\ref{fig:pp_embeddatacomppbar2}. Complete set of the $dca$ and $N_{fit}$ plots can be found in Appendix~\ref{app:dataembed}.

These distributions are also plotted for central (0-5\%) and peripheral (70-80\%) 62.4 GeV Au-Au collisions. The $dca_{\pi^{-}}$ is shown in  Fig.~\ref{fig:auau62_embeddatacomppim1} and Fig.~\ref{fig:auau62_embeddatacomppim2}. The $Nfit_{\pi^{-}}$ is shown in Fig.~\ref{fig:auau62_embeddatacomppim3} and Fig.~\ref{fig:auau62_embeddatacomppim4}.

$Dca$ extracted from real data shows wider a distribution compared to embedding, especially at low transverse momentum. This is due to secondary contaminations, especially at low momentum. The secondary contaminations is most pronounced in the real proton $dca$ distribution at low momentum. 

The number of fit points cut is important to avoid merging and splitting tracks in charged multiplicity (number of fit points $\geq$ 15) and spectra (number of fit points $\geq$ 25) measurements. For each colliding set the number of fit points distributions extracted from embedding and real data agree well for number of fit points 10 and higher. 

The overall agreement of the embedding and real data ensures that corrections extracted from embedding reflect realistic calculations.

\subsection{Corrections} 

Raw spectra are corrected for detector acceptance, tracking inefficiency, hadronic interactions and resonance particle decays.
The following subsections provide detailed overviews of these corrections.

Since detector parameters (gas pressure in the TPC, temperature) can change over
the run, a minimum uncertainty ($\sim$ 5\%) is assigned to the obtained correction factors. 
Errors on efficiencies are binomial and calculated as: 
\begin{equation}
Err(p)=\sqrt{\frac{p(1-p)}{N}}
\end{equation}
where $p$ is the efficiency in a given bin and $N$ is the number of entries in the bin.

\subsection{Energy loss correction}

Low momentum particles lose a significant amount of energy traveling 
through the detector material. The track reconstruction algorithm 
takes into account the Coulomb scattering and the energy loss, but 
assumes $pion$ $mass$ for each particle. Therefore, the reconstructed momentum for heavier particles (in our case: kaons and protons/antiprotons) is biased, lower than the original momentum. 

\begin{figure}[thbp]
	\centering
		\includegraphics[width=0.9\textwidth]{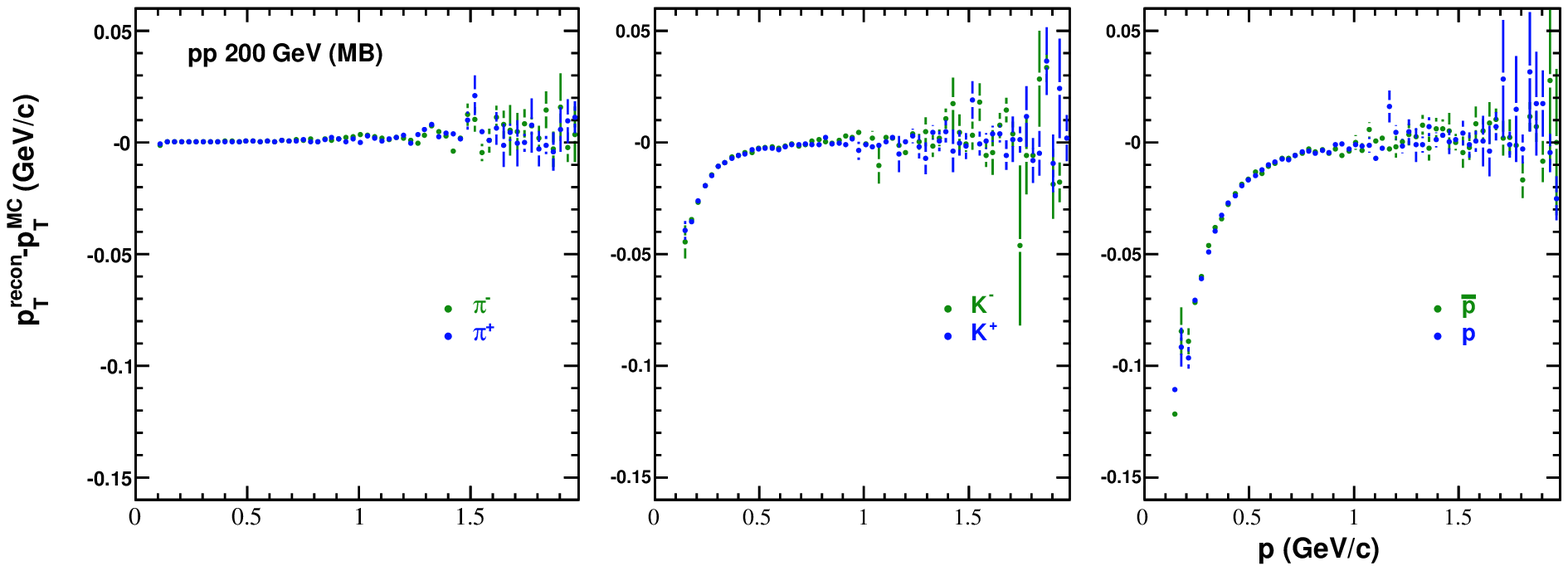}
		\includegraphics[width=0.9\textwidth]{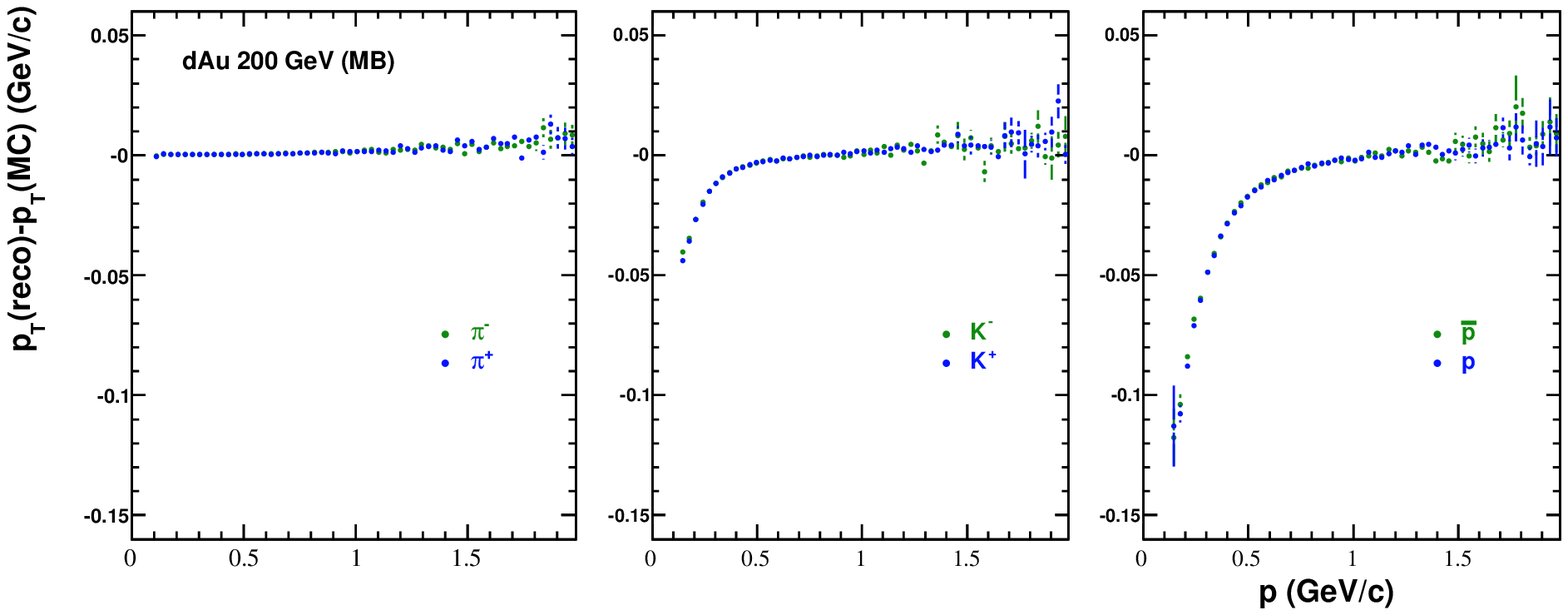}
		\includegraphics[width=0.9\textwidth]{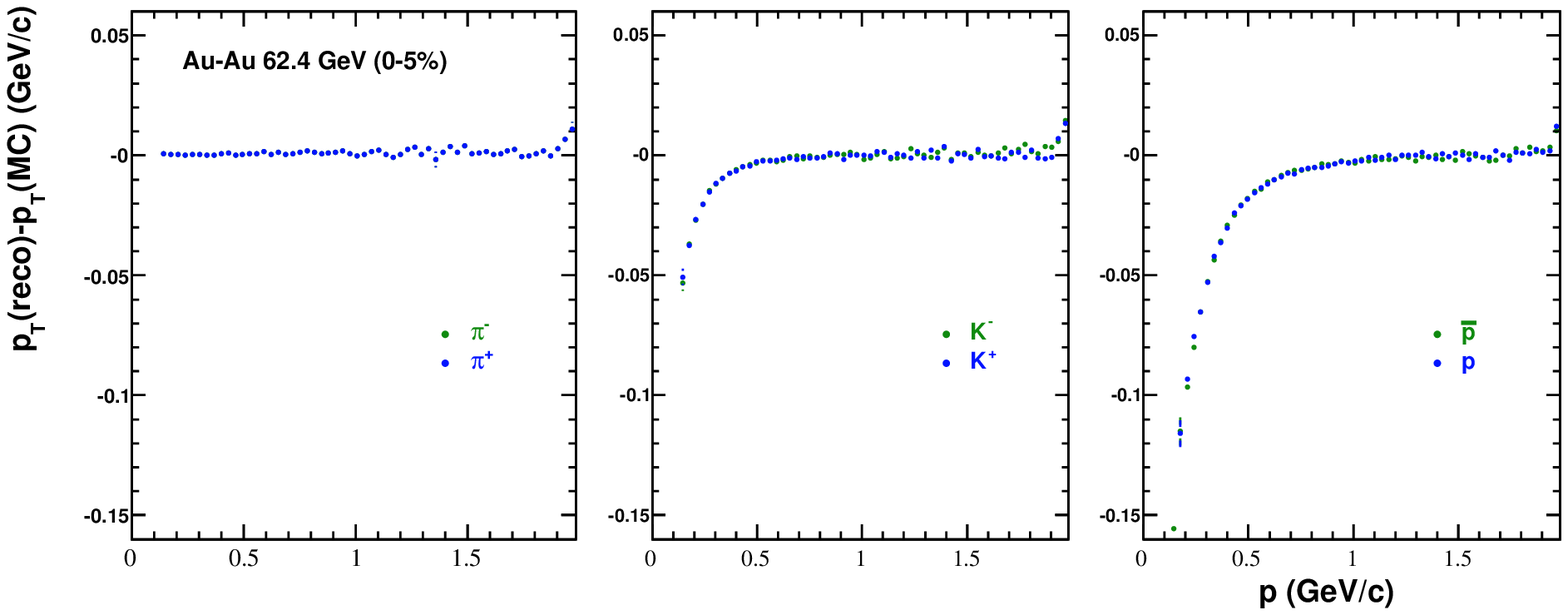}
		\includegraphics[width=0.9\textwidth]{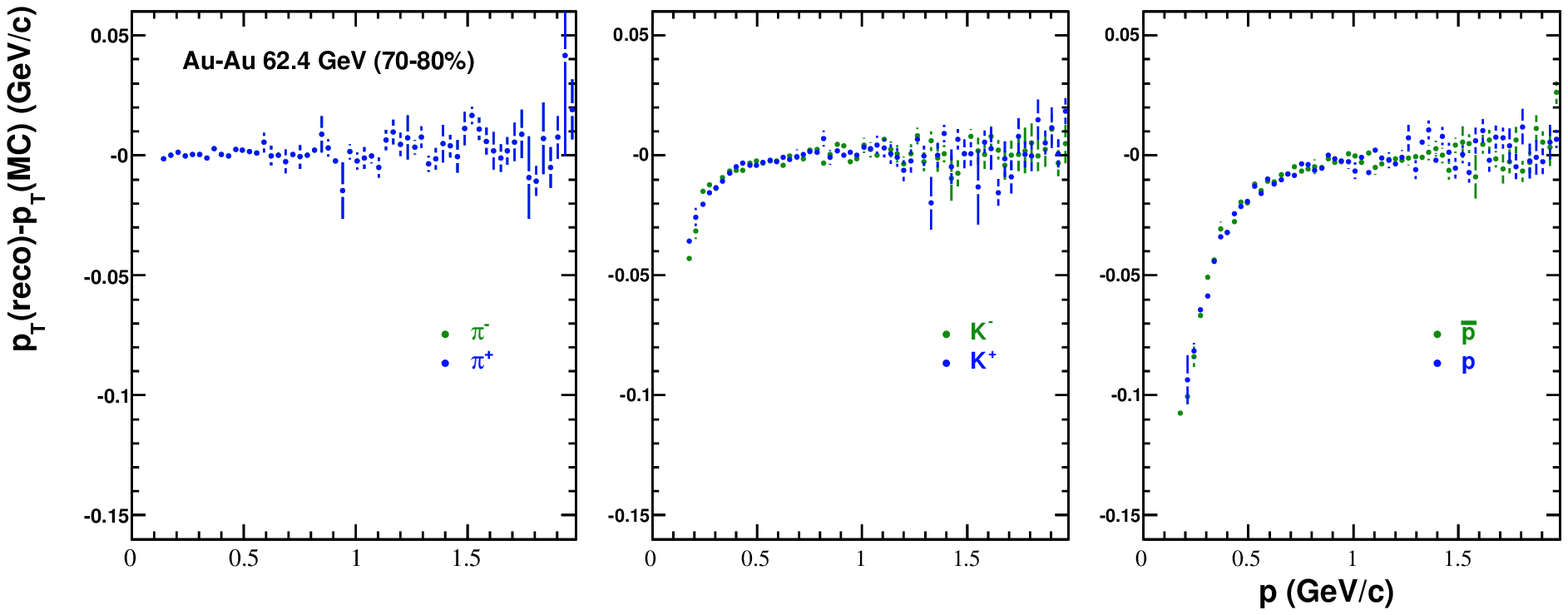}			
\caption{Energy loss correction for $\pi^{\pm}$, $K^{\pm}$, p and $\overline{p}$ 
	as a function of momentum in 200 GeV pp collisions (first row), 200 GeV dAu collisions (second row), central (0-5\%) (third row) and peripheral (70-80\%) (fourth row) 62.4 GeV Au-Au collisions.}
	\label{fig:energyloss}
\end{figure}

The correction is obtained from embedding by comparing the input MC and the reconstructed momentum: p$_{\normalfont{reconstructed}}$-p$_{\normalfont{MC}}$ as a function of p$_{\normalfont{reconstructed}}$ as shown in Fig.~\ref{fig:energyloss}.

Increasing bias can be observed with increasing particle mass at low momentum. 
Furthermore, Fig.~\ref{fig:energyloss} also shows the extracted correction for particles: $\pi^{\pm}$, $K^{\pm}$, $\overline{p}$ and p for 200 GeV pp, 200 GeV dAu and 62.4 GeV Au-Au collisions. At low transverse momenta the difference for protons is $\sim$ 100 - 120 MeV/c and decreases to $<$ 10 MeV at $p_{T}$ = 1 GeV/c. This limits our low $p_{T}$ cut off for protons/antiprotons.

The pion transverse momentum difference is flat through the measured $p_{T}$ range and the correction is smaller than 0.3$\%$ at any $p_{T}$. This is because $\pi$ energy loss is corrected in reconstruction and the remaining small effect is negligible. However, kaons and protons/antiprotons show larger discrepancy between the MC and the reconstructed transverse momentum at low momentum and the deviation from MC input is the same for particles and antiparticles.

Energy loss for a specific particle type is independent of collision type (pp - dAu - peripheral Au-Au) as expected.
The slight difference between collision types is due to the changing detector setup between different runs (pp: Run II., dAu: Run III., and Au -Au: Run IV.) and only controlled by the amount of absorbing material in the detector itself. Between the runs, as already mentioned, the SVT supporting frame has been modified and a new silicon layer has been installed: SSD.  The GEANT description of the SVT and SSD becomes more refined through  the runs and the corresponding simulation packages, but it also introduces uncertainty on the calculated efficiencies and corrections which is included in the overall 5\% uncertainty. 

The energy loss correction for kaons and protons/antiprotons can be parameterized as:
\begin{equation} 
{p_{T}}_{corrected}={p_{T}}_{measured}+c_{1}+c_{2}\left(1+\frac{m^{2}}{p^{2}}\right)^{c_{3}}
\end{equation}
where $m$ is the mass of the particle and $c_{i}$ (i=1, 2, 3) are the parameters extracted from the fit to the energy loss curve.
The change in the fit parameters is negligible between the collision systems and centralities, therefore the characteristic numbers can be quoted:
$c_1 = 0.006 (0.013)$~GeV/$c$, $c_2 = -0.0038 (-0.0081)$, and $c_3 = 1.10 (0.03)$ for kaons (protons/antiprotons), respectively.

The energy loss correction seems to show a small dependence on centrality in dAu and in Au-Au collisions. Weak dependence can be observed with varying quality cuts. The change in the energy loss correction due to different rapidity selection in $|y| < 0.5$ is negligible. 

The energy loss correction is applied off-line to the raw data upon selecting tracks from the dE/dx distribution to be used for spectra analysis. 
Since individual particle identification is not possible, the energy loss correction of the particular specie of interest is applied to all tracks, eg. when analysing kaons each track (even from the pion and proton/antiproton bands) are corrected for kaon energy loss.
This method does not introduce artificial bias on the extracted raw particle yield, since raw yield is only extracted for a particular particle at one time and it only changes the scale in the direction of the transverse momentum, but leaves the magnitude of the dE/dx unchanged.

\subsection{Vertex correction}

In pp and dAu collisions the average number of tracks per event
is small compared to Au-Au collisions, and the event rate is high. 
Multiple bunch crossing (pile-up) within the same read-out window (complete drift of the triggered event to the read out electronics) is a significant problem of drift detectors, which leads to high background rate in the triggered events.
In higher multiplicity collisions a larger number of tracks defines the vertex more precisely, but in low multiplicity
collisions the vertex finder (ppLMV: proton - proton Low Multiplicity Vertex Finder implemented in STAR) is sensitive to pile-up events. Pile-up can shift the position of the reconstructed primary vertex. In very low multiplicity events ppLMV can fail to find the vertex.
 
To correct for these inefficiencies, the possible problems are identified. In a second step a physical quantity is identified which is insensitive to the pile-up rate and accessible from data (since the pile-up level is not known from data). In the third step data is corrected using the quantity mentioned above.
\begin{figure}[!h]
	\centering
		\includegraphics[width=0.45\textwidth]{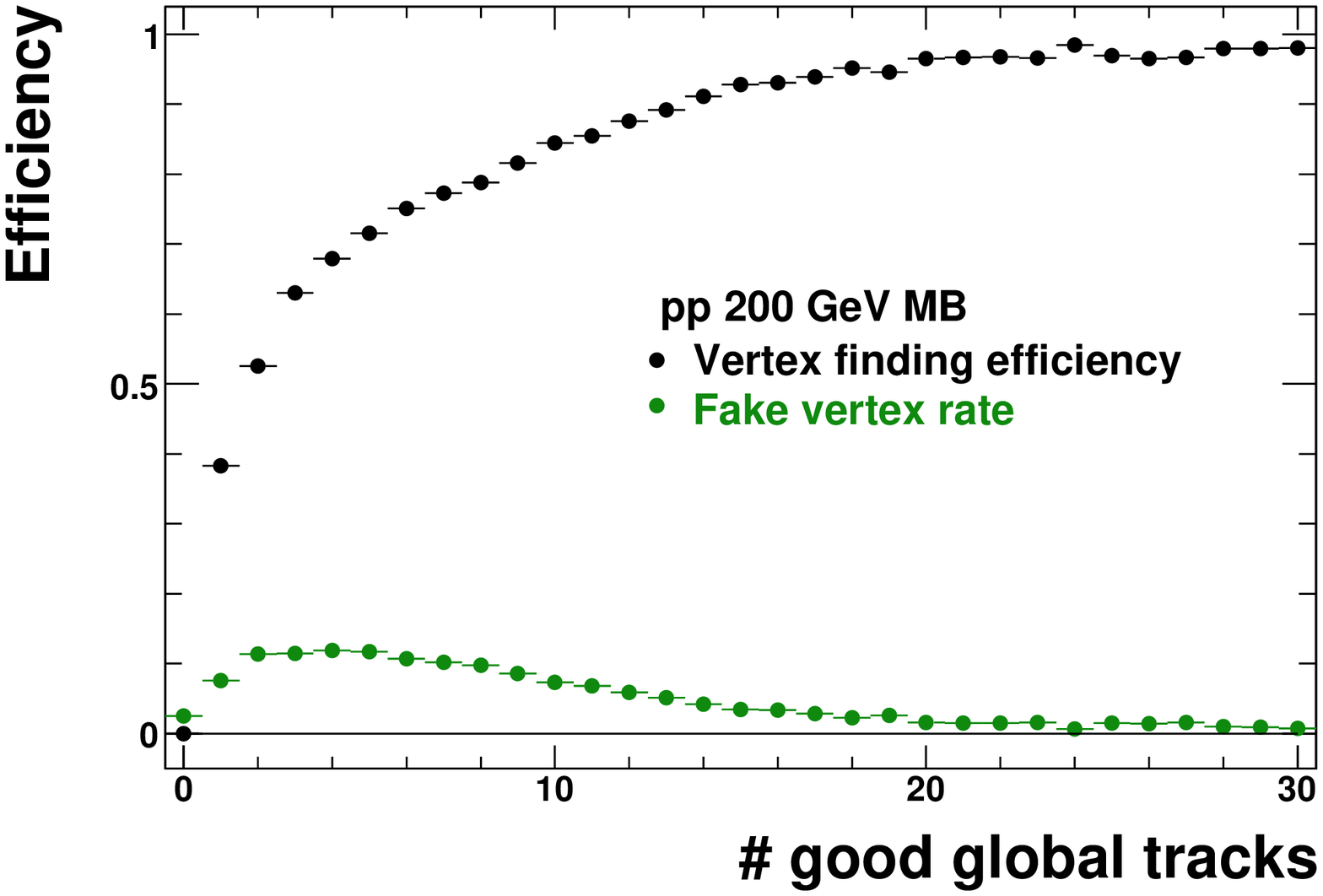}
		\includegraphics[width=0.45\textwidth]{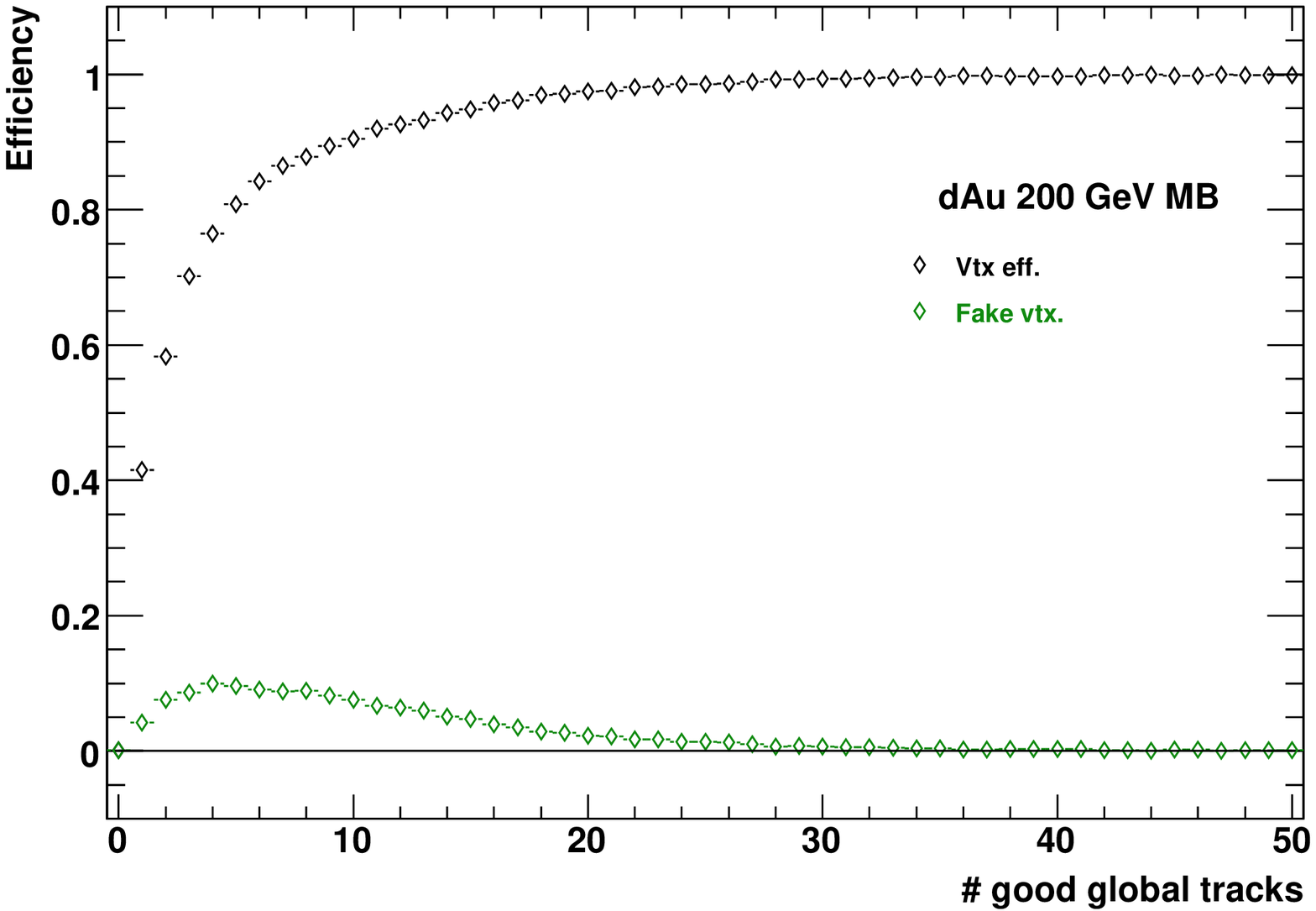}

\caption{Vertex inefficiencies in minimum bias pp and dAu collisions at 200 GeV.}
	\label{fig:vtxineff}
\end{figure}

The ppLMV can fail to find the primary vertex or it can find the vertex at a wrong place.
To correct for these inefficiencies in pp and dAu collisions, MC (HIJING) events were embedded to abort gap events and were 
reconstructed in the full reconstruction chain. (Events triggered and reconstructed at empty bunch crossings are called abort gap events.)
In every MC event there is a well defined primary vertex with well defined coordinates. After reconstructing the embedded events with the MC information in hand the vertex reconstruction efficiency can be studied. 

In data analysis, only those events are taken which satisfy certain quality cuts. The first one is the cut on the primary vertex position. The $x$ and $y$ positions are well defined (and also restricted by the beam pipe), however the $z$ position can vary along the beam direction over a wide range due to the difference in bunch timing of the two collider rings. Good events with $z$ vertex position are selected from $\pm$ 30.0 cm in pp and $\pm$ 50.0 cm in dAu for data analysis. The overall vertex (in)efficiency can be determined as the ratio of the number of reconstructed good events with respect to the number of good MC events satisfying the vertex cut, as shown by Eq.~\ref{eq:vtxineff}.
\begin{equation}
	Efficiency\ =\ \frac{number\ of\ good\ reconstructed\ events}{number\ of\ good\ MC\ events}
	\label{eq:vtxineff}
\end{equation}
This overall correction is applied as a multiplication factor to the extracted raw yield. 
\begin{figure}[!h]
	\centering
		\includegraphics[width=0.45\textwidth]{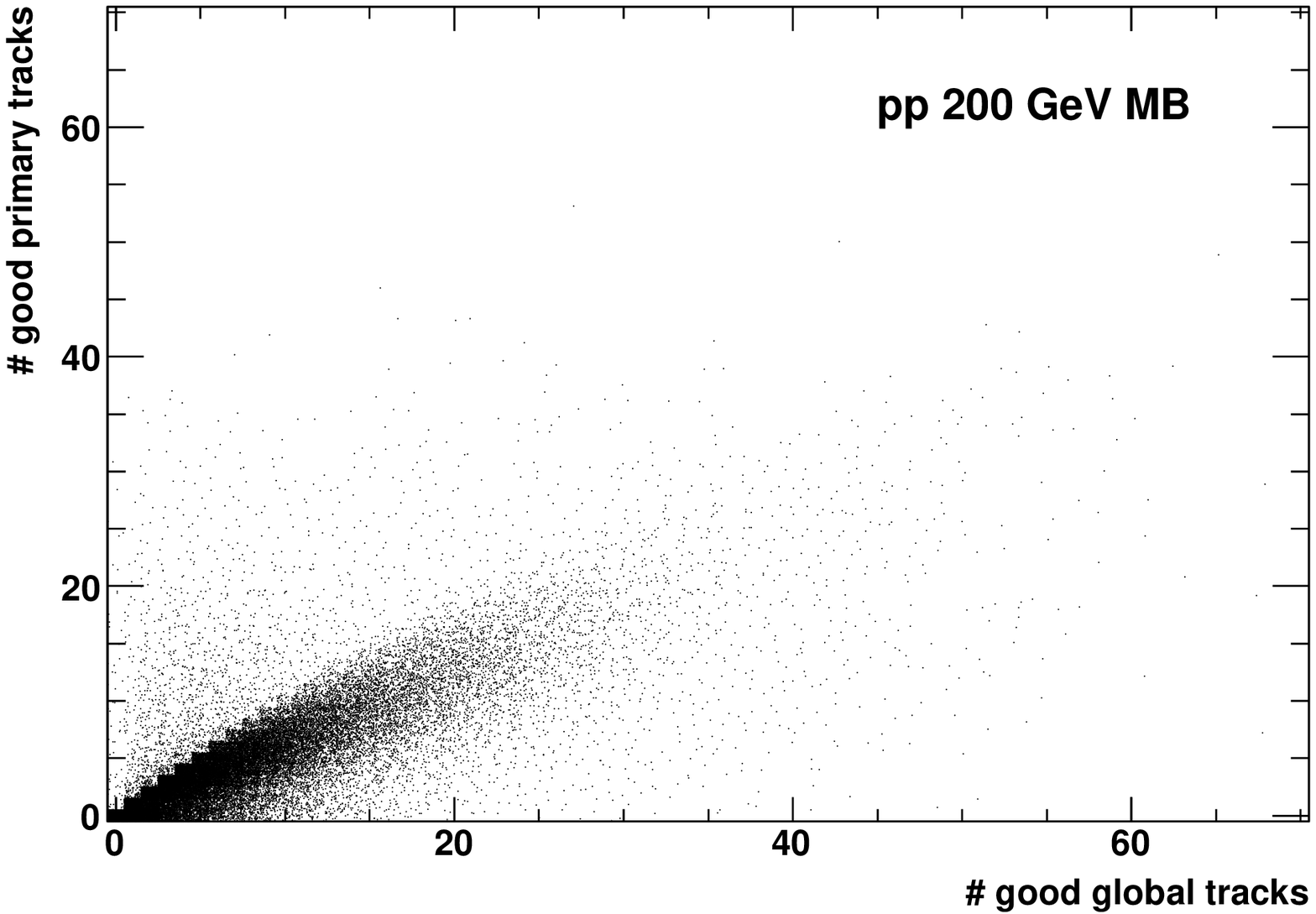}
		\includegraphics[width=0.45\textwidth]{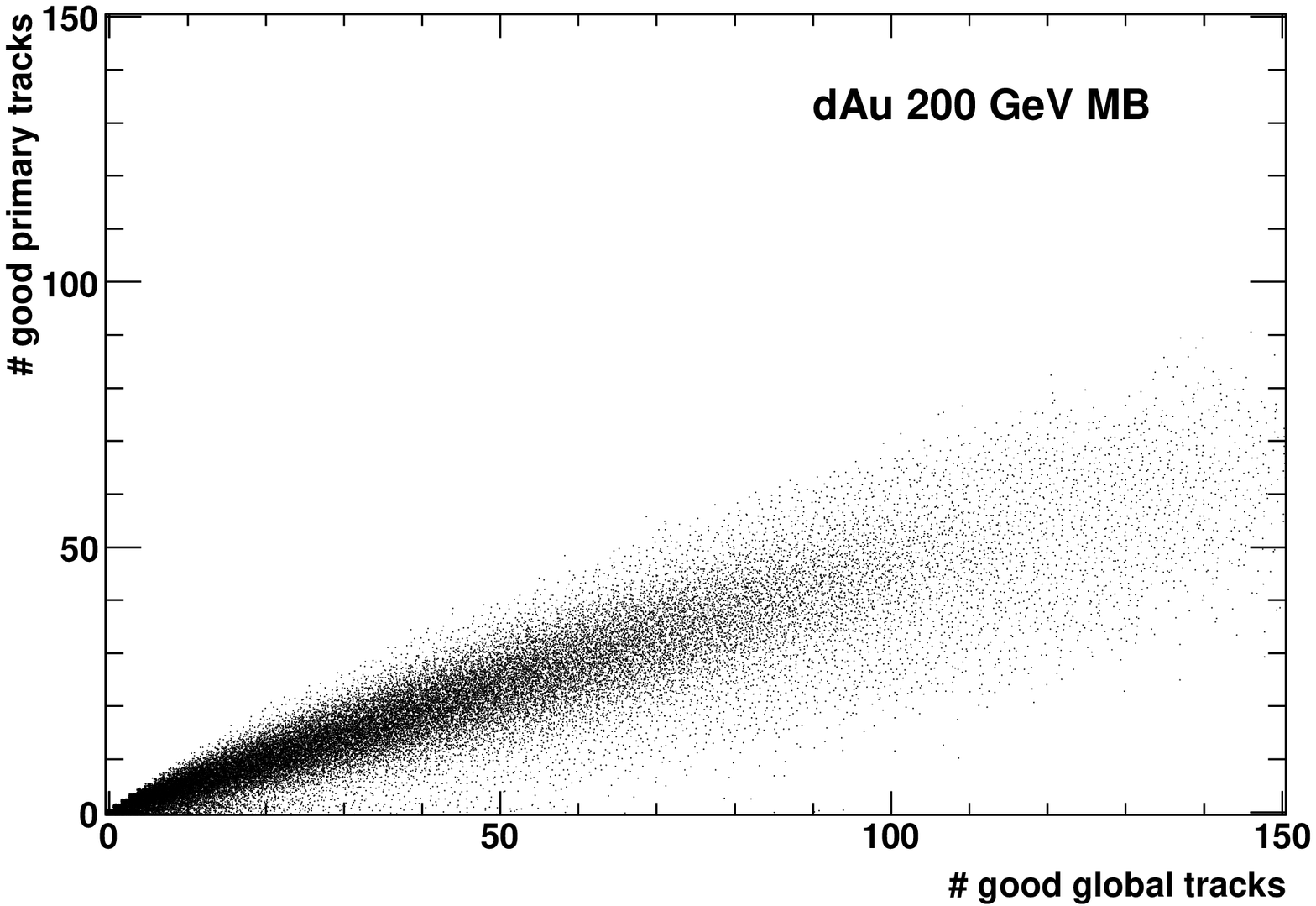}
	
\caption{Map of number of good primary tracks vs. number of good global tracks.}
	\label{fig:vtxmapping}
\end{figure}

To characterize the vertex inefficiency a parameter should be
chosen that can be measured in the data and is not effected 
by the pile-up. In order to study the pile-up effect two simulated
files of pp events are mixed at the raw data level and reconstructed
in the full event reconstruction chain. The first set is considered
the real event and the other set is used as the pile-up background event. 
The pile-up range was varied from 0 - 100 $\%$, where 100$\%$  means
each real event has a pile-up event in it. After mixing, events
were reconstructed and the number of good global and primary tracks
were examined as a function of the pile-up level. (A global or primary track is called good if 
its distance of closest approach is smaller than 3cm and it has at least 15 fit points out of 45 possible).
\begin{figure}[!h]
	\centering
		\includegraphics[width=0.85\textwidth]{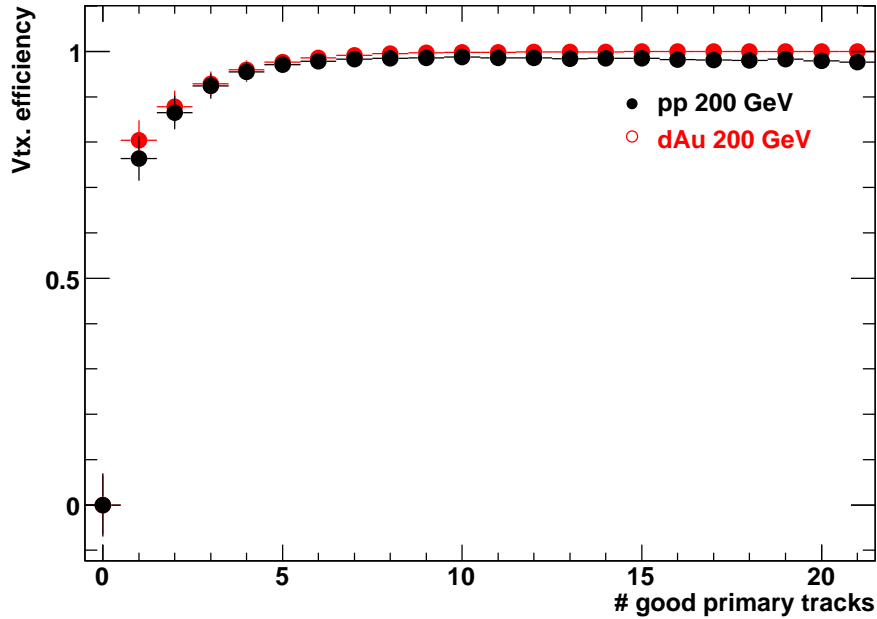}
	\caption{Vertex finding efficiency as a function of good primary tracks in 200 GeV $\bf{pp}$ and $\bf{dAu}$ collisions.}
	\label{fig:ppvtxeff}
\end{figure}
\begin{figure}[!h]
	\centering
		\includegraphics[width=0.85\textwidth]{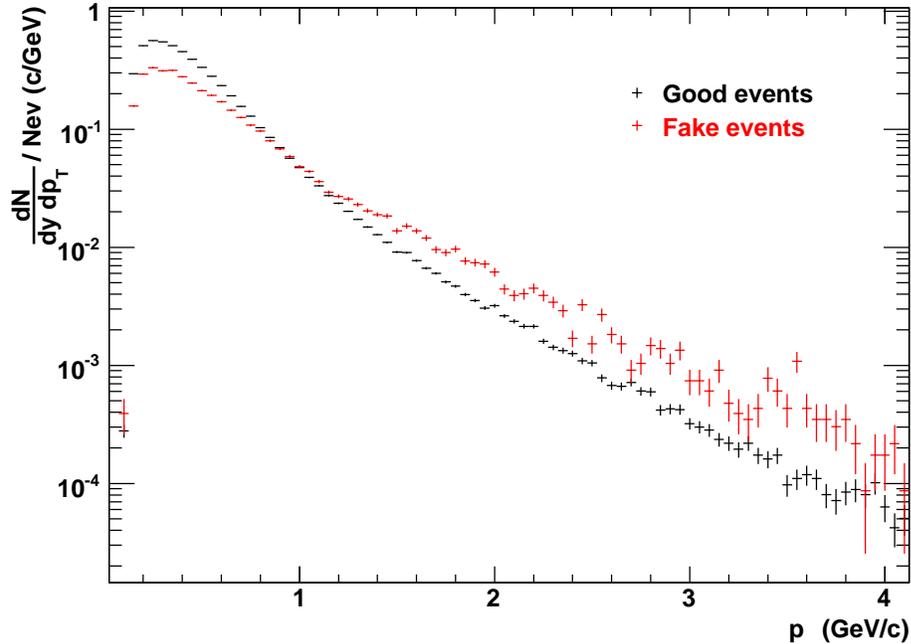}
	\caption{Transverse momentum spectra of good and fake events extracted from embedding in 200 GeV $\bf{pp}$ collisions. }
	\label{fig:ptspectrafake}
\end{figure}

The number of good primary tracks were chosen to characterize the vertex inefficiency
since the number of good primary tracks is independent of the
degree of pile-up. The number of good global tracks increases with increasing
pile-up rate in the event. The vertex inefficiency cannot be described by the
number of good primary tracks directly. Hence, the correction is performed
in two steps. First, the number of lost events the number of fake events are obtained as a function of good global tracks 
(which are required to have at least 15 hit points). 
This is shown in Fig.~\ref{fig:vtxineff}. 
\begin{figure}[!h]
	\centering
		\includegraphics[width=0.85\textwidth]{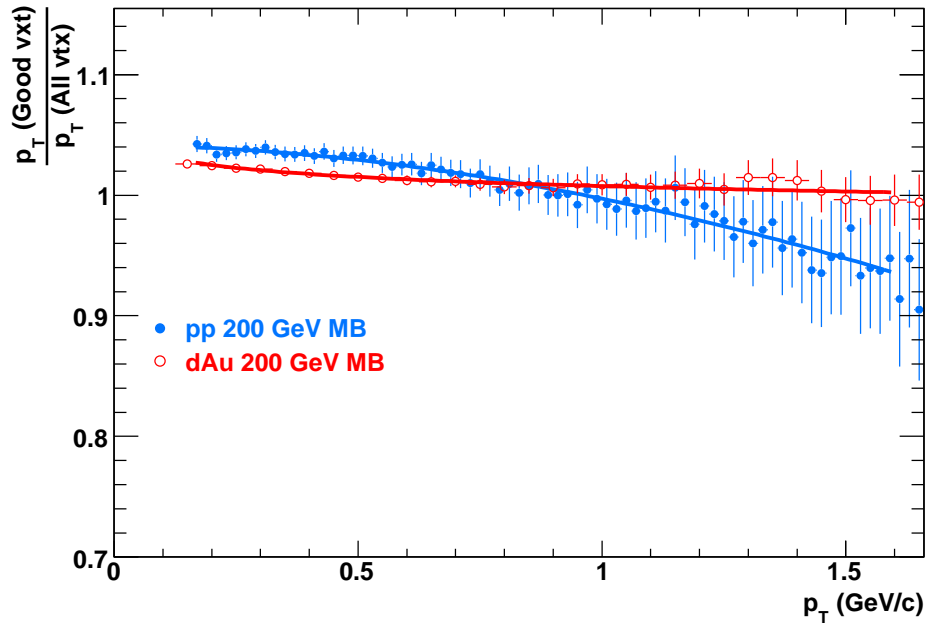}
	\caption{Fake vertex correction as a function of $p_{T}$ in 200 GeV $\bf{pp}$ and $\bf{dAu}$ collisions. }
	\label{fig:FakeVertexCorrection}
\end{figure}
The number of good global track depends on the pile-up as well. 
 In the second step, the efficiency distributions are converted to the function of the number of good primary tracks through 
the mapping of the good primary good global track distribution.
The mapping is shown in Fig.~\ref{fig:vtxmapping}.
For each good primary track bin the lost and fake distributions are convoluted with 
the good global distribution. Finally, the vertex correction is given as a 
function of the number of good primary tracks as shown in Fig.~\ref{fig:ppvtxeff}
and applied in each event to the raw particle spectra. Each event and each track is weighted by
the inverse of the vertex efficiency in pp, minimum bias dAu and peripheral dAu events. 

\subsection{Fake vertex correction}

As we mentioned in the previous section pile-up can affect the vertex reconstruction. One typical problem is the shifted vertex.

In the low multiplicity events the reconstruction software can be biased by the pile-up and the vertex may be reconstructed away from the real vertex. This can be studied via embedding. If the reconstructed vertex in the embedding is farther than 2 cm from the corresponding MC vertex, the reconstructed vertex is labeled as a fake vertex.
\begin{figure}[!h]
	\centering
		\includegraphics[width=0.45\textwidth]{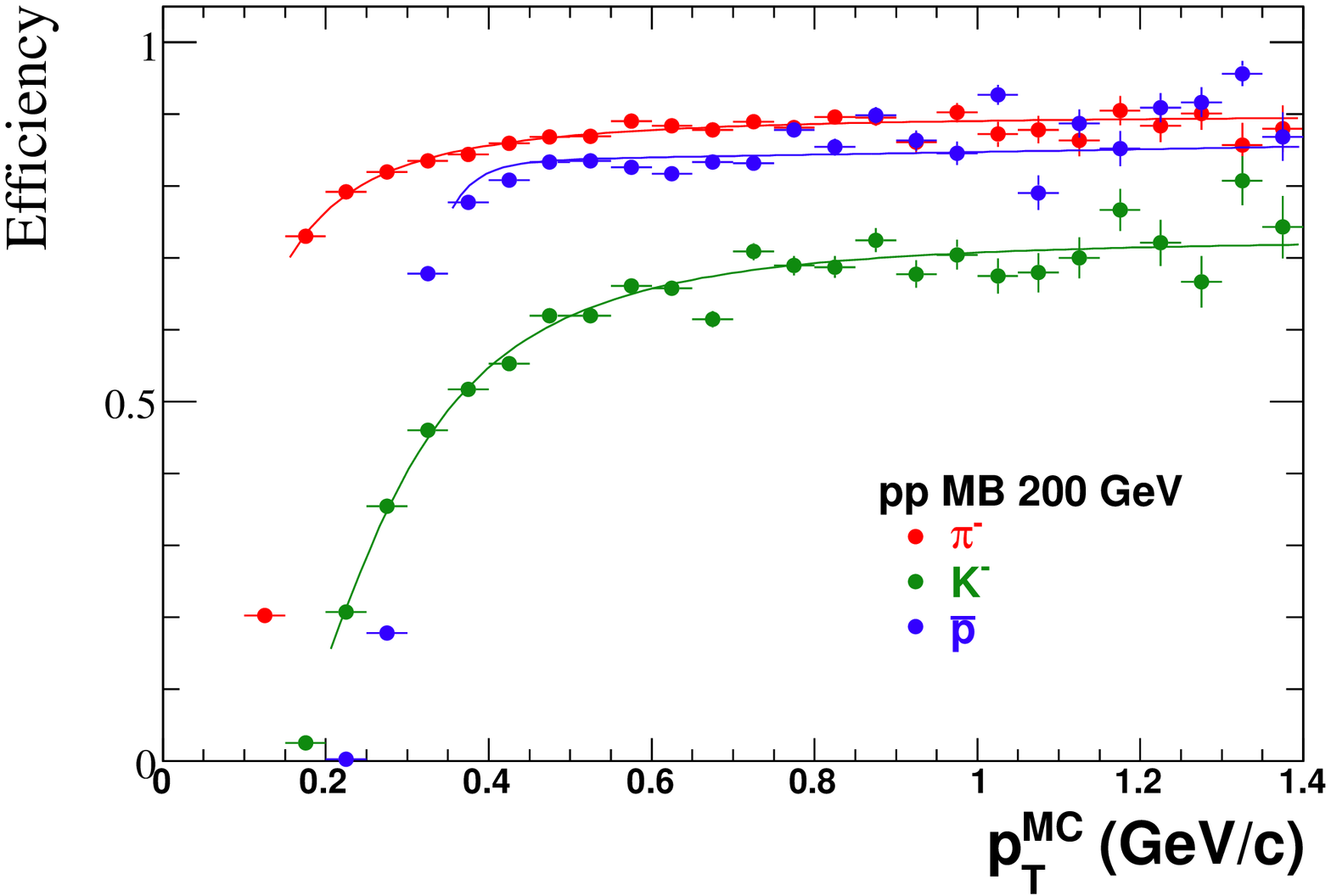}
		\includegraphics[width=0.45\textwidth]{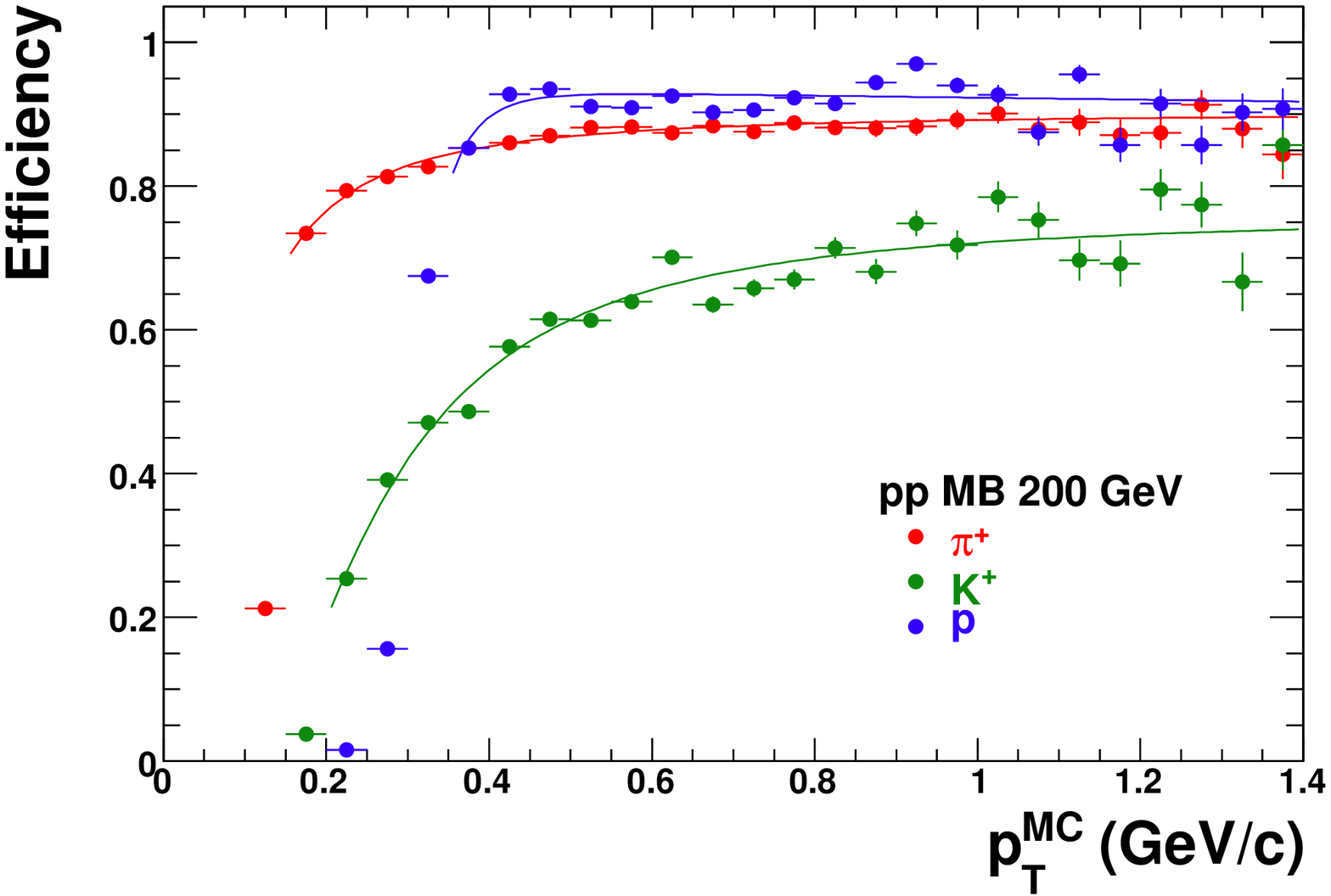}
			\caption{Tracking efficiency of $\pi^{-}$, $K^{-}$, $\overline{p}$ (left panels) and $\pi^{+}$, $K^{+}$ and $p$ (right panels) in 200 GeV minimum bias $\bf{pp}$ collisions as a function of transverse momentum.}
	\label{fig:treffpp}
\end{figure}
\begin{figure}[!h]
	\centering
		\includegraphics[width=0.45\textwidth]{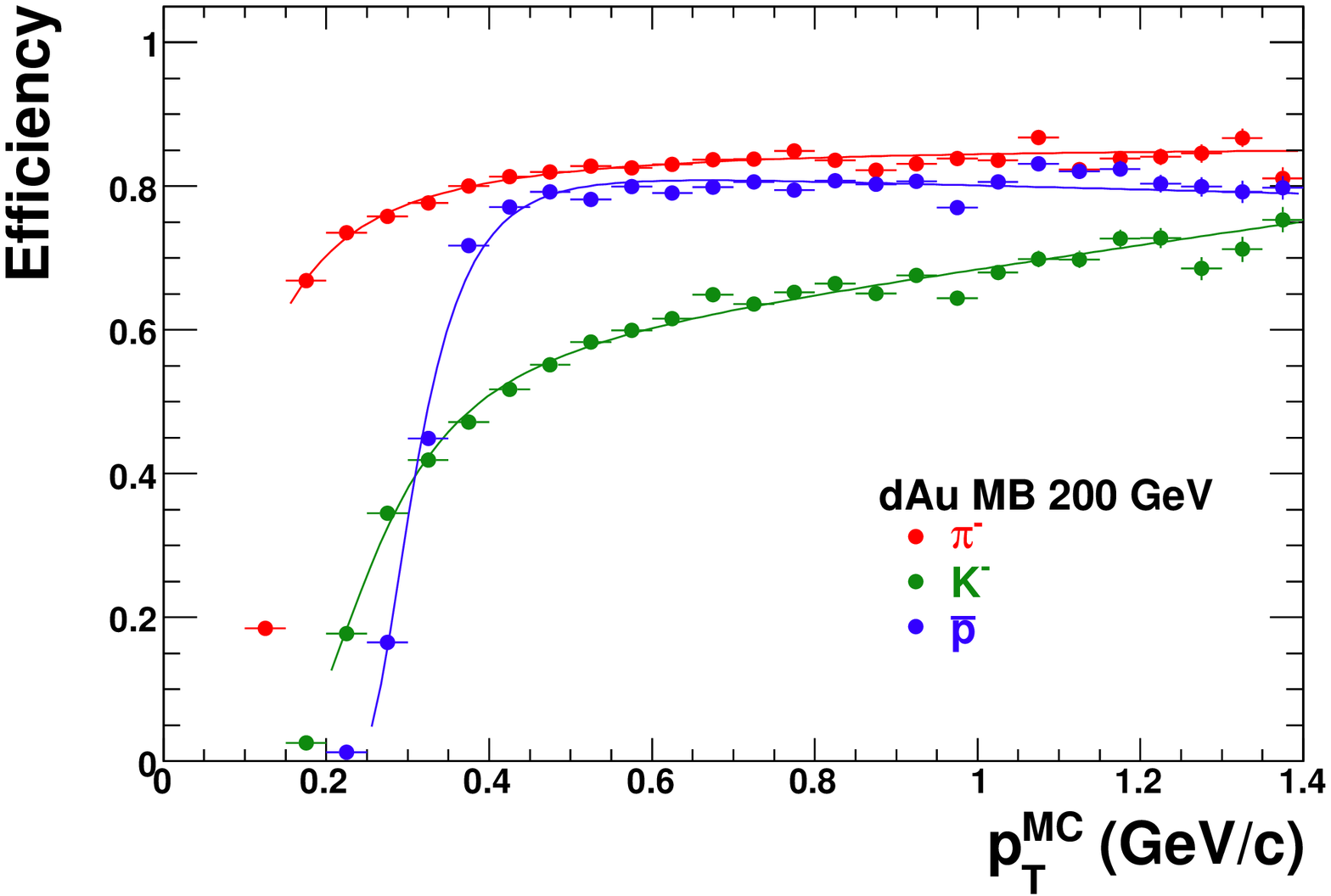}
		\includegraphics[width=0.45\textwidth]{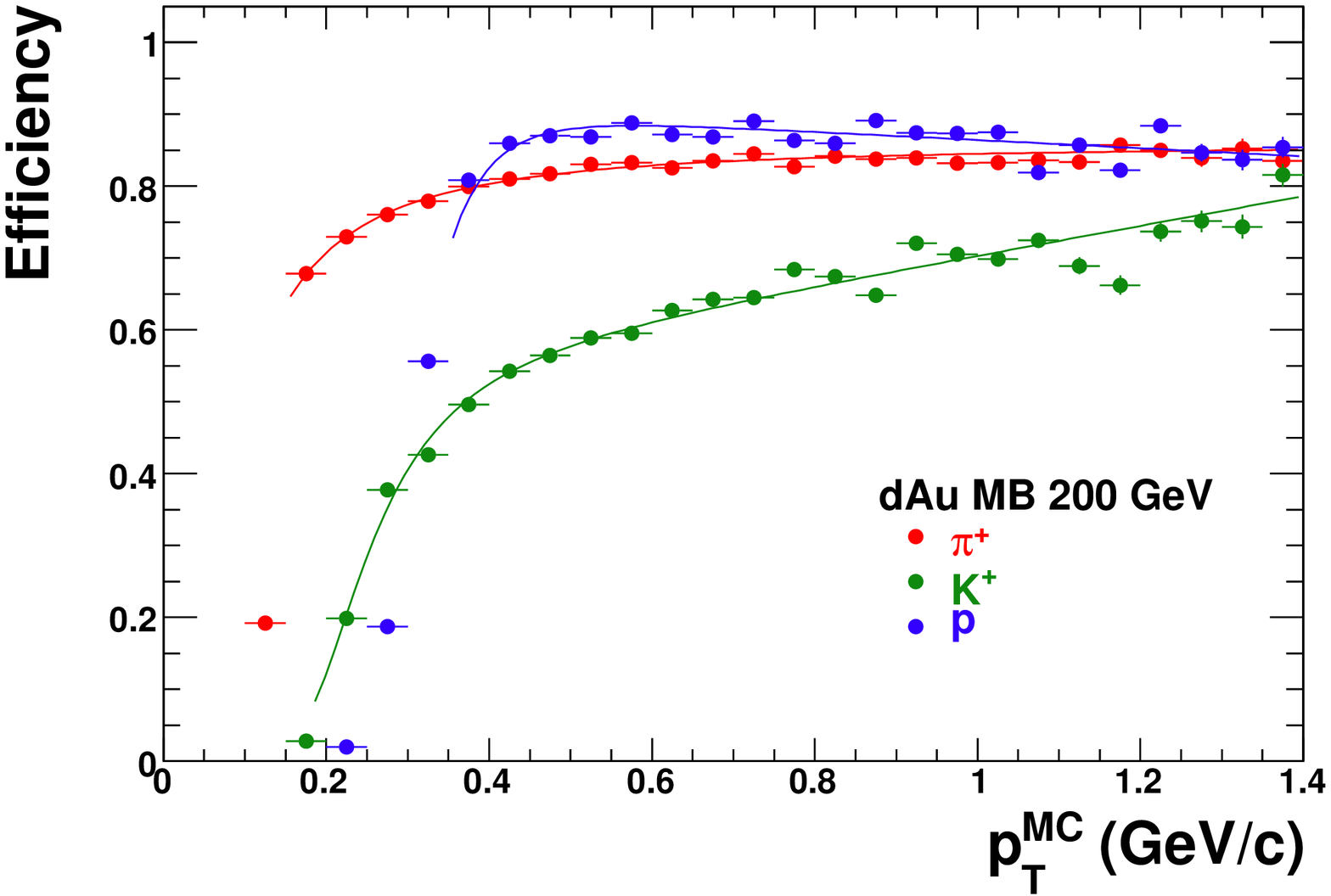}
			\caption{Tracking efficiency of $\pi^{-}$, $K^{-}$, $\overline{p}$ (left panels) and $\pi^{+}$, $K^{+}$ and $p$ (right panels) in 200 GeV $\bf{dAu}$ collisions as a function of transverse momentum.}
	\label{fig:treffdau}
\end{figure}
\begin{figure}[!t]
	\centering
		\includegraphics[width=0.45\textwidth]{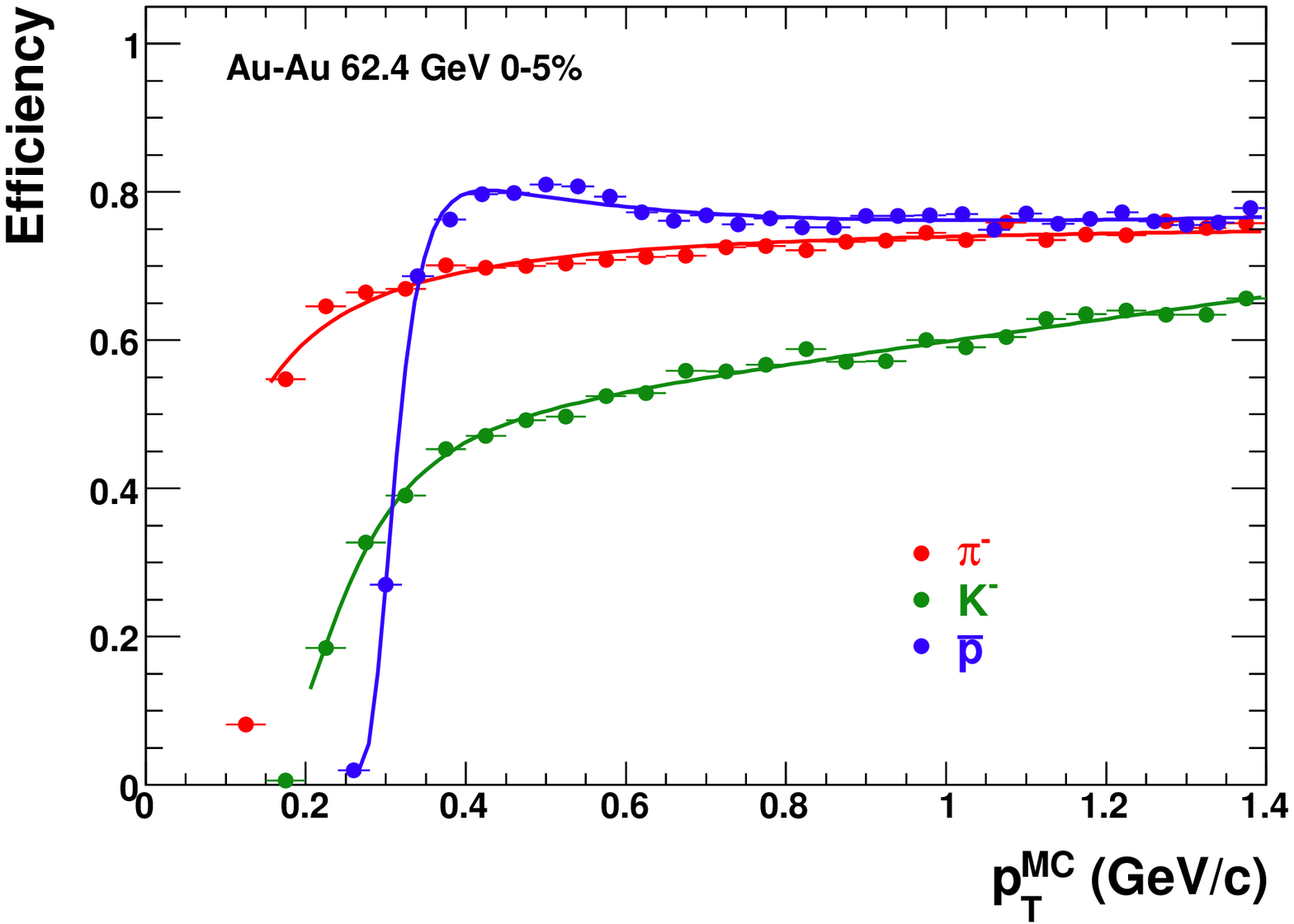}
		\includegraphics[width=0.45\textwidth]{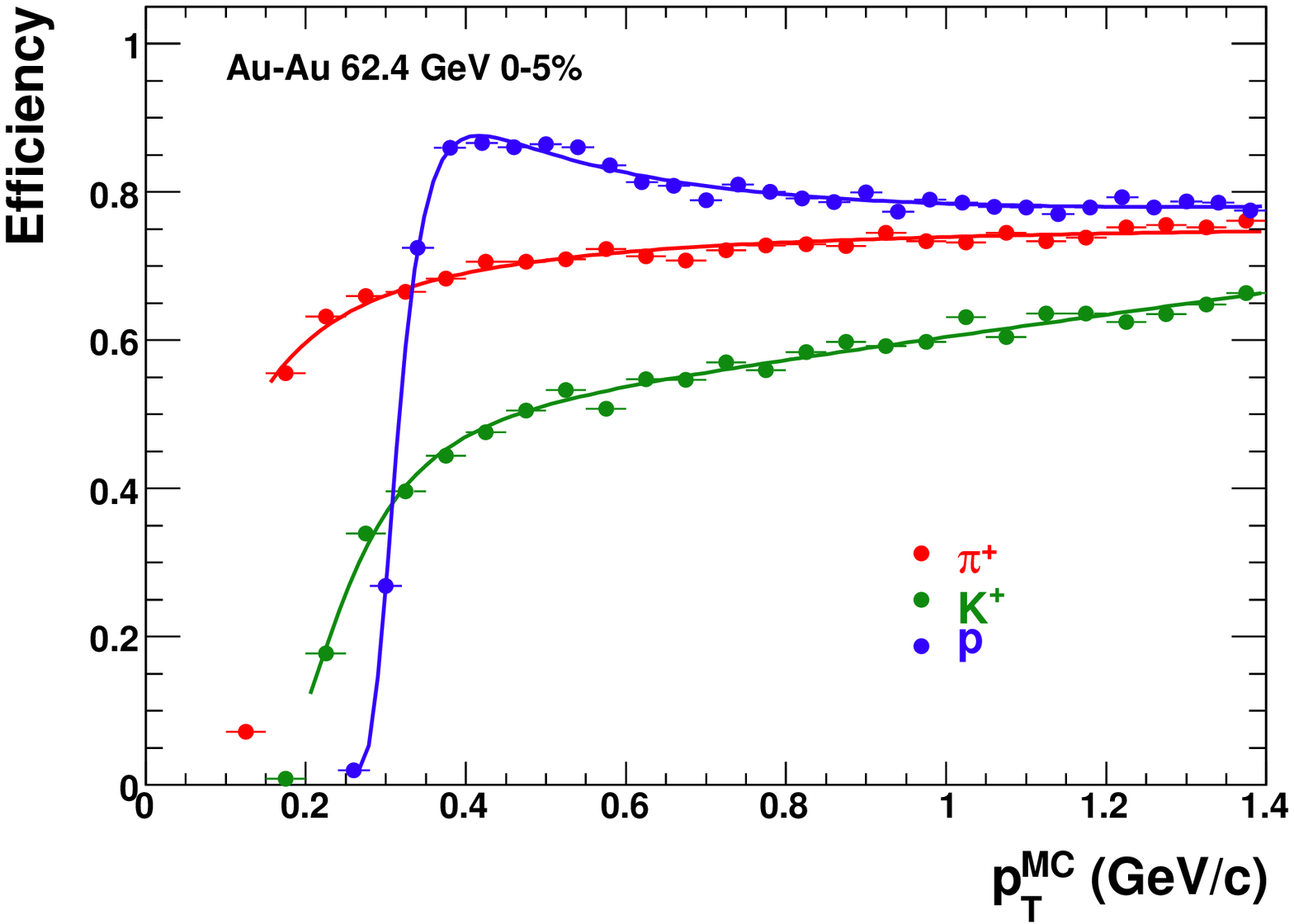}
		
		\includegraphics[width=0.45\textwidth]{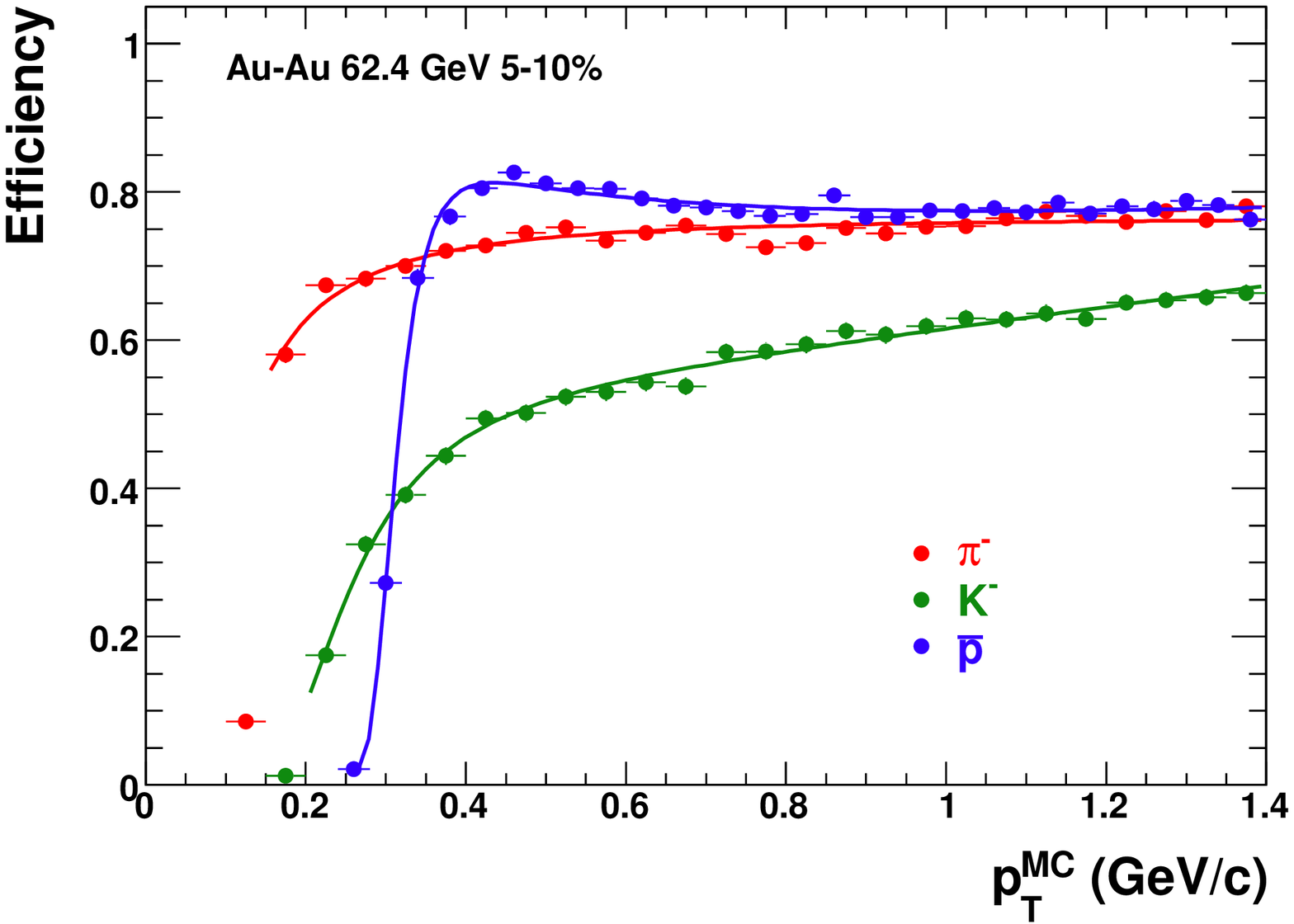}
		\includegraphics[width=0.45\textwidth]{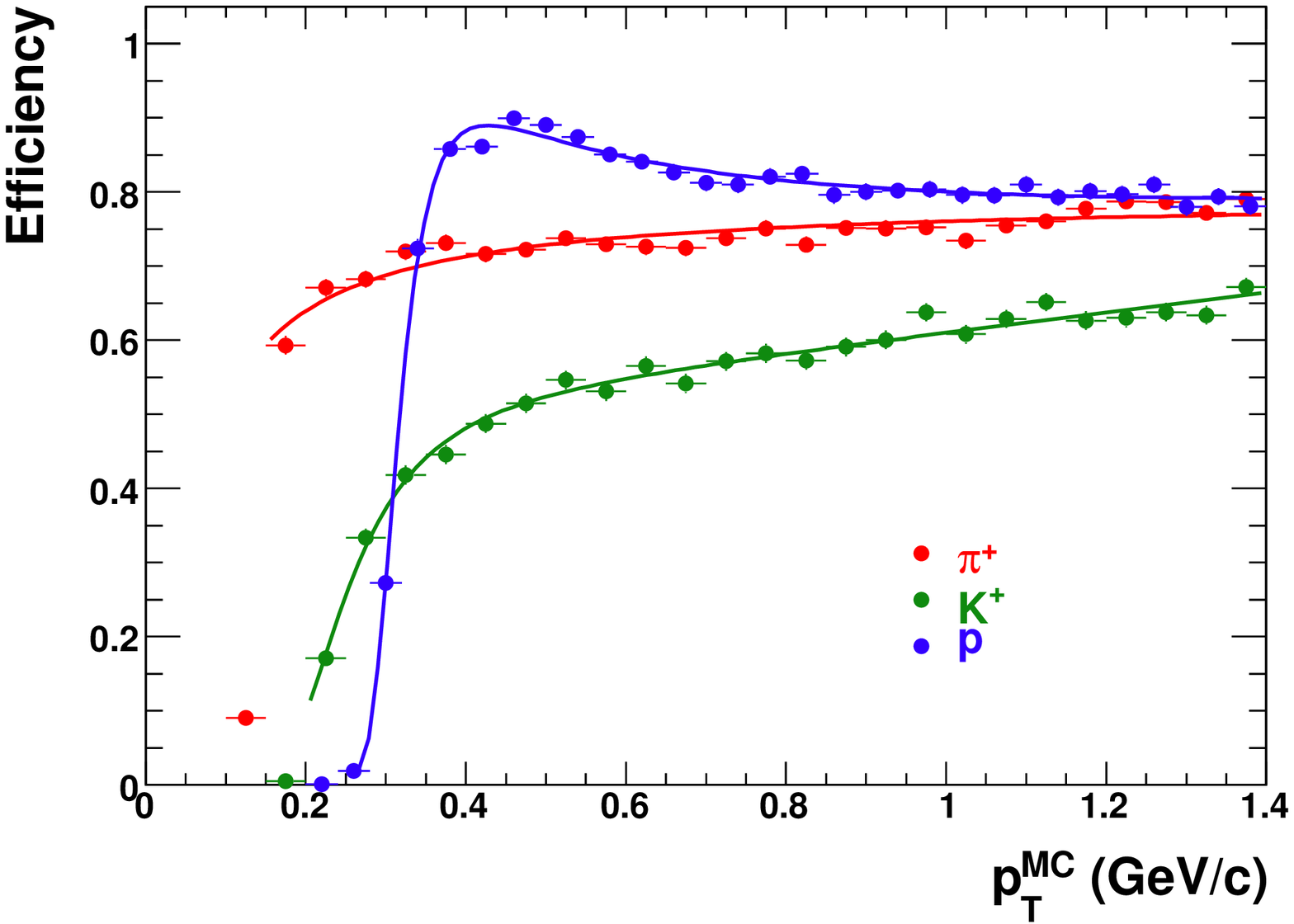}
			
		\includegraphics[width=0.45\textwidth]{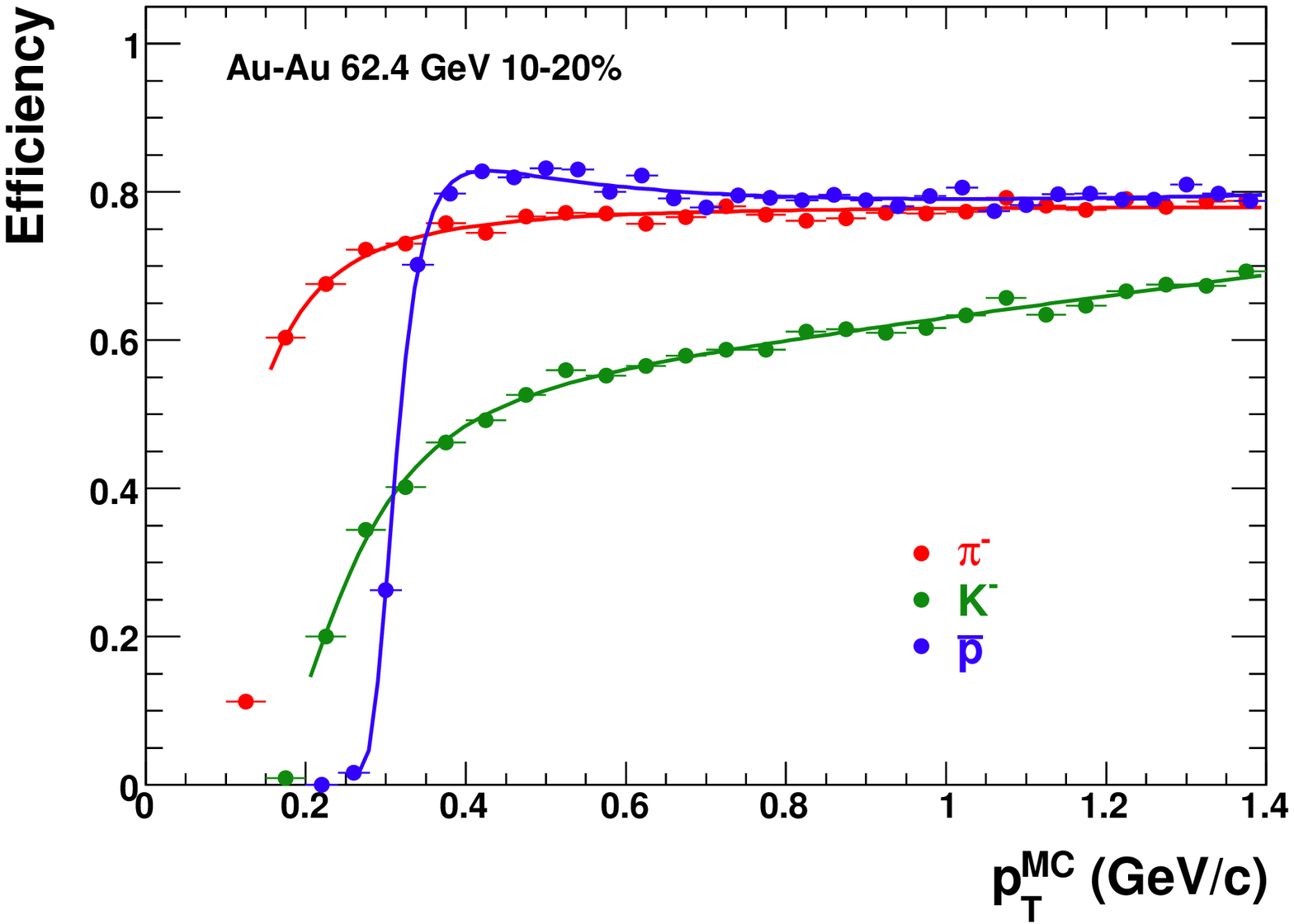}
		\includegraphics[width=0.45\textwidth]{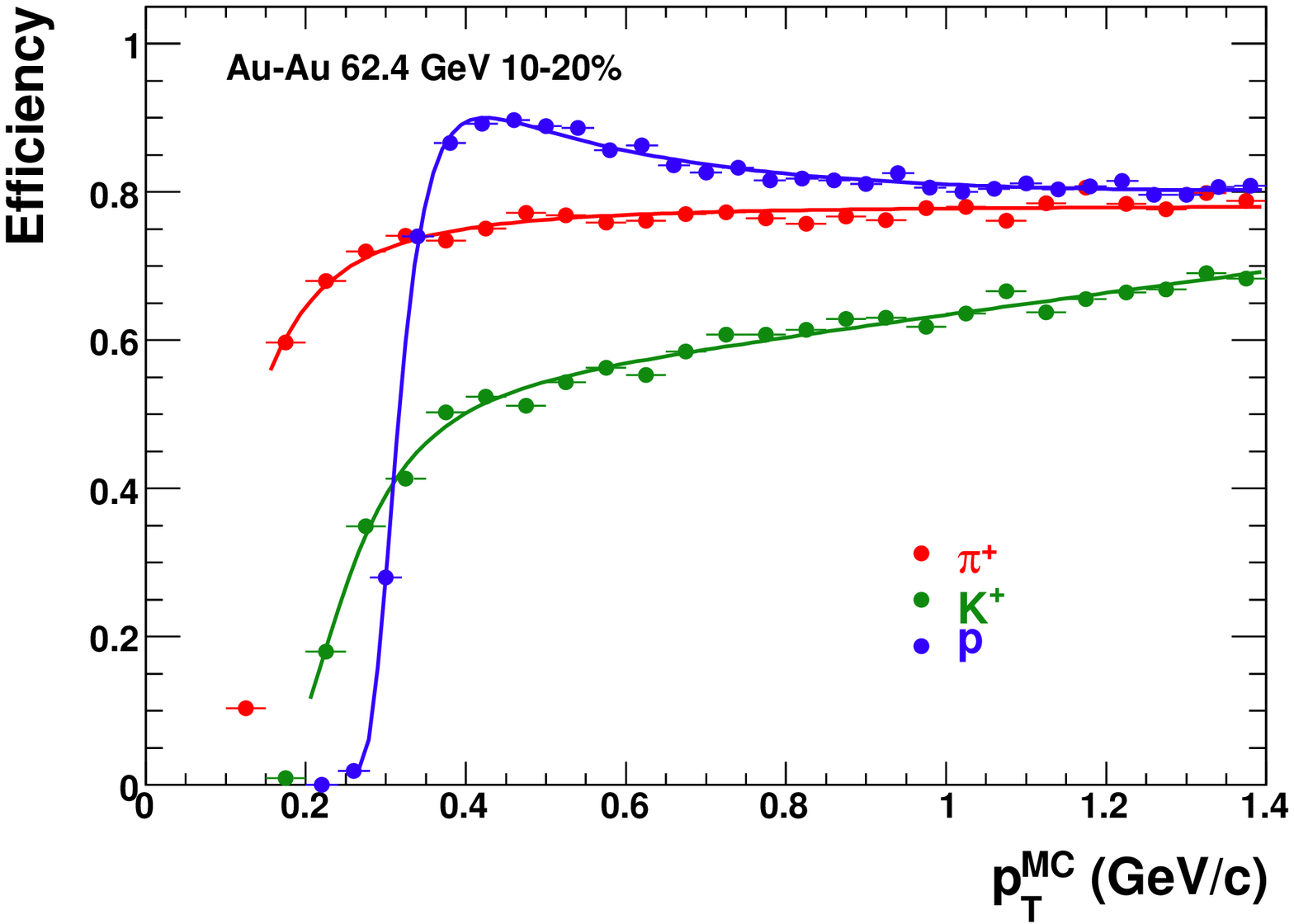}
		
		\includegraphics[width=0.45\textwidth]{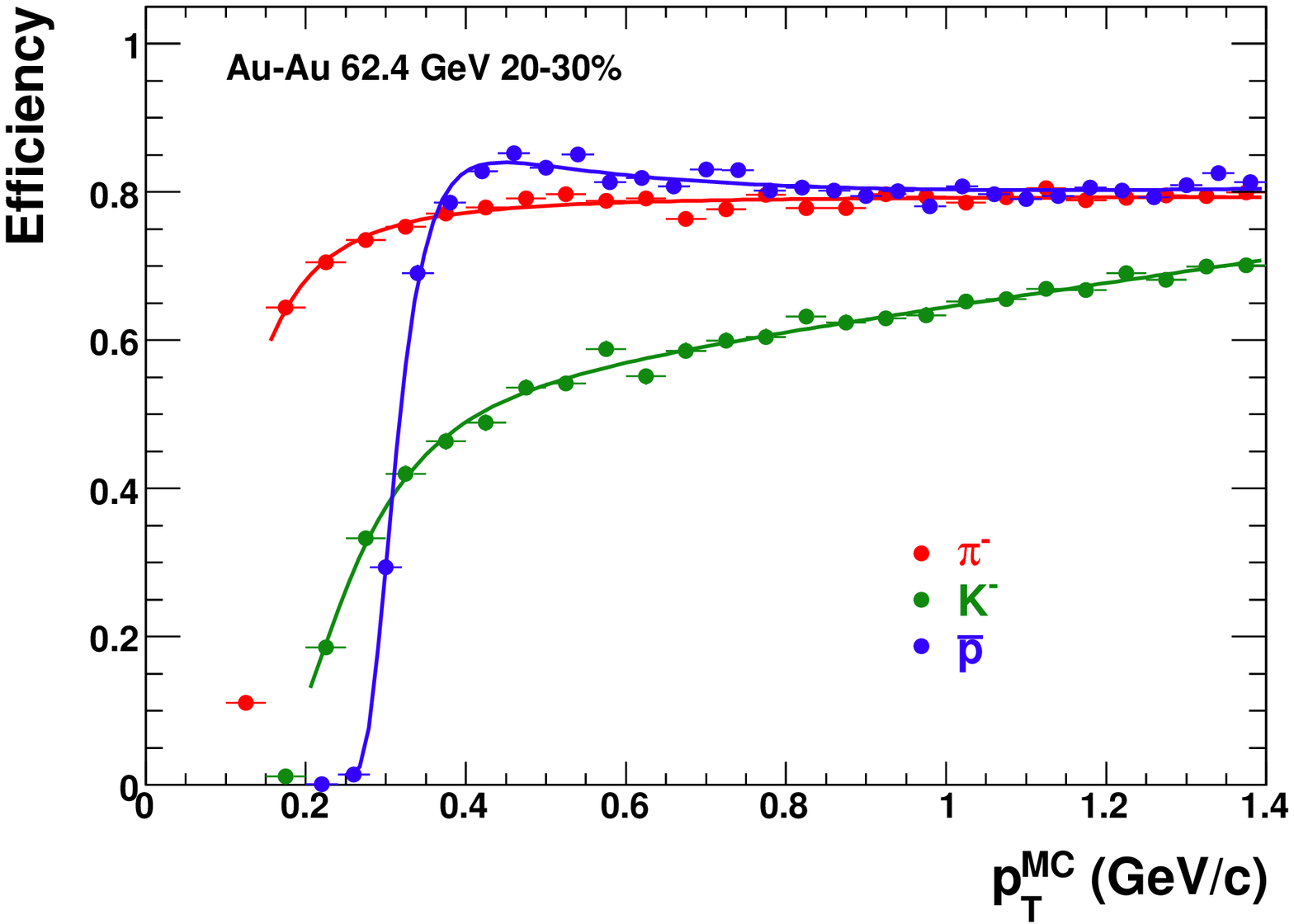}
		\includegraphics[width=0.45\textwidth]{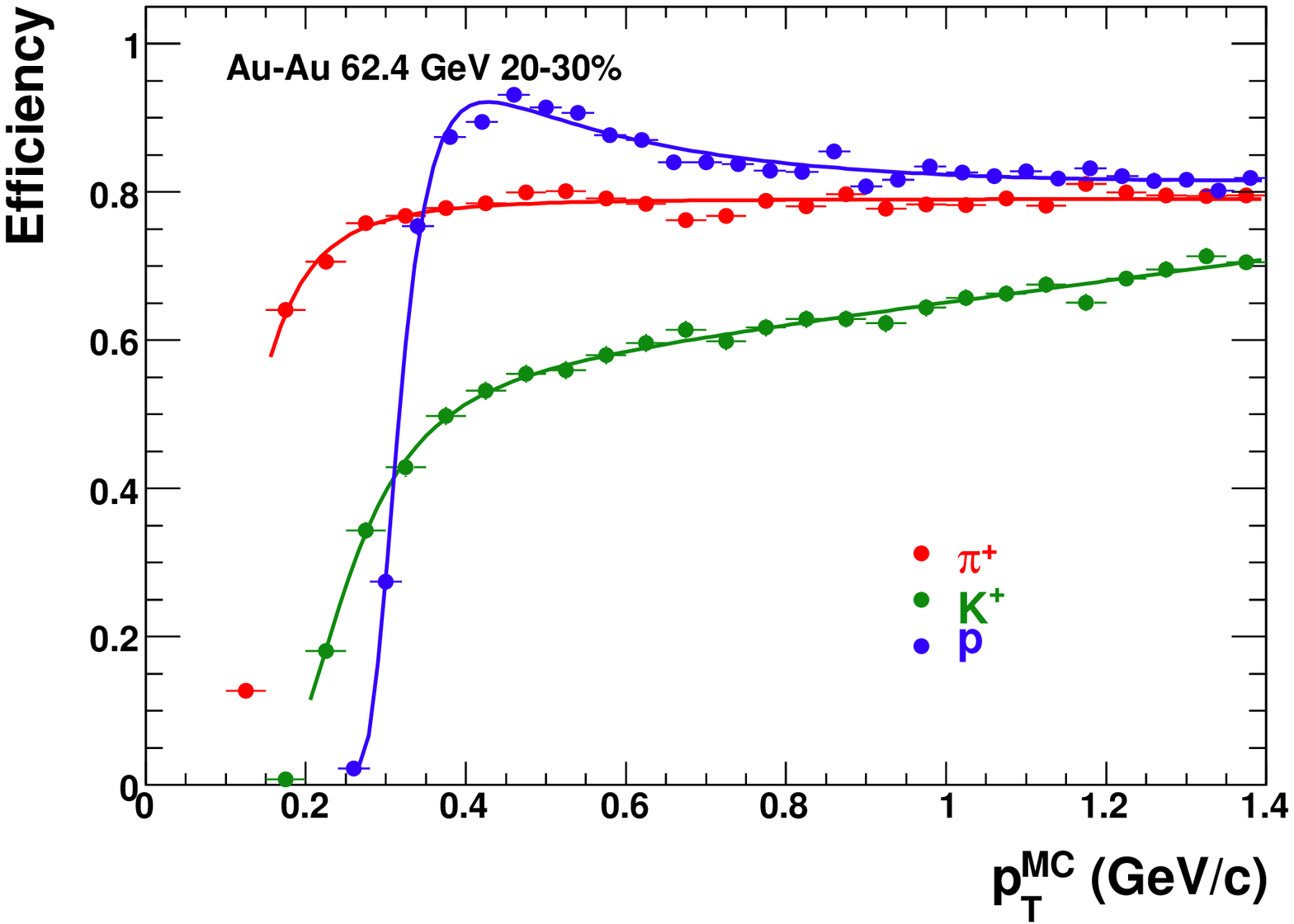}
		
		\includegraphics[width=0.45\textwidth]{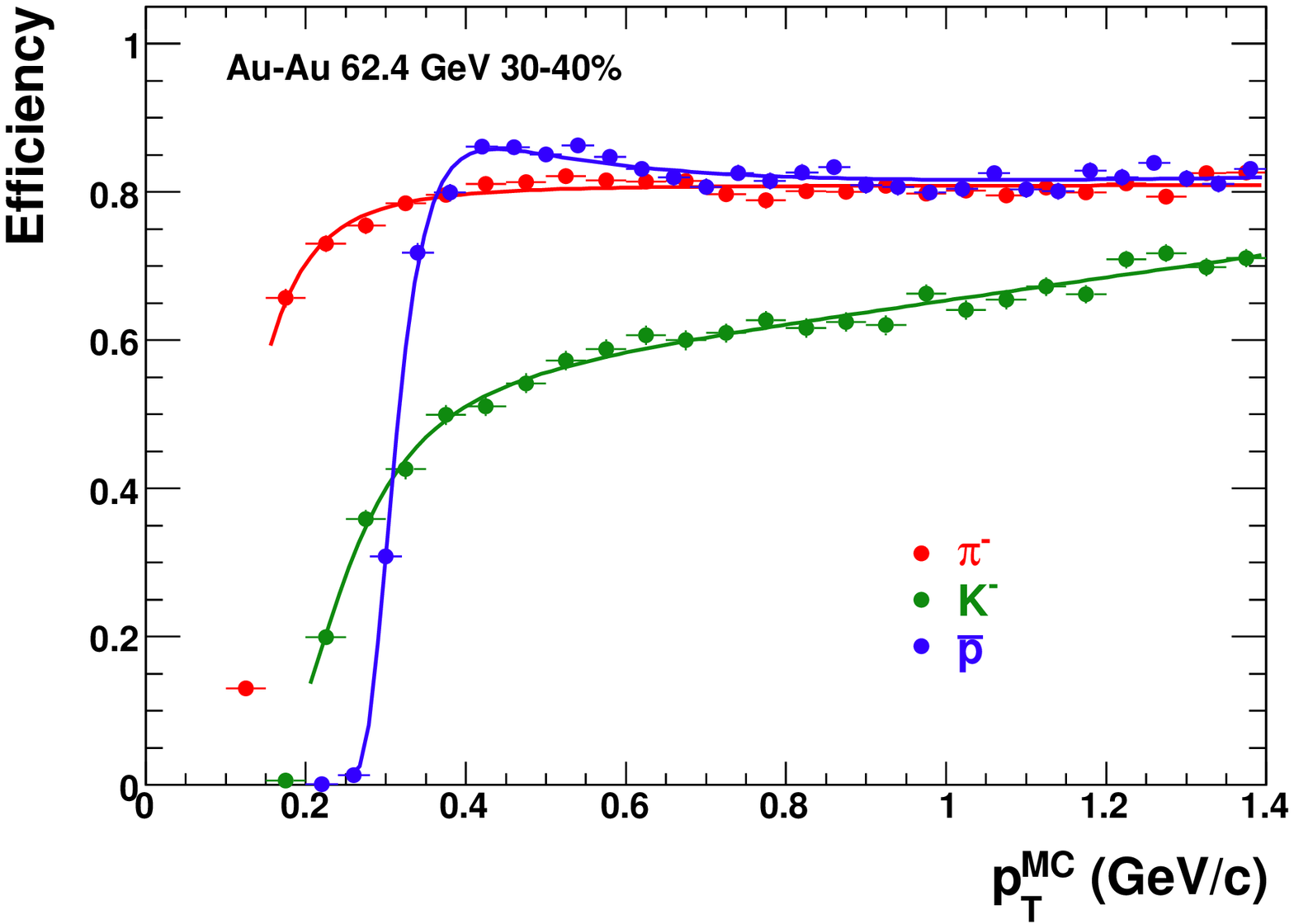}
		\includegraphics[width=0.45\textwidth]{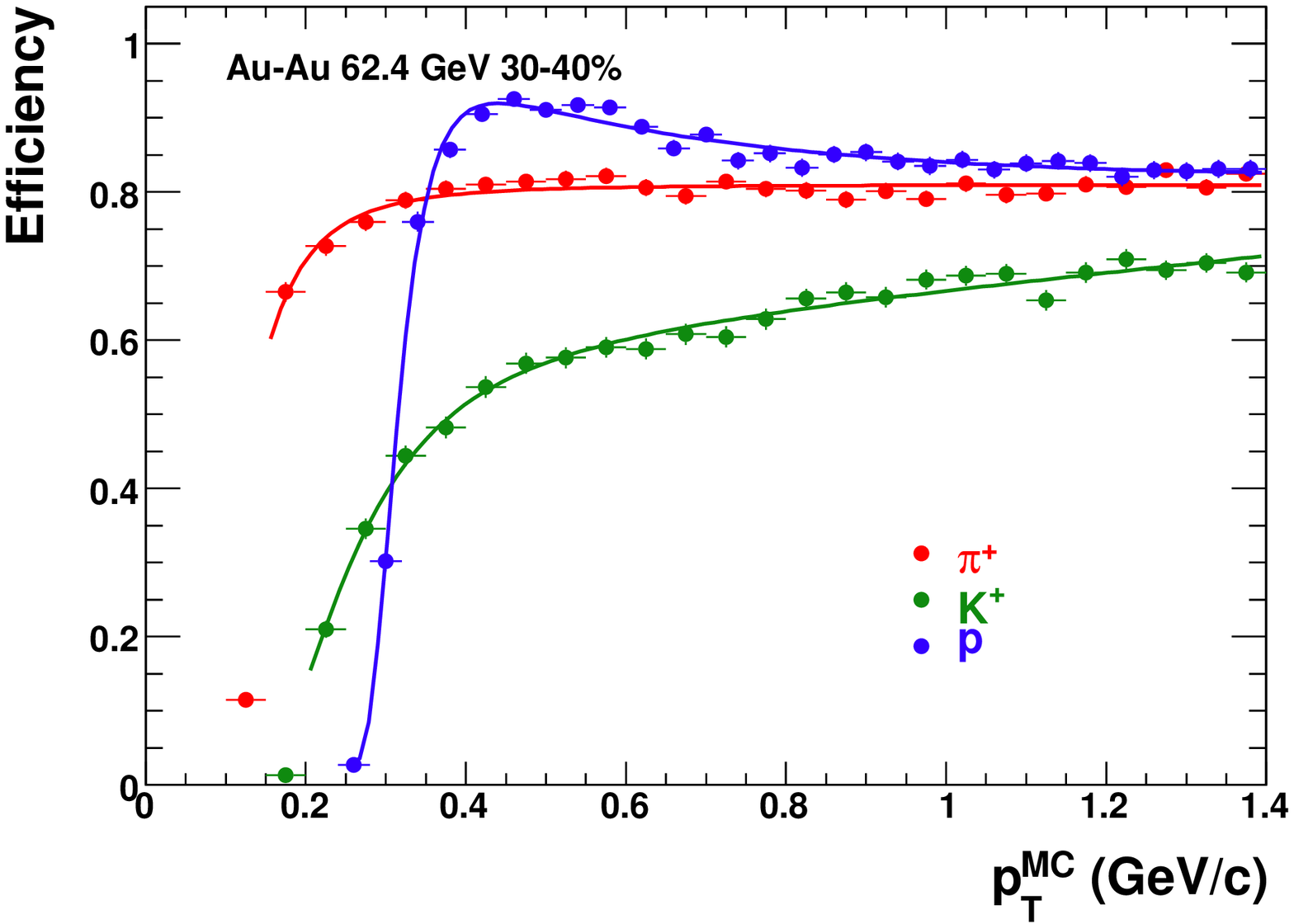}

			\caption{Tracking efficiency of $\pi^{-}$, $K^{-}$, $\overline{p}$ (left panels) and $\pi^{+}$, $K^{+}$ and $p$ (right panels) in 62.4 GeV $\bf{Au-Au}$ collisions as a function of transverse momentum and centrality.}
	\label{fig:treffauau1}
\end{figure}
In the real data it is not possible to determine whether a vertex is fake or not. Therefore,
one has to apply a different approach. It is found that the transverse momentum spectrum of the
fake events is different than the spectrum of good events as shown in Fig.~\ref{fig:ptspectrafake}.
Spectrum from fake events is flatter than from good events, presumably due to the refitting of tracks including the wrong vertex position.

Figure~\ref{fig:FakeVertexCorrection} shows the ratio of charged hadron $p_{T}$ spectrum in good vertex events to that in all events with a reconstructed vertex (i.e. sum of good and fake vertex events) for minimum bias pp and d-Au collisions. The spectra are normalized per event before the ratio is taken. Therefore, a single function:
\begin{equation}
\frac{p_{T}^{good vertex}}{p_{T}^{all vertex}}\ =\ c_{0}\cdot e^{c_{1}\cdot p_{T}^{ c_{2}} }-1
\label{eq:fakecorr}
\end{equation}
is used for the $p_{T}$-dependence of the fake vertex correction. The correction is done by multiplying the $p_{T}$ spectra by the function described in Eq.~\ref{eq:fakecorr}.
In low multiplicity and minimum bias pp events and in peripheral
and minimum bias dAu events the correction is $\sim$ 3 - 5 \% and vanishes around $p_{T} \sim$ 1 GeV as shown in Fig.~\ref{fig:FakeVertexCorrection}.


\subsection{Tracking efficiency and acceptance}

Raw particle spectra have to be corrected for detector and tracking efficiency. The corrections are obtained from MC embedding. The obtained correction includes the net effect of detector acceptance, tracking efficiency interaction losses, decays, etc.

For each investigated particle specie ($\pi^{\pm}, K^{\pm}, p$ and $\overline{p}$) particles are embedded with flat distributions in $p_{T}$ and $y$ to have uniform statistics.
\begin{figure}[!h]
	\centering
	
		\includegraphics[width=0.45\textwidth]{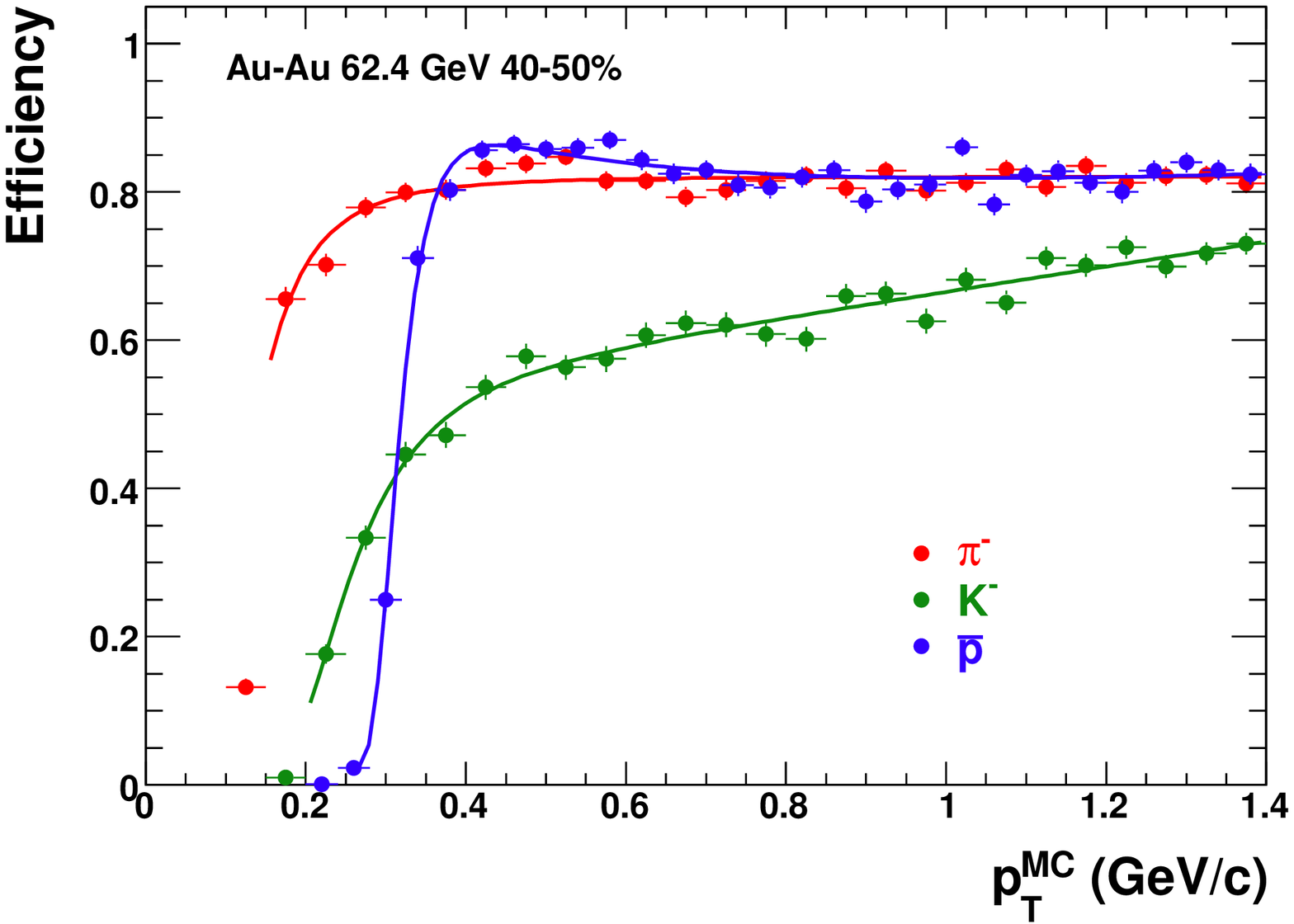}
		\includegraphics[width=0.45\textwidth]{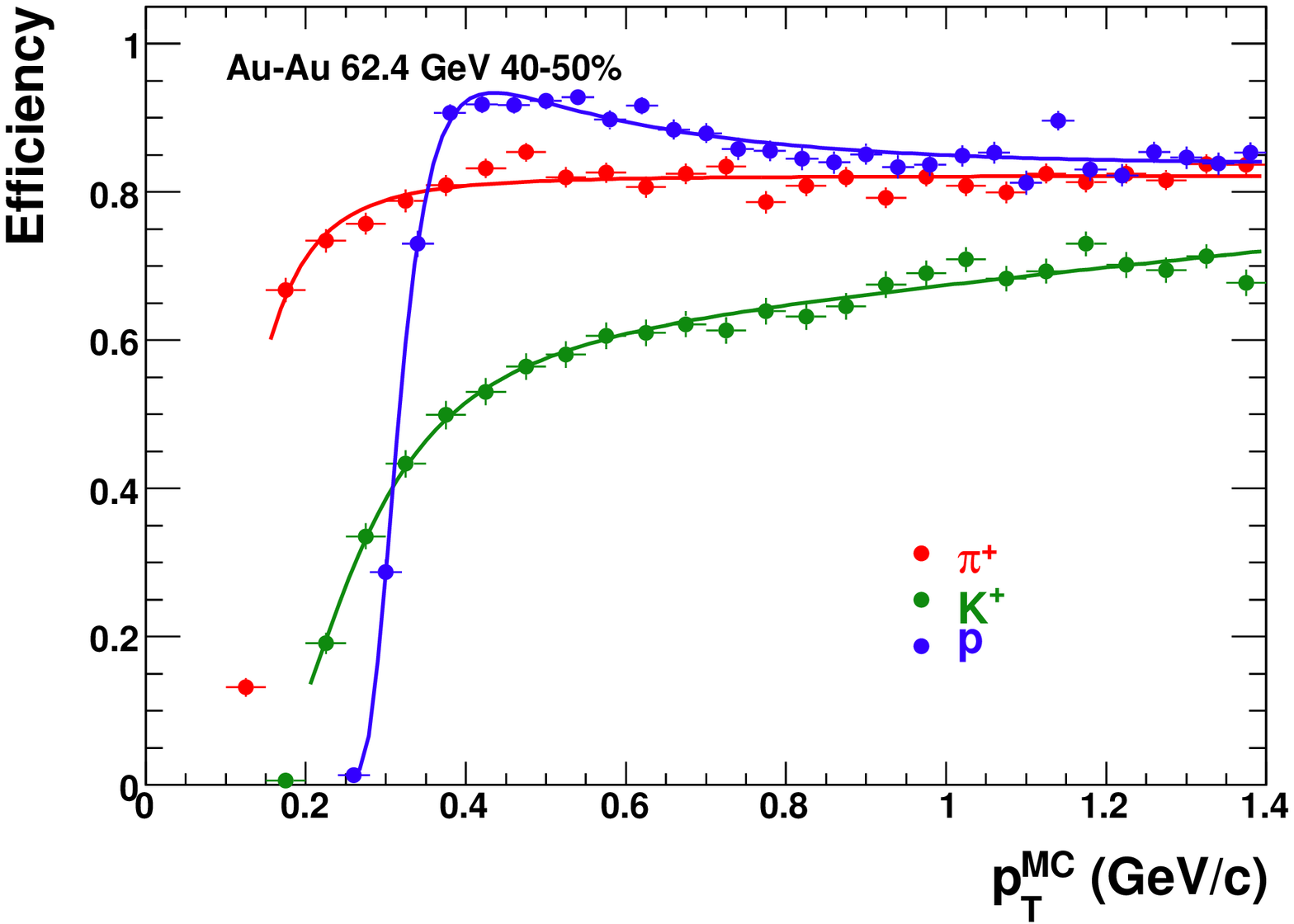}
		\includegraphics[width=0.45\textwidth]{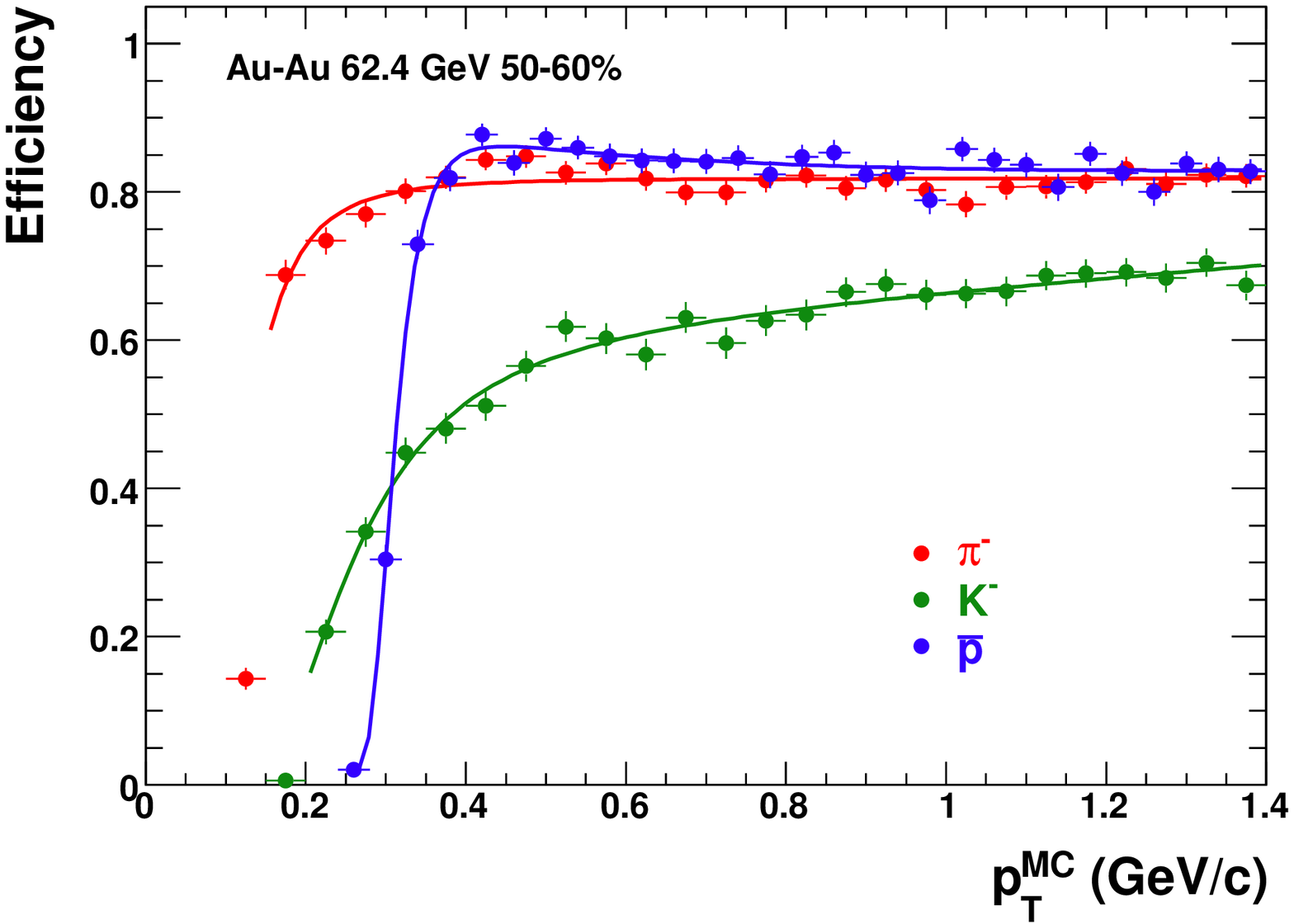}
		\includegraphics[width=0.45\textwidth]{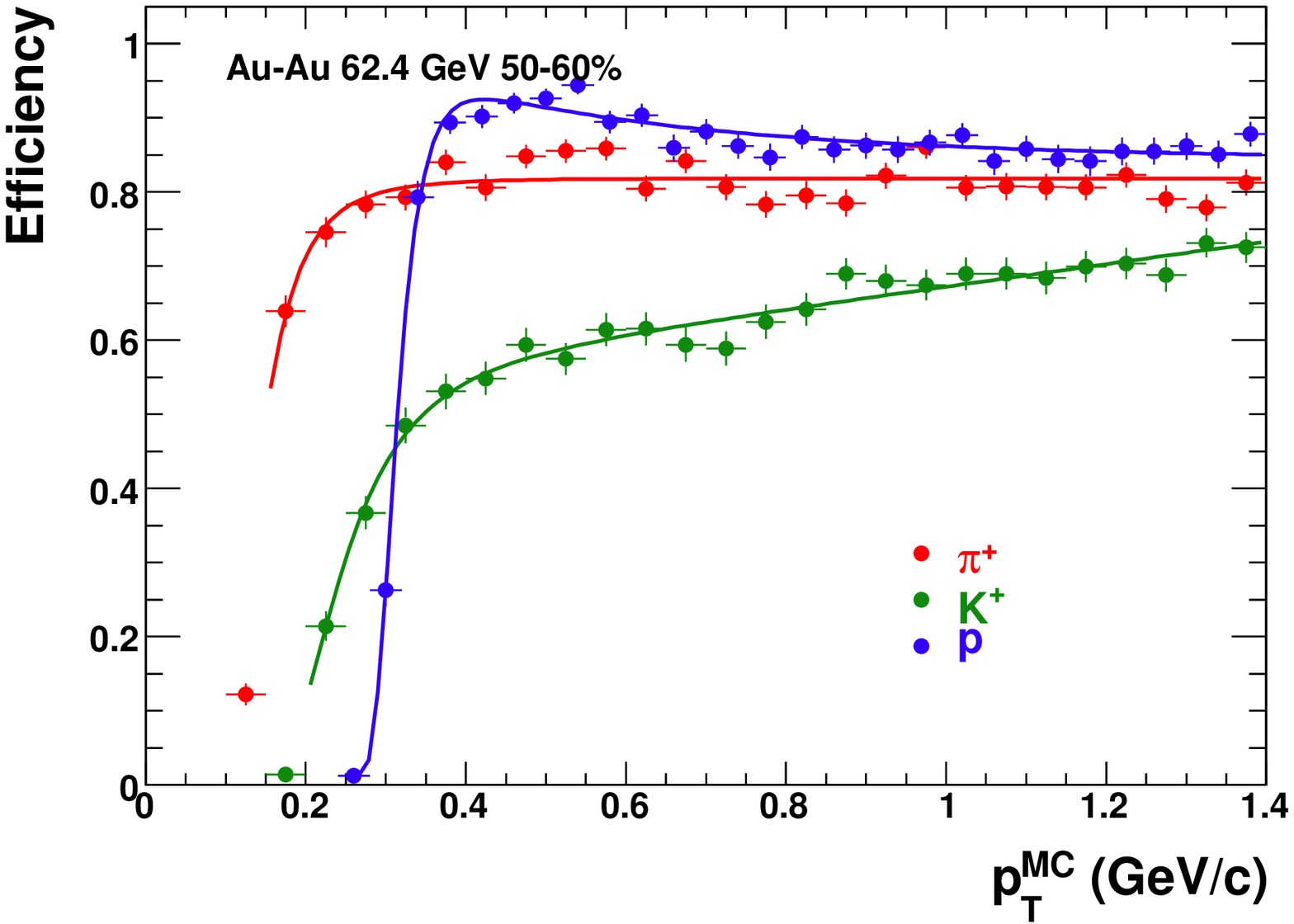}
		\includegraphics[width=0.45\textwidth]{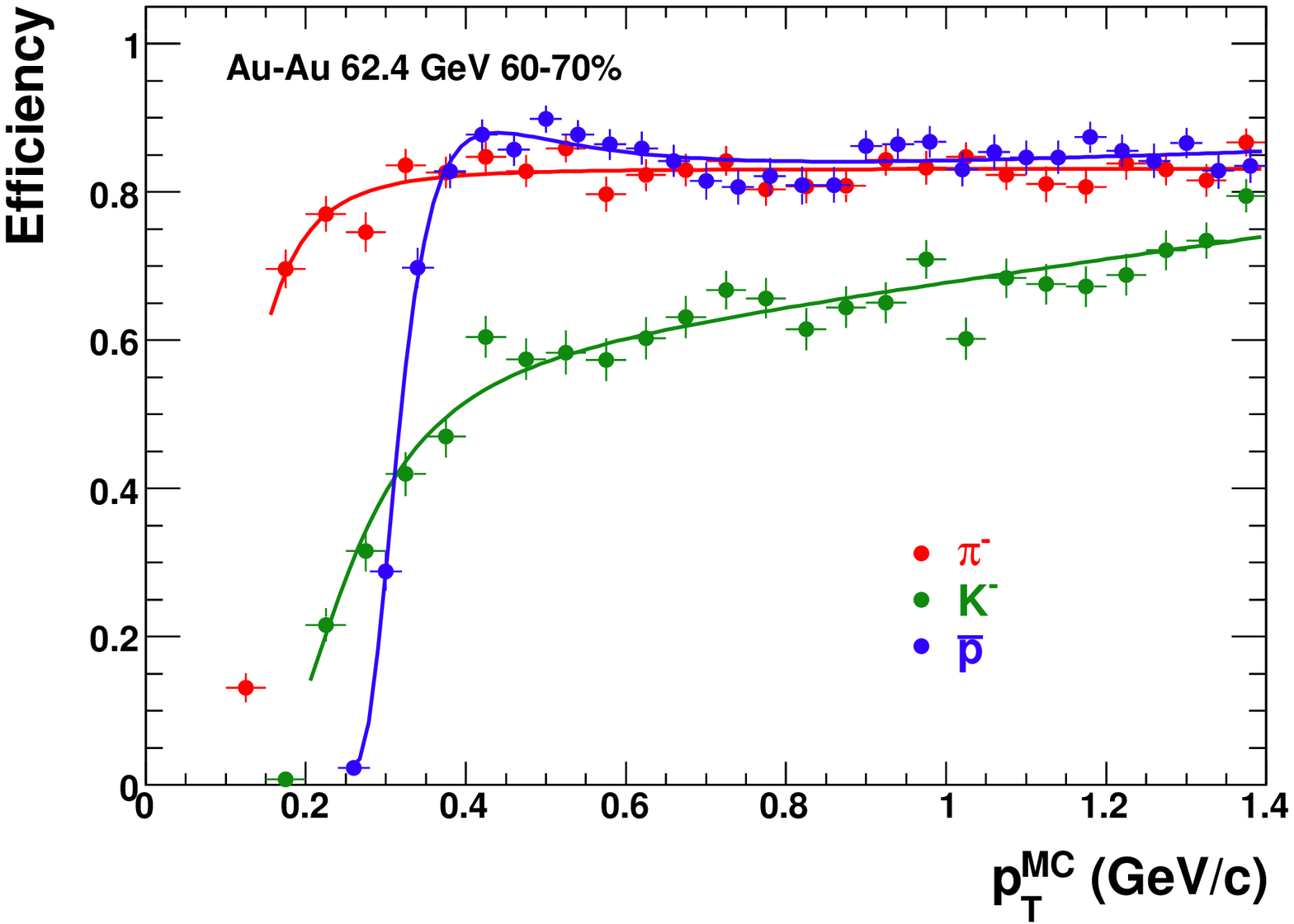}
		\includegraphics[width=0.45\textwidth]{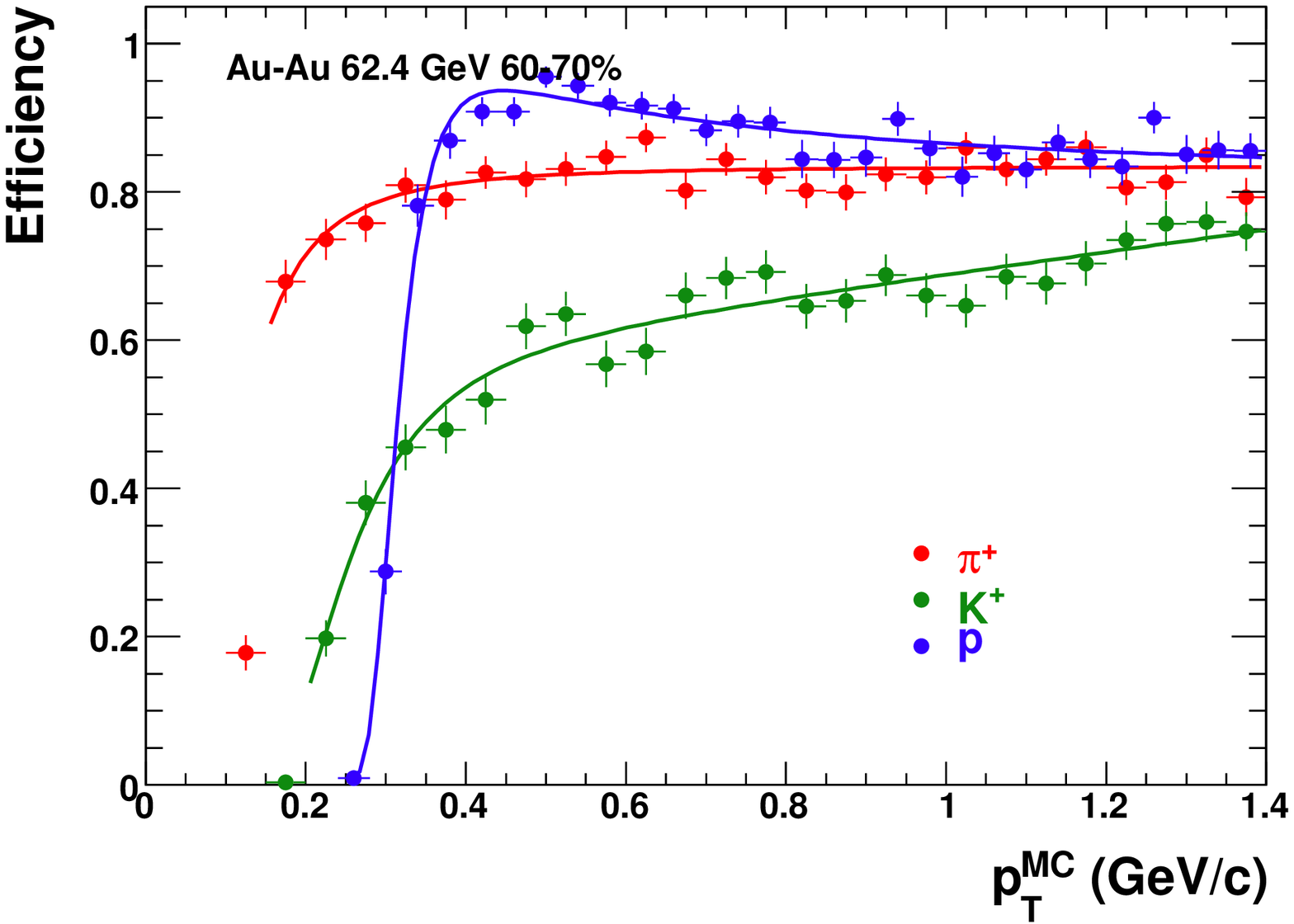}
	  \includegraphics[width=0.45\textwidth]{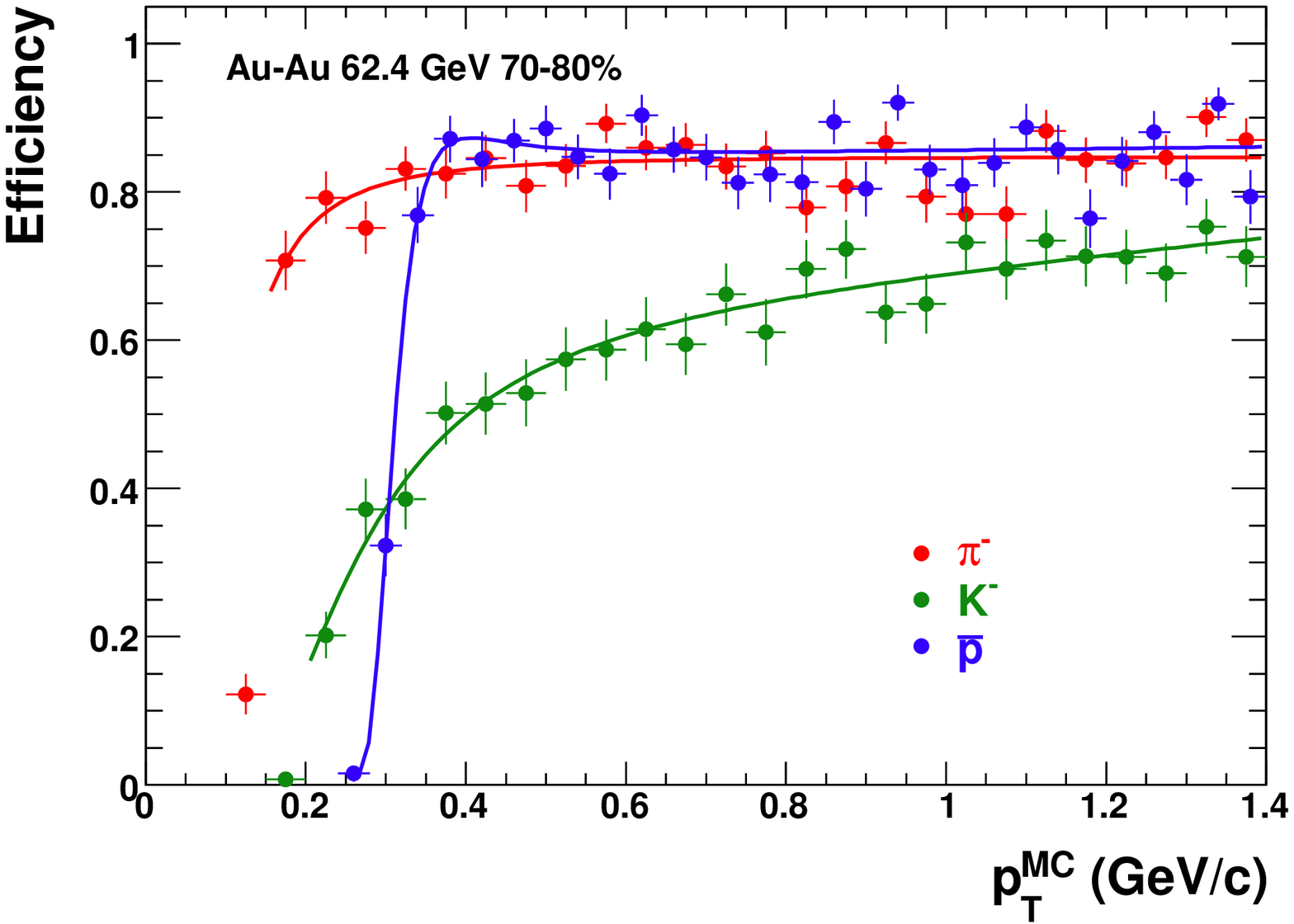}
		\includegraphics[width=0.45\textwidth]{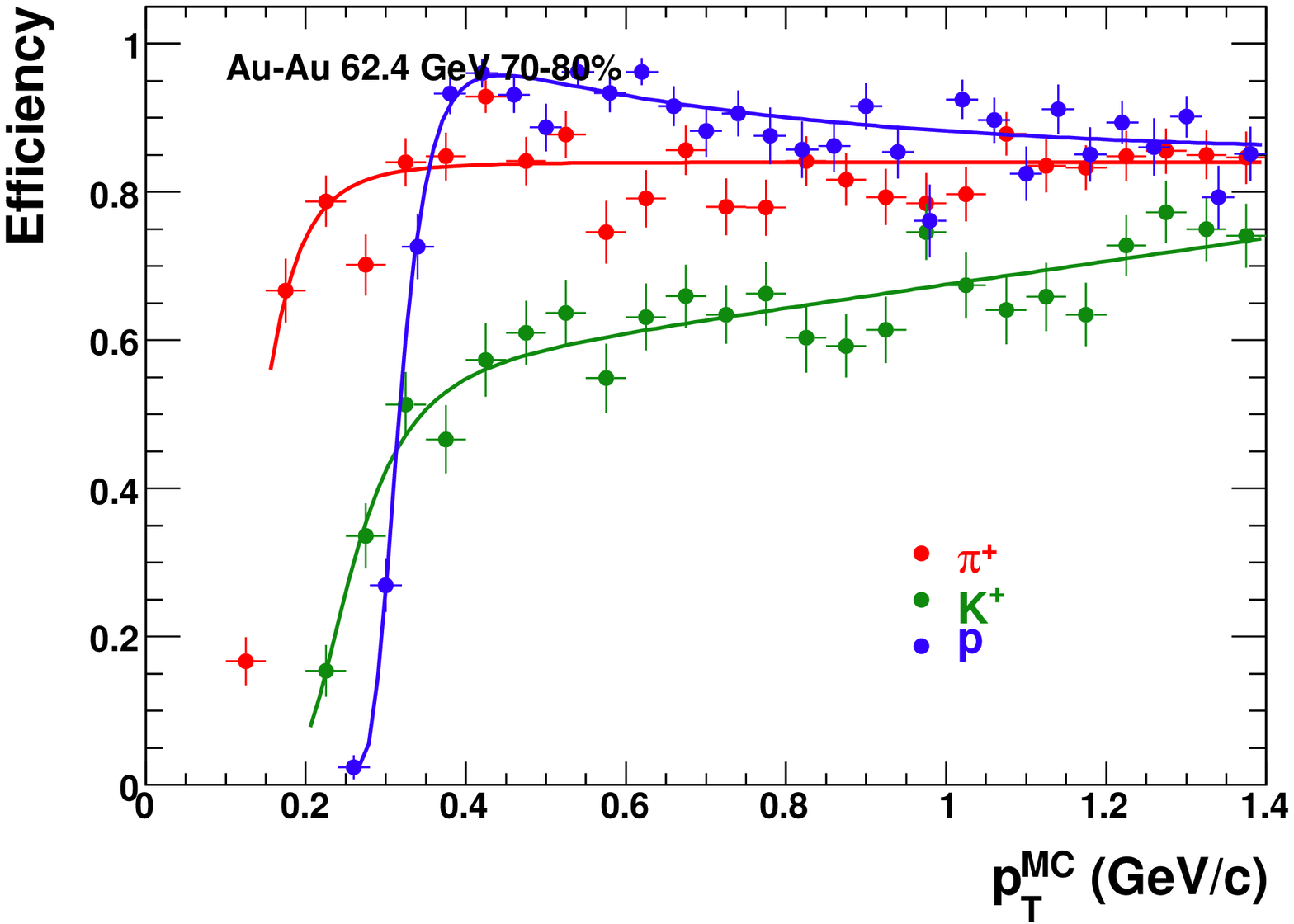}

			\caption{Tracking efficiency of n$\pi^{-}$, $K^{-}$, $\overline{p}$ (left panels) and $\pi^{+}$, $K^{+}$ and $p$ (right panels) in 62.4 GeV $\bf{Au-Au}$ collisions as a function of transverse momentum and centrality.}
	\label{fig:treffauau2}
\end{figure}

Tracking efficiency for 200 GeV pp collisions is shown in Fig.~\ref{fig:treffpp}, for 200 GeV dAu collisions in Fig.~\ref{fig:treffdau} and for 62.4 GeV Au-Au collisions in Fig.~\ref{fig:treffauau1} and Fig.~\ref{fig:treffauau2}.
\begin{table}
\scriptsize
\begin{center}
\caption{Tracking efficiency parameterization for pions.\label{tab:treffpion}}  
\begin{scriptsize}
\begin{tabular}{|c|c|c|c|c|c|c|} 
\hline 
Collision & \multicolumn{3}{|c|}{$\pi^{-}$ ($c_{0,1,2} $)} & \multicolumn{3}{|c|}{$\pi^{+}$ ($c_{0,1,2}$)}\\ \hline
pp MB & 										0.89905 & 0.06974 & 1.73860 & 0.90280 & 0.06526 & 1.60725 \\ \hline
Au-Au 62.4 GeV 70-80\% 		& 0.84839 & 0.08926 & 2.55107 & 0.83915 & 0.12901 & 4.70611 \\ \hline
Au-Au 62.4 GeV 0-5\% 			& 0.76833 & 0.06332 & 1.21189 & 0.76969 & 0.06244 & 1.18241\\ \hline
\end{tabular} 
\end{scriptsize}
\end{center} 
\end{table}  
\begin{table}
\scriptsize
\begin{center}
\caption{Tracking efficiency parameterization for kaons.\label{tab:treffkaon}}  
\begin{scriptsize}
\begin{tabular}{|c|c|c|c|c|c|c|} 
\hline 
Collision & \multicolumn{3}{|c|}{$K^{-}$ ($c_{0,1,2} $)} & \multicolumn{3}{|c|}{$K^{+}$ ($c_{0,1,2}$)}\\ \hline
pp MB 										& 0.73584 & 0.24398 & 2.49097 & 0.75943 & 0.23179 & 2.03419\\ \hline
Au-Au 62.4 GeV 70-80\% 		& 0.60802 & 0.23787 & 2.42484 & 0.51912 & 0.24281 & 5.39325\\ \hline
Au-Au 62.4 GeV 0-5\% 			& 0.45012 & 0.22891 & 3.92516 & 0.45835 & 0.23128 & 4.02922\\ \hline
\end{tabular} 
\end{scriptsize}
\end{center} 
\end{table}  
\begin{table}
\scriptsize
\begin{center}
\caption{Tracking efficiency parameterization for protons/antiprotons.\label{tab:treffproton}}  
\begin{scriptsize}
\begin{tabular}{|c|c|c|c|c|c|c|} 
\hline 
Collision: $\overline{p}$ and $p$ & \multicolumn{2}{|c|}{pp MB } &  \multicolumn{2}{|c|}{Au-Au 62.4 GeV 70-80\%} & \multicolumn{2}{|c|}{Au-Au 62.4 GeV 0-5\%} \\ \hline
$p_{0}$ &  0.76733  & 0.930071  & 7.76198   & 32.7826& 0.35058 & 0.33582  \\ \hline
$p_{1}$ &  0.28580  & 0.290520  & 0.30947   & 0.32498 & 0.30266 & 0.33119  \\ \hline
$p_{2}$ &  13.08348 & 10.57888  & 20.65861  & 16.3731& 11.2653 & 9.44440  \\ \hline
$p_{3}$ &  0.11409  & -0.00805  & 1.71647   & 5.65292 & 0.02679 & 0.01084  \\ \hline
$p_{4}$ &  -        & -         & 40.61263  & 363.1   & 0.53201 & 0.53312  \\ \hline
$p_{5}$ & - 				& -         & 3.29003   &  732.8  & 0.09304 & 0.22337  \\ \hline
$p_{6}$ & - 				& -         & 0.13745   & 0.09921 & 1.11354 & 2.66341  \\ \hline

\end{tabular} 
\end{scriptsize}
\end{center} 

\end{table}  

Tracking efficiencies for $\pi^{+}$ and $\pi^{-}$ are similar. Pion efficiency quickly rises from $p_{T}$ = 0.1 GeV/c to $p_{T} \approx$ 0.3 GeV/c and levels off at $\sim$ 90\%. Small difference can be observed at low transverse momenta for $K^{+}$ and $K^{-}$ and for $p$ and $\overline{p}$ due to the absorption effect in the detecor material. Kaon efficiencies rise monotonicaly with increasing transevse momentum. Proton/antiproton efficiencies rise sharply around $p_{T}$ = 0.3 GeV/c and flatten out with an increasing damping in mid-central and central Au-Au collisions. 

Calculated tracking efficiencies depend on the measured particle
multiplicity, which is clearly shown in Au-Au collisions, comparing the most central (0-5 \%) and the peripheral bins (eg. 60-70 \%). The change in the pion efficiency at $p_{T} = 400$ MeV between most central and most peripheral Au-Au collisions is $\sim$ 20 \%, which correspondes to a change of 350 between the average charged particle multiplicity of the two centralities.
In pp and dAu collisions the variation of event multiplicity is small (in the highest multiplicity pp bin the average multiplicity is $\sim$ 10, and $\sim$ 19 in central dAu collisions), therefore tracking efficiency has no dependence on event classes.

Tracking efficiency also depends on the quality cuts as expected. The final corrected spectra should be the same for different quality cuts.
Systematic uncertainties in efficiency can therefore be assessed by different quality cuts.
\begin{figure}[!h]
	\centering
		\includegraphics[width=0.45\textwidth]{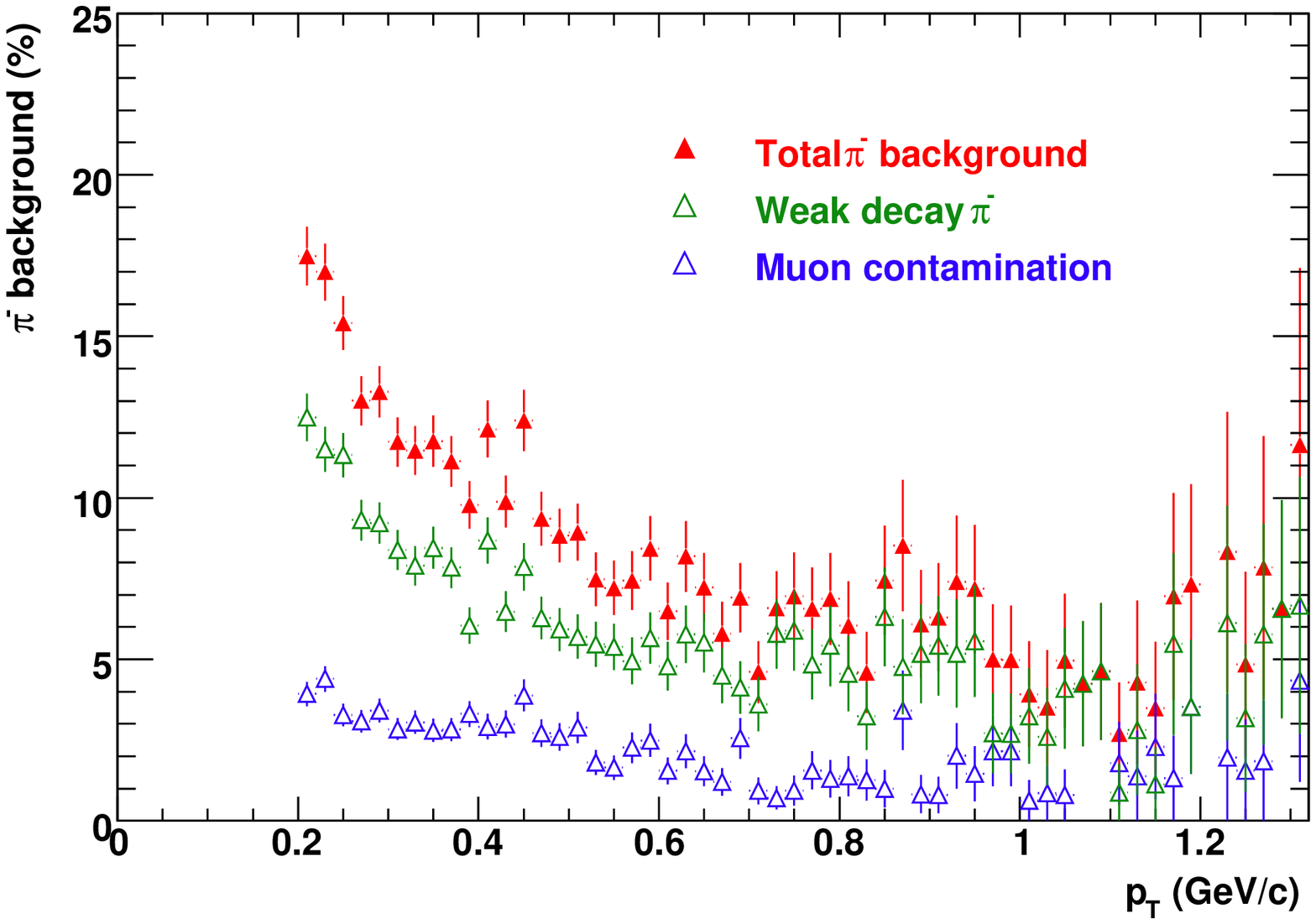}
		\includegraphics[width=0.45\textwidth]{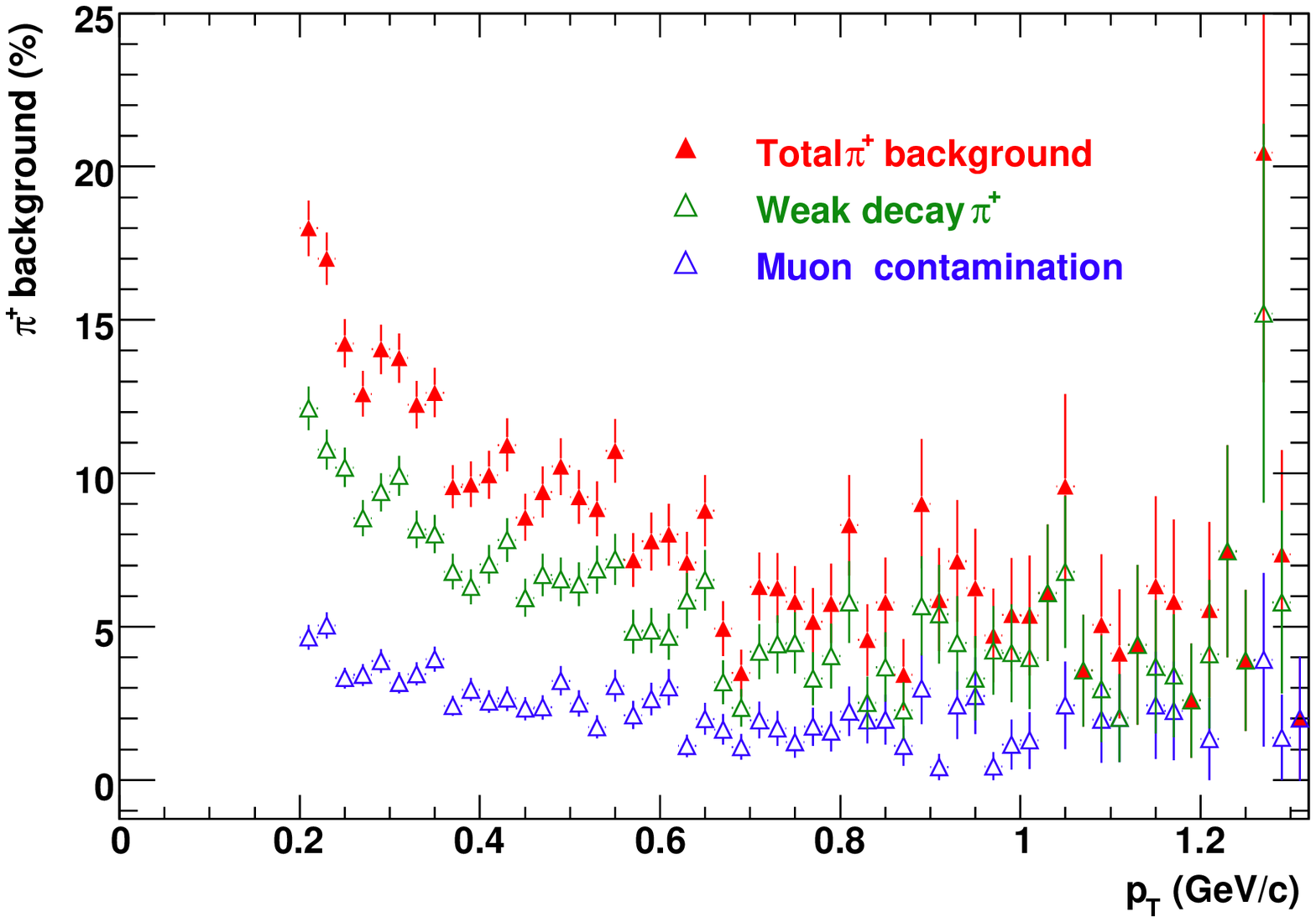}
				\caption{Correction to pion spectra for weak decays ($\Lambda$,$K^{0}_{S}$ and muons) as a function of $p_{T}$ in 200 GeV $\bf{pp}$ collisions. }
	\label{fig:pibgpp}
\end{figure}
\begin{figure}[!h]
	\centering
		\includegraphics[width=0.45\textwidth]{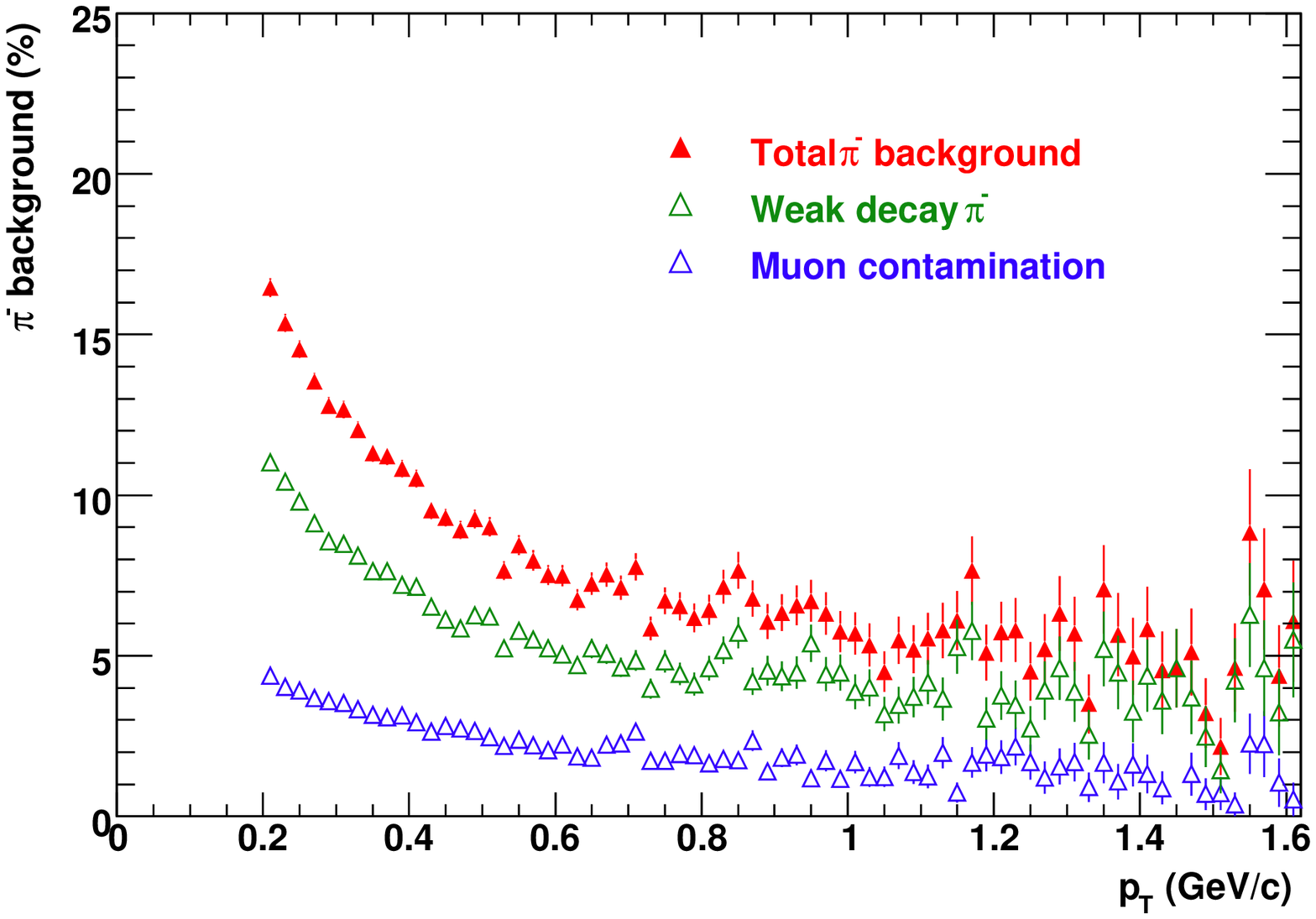}
		\includegraphics[width=0.45\textwidth]{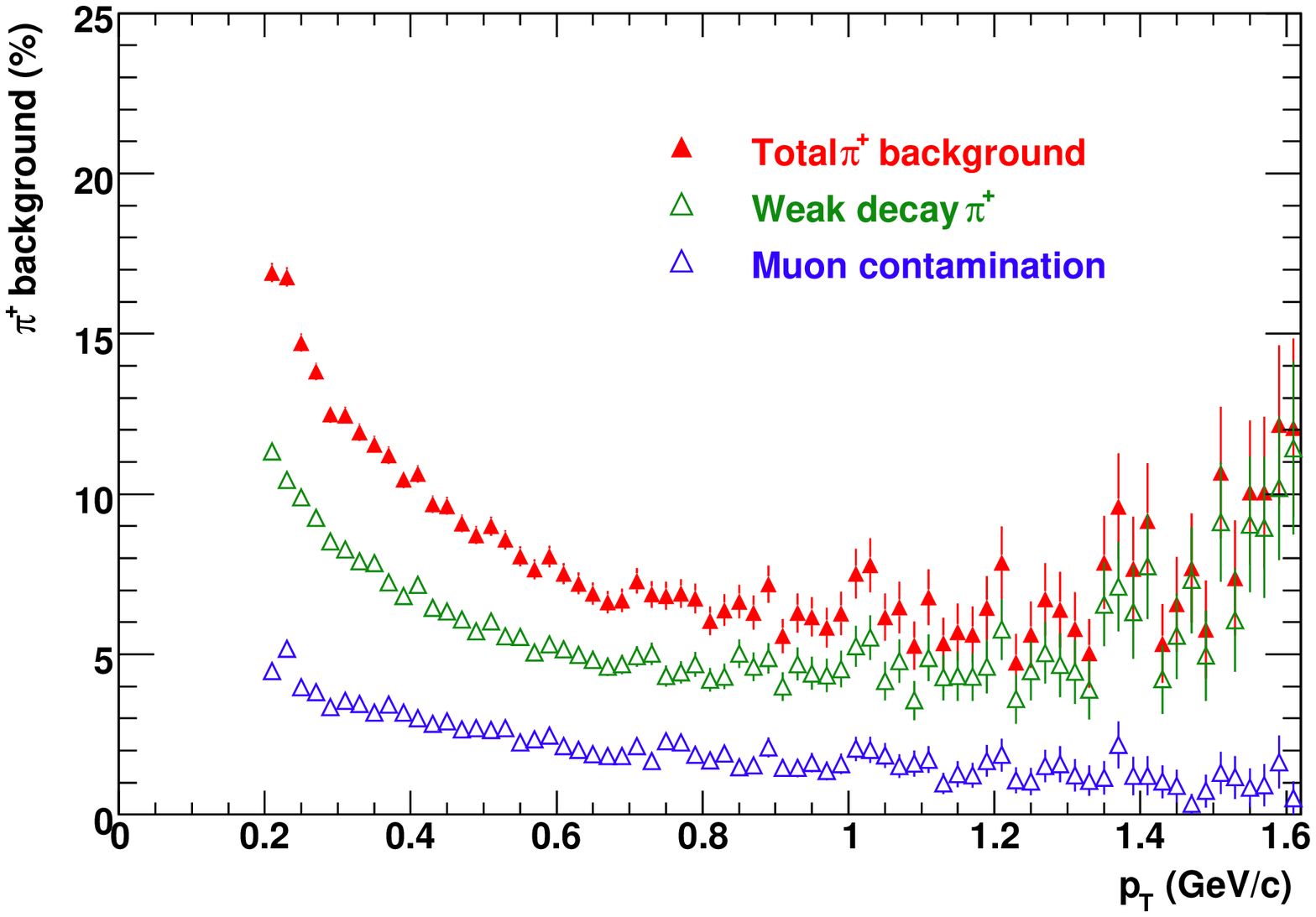}
				\caption{Correction to pion spectra for weak decays ($\Lambda$,$K^{0}_{S}$ and muons) as a function of $p_{T}$ in 200 GeV $\bf{dAu}$ collisions. }
	\label{fig:pibgdAu}
\end{figure}
\begin{figure}[!h]
	\centering
		\includegraphics[width=0.45\textwidth]{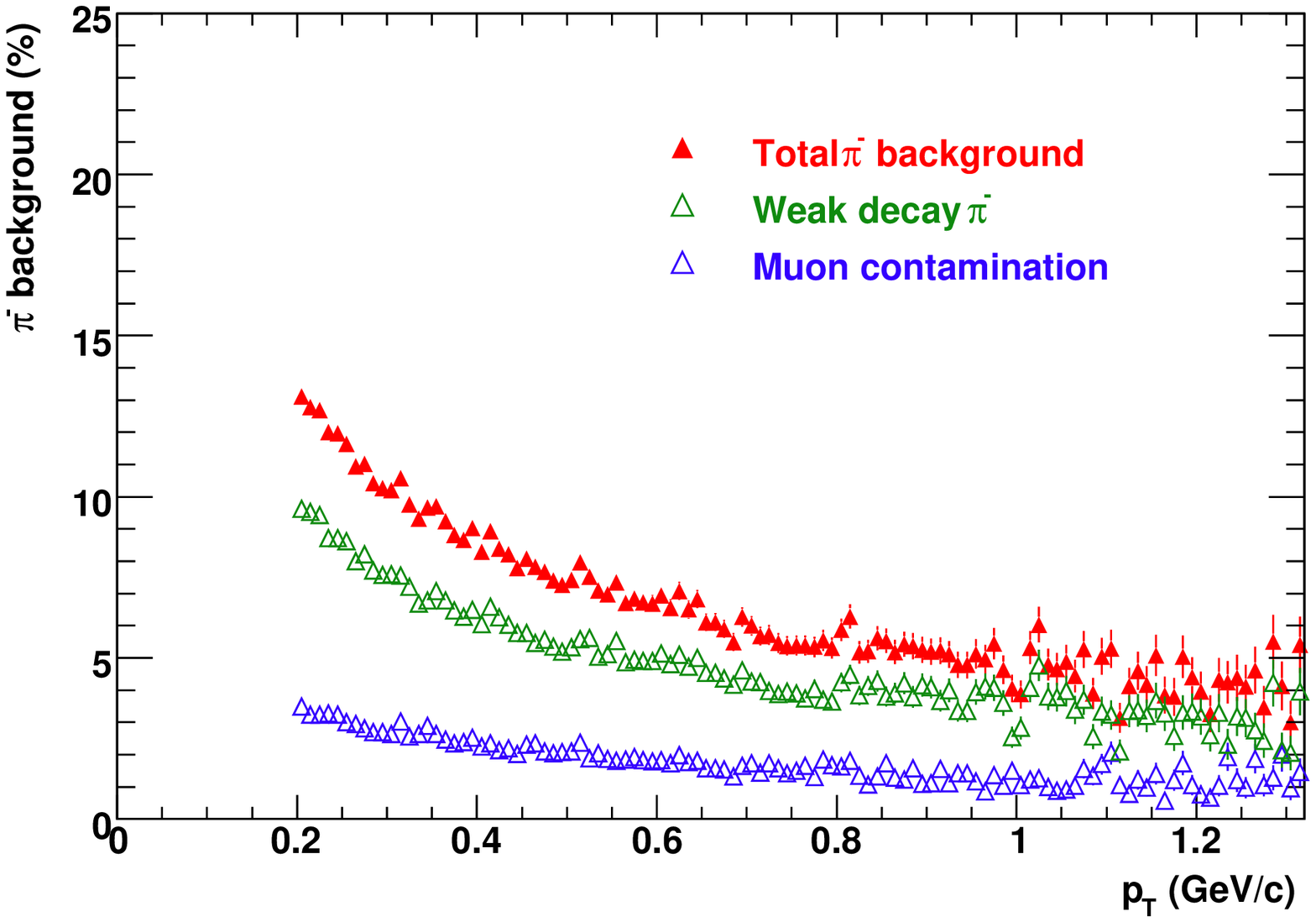}
		\includegraphics[width=0.45\textwidth]{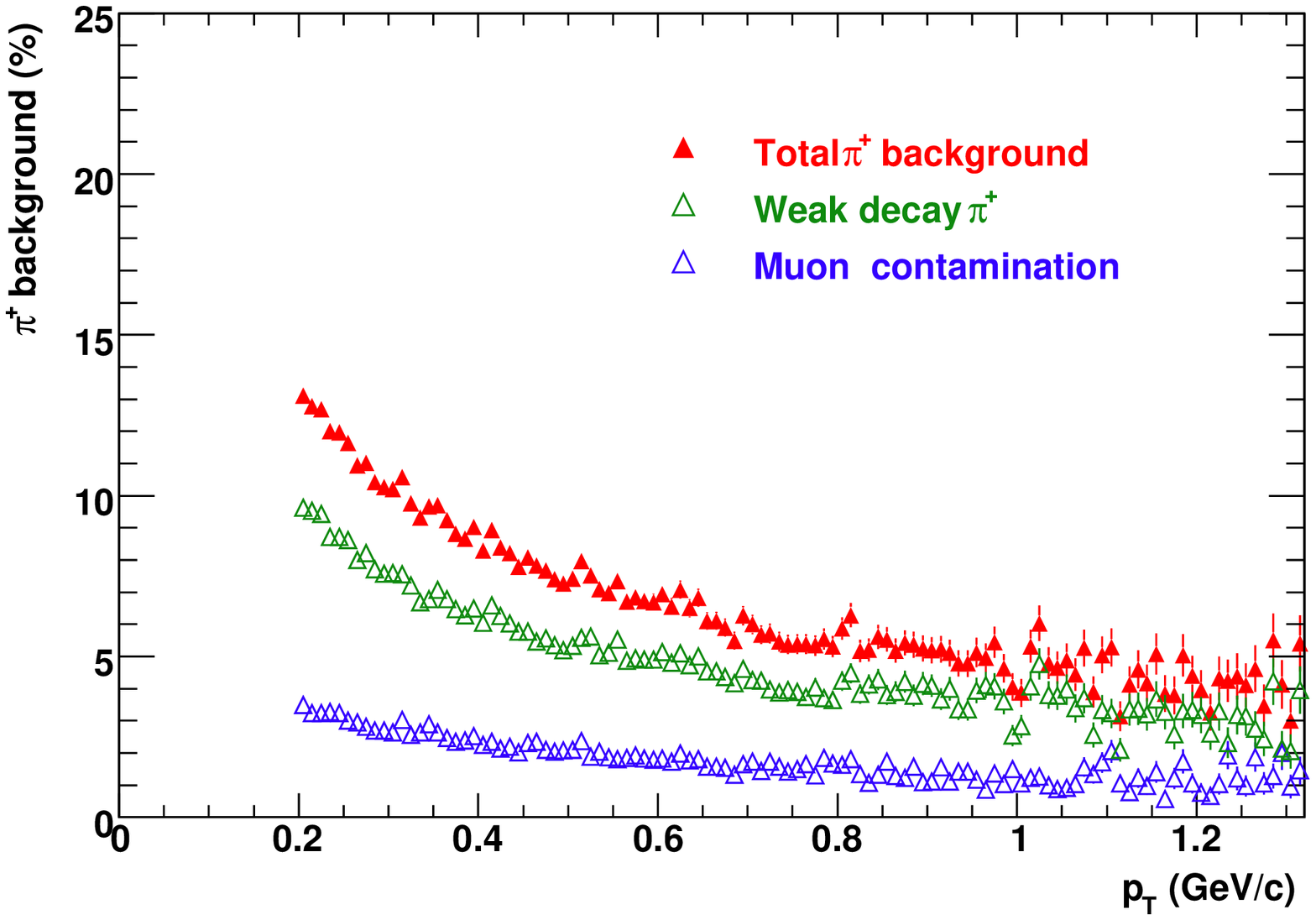}
				\caption{Averaged correction to pion spectra for weak decays ($\Lambda$,$K^{0}_{S}$ and muons) as a function of $p_{T}$ in 62.4 GeV $\bf{Au-Au}$ collisions. }
	\label{fig:pibgAuAu62}
\end{figure}

The extracted raw identified particle spectra are corrected for tracking efficiencies through the following parameterizations.
Pion tracking efficiency is fitted to the following function:
%
\begin{equation}
Efficiency\ (\pi)=c_{0}\cdot e^{-\left(\frac{c_{1}}{p_{T}}\right)^{c_{2}}}
\end{equation}
where $c_{0,1,2}$ are the parameters.
To account for the flat rising part of the kaon efficiency the following function can be used:
\begin{equation}
Efficiency\ (K)=c_{0}\cdot e^{-\left(\frac{c_{1}}{p_{T}}\right)^{c_{2}}}+c_{3}\cdot p_{T}
\label{eq:kaoneff}
\end{equation}
where $c_{0,1,2,3}$ are the parameters.
To characterize the proton/antiproton efficiency two different functions are implemented. For saturating efficiency in pp and dAu collisions  a similar form can be used as for kaons.
To account for the small decrease in the proton/antiproton efficiency a simple linear combination of the previous functions can be used in mid-central and central Au-Au collisions. Table~\ref{tab:treffpion}, Tab.~\ref{tab:treffkaon} and Tab.~\ref{tab:treffproton} show a representative set of the extracted parameters.


As Figs.~\ref{fig:treffpp},~\ref{fig:treffdau},~\ref{fig:treffauau1},~\ref{fig:treffauau2} show, these functions describe the tracking efficiencies well.

\subsection{Pion background corrections}

The pion spectra are corrected for weak decays, muon contamination and background pions from the detector materials. 
Corrections are extracted from HIJING and PYTHIA simulations propagated through the STAR geometry and reconstructed as real data.
For each simulated particle, the origin, the parent particle and the decay particle type are known. From this information, we can select pions created in the simulated collision (primary particles) from the ones created in the detector material or produced from resonance decay.

The weak-decay daughter pions are mainly from $K_{0}$ and $\Lambda$ and are identified by the parent particle information accessible from the simulation. In real data, pion decay muons can be mis-identified as primordial pions because of the similar masses of muon and pion. By selecting the parent particle information in the simulation, the muon contamination can be extracted.

Once these selections are applied, the amount of background contamination can be extracted for each transverse momentum bin as shown in Fig.~\ref{fig:pibgpp} for 200 GeV pp and in Fig.~\ref{fig:pibgdAu} for 200 GeV dAu collisions.
%
%
The magnitude of the background pion contamination falls steeply. At low transverse momentum ($p_{T}$ = $\sim$ 0.3 GeV/c) it is in the order of $\sim$ 15\% and decreases to $\sim$ 5\% at $p_{T}$ = 1 GeV/c. 

The pion background is independent of event multiplicity in 200 GeV $pp$ and d-Au collisions; therefore, a single correction is applied. In 62.4 GeV Au-Au collisions the multiplicity dependence of the pion background is week, within 1.5\% over the entire centrality range. Therefore, a single averaged correction is applied to all centralities, similarly to~\cite{Adams:2003xp}. The correction is shown in Fig.~\ref{fig:pibgAuAu62}.

\subsection{Proton background corrections}
A particle, created in the primary collision, has to travel through  the beam pipe, the SVT (and SSD) layers and the support structure and finally the TPC Inner Field Cage (IFC) before it can be detected in the TPC active drift volume. Particles traversing through the detector material create secondary particles due to the interaction with the nuclei of the detector material~\cite{Ashery:1987nt}. Radiation lengths are small for the subsystems (SVT: $<$ 6\%~\cite{Bellwied:2002ag}, SSD: 1\%~\cite{Arnold:2002wx} and TPC IFC: 0.5\%~\cite{Anderson:2003ur}) but the level of background protons is considerable at low $p_{T}$.
\begin{figure}[!h]
	\begin{center}
		\includegraphics[width=0.3\textwidth]{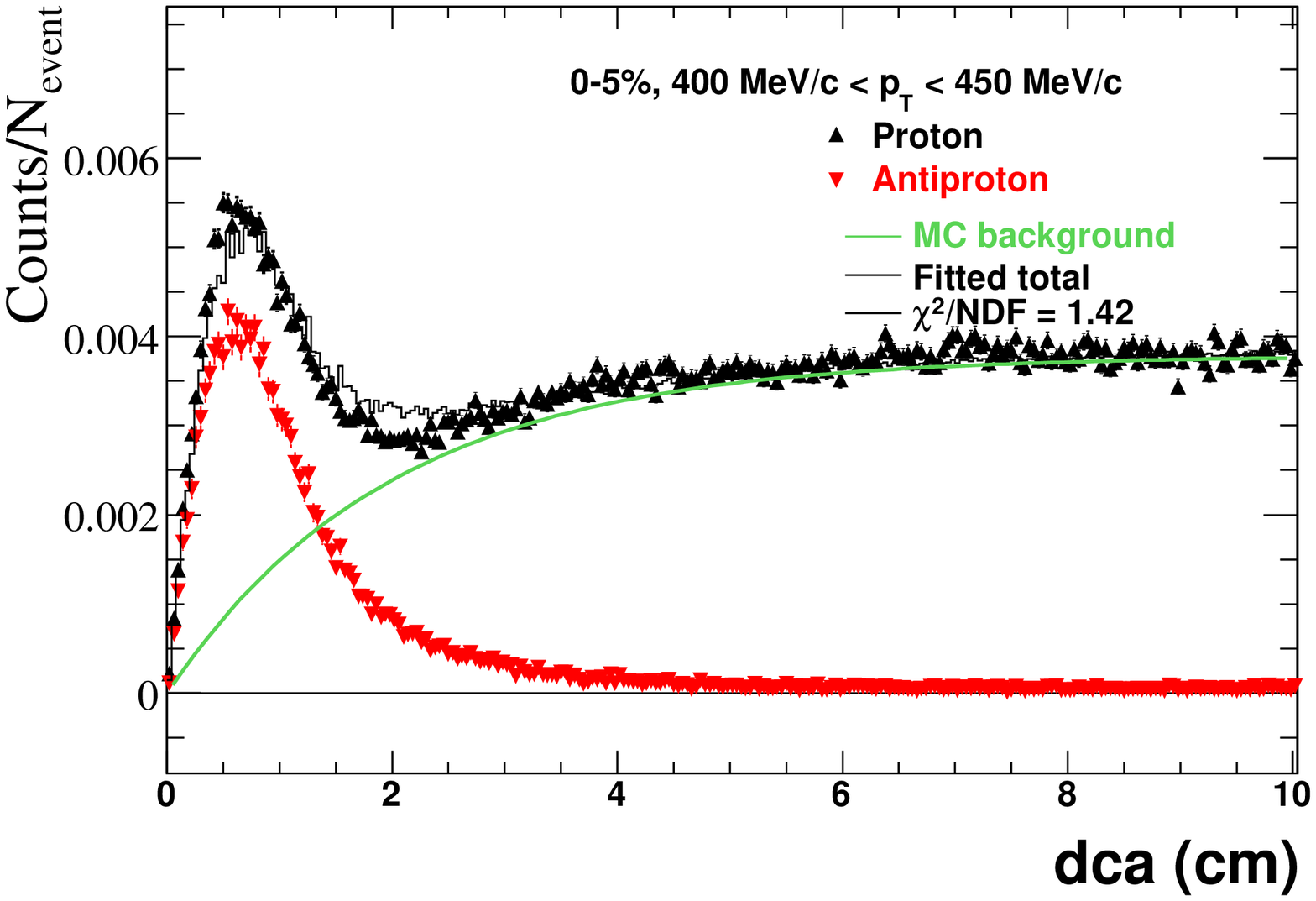}
		\includegraphics[width=0.3\textwidth]{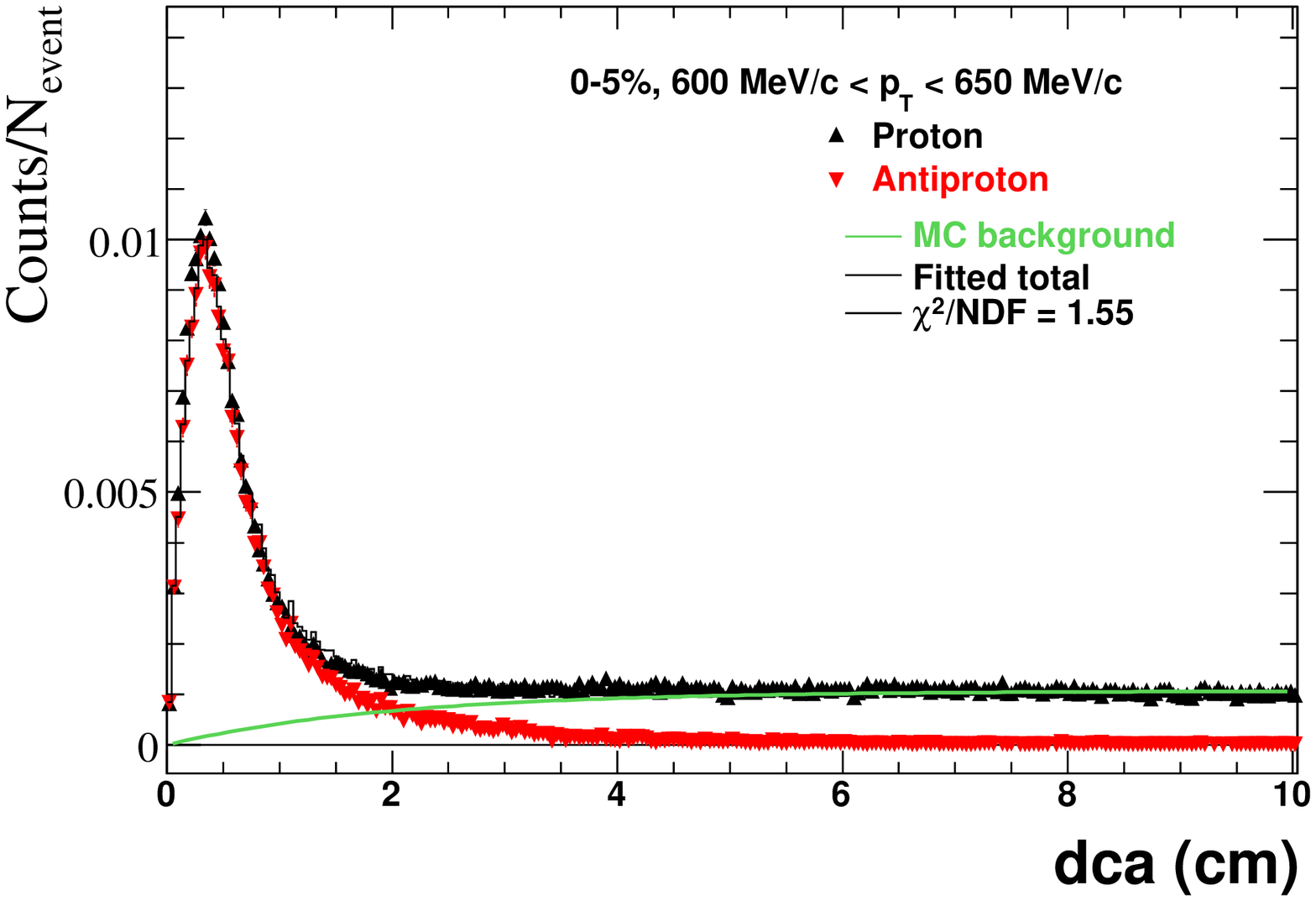}
		\includegraphics[width=0.3\textwidth]{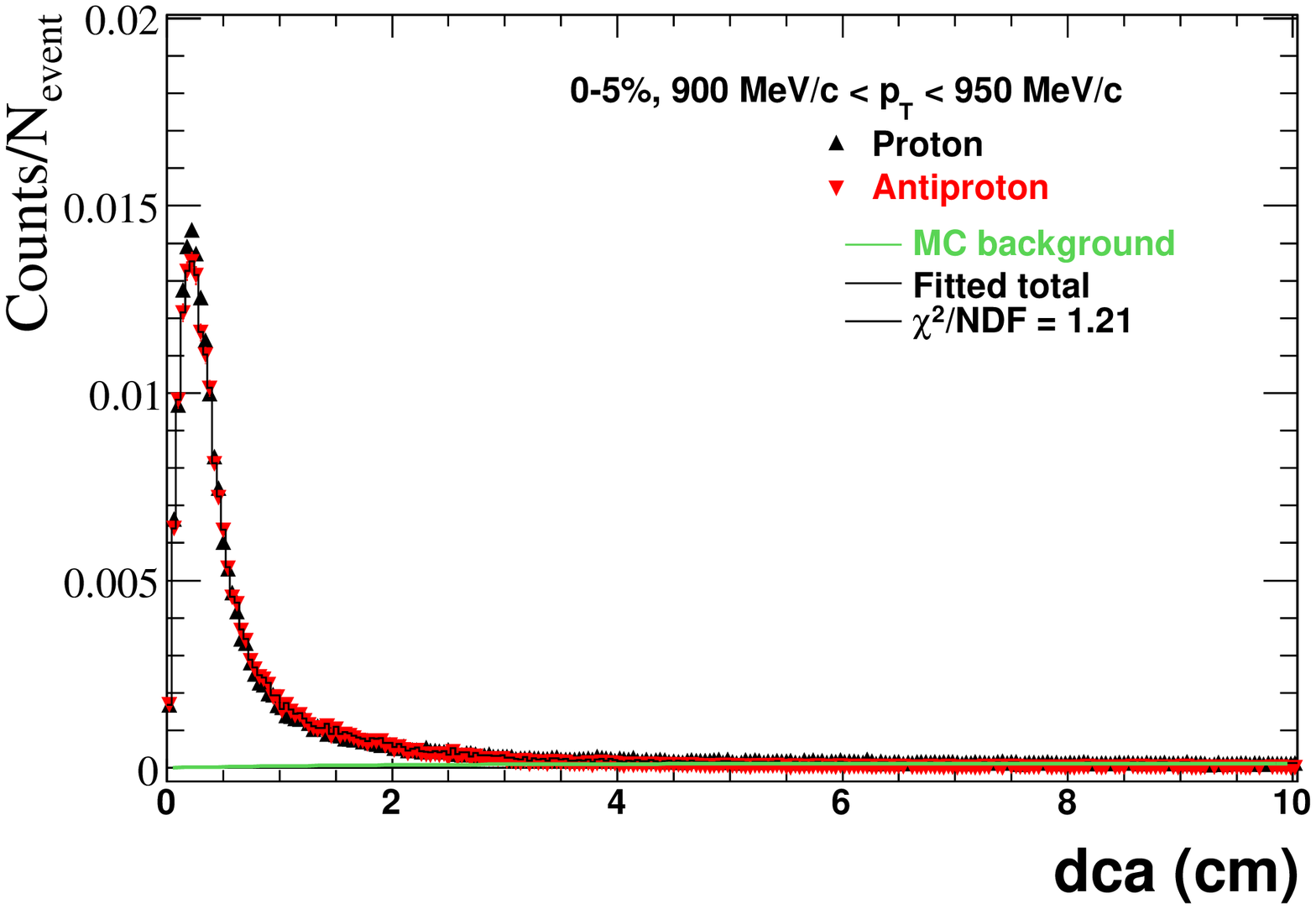}
		\includegraphics[width=0.3\textwidth]{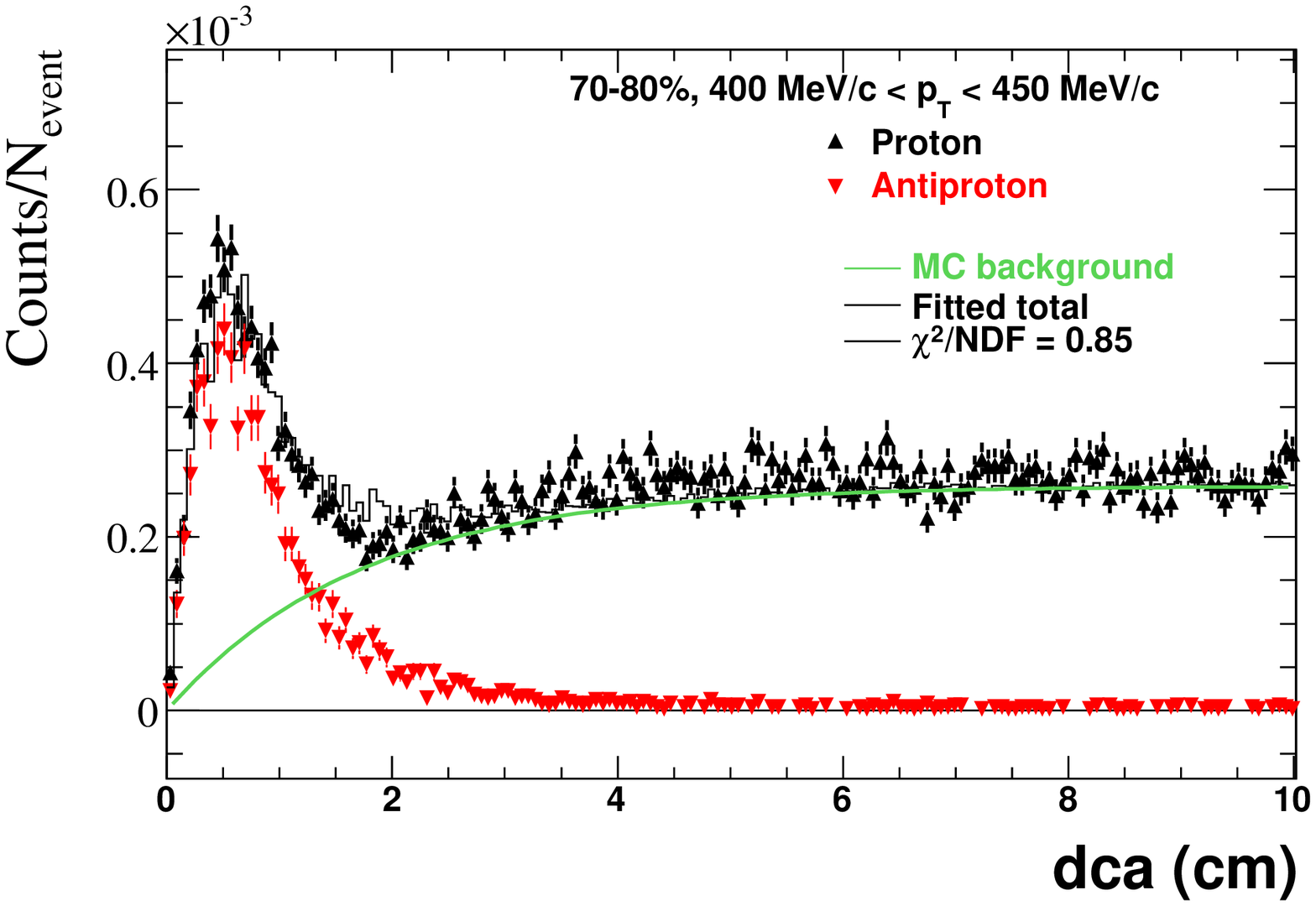}
		\includegraphics[width=0.3\textwidth]{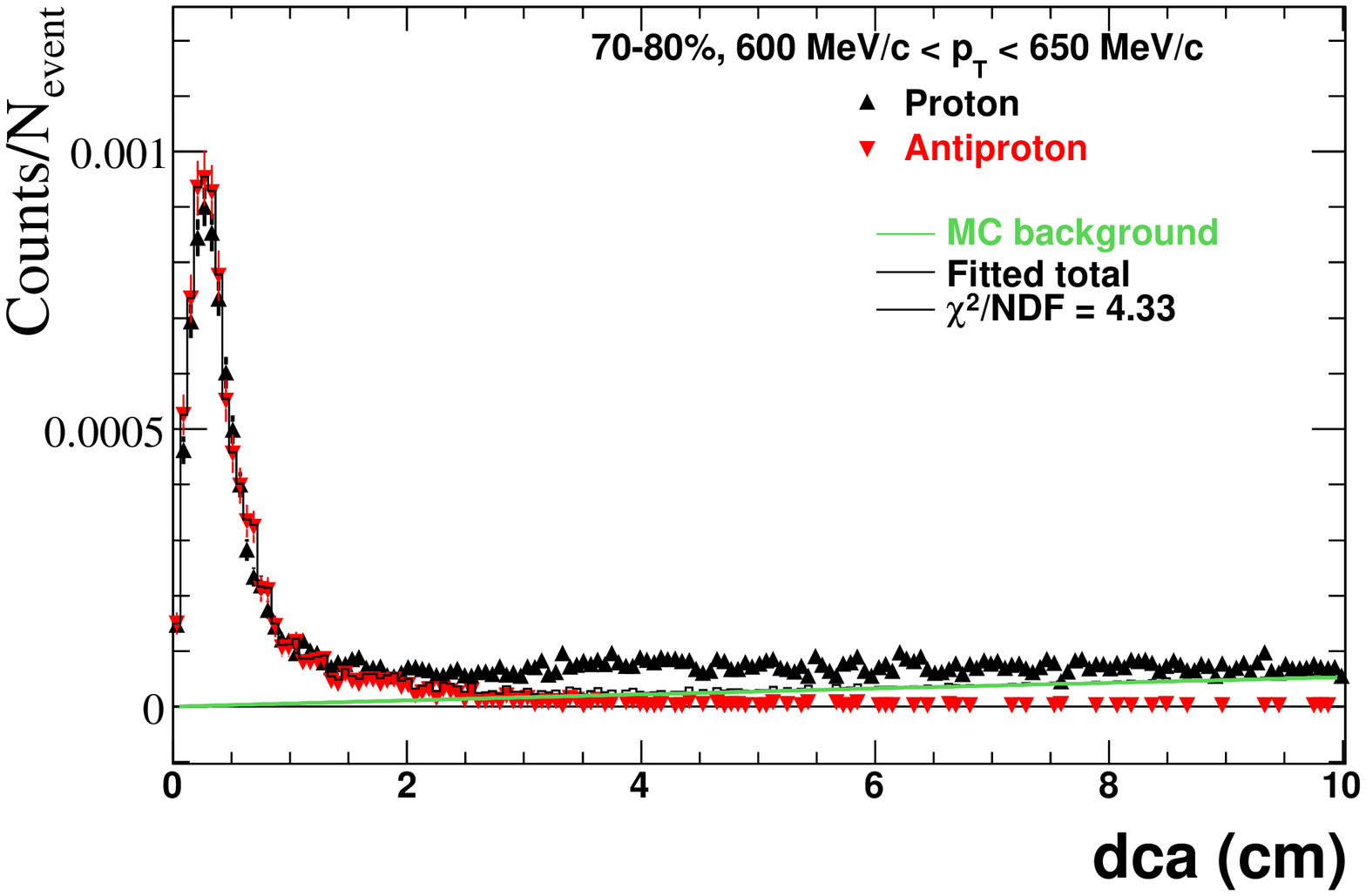}
		\includegraphics[width=0.3\textwidth]{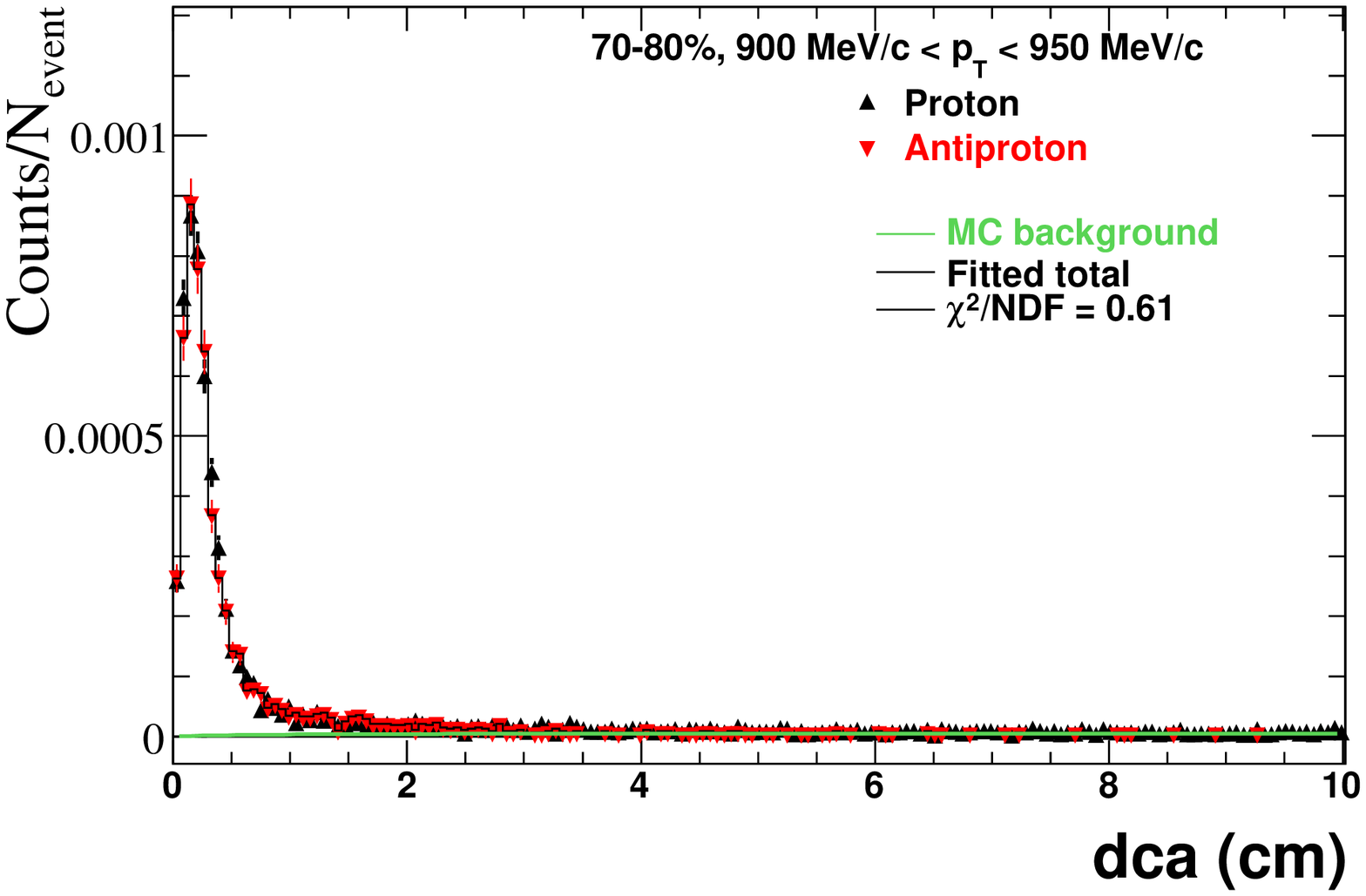}

				\caption{Sample proton/antiproton dca distributions for various transverse momenta in central (0-5\%) and in peripheral (70-80\%) 62.4 GeV Au-Au collisions.
	\label{fig:pbg_auau62_bins}}
	\end{center}
\end{figure}

\begin{figure}[!h]
	\begin{center}

\includegraphics[width=0.3\textwidth]{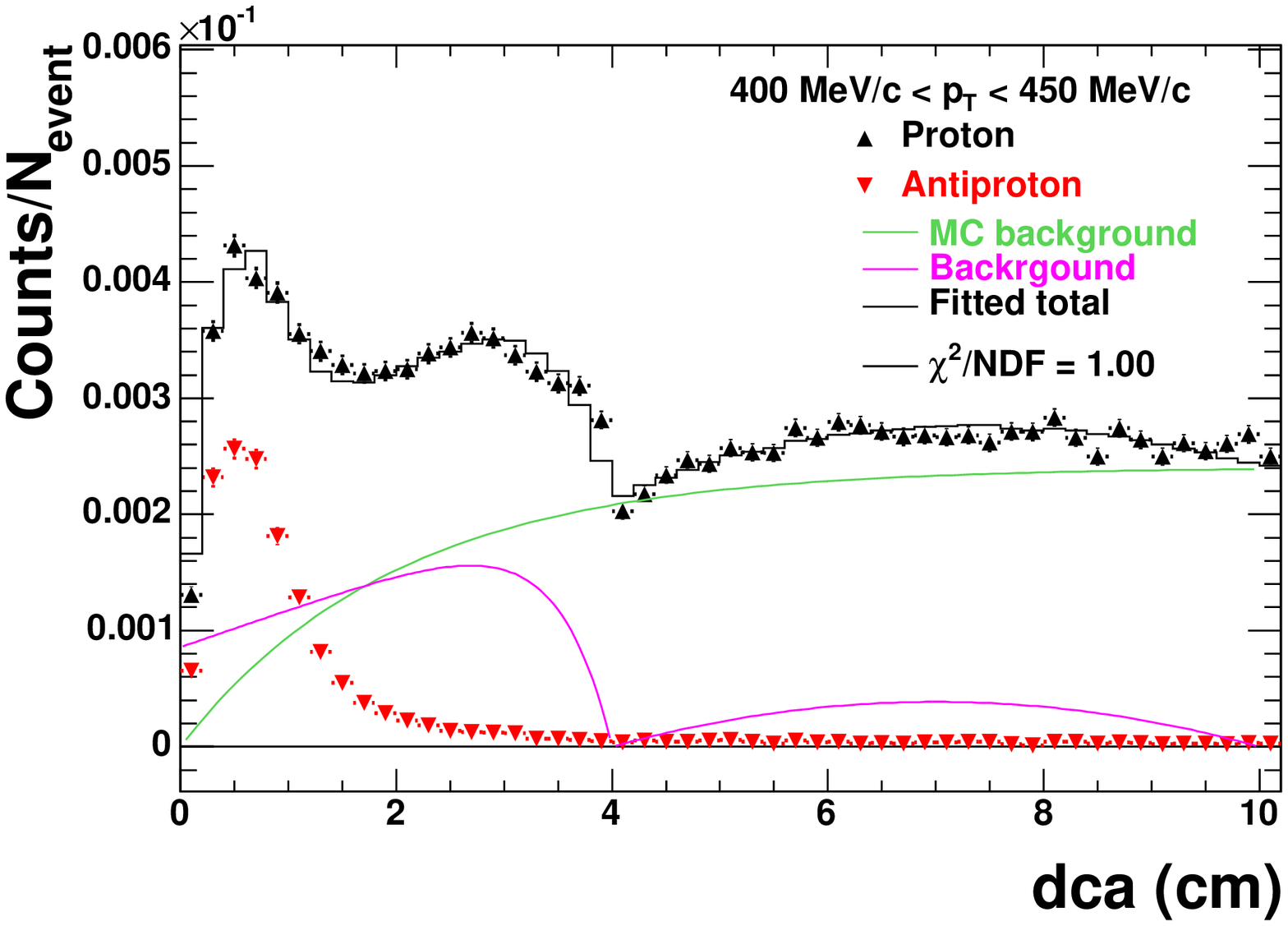}
\includegraphics[width=0.3\textwidth]{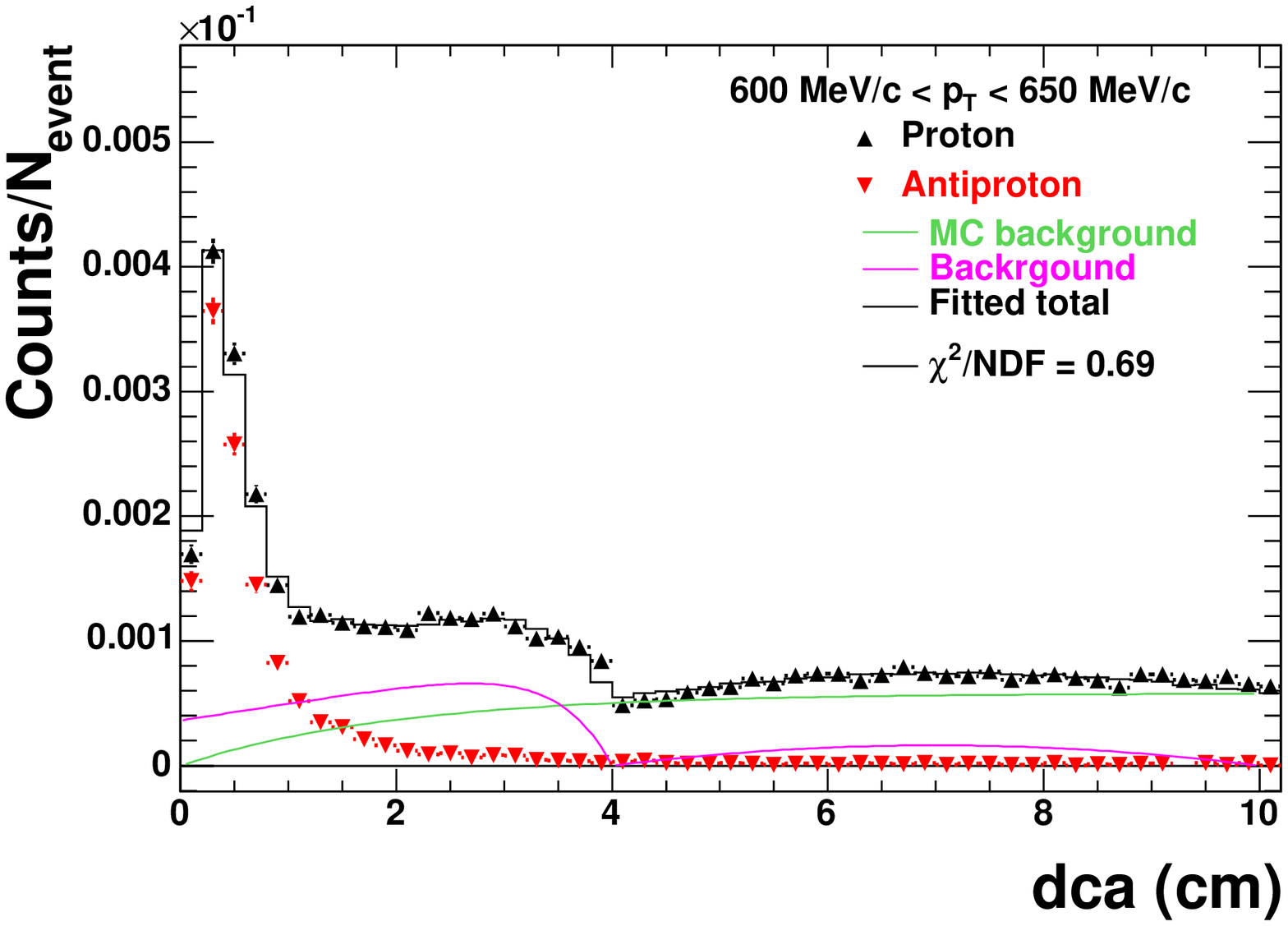}
\includegraphics[width=0.3\textwidth]{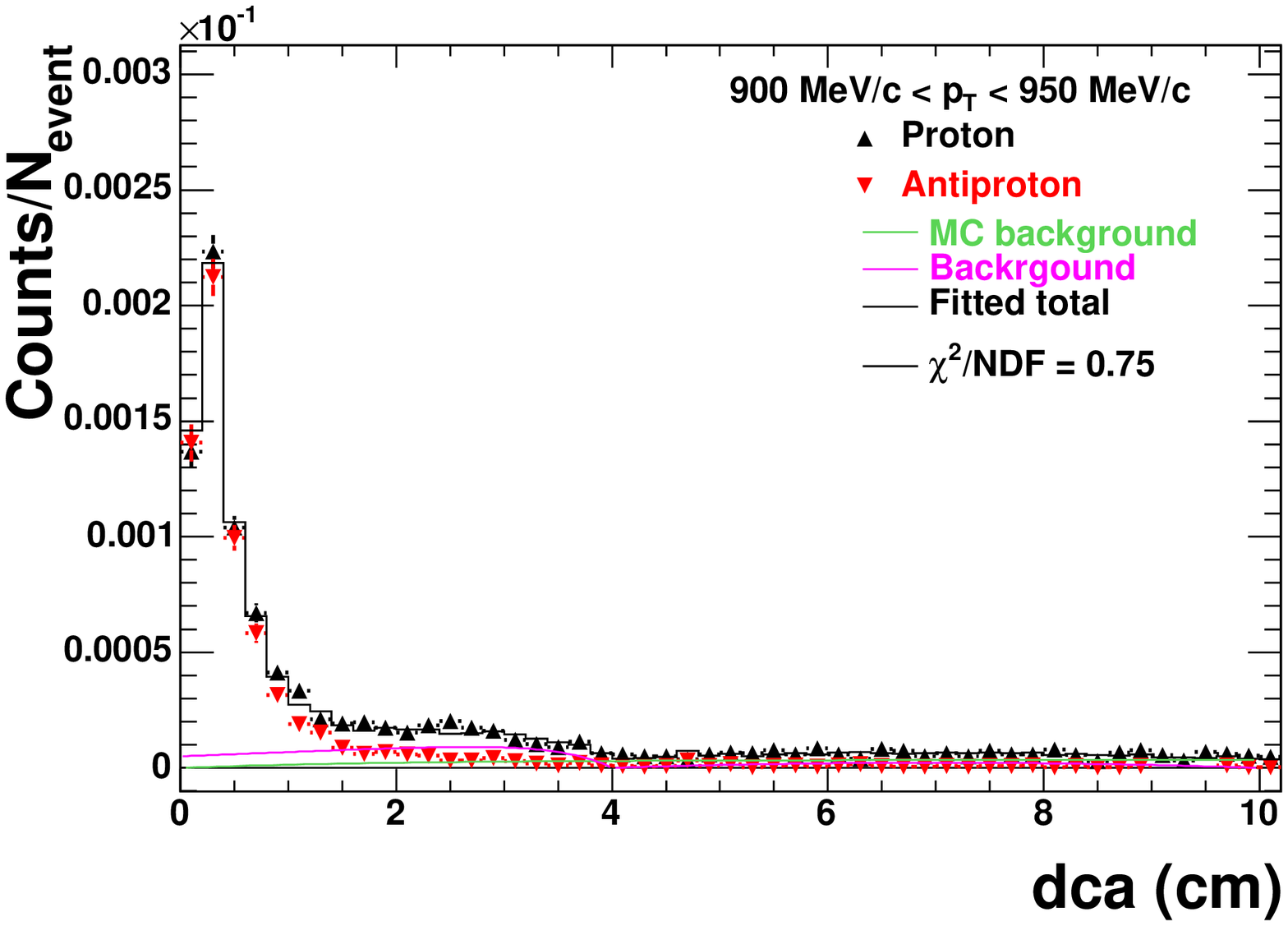}

				\caption{Sample proton/antiproton dca as a function of transverse momentum in Minimum Bias 200 GeV pp collisions.}
				\label{fig:pbg_pp_mb}
				\end{center}
	
\end{figure}
\begin{figure}[!h]
	\begin{center}

\includegraphics[width=0.3\textwidth]{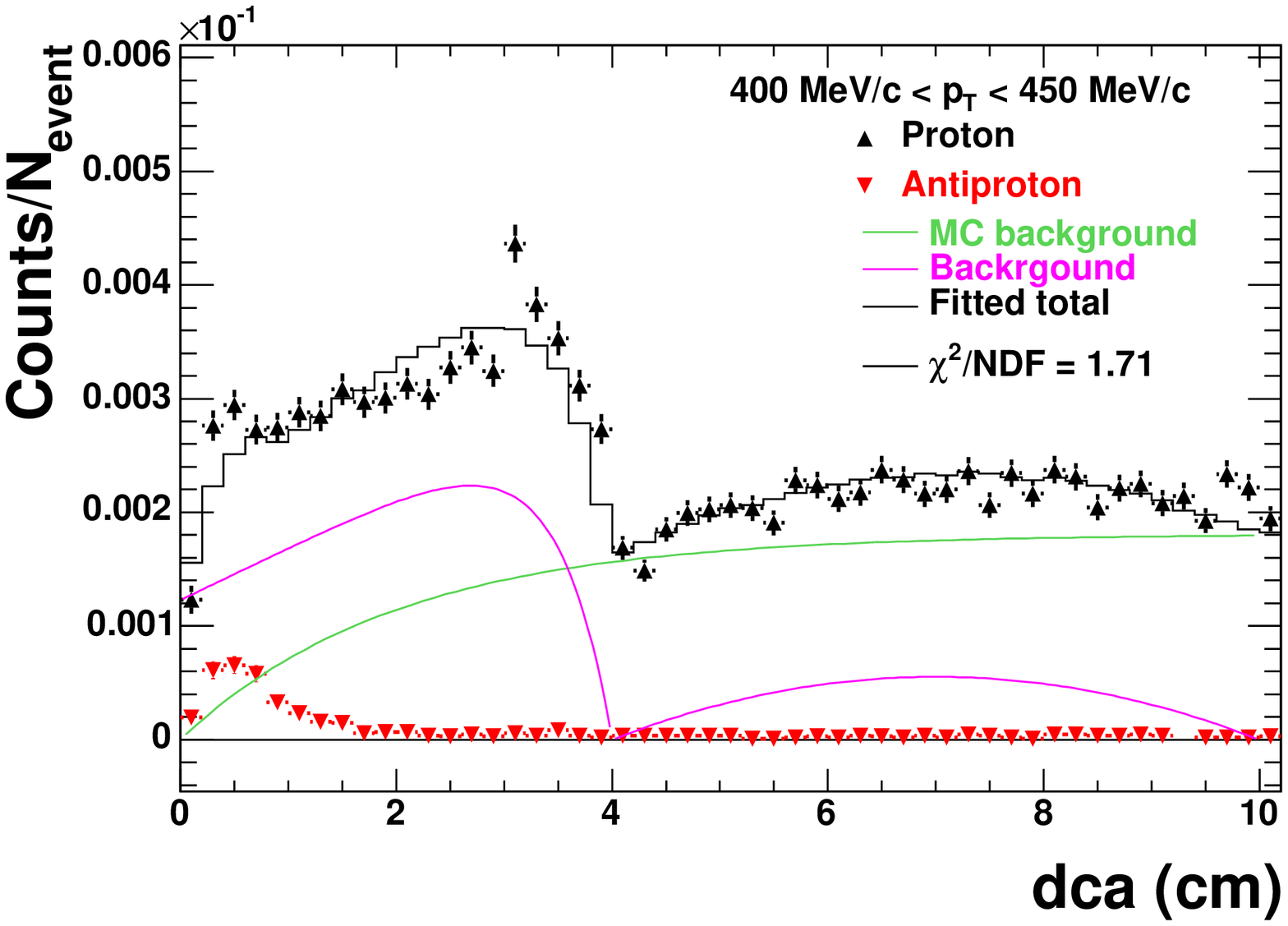}
\includegraphics[width=0.3\textwidth]{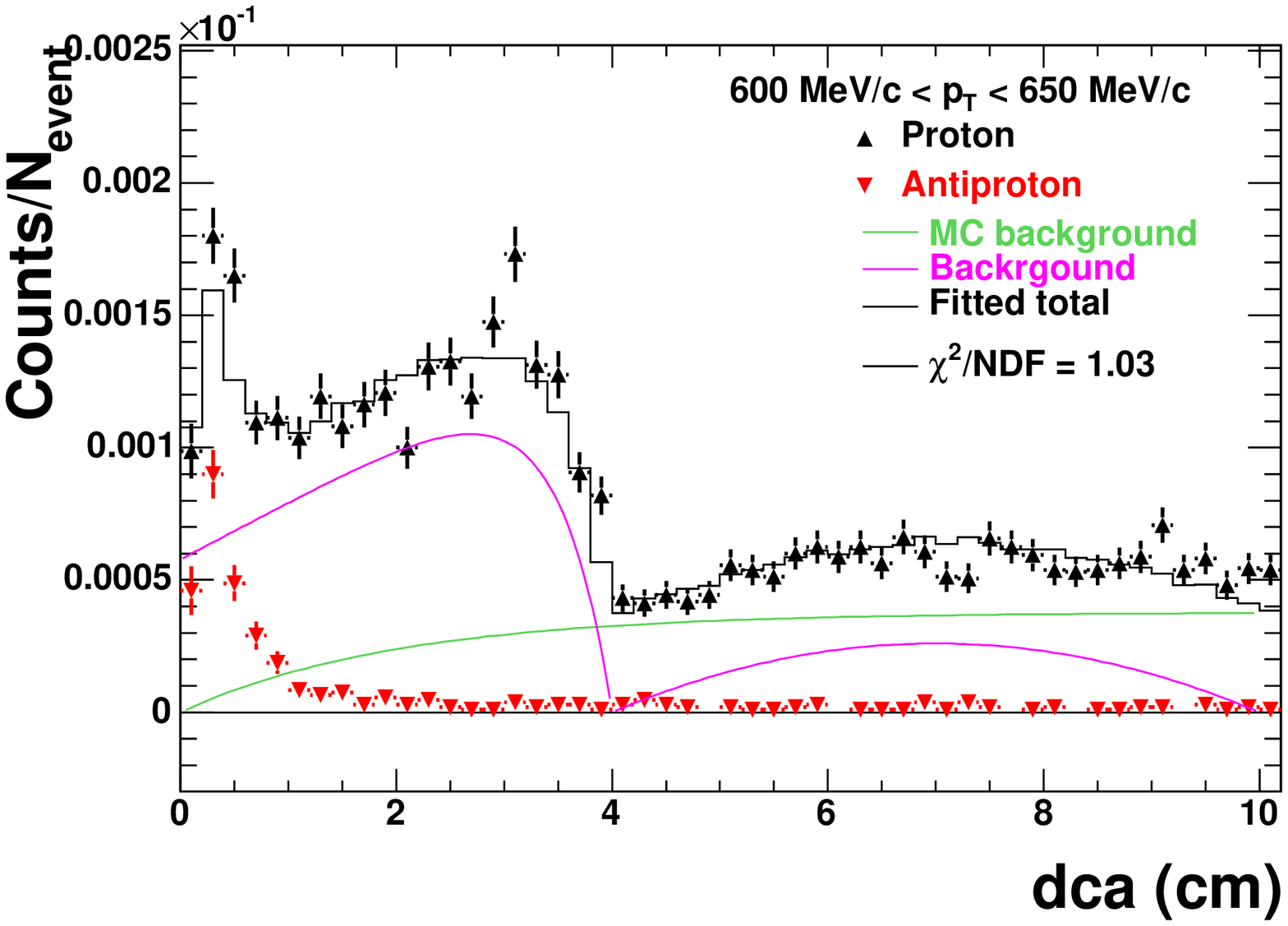}
\includegraphics[width=0.3\textwidth]{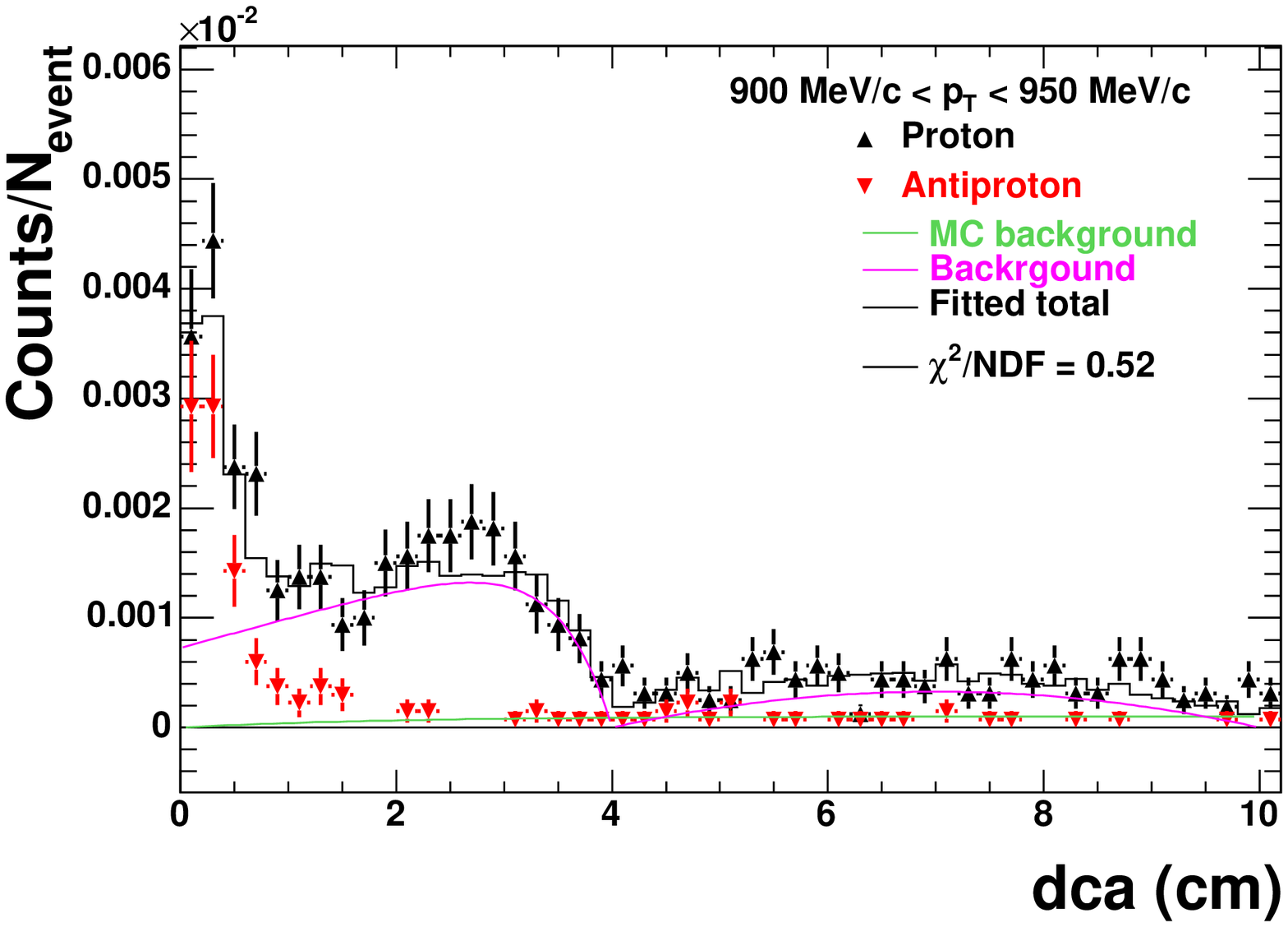}

				\caption{Sample proton/antiproton dca as a function of transverse momentum in $Nch_{0-2}$ 200 GeV pp collisions.\label{fig:pbg_pp_nch02}}
				\end{center}
	
\end{figure}

To correct for background protons, the $dca$ distributions of protons and antiprotons are extracted and compared from real data. Since the $dca$ distribution of the background protons cannot be obtained from real data, the $dca$ distribution of the background protons is obtained from embedding. The method presented here is the same as can be found in earlier STAR publications~\cite{Adler:2001bp,Adler:2001aq}. 

Due to the geometry of the detector structure, secondary protons are created far from the primary vertex (couple cm away, mainly in the beam pipe), hence their global $dca$ will be larger than for primary protons. Since antiprotons do not create secondaries their $dca$ distribution should be the same as that primary protons, hence the background contribution can be extracted as follows:
\begin{equation} 
dca_{proton}\ =\ p_{1}\cdot dca_{background\ protons}\ +\ p_{2}\cdot dca_{antiprotons}\ +\ p_{3}\cdot dca_{primary\ protons}
\label{eq:protonbggeneral}
\end{equation}
\begin{figure}[!h]
	\begin{center}

\includegraphics[width=0.3\textwidth]{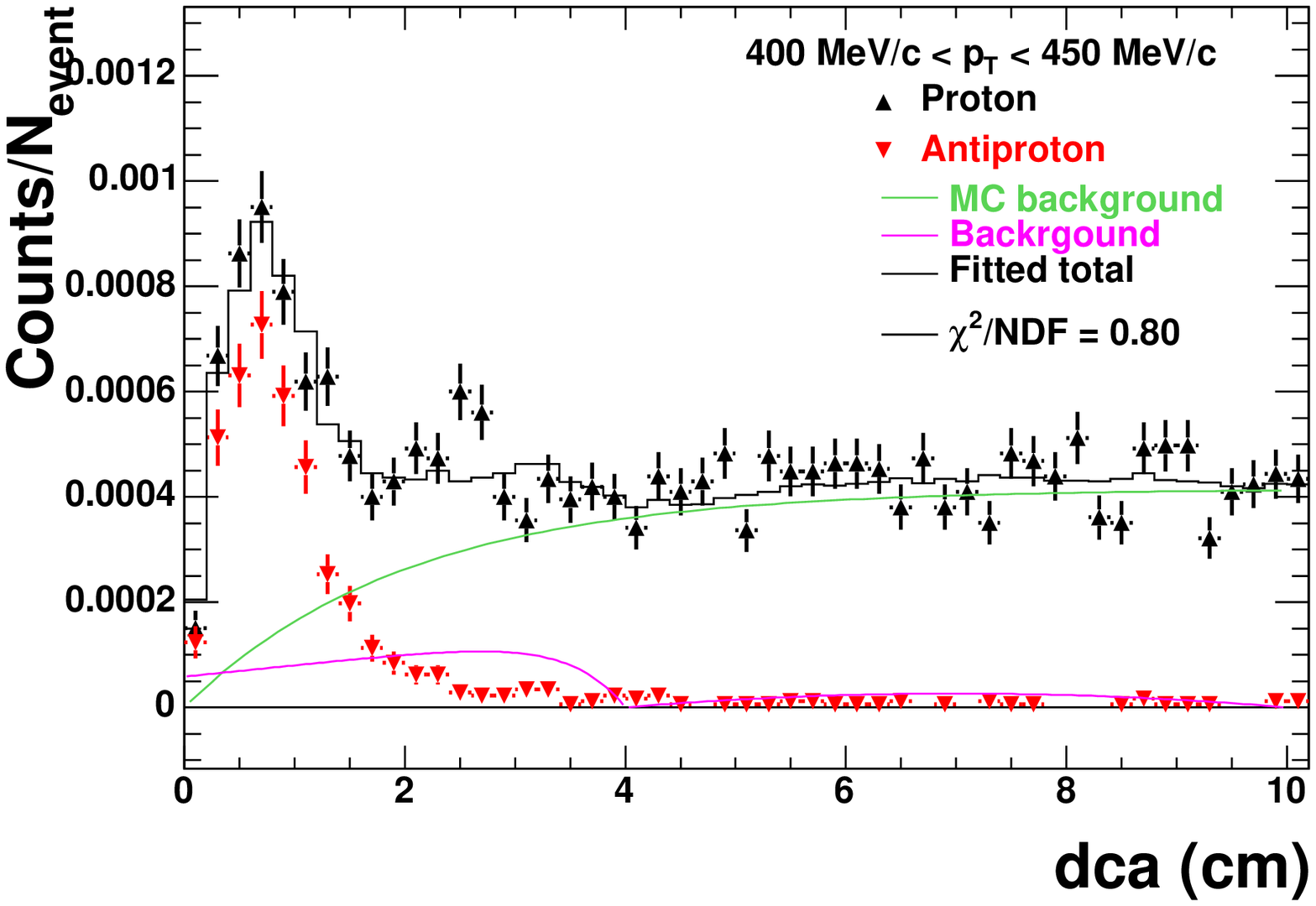}
\includegraphics[width=0.3\textwidth]{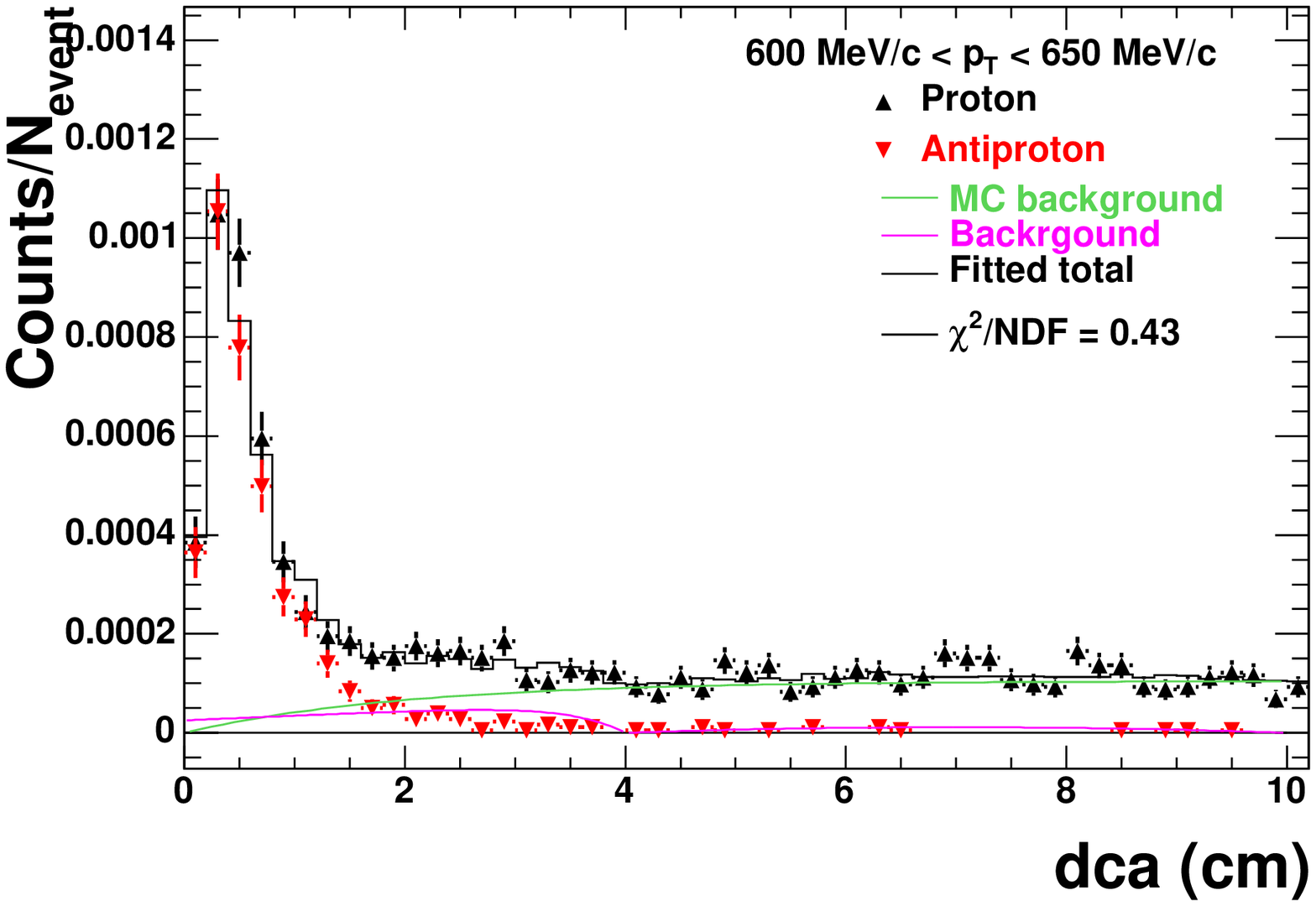}
\includegraphics[width=0.3\textwidth]{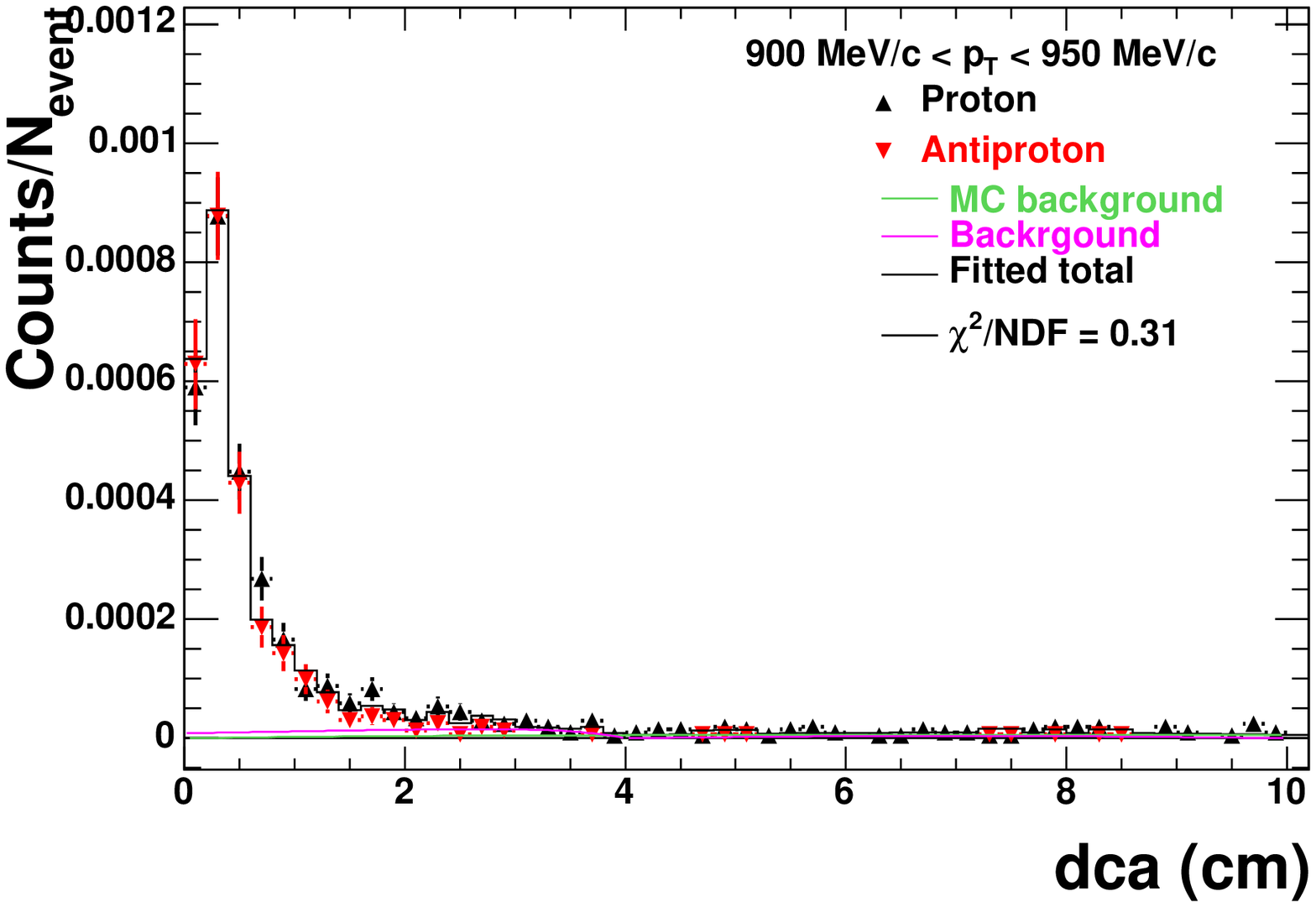}

				\caption{Sample proton/antiproton dca as a function of transverse momentum in $Nch_{9-...}$ 200 GeV pp collisions.	\label{fig:pbg_pp_nch9100}}
				\end{center}

\end{figure}
\begin{figure}[!h]
	\begin{center}

\includegraphics[width=0.3\textwidth]{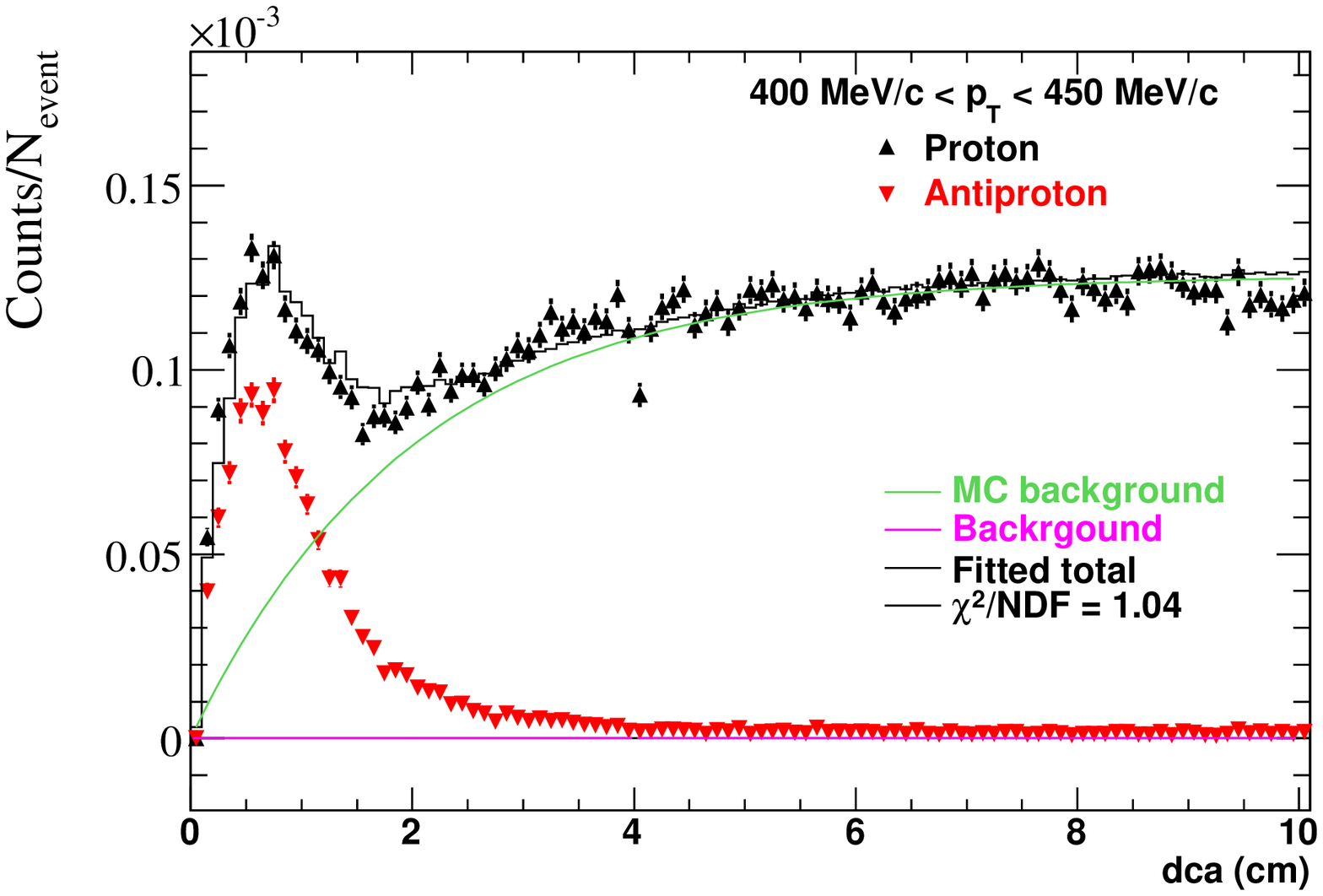}
\includegraphics[width=0.3\textwidth]{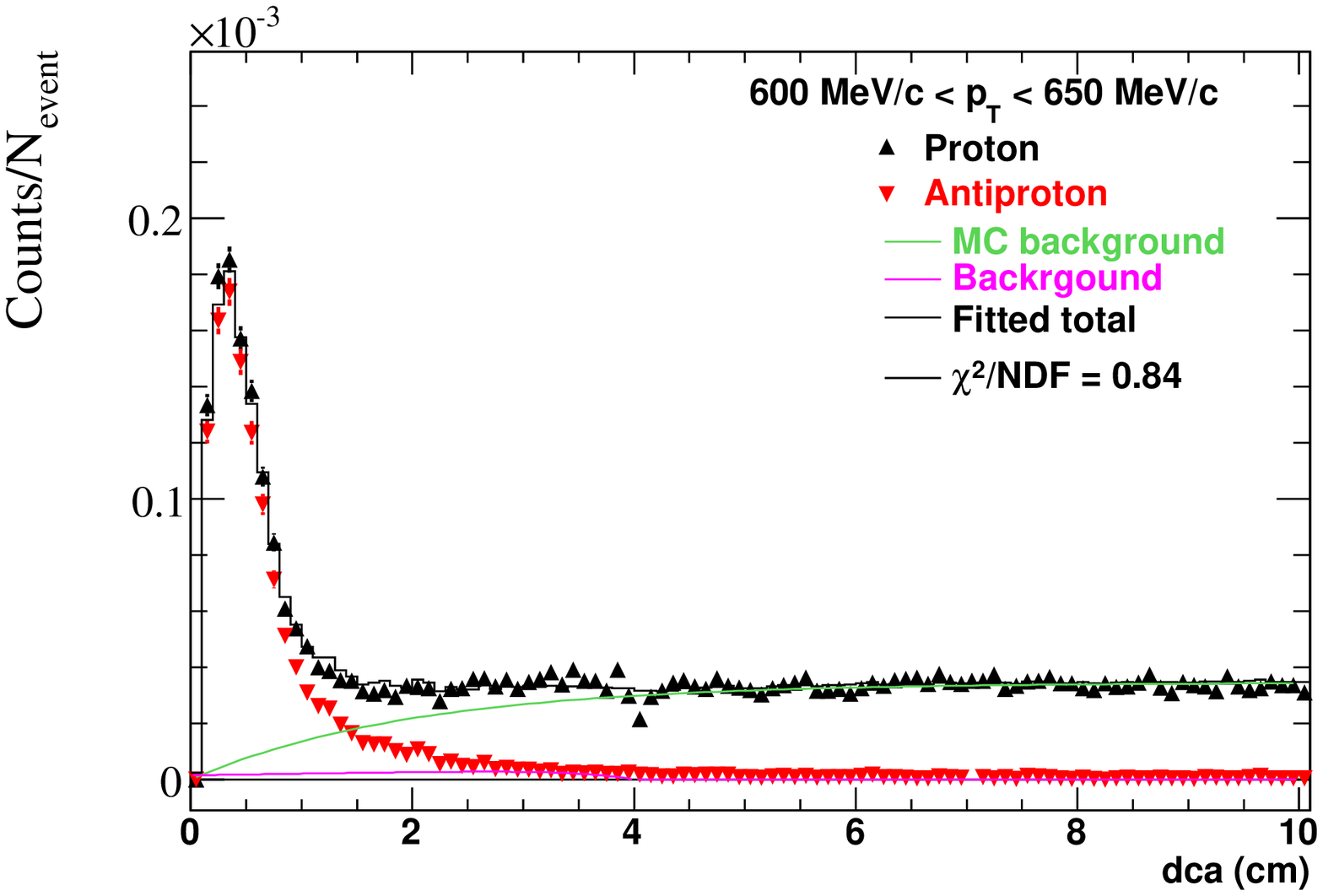}
\includegraphics[width=0.3\textwidth]{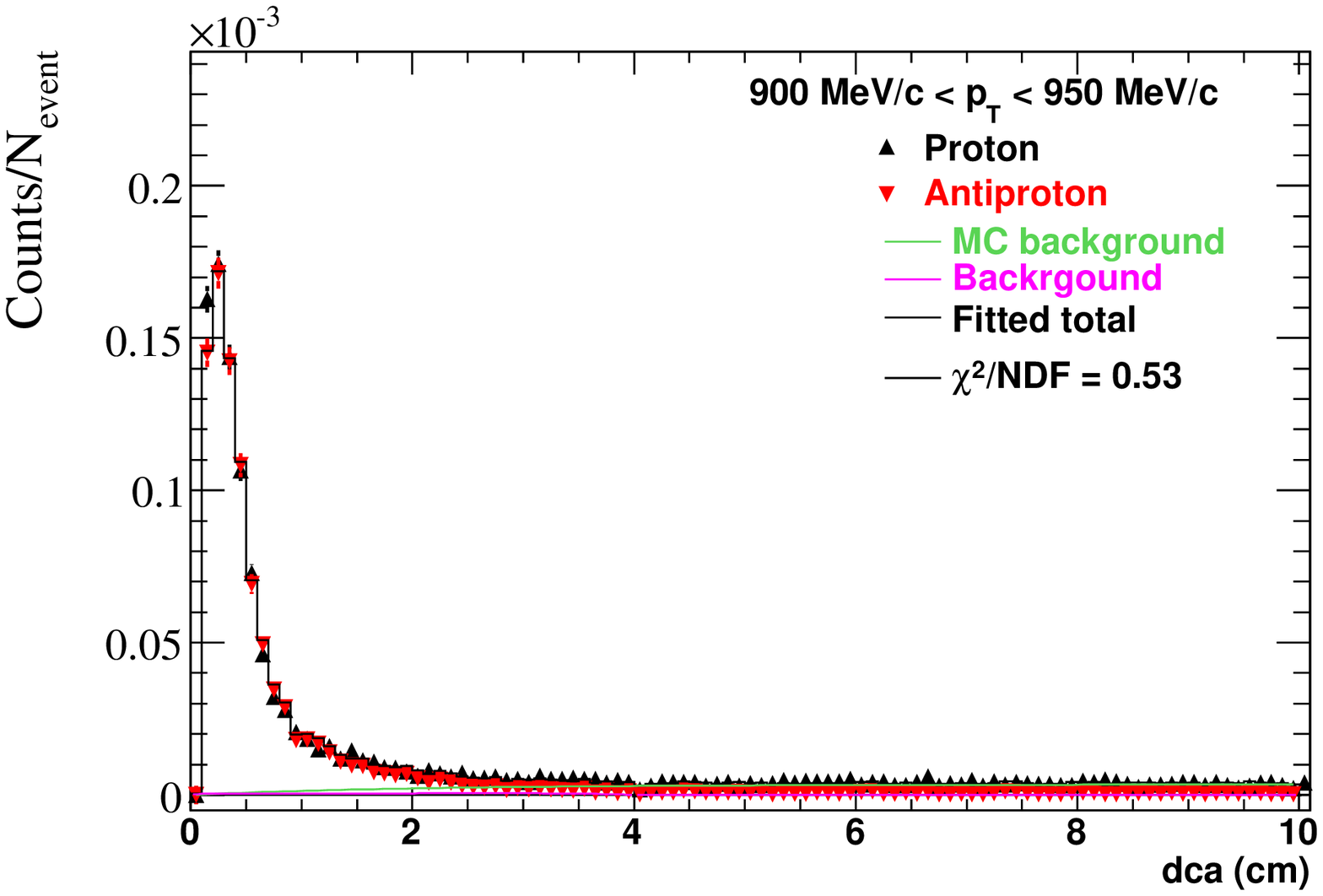}

				\caption{Sample proton/antiproton dca as a function of transverse momentum in minimum bias 200 GeV dAu collisions.}
				\end{center}
	\label{fig:pbg_dau_mb}
\end{figure}
\begin{figure}[!h]
	\begin{center}
\includegraphics[width=0.45\textwidth]{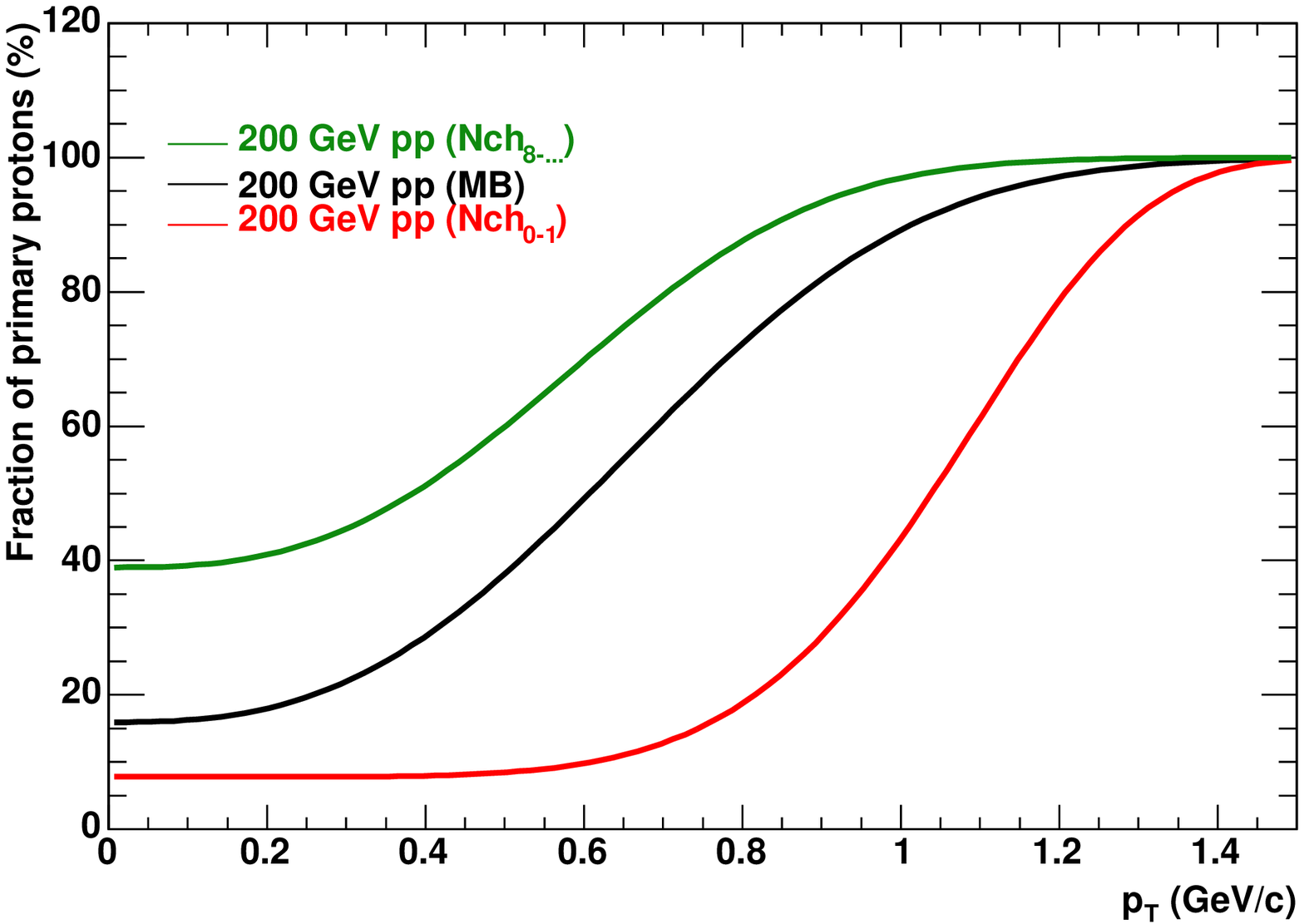}
\includegraphics[width=0.45\textwidth]{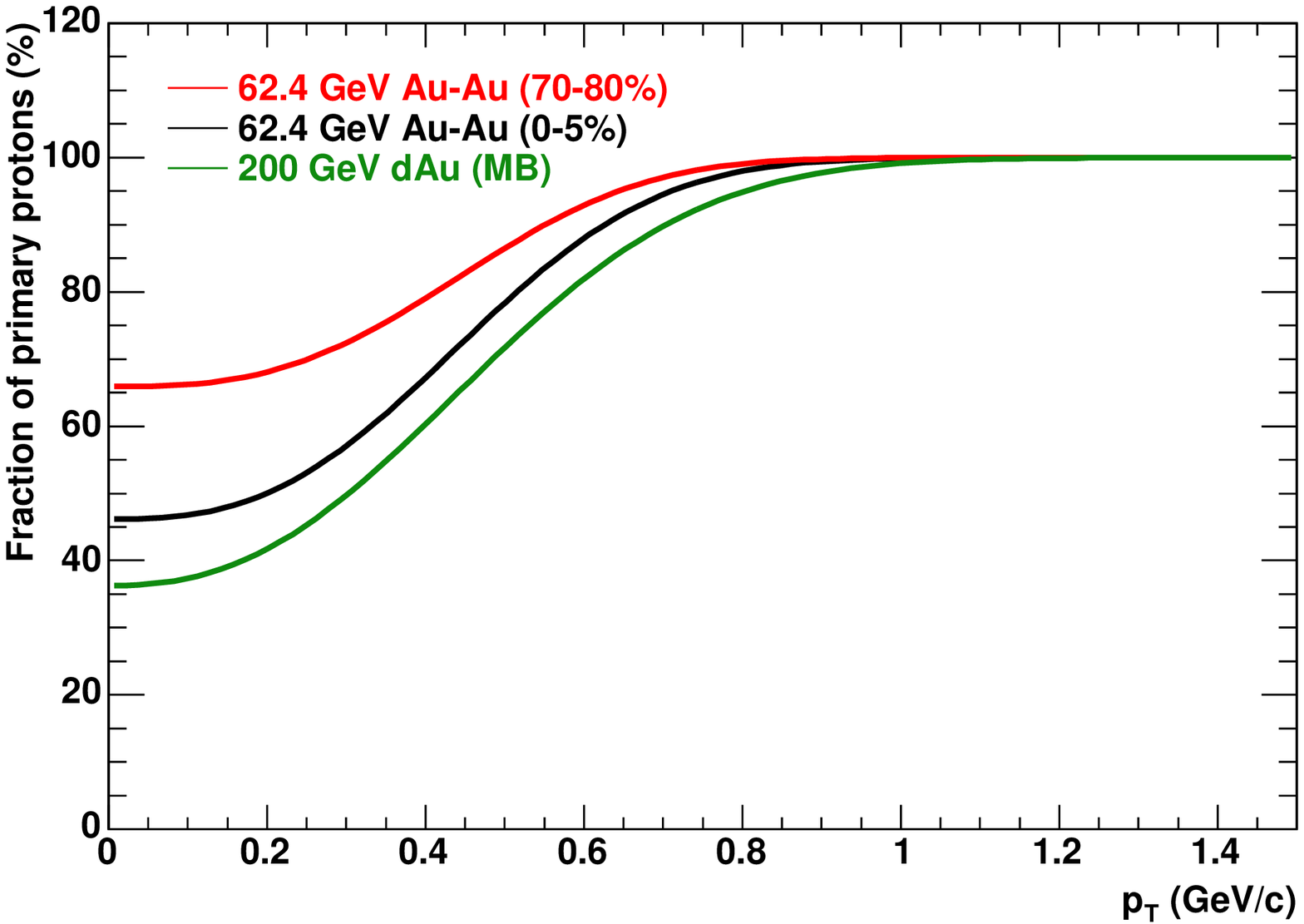}

				\caption{Fraction of primary protons as a function of transverse momentum in 200 GeV pp collisions (left panel) and in 200 GeV dAu and 62.4 GeV Au-Au collisions (right panels).\label{fig:pbg_corr_curves}}
				\end{center}
	
\end{figure}
\begin{figure}[!h]
\end{figure}

Primary tracks are selected within 3$\sigma$ of the Bethe-Bloch parameterization of the protons/anitproton energy loss bands. The $dca$ of primary tracks is defined up to 3 cm, therefore to extract proper corrections, one has to access the dca region up to 10 cm or so.
This is achieved through the mapping of global tracks to primary tracks.

In real data there exists a map between primary and global tracks, so the momentum of global tracks can be associated to the primary one (global and primary momentum can be different for the same track due to refitting) and can be corrected for energy loss of protons/antiprotons (which is obtained for primary tracks). By mapping the global tracks to the energy loss corrected primary tracks, the global dca of global tracks corrected global momentum map is created. This map is obtained for each multiplicity/centrality bin. By this mapping, proton and antiproton $dca$ distributions can be compared up to $dca \sim$ 20 cm, in each $p_{T}$ bin, as shown in Fig.~\ref{fig:pbg_auau62_bins}.

The long, nearly flat $dca$ tail in the proton distribution comes mainly from knock-out background proton, due to interactions of produced particles with detector materials. Antiprotons do not have knock-out background and the flat $dca$ tail is absent from their dca distribution. In order to correct for the knock-out background protons, the $dca$ dependence at $dca<3$~cm is needed for the knock-out protons. Such a dependence is obtained from MC simulations, and is found to be of the form~\cite{Adler:2001bp}:
\begin{equation}
p_{knocked-out\ proton}(dca) = 1-e^{-\frac{dca}{dca_0}}.
\label{eq:dca_knockout}
\end{equation}
This background contribution is indicated by the green curve in Fig.~\ref{fig:pbg_auau62_bins} and for the rest of the proton background plots in this subsection.

Assuming that the background subtracted proton $dca$ distribution is identical in shape to the anti-proton $dca$ distribution, Eq.~\ref{eq:protonbggeneral} can be written as:
\begin{equation}
p(dca) = \overline{p}(dca)/r_{\overline{p}/p} + A\cdot p_{knocked-out proton}(dca).
\label{eq:dca_fit}
\end{equation}
Data can be fitted treating the magnitude of the background protons $A$, the parameter $dca_0$, and the antiproton-to-proton ratio $r_{\overline{p}/p}$ as free parameters.

This assumption is, however, not strictly valid because the weak decay contributions to proton and anti-proton are in principle different and the $dca$ distribution of the weak decay products differs from that of the primordial (anti-)protons. However, the measured anti-lambda/lambda ratio is close to the anti-proton/proton ratio~\cite{Adler:2002uv}. The difference in $dca$ distributions between protons and anti-protons arising from weak decay contaminations is small.

The $dca$ distributions of protons and antiprotons in Au-Au collisions are fitted with Eq.~(\ref{eq:dca_fit}) in each $p_{T}$ and centrality bin, as shown in Fig.~\ref{fig:pbg_auau62_bins}.
The obtained fraction of knock-out background protons is approximately 60\% at $p_{T} = 0.35$~GeV/c and less than 5\% at $p_{T} = 1$~GeV/c. The amount of knock-out background protons depends directly on the total particle multiplicity and their kinetic energies produced in the collisions. Since the proton multiplicity over total particle multiplicty varies somewhat with centrality and the particle kinematics change with centrality, the background fraction depends on centrality. The variation from peripheral to central collisions is on the order of 10\% in Au-Au collisions. 

In pp collisions, the proton background strongly depends on the multiplicity, as shown in Fig.~\ref{fig:pbg_pp_mb} and Fig.~\ref{fig:pbg_pp_nch02}, Fig.~\ref{fig:pbg_pp_nch9100}. In large multiplicity events the evolution of the dca distributions are the same as observed in Au-Au collisions, but in low multiplicity events, even in minimum bias pp collisions, significant background excess develops. This background excess is smaller in minimum bias and peripheral dAu collisions, than in pp collisions and disappears in central dAu collisions as shown in Fig.~\ref{fig:pbg_dau_mb}. Looking at pp and peripheral and minimum bias dAu collisions in detail, on top of the long, flat $dca$ tail due to knock-out protons, there are two bumps in the proton dca distribution, one at $dca<4$~cm and the other at $4<dca<10$~cm. These bumps, absent from the antiproton dca distributions, come from other sources of proton background. The sharp drop at 4~cm suggests that they come from interactions in the beam pipe which is at radius of 4~cm. For straight tracks originating from the beam pipe and uniform in azimuth, the distribution in $dca$ (from the primary vertex of the real event) would be of the form of $1/\sqrt{1-(dca/4 {\rm cm})^2}$. This form gives not an unreasonable description of the bump at $dca<4$~cm considering the finite $dca$ resolution and curving of low momentum particles in the magnetic field. 

One possible source of such interactions is those between beam protons and the beryllium beam pipe. Such interactions are asymmetric and have a nucleon-nucleon center-of-mass energy of $\sqrt{S_{NN}}\approx 14$~GeV, which is smaller than that at the SPS. The proton-antiproton pair production rate is small compared to the number of protons transported from the initial baryons; the antiproton/proton ratio at mid-rapidity ($y\approx 2.5$) should be smaller than that at the SPS, which is on the order of a few percent. While the produced particles (pions, antiprotons, etc.) are symmetric in rapidity, the proton rapidity distributions should be peaked significantly toward the target rapidity ($y=0$) because a large fraction of the protons come from the beryllium target. The number of these protons may not be small compared to protons produced in 200 GeV pp collisions though the relative magnitudes of the bump and produced protons in a $pp$ collision depend on the rate of the background interaction per pp event. At $y=0$, the antiproton/proton ratio should be significantly smaller than that at $y\approx 2.5$, which could explain that such a background is not observable in the antiproton $dca$ distributions.

In principle, such background interactions should also be present in Au-Au running, depending on the quality of beam focusing. However, the center-of-mass of Au-beryllium interactions is shifted significantly to the Au beam rapidities. The proton (and antiproton) yield at rapidity $y=0$ should be very small. This is especially so when compared to the large multiplicity in Au-Au collisions.

The bump at $4<dca<10$~cm in the proton distribution in $pp$ could be due to interactions between the beam and the SVT and SSD materials which are located outside the beam pipe. The two-bump structure is also observed in minimum bias d-Au and peripheral d-Au collisions. The effect is not significant in the other two centralities of d-Au.

In order to correct for those background protons, the bumps are parameterized by
\begin{equation}
p_{bump,\ 0<dca<4\ cm} =\left(1.2+1.8\frac{dca}{4{ cm}}\right)\left(1-\exp \left[-7.2\left(1-\frac{dca}{4{ cm}}\right)\right]\right) \\
\end{equation}
\begin{equation}
p_{bump,\ 4<dca<10\ cm} =0.54\left[1-\left(\frac{dca}{3{ cm}}-\frac{7}{3}\right)^2\right] 
\label{eq:dca_bumps}
\end{equation}
The $dca$ distributions from $pp$ (minimum-bias and various multiplicity classes) and d-Au (minimum bias and the peripheral centrality) are fitted with
\begin{equation}
p(dca) = \overline{p}(dca)/r_{\overline{p}/p} + A\cdot p_{knock-out} + B\cdot p_{bump},
\label{eq:dca_fit_pp}
\end{equation}
treating $A$, $dca_0$, $B$, and $r_{\overline{p}/p}$ as free parameters. For the two non-peripheral d-Au centralities, the same fit procedure using Eq.~\ref{eq:dca_fit}, as for the Au-Au data, is used.
Different curves in Fig.~\ref{fig:pbg_pp_mb} represent the various background contributions. The green curve is the knock-out proton background described by Eq.~\ref{eq:dca_knockout}. The pink curve is the additional proton background described by Eq.~\ref{eq:dca_bumps}. The black histogram is the fit results by Eq.~\ref{eq:dca_fit_pp}. The fitted background protons are subtracted from the proton data. 

The knock-out proton background should scale with the event multiplicity, as the fit results indicate. If the beam-material interaction is responsible for the two-bump structure, then the magnitude of such background should be independent of the event multiplicity. This seems to be supported by the fit results.
The parameter $dca_0$ is found to be approximately 2.0 in pp, d-Au and varies between 2.5 - 1.4 in Au-Au collisions with increasing $p_{T}$ and shows almost no (weak) centrality dependence.

Figure~\ref{fig:pbg_corr_curves} shows the fits to the obtained primary proton fractions as a function of $p_{T}$ in minimum bias pp and dAu, and for two multiplicities/centralities in pp and Au-Au. In higher multiplicity and minimum bias collisions the primary proton fraction rises quickly from $\sim$ 30-40 \% to 100 \% in the measured proton spectra range. The lowest multiplicity pp bin carries significant proton background even at larger transeverse momentum.  In dAu and Au-Au collisions the primary proton fraction is larger and steeply increases to 100 \%. Corrections depend on multiplicity/centrality therefore the primary proton fraction is calculated for each multiplicity/centrality selection.

\chapter{Results and discussion}

In this section we present the transverse momentum spectra of the invariant yield $\left(\frac{1}{2\pi p_{T}}\frac{dN}{dydp_{T}}\right)$ for  identified charged particles: $\pi^{\pm}$, $K^{\pm}$, p and $\bar{p}$ in 200 GeV pp, 200
GeV dAu and in 62.4 GeV Au-Au collisions. Their $\left\langle p_{T}\right\rangle$, dN/dy, and particle ratios are studied.
Bulk properties extracted from chemical and kinetic model fit results are discussed.

\section{Identified particle spectra}
\begin{figure}[!h]
	\begin{center}
		\includegraphics[width=0.95\textwidth]{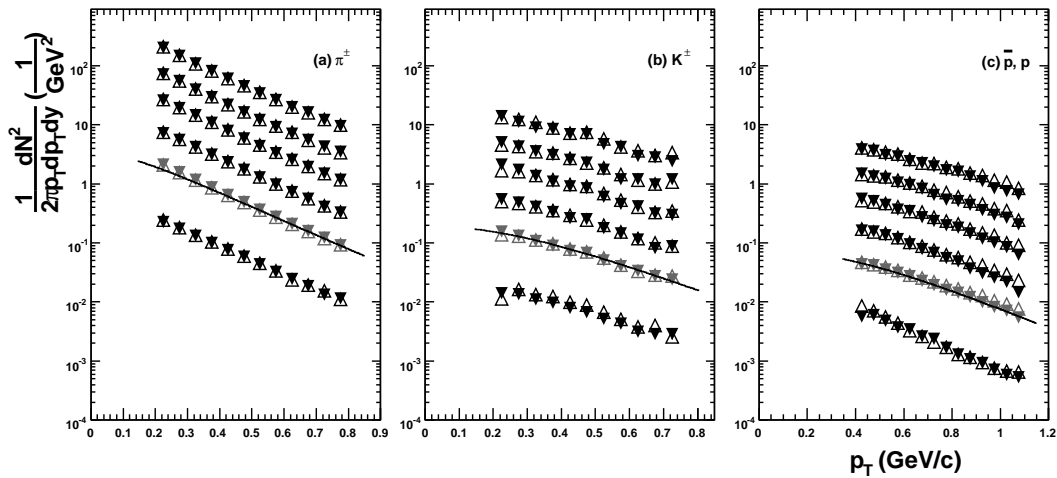}

\caption{ Inclusive identified particle spectra (negative particles: filled symbols, positive particles: empty symbols) measured in 
$\left|y\right|<$ 0.1 in 200 GeV \textbf{pp} collisions. Spectra are plotted
 for five multiplicity bins and the minimum bias spectra are also shown (grey markers). Curves represent Bose-Einstein (pions) and Blast-wave fits (kaons and protons/antiprotons).
  Errors are statistical and point-to-point systematic errors added in quadrature. Spectra are scaled, see text for details.}
	\label{fig:ppspectra}\end{center}
\end{figure}
%

%
Fig.~\ref{fig:ppspectra} shows the minimum bias $\pi^{\pm}$,
$K^{\pm}$, p and $\bar{p}$ spectra in 200 GeV pp collisions. 
Spectra are measured at mid-rapidity in $\left|y\right|<$ 0.1 
in minimum bias collisions and for five multiplicity bins. 
For each particle species, spectra are scaled for clarity, 
except the minimum bias ones. 
%
%
%
\begin{figure}[!t]
	\begin{center}

		\includegraphics[width=0.95\textwidth]{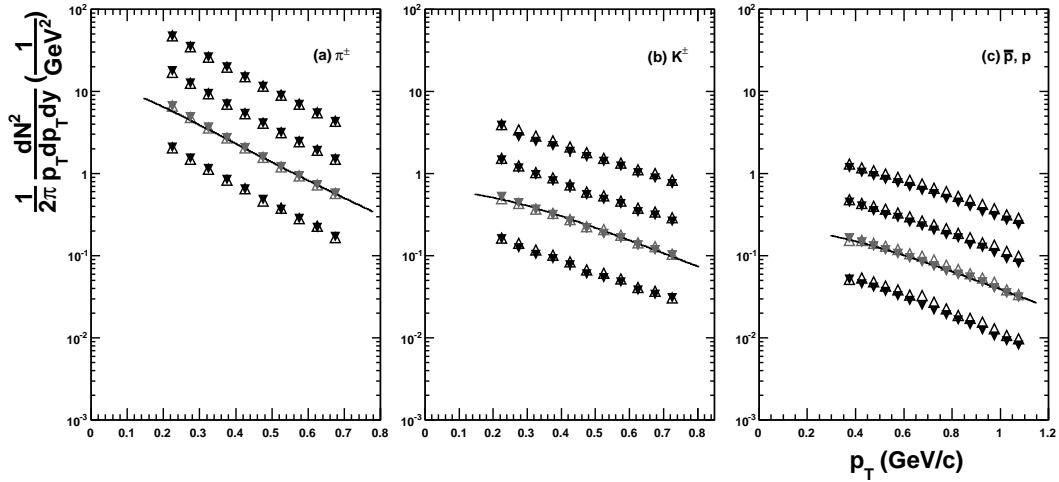}
\caption{ Inclusive identified particle spectra (negative particles: filled symbols, positive particles: empty symbols) measured at  mid-rapidity $\left|y\right|<$ 0.1 in 200 GeV \textbf{dAu} collisions. Grey spectra points represent minimum bias collisions.
Curves represent Bose-Einstein (pions) and Blast-wave fits (kaons and protons/antiprotons).
Errors are statistical and point-to-point systematic errors added in quadrature. Spectra are scaled, see text for reference.}
	\label{fig:dauspectra}\end{center}
\end{figure}
%
%
%
%
The top spectra correspond to 8 $\leq$ Nch and are scaled by a factor of 
32. The next spectra represent 6 $\leq$ Nch $\leq$ 7 and are scaled
by a factor of 16. The third spectra from top represent 4 $\leq$ Nch $\leq$
5 and are scaled by a factor of 8. The next spectra represent 2 $\leq$ Nch 
$\leq$ 3 and are scaled by a factor of 4. The next spectra are the minimum
bias spectra and not scaled. The bottom spectra represent 0 $\leq$ Nch
$\leq$ 1 and are scaled by a factor of 0.5. 
\begin{figure}[!t]
	\begin{center}
	\includegraphics[width=0.95\textwidth]{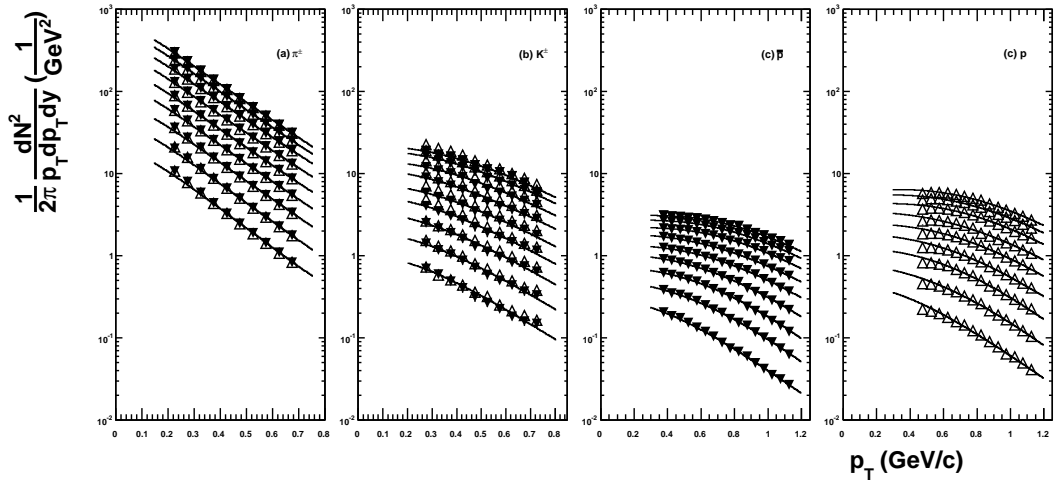}
	
\caption{Centrality dependence of identified particle spectra in 62.4 GeV 
\textbf{Au-Au} collisions (negative particles: filled symbols, positive particles: empty symbols). Spectra from top to bottom are: 0-5\%, 5-10\%, 10-20\%, 20-30\%, 30-40\%, 40-50\%, 50-60\%, 60-70\% and 70-80\%. Curves represent Bose-Einstein (pions) and Blast-wave fits (kaons and protons/antiprotons). Errors are statistical and point-to-point systematic errors added in quadrature.}
	\label{fig:auau62spectra}\end{center}
\end{figure}
\begin{figure}[!h]
	\begin{center}
			\includegraphics[width=0.9\textwidth]{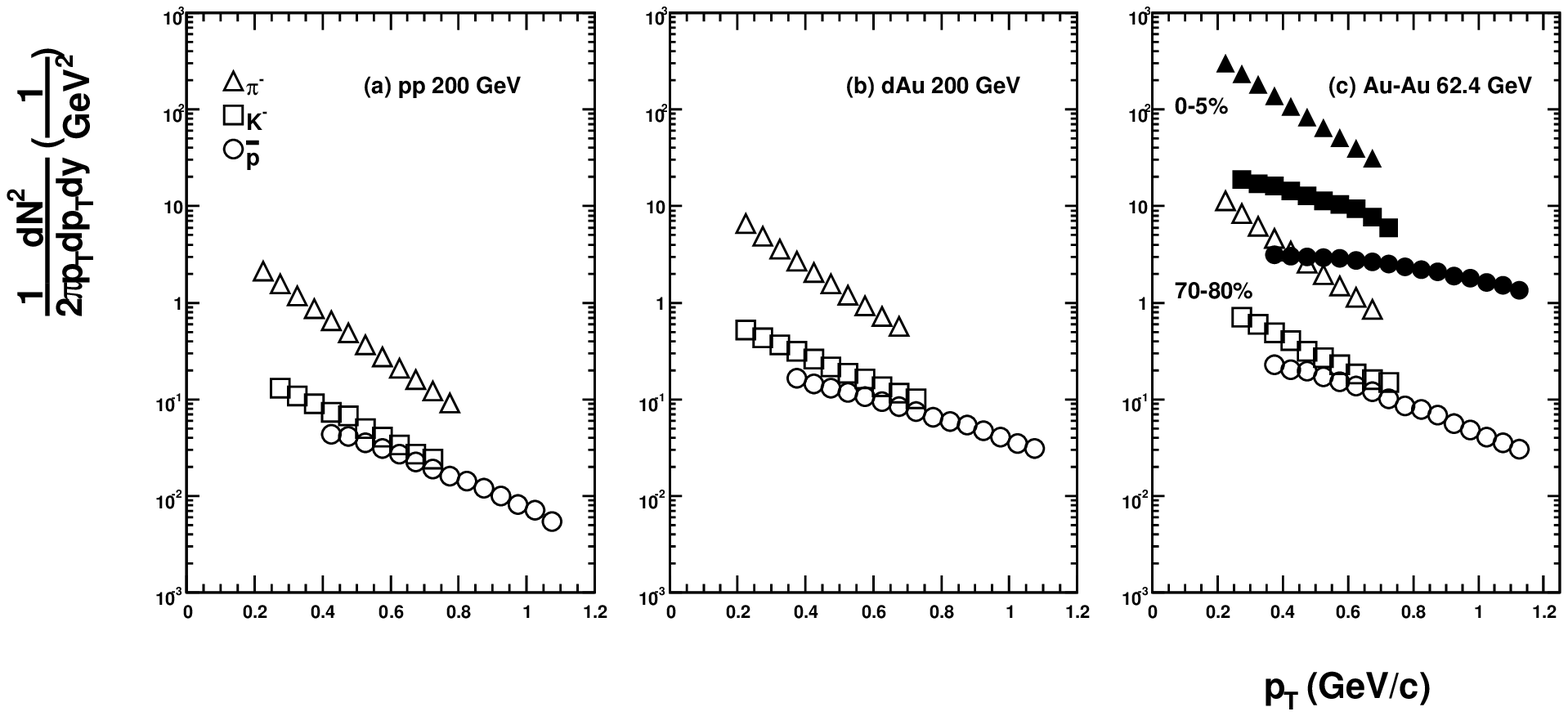}		
\caption{Transverse momentum spectra in 200 GeV pp, dAu and 62.4 GeV Au-Au collisions. In panel (c): filled spectra points represent 0 - 5 \% central collisions, empty spectra points represent 70 - 80 \% peripheral collisions. Errors are statistical and point-to-point systematic errors added in quadrature.}
	\label{fig:allspectra}
		\end{center}
\end{figure}

Minimum bias pp spectra are in good agreement with the previously 
published pp results from STAR~\cite{Adams:2003xp}.
Pion spectra shapes are similar in each multiplicity bin. However, kaons and 
protons/antiprotons show a small gradual flattening with increasing multiplicity
and this effect is more pronounced for protons/antiprotons. 
\begin{figure}[!h]
	\begin{center}
		\includegraphics[width=0.85\textwidth]{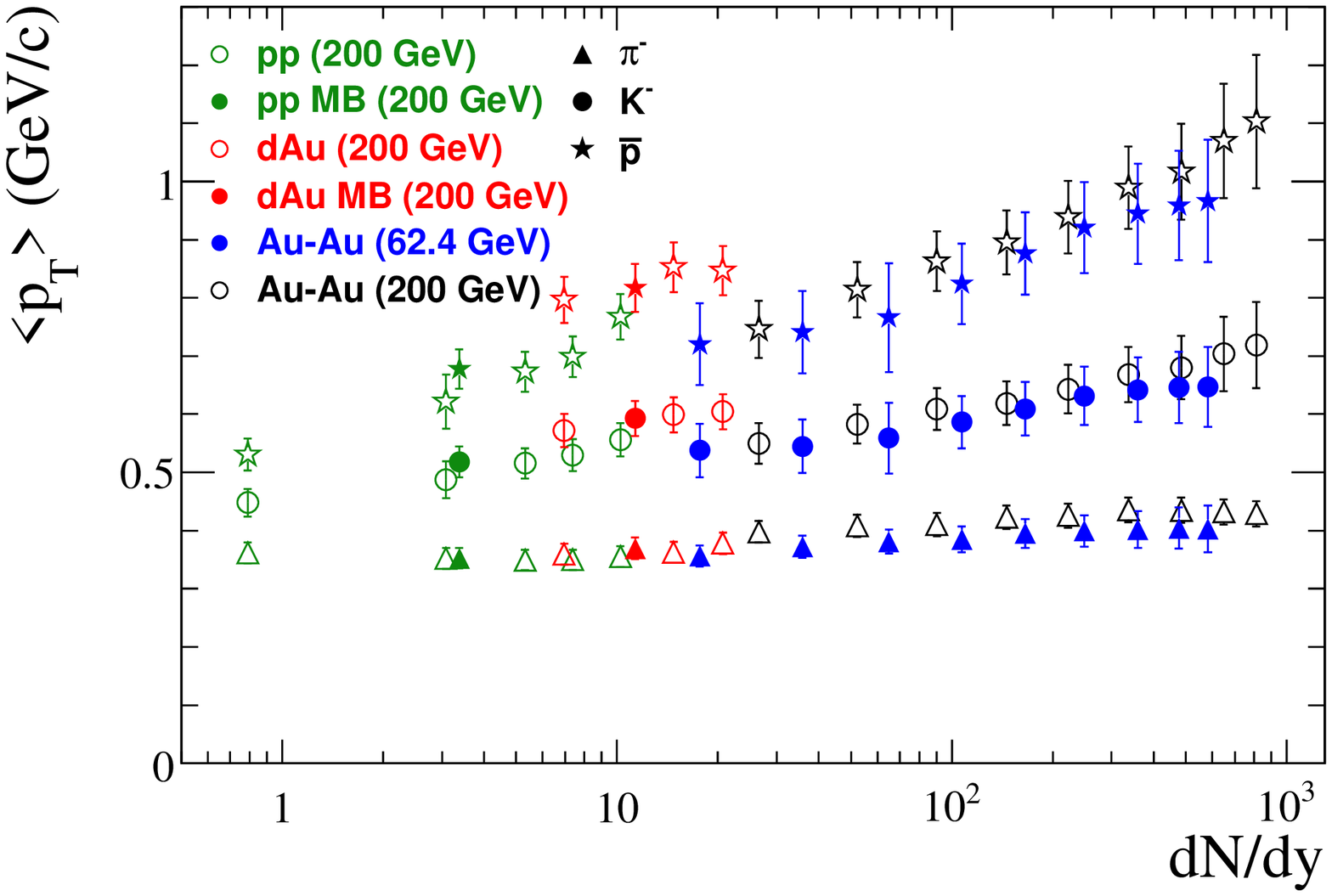}	
		\includegraphics[width=0.85\textwidth]{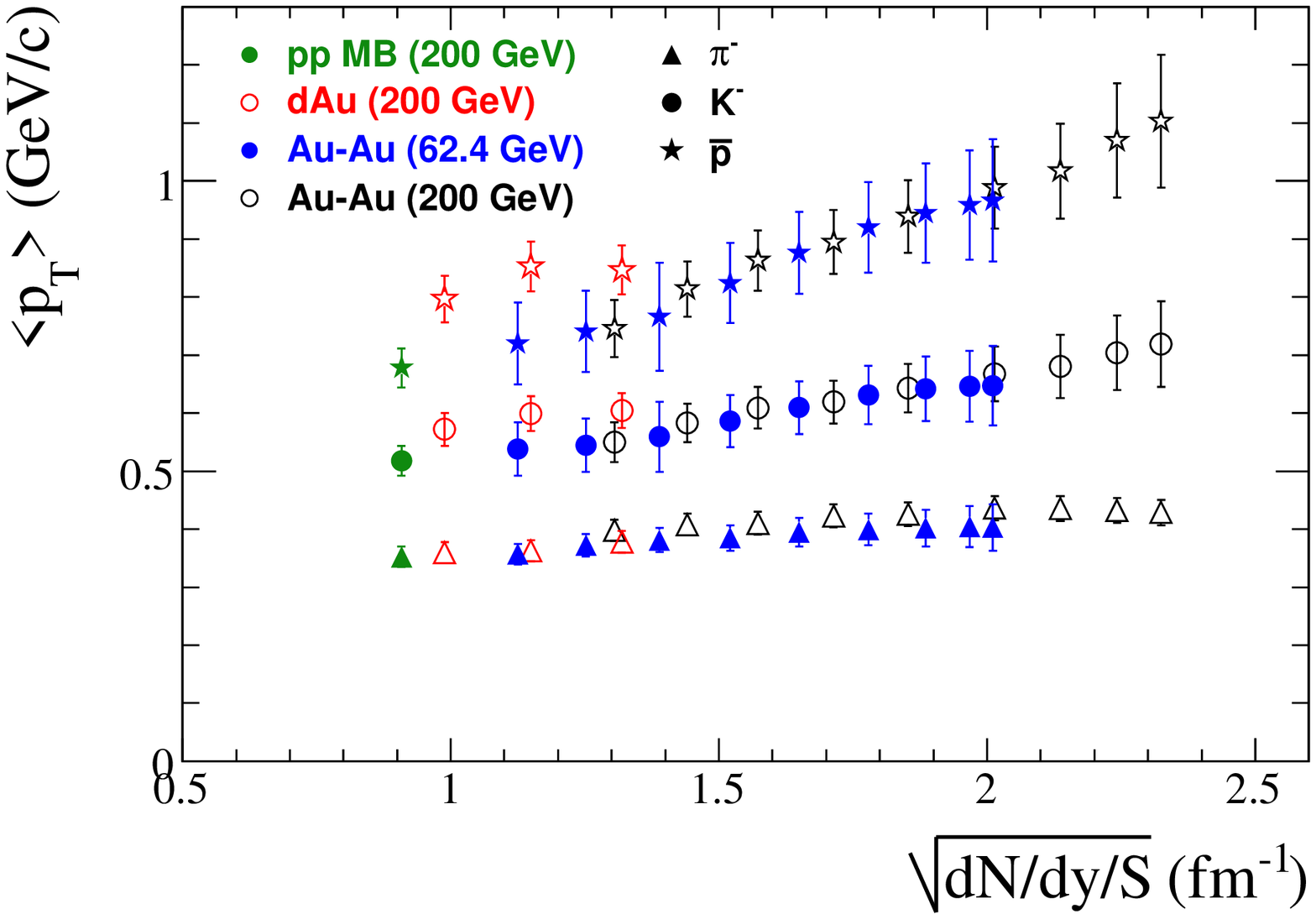}		
		
\caption{Average transverse momenta are shown as a function of corrected $dN/dy$ and $\sqrt{dN/dy/S}$ in 200 GeV pp, dAu and 62.4 and 200 GeV Au-Au collisions. Errors are statistical and systematic errors added in quadrature.}
	\label{fig:meanpt2}\end{center}
\end{figure}
Fig.~\ref{fig:dauspectra} shows the minimum bias and centrality 
dependent $\pi^{\pm}$, $K^{\pm}$, p and $\bar{p}$ spectra in 200 
GeV dAu collisions. The top spectra represent centrality 0-20$\%$ 
and are scaled by a factor of 4. The next spectra show the mid-central 
bin 20-40$\%$ and are scaled by a factor of 2. The third spectra from 
top represent the minimum bias spectra and are not scaled. The bottom
spectra represent the peripheral 40-100$\%$ and are scaled by factor
of 0.5. 
The minimum bias dAu spectra are in good agreement with 
the previously published STAR results~\cite{Adams:2003qm}.

The flattening in the dAu spectra are less pronounced than in pp.
This is presumably due to the centrality selection by the FTPC.
The relative difference in average multiplicity between the most central and most 
peripheral events is smaller than in pp collisions and the underlying physical processes might be different as well.

Fig.~\ref{fig:auau62spectra} shows the centrality dependence of the $\pi^{\pm}$, $K^{\pm}$, p and $\bar{p}$
spectra measured in 62.4 GeV Au-Au collisions. 
Pion spectra shapes show small variation with centrality.
But the shapes of kaon and proton/anitproton spectra exhibit strong dependence
on centrality. The evolution is more pronounced than in pp or dAu.

In Au-Au collisions the change in the spectral shapes is clearly observable for protons/antiprotons (less for kaons). 
The same spectra evolution is observed in 200 GeV Au-Au collisions as well~\cite{Adams:2003xp}.
Detailed discussion of the observed spectra behavior will follow in Sec.~\ref{sec:meanpt} and Sec~\ref{sec:freezeout}.
\begin{figure}[!t]
	\begin{center}
		\includegraphics[width=0.9\textwidth]{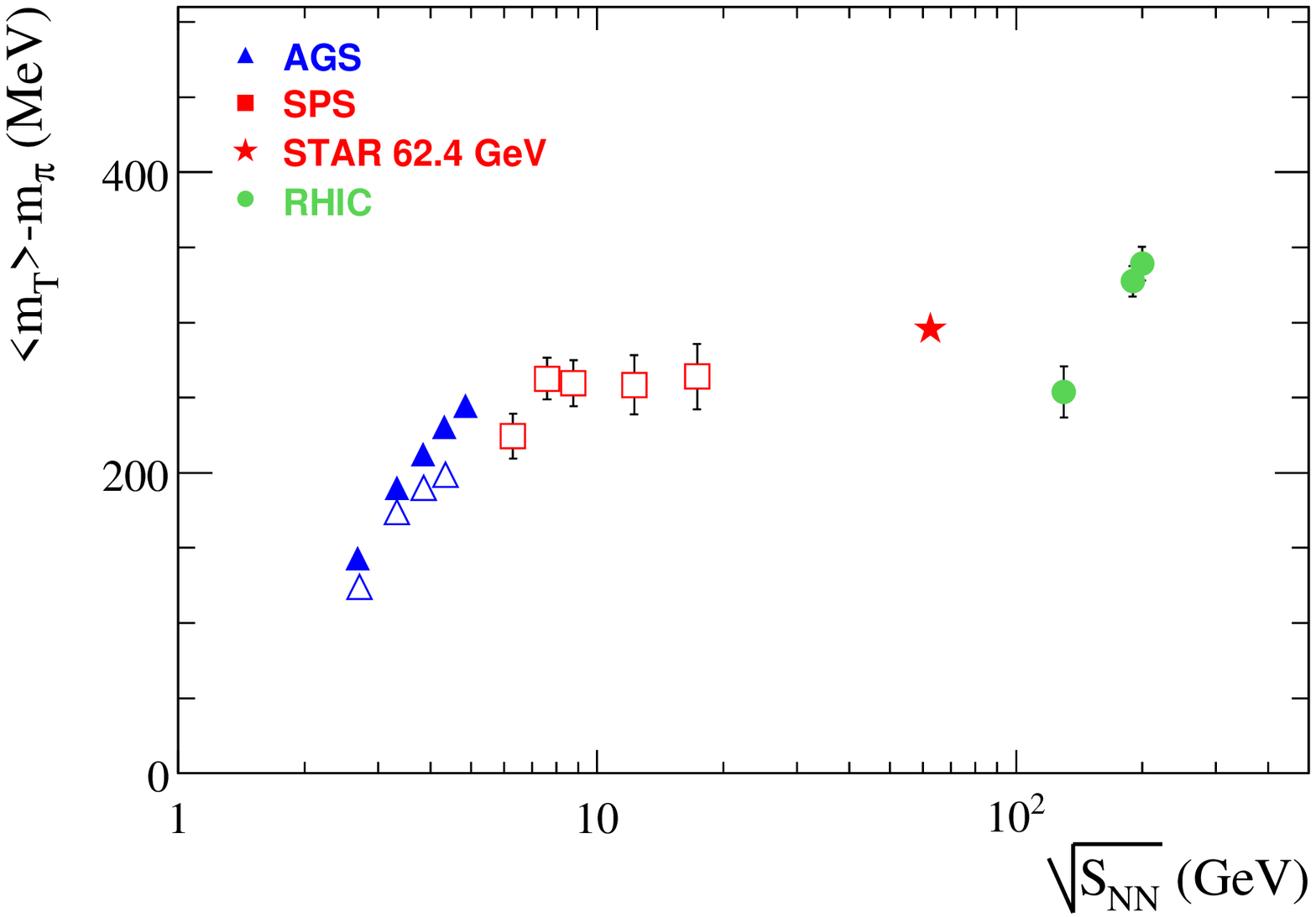}
			\caption{Average transverse mass for pions from AGS to SPS to RHIC energies measured in 0 - 5 \% Pb-Pb and Au-Au collisions. See text for references.}
	\label{fig:tvmasspi}\end{center}
\end{figure}

Fig.~\ref{fig:allspectra} shows the relative magnitude of the minimum bias 200 GeV pp, dAu and the most peripheral and central 62.4 GeV Au-Au spectra. Magnitudes of spectra show modest increase from pp to dAu to peripheral Au-Au collisions and a significant enhancement from peripheral to central Au-Au collisions. Note that the separation of proton and antiproton spectra in 62.4 GeV Au-Au collisions will be discussed later.
\begin{figure}[!t]
	\begin{center}
		\includegraphics[width=0.9\textwidth]{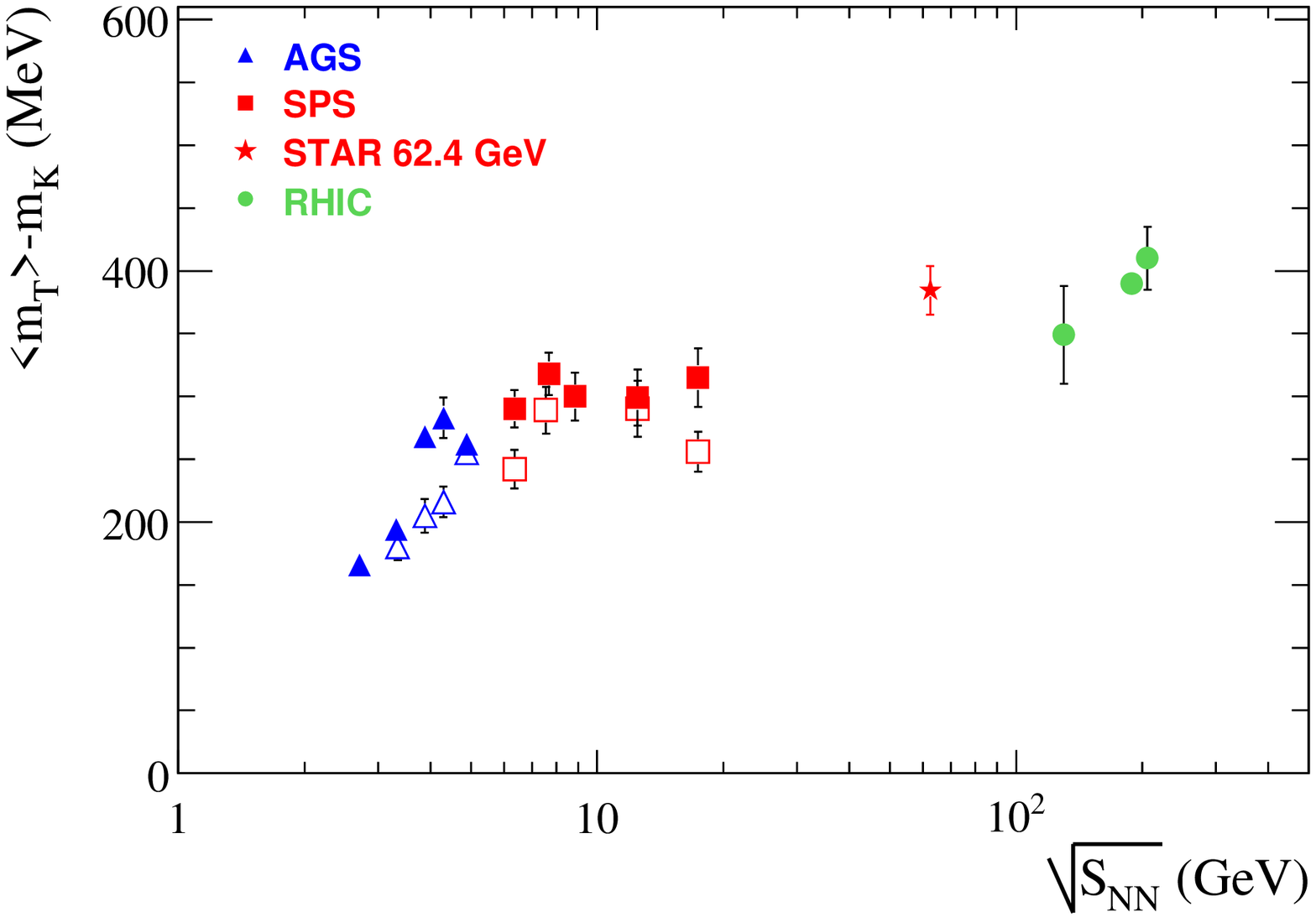}
	\caption{Average transverse mass for kaons from AGS to SPS to RHIC energies measured in 0 - 5 \% Pb-Pb and Au-Au collisions. See text for references.}
	\label{fig:tvmassk}\end{center}
\end{figure}
Spectra points and point-to-point systematic errors are listed in Spectra Tables in Appendix C.
\begin{figure}[!t]
	\begin{center}
		\includegraphics[width=0.9\textwidth]{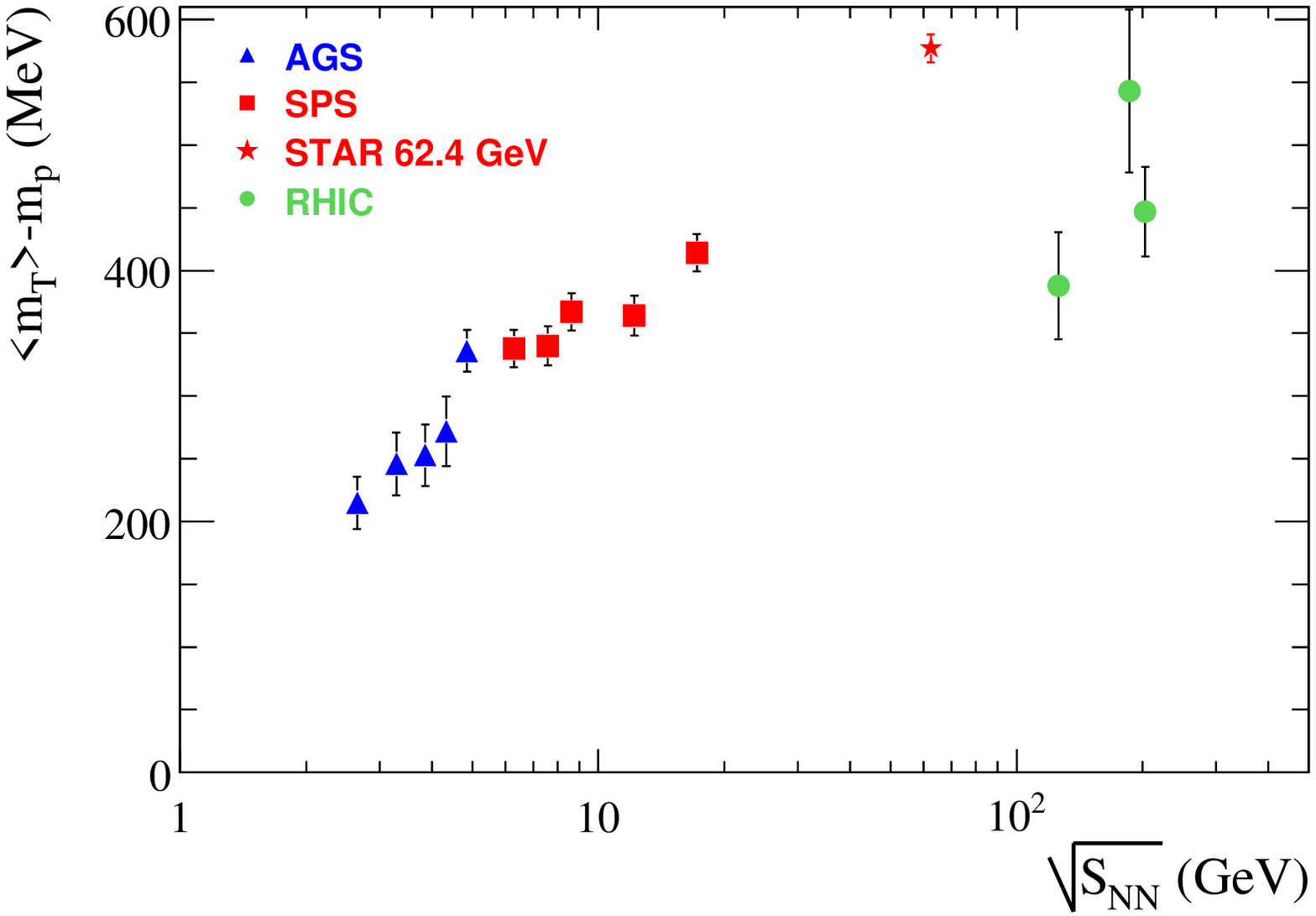}
	\caption{Average transverse mass for protons from AGS to SPS to RHIC energies measured in 0 - 5 \% Pb-Pb and Au-Au collisions. See text for references.}
	\label{fig:tvmassp}\end{center}
\end{figure}
\begin{sidewaysfigure}[!h]
	\begin{center}
	\includegraphics[width=0.9\textwidth]{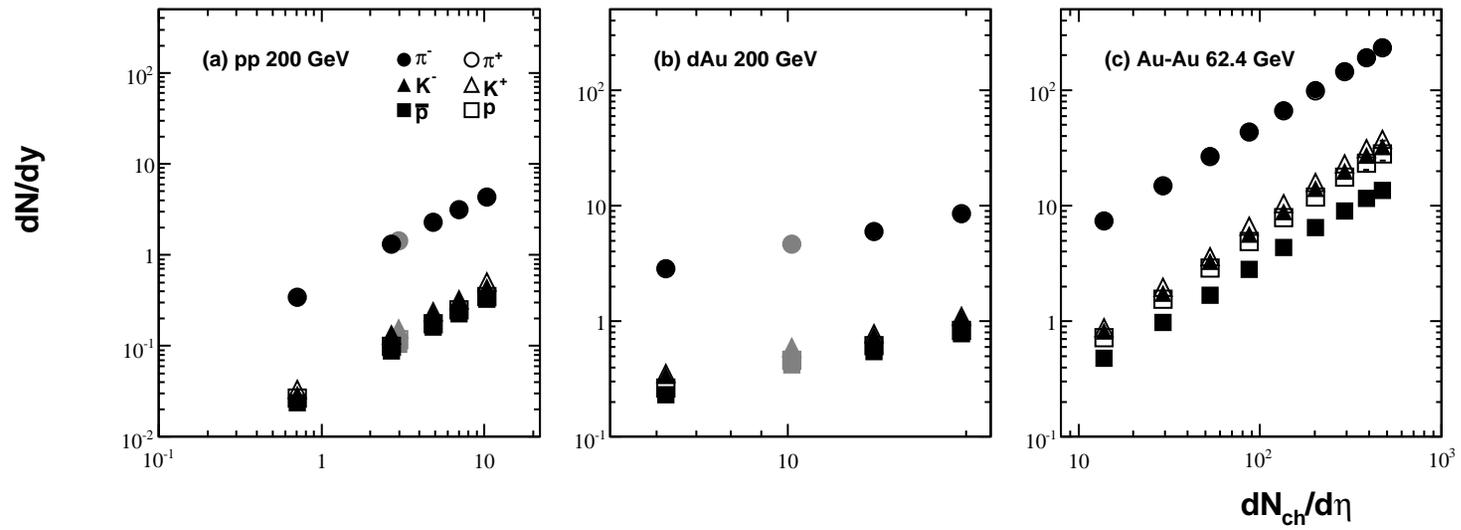}
\caption{Particle yields as a function of uncorrected charged particle multiplicity in 200 GeV pp collisions and as a function of centrality in 200 GeV dAu and 62.4 GeV Au-Au collisions.}
	\label{fig:alldNdy}\end{center}
\end{sidewaysfigure}
\begin{figure}[!h]
	\begin{center}
		\includegraphics[width=0.6\textwidth]{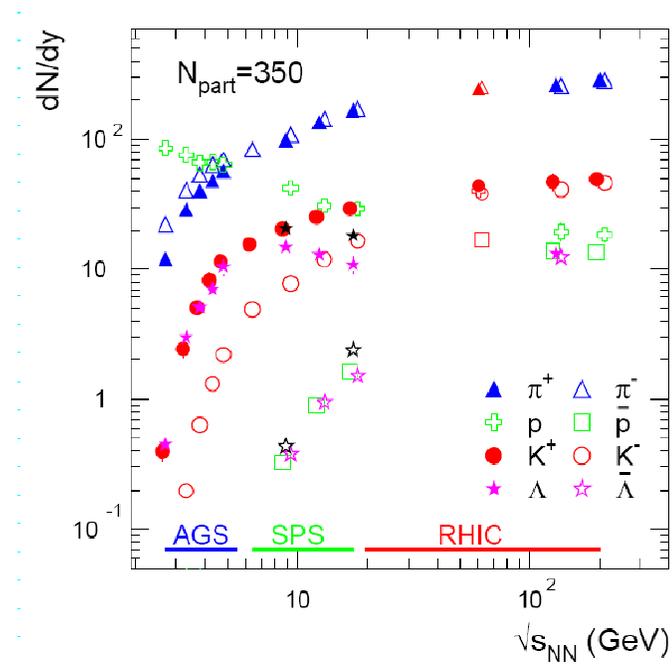}
\caption{Various particle yields measured in central Au-Au/Pb-Pb collisions as a function of center of mass energy. See text for references.}
	\label{fig:alldNdy_thermalpaper}\end{center}
\end{figure}
\begin{figure}[!h]
	\begin{center}
		\includegraphics[width=0.8\textwidth]{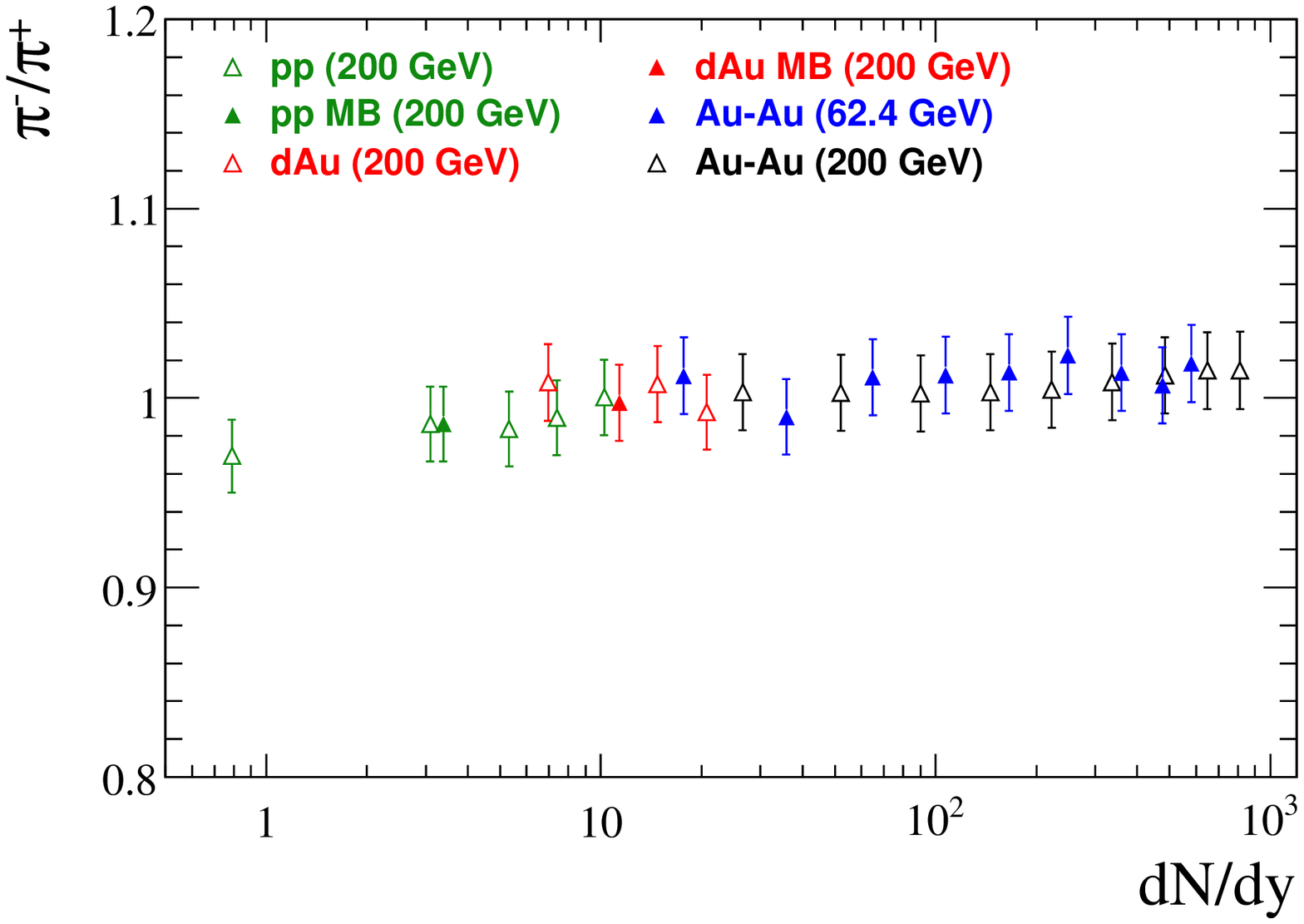}
\caption{$\pi^{-}/\pi{+}$ ratios are shown as a function of corrected dN/dy in pp, dAu and Au-Au collisions.}
	\label{fig:papratiospion}\end{center}
\end{figure}

\section{Systematic uncertainties}

Point-to-point systematic errors are estimated from the variation of the fit parameters, variation of the event and track selection. Since the pion energy loss band can be cleanly separated, the pion average point-to-point systematic error is $\sim$ 4\%. Kaon point-to-point errors in the overlaping electron region are larger, $\sim$ 15 - 20 \%, and $\sim$ 5\% otherwise.  Protons and antiprotons are well separated up to high transverse momenta. Point-to-point uncertainty is $\sim$ 5\% for the protons spectra overall. In the region where the electrons merge they represent a small fraction of the extracted proton yield, hence they do not significantly increase the point-to-point uncertainties in this region. 
The proton background correction is estimated to contribute to the systematic uncertainty on the level of $\sim$ 5\%.
Further overall uncertainties are introduced by embedding. Since selected runs represent all measured events for a specific collision, a $\sim$ 5\% uncertainty is assigned to tracking efficiencies.

\subsection{Extraction of spectra properties}\label{sec:extractspectraparas}
Particle yield (dN/dy), inverse slope parameter (T) and average transverse momentum ($\left\langle p_{T}\right\rangle$) are extracted 
from the measured spectra and extrapolated outside of the measured transverse momentum region.
Extrapolation is based on different functional forms presented in detail in Appendix B. 

The low momentum part of the particle spectra ($p_{T} <$ 2 GeV/c) has been considered exponential
because of the thermal production mechanism. However, as shown in Fig.~\ref{fig:auau62spectra} kaon and proton/antiproton spectra differ from exponential, especially in central Au-Au collisions.

In elementary collisions (eg. proton-proton), particle production models describe a static, thermal source that leads to exponential behavior of the low momentum particle spectra. As it was known from the lower energy heavy-ion collisions, pressure generated during the collision process boosts the produced particle away from the center of the collision. This mechanism leads to an expanding source, which might be thermalized. This pressure generated boost manifests itself in the change of the shapes of particle spectra, depending on the mass of the measured particle~\cite{Schnedermann:1994gc}. 

The blast-wave model~\cite{Schnedermann:1994gc} describes the hydrodynamical evolution of the heavy-ion collisions. 
Due to the geometry of the collision, hydrodynamic equations are set for a cylindrically symmetric system. The blast-wave model is only a parameterization of the non-viscous, ideal hydrodynamical equations. The blast-wave model does not account for particle production, only for their propagation through the hydrodynamical expansion of the system. The blast-wave model assumes local thermal equilibrium, which is expected to develop early in the heavy-ion collisions, therefore the system evolution starts from a thermalized state in the model.

Bulk particle spectra ($\pi^{\pm}, K^{\pm}, \overline{p}, p, \Lambda, \overline{\Lambda}$) in heavy-ion collisions can be simultaneously described by the blast-wave model with three parameters: the kinetic freeze-out temperature ($T_{kin}$), the average transverse flow velocity (or radial flow, $\left\langle \beta \right\rangle$) and the exponent of the flow profile ($n$). Since the blast-wave model describes the spectra of the primordial particles (those particles which are created in the collisions process and not from resonance decays) and pion spectra are expected to carry significant contribution from resonance decays at low transverse momenta ($p_{T}<$ 0.5 GeV/c), pion spectra points below $p_{T}<$ 0.475 GeV/c are excluded from the blast-wave fit. Detailed investigation of the resonance contribution to extracted blast-wave parameters is described in Section~\ref{sec:resonances}. 

Since the low momentum part of the pion spectra is not described well by the blast-wave model, the Bose-Einstein functional form is chosen to extract spectra properties. For each collision system the Bose-Einstein functional form provides the best description of the measured pion spectra in transverse momentum region accessible by the dE/dx technique. Kaon and proton spectra properties are extracted from the blast-wave model fit, which gives the best description of the measured spectral shapes in Au-Au collisions. Results presented in the following sections are extracted in the way described above, unless stated otherwise.
\begin{figure}[!h]
	\begin{center}
	\includegraphics[width=0.8\textwidth]{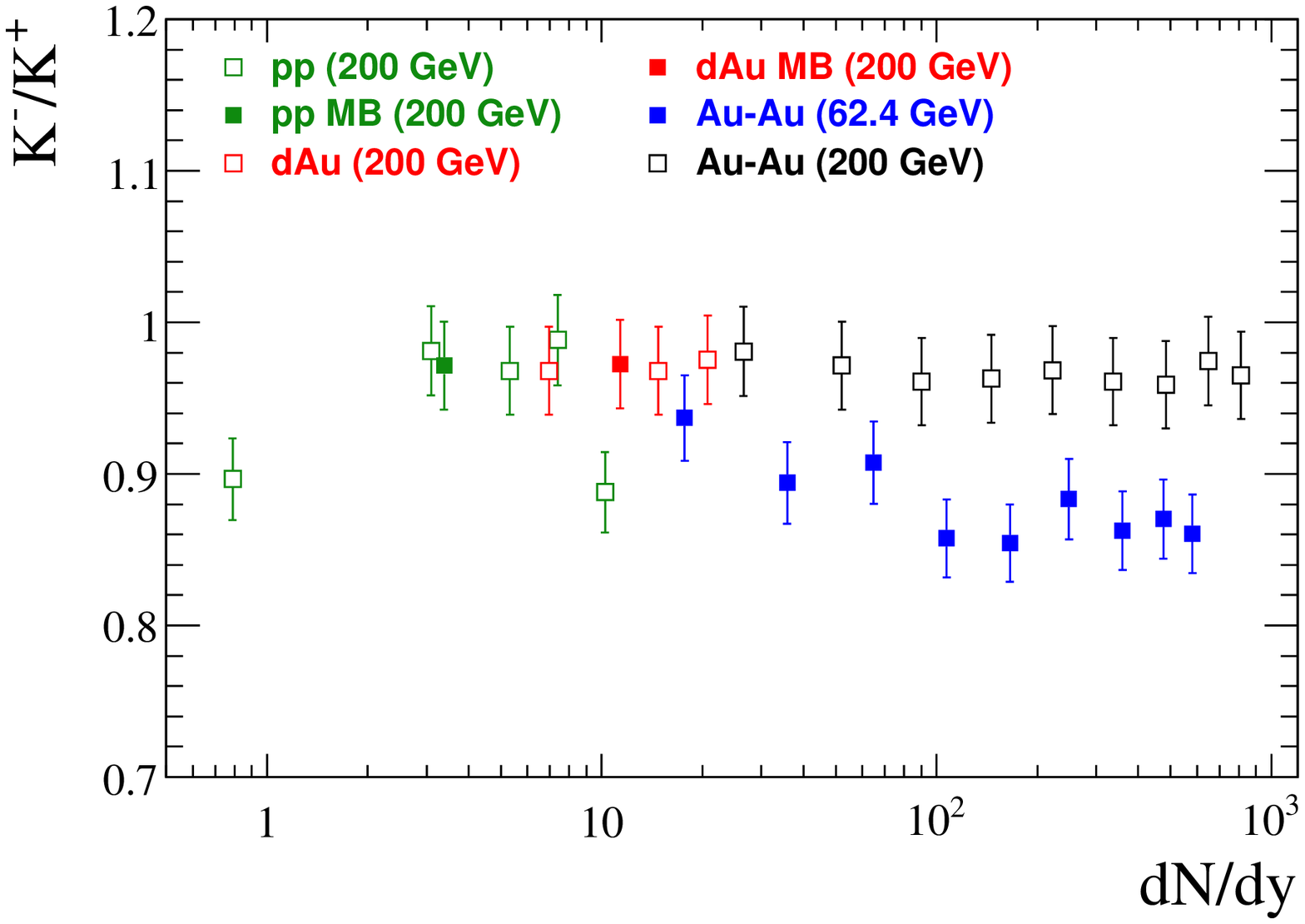}
\caption{$K^{-}/K{+}$ ratios are shown as a function of corrected dN/dy in pp, dAu and Au-Au collisions. }
	\label{fig:papratioskaon}\end{center}
\end{figure}

%
\begin{figure}[!h]
	\begin{center}
	\includegraphics[width=0.8\textwidth]{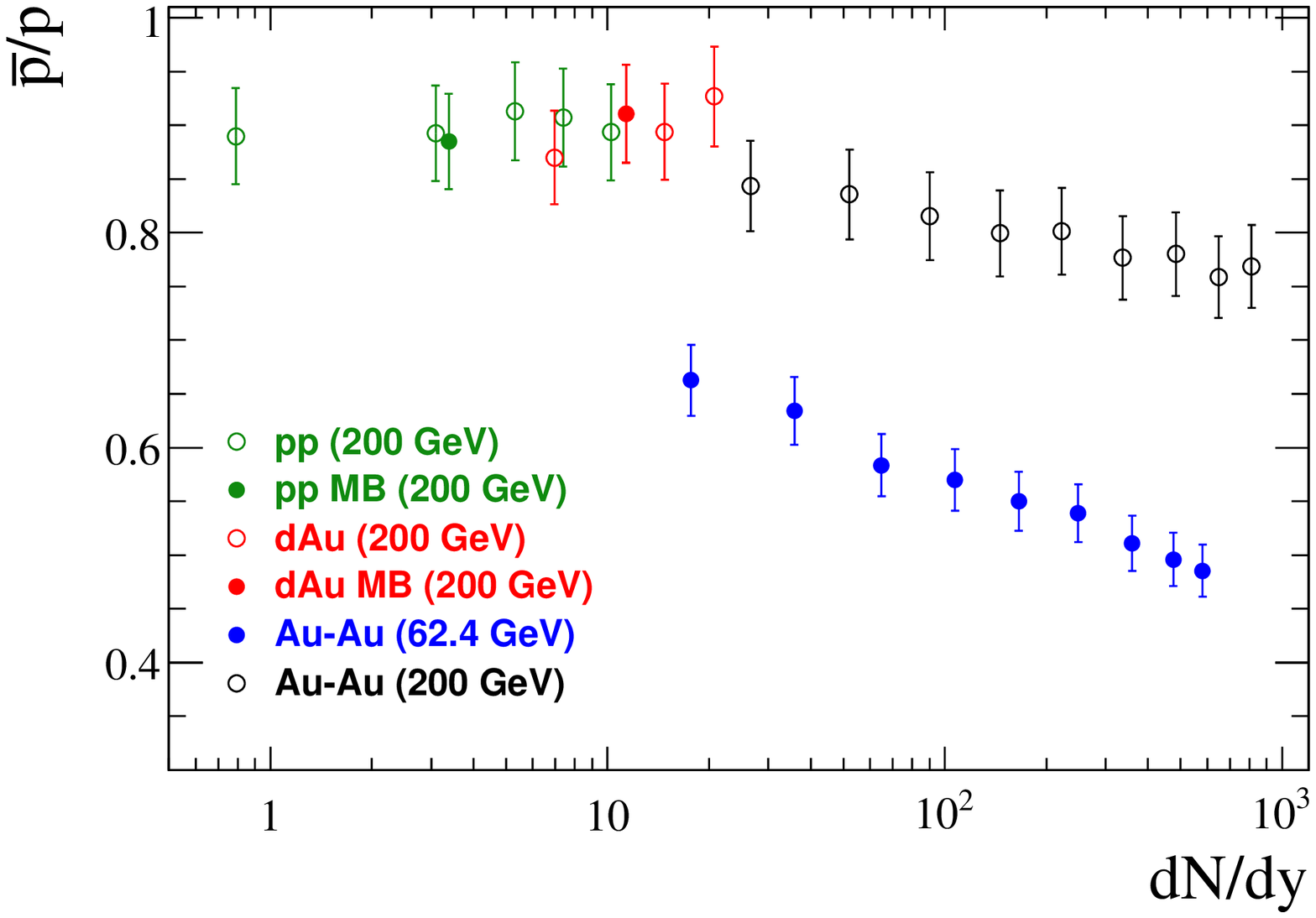}
\caption{$\overline{p}/p$ ratios are shown as a function of corrected dN/dy in pp, dAu and Au-Au collisions.}
	\label{fig:papratiosp}\end{center}
\end{figure}
\begin{figure}[!h]
	\begin{center}
	\includegraphics[width=0.8\textwidth]{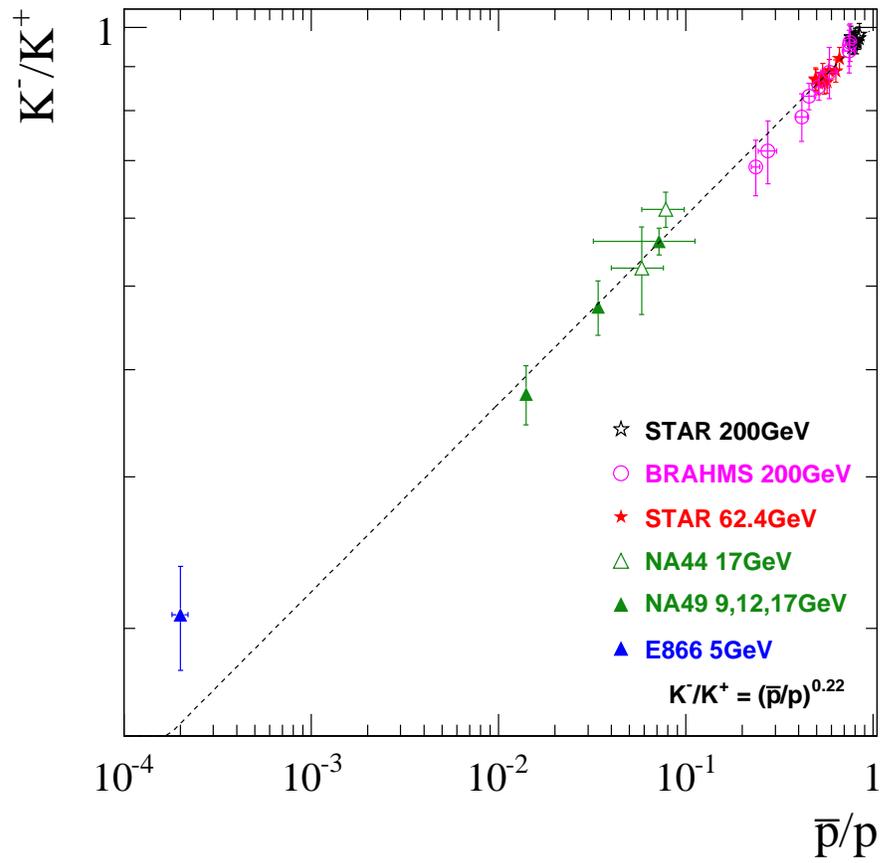}
\caption{Ratio of charged kaons vs. the ratio of antiprotons to protons in various heavy-ion collisions. See text for references.}
	\label{fig:kktopbarp}\end{center}
\end{figure}
\begin{figure}[!h]
	\begin{center}
	\includegraphics[width=0.8\textwidth]{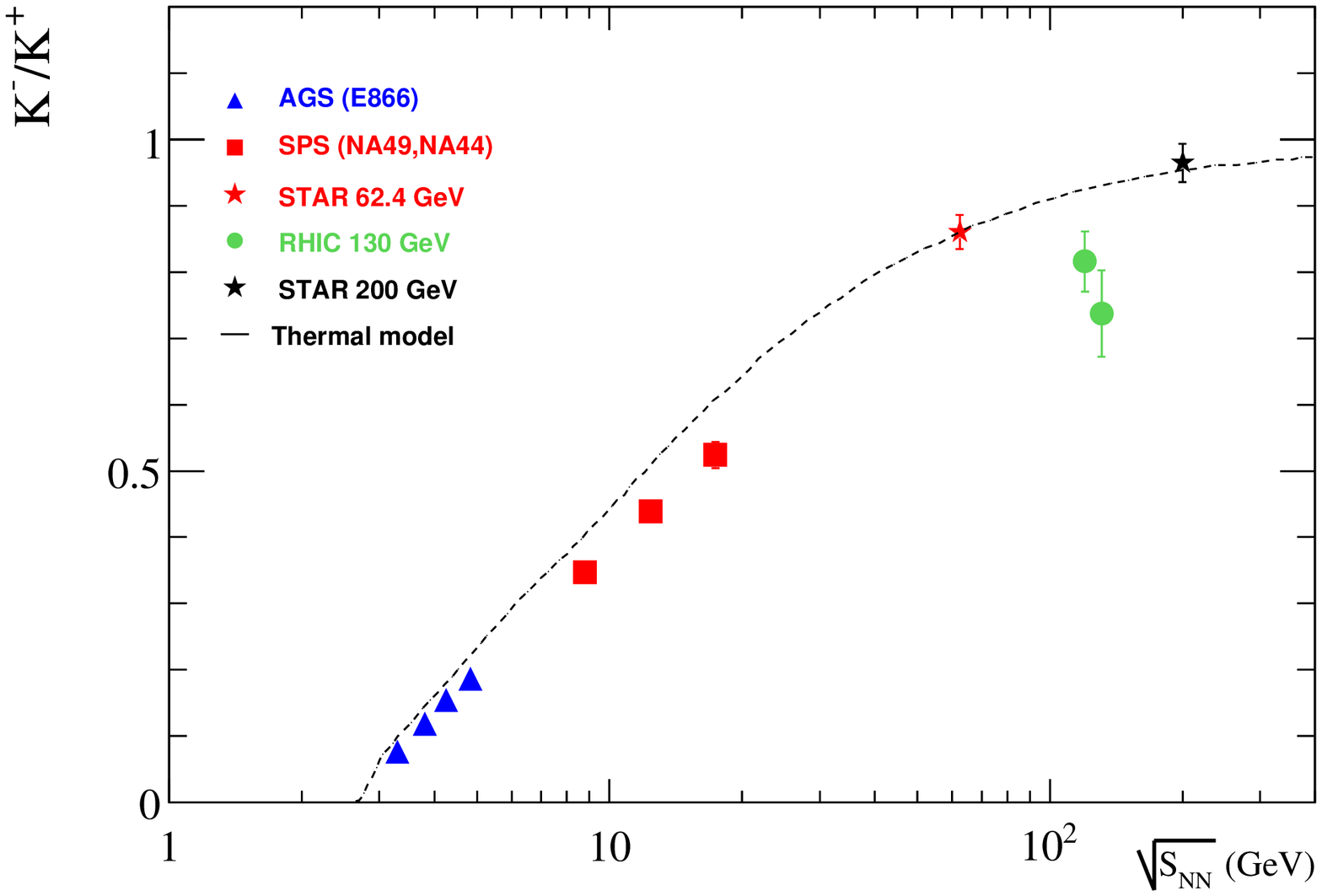}
	\caption{Particle antiparticle ratios ($K^{-}/K^{+}$) are shown as a function of center of mass energy for central (0-5\%) heavy-ion (Pb-Pb/Au-Au) collisions. See text for references. }
	\label{fig:papratios_S1}\end{center}
\end{figure}

\begin{figure}[!h]
	\begin{center}
\includegraphics[width=0.8\textwidth]{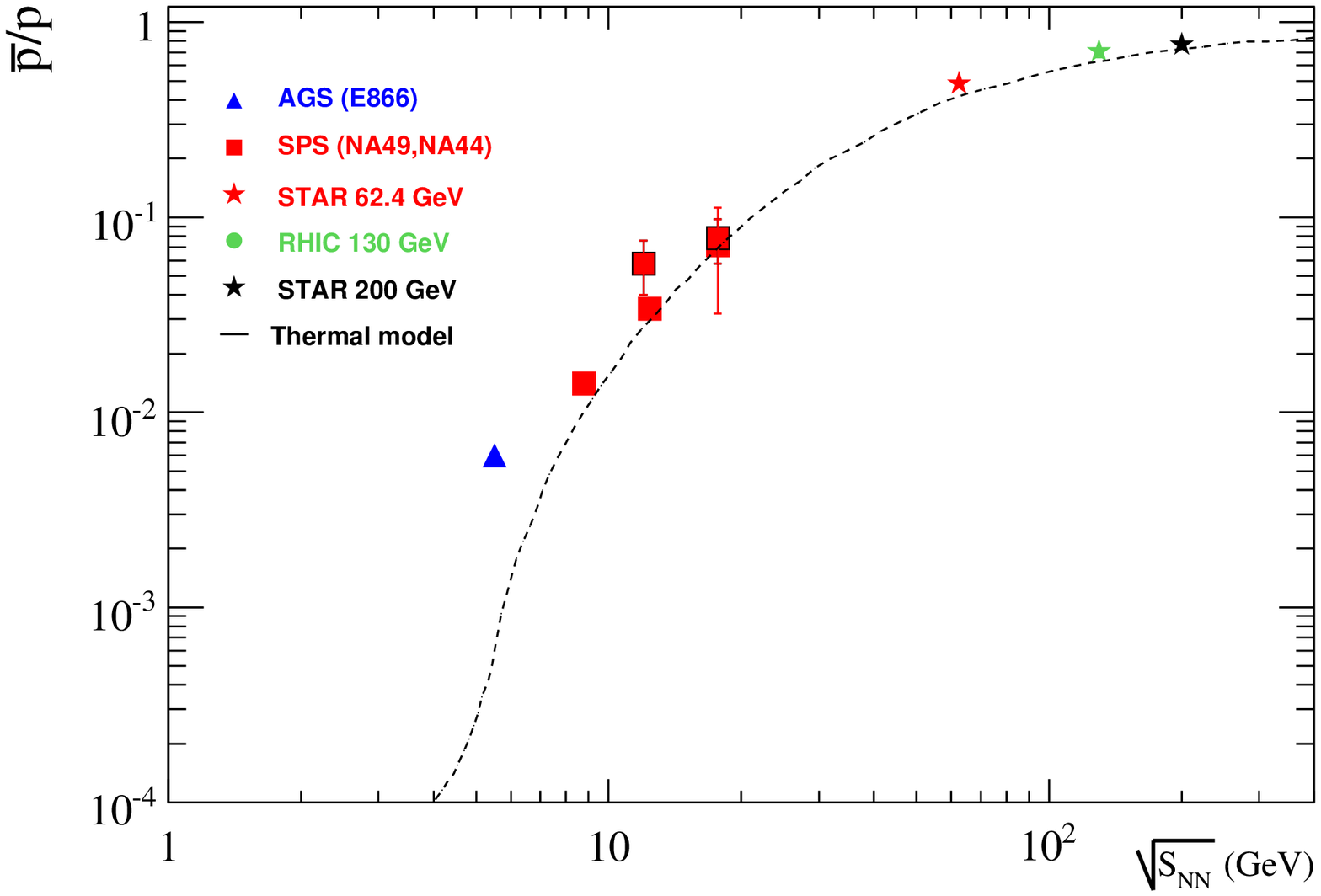}
	\caption{Particle antiparticle ratios ($\overline{p}$/p) are shown as a function of center of mass energy for central (0-5\%) heavy-ion (Pb-Pb/Au-Au) collisions. See text for references.}
	\label{fig:papratios_S2}\end{center}
\end{figure}
\clearpage
\begin{figure}[!h]
	\begin{center}
	\includegraphics[width=0.8\textwidth]{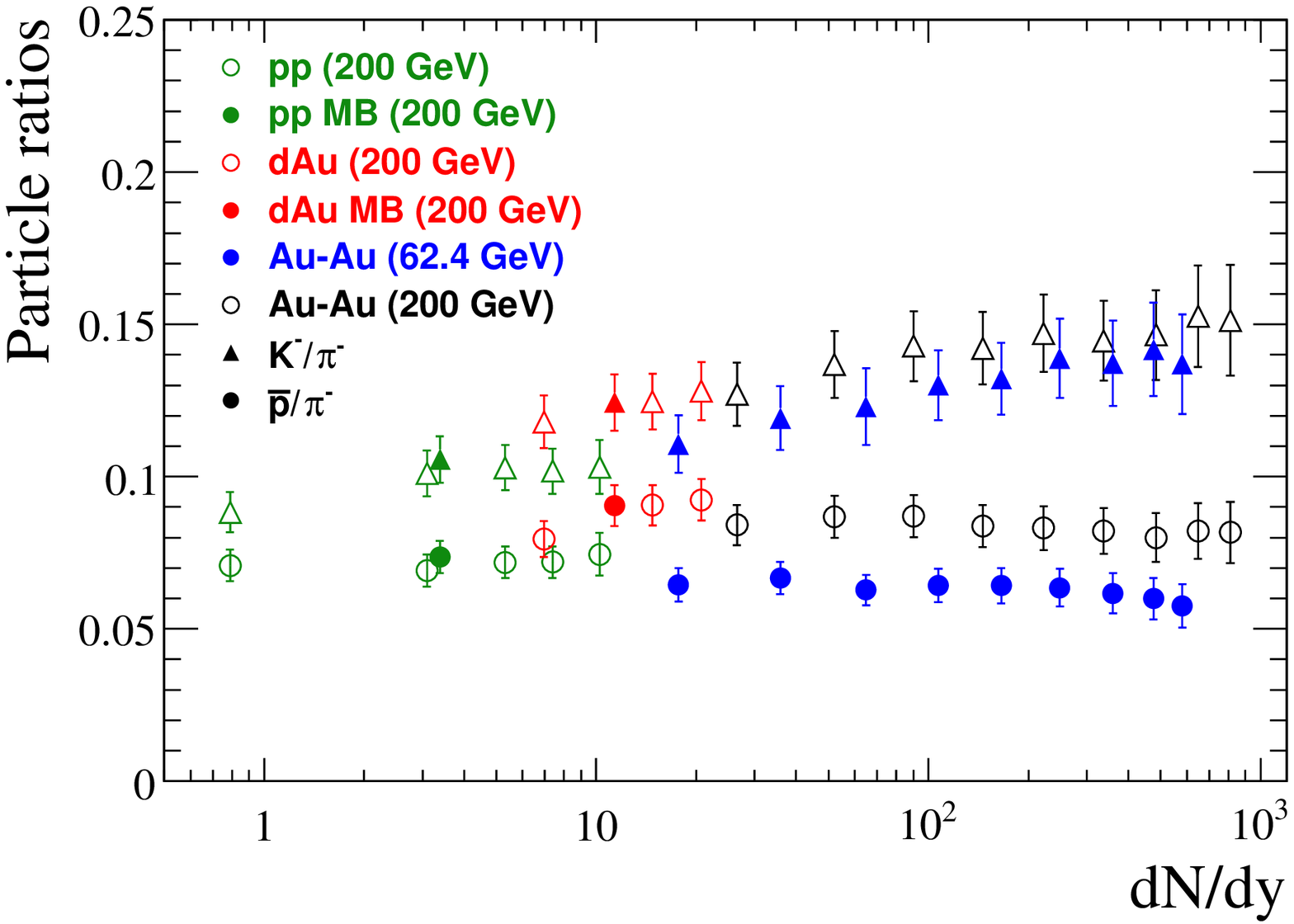}
 \includegraphics[width=0.8\textwidth]{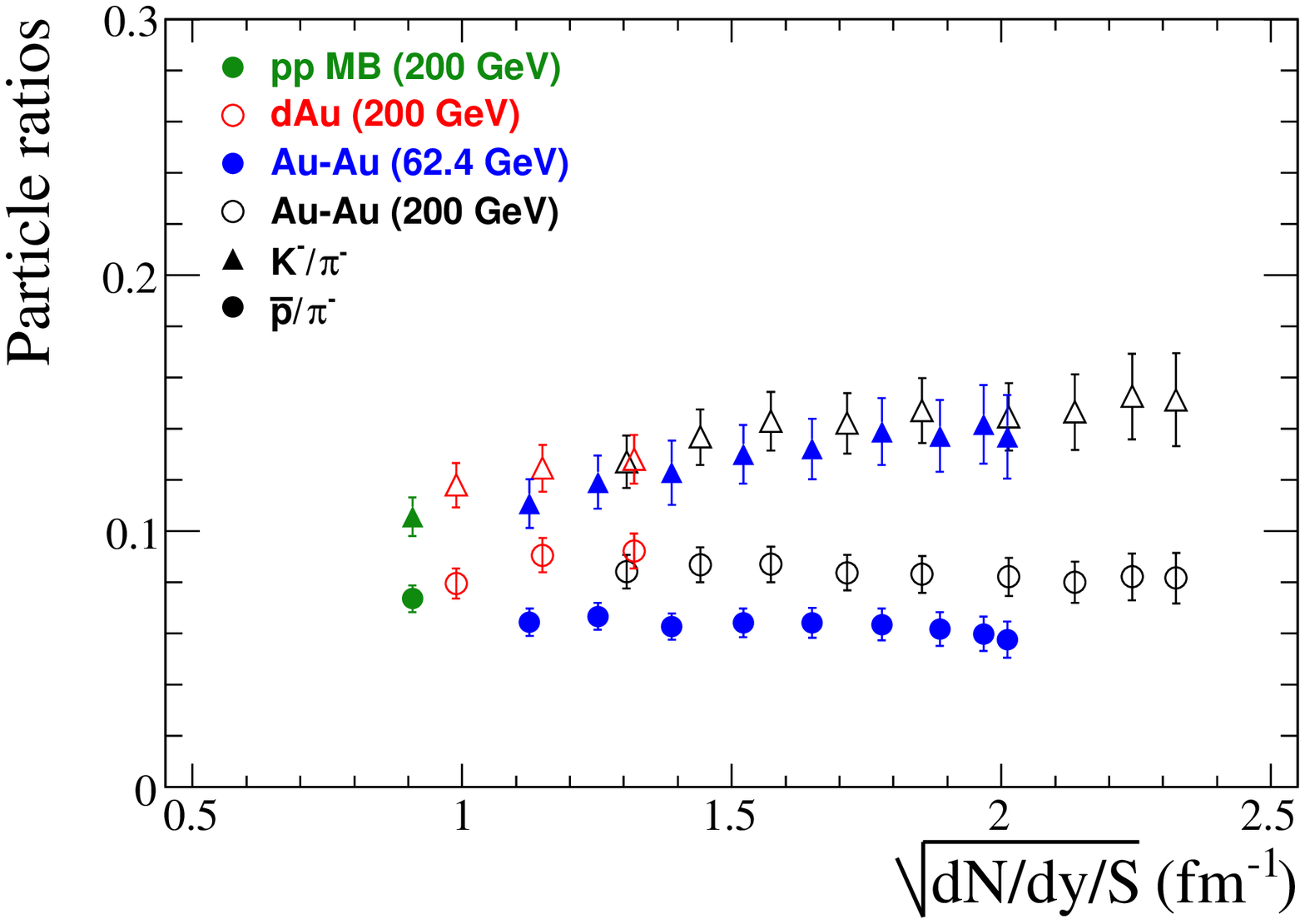}

\caption{Unlike particle ratios: $K^{-}/\pi^{-}$ and $\overline{p}/\pi^{-}$ are shown as a function of corrected dN/dy and $\sqrt{dN/dy/S}$ in 62.4 GeV and in 200 GeV Au-Au collisions. Errors are statistical and systematic added in quadrature.}
	\label{fig:unlikeratios1}\end{center}
\end{figure}
\begin{figure}[!h]
	\begin{center}
		\includegraphics[width=0.8\textwidth]{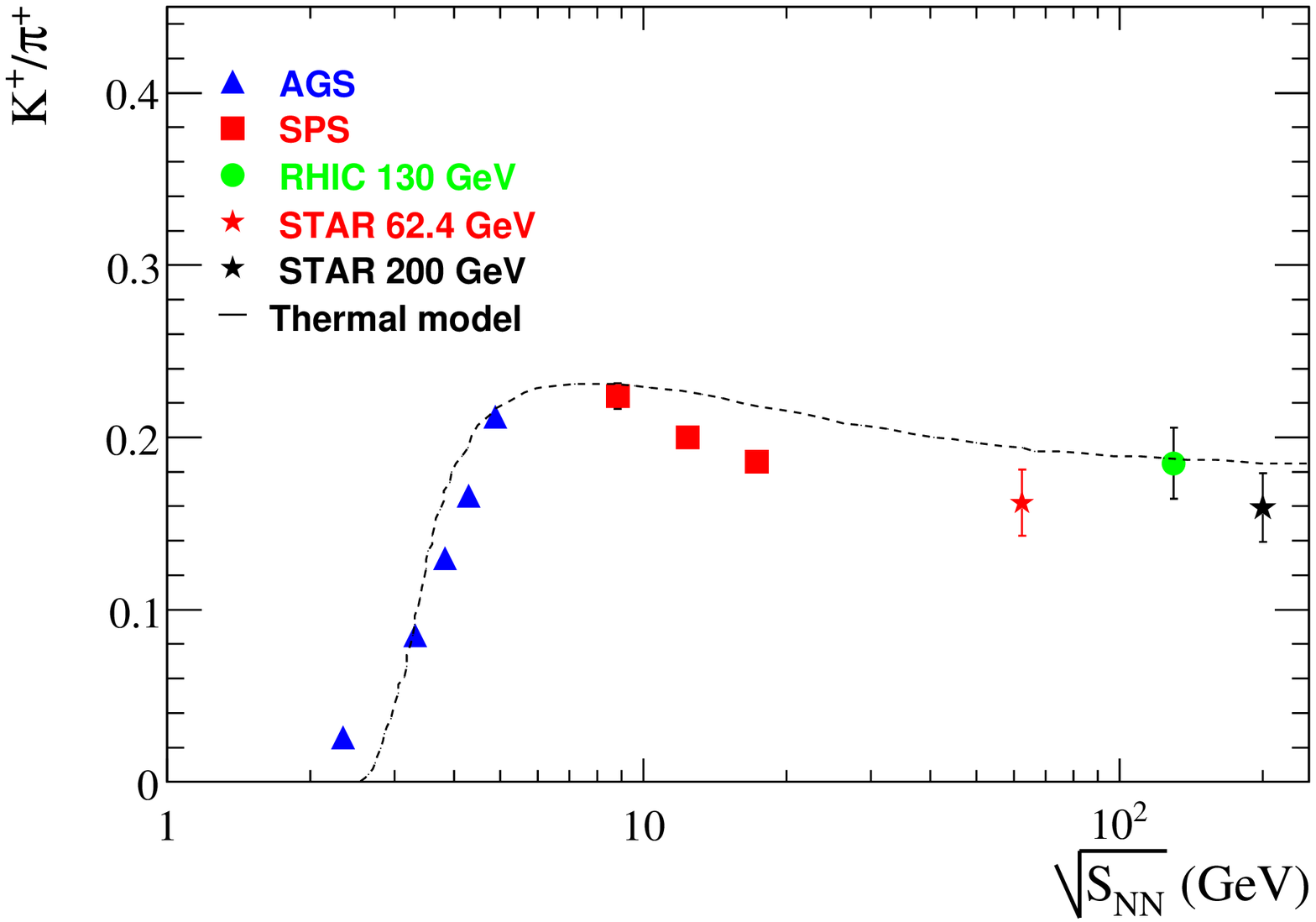}
	\caption{Mid-rapidity $K^{+}/\pi^{+}$ ratio as a function of collision energy for central (0-5\%) heavy-ion (Pb-Pb/Au-Au) collisions. See text for references.}
	\label{fig:kaon2pion_S1}\end{center}
\end{figure}
\begin{figure}[!h]
	\begin{center}
	\includegraphics[width=0.8\textwidth]{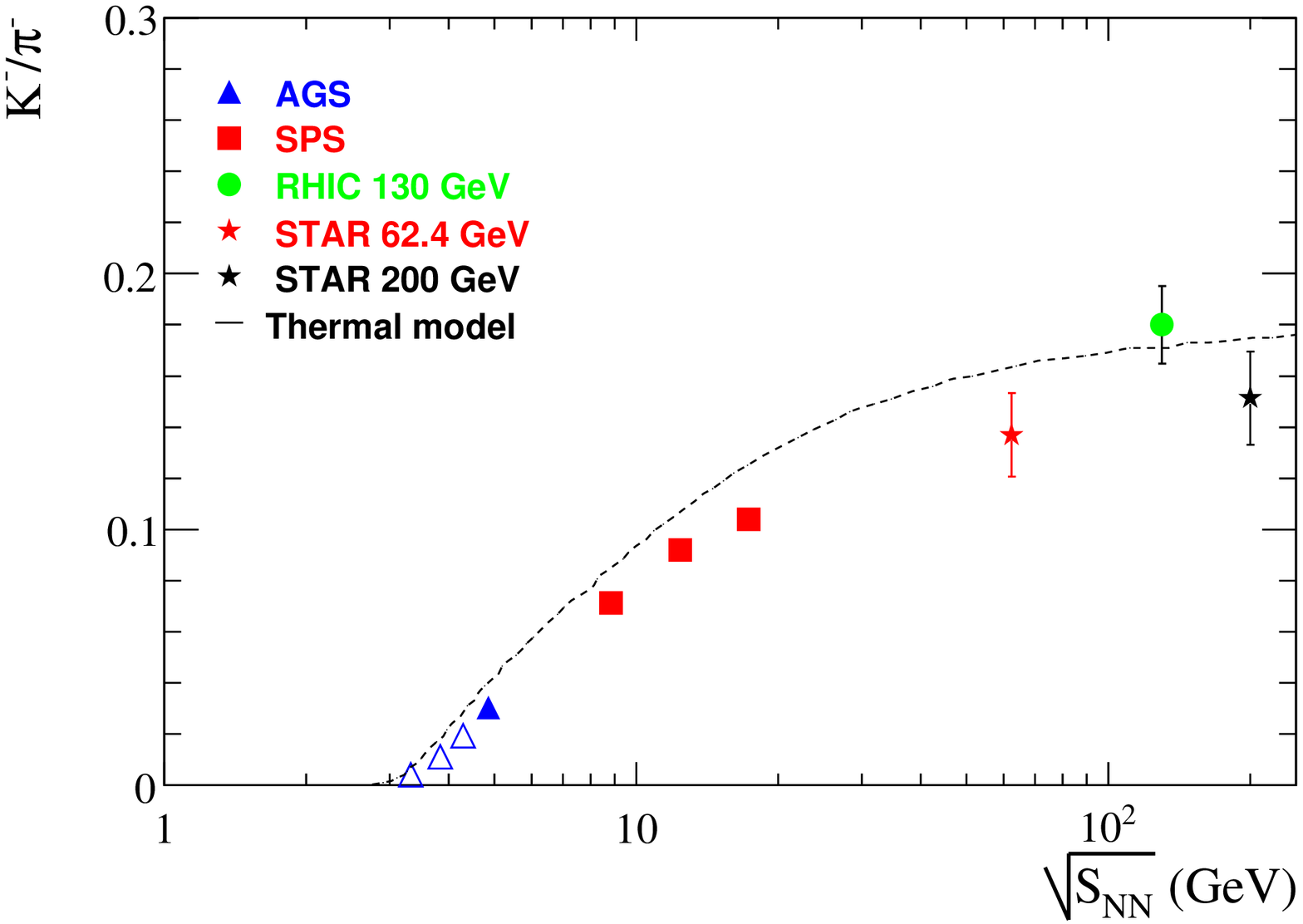}
	\caption{Mid-rapidity $K^{-}/\pi^{-}$ ratio as a function of collision energy for central (0-5\%) heavy-ion (Pb-Pb/Au-Au) collisions. See text for references.}
	\label{fig:kaon2pion_S2}\end{center}
\end{figure}

\section{Average transverse momenta}\label{sec:meanpt}

To characterize the change in the measured spectral shapes by collision type, energy and centrality, the average transverse momentum is investigated. 
\begin{equation}
\left\langle p_{T} \right\rangle\ =\ \frac{\int \frac{1}{2\pi p_{T}}\frac{dN}{dydp_{T}}\cdot  2\pi\cdot p_{T}^{2}\cdot dp_{T}}{\int \frac{1}{2\pi p_{T}}\frac{dN}{dydp_{T}}\cdot2\pi\cdot p_{T}\cdot dp_{T}}\
\label{eq:meanpt}
\end{equation}
The definition of the average transverse momentum is given in Eq.~\ref{eq:meanpt}. Numerical integration goes from 0 to 10 GeV/c in the $\left\langle p_{T} \right\rangle$ calculations.

Investigation of the average transverse momentum as a function of charged hadron multiplicity was an important analysis of previous experiments, since the anomalous behavior of the average transverse momentum as a function of the measured charged particle multiplicity can indicate the phase transition from the quark gluon plasma to the hadronic phase~\cite{VanHove:1982vk}. Following van Hove's approach: charged particle multiplicity is proportional to the entropy. The entropy is created early in the collision at thermalization, and followed by hydrodynamical adiabatic expansion with conserved entropy. The shape of the transverse momentum spectrum carries the combined effect of the temperature in the collision and the expansion of the system. The average transverse momentum increases as a function of charged particle multiplicity. In the case of a phase transition the entropy density is expected to increase but the temperature is expected to remain nearly constant. Therefore, the average transverse momentum is expected to reach a plateau at large charged particle multiplicities. In 200 GeV pp collisions the multiplicity selection allows a selection of events with a few times the average multiplicity, and Au-Au collisions provide an even wider multiplicity range. 

Fig.~\ref{fig:meanpt2} also shows the evolution of the average transverse momentum for 200 GeV pp, dAu and Au-Au collisions as a function of charged particle multiplicity and $\sqrt{dN/dy/S}$.

The average transverse momentum of pions is flat through  the collision systems. The average transverse momenta of kaons and protons/antiprotons follow the same increasing trend for pp and dAu collisions. In the highest multiplicity pp bin and in the most central dAu collisions the average transverse momenta of kaons and protons/antiprotons are larger than in peripheral Au-Au collisions, although the systematic errors overlap. The average transverse momenta of kaons and protons/antiprotons increase from peripheral to central Au-Au collisions and exceed that measured in the highest multiplicity the pp bin and in the most central dAu collisions. In Au-Au collisions the increase is predominantly driven by the collective expansion of the system. In pp and dAu collisions the increase of the average transverse momentum, measured from the azimuthally averaged spectrum, is expected to reflect the contribution from semi-hard scatterings and multi-parton collisions ($k_{T}$ broadening). These different physical processes might lead to the observed departure of the average transverse momenta in pp and dAu compared to Au-Au collisions. 

Fig.~\ref{fig:meanpt2} shows the evolution of the average transverse momentum for 62.4 GeV and 200 GeV Au-Au collisions. First we notice that the average transverse momentum for each particle species are the same within errors at both energies despite the factor of 3 difference in the collision energies. This implies similar system evolution at both energies despite the different initial conditions (energy density and baryon constant of the collisions zone).
The average transverse momenta of pions are flat from peripheral to central collisions. The average transverse momenta of kaons and protons/antiprotons increases quickly from peripheral to mid-central collisions and seem to saturate for kaons and increase less steeply for protons/antiprotons. 
The 200 GeV Au-Au data are presented in~\cite{Adams:2003xp}. Systematic errors on the $\left\langle p_{T}\right\rangle$ are also estimated by using the various functional forms mentioned above for extrapolation of the spectra, as can be found in~\cite{Adams:2003xp}.

\subsection{Average transverse mass}

Kaon transverse mass spectra in nucleon-nucleon and in nucleus-nucleus collisions from AGS to RHIC have exponential shape. Based on van Hove's argument and to seek a better understanding of the spectral shape evolution, instead of the average transverse momentum, the average transverse mass is investigated~\cite{Gorenstein:2003cu}. Particle spectra in 0-5 \% central Pb-Pb/Au-Au collisions are fitted with the exponential function:
\begin{equation}
\frac{d^{3}N}{2\pi m_{T}dm_{T}dy}\approx C\cdot e^{-m_{T}/T}
\end{equation}
where $T$ is the inverse slope parameter and the average transverse mass is given by:
\begin{equation}
\left\langle m_{T}\right\rangle\ =\ T\ + m\ + \frac{T^{2}}{m+T}.
\end{equation}

Fig.~\ref{fig:tvmasspi}, Fig.~\ref{fig:tvmassk} and Fig.~\ref{fig:tvmassp} present a compilation of the available data for 0-5 \% central Pb-Pb/Au-Au collisions from AGS to RHIC energies~\cite{Blume:2004ci}. It was argued that a plateau structure is developed at SPS energies, where the transition between confined and deconfined matter is expected to be located~\cite{Gorenstein:2003cu}. The plateau structure, in van Hove's picture, similarly to the average transverse momentum - charged particle multiplicity correlation, might indicate the onset of the phase transition. 
It is interesting to place the 62.4 GeV STAR measurement between the top SPS and the previously available RHIC energies. The flat trend as observed for pions at the SPS seems to continue or there might be a hint for a small rise in the average transverse mass. For kaons there is a small increase in the average transverse mass toward RHIC energies. The average transverse mass of protons exhibit the steepest trend with increasing energy. 

We argue in light of the 62.4 GeV data points that a definite change can be observed in the trend of $\left\langle m_{T}\right\rangle$ at the AGS-SPS energies for pions which is smoothened out for kaons and protons/antiprotons toward RHIC energies. At this point a lower energy RHIC run with high statistics would be necessary to test the above arguments within the same experimental framework in the SPS energy range. 
One might conclude from these excitation functions, that the initial conditions of the systems created in central (0-5 \%) Pb-Pb/Au-Au collisions are similar from SPS to RHIC energies and the increase of the collision energy is translated to the system expansion.

\begin{figure}[!h]
	\begin{center}
	\includegraphics[width=0.9\textwidth]{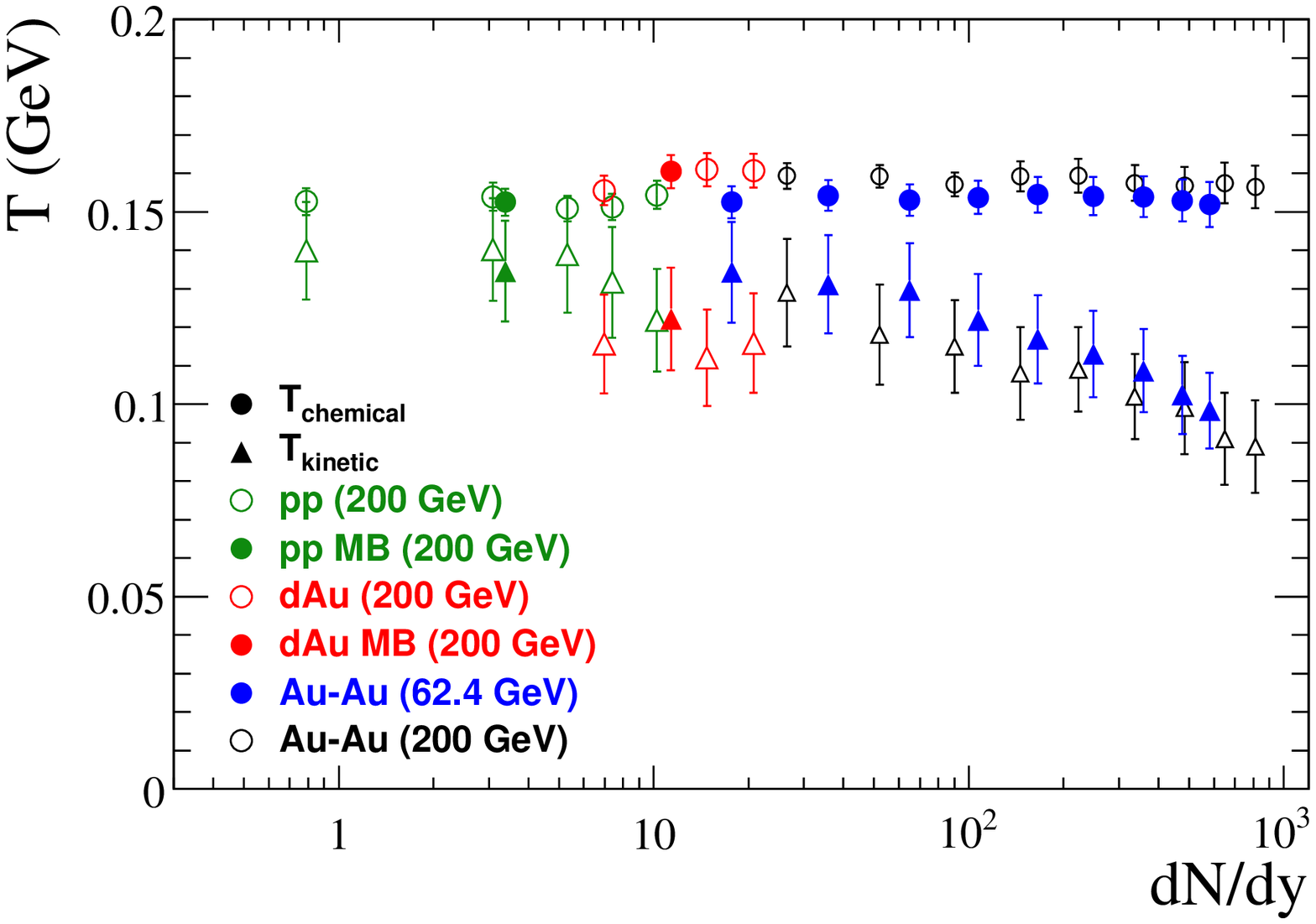}
	\includegraphics[width=0.9\textwidth]{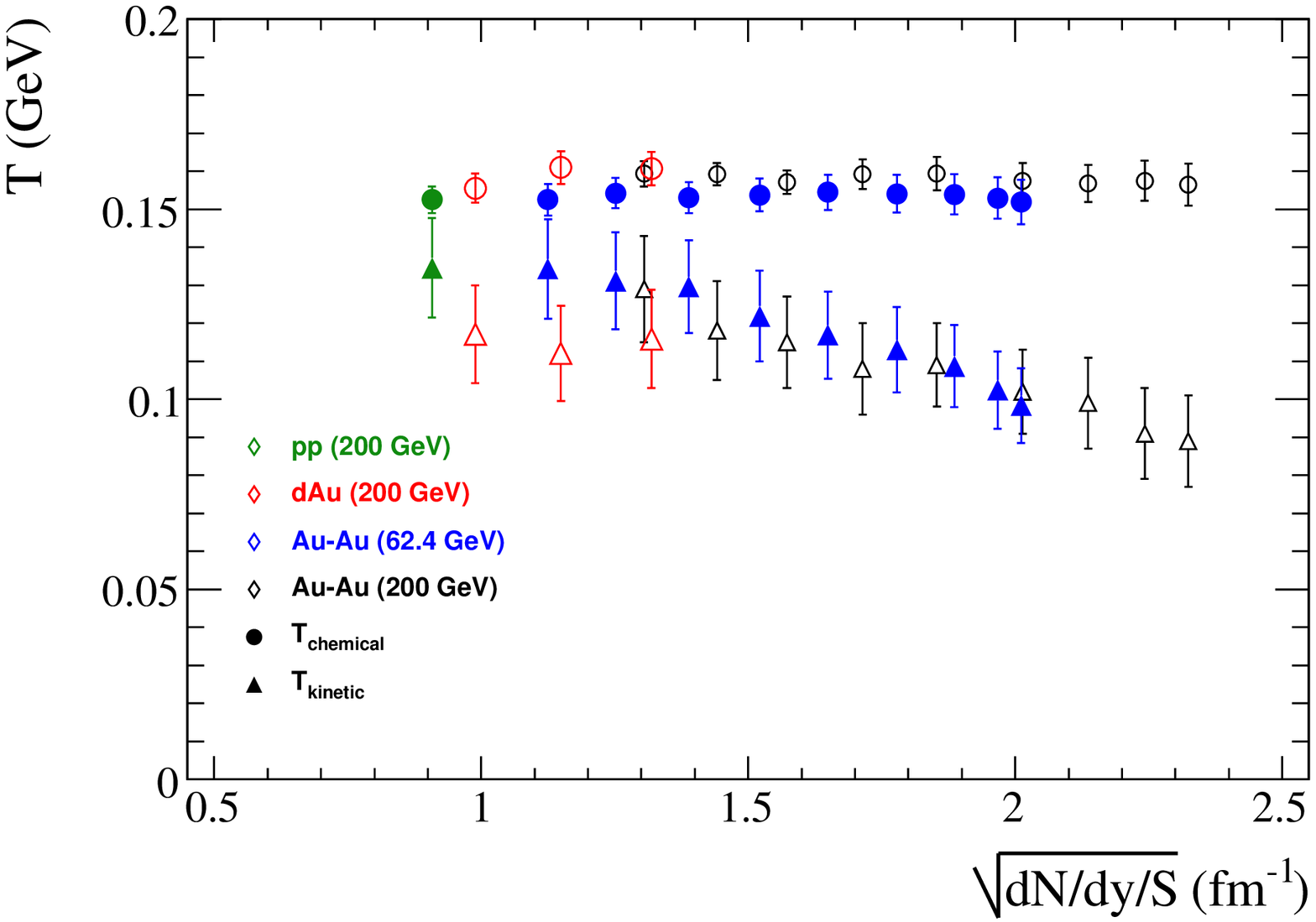}
\caption{ Chemical and kinetic freeze-out temperatures as a function of corrected dN/dy and $\sqrt{dN/dy/S}$.}
	\label{fig:tfo}\end{center}
\end{figure}

\section{Particle production}

\subsection{Total dN/dy}\label{totaldNdy}

The inclusive particle yield measured at mid-rapidity ($|y|<0.1$) for each identified particle spectrum is calculated from the available measured $p_{T}$ range and extrapolated outside. For extrapolation two methods are used: Bose-Einstein fit function for pions and blast-wave model calculation for kaons and protons. Table~\ref{tab:yields} and Tab.~\ref{tab:yieldsextrapolation} show the extracted dN/dy and the amount of extrapolation for each collision system.

Fig.~\ref{fig:alldNdy} shows the evolution of the extracted particle yields as a function of uncorrected charged multiplicity. For each collision system, the extracted yield shows nearly linear evolution with charged multiplicity. 
Fig.~\ref{fig:alldNdy_thermalpaper} shows a compilation of extracted particle yields in central ($\sim$ 0-5\%) Au-Au/Pb-Pb collisions from AGS to RHIC energies~\cite{Andronic:2005yp}, including the STAR 62.4 GeV Au-Au data points as well (red markers but with the same shape coding). One should note that data points presented for higher RHIC energies are measured by the PHENIX experiment. Weak decay corrections are applied to proton/antiproton spectra but not to pion spectra. Fig.~\ref{fig:alldNdy_thermalpaper} also depicts the particle composition of the collision products. Pion production dominates at all energies above a center of mass energy $\sim$ 5 GeV kaon production shows similar evolution to pions. One can observe a change in the trend around $\sqrt{S_{NN}}$= $\sim$ 5 GeV. Below this energy the collisions are dominated by the incoming nuclei, which undergo a significant stopping in the collision zone, also shown by the proton yield. At higher energies, nuclei become more transparent, at the SPS region production of particles with strange quark content starts to incerase and above the top SPS energy the chemical composition evolves smoothly to RHIC energies. The 62.4 GeV data set is situated in the transition region between collisions dominated by stopping and the transparent collisions, therefore contributes significanlty to the systematic mapping of particle production in heavy-ion collisions. (Note that in top RHIC energy collisions there is a finite number of net-baryon present in the collision zone~\cite{Adams:2003xp}.)
\begin{figure}[!h]
	\begin{center}
			\includegraphics[width=0.9\textwidth]{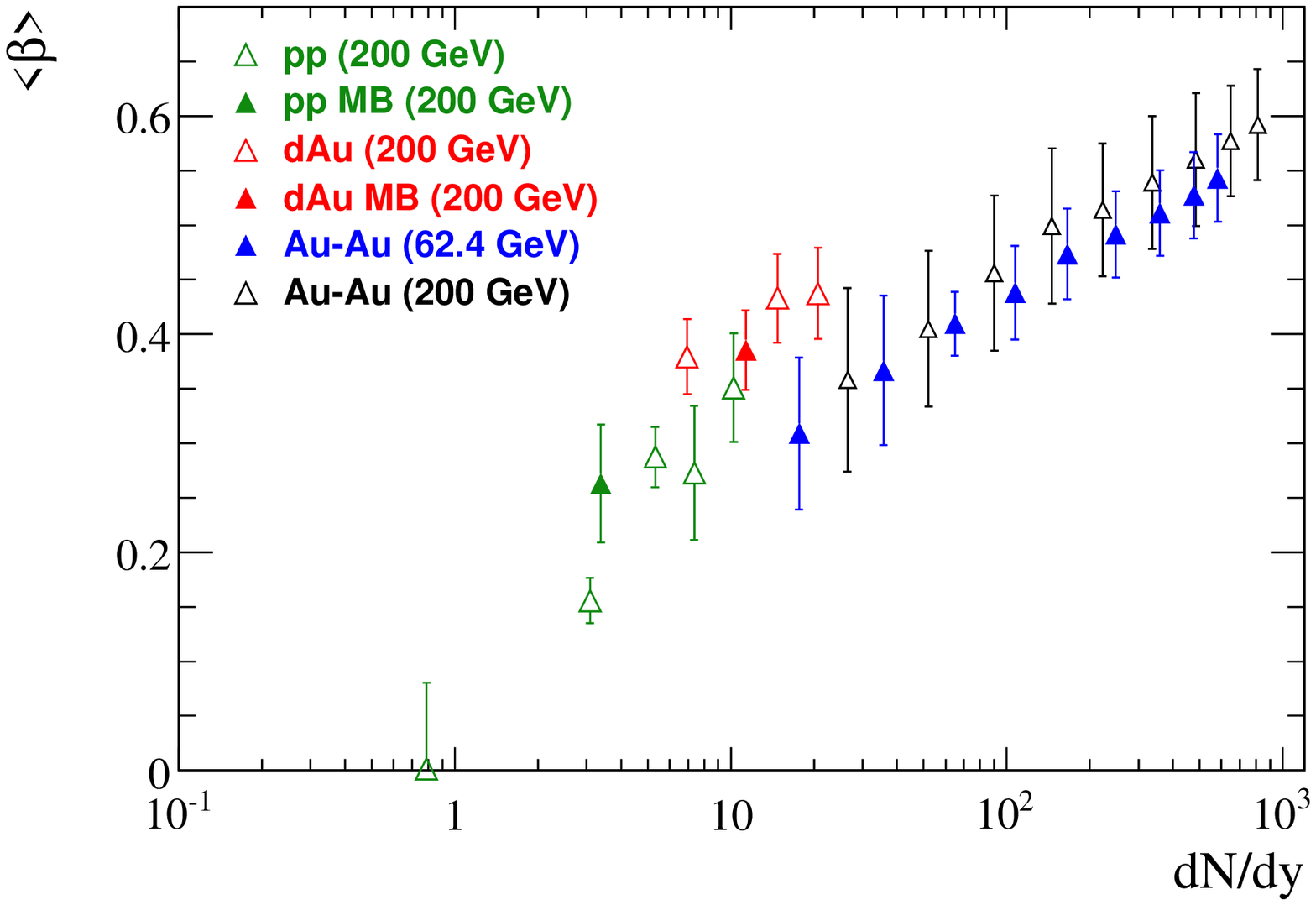}
		\includegraphics[width=0.9\textwidth]{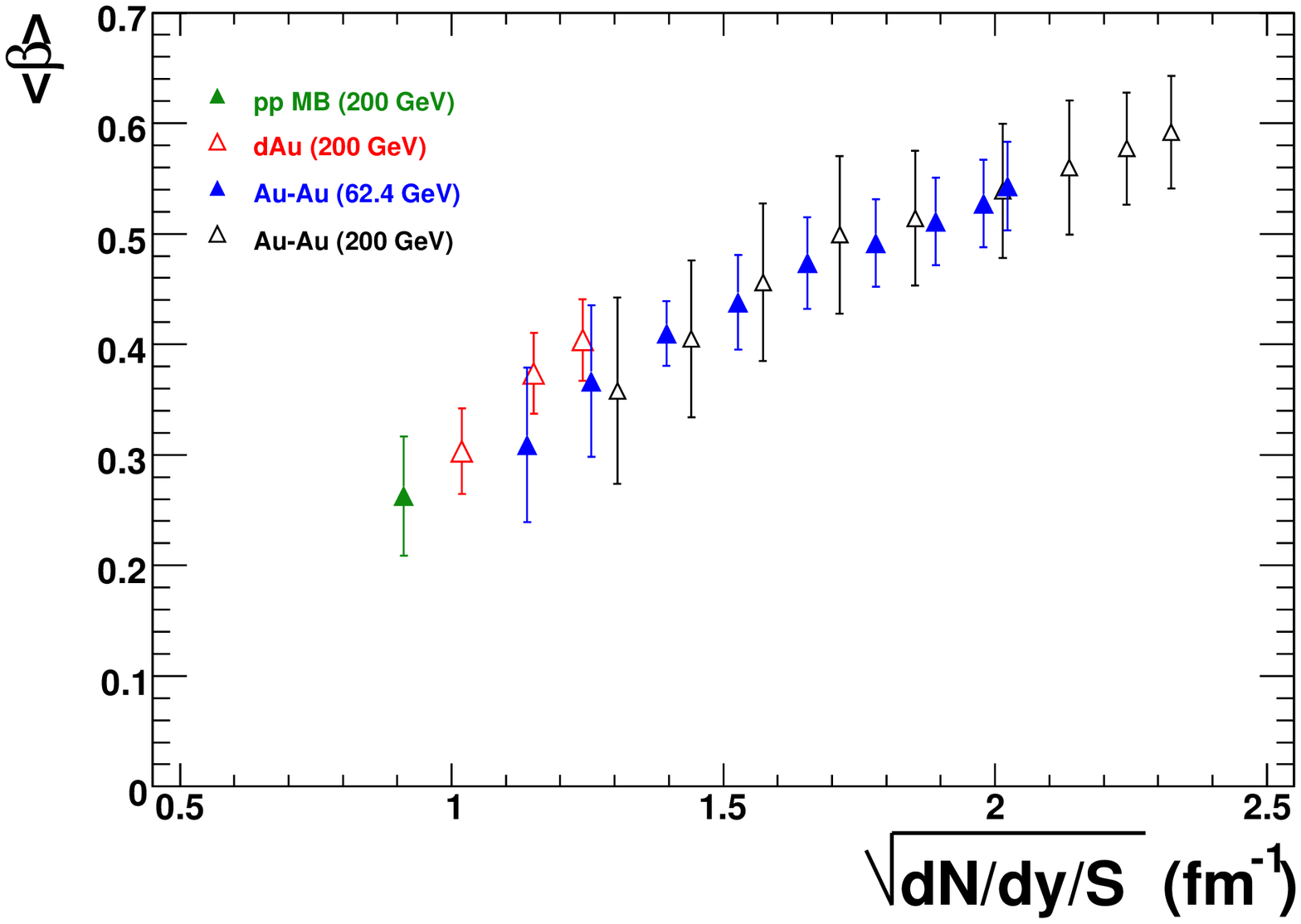}
\caption{Average transverse flow velocity extracted from blast-wave fits as a function of corrected dN/dy and $\sqrt{dN/dy/S}$.}
	\label{fig:beta}\end{center}
\end{figure}

\subsection{Particle ratios}\label{sec:ratios}

The particle-antiparticle ratios ($\pi^{-}/\pi{+}$, $K^{-}/K^{+}$, $\overline{p}/p$),
and unlike particle ratios ($K^{\pm}/\pi^{\pm}$, $p(\overline{p})/\pi^{\pm}$) are presented
as a function of multiplicity/centrality in this section.
Particle ratios are calculated from the integrated inclusive particle yields as described in Section~\ref{sec:extractspectraparas}. 

Fig.~\ref{fig:papratiospion}, Fig.~\ref{fig:papratioskaon} and Fig.~\ref{fig:papratiosp} show the dependence of particle ratios for
$\pi^{-}/\pi^{+}$, $K^{-}/K^{+}$ and $\overline{p}/p$ as a function of charged multiplicity in pp, dAu and Au-Au collisions at 200 GeV and in 62.4 - 130 - 200~\cite{Adams:2003xp} GeV Au-Au collisions. 

The $\pi^{-}/\pi^{+}$ ratio is $\approx$ 1 at each collision energy and collision species. The ratios are flat as the function of multiplicity/centrality. Similar behavior has been observed at lower collision energies as well.

The $K^{-}/K^{+}$ ratios are close to 90 \% in pp, dAu and Au-Au collisions
at 200 GeV. The same ratios show slight decrease from 200 GeV to 130 to 62.4 GeV. At lower energies due to the non-zero net baryon density in the collisions zone the associated production of kaons with hyperons will be different from these produced with antihyperons. 

\begin{figure}[!h]
	\begin{center}
		\includegraphics[width=0.9\textwidth]{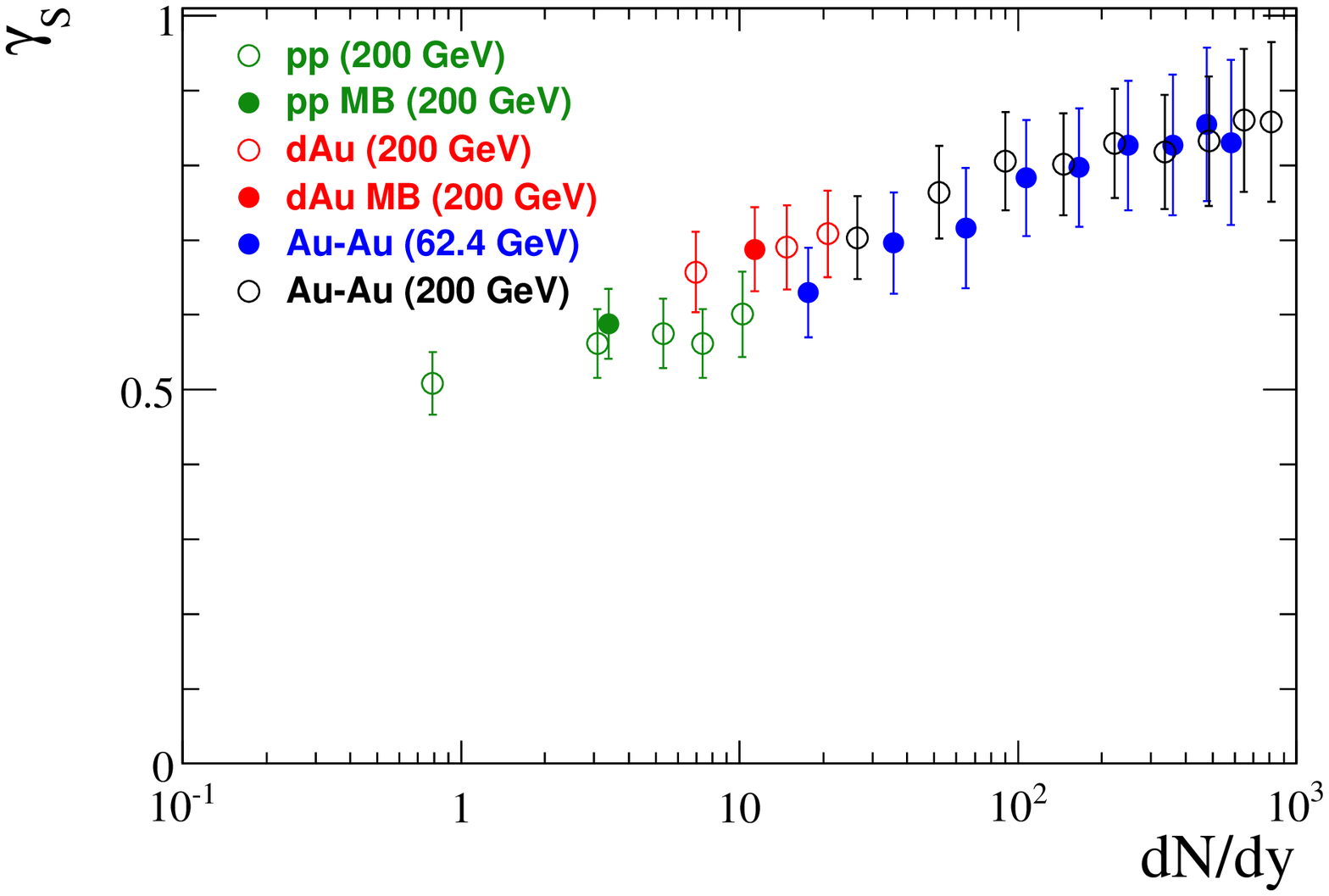}
		\includegraphics[width=0.9\textwidth]{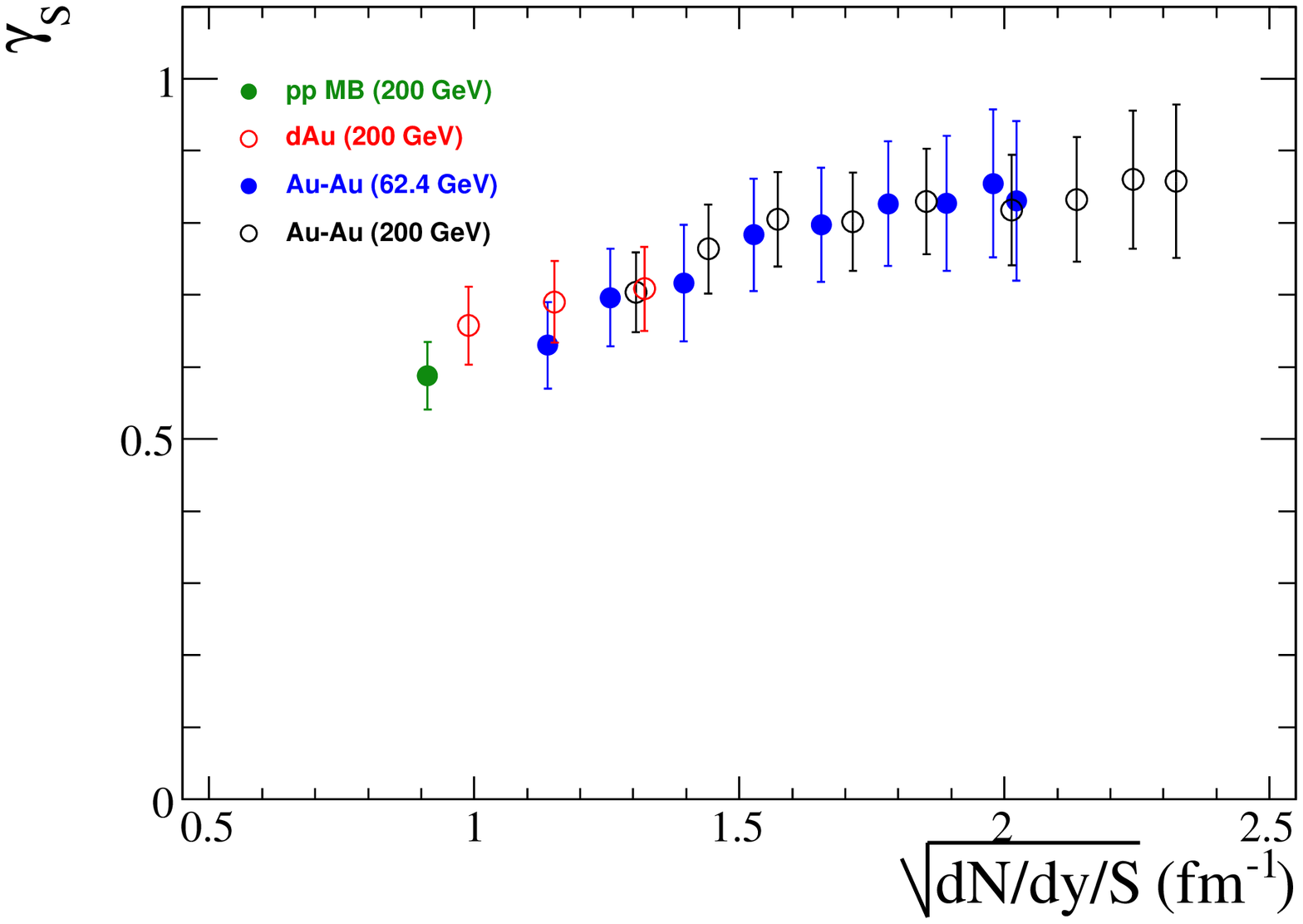}
\caption{Strangeness suppression factor extracted from chemical model fit in 200 GeV pp, dAu and Au-Au and 62.4 GeV Au-Au collisions corrected dN/dy and $\sqrt{dN/dy/S}$.}
	\label{fig:chemgamma}\end{center}
\end{figure}
The $\overline{p}/p$ ratio seems to be flat through  the various collision system at 200 GeV within errors, although a slight increase might be observed from pp to dAu to peripheral Au-Au and there seems to be a small drop in the ratio starting from mid central to central Au-Au collisions. 
The $\overline{p}/p$ ratio is similar at 130 GeV and 200 GeV, but 
shows significant drop at 62.4 GeV. The $\overline{p}/p$ ratio is 
flat at higher collision energies although there seems to be a hint of a drop in 200 GeV Au-Au as well.
More pronounced drop of the ratio toward central collisions at 62.4 GeV, is consistent with larger 
stopping at central collisions.    

The $K^{-}/K^{+}$ and $\overline{p}/p$ ratios are shown in Fig.~\ref{fig:papratios_S1} and Fig.~\ref{fig:papratios_S1} as a function of $\sqrt{S_{NN}}$. Lower energy data points (AGS, SPS) are from~\cite{Afanasiev:2002mx,Ahle:2000wq}, RHIC points are from~\cite{Adler:2003cb,Adler:2002wn,Adler:2001bp} and the thermal model calculation is presented in~\cite{Andronic:2005yp}. Ratios smoothly evolve from AGS to RHIC energies approaching 1. The $\pi^{-}/\pi^{+}$ ratios are flat from SPS to RHIC energies~\cite{Andronic:2005yp}, although not plotted here. The connection between kaon and proton/anitproton production is clearly seen in Fig.~\ref{fig:kktopbarp}, where $K^{-}/K^{+}$ ratios are plotted as a function of $\overline{p}/p$ for the energies and rapidities cited in~\cite{Bearden:2003fw}. Implication of this behavior to the chemical freeze-out will be discussed in details in Sec.~\ref{sec:freezeoutprop}.

\begin{figure}[!h]
	\begin{center}
		\includegraphics[width=0.9\textwidth]{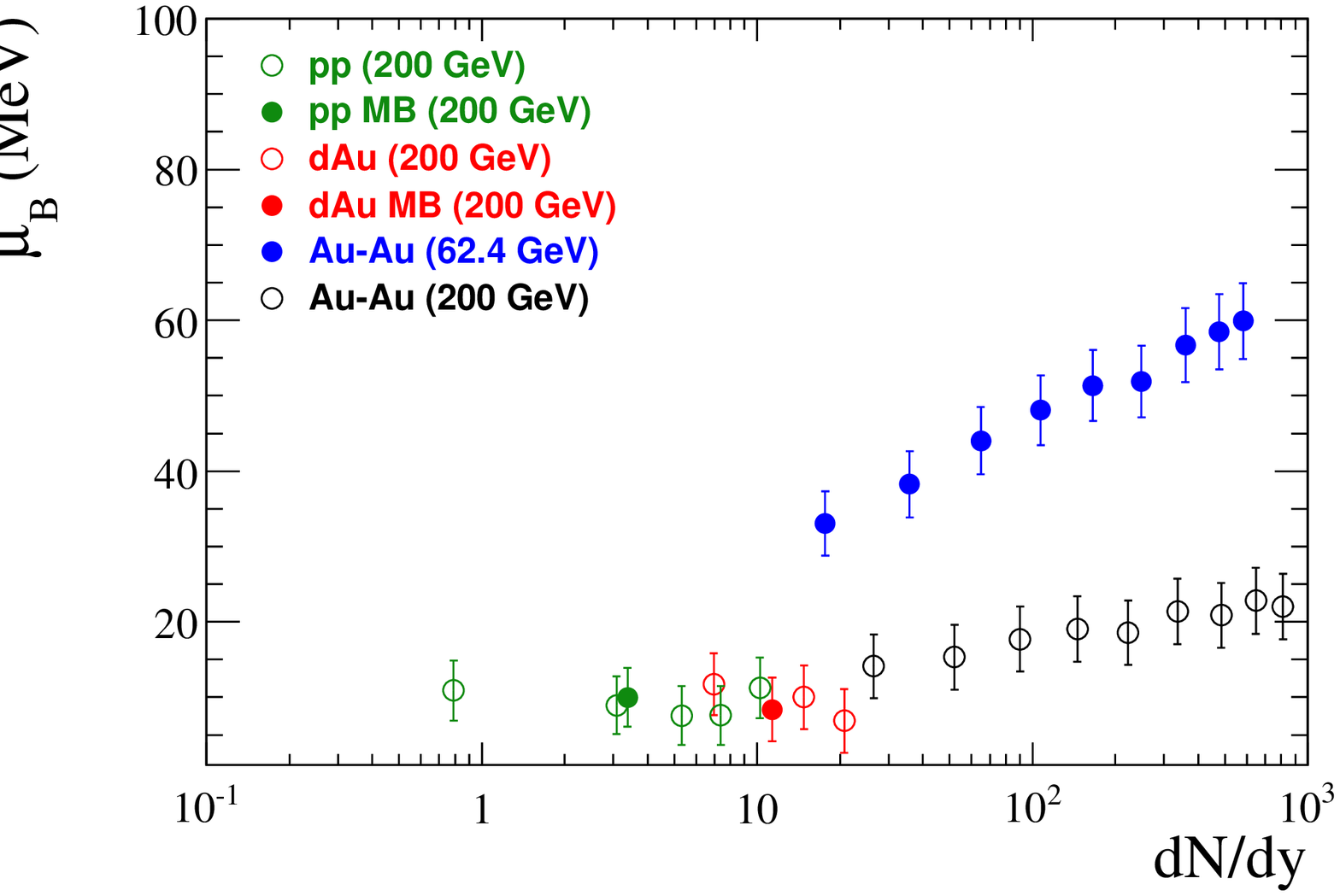}
		\includegraphics[width=0.9\textwidth]{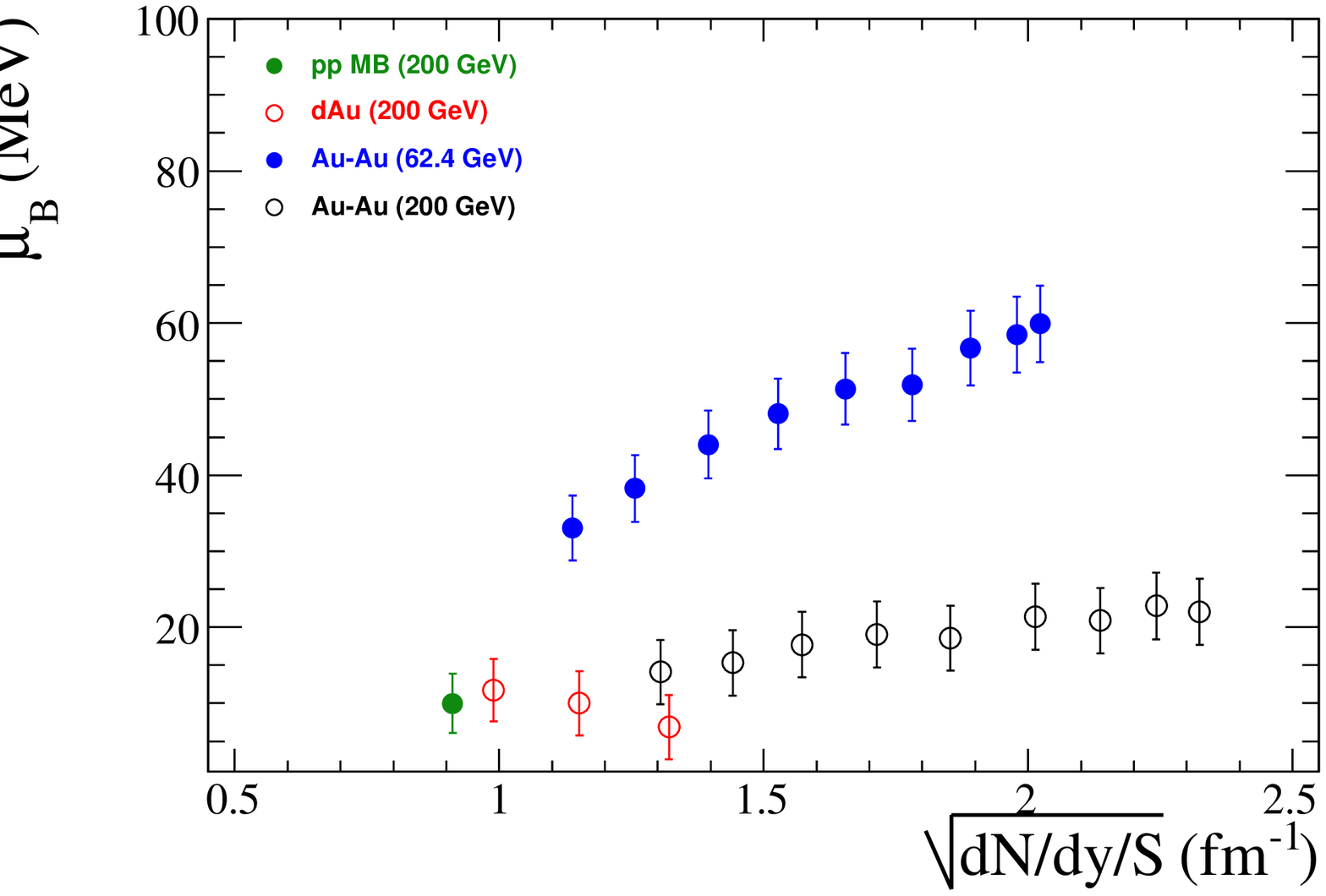}
\caption{Baryon chemical potential extracted from chemical model fit in 200 GeV pp, dAu and Au-Au and 62.4 GeV Au-Au collisions corrected dN/dy and $\sqrt{dN/dy/S}$.}
	\label{fig:chemmuB}
	\end{center}
\end{figure}

Fig~\ref{fig:unlikeratios1} shows the unlike particle ratios $K^{-}/\pi^{-}$ and
$\overline{p}/\pi^{-}$ in pp, dAu and Au-Au collisions as a function of charged hadron multiplicity and $dN/dy/S$. 
These ratios represent the bulk strangeness and baryon production. Strangeness enhancement in heavy-ion collisions with respect to pp collisions is considered one of the possible signatures of the QGP formation. In elementary collisions (eg. pp) or in hadron gas (heavy-ion collisions without QGP formation) strangeness production leads to strange particle (hadron) pairs requiring large amount of energy. In QGP strange quark - antiquark pair can be produced which are energetically favored with respect to strange hadron pair production. Most of the higher mass antibaryons decay into antiprotons, therefore the $\overline{p}/\pi^{-}$ ratio is a good measure of overall antibaryon production.

The ratios: $K^{-}/\pi^{-}$ and $\overline{p}/\pi^{-}$ gradually increase from pp to peripheral, mid-peripheral Au-Au
 collisions and saturate in mid-central and central collisions. Kaon production in central Au-Au collisions is increased by $\approx$
50 \% with respect to minimum bias pp collisions. 
Ratios at 62.4 GeV are lower than at 200 GeV and $\overline{p}/\pi^{-}$ show more pronounced drop at 62.4 GeV than at 200 GeV.

Fig.~\ref{fig:kaon2pion_S1} and Fig.~\ref{fig:kaon2pion_S2} show the $K^{+}/\pi^{+}$ and $K^{-}/\pi^{-}$ ratios from AGS~\cite{Ahle:1999uy,Klay:2003zf,Bearden:2002ib} to SPS to RHIC energies. 
$K^{+}/\pi^{+}$ shows a sharp increase at lower energies and drops at energies higher than $\sim$ 10 GeV, whereas $K^{-}/\pi^{-}$ ratios monotonically increase. Behavior of the $K/\pi$ ratios can be explained with the different energy dependence of the kaon production rate and the net baryon density. The 62.4 GeV STAR measurements follow the smooth trend for both ratios and are close to the thermal model description~\cite{Andronic:2005yp}. 

Systematic uncertainties on particle ratios come from those on the extrapolated yields using the various functional forms discussed above. The systematic uncertainties on the extrapolated yields are somewhat correlated, thus are partially canceled in the ratios.

\section{Freeze-out properties}\label{sec:freezeoutprop}

In this section we present the freeze-out properties of the various systems, treating the chemical and kinetic freeze-outs separately.
Chemical freeze-out is investigated in terms of thermal models~\cite{Xu:2001zj,Braun-Munzinger:1999qy,Andronic:2005yp}. Kinetic freeze-out happens later than the chemical freeze-out, meanwhile the system expands and cools. Kinetic freeze-out is described by the blast-wave model fits to the measured data.

\begin{sidewaysfigure}[!h]
	\begin{center}

		\includegraphics[width=0.48\textwidth]{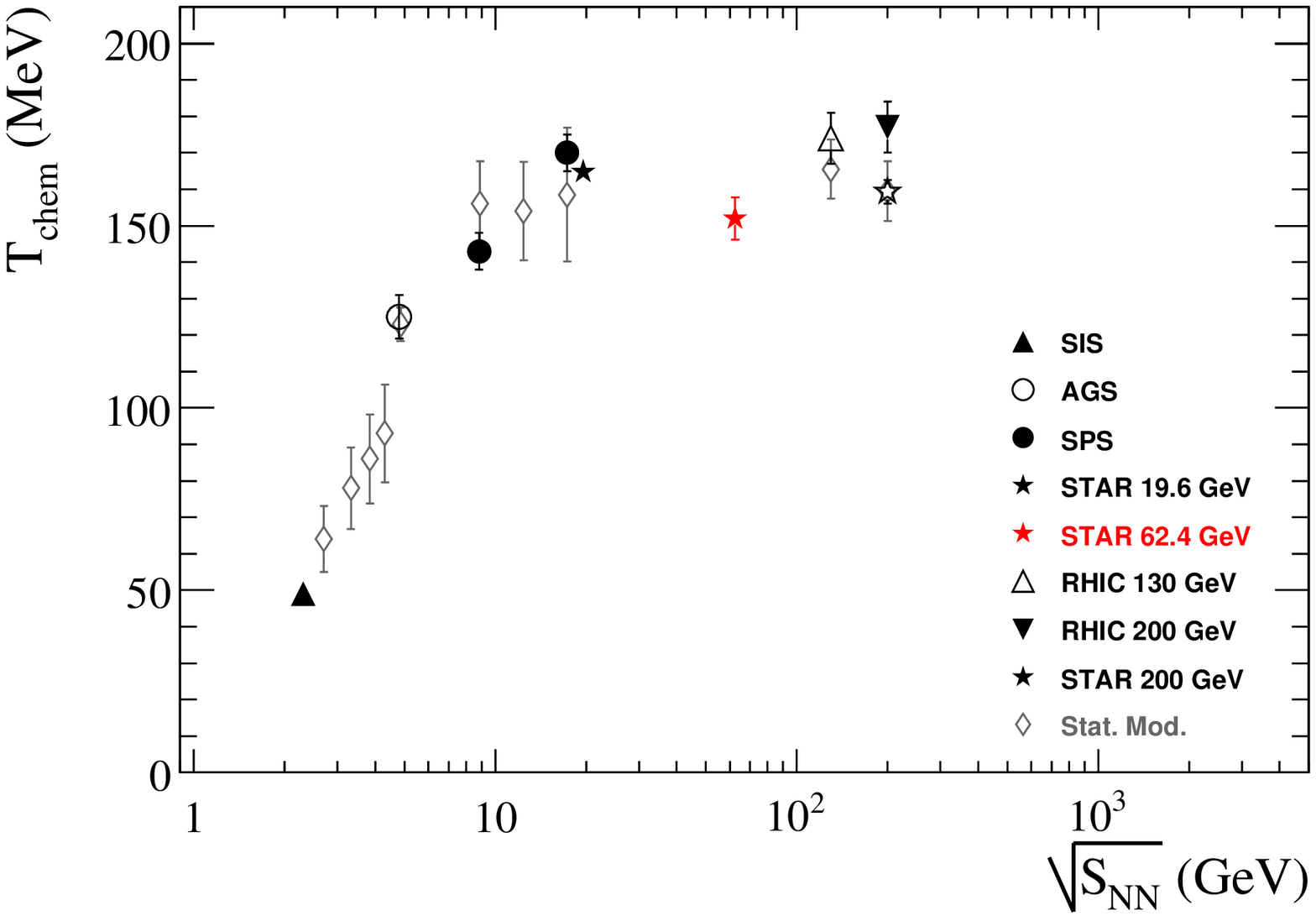}
		\includegraphics[width=0.48\textwidth]{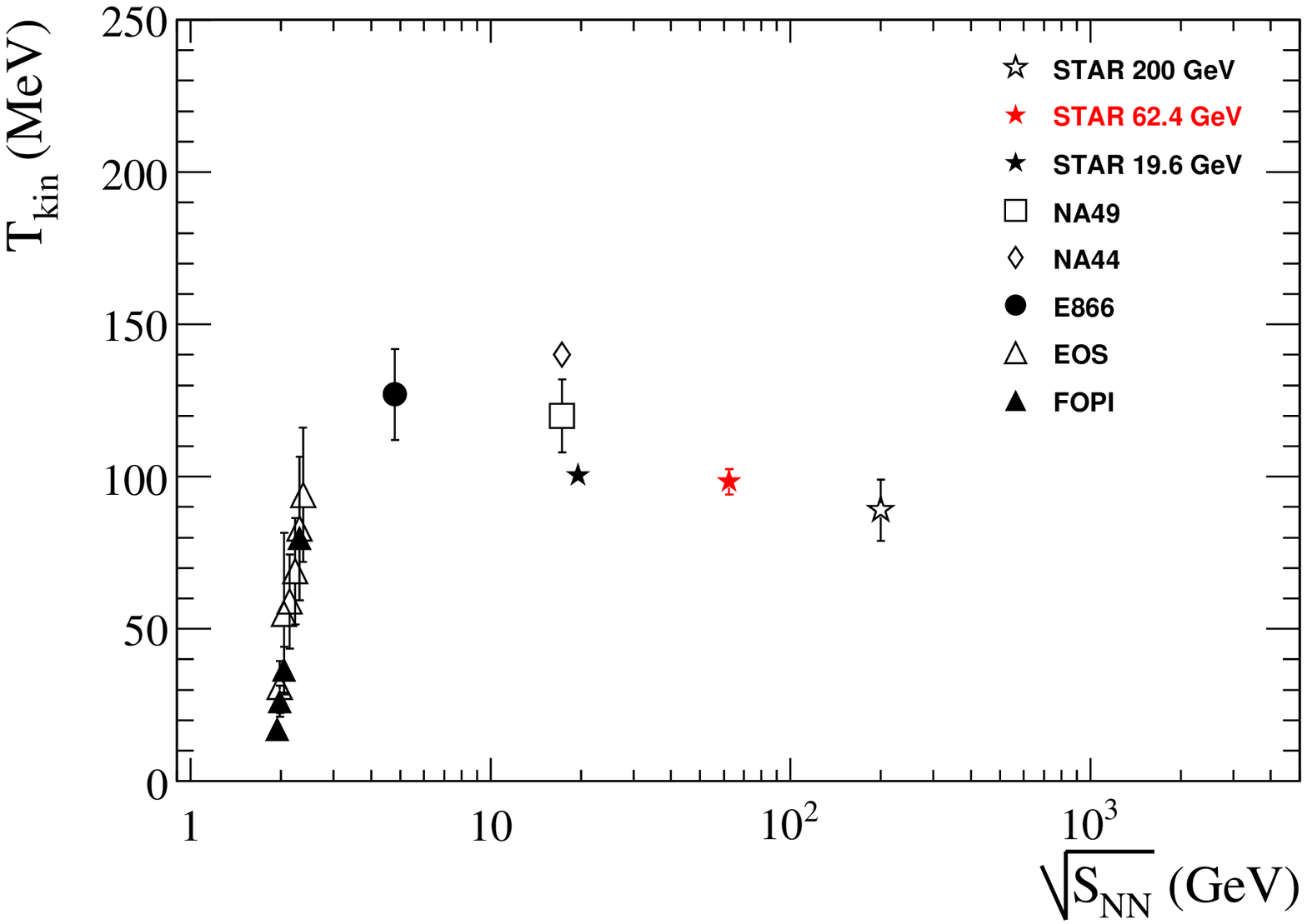}
		\includegraphics[width=0.48\textwidth]{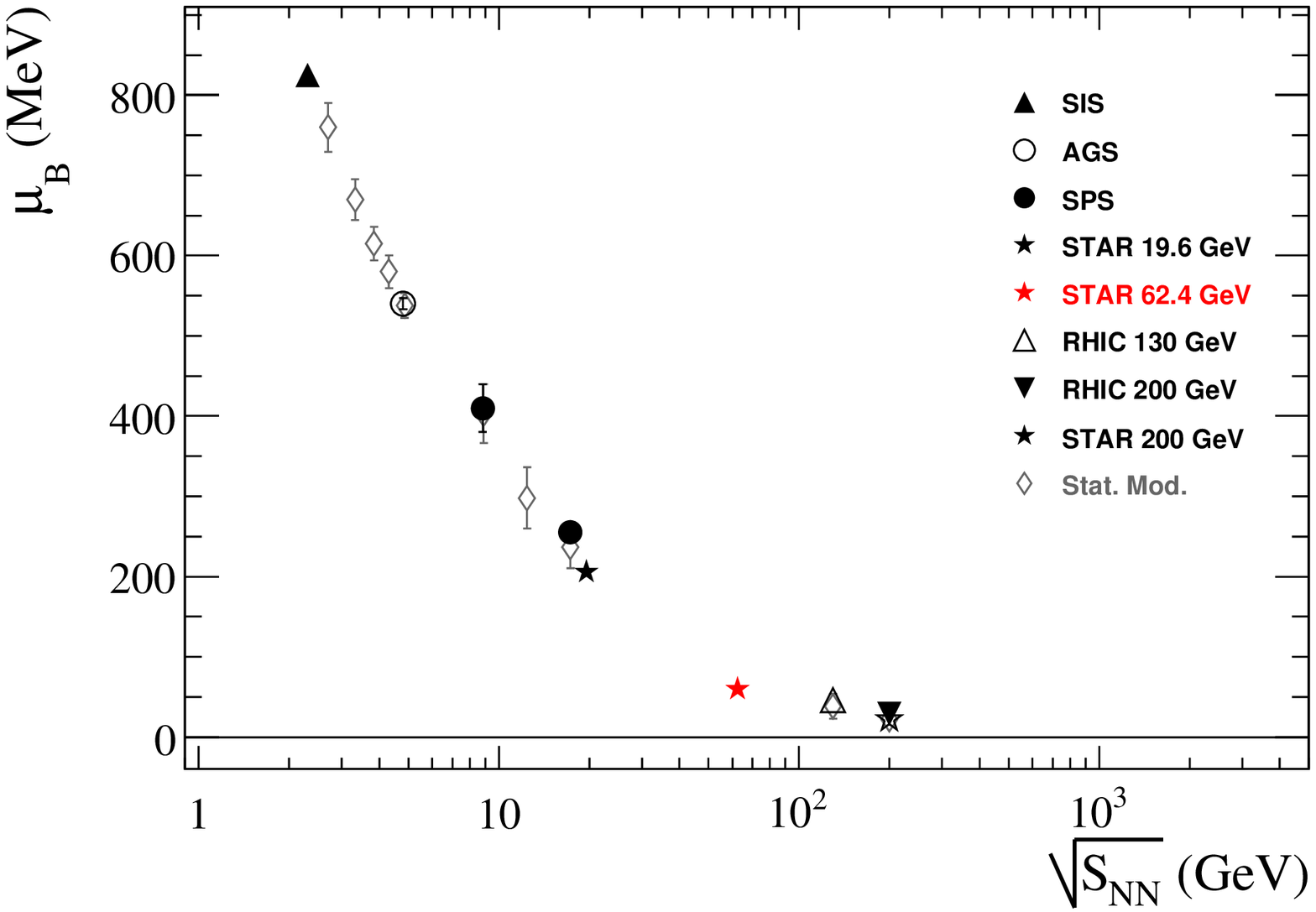}
		\includegraphics[width=0.48\textwidth]{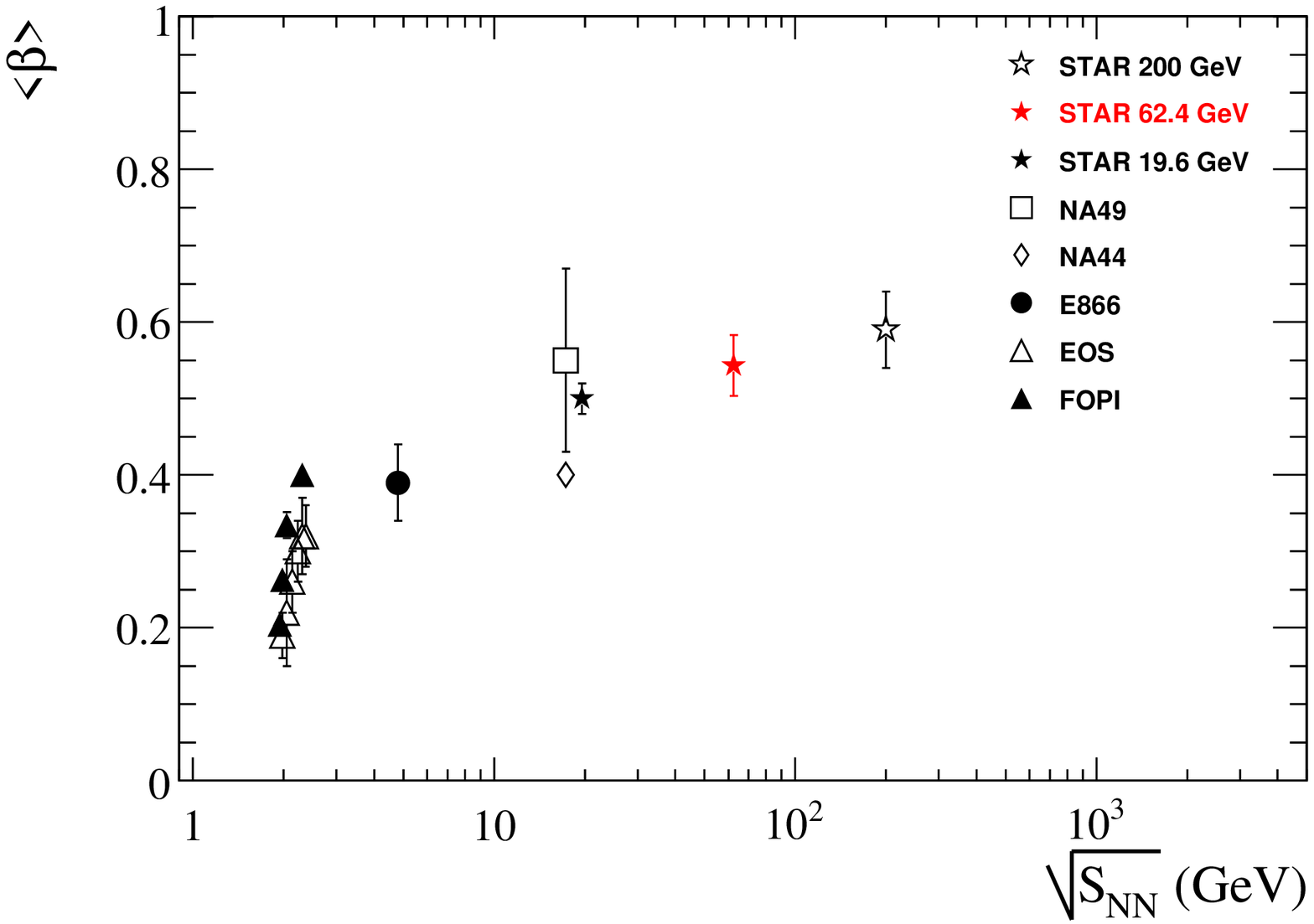}
\caption{Excitation function of chemical and kinetic freeze-out parameters in central heavy-ion collisions. See text for references.}
	\label{fig:chemprop}
	\end{center}
\end{sidewaysfigure}
\subsection{Chemical and kinetic freeze-out}\label{sec:freezeout}
In our framework the measured particle ratios are fitted with a four parameter chemical freeze-out
model, where the free parameters are the chemical freeze-out temperature 
($T_{chem}$), the baryon and strangeness chemical potentials ($\mu_{B}$,
 $\mu_{S}$) and the strangeness suppression factor ($\gamma_{S}$). 
The following bulk ratios are included in the fit: $\pi^{+}/\pi^{-}$, 
$K^{-}/K^{+}$, $\overline{p}/p$, $K^{-}/\pi^{-}$, 
$\overline{p}/\pi^{-}$. The fit is performed for each collision systems and each multiplicity/centrality class. 

The first observation is the independence of the chemical freeze-out temperature of collision system and multiplicity/centrality. In each system investigated the chemical freeze-out temperature is around $\approx$ 150 - 156 MeV which is close to the LQCD calculation for three flavors (154 MeV $\pm$ 8 MeV)~\cite{Karsch:2004ti}. For Au-Au collisions at the two investigated collision energies the chemical freeze-out temperatures are the same. 
Chemical freeze-out temperature as a function of $\sqrt{S_{NN}}$ shows a sharp rise in the AGS range and seems to be flat from SPS to RHIC (starting at $\sim$ 10 GeV)~\cite{Andronic:2005yp}.

Further chemical freeze-out parameters can be extracted, such as the strangeness suppression factor, shown in Fig.~\ref{fig:chemgamma}. In central Au-Au collisions $\gamma_{S}$ approaches 0.9 from the pp value $\sim$ 0.5. This infers that the system created in the collision evolves toward chemical equilibrium even for the strangeness sector. Perfect chemical equilibrium is achieved if $\gamma_{S}$ = 1. Including only charged kaons in the chemical model fit leads to a smaller saturation value of the strangeness suppression factor. Chemical model fits including strange baryons ($\Lambda, \Lambda_{1520}, \Xi, \Omega$) and mesons ($\phi, K^{*}, K_{S}$) with the bulk particles measured leads to $\gamma_{S} \approx$ 1 in central Au-Au collision at RHIC energies.

As shown in Fig.~\ref{fig:kktopbarp}, $K^{-}/K^{+}$ ratios are plotted as a function of $\overline{p}/p$. From naive quark counting the correlation between these ratios is expected to exert a power law relation with an exponent of 1/3, but the measured ratios seem to differ from that as shown in Fig.~\ref{fig:kktopbarp}. Chemical model fits at fixed chemical freeze-out temperature (170 MeV) are able to describe the measured correlation with varying baryon chemical potential~\cite{Becattini:2000jw}. The good description of the correlation between these particle ratios infers chemical equilibrium in the measured collisions in a rapidity range $|y| < \sim$ 3.0. 

Finally, the baryon chemical potential is presented in Fig.~\ref{fig:chemmuB}, which can characterize the transparency and/or the net baryon content of the collision zone. At 200 GeV the baryon chemical potential is small $\sim$ 10 - 22 MeV, but increases to $\sim$ 40 - 80 MeV at 62.4 GeV. The observed decreasing trend of $\overline{p}$ ratios at 62.4 and 200 GeV are due to the increasing stopping toward central Au-Au collisions.

Kinetic freeze-out parameters are extracted from the blast-wave model fits. Figure~\ref{fig:tfo} summarizes the extracted freeze-out temperatures for 200 GeV pp, dAu and Au-Au~\cite{Adams:2003xp} collisions and for Au-Au collisions at 62.4 GeV. 
As opposed to the chemical freeze-out temperature, the kinetic freeze-out temperature shows strong evolution with collision system and multiplicity/centrality. At very low multiplicity pp collisions kinetic freeze-out temperature is close to the chemical freeze-out temperature. As the multiplicity increases kinetic freeze-out temperature decreases. This trend continues through dAu centralities and through the Au-Au centralities. Au-Au collisions at 62.4 GeV and 200 GeV show similar freeze-out.

The apparent average transverse flow velocity shows the same increasing trend, as depicted in Fig.~\ref{fig:beta}.
In Au-Au collisions the average transverse flow velocity is interpreted as the result of collective expansion of the system created at collision. In pp and dAu collisions the system is considered too small for expansion, but with increasing multiplicity in pp and increasing centrality in dAu due to the azimuthally averaged spectra through several collisions, (mini)jets and the increasing contribution from $k_{T}$ broadening can mimic the expected collective behavior, which can be measured on the event-by-event basis in Au-Au collisions but not in pp or dAu. Despite the different physical processes the extracted average flow velocity evolves smoothly. Now, concentrating on Au-Au collision only, the decreasing/increasing trend of the kinetic freeze-out temperature/average flow velocity is the same within errors at 62.4 GeV and at 200 GeV. The magnitude of the freeze-out parameters seem to be only governed by the charged particle multiplicity. This might suggest that a given collision system (in our case Au-Au) above the expected phase transition center of mass energy (independent of the initial conditions, eg. baryon content of the collision zone, but not the initial energy density) follows a general evolution: after chemical freeze-out the system expands and cools to kinetic freeze-out determined by the charged particle multiplicity (i.e. the initial energy density of the collision).

It would be an interesting continuation of this work with improved statistics of the pp and dAu data to study the freeze-out properties of high multiplicity but not jet like events. That might lead to a better understanding of gluon saturation in high energy collisions. Unfortunately this kind of event selection with the current statistics of 200 GeV pp and 200 GeV dAu data is not possible.

Over the last decades several experiments have measured the freeze-out properties. Fig.~\ref{fig:chemprop} shows the excitation function of the chemical freeze-out temperature (top left), the baryon chemical potential (bottom left), the kinetic freeze-out temperature (top right) and the average transverse flow velocity (bottom right) from a collection of experimental and theoretical works for central Pb-Pb/Au-Au collisions. A collection of numerical values can be found in~\cite{Andronic:2005yp} and references therein. The 20 GeV Au-Au data points are from~\cite{Picha:2005pq}.

Each freeze-out parameter follows a smooth trend with increasing collision energy, and the 62.4 GeV Au-Au measurements fit well in this trend. A distinct change can be observed at $\sim$ 10 GeV for the chemical and kinetic freeze-out temperatures and for the average transverse flow velocity. The chemical freeze-out temperature seems to be independent of the increasing collision energy above $\sim$ 10 GeV. The kinetic freeze-out temperature shows a slight decrease and the average transverse flow velocity shows a monotonic increase $\sim$ 10 GeV. 
The baryon chemical potential rapidly drops from $\sim$ 400 MeV at 10 GeV to $\sim$ 10 - 20 MeV at 200 GeV, reaching the nearly transparent regime of heavy-ion collisions. 

Systematic uncertainties on the kinetic freeze-out parameters are also assessed by excluding the kaon spectra or the proton spectra from the Blast-wave fit. An overall 10\% uncertainty can be applied on the kinetic freeze-out temperature and 8\% on the average transverse flow velocity. The errors on the chemical freeze-out parameters include the systematic error from the particle ratios, can be found in~\cite{Adams:2003xp}.

\section{Effect of resonance decays on the kinetic freeze-out temperature\label{sec:resonances}}

\subsection{Extraction of freeze-out properties in heavy-ion collisions}

Kinetic freeze-out parameters are extracted from low momentum identified particle spectra, where the effect of the collective flow field on the particles is the strongest. However, as long standing common knowledge, the low momentum (especially bulk particles: $\pi$, K, p and $\overline{p}$) particles carry significant contribution from resonance particle decays. 

One possible way to reduce the effect of resonance contribution is to treat the identified particle spectra as primordial and exclude the low momentum part ($p_{T} <$ 475 MeV) of the pion spectra. The kinetic freeze-out temperature extracted this way is significantly lower than the chemical freeze-out temperature in central Au-Au collisions at RHIC energies (except 20 GeV). There are arguments to include all resonance particles in the description of the freeze-out, hence the chemical and kinetic freeze-out can be described with a single freeze-out temperature~\cite{Broniowski:2001we}, though chemical models generally fail to reproduce the multi strange particle ratios, which is the input for the single freeze-out model. Single freeze-out models can describe identified particle spectra in the level of 20-30 \%, which is significant if we consider the low momentum part of the spectra which contributes in the order of $\sim$ 90\% to the total particle yield.

\subsection{Motivation of the study}

Our goal is to provide a good description of the measured identified particle spectra ($\pi^{\pm}$, $K^{\pm}$, p and $\overline{p}$) in central (0-5\%) 200 GeV Au-Au collisions and investigate the effect of resonances on the extracted kientic freeze-out parameters.
This rigorous study is carried out with the help of a model combined from chemical and thermal freeze-out models, in which primordial and resonance spectra can be calculated and fitted to the measured identified particle spectra and the freeze-out properties. Kinetic freeze-out temperature, the average transverse flow velocity and the $shape$ of the flow distribution can be extracted. 

\subsection{Model description}

STAR has measured a wide variety of particles in 200 GeV Au-Au collisions: $\rho$, $\omega$, $\eta$, $\eta'$, $K^{*0}$, $K^{*\pm}$, $\phi$ and $\Lambda$, $\Delta$, $\Sigma$, $\Xi$, $\Lambda_{1520}$, $\Sigma_{1385}$, $\Omega$. The experimental knowledge of strange particle yields can constrain the resonance ratios in the chemical model description. The relative proportions of particles and resonances in our study are determined by chemical freeze-out parameters that are fixed. We used $T_{ch}$= 160 MeV, $\mu_{B}$= 22 MeV, $\mu_{S}$= 1.4 MeV, and $\gamma$=0.98, which were obtained from the fit to the identified particle spectra ($\pi^{\pm}$, $K^{\pm}$, p and $\overline{p}$) in central (0-5\%) 200 GeV Au-Au collisions. 

The kinetic component of our model is based on the model by Wiedemann and Heinz~\cite{Wiedemann:1996ig}, which calculates the $p_{T}$ spectra of thermal particles and decay products from resonances in an analytical framework based on the source function of the collision. This model has been modified to accommodate the assumptions in the data description ~\cite{Adams:2003xp}:
\begin{enumerate}
	\item{We assumed two distinct freeze-out temperatures. Chemical freeze-out is fixed by the measured particle ratios. Kinetic freeze-out temperature is a free parameter which can be extracted from the model fit to the measured spectra. The original model~\cite{Wiedemann:1996ig} uses single temperature and sets the chemical potential to zero.}

	\item{We included a more extensive list of resonances, namely: $\rho$, $\omega$, $\eta$, $\eta'$, $K^{*0}$, $K^{*\pm}$, $\phi$ and $\Lambda$, $\Delta$, $\Sigma$, $\Xi$, $\Lambda_{1520}$, $\Sigma_{1385}$, $\Omega$. Each charged decay mode of the resonance particles is taken into account according to the Particle Data Book~\cite{Hagiwara:2002fs}.}
	
\item{In the original code~\cite{Wiedemann:1996ig} the Gaussian source function is implemented, which is changed to box profile to work with the flow profile: $\beta=\beta_{S}\left(r/R\right)^{n}$, as in the data description~\cite{Adams:2003xp}. This assumption is valid in the mid-rapidity region at RHIC for 200 GeV Au-Au collisions.}

\item{Constant $dN/dy$ distributions are implemented instead of Gaussian as in the original code~\cite{Wiedemann:1996ig}. Rapidity distributions are needed for resonances, which can decay into particles at mid-rapidity where our measurements are made.}
\end{enumerate}

Within this modified kinetic freeze-out model primordial and resonance spectra are calculated at the same kinetic freeze-out temperature. 
To limit the necessary computing time required by the calculation, for each spectrum only spectra points between 0.0 GeV/c $< p_{T} <$ 5.0 GeV/c are calculated with 50 MeV steps. This gives a good estimate of the spectrum shape, which is necessary to assign proper yield to the spectrum.
Spectra of resonance decay daughters are calculated through each decay they undergo and are combined according to the proper spin, isospin degeneracies and decay branching ratios. 
\subsection{Calculated spectra}
\begin{figure}[!h]
\begin{center}
\includegraphics[width=0.45\textwidth]{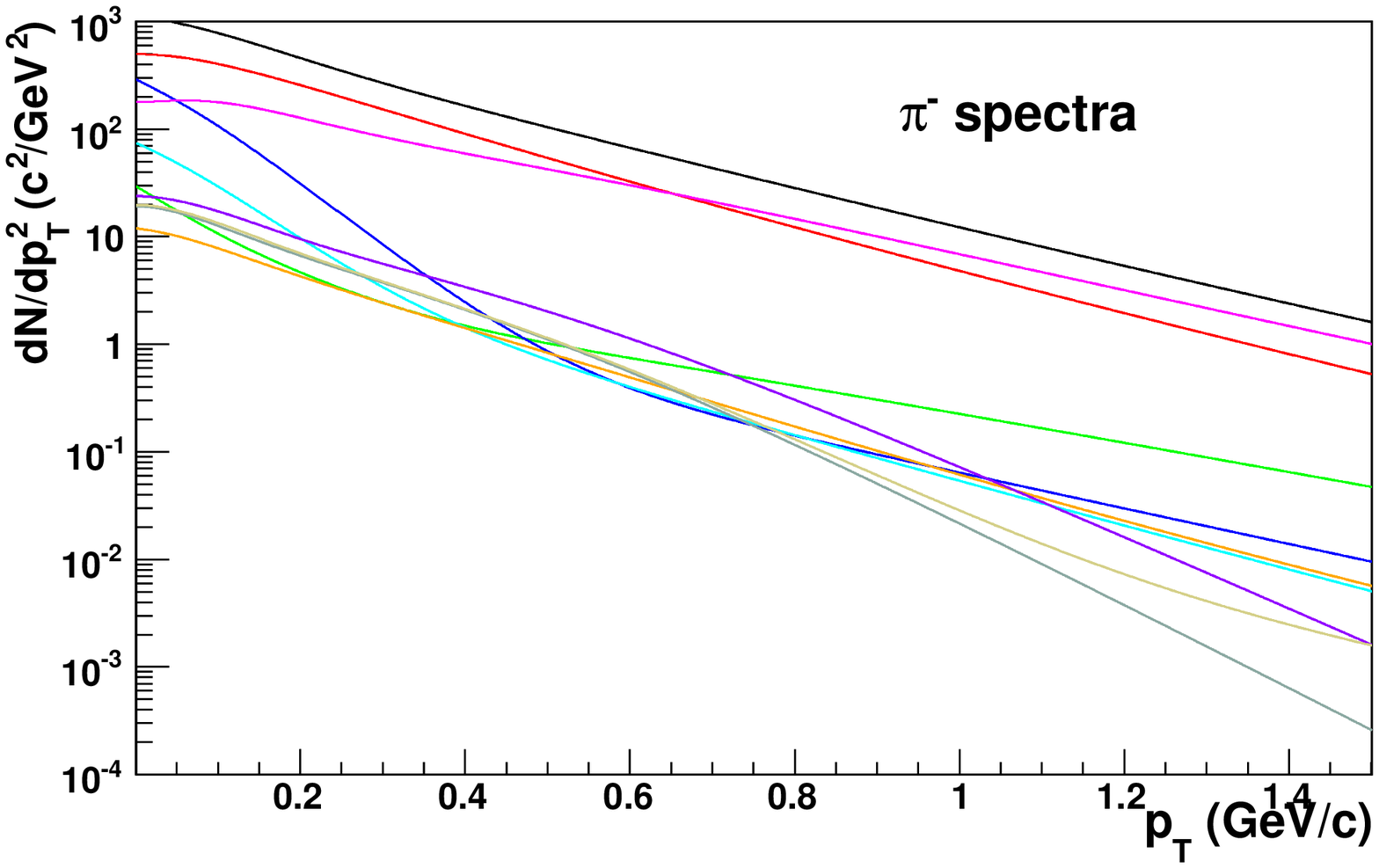} 
\includegraphics[width=0.45\textwidth]{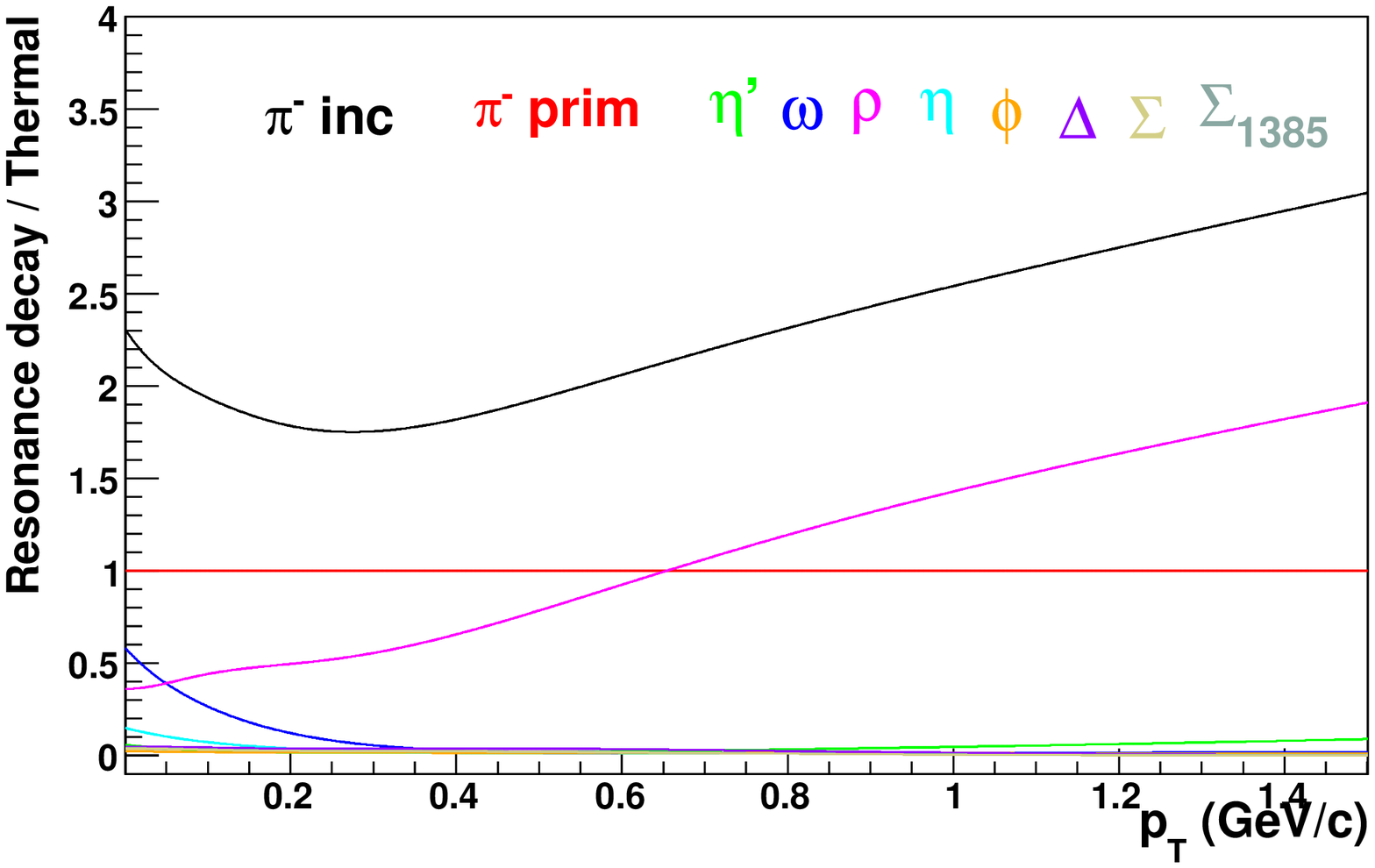} 

\includegraphics[width=0.45\textwidth]{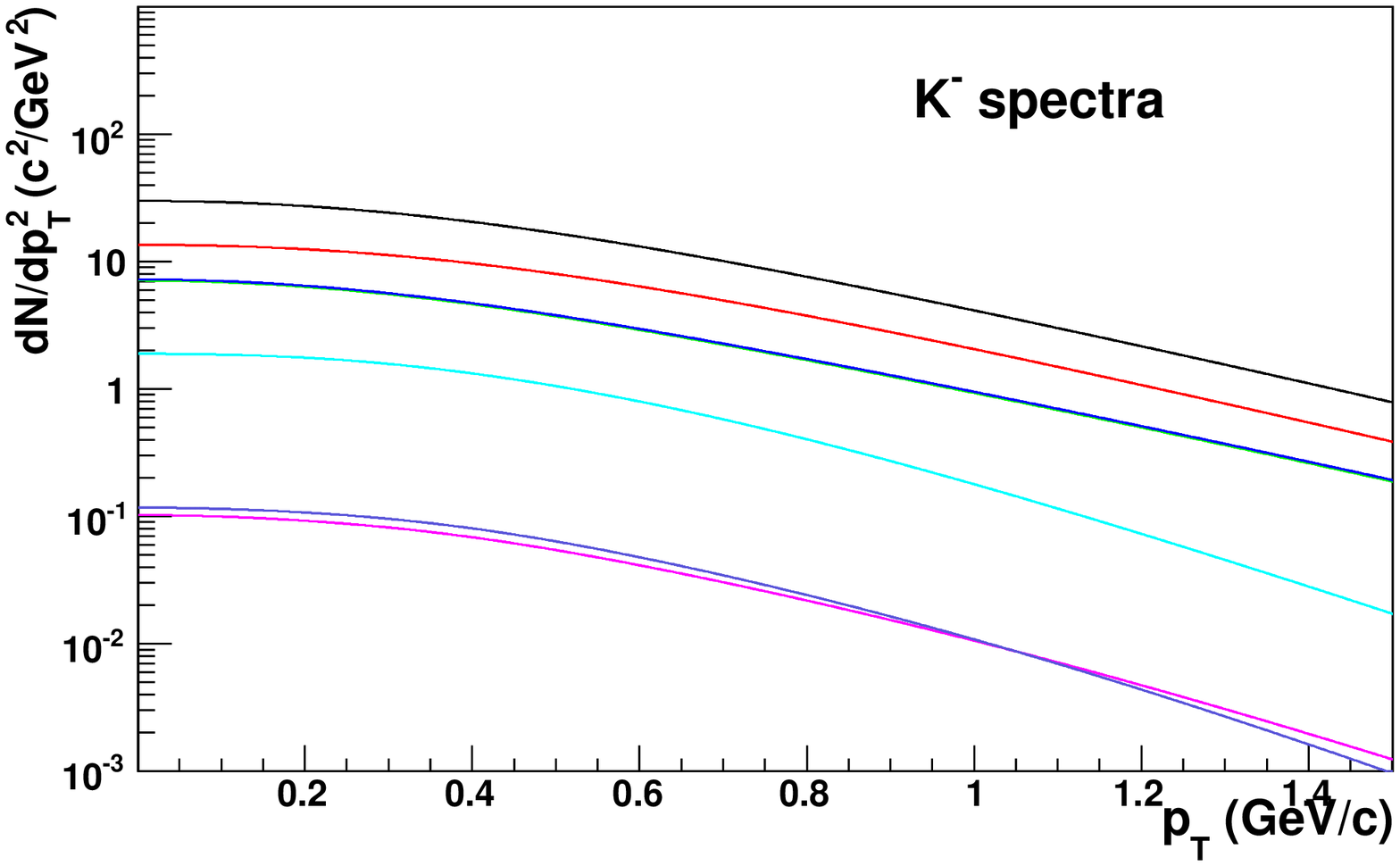} 
\includegraphics[width=0.45\textwidth]{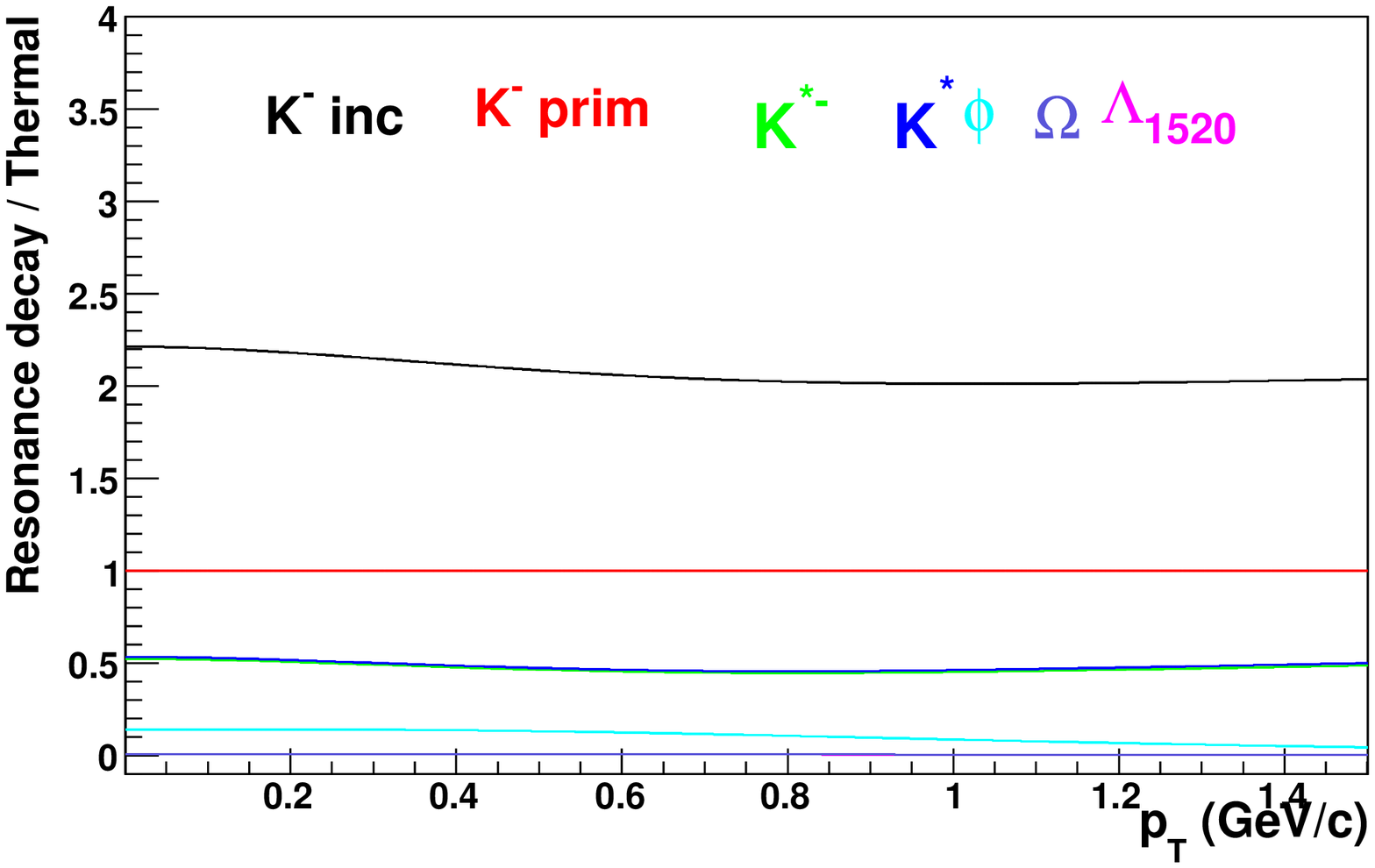}
\\
\includegraphics[width=0.45\textwidth]{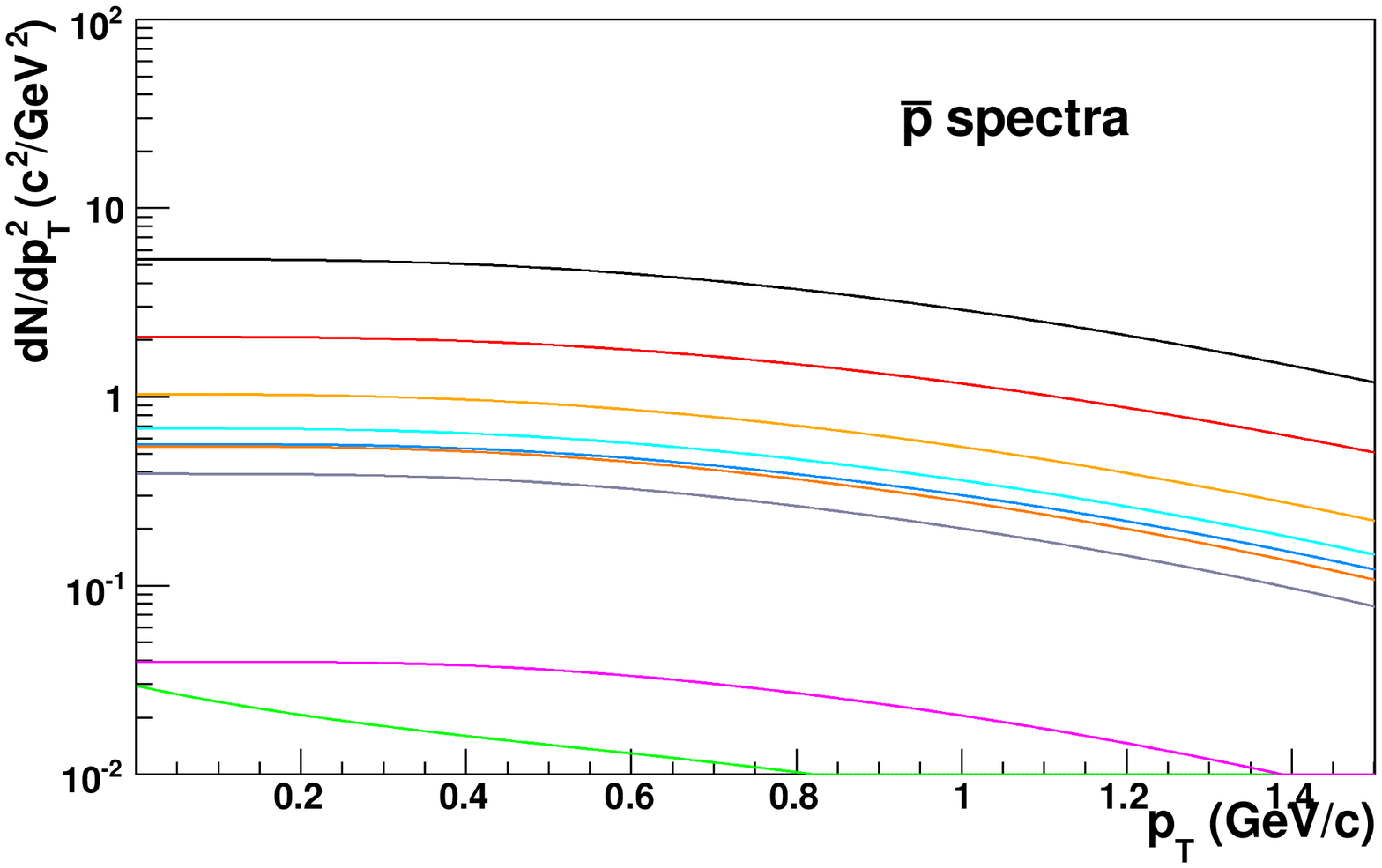}
\includegraphics[width=0.45\textwidth]{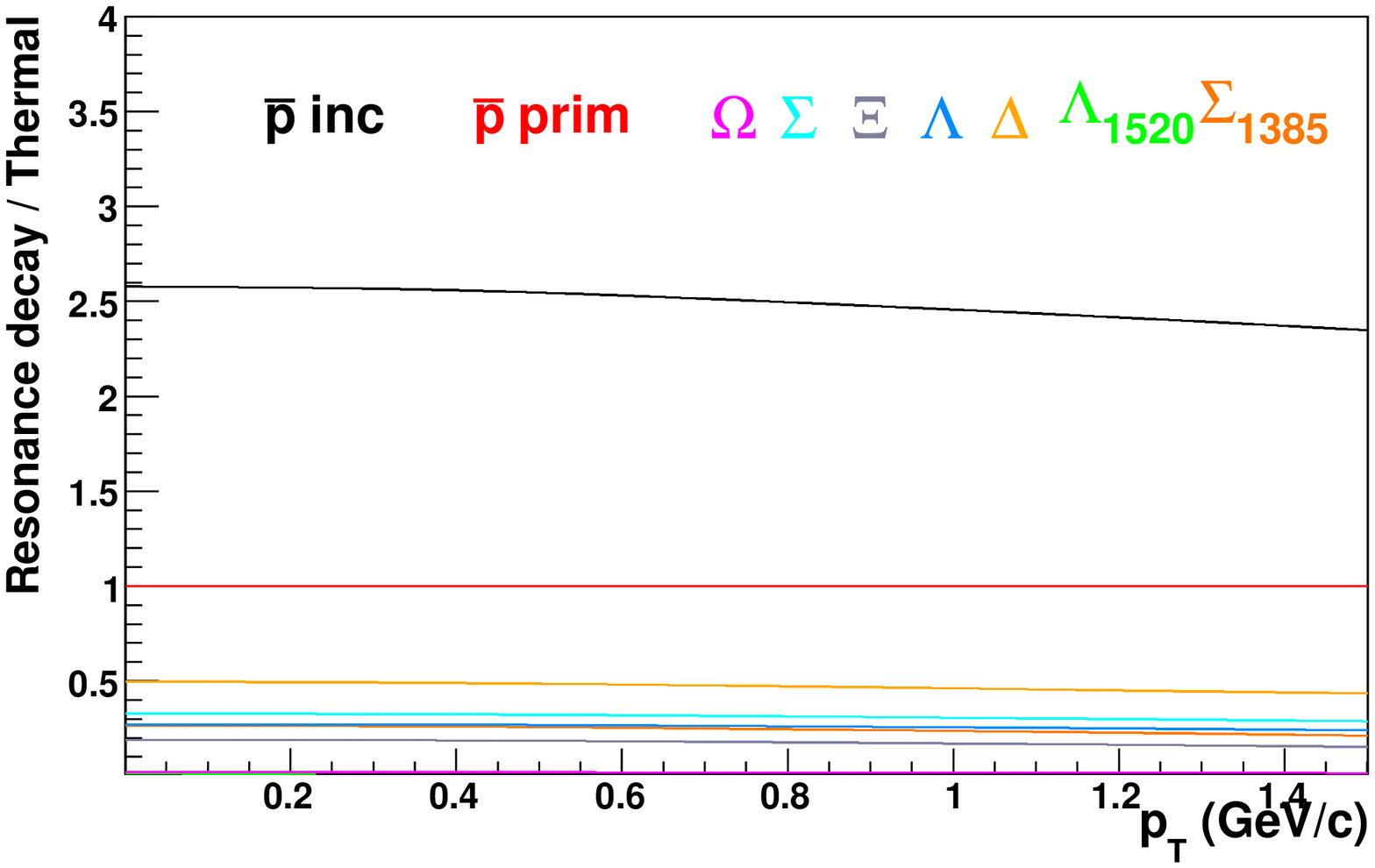} 

\end{center}
\vspace*{-0.1cm}
\caption{Calculated primordial and resonance decay daughter spectra and their ratio with respect to the thermal production in 0-5\% Au-Au collisions at 200 GeV. Spectra are calculated with the following parameters: $T_{kin}$ = 89 MeV, $\beta$ = 0.59, n = 0.82. }
\label{fig:calc_specAuAu}
\end{figure}
\begin{figure}[!h]
\begin{center}
\resizebox{.45\textwidth}{!}{\includegraphics{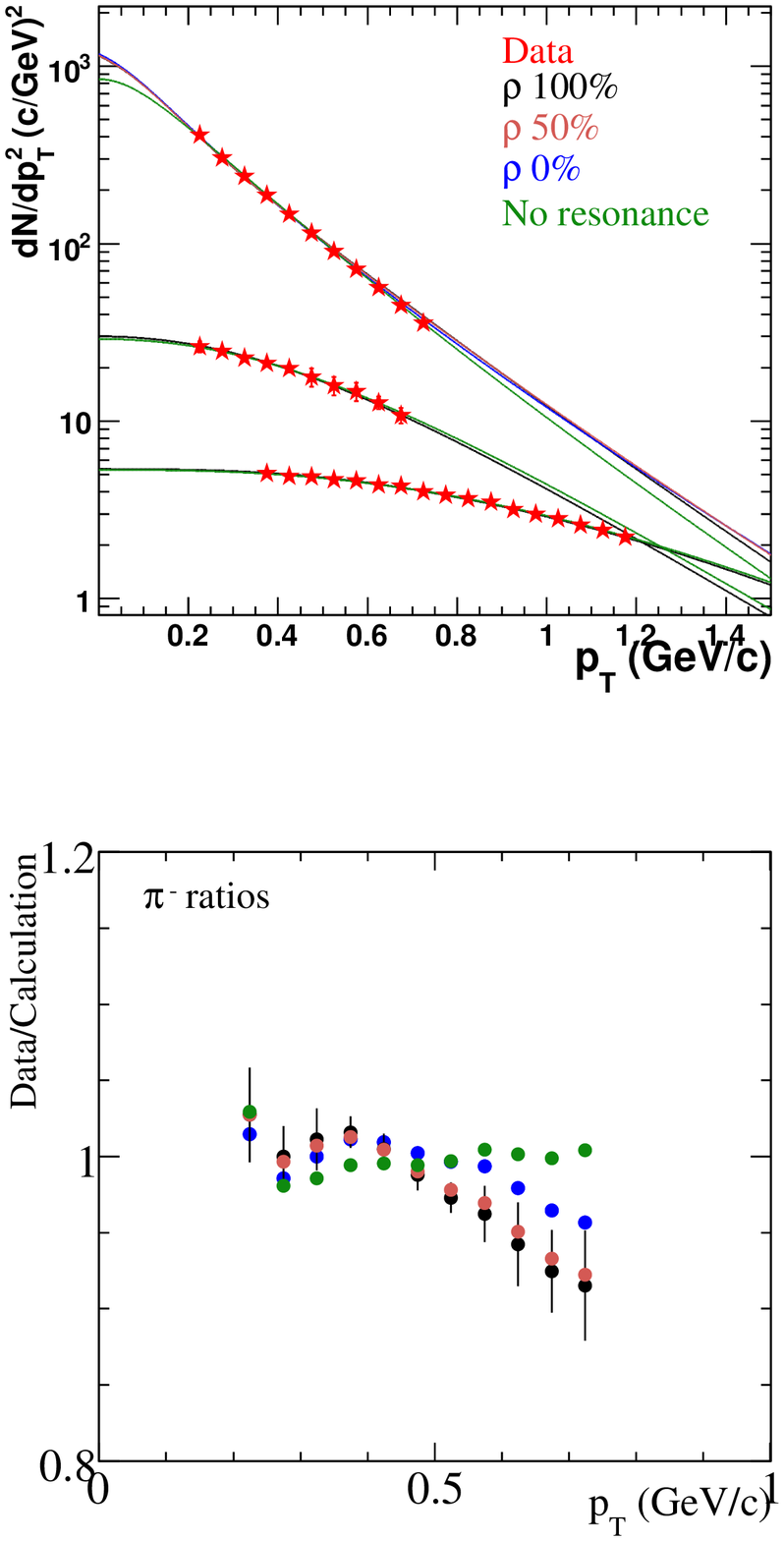}}
\resizebox{.45\textwidth}{!}{\includegraphics{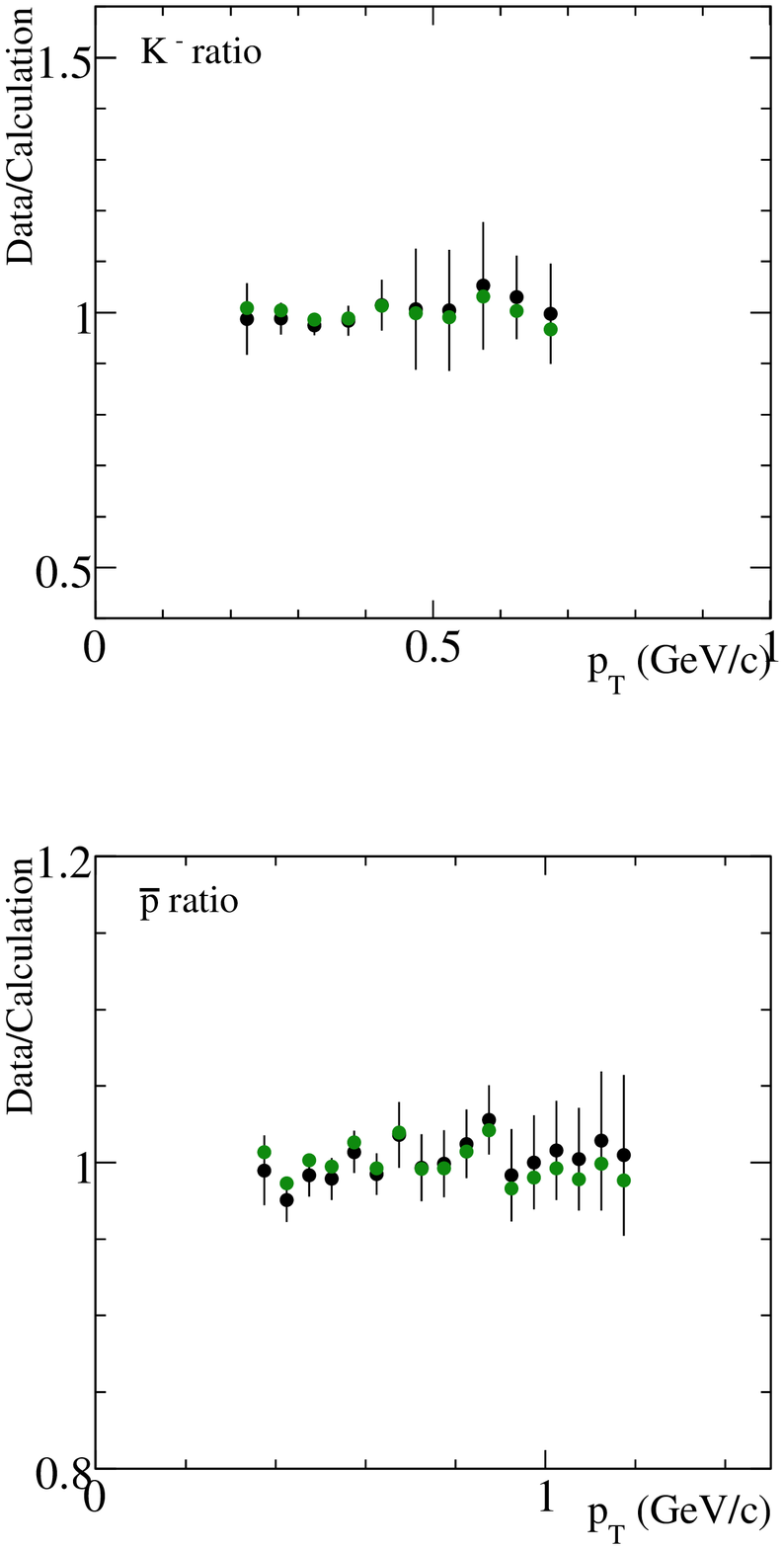}}

\end{center}
\vspace*{-0.1cm}
\caption{Top left panel: Fit of the calculated spectra to the measured ones in top 5\% central Au-Au collisions at 200 GeV ~\cite{Adams:2003xp}. Four calculated spectra are shown for $\pi^-$ (upper curves): including resonances with three different $\rho$ contributions and excluding resonances. Only two calculated curves are shown for $K^-$ (middle curves) and $\bar{p}$ (lower curves): including resonances with 100$\%$ $\rho$ and excluding resonances. Other panels: data / calculation ratios. Error bars are from statistical and point-to-point systematic errors on the data, and are shown for only one set of the data points.}
\label{fig:calc_spec_fitAuAu}
\end{figure}
The calculated particle spectra of $\pi^{-}$, $K^{-}$ and $\overline{p}$ are shown in Fig.~\ref{fig:calc_specAuAu} (left panels). Since the measured pion spectra are corrected for weak decays, the calculated inclusive pion spectra do not contain weak decay pions. Resonance contributions are labeled by the initial resonance particle, e.g. a $\pi$ emerging from the $\eta' \rightarrow \eta \rightarrow \pi$ decays is labeled as $\pi_{\eta'}$ and summed over each possible decay mode of $\eta'$. The calculated inclusive pion spectra include contributions from $\Lambda_{1520}$ which are not plotted.
The right panels of Fig.~\ref{fig:calc_specAuAu} show the resonance contributions to the inclusive spectra relative to the primordial one. 

The low $p_{T}$ pion enhancement is the counter play of $\rho$, $\omega$ and $\eta$; at higher $p_{T}$ the $\rho$ contribution dominates.
The inclusive kaon and antiproton spectra do not show significant changes in the spectral shapes compared to the primordial ones. The largest contributions to the inclusive kaon spectra are from $K^{*0}$ and $K^{*-}$ and the largest contributions to the inclusive $p$ and $\overline{p}$ spectra are from $\Lambda$, $\Delta$, and $\Sigma$'s. 

One should note here that different particles freeze-out at different temperatures, which would alter the final calculated inclusive spectrum. Within our model it is not possible to accommodate individual kinetic freeze-out temperatures, but would be an interesting task for model calculations such as RQMD or other parton cascade models.

\subsection{Short lived resonances and fit results}

It is an open question what flow velocity and temperature should be assigned to the short lived resonances, such as $\rho$ and $\Delta$. These resonance decays are expected to be constantly regenerated during the system evolution, since their life-times are shorter ($c\tau_{\rho} =$ 1.3 fm, $c\tau_{\Delta} =$ 1.6 fm) than the expected system evolution time (c$\tau \approx$ 10 fm).

Processes of $\rho \rightarrow \pi\pi$ and $\pi\pi \rightarrow \rho$, for example, constantly occur along the dynamical evolution of the system. Thus, it is reasonable to expect that the final $\rho$ decay pions carry the same flow information as the primordial pions do. In other words, the regenerated $\rho$ gain negligible flow velocity during its short life span except the inherited flow from the two resonant pions.

To gain better insights, three cases are considered for $\rho$, which gives the largest contribution to the measured $\pi$ spectra:
\begin{enumerate}
	\item{The $\rho$ decay pions have the same $p_{T}$ spectra shape as the primordial pions.}
	\item{The $\rho$ acquires flow as given by kinetic freeze-out temperature and transverse flow velocity, and the decay pions are calculated from decay kinematics.}
	\item{Half of the $\rho$ contribution is taken like in (1) and the other half as in (2). }
	\end{enumerate}
Case (2) has the largest flow for decay pions because the $\rho$, being heavy, acquires flow more efficiently than pions.

In the fit of the calculated spectra to the measured the free parameters are the kinetic freeze-out temperature ($T_{kin}$), the average transverse flow velocity ($\beta$) and the exponent flow profile ($n$). Two different fit methods are implemented based on the Minuit package of the ROOT~\cite{Brun:1997pa}. In the first fit, finite parameter values are used with fixed $n$ = 0.82; results are summarized in Table~\ref{tab:reso_fits1}. In the second method all three parameter values are set to free and to be able to vary with 6 decimal point precision; results are summarized in Table~\ref{tab:reso_fits2}. The first method allows a fast mapping and minimization to find the best fit for the $T_{kin}$, $\beta$. The second method provides a better estimate of the extracted parameters and the errors. In the minimization the point-to-point systematic errors on the measured spectra points are taken into account Due to the conservatively estimated point-to-point systematic errors, the extracted $\chi^{2}$/ndf values are below unity for the best fits.
\begin{table}[!h] 
\begin{center}
\caption{Extracted kinetic freeze-out parameters and fit $\chi^{2}$ in 0-5\% central Au-Au collisions at 200 GeV. The flow profile $n$ parameter is fixed to be 0.82.\label{tab:reso_fits1}}
\begin{tabular}{|c|c|c|c|}
\hline
Set & $T_{kin}$ $(MeV)$ & $\langle\beta\rangle$ & $\chi^{2}/ndf$\\ \hline
No resonances & $86.8^{ +0.7}_{ -0.6}$ & $0.595^{ +0.002}_{ -0.003}$  & 0.26\\
$\rho$ 0 $\%$ & $94.6^{ +0.9}_{ -1.0}$ & $0.603^{ +0.004}_{ -0.002}$  & 0.37\\
$\rho$ 50 $\%$& $87.4^{ +0.9}_{ -1.1}$ & $0.605^{ +0.002}_{ -0.002}$ &  0.45 \\
$\rho$ 100 $\%$& $77.2^{ +0.8}_{ -0.9}$ & $0.604^{ +0.004}_{ -0.003}$ &  0.60\\
\hline
\end{tabular}
\end{center}
\end{table}
\begin{table}[!h] 
\begin{center}
\caption{Extracted kinetic freeze-out parameters and fit $\chi^{2}$ in 0-5\% central Au-Au collisions at 200 GeV. All three parameters are free.}\label{tab:reso_fits2}
\begin{tabular}{|c|c|c|c|c|}
\hline
Set & $T_{kin}$ $(MeV)$ & $\langle\beta\rangle$ & $n$ &$\chi^{2}/ndf$\\ \hline
No resonances & 87.455 $\pm$ 2.545 & 0.604 $\pm$ 0.003 & 0.835 $\pm$ 0.020 & 0.695 \\ 
$\rho$ 0 $\%$ &  	91.350 $\pm$ 1.350 & 0.619 $\pm$ 0.001 & 0.820 $\pm$ 0.001 & 0.788 \\ 
$\rho$ 50 $\%$&  	87.555 $\pm$ 0.886 & 0.598 $\pm$ 0.003 & 0.820 $\pm$ 0.008 & 0.424 \\ 
$\rho$ 100 $\%$&  78.287 $\pm$ 0.966 & 0.599 $\pm$ 0.005 & 0.801 $\pm$ 0.019 & 0.558 \\
\hline
\end{tabular}
\end{center}
\end{table}
%
%
%
\begin{figure}[!h]
\begin{center}
\resizebox{.3\textwidth}{!}{\includegraphics{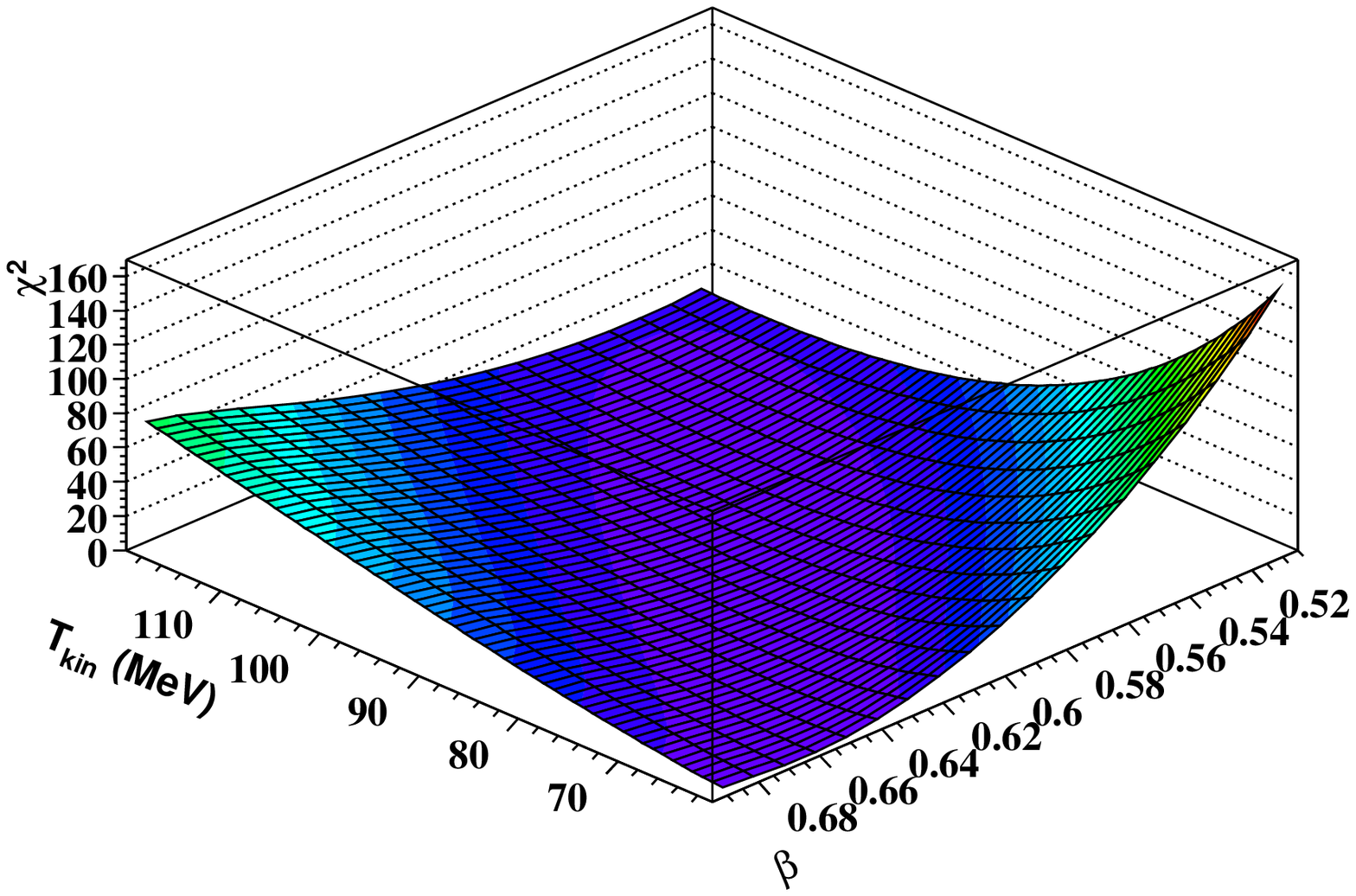}}
\resizebox{.3\textwidth}{!}{\includegraphics{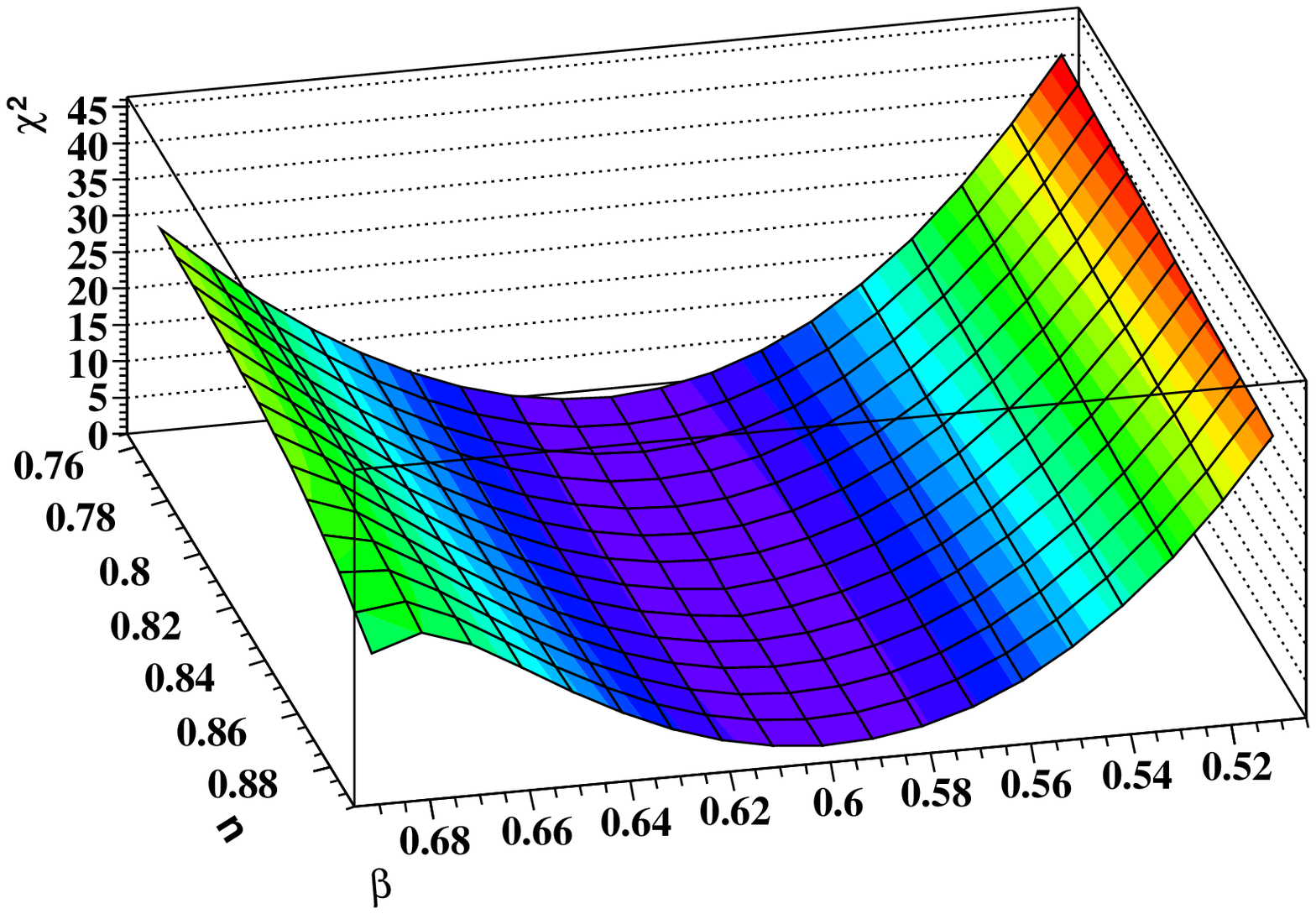}}
\resizebox{.3\textwidth}{!}{\includegraphics{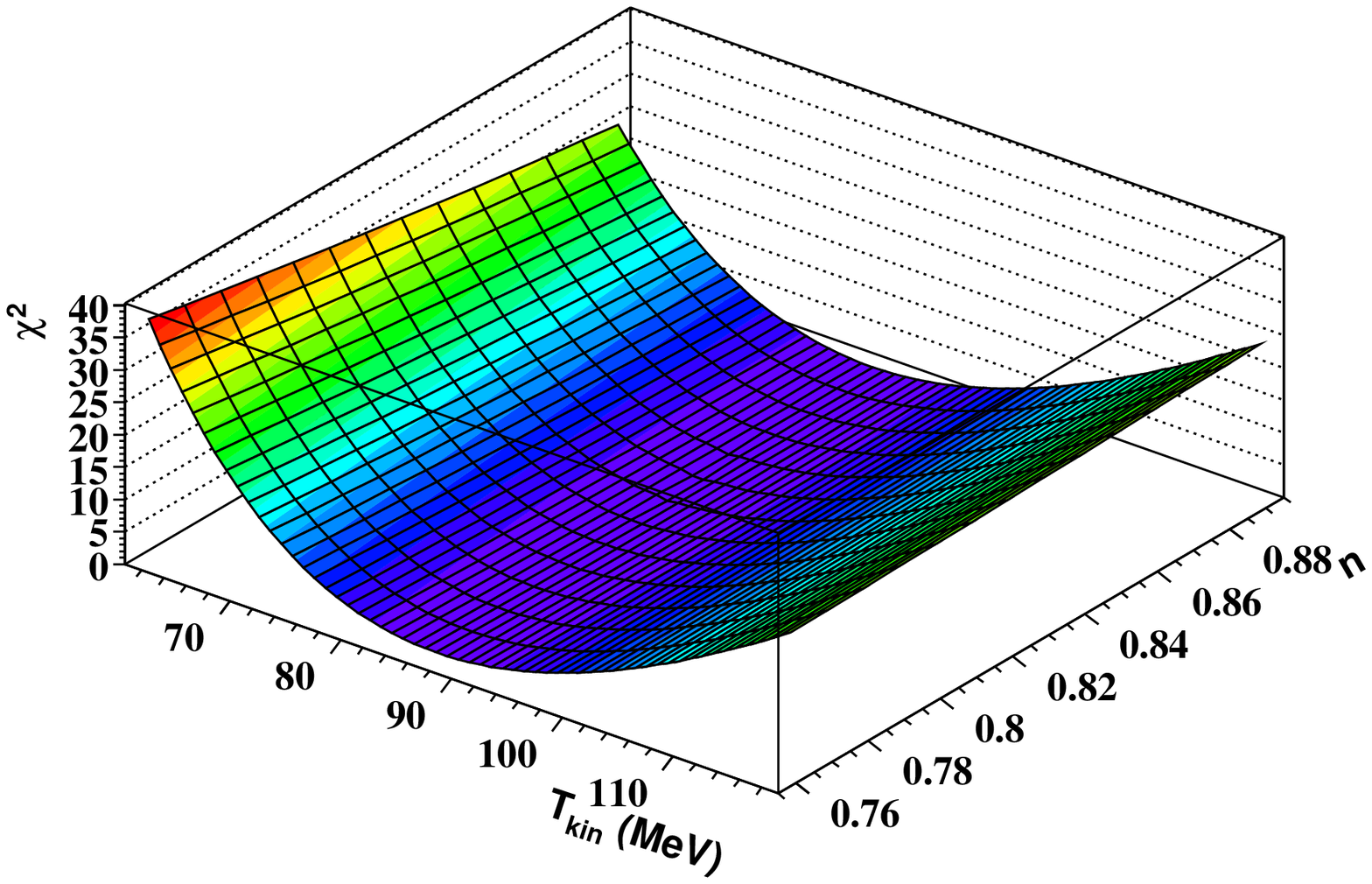}}
\end{center}
\vspace*{-0.1cm}
\caption{ Fine mapping of $\chi^{2}$ as a function of $T_{kin}$ and $\langle\beta\rangle$ and $n$ for the no resonances case. Number of degrees of freedom is $NDF=75$. The $\chi^{2}$ distributions are similar for the resonance cases.}
\label{fig:chi2s}
\end{figure}
Table~\ref{tab:reso_fits1} and Table~\ref{tab:reso_fits2} show the fit results for the three short lived resonance cases and for the two fit methods. Also listed are the fit results without including resonances. The spectral shapes are found to be less sensitive to the kinetic freeze-out temperature than the flow velocity. Fig.~\ref{fig:chi2s} shows, as an example, the fitted $\chi^2$ versus fit parameters $T_{kin}$, $\langle\beta\rangle$ and $n$. It can be seen from the figure that $\langle\beta\rangle$ is better constrained than $T_{kin}$. 

Fig.~\ref{fig:calc_spec_fitAuAu} shows the fits of the calculated inclusive spectra to the measured ones. Fits are performed to the six measured spectra simultaneously, but only negatively charged particles are shown. For $K^-$ and $\overline{p}$, results from the 100 \% $\rho$ case fit and the fit excluding resonances are plotted, while all fits are plotted for pions.
In the case of 100$\%$ $\rho$, the calculated spectrum starts to deviate from the data above $p_{T}$ $\sim$ 400 MeV. In the case of 0$\%$ $\rho$, below $p_{T}$ $\sim$ 400 MeV  the calculated inclusive spectrum is enhanced by $\omega$ and $\eta$, which become more important without the $\rho$.
The model, with all the three cases of $\rho$ contributions, seems to describe the data well. Nonetheless, the fitted $T_{kin}$ values with all three cases of $\rho$ contributions seem to agree with that obtained without including resonances within the systematic error of $\pm$ 10 MeV ~\cite{Adams:2003xp}. In other words, resonance decays appear to have no significant effect on the extracted kinetic freeze-out parameters as shown in Table~\ref{tab:reso_fits1} and Table~\ref{tab:reso_fits2}. This is primarily due to the limited $p_{T}$ ranges of our data where resonance decay products have more or less similar spectral shapes as the primordial particles do.
\begin{figure}[!h]
\begin{center}
\resizebox{.45\textwidth}{!}{\includegraphics{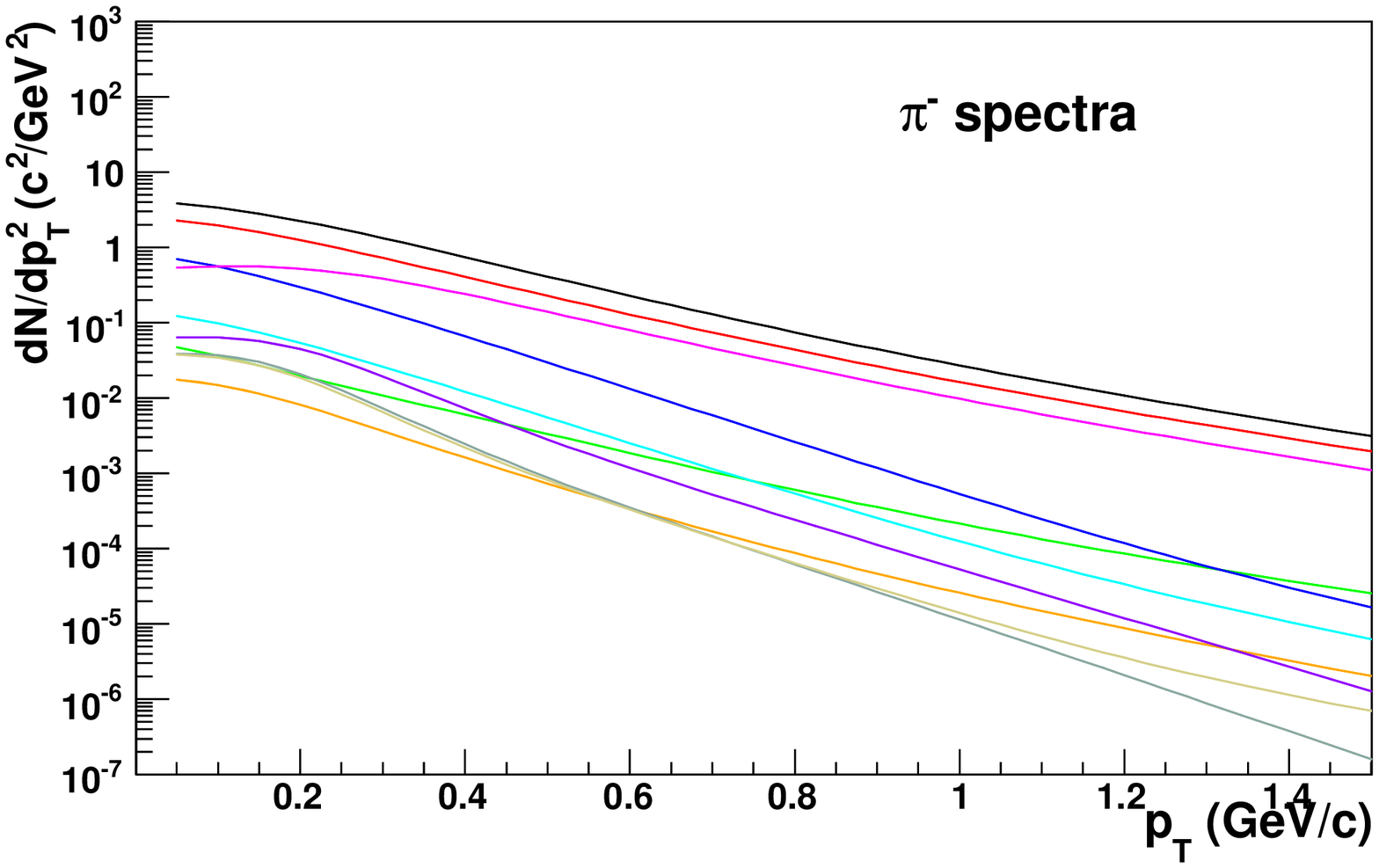}}
\resizebox{.45\textwidth}{!}{\includegraphics{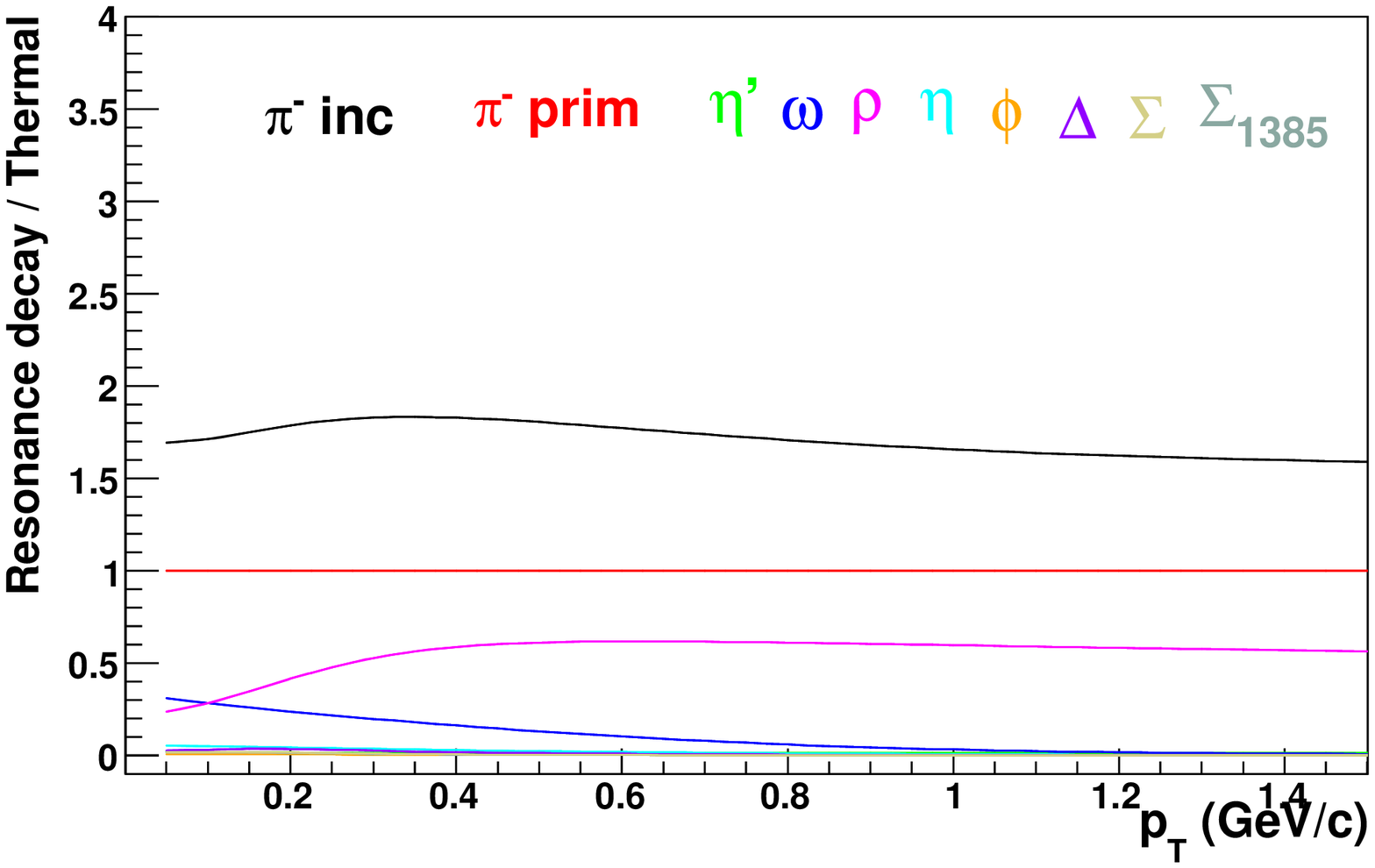}}
\resizebox{.45\textwidth}{!}{\includegraphics{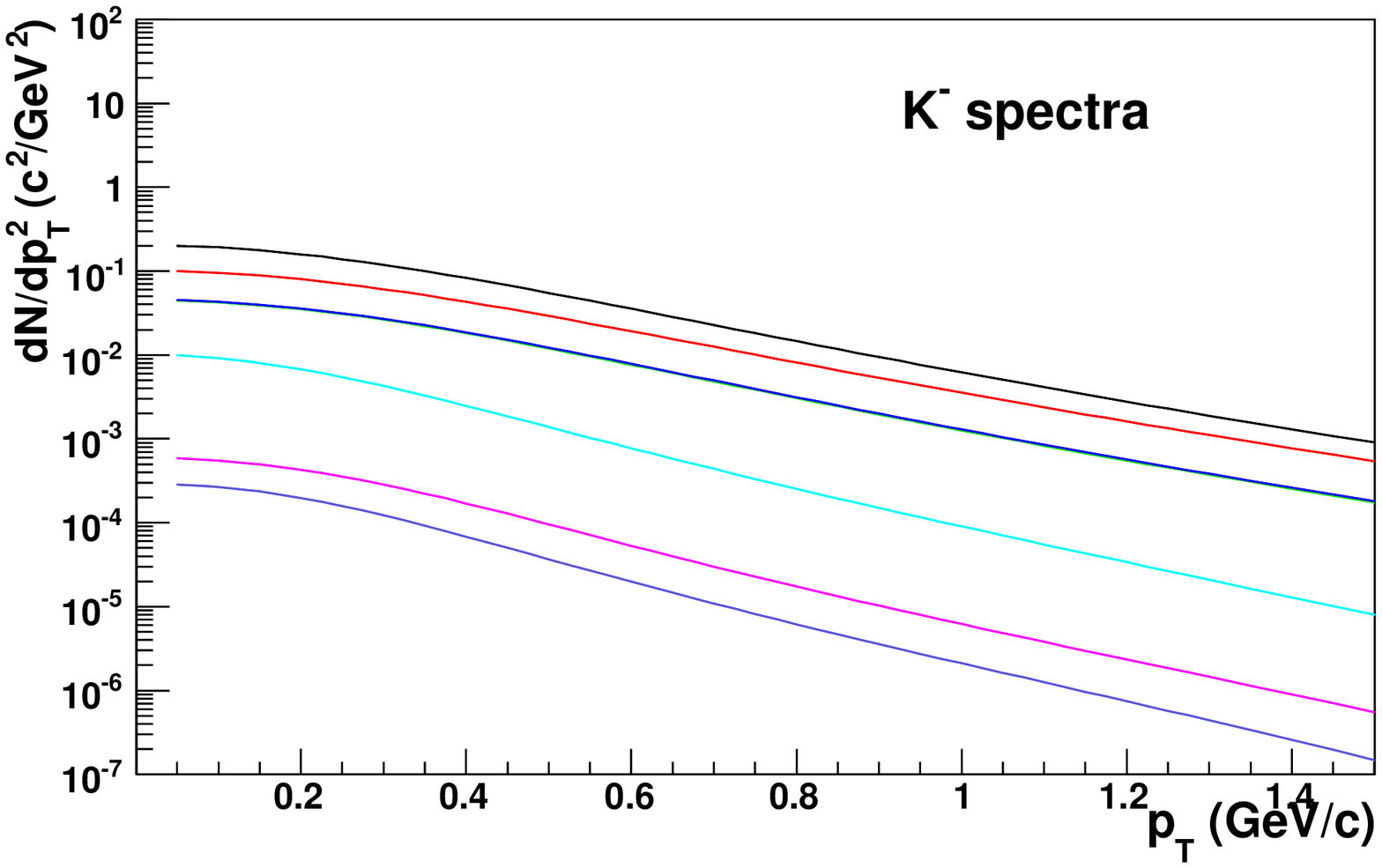}}
\resizebox{.45\textwidth}{!}{\includegraphics{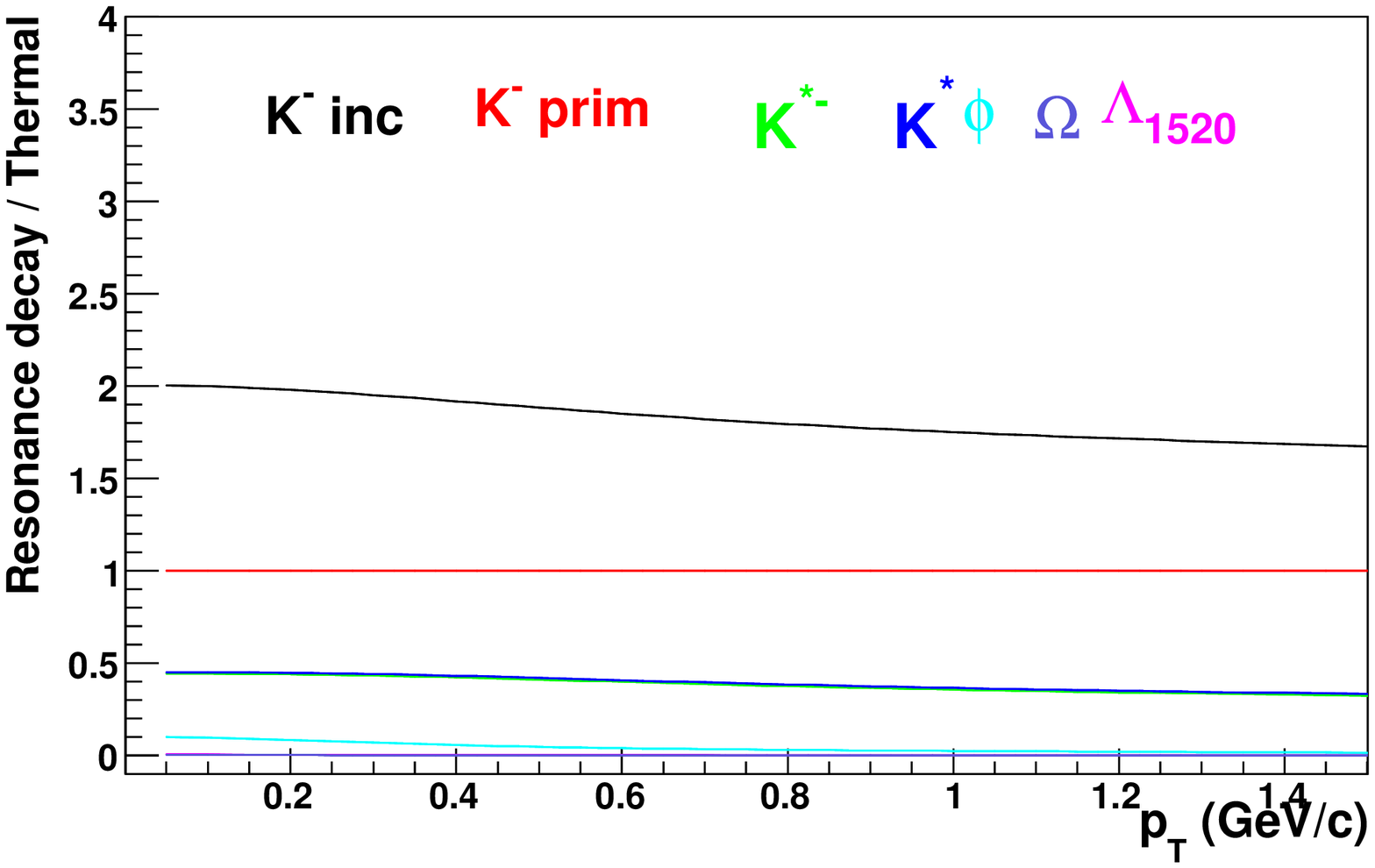}}
\resizebox{.45\textwidth}{!}{\includegraphics{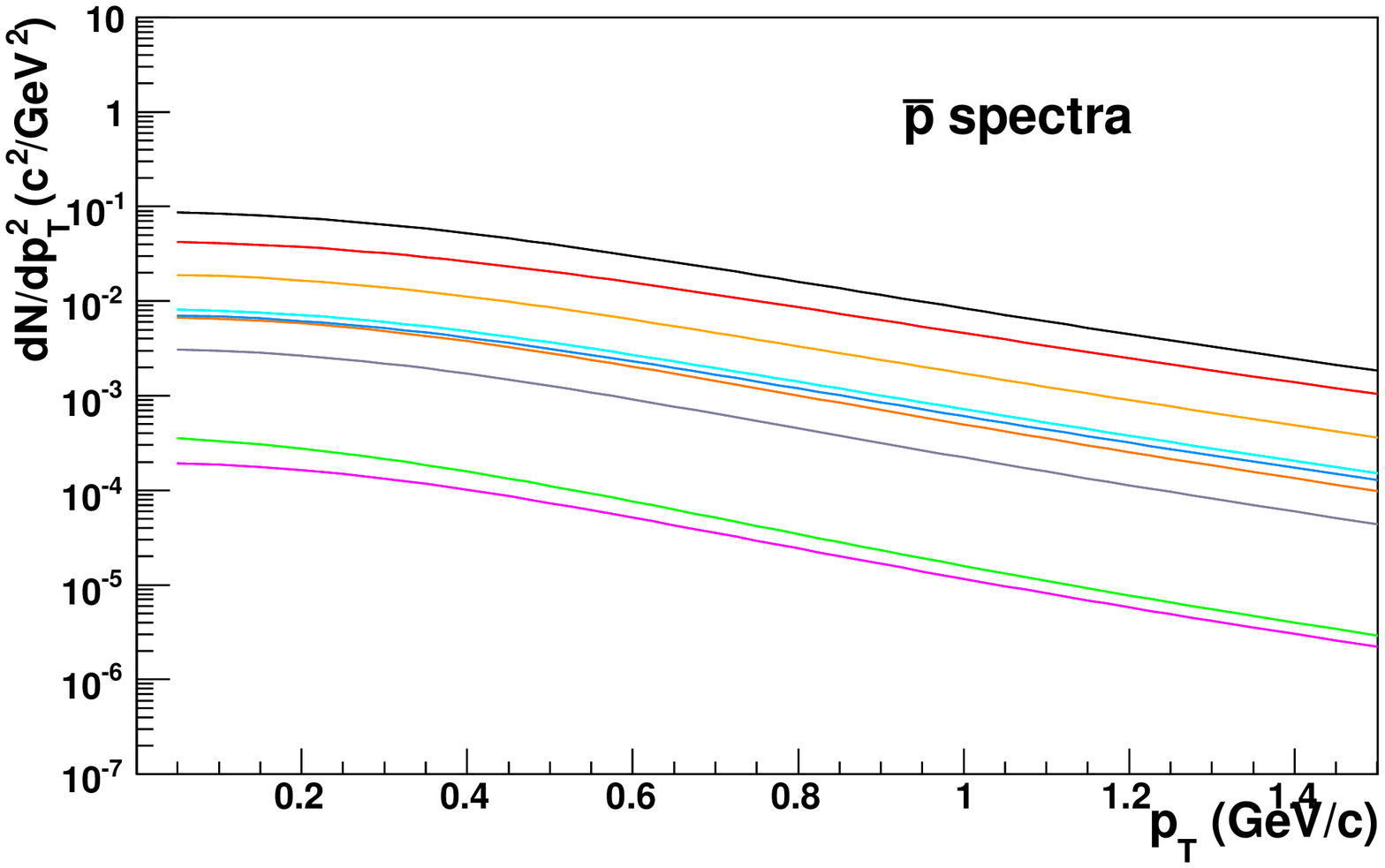}}
\resizebox{.45\textwidth}{!}{\includegraphics{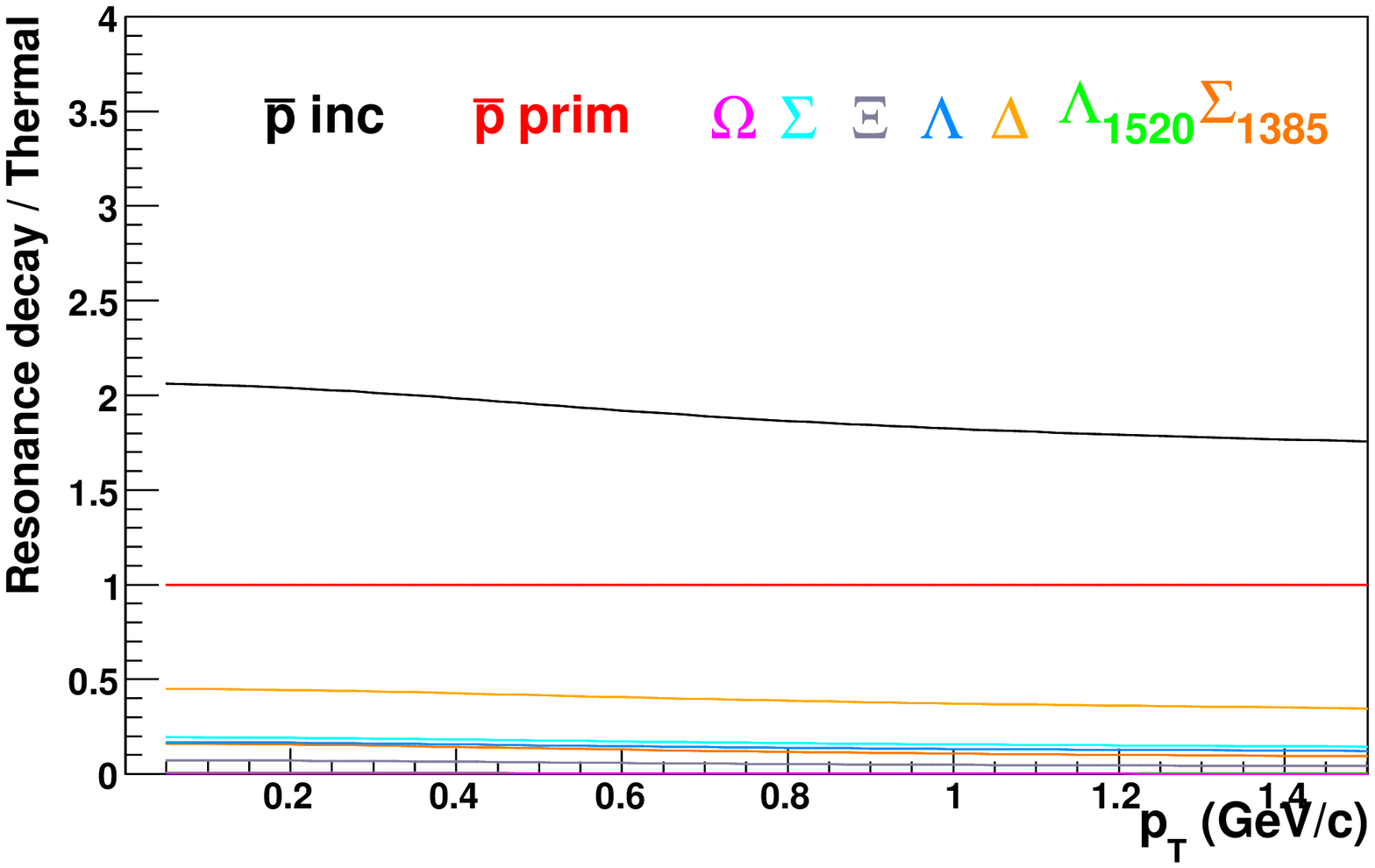}}

\end{center}
\vspace*{-0.1cm}
\caption{Calculated primordial and resonance decay daughter spectra and their ratio with respect to the thermal production in 200 GeV pp collisions. Spectra are calculated with the following parameters: $T_{kin}$ = 118 MeV, $\beta$ = 0.29 and n = 3.1.}
\label{fig:calc_specpp}
\end{figure}
\begin{figure}[!h]
\begin{center}
\resizebox{.45\textwidth}{!}{\includegraphics{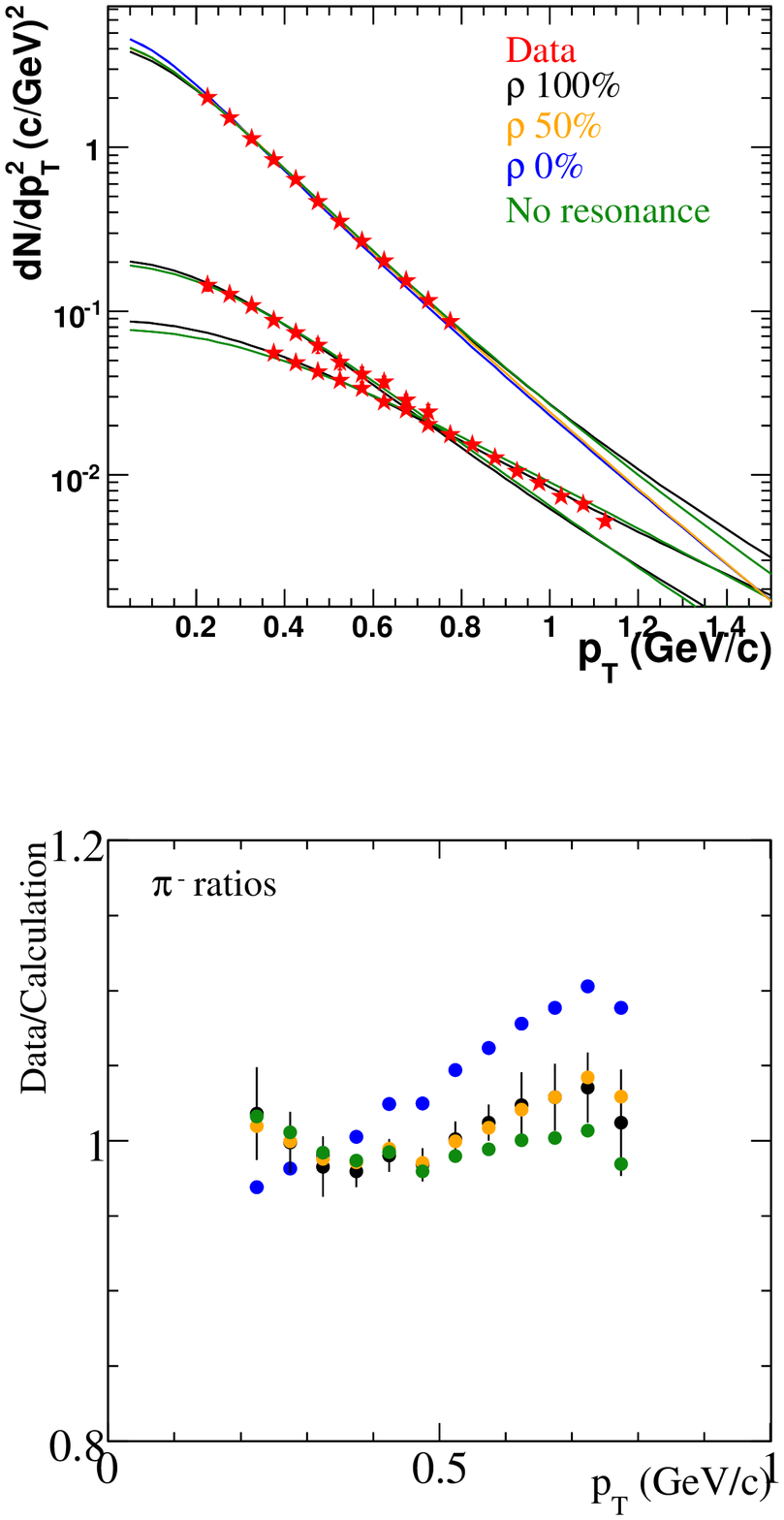}}
\resizebox{.45\textwidth}{!}{\includegraphics{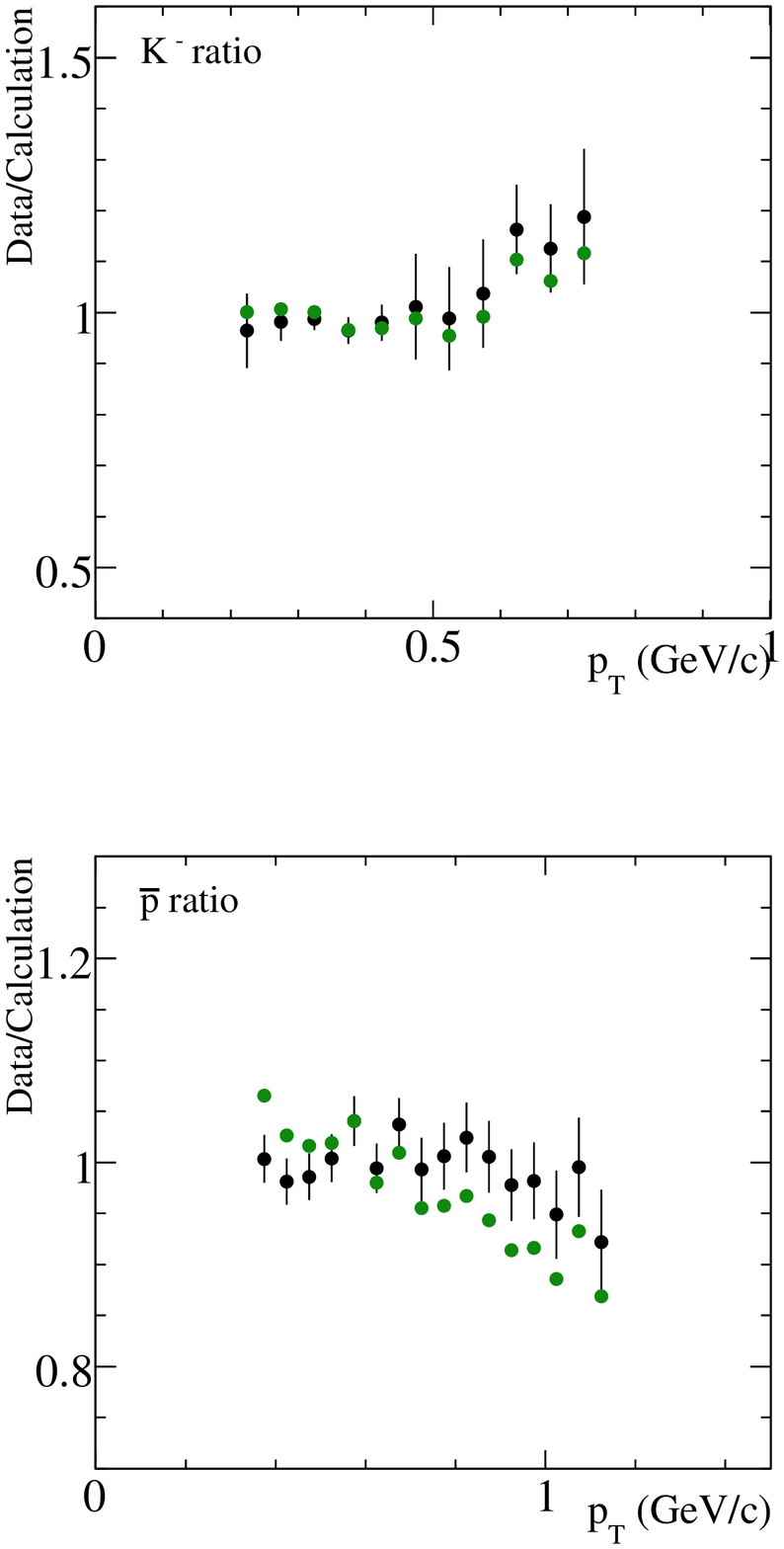}}
\end{center}
\vspace*{-0.1cm}
\caption{Top left panel: Fit of the calculated spectra to the measured ones in minimum bias pp collisions at 200 GeV~\cite{Adams:2003xp}. Four calculated spectra are shown for $\pi^-$ (upper curves): including resonances with three different $\rho$ contributions and excluding resonances. Only two calculated curves are shown for $K^-$ (middle curves) and $\bar{p}$ (lower curves): including resonances with 100$\%$ $\rho$ and excluding resonances. Other panels: data / calculation ratios. Error bars are from statistical and point-to-point systematic errors on the data, and are shown for only one set of the data points.}
\label{fig:calc_specpp_fit}
\end{figure}
\begin{table}[!h] 
\begin{center}
\caption{Extracted kinetic freeze-out parameters and fit $\chi^{2}$ in minimum bias pp collisions at 200 GeV. All three parameters are free.\label{Tab:reso_fitspp}}
\begin{tabular}{|c|c|c|c|c|}
\hline
Set & $T_{kin}$ $(MeV)$ & $\langle\beta\rangle$ & $n$ &$\chi^{2}/ndf$\\ \hline
No resonances & 119.497 $\pm$ 1.503  & 0.298 $\pm$ 0.027 & 2.475 $\pm$ 1.163 & 1.395 \\ 
 $\rho$ 0 $\%$ & 121.858 $\pm$ 0.858  & 0.345 $\pm$ 0.002 & 1.190 $\pm$ 0.034 & 4.400 \\ 
 $\rho$ 50 $\%$ & 122.154 $\pm$ 1.154  & 0.350 $\pm$ 0.010 & 0.996 $\pm$ 0.312 & 2.464 \\ 
 $\rho$ 100 $\%$ & 117.813 $\pm$ 3.187  & 0.293 $\pm$ 0.012 & 3.124 $\pm$ 0.579 & 1.061 \\ 
\hline 
\end{tabular}
\end{center}
\end{table}
\begin{table}[!h] 
\begin{center}
\caption{Extracted kinetic freeze-out parameters and fit $\chi^{2}$ in minimum bias pp collisions at 200 GeV. The $n$ parameter is fixed to be 2.0.\label{Tab:reso_fitsppfx}}
\begin{tabular}{|c|c|c|c|c|}
\hline
Set & $T_{kin}$ $(MeV)$ & $\langle\beta\rangle$ & $n$ &$\chi^{2}/ndf$\\ \hline
No resonances & 122.237 $\pm$ 1.237  & 0.292 $\pm$ 0.003 & 2.000 $\pm$ 0.000 & 1.306 \\ 
$\rho$ 0 $\%$ & 121.560 $\pm$ 0.560  & 0.321 $\pm$ 0.002 & 2.000 $\pm$ 0.000 & 3.134 \\  
$\rho$ 50 $\%$ & 121.974 $\pm$ 0.974  & 0.320 $\pm$ 0.003 & 2.000 $\pm$ 0.000 & 1.549 \\
$\rho$ 100 $\%$ & 119.853 $\pm$ 1.147  & 0.302 $\pm$ 0.003 & 2.000 $\pm$ 0.000 & 1.826 \\ 
\hline 
\end{tabular}

\end{center}
\end{table}
We also fitted the spectra data with a single, fixed kinetic freeze-out temperature $T_{kin}=T_{ch}=160$ MeV including resonances and with 100 $\%$ $\rho$ contribution. The fitted $\langle\beta\rangle$ is $0.520^{ +0.001}_{ -0.002}$ with $\chi^{2}/NDF$=19.56. A single temperature scenario is therefore ruled out by the data.

The $\chi^{2}/NDF$ is smaller than unity because we included in the fit the point-to-point systematic errors (dominate over statistical ones), which were estimated on the conservative side and might not be completely random. If we scale the $\chi^{2}/NDF$ such that the minimum is unity, then we get somewhat smaller statistical errors on the fit parameters.

\section{Model fit to 200 GeV pp collisions}

It was suggested that resonance contribution to the low momentum bulk particle spectra can be more significant in pp collision than in central Au-Au collisions. The same method we presented earlier can be repeated for minimum bias pp collisions as well.

Fig.~\ref{fig:calc_specpp} shows the calculated spectra for negative particles in minimum bias pp collisons at 200 GeV. The inclusive calculated kaon and antiproton spectra are slightly modified compared to the primordial ones. The inclusive calculated pion spectrum is less modified compared to the central Au-Au case. The largest contribution can be attributed to the $\rho$ meson. The $\eta$ and $\omega$ mesons are less significant than in central Au-Au collisions at 200 GeV. 

Fig.~\ref{fig:calc_specpp_fit} shows the fit to the measured particle spectra in 200 GeV pp collisions. Fits are performed simultaneously to the six particle spectra but only the negatives are shown. The calculated spectra can describe the measured particle spectra well in the measured transverse momentum region. The numerical results are listed in Tab.~\ref{Tab:reso_fitspp} and in Tab.~\ref{Tab:reso_fitsppfx}.

In summary, we do not observe significant modification in the extracted kinetic freeze-out parameters including resonances in the measured transverse momentum range for both minimum bias pp and central (0 - 5\%) Au-Au collisions.

\section{Conclusions and Summary}

In this thesis multiplicity/centrality dependent identified particle spectra of $\pi^{\pm}$, $K^{\pm}$, $p$ and $\overline{p}$ are presented from $\sqrt{S_{NN}} =$ 200 GeV pp, dAu and $\sqrt{S_{NN}} =$ 62.4 GeV Au-Au collisions. Measurements were carried out by the STAR experiment at RHIC. The main detector used in the measurements is the TPC. Charged particles are identified by the specific energy loss in the TPC gas with the dE/dx technique. 

Transverse momentum spectra of kaons and protons/antiprotons show hardening with increasing multiplicity/centrality.
The average transverse momenta in 62.4 and 200 GeV Au-Au collisions follow the same trend and seem to scale with multiplicity. The average transverse momenta of pions are flat over wide range of multiplicity, and show monotonic increase those of kaons, protons/antiprotons. The average transverse momenta as a function of the multiplicity density per transverse area ($\sqrt{dN/dy/S}$) are the same within errors for 62.4 and 200 GeV Au-Au collisions.

The average transverse momenta of kaons and protons/antiprotons in 200 GeV pp and dAu collision show departing trend from that of 62.4 and 200 GeV Au-Au. The average transverse momenta of kaons and protons/antiprotons in 200 GeV pp and dAu collision are larger than in 200 GeV Au-Au collisions at similar multiplicities. The average transverse momenta of pions show almost no dependence.

Particle/antiparticle ratios are independent of multiplicity/centrality in 200 GeV collisions. The $\overline{p}/p$ ratio shows significant decrease from peripheral to central 62.4 GeV Au-Au collisions. In 200 GeV this drop is not well pronounced. In 62.4 GeV Au-Au collisions nuclear stopping increases compared to the almost transparent 200 GeV Au-Au collisions and the amount of net baryons present in the collisions zone is significant.

Increase in the nuclear stopping is also reflected in the decrease of the $K^{-}/K^{+}$ ratio at 62.4 GeV in contrast to 200 GeV Au-Au collisions. The connection of these particle ratios ($K^{-}/K^{+}$ versus $\overline{p}/p$ and others) is well reproduced by chemical model calculations over wide ranges of collision energy, centrality, rapidity. This might indicate (local) equilibration of the system. 

The $K^{-}/\pi^{-}$ ratios at RHIC energies (62.4 GeV, 130 GeV, 200 GeV) from mid-peripheral to central Au-Au collisions are independent of centrality, while in lower energies they show a steep rise. This implies that strangeness production is the same at these RHIC energies.

Chemical freeze-out properties are investigated from the measured particle ratios. The chemical freeze-out temperature is $\sim$ 150 - 156 MeV (which is close to the lattice QCD calculation with three flavors: 154 $\pm$ 8 MeV) and is independent of multiplicity / centrality. The baryon chemical potential is $\sim$ 7 - 18 MeV at 200 GeV and $\sim$ 40 - 75 MeV at 62.4 GeV. The ad-hoc strangeness suppression factor increases from 0.5 at pp, dAu and saturates at mid-central, central Au-Au collisions at 0.85. The strangeness suppression factor shows the same trend at 62.4 and at 200 GeV. 

The kinetic freeze-out properties are extracted from simultaneous blast-wave model fits to the $\pi^{\pm}$, $K^{\pm}$, $p$ and $\overline{p}$ spectra. The kinetic freeze-out temperature decreases with increasing multiplicity/centrality while the average transverse flow velocity increases.  

Investigation of the excitation functions of the freeze-out parameters indicate that above $\sqrt{S_{NN}} \sim$10 GeV further increase in the collision energy only creates a larger flow, but the chemical freeze-out properties are the same. Particle production is governed by the baryon chemical potential as shown from the $K^{-}/K^{+}$ vs. $\overline{p}/p$ correlation. The kinetic freeze-out properties of the system are determined by the collision energy and centrality. 

Since the bulk $\pi^{\pm}$, $K^{\pm}$, $p$ and $\overline{p}$ spectra carry contribution from resonance decays, the kinetic freeze-out properties are investigated including resonance particles in the blast-wave model calculation. The inclusive $\pi^{\pm}$, $K^{\pm}$, $p$ and $\overline{p}$ spectra are calculated from thermal and blast-wave models with resonances for central (0-5\%) Au-Au collisions and for Minimum Bias pp collisions at 200 GeV. The extracted kinetic freeze-out temperature and the average transverse flow velocity do not show significant effect from resonances. This is because in the measured STAR transverse momentum range resonance decays do not significantly alter the shapes of the final inclusive spectra.

The main contributions of this thesis to the search of the Quark Gluon Plasma are the following. From the measured particle ratios, there is a strong indication that the chemical freeze-out happens close to the universal hadronization or phase transition. The chemical freeze-out temperature is unique and independent of the collision energy from the top SPS to RHIC energies. Kinetic freeze-out happens later; the kinetic freeze-out temperature and the average flow velocity extracted from the identified particle spectral shapes suggest significant cooling and expansion from chemical to kinetic freeze-out. Rigorous investigation of the effect of resonances on the kinetic freeze-out parameters shows that, given the STAR measured transverse momentum range, resonance decays have no significant effect on the extracted kinetic freeze-out parameters. This measurement also rules out the single freeze-out scenario.




%
%
%


%
%
%

\appendix

\chapter{Kinematic variables}

We introduce the basic kinetic variables used in high energy heavy-ion collisions.
\\
First, we use a Cartesian coordinate system; the $z$ direction of the coordinate system in the laboratory frame is set to be parallel to the beam direction, and the $x$ and $y$ components are perpendicular to the beam direction (they span the $transverse\ plane$). The azimuthal angle $\phi$ is measured in the transverse plane with respect to the $x$ axis. 
\\
The three particle momentum can be decomposed to longitudinal ($p_z$) component and transverse component:
\begin{equation}
p_T = \sqrt{p_{x}^{2}+p_{y}^{2}}.
\end{equation}
The transverse mass is defined as:
\begin{equation}
m_T = \sqrt{m_{0}^{2}+p_{T}^{2}}
\end{equation}
where $m_{0}$ is the rest mass of the particle.
Energy of a given particle is defined through the relativistic formula:
\begin{equation}
E\ =\ \sqrt{p_{\mu}p^{\mu}+m^{2}}\ =\ \sqrt{m_{T}^{2}+p_{z}^{2}}.
\end{equation}
From the measured energy of the particle and the measured momentum we can derive two useful quantities: the rapidity and the pseudo-rapidity variables. Rapidity is defined as: 
\begin{equation}
y\ =\ \frac{1}{2}\frac{E+p_{z}}{E-p_{z}},
\end{equation}
and the pseudorapidity is defined as:
\begin{equation}
\eta= -\ \normalfont{ln(tan}(\theta/2)).
\end{equation}

The differential invariant yield section is defined as the number of particles in a phase space segment, which is commonly described in cylindrical coordinates.

\begin{equation}
E\frac{d^{3}N}{dp^{3}}\ =\ \frac{d^{3}N}{p_{T}dp_{T}d\phi dy}\ =\ \frac{d^{3}N}{m_{T}dm_{T}d\phi dy}.
\end{equation}
In our studies we investigate the azimuthally averaged particle spectra, hence the invariant cross section can be written as:
\begin{equation}
E\frac{d^{3}N}{dp^{3}}\ =\ \frac{d^{3}N}{2\pi p_{T}dp_{T}dy}\ =\ \frac{d^{3}N}{2\pi m_{T}dm_{T}dy}.
\end{equation}

Various fitting functions are used to fit the invariant cross section and extract particle yields and spectra properties.
Properties of pion spectra are extracted with the Bose - Einstein function:
\begin{equation} 
E\frac{d^{3}N}{dp^{3}}\ \propto \frac{1}{e^{m_T/T}-1}.
\end{equation}
To estimate the systematic uncertainties two other function forms are used to fit the particle spectra, the $m_{T}$ exponential function:
\begin{equation} 
E\frac{d^{3}N}{dp^{3}}\ \propto e^{-m_T/T},
\end{equation}
and the Boltzmann function:
\begin{equation} 
E\frac{d^{3}N}{dp^{3}}\ \propto m_{T} e^{-m_T/T}.
\end{equation}
\clearpage
\chapter{Extraction of yield}

\begin{sidewaysfigure}[!t]
\begin{center}
\resizebox{.225\textwidth}{!}{\includegraphics{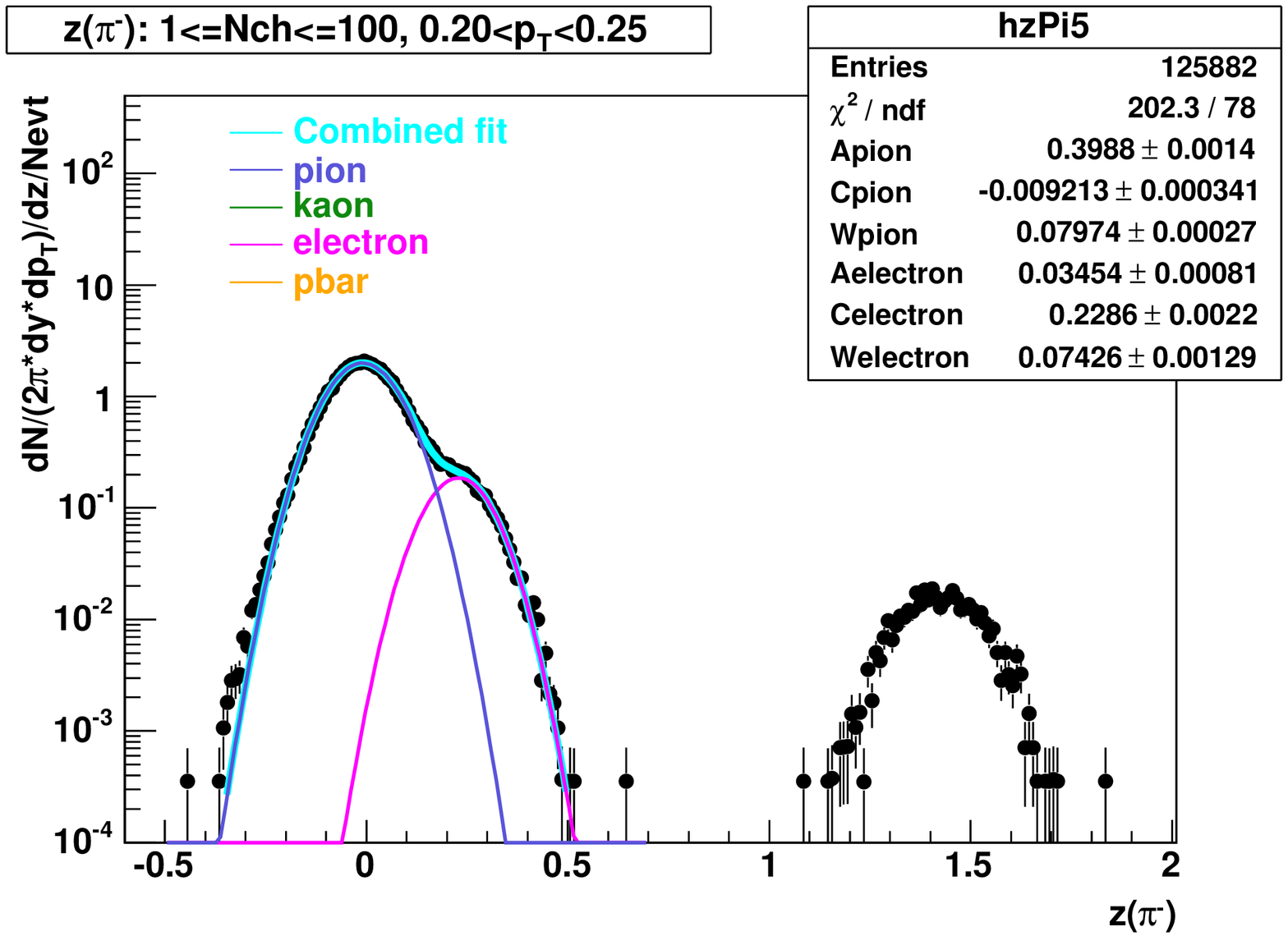}}
\resizebox{.225\textwidth}{!}{\includegraphics{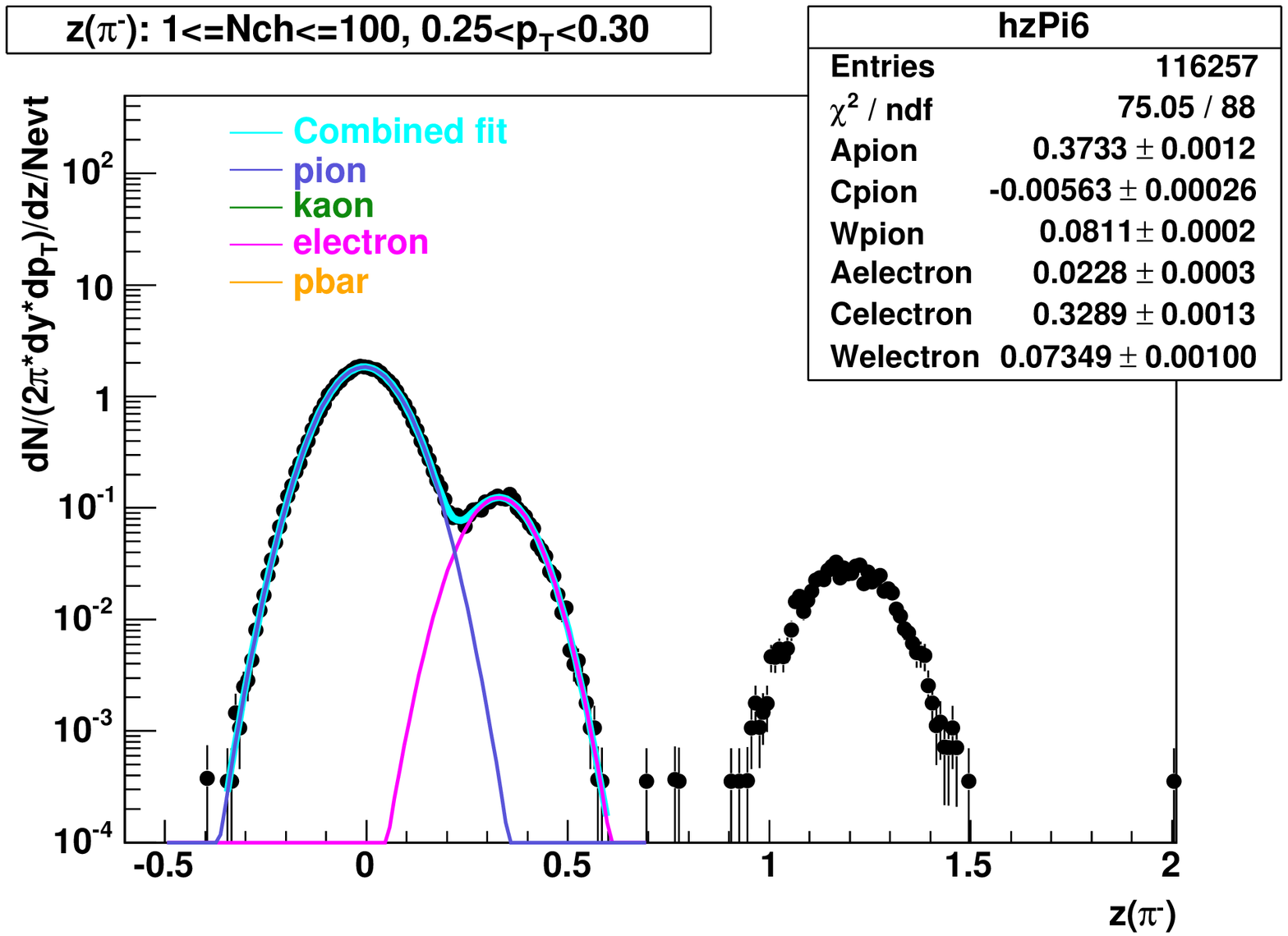}}
\resizebox{.225\textwidth}{!}{\includegraphics{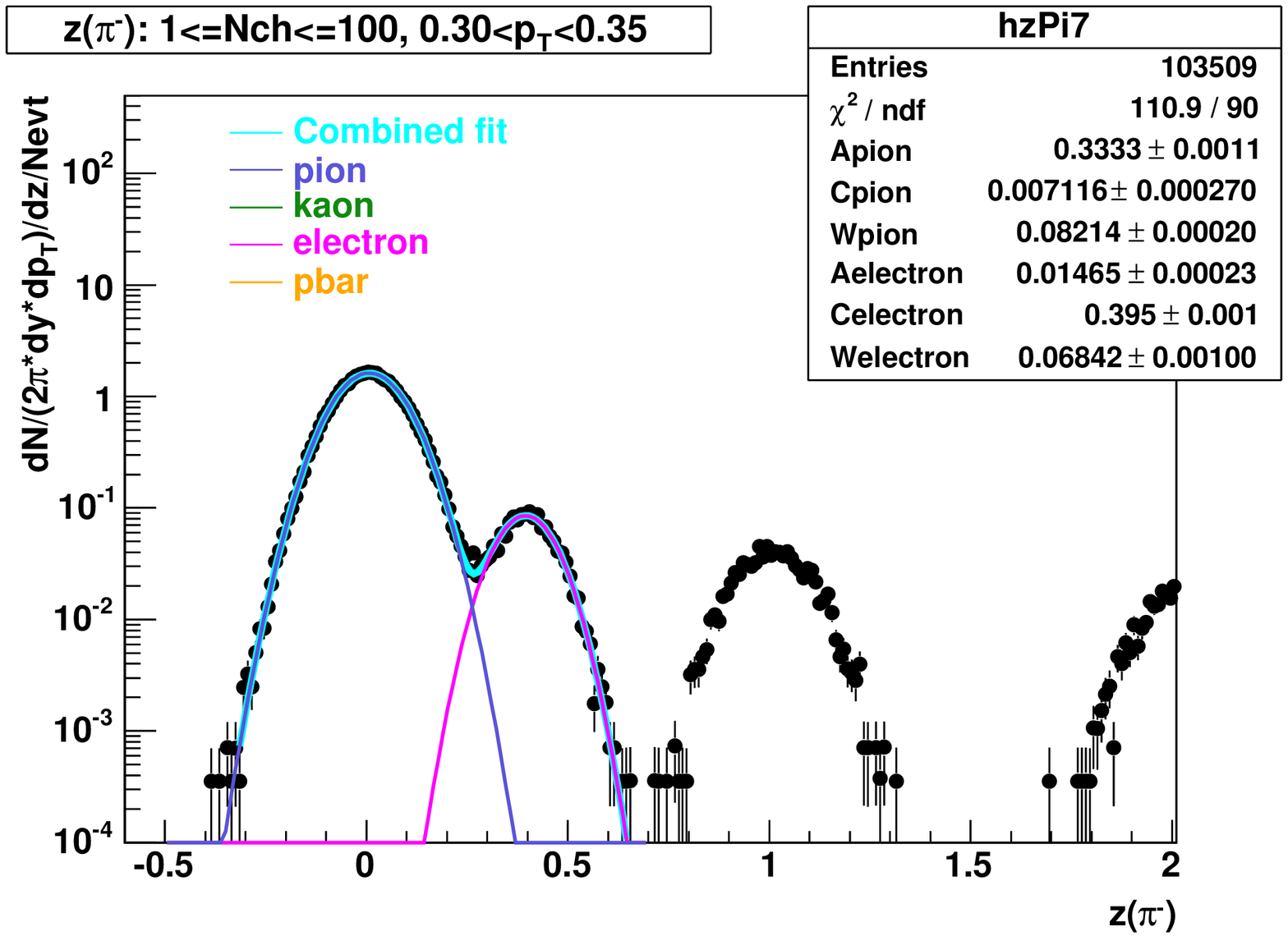}}
\resizebox{.225\textwidth}{!}{\includegraphics{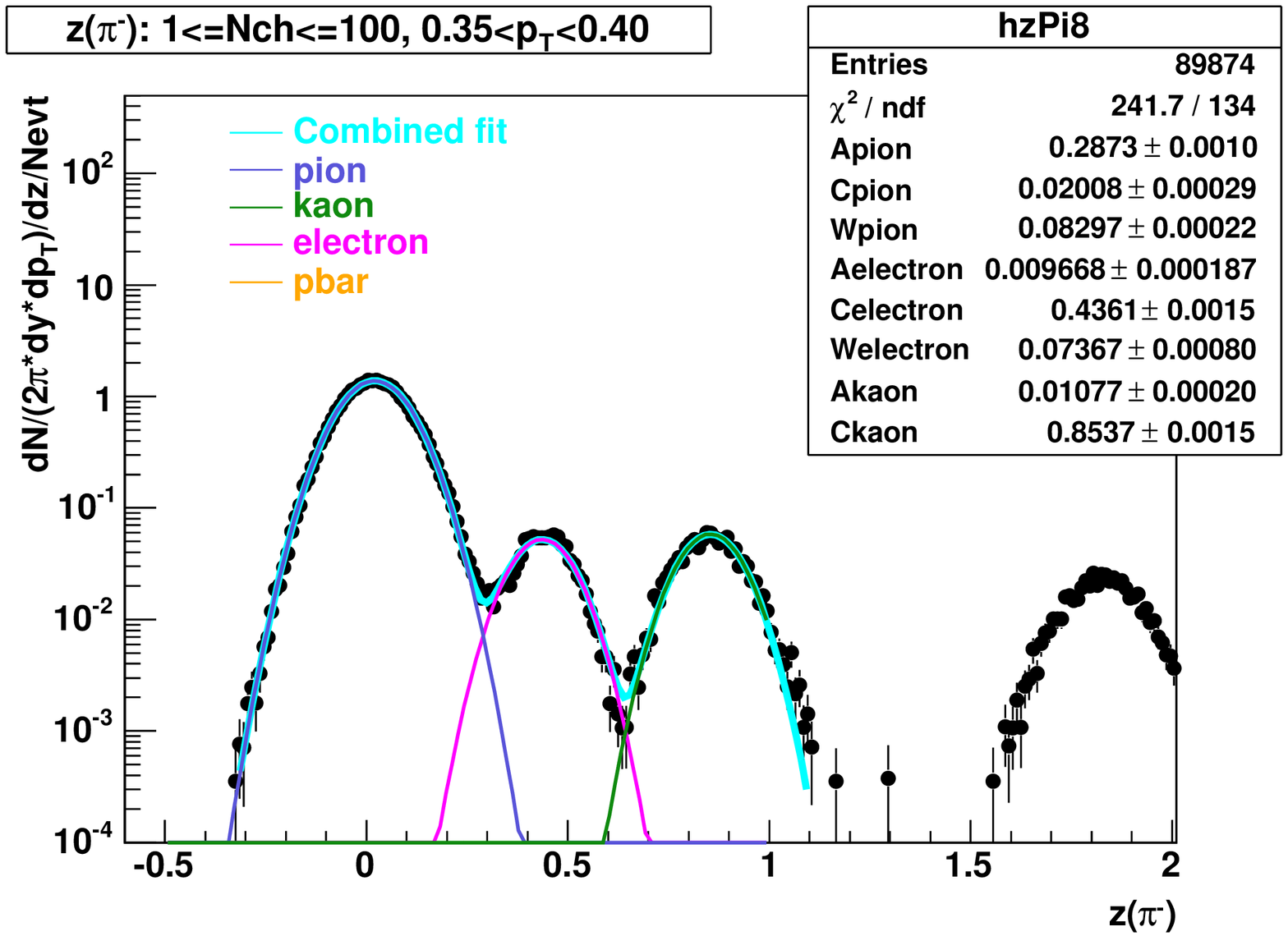}}\\
\resizebox{.225\textwidth}{!}{\includegraphics{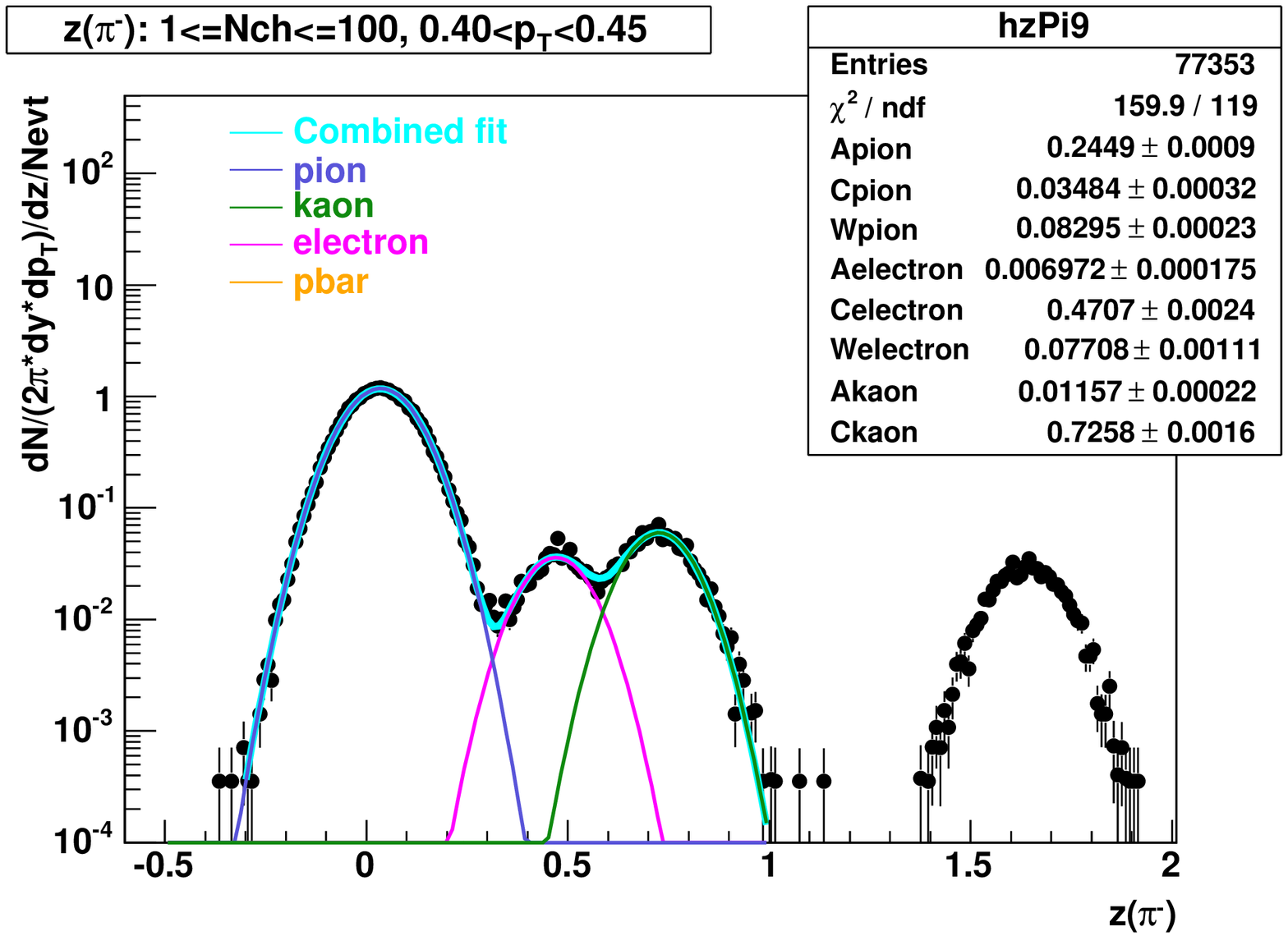}}
\resizebox{.225\textwidth}{!}{\includegraphics{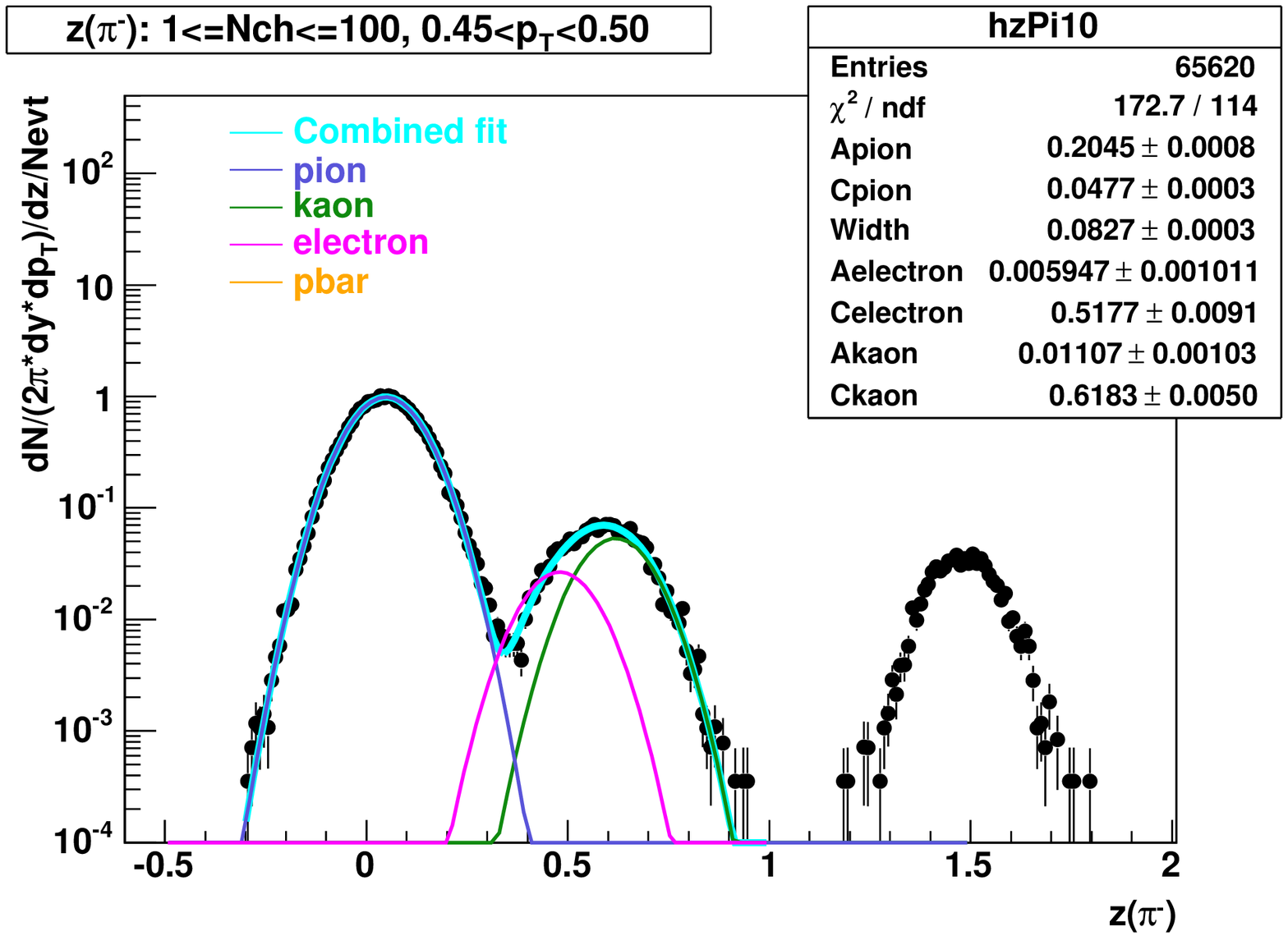}}
\resizebox{.225\textwidth}{!}{\includegraphics{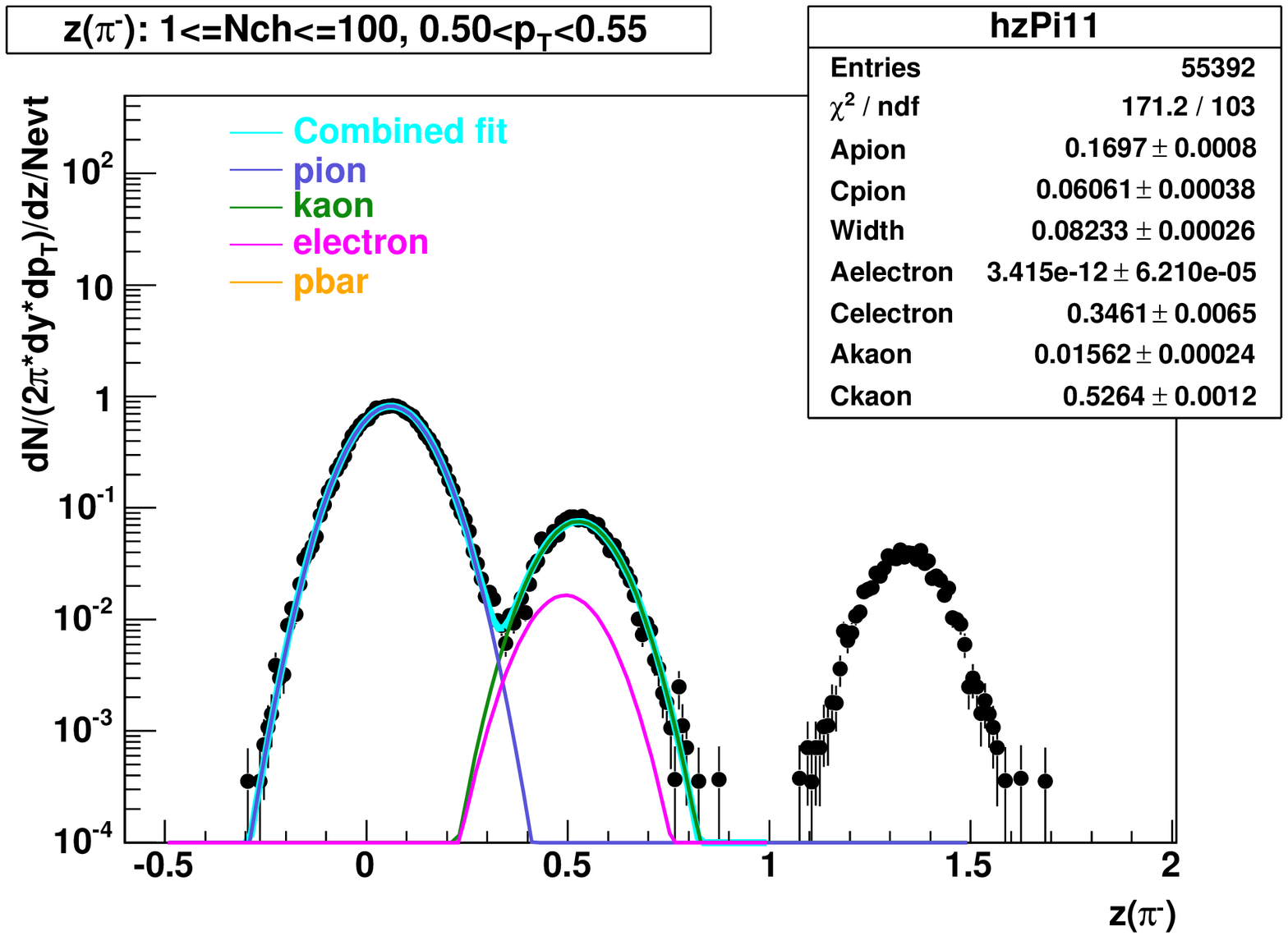}}
\resizebox{.225\textwidth}{!}{\includegraphics{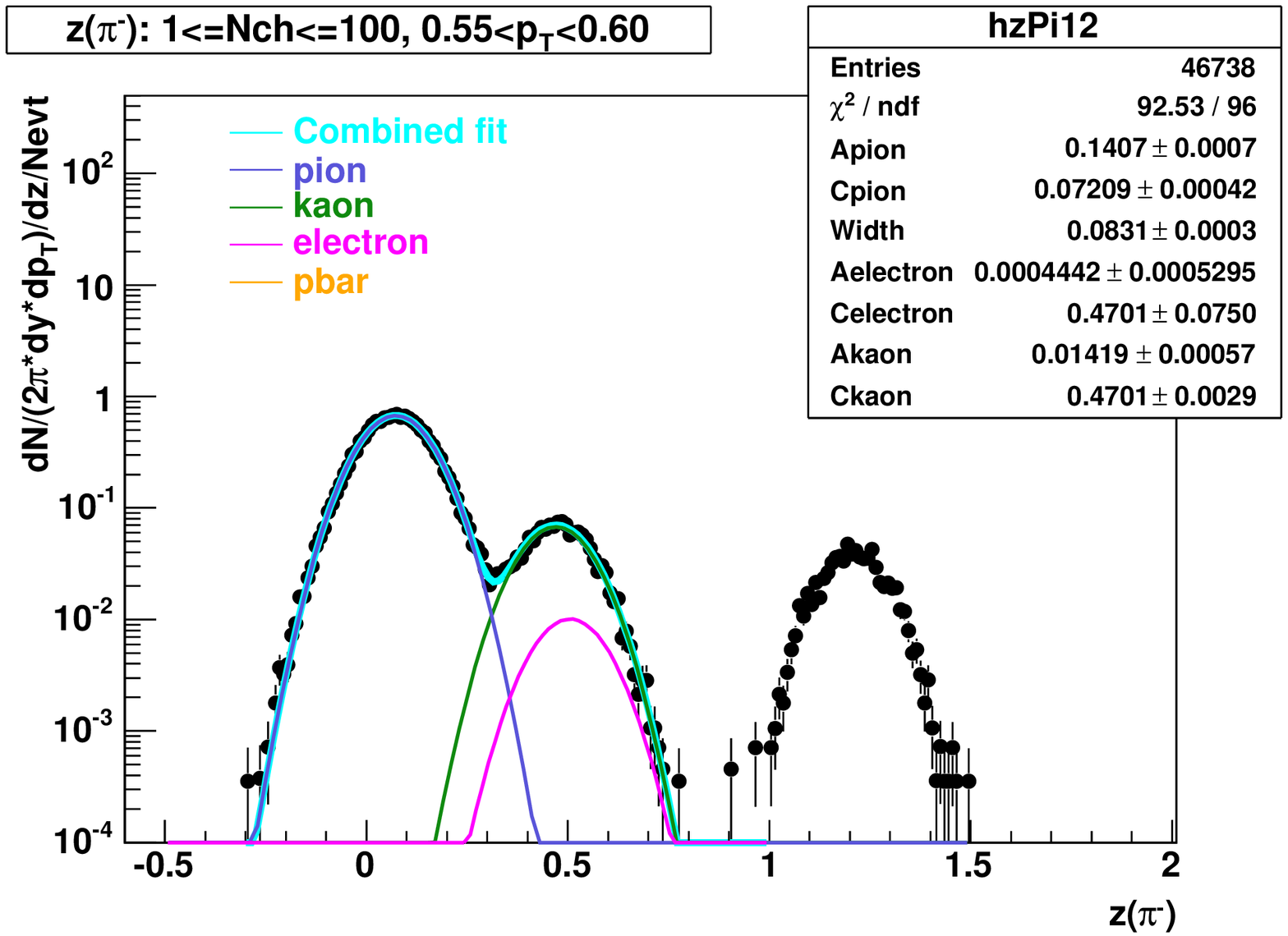}}\\
\resizebox{.225\textwidth}{!}{\includegraphics{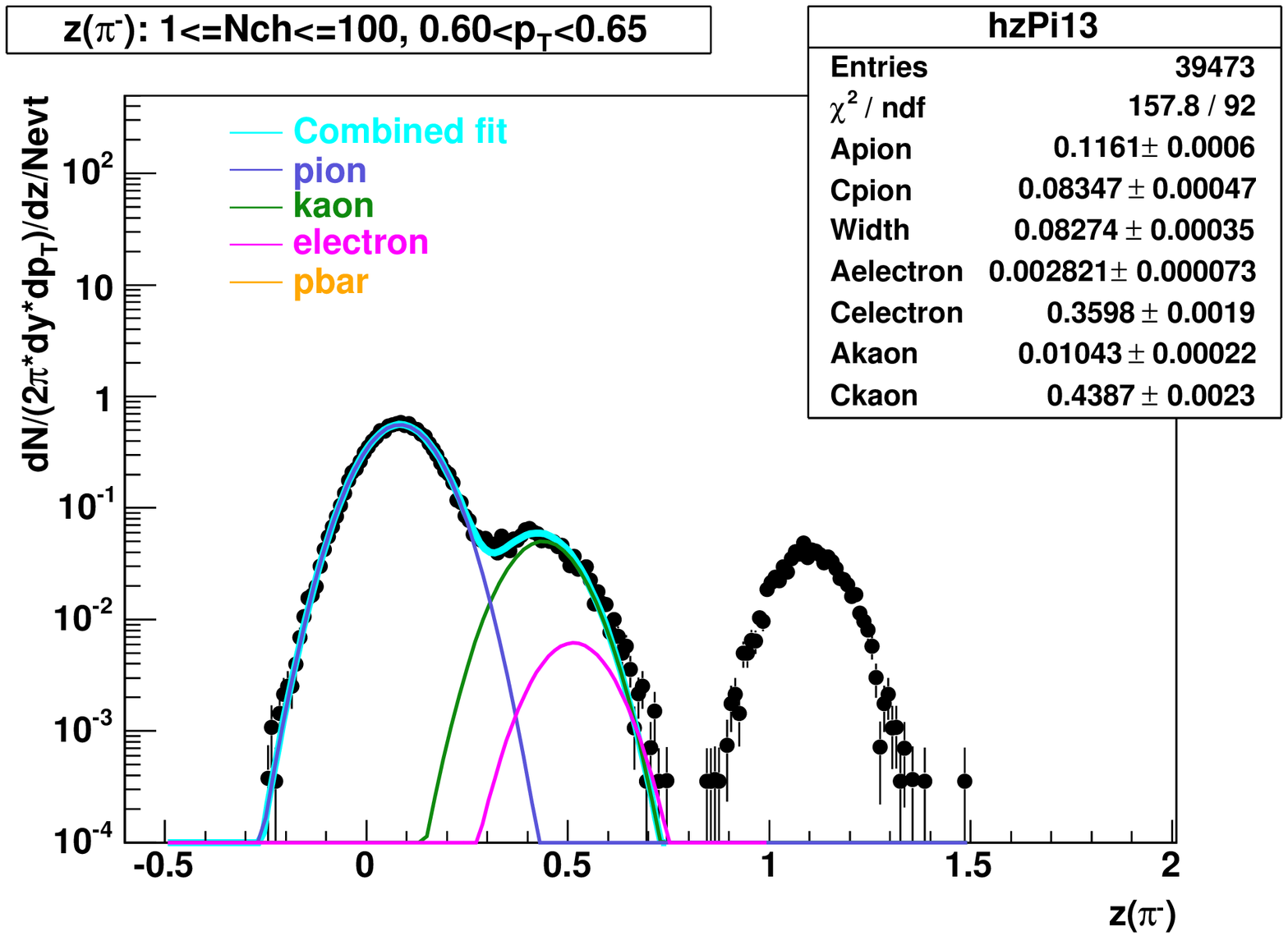}}
\resizebox{.225\textwidth}{!}{\includegraphics{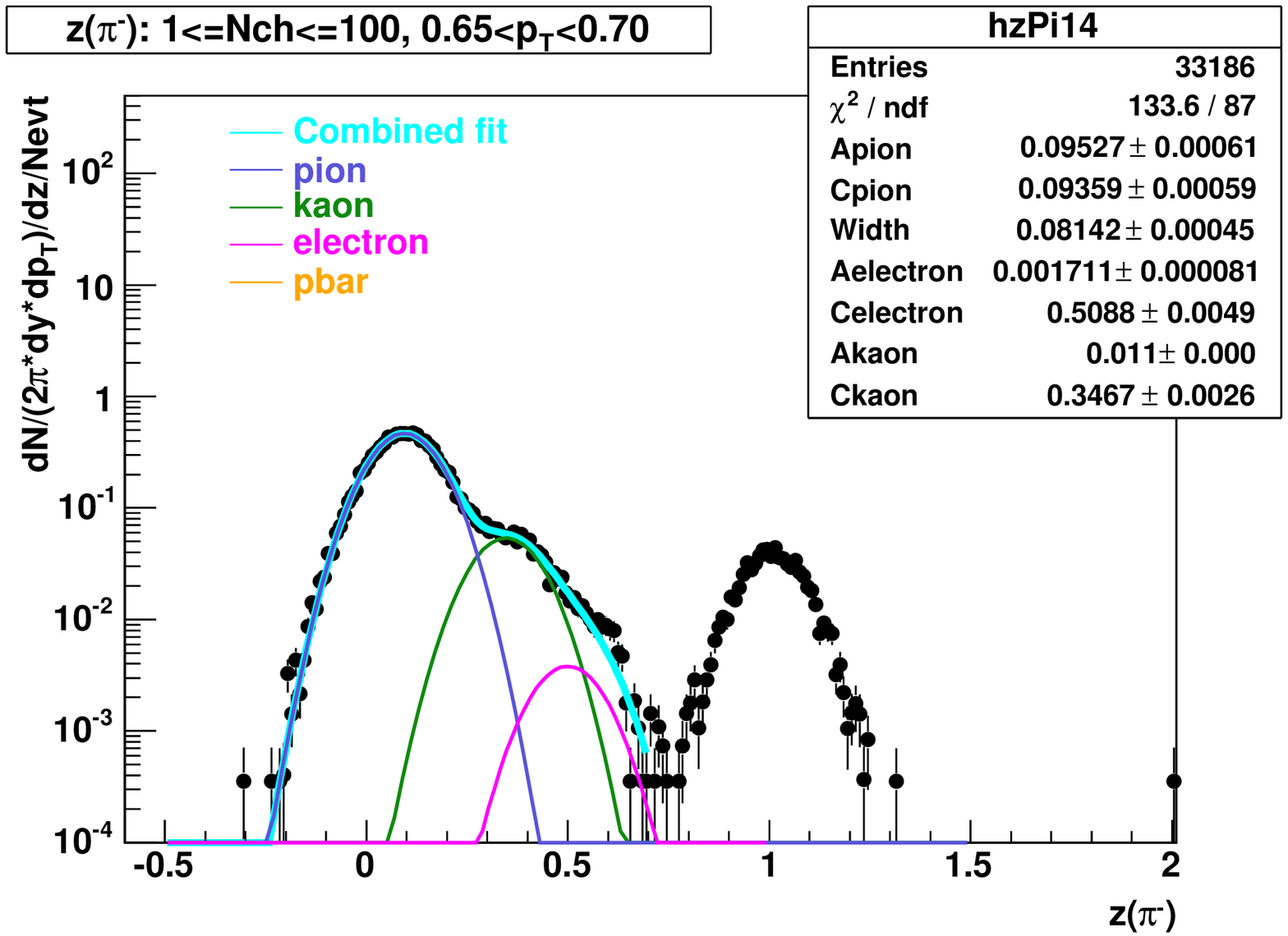}}
\resizebox{.225\textwidth}{!}{\includegraphics{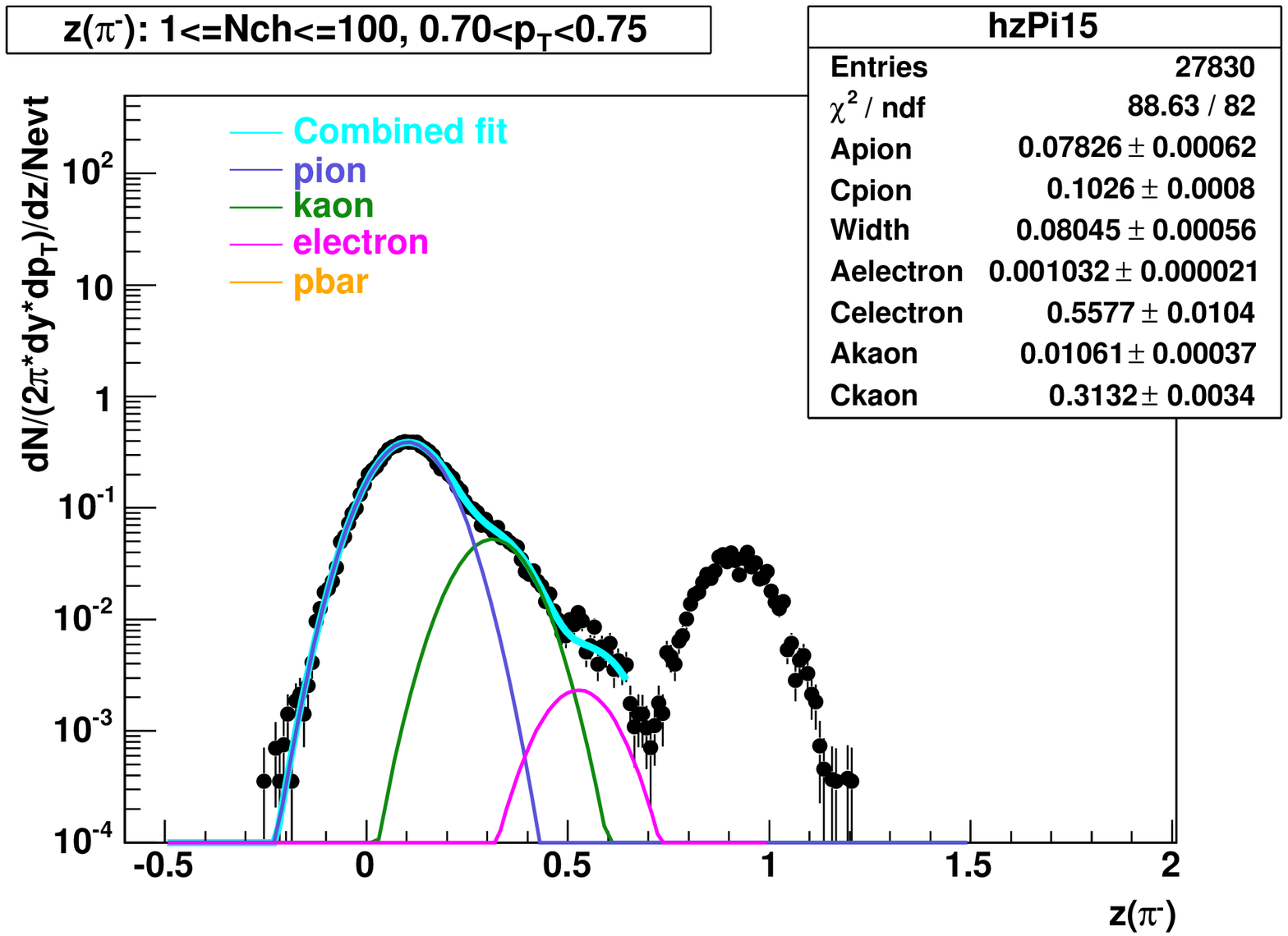}}
\resizebox{.225\textwidth}{!}{\includegraphics{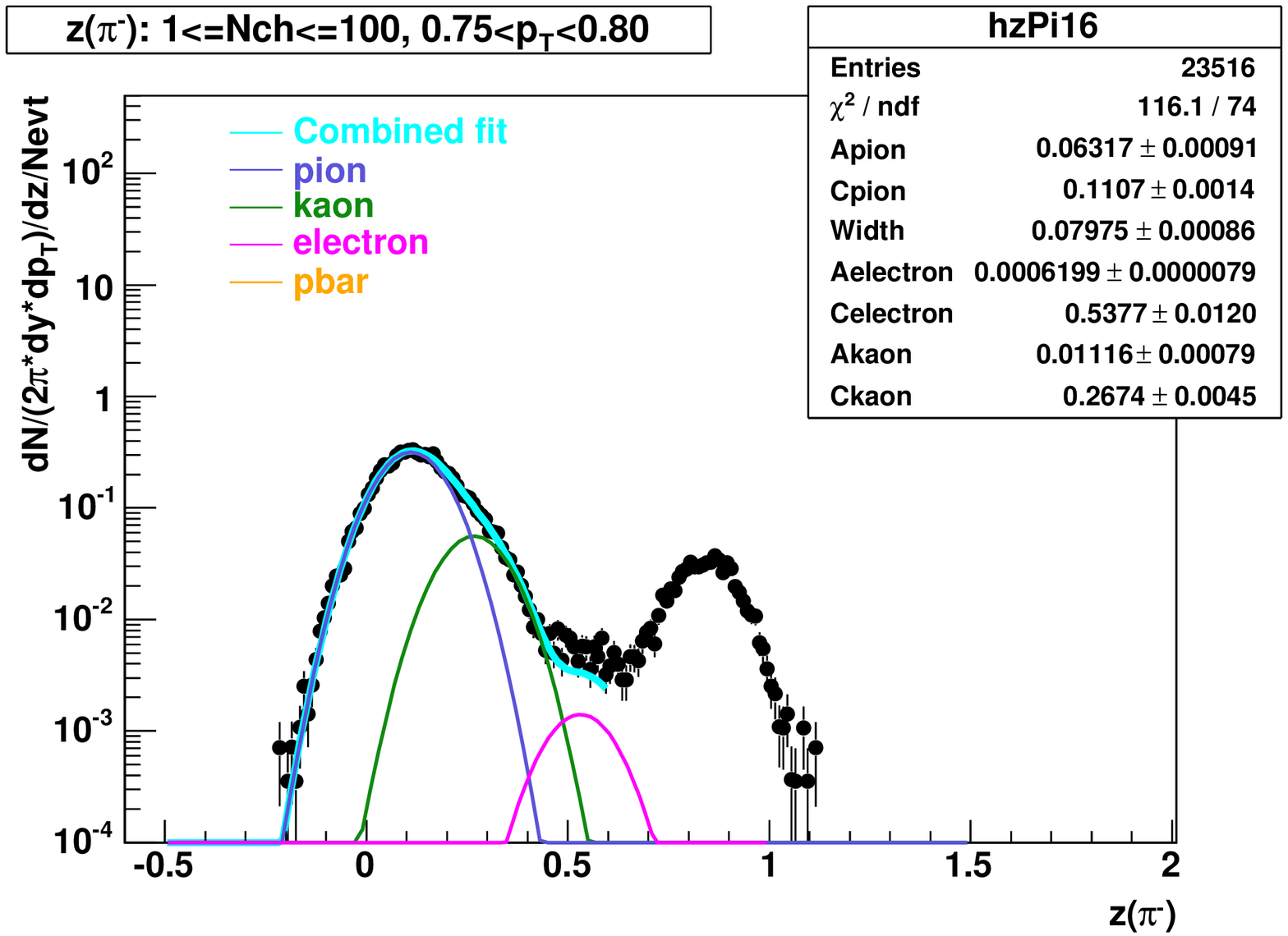}}
	 \caption{Gaussian fits to the $z$ distribution of pions in 200 GeV pp collisions.\label{fig:ppGaussianFitsPion}}
\end{center} 
\end{sidewaysfigure}
\begin{sidewaysfigure}[!h]
\begin{center}
\resizebox{.225\textwidth}{!}{\includegraphics{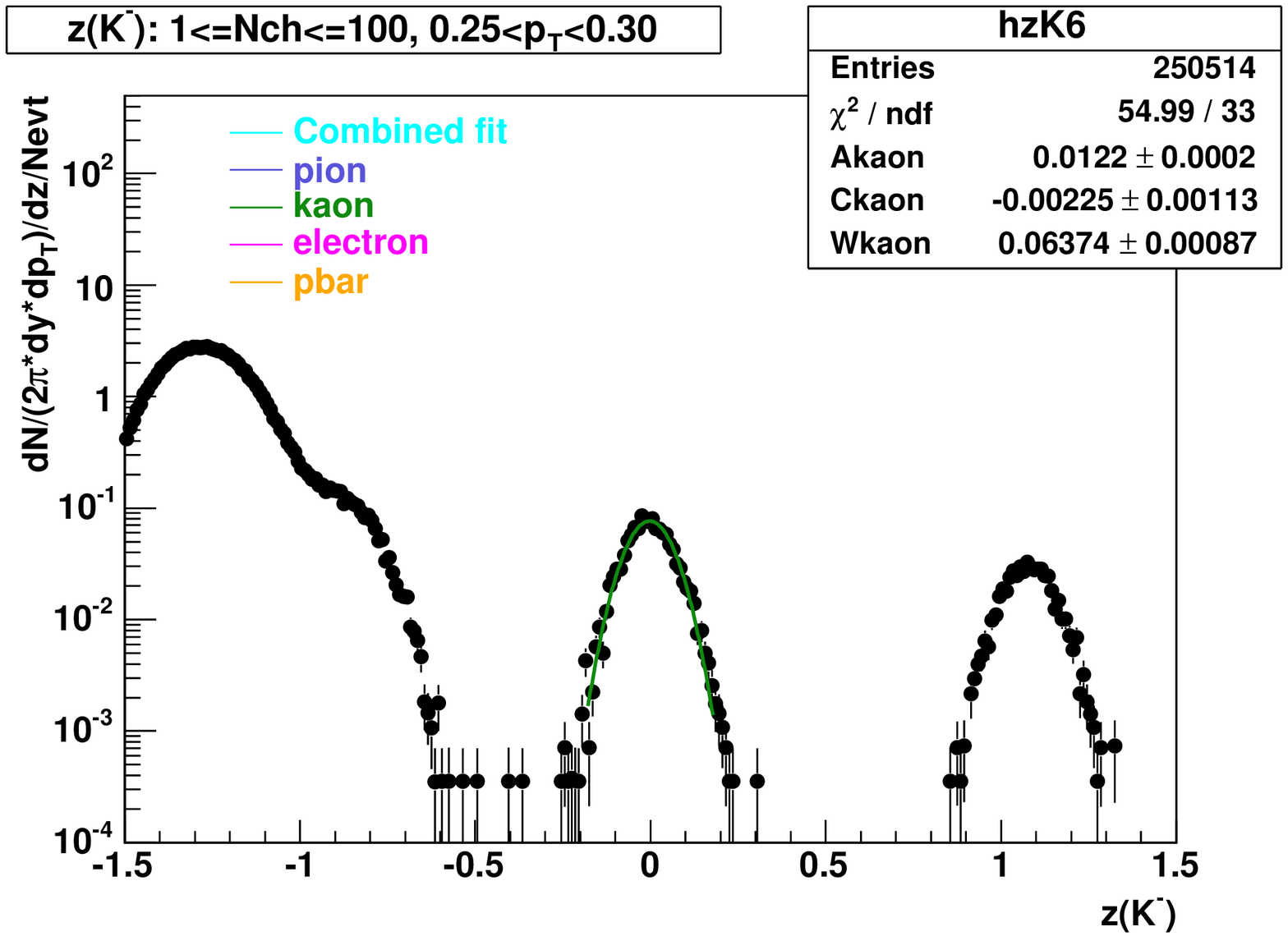}}
\resizebox{.225\textwidth}{!}{\includegraphics{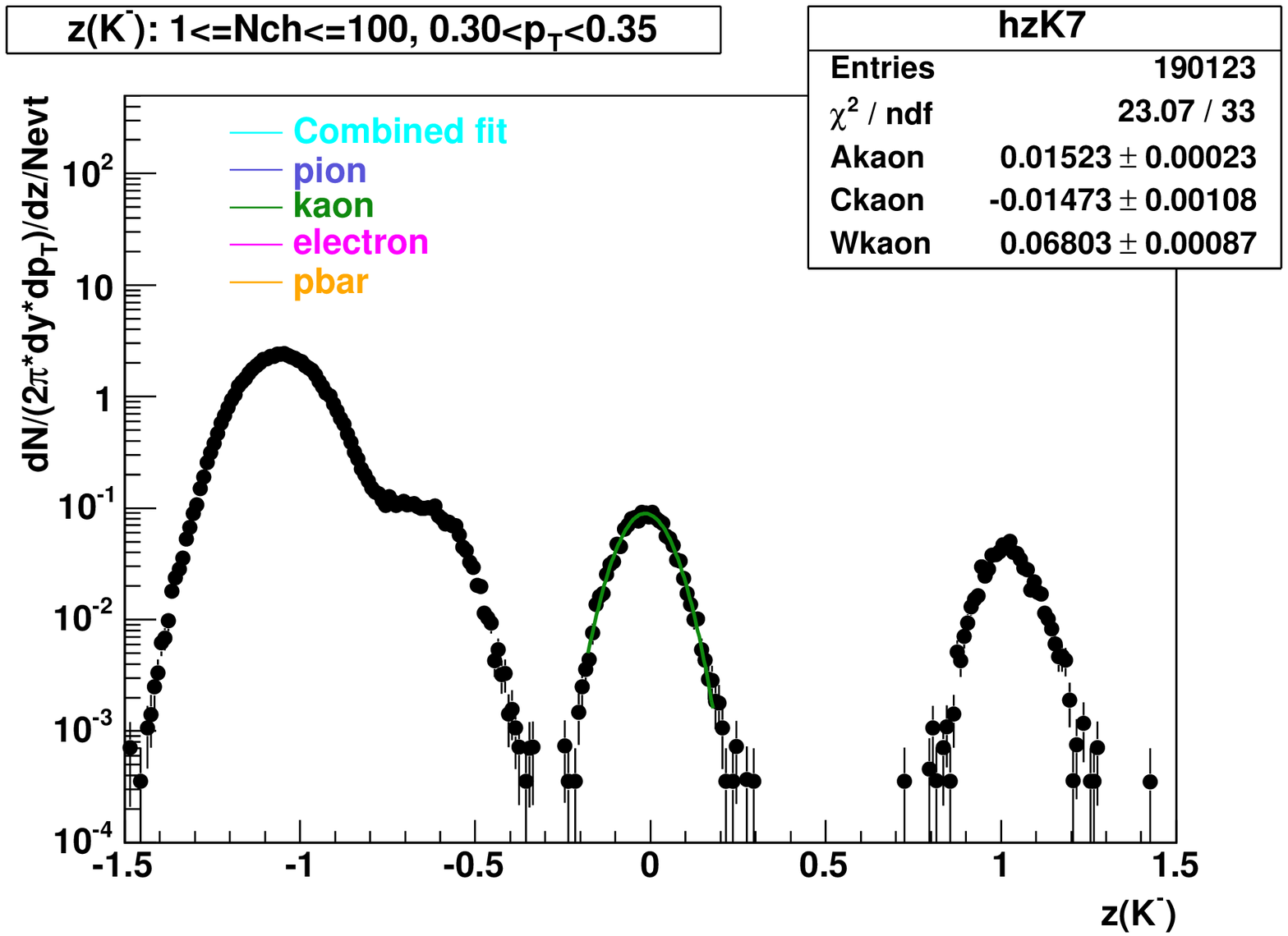}}
\resizebox{.225\textwidth}{!}{\includegraphics{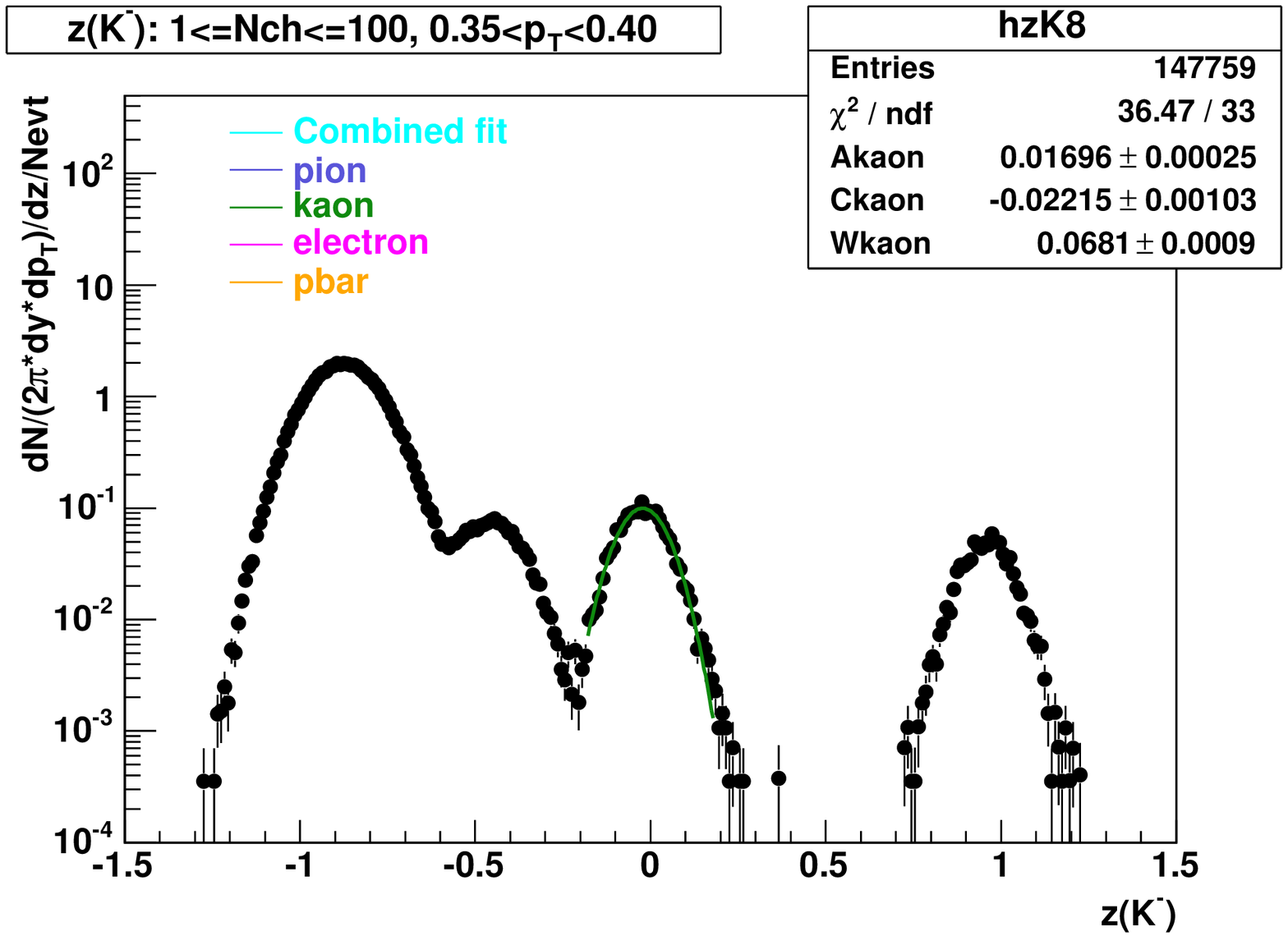}}
\resizebox{.225\textwidth}{!}{\includegraphics{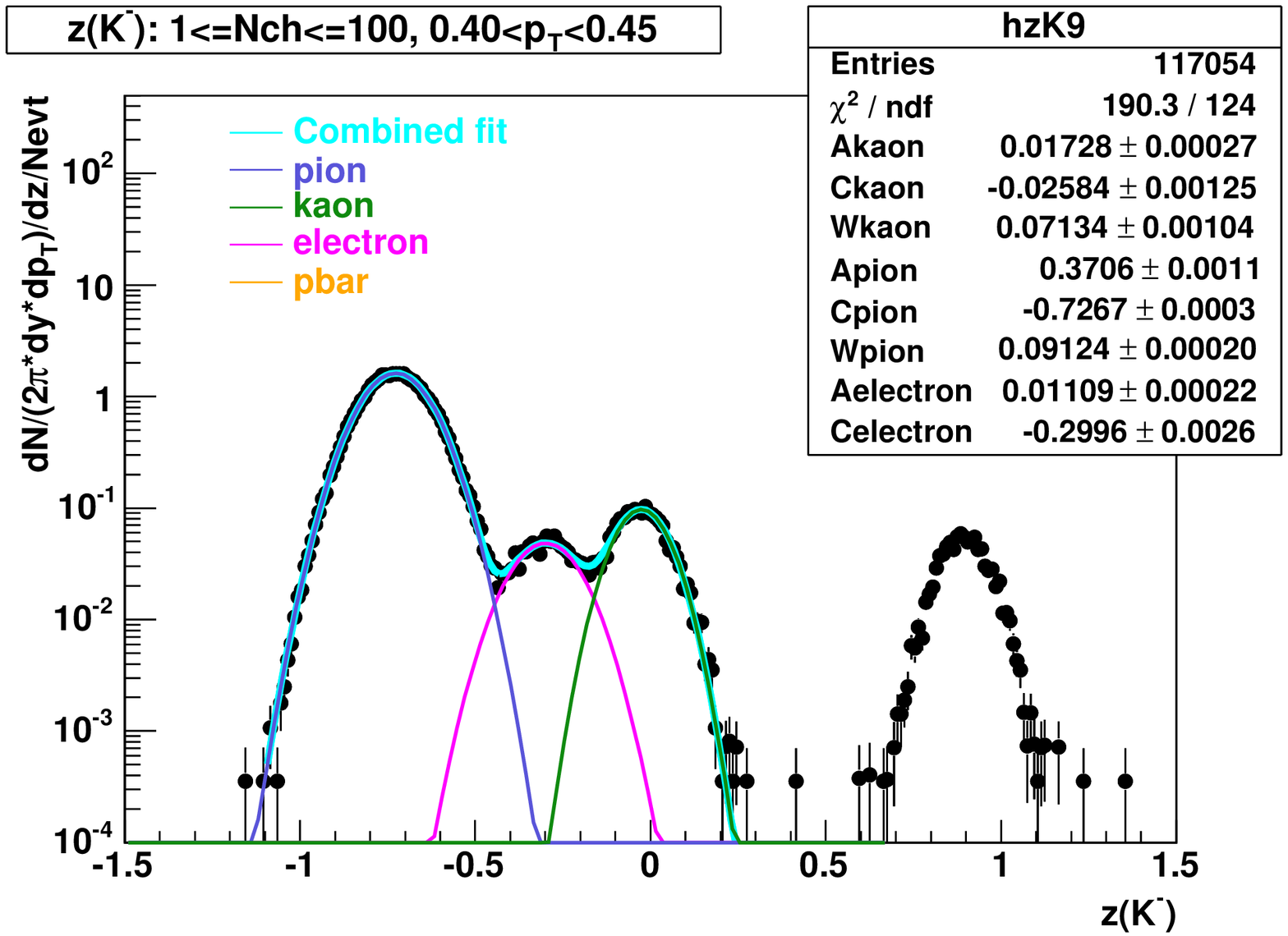}}\\
\resizebox{.225\textwidth}{!}{\includegraphics{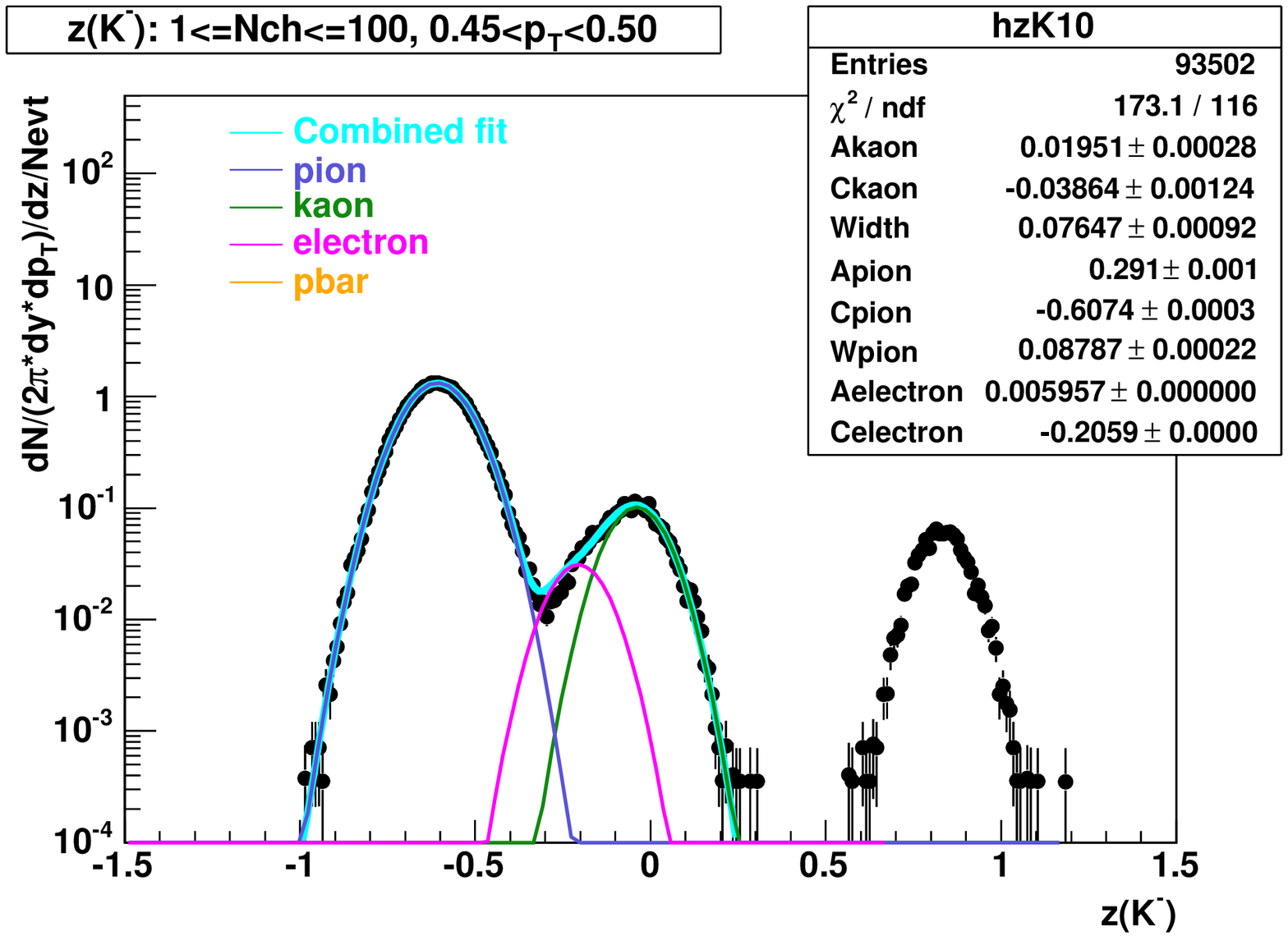}}
\resizebox{.225\textwidth}{!}{\includegraphics{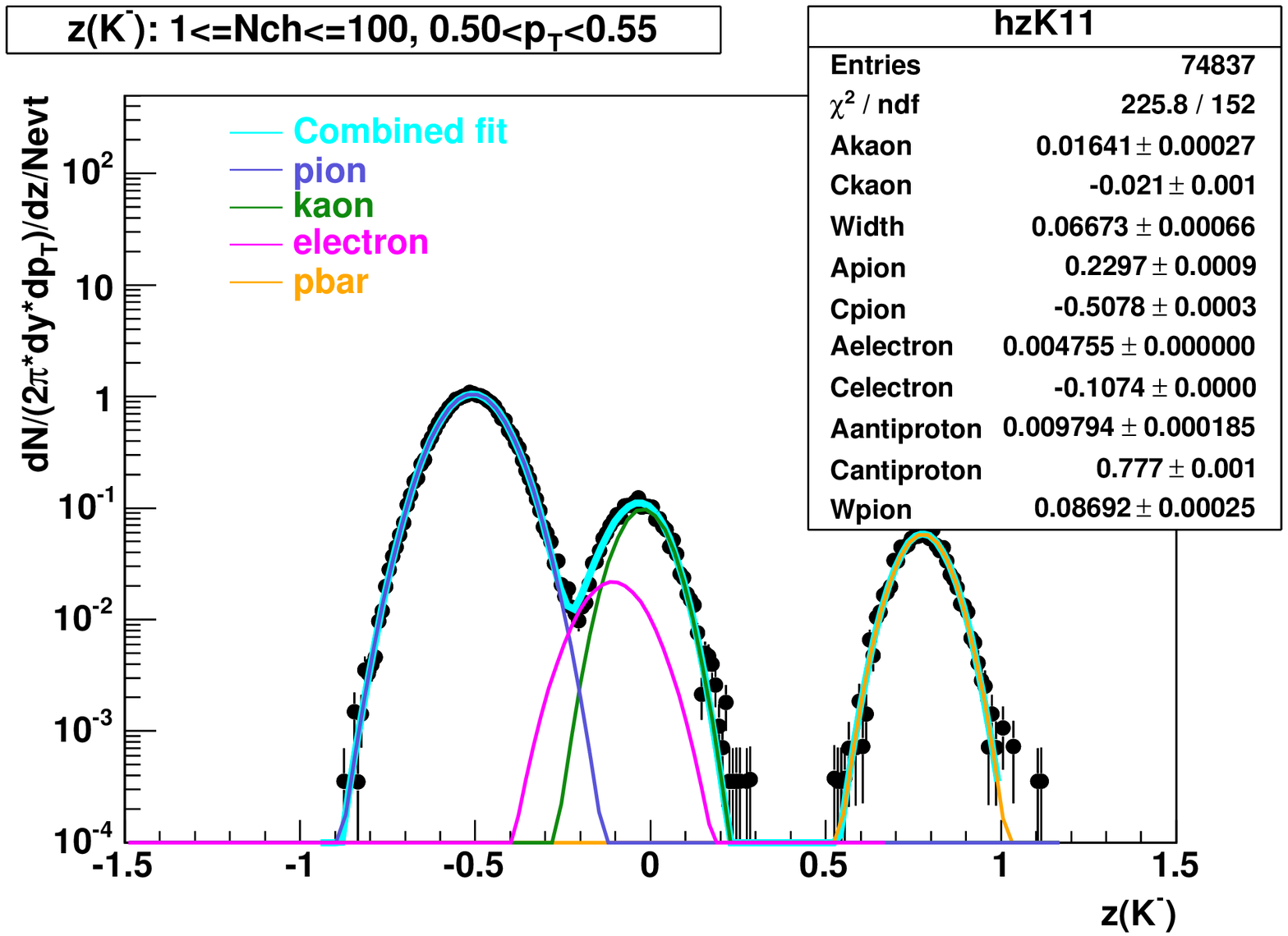}}
\resizebox{.225\textwidth}{!}{\includegraphics{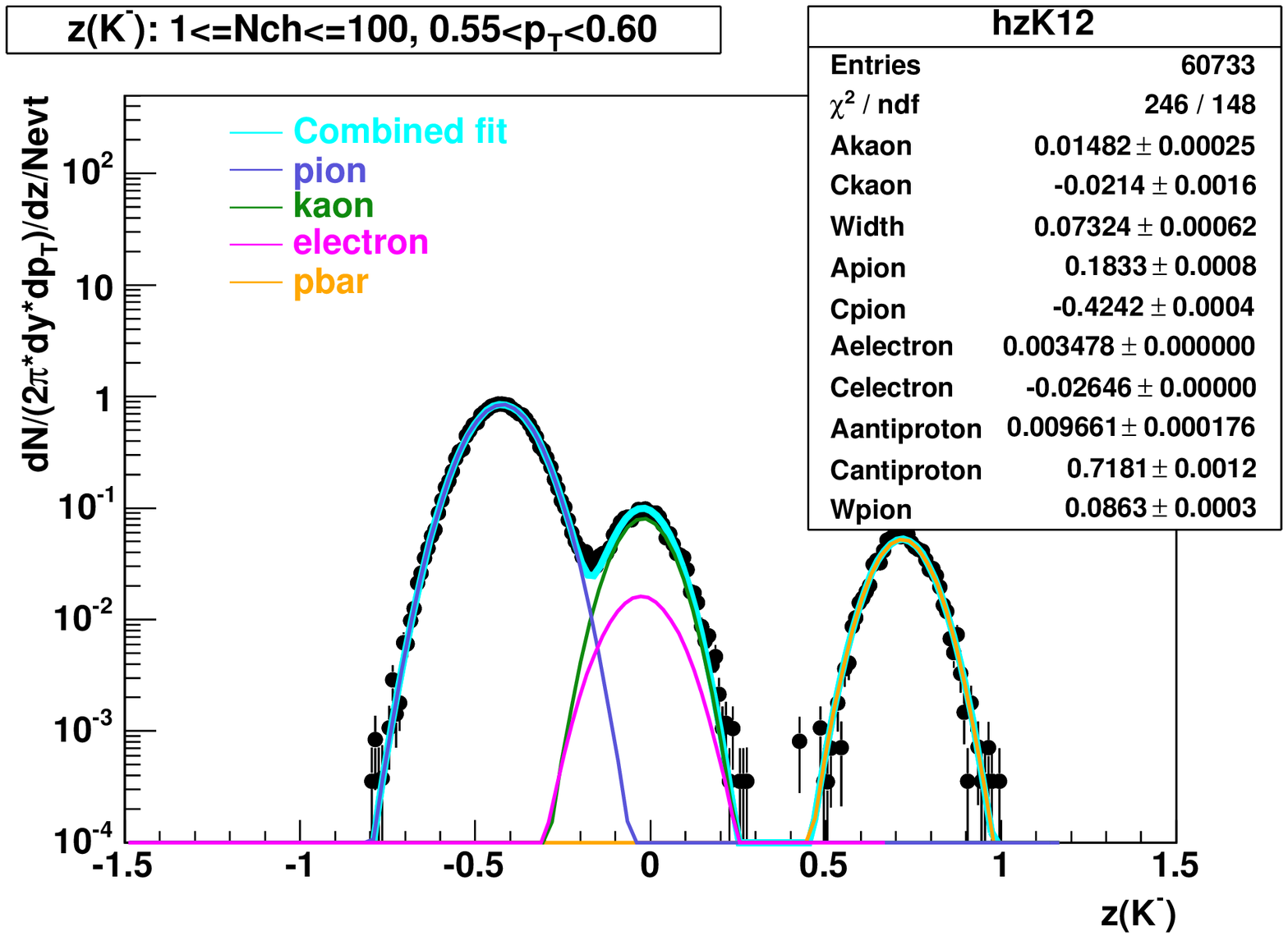}}
\resizebox{.225\textwidth}{!}{\includegraphics{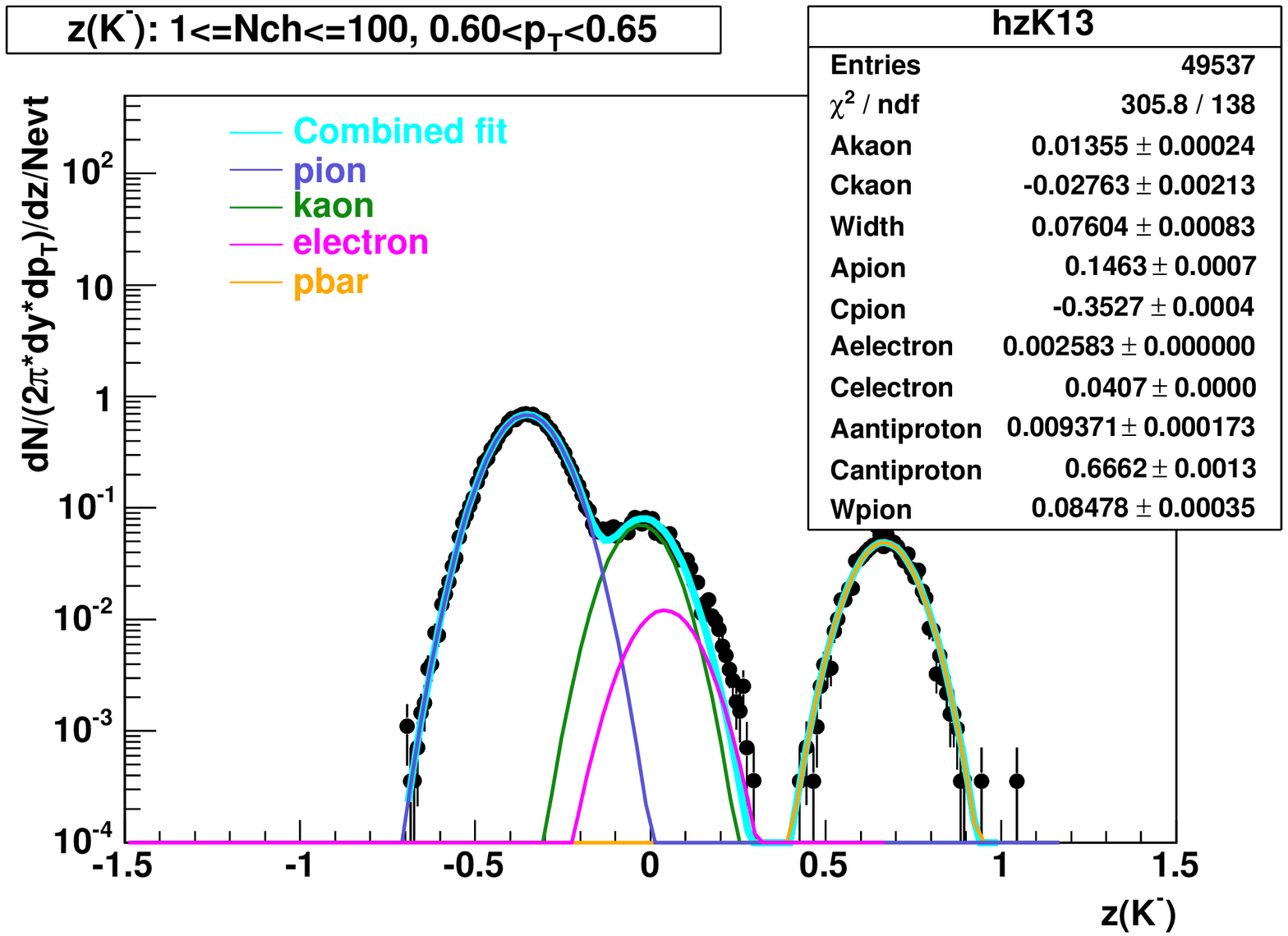}}\\
\resizebox{.225\textwidth}{!}{\includegraphics{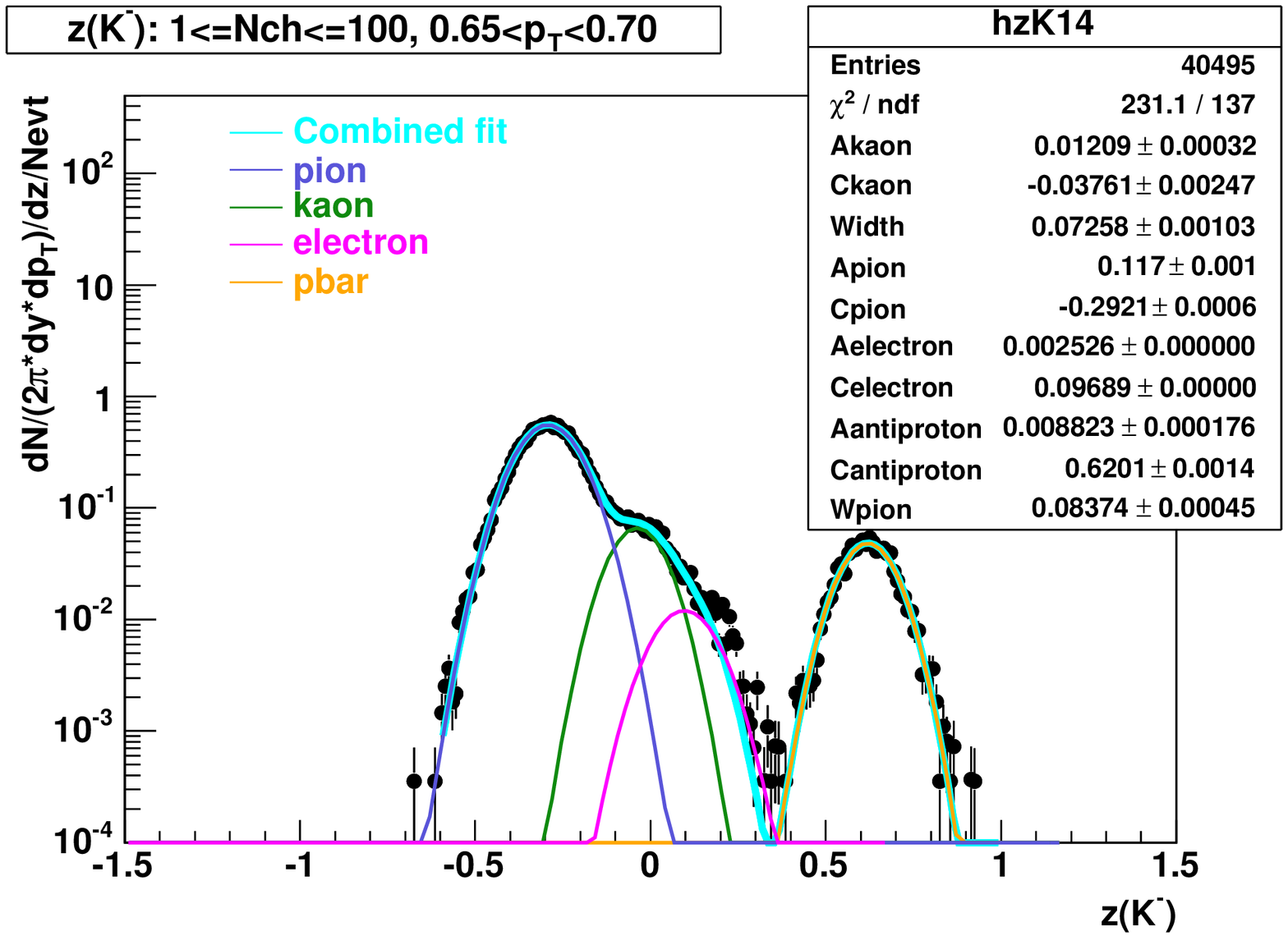}}
\resizebox{.225\textwidth}{!}{\includegraphics{Plots/pid/SpFits/km_Nch_1_100_pT_0.70_0.75.eps}}
	 \caption{Gaussian fits to the $z$ distribution of kaons in 200 GeV pp collisions.\label{fig:ppGaussianFitsKaon}}
\end{center} 
\end{sidewaysfigure}
\begin{sidewaysfigure}[!h]
		\begin{center}																											
\resizebox{.225\textwidth}{!}{\includegraphics{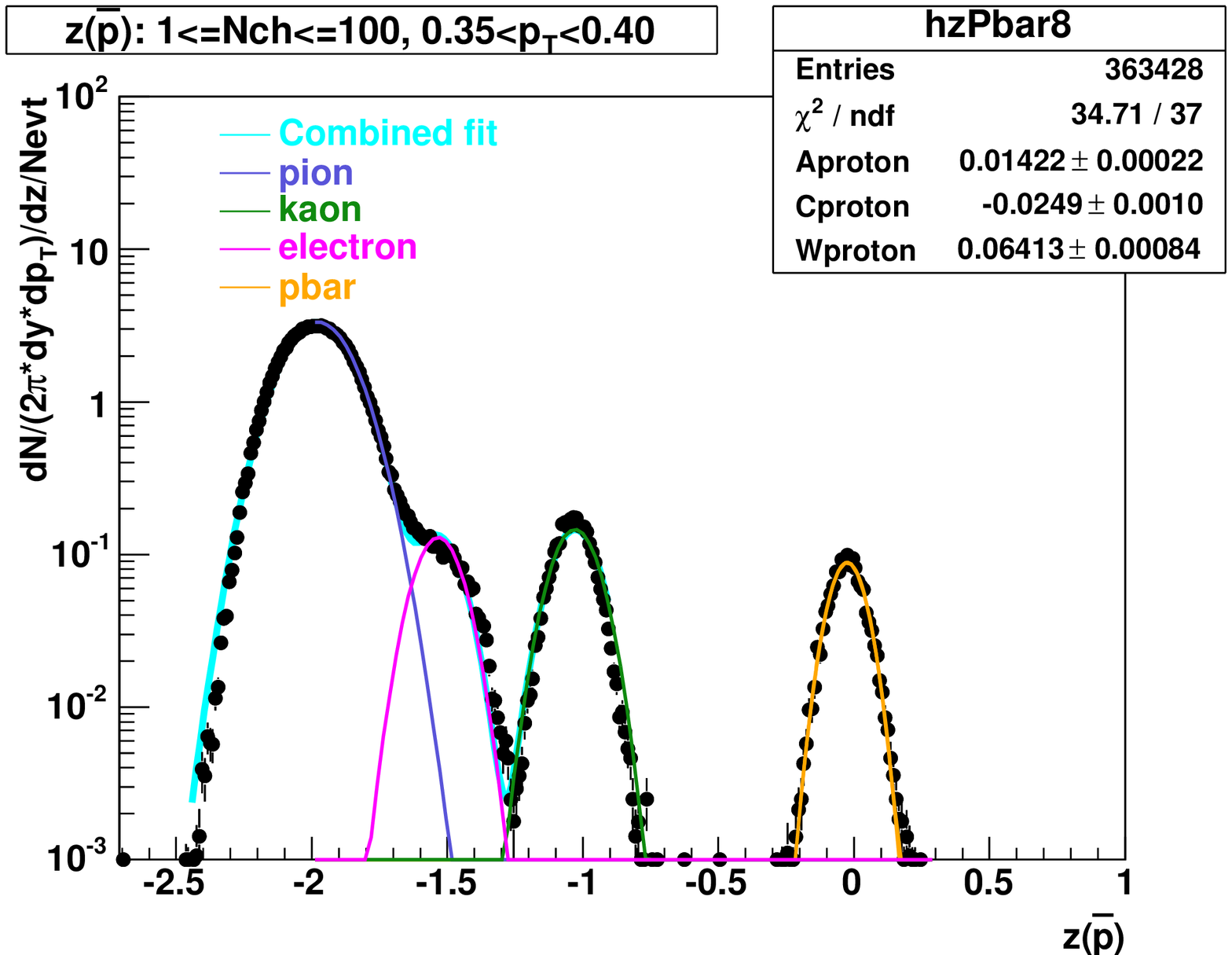}}
\resizebox{.225\textwidth}{!}{\includegraphics{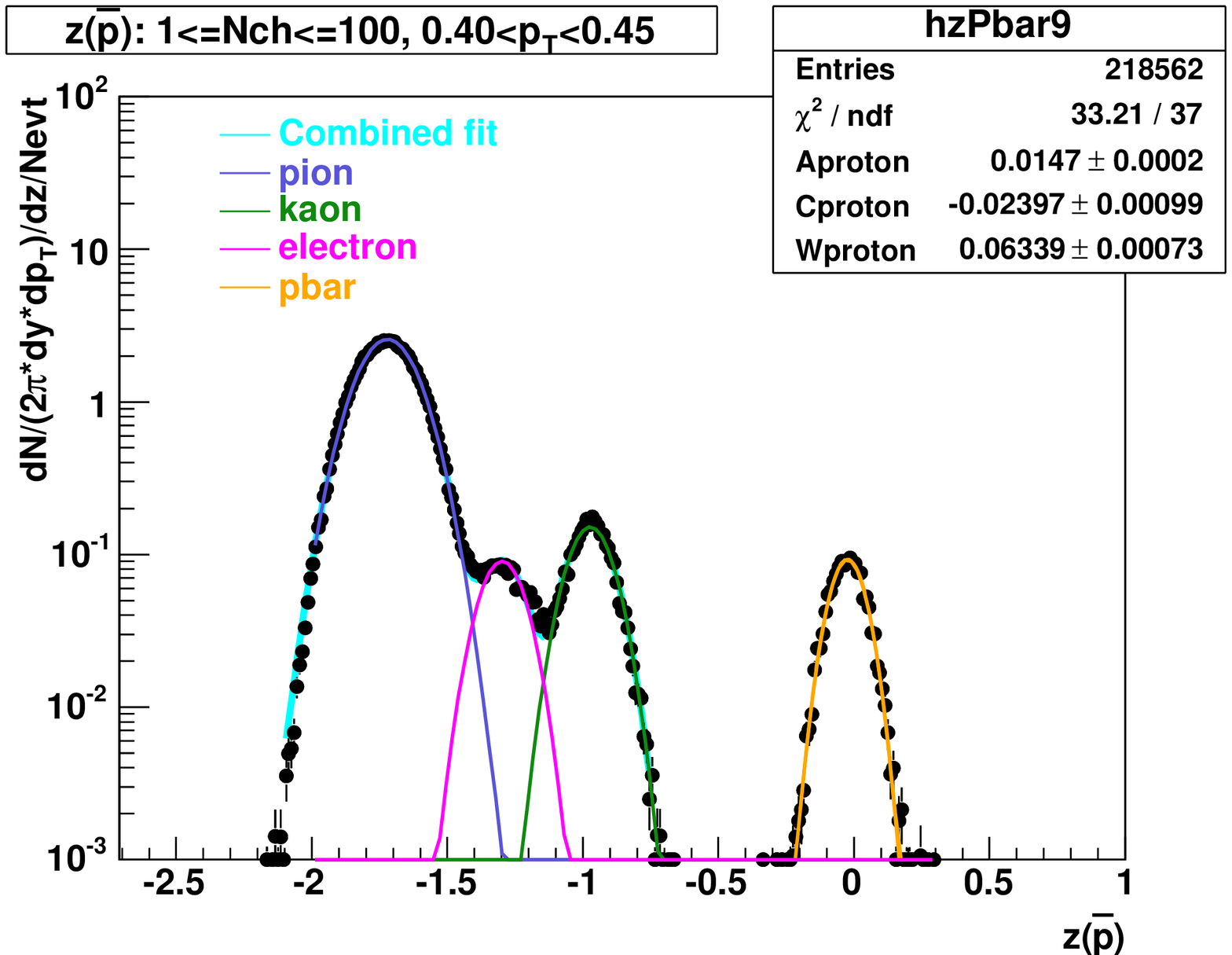}}
\resizebox{.225\textwidth}{!}{\includegraphics{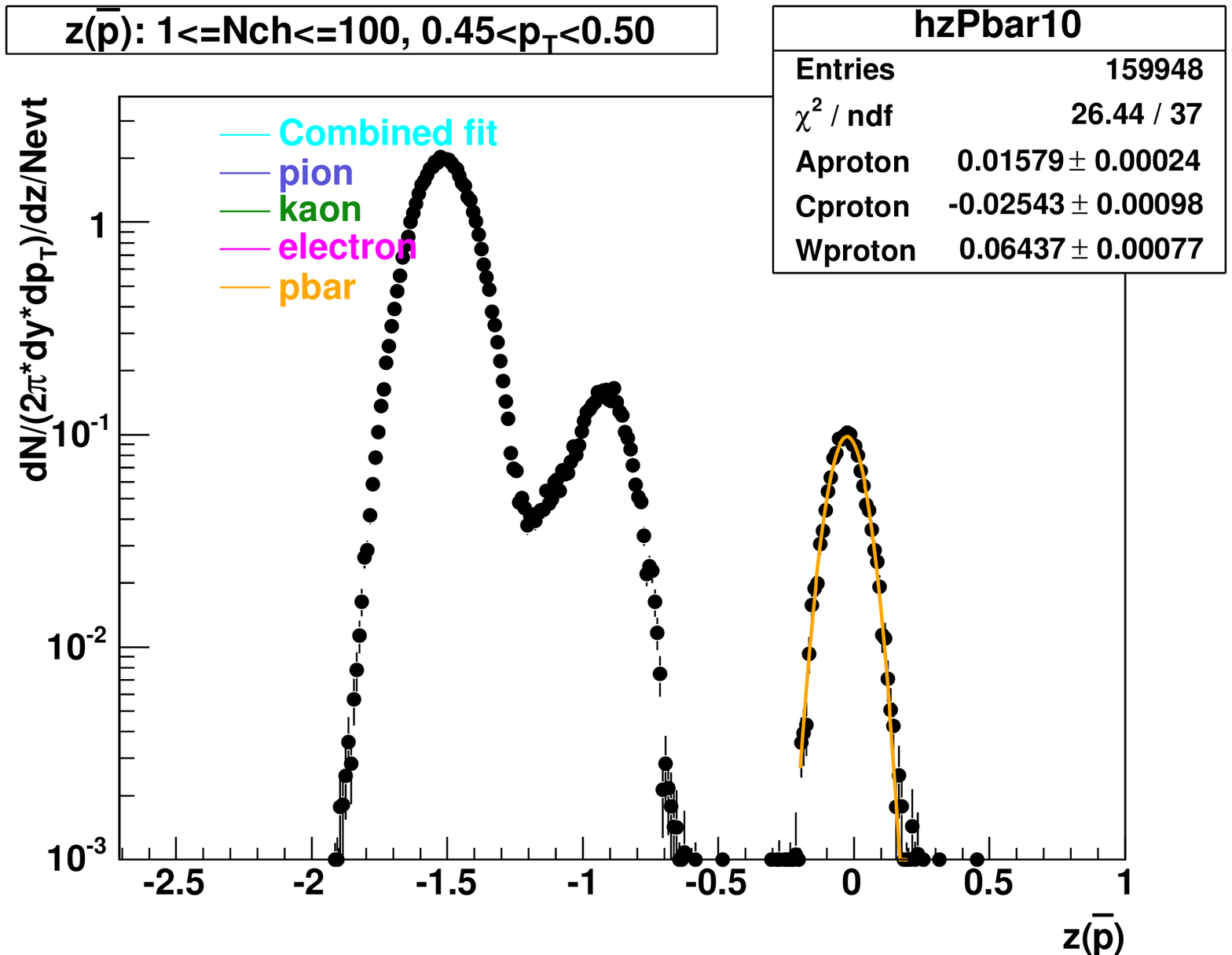}}
\resizebox{.225\textwidth}{!}{\includegraphics{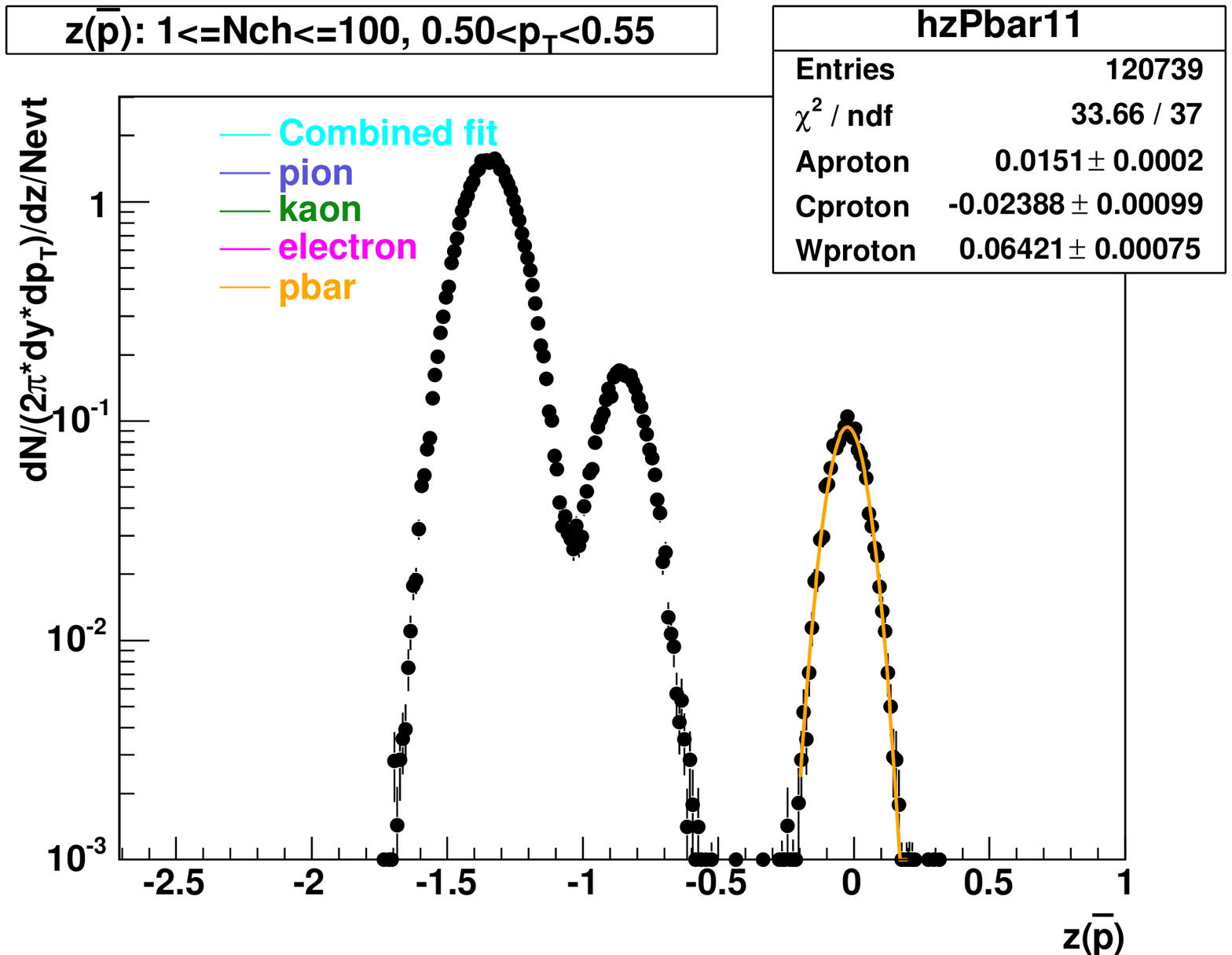}}\\
\resizebox{.225\textwidth}{!}{\includegraphics{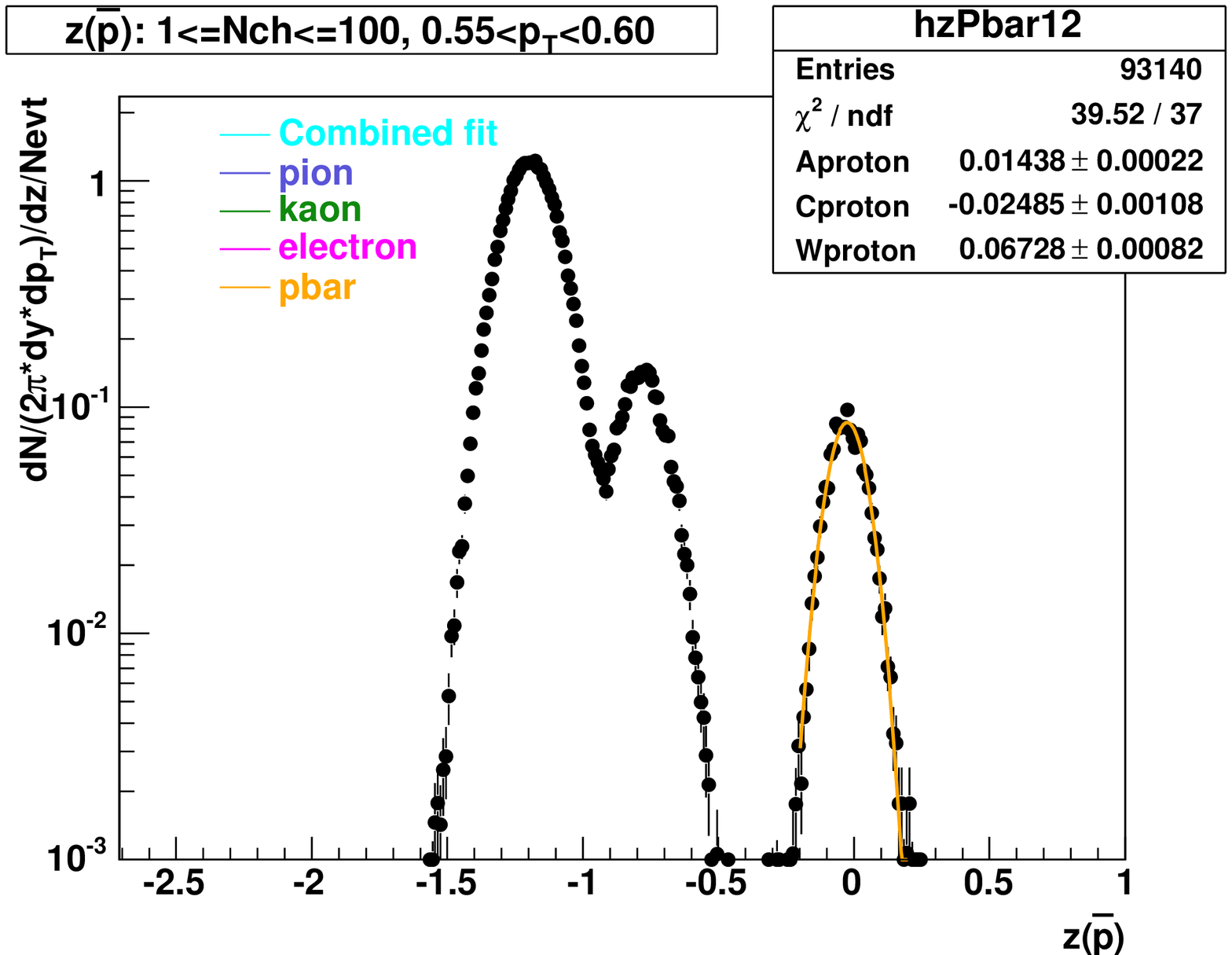}}
\resizebox{.225\textwidth}{!}{\includegraphics{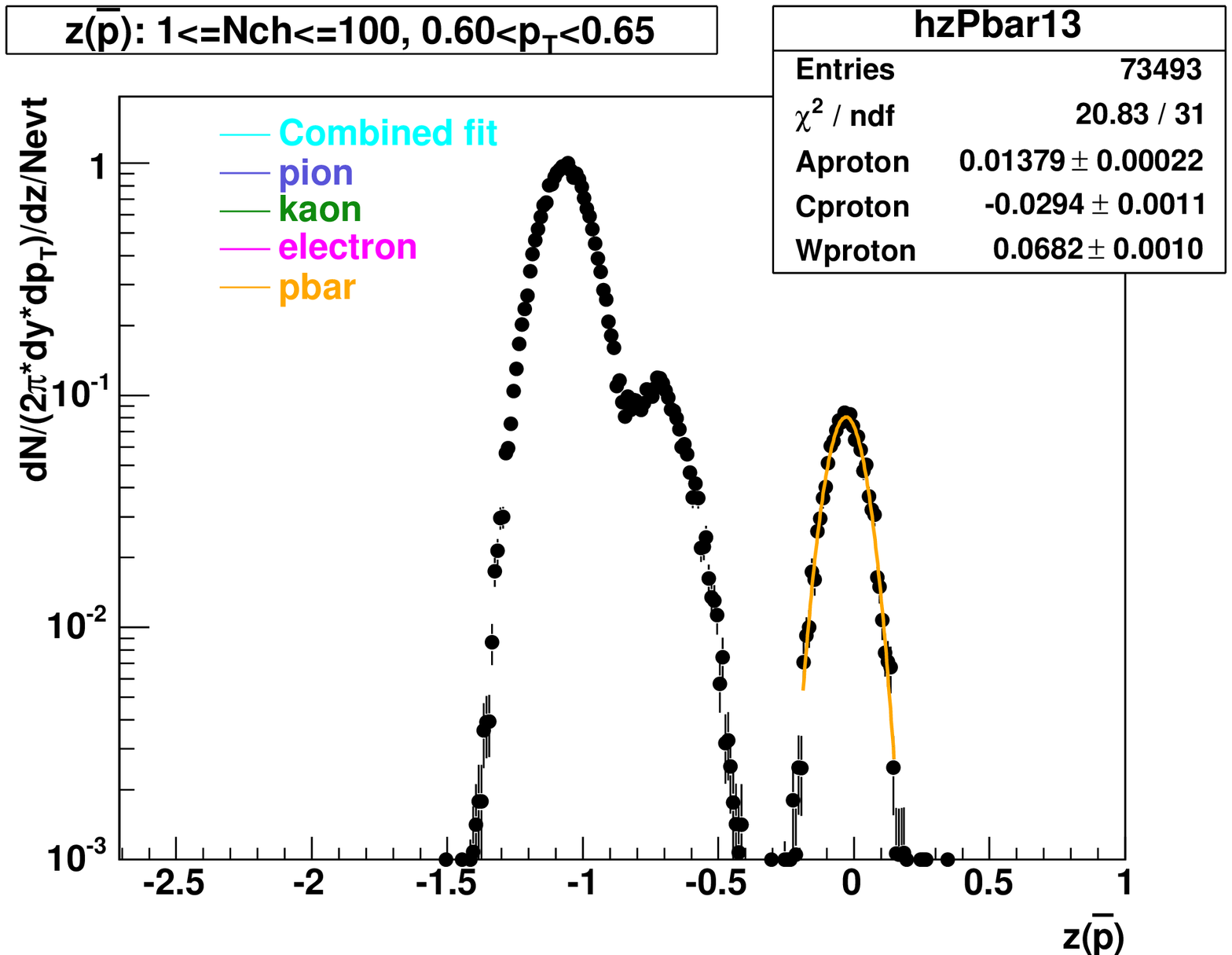}}
\resizebox{.225\textwidth}{!}{\includegraphics{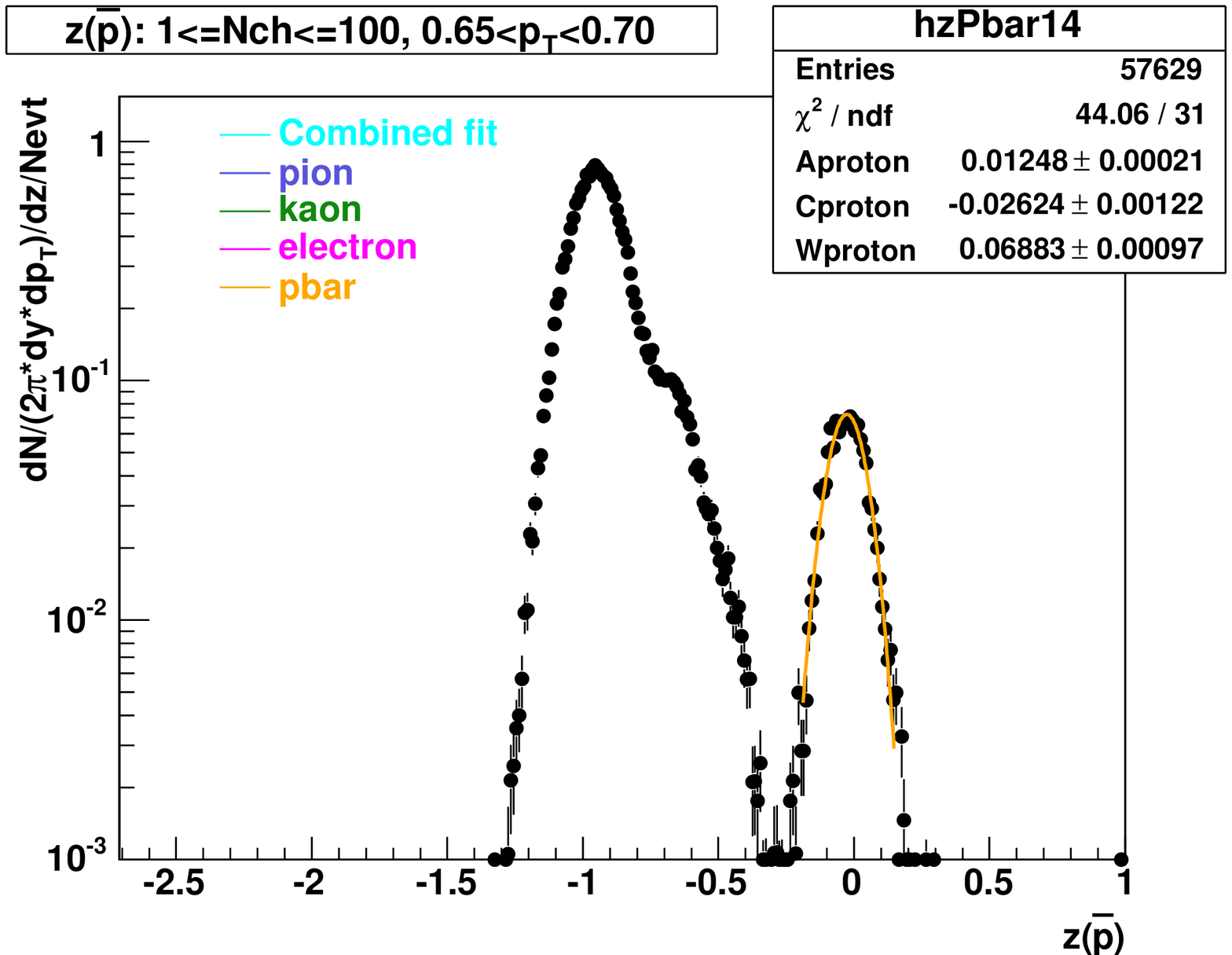}}
\resizebox{.225\textwidth}{!}{\includegraphics{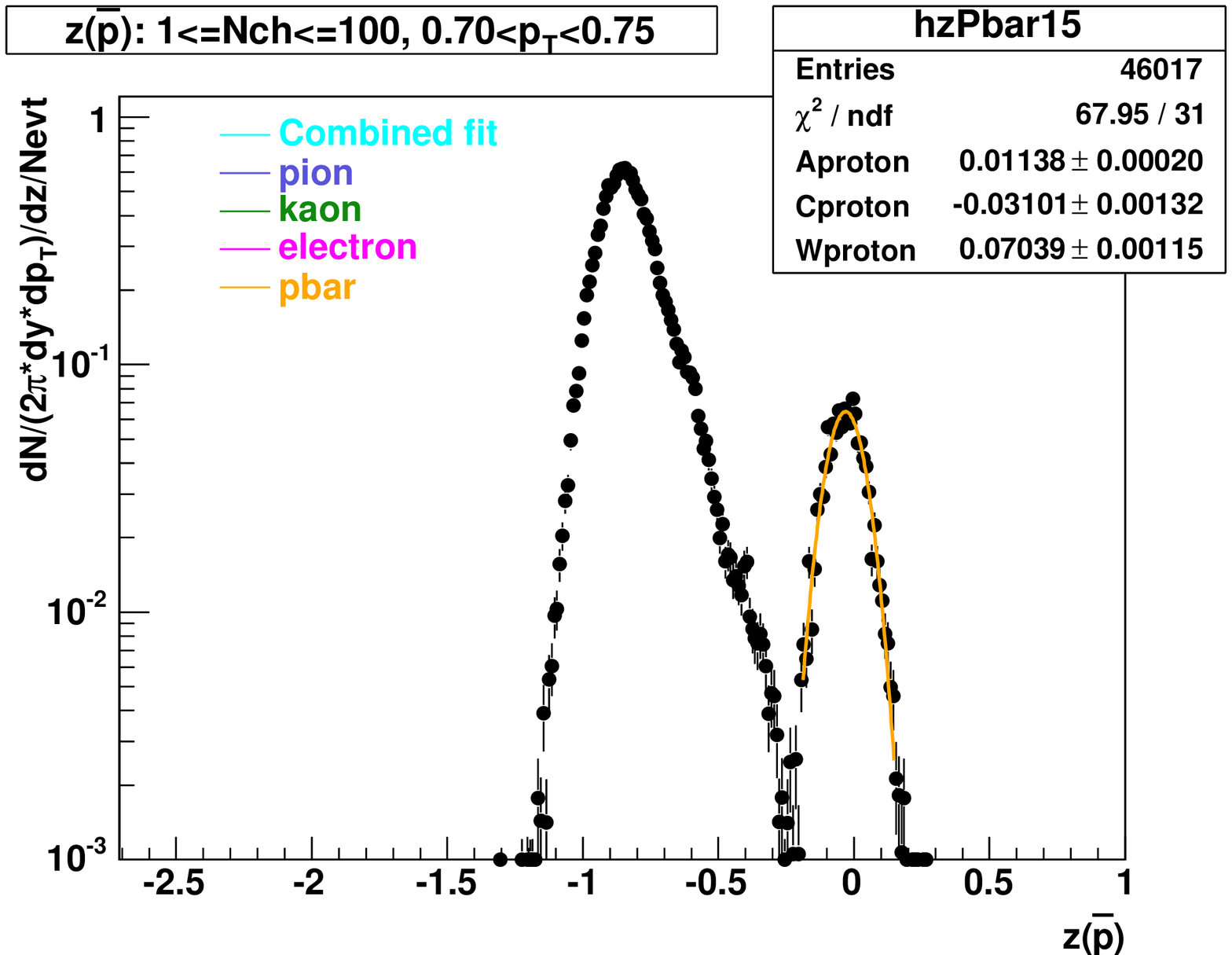}}\\
\resizebox{.225\textwidth}{!}{\includegraphics{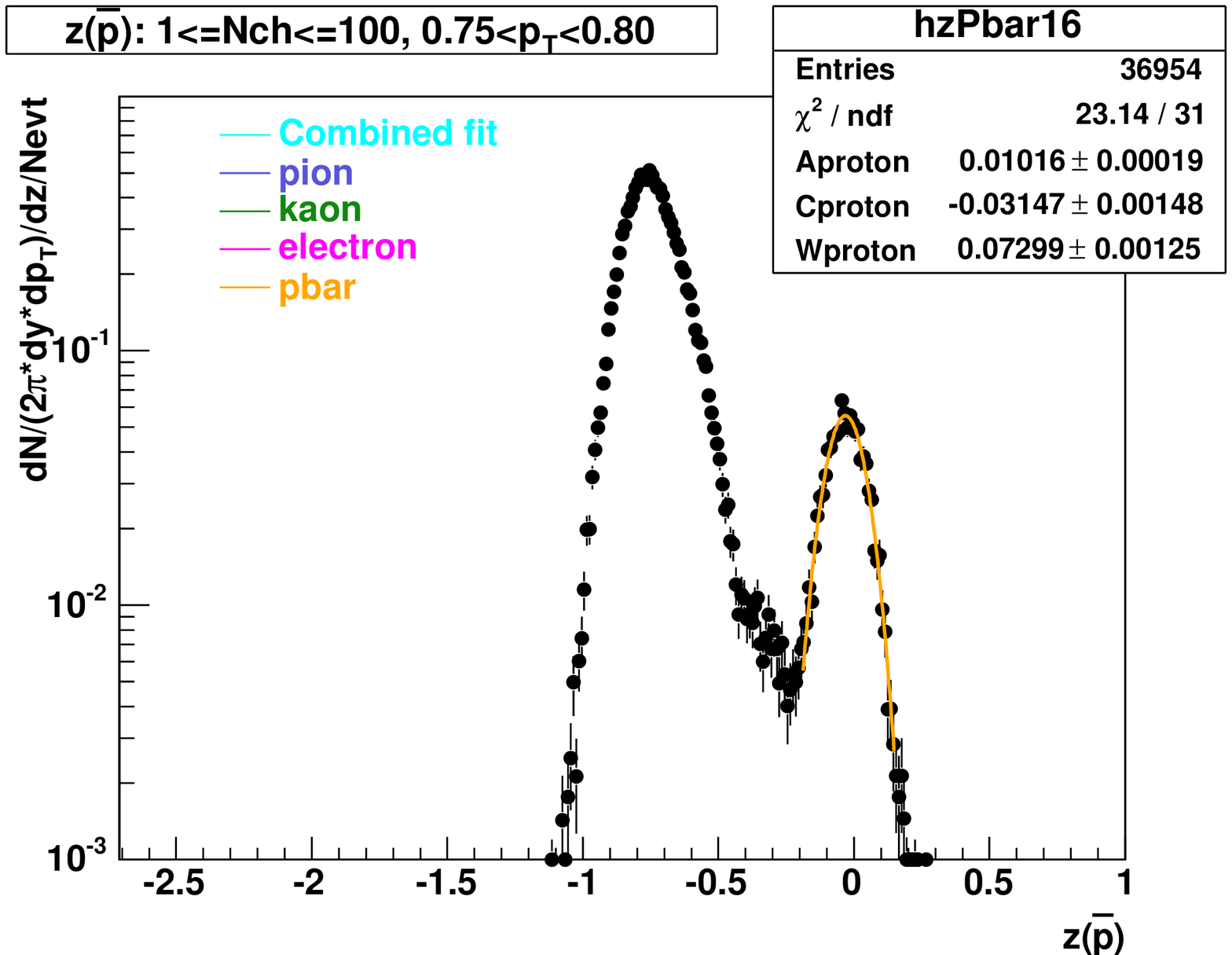}}
\resizebox{.225\textwidth}{!}{\includegraphics{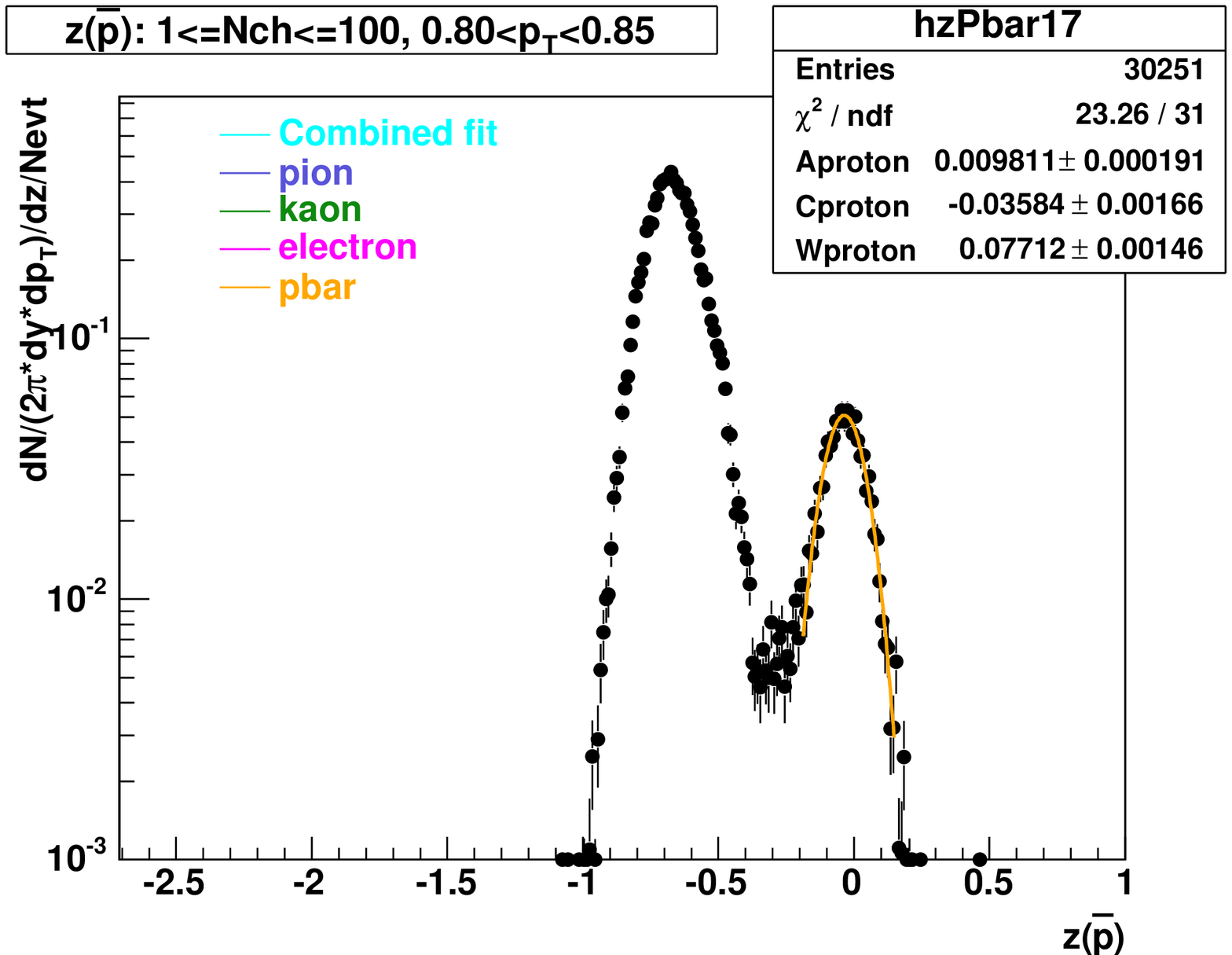}}
\resizebox{.225\textwidth}{!}{\includegraphics{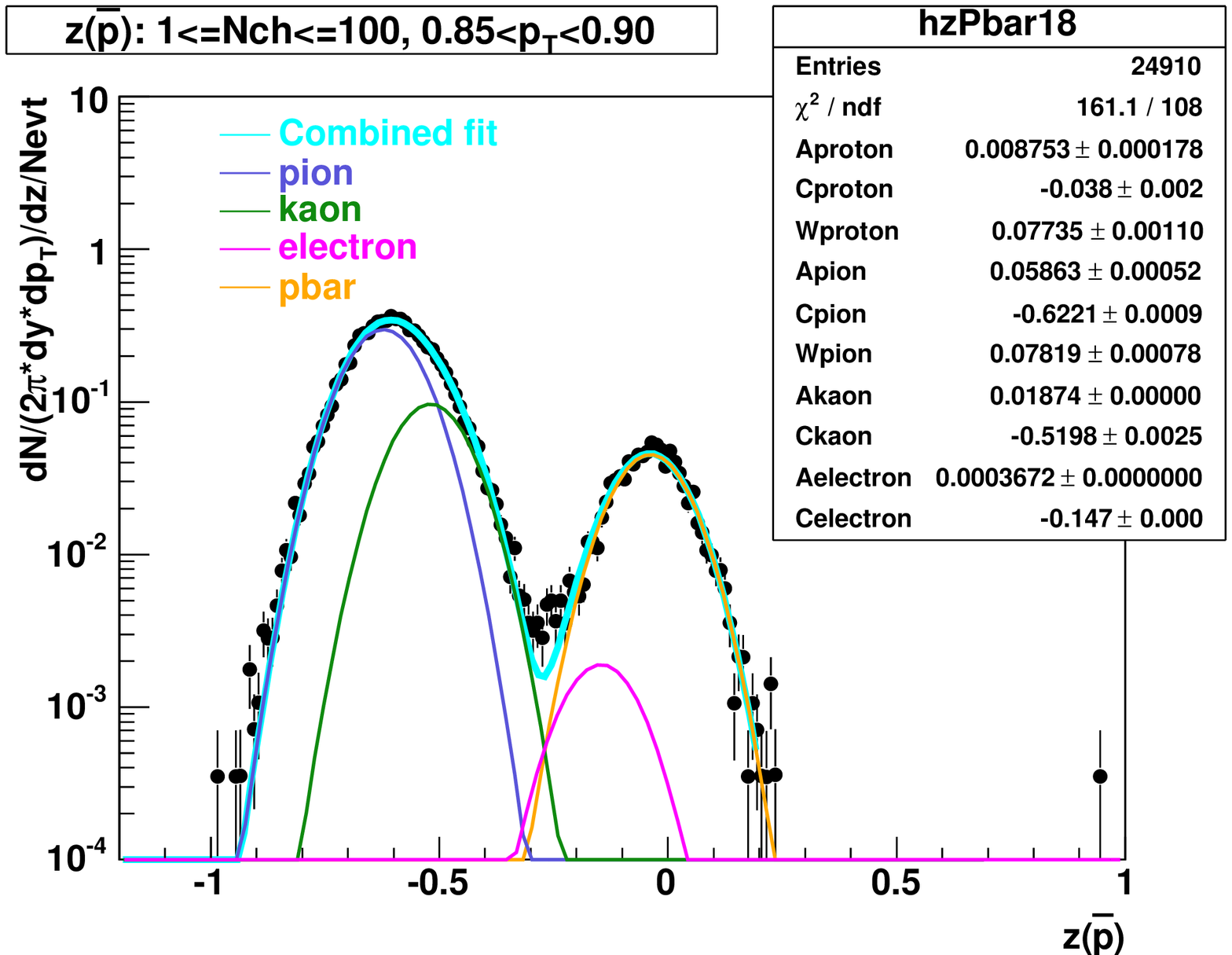}}
\resizebox{.225\textwidth}{!}{\includegraphics{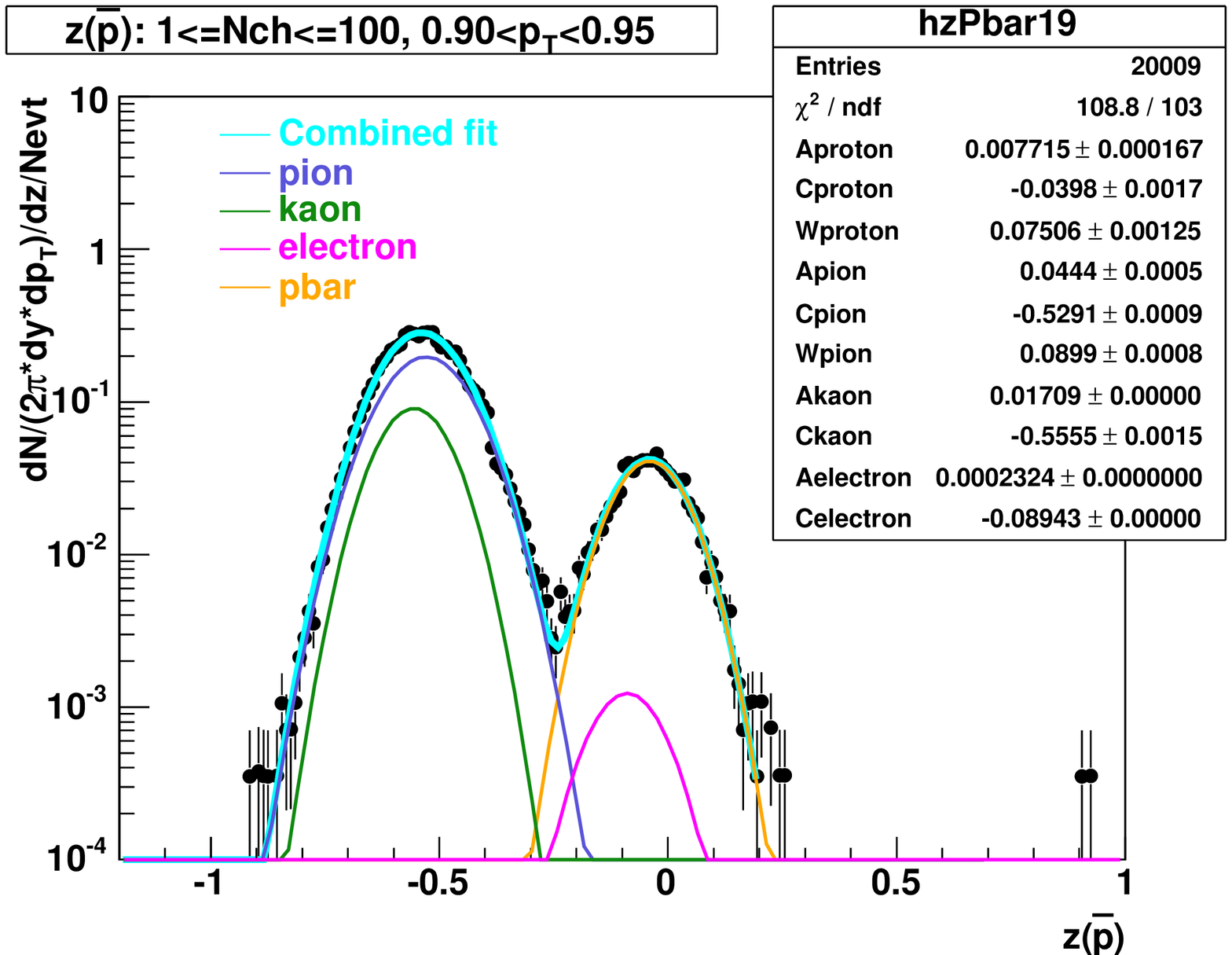}}\\
\resizebox{.225\textwidth}{!}{\includegraphics{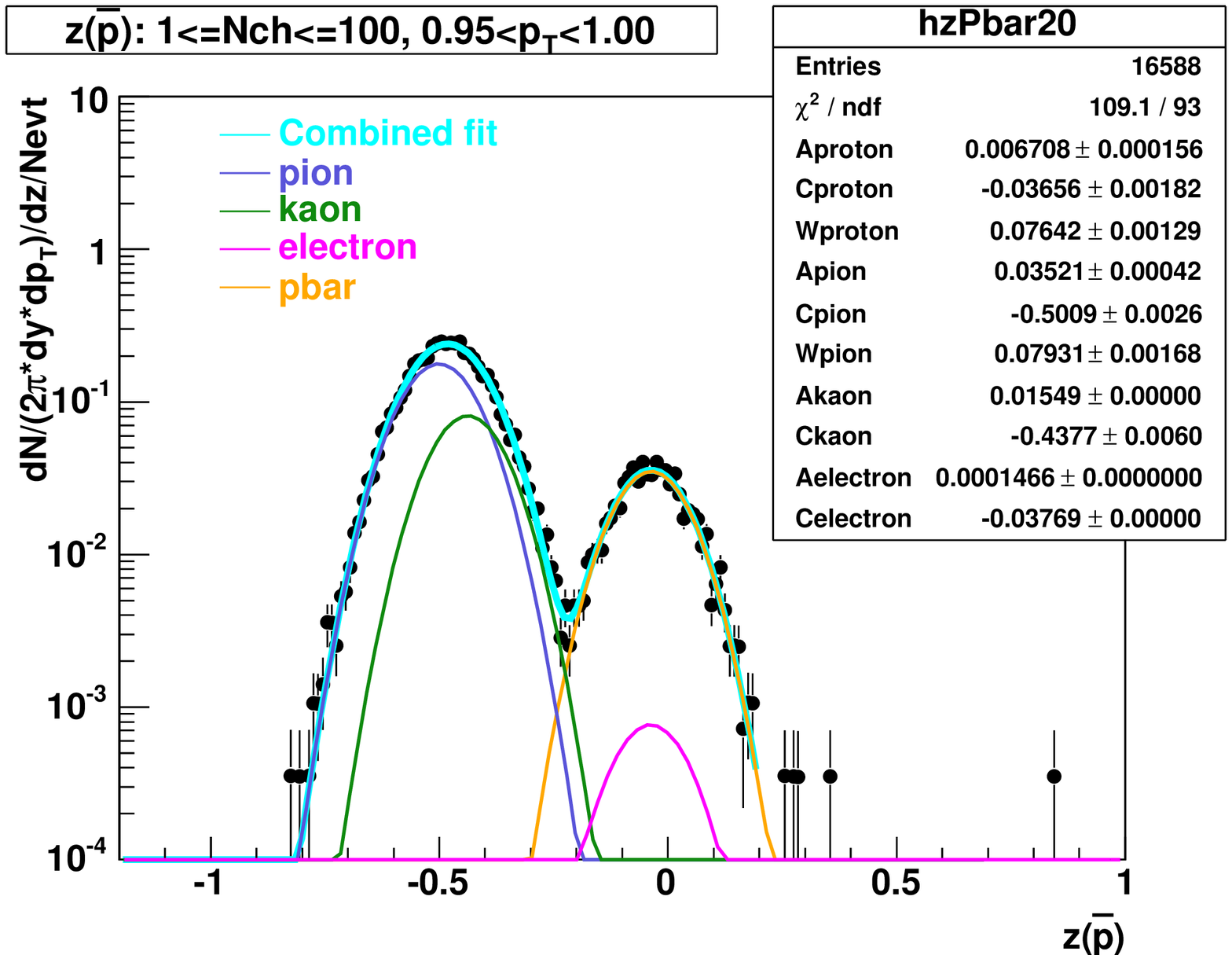}}
\resizebox{.225\textwidth}{!}{\includegraphics{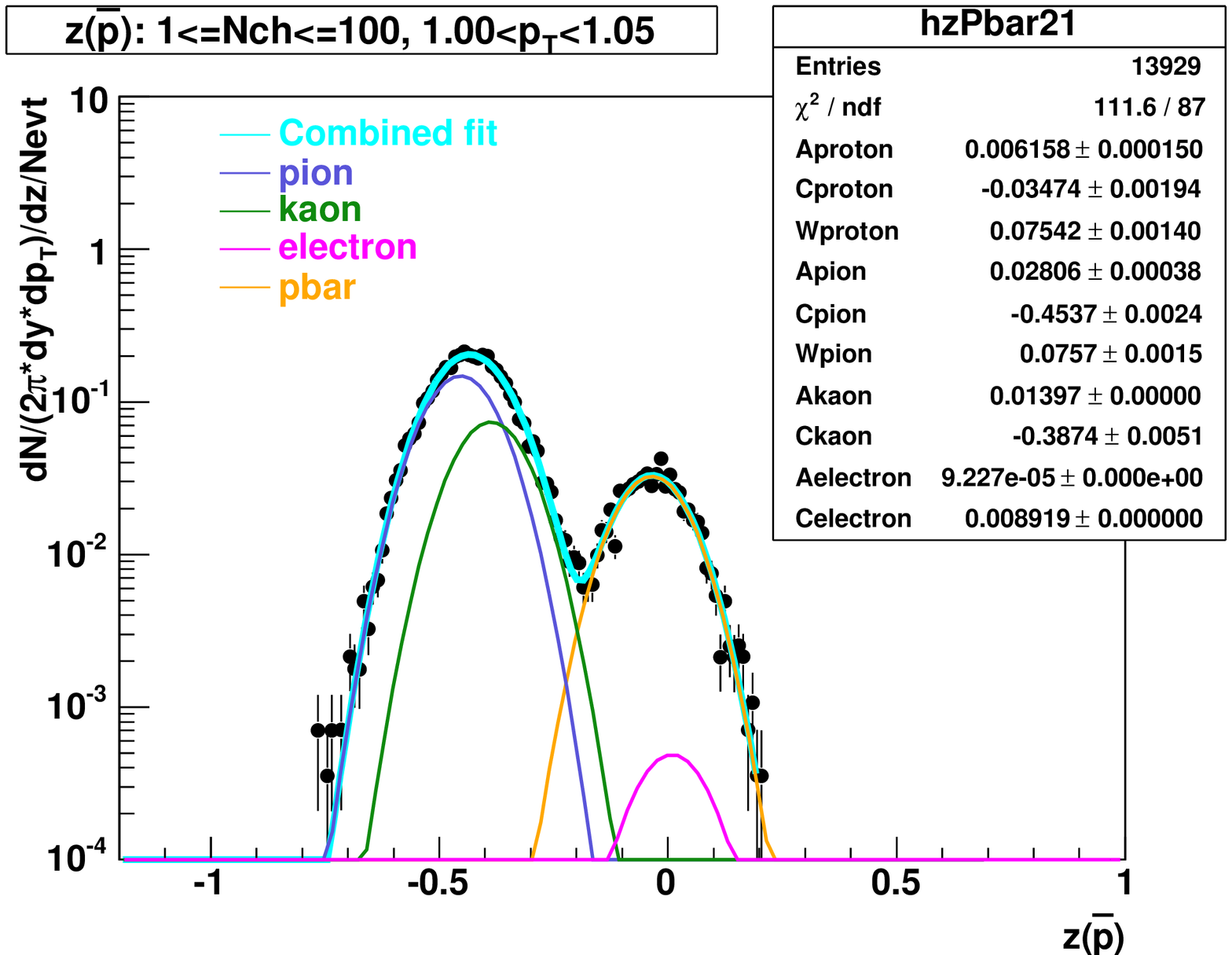}}
\resizebox{.225\textwidth}{!}{\includegraphics{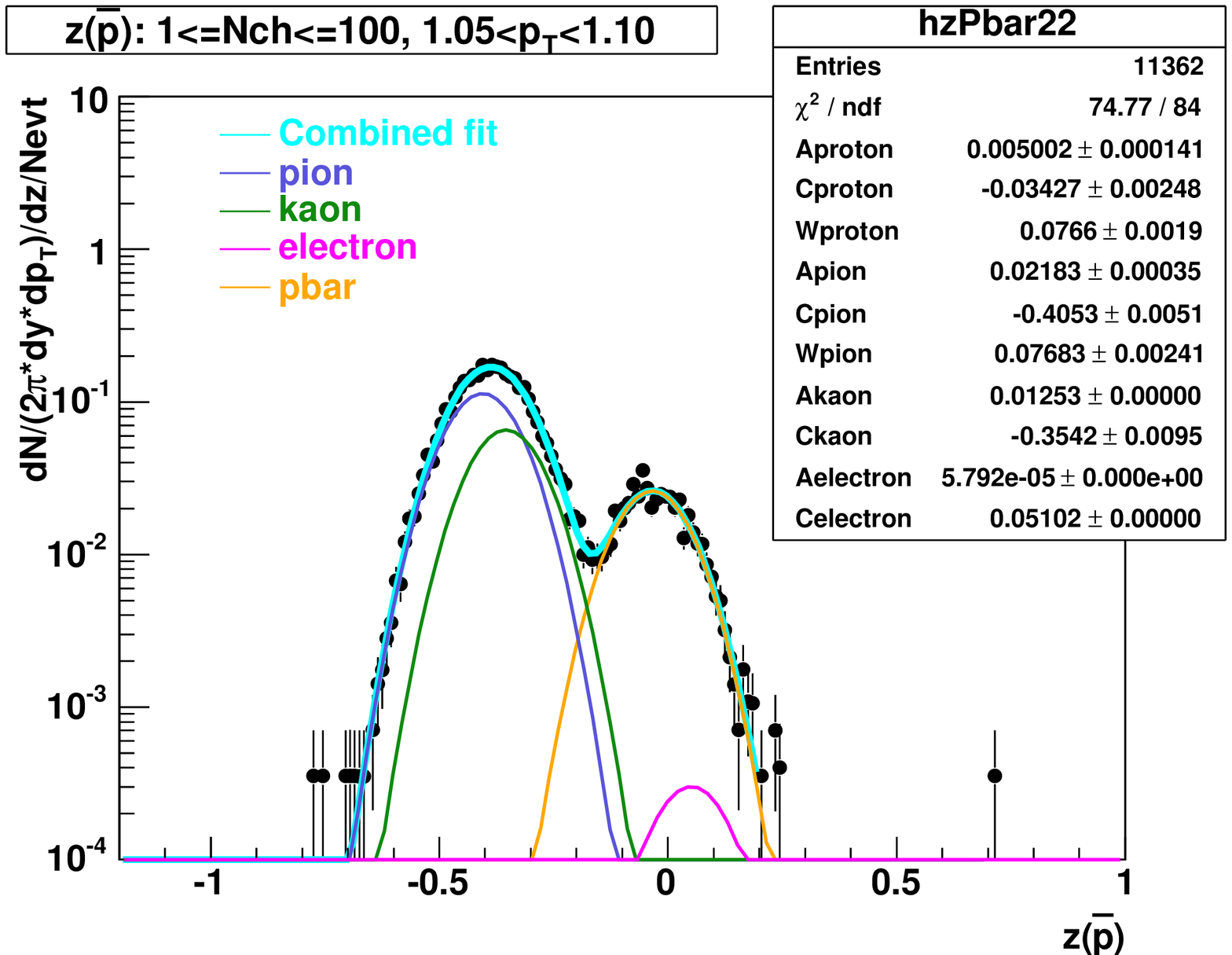}}

	 \caption{Gaussian fits to the $z$ distribution of antiprotons in 200 GeV pp collisions.\label{fig:ppGaussianFitsPbar}}
\end{center} 
\end{sidewaysfigure}
%

%

\chapter{Tables}
\begin{sidewaystable}{}
\begin{center} 
\begin{scriptsize}
\caption{Extrapolated average transverse momenta, $\meanpt$ in $GeV/c$, of identified particles for various collision systems and centralities. Errors are the quadratic sum of statistical and systematic errors, but dominated by systematic errors.}
\begin{tabular}{cc|c|c|c|c|c|c} 
\hline \hline
System & Centrality & $\pi^{-}$ & $\pi^{+}$ & $K^{-}$ & $K^{+}$ &  $\overline{p}$ & $p$  \\ \hline  
  & min-bias & 0.35 $\pm$ 0.02 & 0.35 $\pm$ 0.02  &  0.52 $\pm$ 0.03  &  0.52 $\pm$ 0.03 &  0.68 $\pm$ 0.03  &  0.68 $\pm$ 0.03  \\  
  & bin 1 & 0.36 $\pm$ 0.02 & 0.36 $\pm$ 0.02  &  0.45 $\pm$ 0.02  &  0.45 $\pm$ 0.02 &  0.53 $\pm$ 0.03  &  0.52 $\pm$ 0.03  \\  
 pp & bin 2 & 0.35 $\pm$ 0.02 & 0.35 $\pm$ 0.02  &  0.50 $\pm$ 0.03  &  0.50 $\pm$ 0.03 &  0.63 $\pm$ 0.04  &  0.64 $\pm$ 0.04  \\  
 200 GeV & bin 3 & 0.35 $\pm$ 0.02 & 0.35 $\pm$ 0.02  &  0.52 $\pm$ 0.03  &  0.52 $\pm$ 0.03 &  0.67 $\pm$ 0.03  &  0.67 $\pm$ 0.03  \\  
  & bin 4 & 0.35 $\pm$ 0.02 & 0.35 $\pm$ 0.02  &  0.53 $\pm$ 0.03  &  0.53 $\pm$ 0.03 &  0.70 $\pm$ 0.04  &  0.70 $\pm$ 0.04  \\  
  & bin 5 & 0.36 $\pm$ 0.02 & 0.36 $\pm$ 0.02  &  0.55 $\pm$ 0.05  &  0.56 $\pm$ 0.05 &  0.76 $\pm$ 0.08  &  0.77 $\pm$ 0.08  \\  

\hline 
  & min-bias & 0.37 $\pm$ 0.02 & 0.37 $\pm$ 0.02  &  0.59 $\pm$ 0.03  &  0.59 $\pm$ 0.03 &  0.82 $\pm$ 0.04  &  0.82 $\pm$ 0.04  \\  
 d-Au & 40 - 100\% & 0.36 $\pm$ 0.02 & 0.36 $\pm$ 0.02  &  0.57 $\pm$ 0.03  &  0.57 $\pm$ 0.03 &  0.80 $\pm$ 0.04  &  0.80 $\pm$ 0.04  \\  
 200 GeV & 20 - 40\% & 0.36 $\pm$ 0.02 & 0.37 $\pm$ 0.02  &  0.60 $\pm$ 0.03  &  0.60 $\pm$ 0.03 &  0.85 $\pm$ 0.04  &  0.86 $\pm$ 0.04  \\ 
  & 0 - 20\% & 0.38 $\pm$ 0.02 & 0.38 $\pm$ 0.02  &  0.60 $\pm$ 0.03  &  0.60 $\pm$ 0.03 &  0.85 $\pm$ 0.04  &  0.85 $\pm$ 0.04  \\  
\hline 
  & 70 - 80\% & 0.36 $\pm$ 0.02  & 0.36 $\pm$ 0.02  &  0.54 $\pm$ 0.05  &  0.54 $\pm$ 0.05 & 0.72 $\pm$ 0.07  &  0.73 $\pm$ 0.07  \\  
  & 60 - 70\% & 0.37 $\pm$ 0.02  & 0.36 $\pm$ 0.02  &  0.54 $\pm$ 0.05  &  0.55 $\pm$ 0.05 & 0.74 $\pm$ 0.07  &  0.75 $\pm$ 0.07  \\  
  & 50 - 60\% & 0.38 $\pm$ 0.02  & 0.38 $\pm$ 0.02  &  0.56 $\pm$ 0.06  &  0.56 $\pm$ 0.06 & 0.77 $\pm$ 0.09  &  0.77 $\pm$ 0.09  \\  
 62.4 GeV & 40 - 50\% & 0.39 $\pm$ 0.02  & 0.38 $\pm$ 0.02  &  0.59 $\pm$ 0.05  &  0.59 $\pm$ 0.05 & 0.82 $\pm$ 0.07  &  0.83 $\pm$ 0.07  \\  
  & 30 - 40\% & 0.40 $\pm$ 0.02  & 0.39 $\pm$ 0.02  &  0.61 $\pm$ 0.05  &  0.61 $\pm$ 0.05 & 0.88 $\pm$ 0.07  &  0.88 $\pm$ 0.07  \\  
 Au - Au & 20 - 30\% & 0.40 $\pm$ 0.03  & 0.40 $\pm$ 0.03  &  0.63 $\pm$ 0.05  &  0.63 $\pm$ 0.05 & 0.92 $\pm$ 0.08  &  0.92 $\pm$ 0.08  \\  
  & 10 - 20\% & 0.40 $\pm$ 0.03  & 0.40 $\pm$ 0.03  &  0.64 $\pm$ 0.06  &  0.64 $\pm$ 0.06 & 0.94 $\pm$ 0.09  &  0.95 $\pm$ 0.08  \\  
  & 5 - 10\% & 0.40 $\pm$ 0.04  & 0.41 $\pm$ 0.04  &  0.65 $\pm$ 0.06  &  0.65 $\pm$ 0.06 & 0.96 $\pm$ 0.09  &  0.96 $\pm$ 0.09  \\  
  & 0 - 5\% & 0.40 $\pm$ 0.04  & 0.41 $\pm$ 0.04  &  0.65 $\pm$ 0.07  &  0.65 $\pm$ 0.07 & 0.97 $\pm$ 0.11  &  0.97 $\pm$ 0.10  \\

\hline \hline
\end{tabular}  
\end{scriptsize}
\end{center}  
\label{tab:meanpt}
\end{sidewaystable}


\begin{sidewaystable}{}
\begin{center} 
\begin{scriptsize}
\caption{Integrated multiplicity density, $dN/dy$, of identified particles for various collision systems and centralities. Errors are the quadratic sum of statistical and systematic errors, but dominated by systematic errors.}
\begin{tabular}{cc|c|c|c|c|c|c} 
\hline \hline
System & Centrality & $\pi^{-}$ & $\pi^{+}$ & $K^{-}$ & $K^{+}$ &  $\overline{p}$ & $p$  \\ \hline  
     & min-bias & 1.42 $\pm$ 0.12 & 1.44 $\pm$ 0.13  &  0.15 $\pm$ 0.01  &  0.15 $\pm$ 0.01 &  0.10 $\pm$ 0.01  &  0.12 $\pm$ 0.01  \\  
        & bin 1 & 0.33 $\pm$ 0.03 & 0.35 $\pm$ 0.03  &  0.030 $\pm$ 0.01  &  0.03 $\pm$ 0.01 &  0.02 $\pm$ 0.01  &  0.03 $\pm$ 0.02  \\  
 pp & bin 2 & 1.29 $\pm$ 0.11 & 1.31 $\pm$ 0.11  &  0.13 $\pm$ 0.01  &  0.13 $\pm$ 0.01 &  0.09 $\pm$ 0.01  &  0.10 $\pm$ 0.01  \\  
 200 GeV & bin 3 & 2.24 $\pm$ 0.20 & 2.28 $\pm$ 0.20  &  0.231 $\pm$ 0.02  &  0.24 $\pm$ 0.02 &  0.16 $\pm$ 0.01  &  0.18 $\pm$ 0.02  \\  
        & bin 4 & 3.12 $\pm$ 0.27 & 3.15 $\pm$ 0.28  &  0.32 $\pm$ 0.03  &  0.32 $\pm$ 0.03 &  0.22 $\pm$ 0.02  &  0.25 $\pm$ 0.02  \\  
        & bin 5 & 4.31 $\pm$ 0.38 & 4.31 $\pm$ 0.38  &  0.44 $\pm$ 0.04  &  0.50 $\pm$ 0.05 &  0.32 $\pm$ 0.03  &  0.36 $\pm$ 0.04  \\  

\hline 
 & min-bias & 4.64 $\pm$ 0.41 & 4.65 $\pm$ 0.41  &  0.58 $\pm$ 0.05  &  0.59 $\pm$ 0.05 &  0.42 $\pm$ 0.04  &  0.46 $\pm$ 0.04  \\  
 d-Au & 40 - 100\% & 2.89 $\pm$ 0.26 & 2.87 $\pm$ 0.27  &  0.34 $\pm$ 0.03  &  0.35 $\pm$ 0.03 &  0.23 $\pm$ 0.02  &  0.26 $\pm$ 0.02 \\ 
 200 GeV & 20 - 40\% & 6.06 $\pm$ 0.54 & 6.01 $\pm$ 0.54  &  0.76 $\pm$ 0.07  &  0.78 $\pm$ 0.07 &  0.55 $\pm$ 0.05  &  0.61 $\pm$ 0.05 \\
  & 0 - 20\% & 8.43 $\pm$ 0.75 & 8.49 $\pm$ 0.76  &  1.08 $\pm$ 0.09  &  1.11 $\pm$ 0.09 &  0.78 $\pm$ 0.07  &  0.84 $\pm$ 0.07  \\     
\hline 
           & 70 - 80\% & 7.43 $\pm$ 0.66  & 7.35 $\pm$ 0.65  &  0.82 $\pm$ 0.08  &  0.88 $\pm$ 0.09 & 0.48 $\pm$ 0.05  &  0.72 $\pm$ 0.07  \\  
          & 60 - 70\% & 14.7 $\pm$ 1.3  & 14.85 $\pm$ 1.3  &  1.75 $\pm$ 0.17  &  1.96 $\pm$ 0.20 & 0.98 $\pm$ 0.09  &  1.55 $\pm$ 0.14  \\  
           & 50 - 60\% & 26.8 $\pm$ 2.4  & 26.6 $\pm$ 2.4  &  3.30 $\pm$ 0.37  &  3.64 $\pm$ 0.41 & 1.68 $\pm$ 0.16  &  2.89 $\pm$ 0.26  \\  
 62.4 GeV & 40 - 50\% & 43.8 $\pm$ 4.1  & 43.2 $\pm$ 4.0  &  5.69 $\pm$ 0.57  &  6.64 $\pm$ 0.66 & 2.81 $\pm$ 0.27  &  4.93 $\pm$ 0.46  \\  
           & 30 - 40\% & 67.5 $\pm$ 6.4  & 66.6 $\pm$ 6.4  &  8.92 $\pm$ 0.90  &  10.4 $\pm$ 1.1 & 4.33 $\pm$ 0.44  &  7.88 $\pm$ 0.76  \\  
           
 Au - Au & 20 - 30\% & 101 $\pm$ 10 & 98.9 $\pm$ 9.9  &  14.1 $\pm$ 1.5  &  15.9 $\pm$ 1.67 & 6.43 $\pm$ 0.70  &  11.9 $\pm$ 1.2  \\
 
          & 10 - 20\% & 145 $\pm$ 15  & 143 $\pm$ 15  &  19.9 $\pm$ 2.2  &  23.2 $\pm$ 2.6 & 8.99 $\pm$ 1.04  &  17.6 $\pm$ 1.9  \\  
  
        & 5 - 10\% & 192 $\pm$ 22  & 191 $\pm$ 22  &  27.2 $\pm$ 3.2  &  31.3 $\pm$ 3.69 & 11.5 $\pm$ 1.40  &  23.2 $\pm$ 2.76  \\  

          & 0 - 5\% & 236 $\pm$ 29  & 232 $\pm$ 29  &  32.4 $\pm$ 4.2  &  37.7 $\pm$ 4.8 & 13.6 $\pm$ 1.8  &  28.1 $\pm$ 3.6   \\  
\hline \hline
\end{tabular} 
\end{scriptsize} 
\end{center} 
\label{tab:yields}
\end{sidewaystable}


\begin{sidewaystable}{}
\begin{center} 
\begin{scriptsize}
\caption{Chemical and kinetic freeze-out properties in 200 GeV pp, dAu and 62.4 GeV, 200 GeV  Au-Au collisions. Errors are statistical. See text for systematic error estimates.}
\begin{tabular}{cc|ccccc|cccc} 
\hline \hline
System & Centrality & $\Tch$ (MeV) & $\mu_B$ (MeV) & $\mu_S$ (MeV) & $\gamma_S$ & $\chi^{2}/NDF$ & $\Tkin$ (MeV) & $\beta$ & $n$ & $\chi^{2}/NDF$ \\ \hline 
& Minimum Bias & 152.5 $\pm$ 3.5 & 10.0 $\pm$ 3.9 & 0.8 $\pm$ 2.5 & 0.59 $\pm$ 0.05 &  0.48 &  134.5 $\pm$ 4.8 &  0.22 $\pm$ 0.03 & 4.44 $\pm$ 0.78 & 1.23  \\

 & Bin1 & 151.6 $\pm$ 3.5 & 10.9 $\pm$ 4.0 & -5.3 $\pm$ 2.5 & 0.51 $\pm$ 0.04 &  2.48 &  139.8 $\pm$ 1.5 &  0.00 $\pm$ 0.08 & 15.00 $\pm$ 7.47 & 8.51  \\
pp & Bin2 & 149.9 $\pm$ 3.5 & 8.9 $\pm$ 3.8 & 1.3 $\pm$ 2.5 & 0.56 $\pm$ 0.05 &  0.53 &  140.2 $\pm$ 4.4 &  0.16 $\pm$ 0.02 & 7.00 $\pm$ 0.79 & 1.84  \\ 

200 GeV & Bin3 & 150.9 $\pm$ 3.4 & 7.5 $\pm$ 3.9 & -0.1 $\pm$ 2.5 & 0.58 $\pm$ 0.05 &  0.70 &  135.2 $\pm$ 4.2 &  0.29 $\pm$ 0.04 & 6.94 $\pm$ 1.54 & 1.07  \\

 & Bin4 & 151.2 $\pm$ 3.4 & 7.6 $\pm$ 3.9 & 1.5 $\pm$ 2.5 & 0.56 $\pm$ 0.05 &  0.27 &  131.7 $\pm$ 8.2 &  0.27 $\pm$ 0.06 & 3.01 $\pm$ 1.08 & 1.10  \\
 
 & Bin5 & 152.6 $\pm$ 4.5 & 11.3 $\pm$ 4.0 & -6.1 $\pm$ 2.5 & 0.60 $\pm$ 0.06 &  0.00 &  121.8 $\pm$ 7.6 &  0.35 $\pm$ 0.04 & 2.08 $\pm$ 0.43 & 1.51  \\ 

\hline
 
 & Minimum Bias & 160.4 $\pm$ 4.3 & 8.4 $\pm$ 4.2 & 0.2 $\pm$ 2.7 & 0.69 $\pm$ 0.06 &  0.02 &  122.2 $\pm$ 7.6 &  0.39 $\pm$ 0.02 & 1.86 $\pm$ 0.17 & 1.92  \\

dAu & 40-100\% & 155.5 $\pm$ 3.8 & 11.7 $\pm$ 4.1 & 1.0 $\pm$ 2.6 & 0.66 $\pm$ 0.05 &  0.17 &  115.6 $\pm$ 7.5 &  0.38 $\pm$ 0.02 & 2.01 $\pm$ 0.16 & 2.11  \\ 

200 GeV & 20-40\% & 160.9 $\pm$ 4.4 & 10.0 $\pm$ 4.2 & 0.3 $\pm$ 2.7 & 0.69 $\pm$ 0.06 &  0.13 &  112.1 $\pm$ 7.4 &  0.43 $\pm$ 0.02 & 1.34 $\pm$ 0.13 & 1.10  \\  

& 0-20\% & 160.7 $\pm$ 4.4 & 6.9 $\pm$ 4.2 & 0.0 $\pm$ 2.7 & 0.71 $\pm$ 0.06 &  0.14 &  115.9 $\pm$ 7.6 &  0.44 $\pm$ 0.02 & 1.23 $\pm$ 0.14 & 1.84  \\ 
\hline

 & 70-80\% & 152.5 $\pm$ 4.1 & 33.0 $\pm$ 4.3 & 5.2 $\pm$ 2.6 & 0.63 $\pm$ 0.06 &  0.34 &  134.3 $\pm$ 4.9 &  0.31 $\pm$ 0.07 & 1.08 $\pm$ 0.76 & 0.95  \\ 

 & 60-70\% & 154.2 $\pm$ 4.0 & 38.3 $\pm$ 4.4 & 2.8 $\pm$ 2.5 & 0.70 $\pm$ 0.07 &  0.25 &  131.2 $\pm$ 4.9 &  0.37 $\pm$ 0.06 & 0.60 $\pm$ 0.47 & 0.77  \\ 

 & 50-60\% & 153.0 $\pm$ 4.1 & 44.0 $\pm$ 4.5 & 6.0 $\pm$ 2.5 & 0.72 $\pm$ 0.08 &  0.30 &  129.5 $\pm$ 3.6 &  0.41 $\pm$ 0.01 & 0.20 $\pm$ 1.46 & 0.57  \\ 
 
Au - Au & 40-50\% & 153.8 $\pm$ 4.4 & 48.1 $\pm$ 4.6 & 2.4 $\pm$ 2.5 & 0.78 $\pm$ 0.08 &  0.36 &  121.9 $\pm$ 4.8 &  0.44 $\pm$ 0.03 & 0.76 $\pm$ 0.18 & 0.57  \\ 

 & 30-40\% & 154.4 $\pm$ 4.6 & 51.3 $\pm$ 4.7 & 3.0 $\pm$ 2.5 & 0.80 $\pm$ 0.08 &  0.45 &  116.9 $\pm$ 4.6 &  0.47 $\pm$ 0.02 & 0.69 $\pm$ 0.13 & 0.80  \\ 

62 GeV & 20-30\% & 154.0 $\pm$ 5.0 & 51.9 $\pm$ 4.8 & 6.0 $\pm$ 2.5 & 0.83 $\pm$ 0.09 &  1.22 &  113.0 $\pm$ 4.5 &  0.49 $\pm$ 0.02 & 0.78 $\pm$ 0.09 & 1.54  \\ 

 & 10-20\% & 153.9 $\pm$ 5.3 & 56.7 $\pm$ 4.9 & 5.6 $\pm$ 2.5 & 0.83 $\pm$ 0.09 &  0.44 &  108.7 $\pm$ 4.5 &  0.51 $\pm$ 0.02 & 0.80 $\pm$ 0.07
 & 2.08  \\ 
 & 5-10\% & 152.9 $\pm$ 5.4 & 58.5 $\pm$ 5.0 & 7.0 $\pm$ 2.5 & 0.85 $\pm$ 0.10 &  0.11 &  102.4 $\pm$ 4.3 &  0.53 $\pm$ 0.01 & 0.84 $\pm$ 0.05 & 1.79  \\ 

 & 0-5\% & 151.9 $\pm$ 5.8 & 59.9 $\pm$ 5.0 & 6.6 $\pm$ 2.5 & 0.83 $\pm$ 0.11 &  0.82 &  98.3 $\pm$ 4.2 &  0.54 $\pm$ 0.01 & 0.82 $\pm$ 0.04 & 2.19  \\ 

 \hline 
\hline
\end{tabular}  
\end{scriptsize}
\end{center}  
\label{tab:freezeout_prop}
\end{sidewaystable}


\begin{sidewaystable}{}
\centering 
\begin{scriptsize}
\caption{Centralities in $pp$ and d-Au collisions at 200 GeV and in Au-Au collisions at 62.4 GeV, 130 GeV, and 200 GeV. The uncorrected charged particle multiplicity $\dNraw$ for d-Au is measured in the FTPC within $-3.8<\eta<-2.8$, and for all other systems in the TPC within $|\eta|<0.5$. The corrected charged particle multiplicity $\left\la\dNdeta\right\ra$ are from the TPC within $|\eta|<0.5$ for all collision systems.}
\begin{tabular}{cccccc|cccc}
\hline \hline

Collision & Centrality & $\dNraw$ &  $\left\la \dNraw \right\ra$ & $\left\la \dNdeta \right\ra$ & $\left\la \dNdy \right\ra$& $\left\la\Npart^{\rm proj}\right\ra$ & $\left\la\Npart\right\ra$ & $\left\la\Ncoll\right\ra$ & $\left\la S \right\ra$ (fm$^2$) \\ 
\hline 
				& min-bias  &  -     	   & 2.4 & 2.98$\pm$0.34 & 3.41$\pm$0.18  & & 2 & 1 & 4.1$\pm$0.1 \\ 
 				& bin 1     & 0-1   	   & 0.6 & 0.71$\pm$0.55 & 0.79$\pm$0.04  & & & & \\ 
pp 			& bin 2     & 2-3   	   & 2.4 & 2.68$\pm$0.48 & 3.04$\pm$0.13  & & & & \\ 
200 GeV & bin 3 		& 4-5   	   & 4.4 & 4.86$\pm$0.39 & 5.26$\pm$0.23  & & & & \\ 
 				& bin 4     & 6-7   	   & 6.4 & 7.02$\pm$0.48 & 7.32$\pm$0.33  & & & & \\ 
 				& bin 5     & $\geq 8$	 & 9.6 & 10.6$\pm$0.9  & 10.2$\pm$0.4   & & & & \\ 
\hline 
 	      & min-bias &  -	      & 10.2 & 10.2$\pm$0.68 & 11.4$\pm$0.6   & 1.62$\pm$0.02 & 8.31$\pm$0.37   &  7.51$\pm$0.39 &  6.7$\pm$0.4 \\  
d-Au 	  & 40-100\% & 0-9   	  &  6.2 & 6.23$\pm$0.34 & 6.96$\pm$0.37  & 1.46$\pm$0.02 & 5.14$\pm$0.44   &  4.21$\pm$0.79 &  6.7$\pm$0.2 \\  
200 GeV & 20-40\%  & 10-16 	  & 12.6 & 14.1$\pm$1.0  & 14.8$\pm$0.8 & 1.85$\pm$0.03 & 11.2$\pm$1.1      & 10.6$\pm$0.8 & 11.2$\pm$0.9 \\  
      	& 0-20\%   & $\geq$17	& 17.6 & 19.9$\pm$1.6  & 20.8$\pm$1.0 & 1.96$\pm$0.01 & 15.7$\pm$1.2      & 15.1$\pm$1.1 & 15.1$\pm$1.3 \\  
\hline 
    	    & 70-80\%  &  9-19    	&  12.4 & 13.9$\pm$1.1 		&18.1$\pm$1.0 && 12.9$\pm$0.4  & 11.4$\pm$0.7 &  14.0$\pm$1.0 \\
 	        & 60-70\%  &  20-37   	&  26.8 & 29.1$\pm$2.2 		&36.1$\pm$1.9 && 25.3$\pm$0.8  & 26.7$\pm$1.8 &  22.8$\pm$1.4 \\
 	        & 50-60\%  &  38-64   	&  49.1 & 53.1$\pm$4.2 	  &65.5$\pm$3.5  && 44.8$\pm$1.1  & 56.2$\pm$3.8 &  33.6$\pm$1.8 \\
Au-Au	    & 40-50\%  & 65-101   	&  81.0 & 87.2$\pm$7.1  	&108$\pm$6  && 72.4$\pm$1.9  & 107$\pm$ 8   &  46.3$\pm$2.2 \\
    	    & 30-40\%  & 102-153  	& 125.2 & 135$\pm$11  &167$\pm$9  && 110.4$\pm$2.1 & 192$\pm$15   &  60.9$\pm$2.7 \\
62.4 GeV  & 20-30\%  & 154-221  	& 184.8 & 202$\pm$17  &249$\pm$14 && 160.3$\pm$2.4 & 318$\pm$25   &  78.6$\pm$3.3 \\
 	        & 10-20\%  & 222-312  	& 263.6 & 292$\pm$25  &361$\pm$22 && 226.4$\pm$3.2 & 508$\pm$41   & 101.0$\pm$4.1 \\
 	        &  5-10\%  & 313-372  	& 340.5 & 385$\pm$33  &482$\pm$32 && 291.0$\pm$3.6 & 714$\pm$59   & 123.1$\pm$4.8 \\
 	        &  0-5\%   & $\geq$373	& 411.8 & 472$\pm$41  &588$\pm$42 && 346.7$\pm$2.9 & 914$\pm$79   & 143.8$\pm$5.2 \\
\hline \hline
\end{tabular}  
\label{tab:auau_coll_prop}
\end{scriptsize}
\end{sidewaystable}

\begin{table}
\begin{center} 
\begin{scriptsize}
\caption{Measured and extrapolated contributions to the total yield for negatively charged particles in 200 GeV pp, dAu and 62.4 GeV Au-Au collisions. Bose-Einstein fit is used for pions and Blast-wave fit is used for kaons and antiprotons.}
\begin{tabular}{c|ccc}
\hline
	& measured & \multicolumn{2}{c}{extrapolated $dN/dy$} \\
system	& $dN/dy$  & low $\pt$ & high $\pt$ \\ \hline
	& \multicolumn{3}{c}{$\pi$, measured $\pt$ range: 0.225-0.775 GeV/c} \\
pp MB 												& 61.6\% & 32.4\% & 6.0\% \\
dAu MB 												& 58.3\% & 30.6\% & 11.1\% \\
Au-Au 80-70\% 								& 58.4\% & 31.6\% & 10.0\% \\
Au-Au 40-30\% 								& 58.0\% & 28.3\% & 13.7\% \\
Au-Au 5-0\%										& 57.8\% & 27.7\% & 14.5\% \\ \hline
	& \multicolumn{3}{c}{$K$, measured $\pt$ range: 0.225-0.725 GeV/c} \\
pp MB 											& 66.7\% &14.8\% & 18.5\%\\
dAu MB											& 60.5\% & 12.3\% & 27.2\% \\		
Au-Au 80-70\%							 	& 58.9\% & 21.1\% & 20.0\% \\
Au-Au 40-30\% 							& 56.8\%& 15.5\%& 27.7\% \\
Au-Au 5-0\%									& 54.2\% & 12.8\% & 33.0\% \\ \hline
	& \multicolumn{3}{c}{$\pbar$, measured $\pt$ range: 0.425-1.25 GeV/c} \\
pp MB 											& 58.5\% & 12.7\% & 28.8\%  \\
dAu MB 											& 68.3\% & 17.2\% & 24.5\% \\
Au-Au 80-70\% 							& 67.2\% & 11.5\% & 21.3\% \\
Au-Au 40-30\% 							& 65.0\% & 12.8\% & 22.2\%   \\
Au-Au 5-0\%									& 59.2\% & 9.4\% & 31.4\% \\ \hline
\end{tabular}
\end{scriptsize}
\end{center} 
\label{tab:yieldsextrapolation}
\end{table}

\clearpage

%
\begin{table}[!h] 
\begin{scriptsize} 
\begin{center} 
\caption{Invariant yields for $\pi^{-}$ and $\pi^{+}$ at mid-rapidity in {\bf Minimum Bias} ${\bf pp}$ collisions at 200 GeV.} \label{tab:Tables/ppPionSpectrummb.tex} 
\begin{tabular}{|c|c|c|c|} 
\hline   
 $p_{T}$ (GeV/c) &  $\Delta p_{T}$ (GeV/c) & $\pi^{-}$  & $\pi^{+}$  \\ \hline 
2.25e-01 &  5.00e-02 &  2.03e+00  $\pm$ 7.08e-03 $\pm$ 2.03e-02 & 2.07e+00 $\pm$ 6.21e-03 $\pm$ 2.07e-02 \\ 
2.75e-01 &  5.00e-02 &  1.52e+00  $\pm$ 4.72e-03 $\pm$ 1.52e-02 & 1.54e+00 $\pm$ 4.65e-03 $\pm$ 1.54e-02 \\ 
3.25e-01 &  5.00e-02 &  1.14e+00  $\pm$ 3.69e-03 $\pm$ 1.14e-02 & 1.14e+00 $\pm$ 3.67e-03 $\pm$ 1.14e-02 \\ 
3.75e-01 &  5.00e-02 &  8.43e-01  $\pm$ 2.94e-03 $\pm$ 8.43e-03 & 8.58e-01 $\pm$ 2.97e-03 $\pm$ 8.58e-03 \\ 
4.25e-01 &  5.00e-02 &  6.31e-01  $\pm$ 2.37e-03 $\pm$ 6.31e-03 & 6.37e-01 $\pm$ 2.40e-03 $\pm$ 6.37e-03 \\ 
4.75e-01 &  5.00e-02 &  4.70e-01  $\pm$ 1.94e-03 $\pm$ 4.70e-03 & 4.79e-01 $\pm$ 1.97e-03 $\pm$ 4.79e-03 \\ 
5.25e-01 &  5.00e-02 &  3.52e-01  $\pm$ 1.60e-03 $\pm$ 3.52e-03 & 3.60e-01 $\pm$ 1.62e-03 $\pm$ 3.60e-03 \\ 
5.75e-01 &  5.00e-02 &  2.66e-01  $\pm$ 1.33e-03 $\pm$ 5.31e-03 & 2.73e-01 $\pm$ 1.35e-03 $\pm$ 5.45e-03 \\ 
6.25e-01 &  5.00e-02 &  2.02e-01  $\pm$ 1.12e-03 $\pm$ 4.03e-03 & 2.06e-01 $\pm$ 1.11e-03 $\pm$ 4.13e-03 \\ 
6.75e-01 &  5.00e-02 &  1.54e-01  $\pm$ 9.70e-04 $\pm$ 4.61e-03 & 1.56e-01 $\pm$ 9.93e-04 $\pm$ 4.69e-03 \\ 
7.25e-01 &  5.00e-02 &  1.20e-01  $\pm$ 8.15e-04 $\pm$ 3.61e-03 & 1.19e-01 $\pm$ 9.46e-04 $\pm$ 3.57e-03 \\ 
7.75e-01 &  5.00e-02 &  9.80e-02  $\pm$ 7.12e-04 $\pm$ 2.94e-03 & 9.36e-02 $\pm$ 7.50e-04 $\pm$ 2.81e-03 \\ 
\hline  
\end{tabular}  
\end{center}  
\end{scriptsize}  
\end{table}  
\begin{table}[!h] 
\begin{scriptsize} 
\begin{center} 
\caption{Invariant yields for $\pi^{-}$ and $\pi^{+}$ at mid-rapidity for the multiplicity bin: $Nch_{0-2}$ in ${\bf pp}$ collisions at 200 GeV.} \label{tab:Tables/ppPionSpectrum02.tex} 
\begin{tabular}{|c|c|c|c|} 
\hline   
 $p_{T}$ (GeV/c) &  $\Delta p_{T}$ (GeV/c) & $\pi^{-}$  & $\pi^{+}$  \\ \hline 
2.25e-01 &  5.00e-02 &  4.59e-01  $\pm$ 5.07e-03 $\pm$ 4.59e-03 & 4.89e-01 $\pm$ 5.03e-03 $\pm$ 4.89e-03 \\ 
2.75e-01 &  5.00e-02 &  3.54e-01  $\pm$ 3.75e-03 $\pm$ 3.54e-03 & 3.60e-01 $\pm$ 3.73e-03 $\pm$ 3.60e-03 \\ 
3.25e-01 &  5.00e-02 &  2.64e-01  $\pm$ 2.95e-03 $\pm$ 2.64e-03 & 2.69e-01 $\pm$ 2.92e-03 $\pm$ 2.69e-03 \\ 
3.75e-01 &  5.00e-02 &  2.01e-01  $\pm$ 2.37e-03 $\pm$ 2.01e-03 & 2.09e-01 $\pm$ 2.42e-03 $\pm$ 2.09e-03 \\ 
4.25e-01 &  5.00e-02 &  1.53e-01  $\pm$ 1.93e-03 $\pm$ 1.53e-03 & 1.57e-01 $\pm$ 1.95e-03 $\pm$ 1.57e-03 \\ 
4.75e-01 &  5.00e-02 &  1.18e-01  $\pm$ 1.59e-03 $\pm$ 1.18e-03 & 1.20e-01 $\pm$ 1.61e-03 $\pm$ 1.20e-03 \\ 
5.25e-01 &  5.00e-02 &  8.83e-02  $\pm$ 1.31e-03 $\pm$ 8.83e-04 & 9.07e-02 $\pm$ 1.33e-03 $\pm$ 9.07e-04 \\ 
5.75e-01 &  5.00e-02 &  6.46e-02  $\pm$ 1.07e-03 $\pm$ 1.29e-03 & 6.76e-02 $\pm$ 1.10e-03 $\pm$ 1.35e-03 \\ 
6.25e-01 &  5.00e-02 &  4.90e-02  $\pm$ 9.10e-04 $\pm$ 9.80e-04 & 4.72e-02 $\pm$ 9.05e-04 $\pm$ 9.45e-04 \\ 
6.75e-01 &  5.00e-02 &  3.64e-02  $\pm$ 7.52e-04 $\pm$ 1.09e-03 & 3.87e-02 $\pm$ 8.33e-04 $\pm$ 1.16e-03 \\ 
7.25e-01 &  5.00e-02 &  2.68e-02  $\pm$ 6.24e-04 $\pm$ 8.03e-04 & 3.08e-02 $\pm$ 6.71e-04 $\pm$ 9.23e-04 \\ 
7.75e-01 &  5.00e-02 &  2.31e-02  $\pm$ 5.61e-04 $\pm$ 6.93e-04 & 2.21e-02 $\pm$ 5.83e-04 $\pm$ 6.63e-04 \\ 
\hline  
\end{tabular}  
\end{center}  
\end{scriptsize}  
\end{table}  
\begin{table}[!h] 
\begin{scriptsize} 
\begin{center} 
\caption{Invariant yields for $\pi^{-}$ and $\pi^{+}$ at mid-rapidity for the multiplicity bin: $Nch_{3-4}$ in ${\bf pp}$ collisions at 200 GeV.} \label{tab:Tables/ppPionSpectrum34.tex} 
\begin{tabular}{|c|c|c|c|} 
\hline   
 $p_{T}$ (GeV/c) &  $\Delta p_{T}$ (GeV/c) & $\pi^{-}$  & $\pi^{+}$  \\ \hline 
2.25e-01 &  5.00e-02 &  1.82e+00  $\pm$ 1.21e-02 $\pm$ 1.82e-02 & 1.85e+00 $\pm$ 1.01e-02 $\pm$ 1.85e-02 \\ 
2.75e-01 &  5.00e-02 &  1.38e+00  $\pm$ 7.75e-03 $\pm$ 1.38e-02 & 1.41e+00 $\pm$ 7.75e-03 $\pm$ 1.41e-02 \\ 
3.25e-01 &  5.00e-02 &  1.05e+00  $\pm$ 6.14e-03 $\pm$ 1.05e-02 & 1.06e+00 $\pm$ 6.13e-03 $\pm$ 1.06e-02 \\ 
3.75e-01 &  5.00e-02 &  7.76e-01  $\pm$ 4.89e-03 $\pm$ 7.76e-03 & 7.91e-01 $\pm$ 4.94e-03 $\pm$ 7.91e-03 \\ 
4.25e-01 &  5.00e-02 &  5.85e-01  $\pm$ 3.97e-03 $\pm$ 5.85e-03 & 5.90e-01 $\pm$ 3.99e-03 $\pm$ 5.90e-03 \\ 
4.75e-01 &  5.00e-02 &  4.33e-01  $\pm$ 3.23e-03 $\pm$ 4.33e-03 & 4.46e-01 $\pm$ 3.28e-03 $\pm$ 4.46e-03 \\ 
5.25e-01 &  5.00e-02 &  3.23e-01  $\pm$ 2.66e-03 $\pm$ 3.23e-03 & 3.29e-01 $\pm$ 2.70e-03 $\pm$ 3.29e-03 \\ 
5.75e-01 &  5.00e-02 &  2.43e-01  $\pm$ 2.20e-03 $\pm$ 4.85e-03 & 2.47e-01 $\pm$ 2.22e-03 $\pm$ 4.95e-03 \\ 
6.25e-01 &  5.00e-02 &  1.82e-01  $\pm$ 1.85e-03 $\pm$ 3.64e-03 & 1.88e-01 $\pm$ 1.88e-03 $\pm$ 3.75e-03 \\ 
6.75e-01 &  5.00e-02 &  1.38e-01  $\pm$ 1.58e-03 $\pm$ 4.15e-03 & 1.40e-01 $\pm$ 1.60e-03 $\pm$ 4.19e-03 \\ 
7.25e-01 &  5.00e-02 &  1.07e-01  $\pm$ 1.35e-03 $\pm$ 3.20e-03 & 1.06e-01 $\pm$ 1.51e-03 $\pm$ 3.19e-03 \\ 
7.75e-01 &  5.00e-02 &  8.58e-02  $\pm$ 1.16e-03 $\pm$ 2.57e-03 & 8.32e-02 $\pm$ 1.23e-03 $\pm$ 2.49e-03 \\ 
\hline  
\end{tabular}  
\end{center}  
\end{scriptsize}  
\end{table}  
\begin{table}[!h] 
\begin{scriptsize} 
\begin{center} 
\caption{Invariant yields for $\pi^{-}$ and $\pi^{+}$ at mid-rapidity for the multiplicity bin: $Nch_{5-6}$ in ${\bf pp}$ collisions at 200 GeV.} \label{tab:Tables/ppPionSpectrum56.tex} 
\begin{tabular}{|c|c|c|c|} 
\hline   
 $p_{T}$ (GeV/c) &  $\Delta p_{T}$ (GeV/c) & $\pi^{-}$  & $\pi^{+}$  \\ \hline 
2.25e-01 &  5.00e-02 &  3.25e+00  $\pm$ 2.11e-02 $\pm$ 3.25e-02 & 3.31e+00 $\pm$ 1.90e-02 $\pm$ 3.31e-02 \\ 
2.75e-01 &  5.00e-02 &  2.40e+00  $\pm$ 1.43e-02 $\pm$ 2.40e-02 & 2.45e+00 $\pm$ 1.41e-02 $\pm$ 2.45e-02 \\ 
3.25e-01 &  5.00e-02 &  1.81e+00  $\pm$ 1.12e-02 $\pm$ 1.81e-02 & 1.81e+00 $\pm$ 1.12e-02 $\pm$ 1.81e-02 \\ 
3.75e-01 &  5.00e-02 &  1.34e+00  $\pm$ 8.98e-03 $\pm$ 1.34e-02 & 1.36e+00 $\pm$ 9.03e-03 $\pm$ 1.36e-02 \\ 
4.25e-01 &  5.00e-02 &  9.90e-01  $\pm$ 7.22e-03 $\pm$ 9.90e-03 & 1.01e+00 $\pm$ 7.33e-03 $\pm$ 1.01e-02 \\ 
4.75e-01 &  5.00e-02 &  7.32e-01  $\pm$ 5.87e-03 $\pm$ 7.32e-03 & 7.47e-01 $\pm$ 5.94e-03 $\pm$ 7.47e-03 \\ 
5.25e-01 &  5.00e-02 &  5.48e-01  $\pm$ 4.83e-03 $\pm$ 5.48e-03 & 5.67e-01 $\pm$ 4.92e-03 $\pm$ 5.67e-03 \\ 
5.75e-01 &  5.00e-02 &  4.14e-01  $\pm$ 4.02e-03 $\pm$ 8.27e-03 & 4.29e-01 $\pm$ 4.11e-03 $\pm$ 8.59e-03 \\ 
6.25e-01 &  5.00e-02 &  3.18e-01  $\pm$ 3.40e-03 $\pm$ 6.35e-03 & 3.26e-01 $\pm$ 3.46e-03 $\pm$ 6.53e-03 \\ 
6.75e-01 &  5.00e-02 &  2.39e-01  $\pm$ 2.93e-03 $\pm$ 7.16e-03 & 2.44e-01 $\pm$ 2.99e-03 $\pm$ 7.32e-03 \\ 
7.25e-01 &  5.00e-02 &  1.90e-01  $\pm$ 2.48e-03 $\pm$ 5.70e-03 & 1.86e-01 $\pm$ 2.83e-03 $\pm$ 5.58e-03 \\ 
7.75e-01 &  5.00e-02 &  1.55e-01  $\pm$ 2.17e-03 $\pm$ 4.65e-03 & 1.45e-01 $\pm$ 2.81e-03 $\pm$ 4.36e-03 \\ 
\hline  
\end{tabular}  
\end{center}  
\end{scriptsize}  
\end{table}  
\begin{table}[!h] 
\begin{scriptsize} 
\begin{center} 
\caption{Invariant yields for $\pi^{-}$ and $\pi^{+}$ at mid-rapidity for the multiplicity bin: $Nch_{7-8}$ in ${\bf pp}$ collisions at 200 GeV.} \label{tab:Tables/ppPionSpectrum78.tex} 
\begin{tabular}{|c|c|c|c|} 
\hline   
 $p_{T}$ (GeV/c) &  $\Delta p_{T}$ (GeV/c) & $\pi^{-}$  & $\pi^{+}$  \\ \hline 
2.25e-01 &  5.00e-02 &  4.55e+00  $\pm$ 3.80e-02 $\pm$ 4.55e-02 & 4.65e+00 $\pm$ 3.34e-02 $\pm$ 4.65e-02 \\ 
2.75e-01 &  5.00e-02 &  3.44e+00  $\pm$ 2.53e-02 $\pm$ 3.44e-02 & 3.41e+00 $\pm$ 2.50e-02 $\pm$ 3.41e-02 \\ 
3.25e-01 &  5.00e-02 &  2.47e+00  $\pm$ 1.95e-02 $\pm$ 2.47e-02 & 2.51e+00 $\pm$ 1.96e-02 $\pm$ 2.51e-02 \\ 
3.75e-01 &  5.00e-02 &  1.82e+00  $\pm$ 1.55e-02 $\pm$ 1.82e-02 & 1.86e+00 $\pm$ 1.57e-02 $\pm$ 1.86e-02 \\ 
4.25e-01 &  5.00e-02 &  1.37e+00  $\pm$ 1.26e-02 $\pm$ 1.37e-02 & 1.37e+00 $\pm$ 1.26e-02 $\pm$ 1.37e-02 \\ 
4.75e-01 &  5.00e-02 &  1.01e+00  $\pm$ 1.02e-02 $\pm$ 1.01e-02 & 1.04e+00 $\pm$ 1.04e-02 $\pm$ 1.04e-02 \\ 
5.25e-01 &  5.00e-02 &  7.61e-01  $\pm$ 8.43e-03 $\pm$ 7.61e-03 & 7.72e-01 $\pm$ 8.53e-03 $\pm$ 7.72e-03 \\ 
5.75e-01 &  5.00e-02 &  5.73e-01  $\pm$ 6.99e-03 $\pm$ 1.15e-02 & 5.87e-01 $\pm$ 7.01e-03 $\pm$ 1.17e-02 \\ 
6.25e-01 &  5.00e-02 &  4.42e-01  $\pm$ 5.96e-03 $\pm$ 8.85e-03 & 4.63e-01 $\pm$ 6.10e-03 $\pm$ 9.26e-03 \\ 
6.75e-01 &  5.00e-02 &  3.40e-01  $\pm$ 5.21e-03 $\pm$ 1.02e-02 & 3.38e-01 $\pm$ 5.24e-03 $\pm$ 1.01e-02 \\ 
7.25e-01 &  5.00e-02 &  2.72e-01  $\pm$ 4.41e-03 $\pm$ 8.16e-03 & 2.51e-01 $\pm$ 4.94e-03 $\pm$ 7.52e-03 \\ 
7.75e-01 &  5.00e-02 &  2.23e-01  $\pm$ 3.83e-03 $\pm$ 6.68e-03 & 2.11e-01 $\pm$ 4.07e-03 $\pm$ 6.33e-03 \\ 
\hline  
\end{tabular}  
\end{center}  
\end{scriptsize}  
\end{table}  
\begin{table}[!h] 
\begin{scriptsize} 
\begin{center} 
\caption{Invariant yields for $\pi^{-}$ and $\pi^{+}$ at mid-rapidity for the multiplicity bin: $Nch_{9-...}$ in ${\bf pp}$ collisions at 200 GeV.} \label{tab:Tables/ppPionSpectrum9100.tex} 
\begin{tabular}{|c|c|c|c|} 
\hline   
 $p_{T}$ (GeV/c) &  $\Delta p_{T}$ (GeV/c) & $\pi^{-}$  & $\pi^{+}$  \\ \hline 
2.25e-01 &  5.00e-02 &  6.31e+00  $\pm$ 5.61e-02 $\pm$ 6.31e-02 & 6.31e+00 $\pm$ 5.69e-02 $\pm$ 6.31e-02 \\ 
2.75e-01 &  5.00e-02 &  4.60e+00  $\pm$ 3.50e-02 $\pm$ 4.60e-02 & 4.61e+00 $\pm$ 3.48e-02 $\pm$ 4.61e-02 \\ 
3.25e-01 &  5.00e-02 &  3.42e+00  $\pm$ 2.73e-02 $\pm$ 3.42e-02 & 3.36e+00 $\pm$ 2.71e-02 $\pm$ 3.36e-02 \\ 
3.75e-01 &  5.00e-02 &  2.52e+00  $\pm$ 2.17e-02 $\pm$ 2.52e-02 & 2.53e+00 $\pm$ 2.18e-02 $\pm$ 2.53e-02 \\ 
4.25e-01 &  5.00e-02 &  1.88e+00  $\pm$ 1.76e-02 $\pm$ 1.88e-02 & 1.87e+00 $\pm$ 1.76e-02 $\pm$ 1.87e-02 \\ 
4.75e-01 &  5.00e-02 &  1.42e+00  $\pm$ 1.44e-02 $\pm$ 1.42e-02 & 1.41e+00 $\pm$ 1.44e-02 $\pm$ 1.41e-02 \\ 
5.25e-01 &  5.00e-02 &  1.06e+00  $\pm$ 1.19e-02 $\pm$ 1.06e-02 & 1.09e+00 $\pm$ 1.20e-02 $\pm$ 1.09e-02 \\ 
5.75e-01 &  5.00e-02 &  8.25e-01  $\pm$ 1.00e-02 $\pm$ 1.65e-02 & 8.38e-01 $\pm$ 1.02e-02 $\pm$ 1.68e-02 \\ 
6.25e-01 &  5.00e-02 &  6.17e-01  $\pm$ 8.28e-03 $\pm$ 1.23e-02 & 6.26e-01 $\pm$ 8.41e-03 $\pm$ 1.25e-02 \\ 
6.75e-01 &  5.00e-02 &  4.95e-01  $\pm$ 7.56e-03 $\pm$ 1.49e-02 & 4.94e-01 $\pm$ 7.41e-03 $\pm$ 1.48e-02 \\ 
7.25e-01 &  5.00e-02 &  3.87e-01  $\pm$ 6.32e-03 $\pm$ 1.16e-02 & 3.87e-01 $\pm$ 7.74e-03 $\pm$ 1.16e-02 \\ 
7.75e-01 &  5.00e-02 &  3.12e-01  $\pm$ 5.40e-03 $\pm$ 9.37e-03 & 3.04e-01 $\pm$ 5.74e-03 $\pm$ 9.13e-03 \\ 
\hline  
\end{tabular}  
\end{center}  
\end{scriptsize}  
\end{table}  

%
\begin{table}[!h] 
\begin{scriptsize} 
\begin{center} 
\caption{Invariant yields for $K^{-}$ and $K^{+}$ at mid-rapidity in {\bf Minimum Bias} ${\bf pp}$ collisions at 200 GeV. \label{tab:Tables/ppKaonSpectrummb.tex}} 
\begin{tabular}{|c|c|c|c|} 
\hline   
 $p_{T}$ (GeV/c) &  $\Delta p_{T}$ (GeV/c) & $K^{-}$ & $K^{+}$ \\ \hline 
2.25e-01 &  5.00e-02 &  1.59e-01  $\pm$ 3.46e-03 $\pm$ 4.77e-03 & 1.37e-01 $\pm$ 3.96e-03 $\pm$ 4.11e-03 \\ 
2.75e-01 &  5.00e-02 &  1.32e-01  $\pm$ 2.24e-03 $\pm$ 1.32e-03 & 1.30e-01 $\pm$ 2.49e-03 $\pm$ 1.30e-03 \\ 
3.25e-01 &  5.00e-02 &  1.08e-01  $\pm$ 1.65e-03 $\pm$ 2.16e-03 & 1.11e-01 $\pm$ 1.82e-03 $\pm$ 2.23e-03 \\ 
3.75e-01 &  5.00e-02 &  8.99e-02  $\pm$ 1.30e-03 $\pm$ 2.70e-03 & 9.62e-02 $\pm$ 1.46e-03 $\pm$ 2.89e-03 \\ 
4.25e-01 &  5.00e-02 &  7.37e-02  $\pm$ 1.14e-03 $\pm$ 8.84e-03 & 7.95e-02 $\pm$ 1.19e-03 $\pm$ 9.54e-03 \\ 
4.75e-01 &  5.00e-02 &  6.77e-02  $\pm$ 1.02e-03 $\pm$ 8.80e-03 & 7.44e-02 $\pm$ 1.35e-03 $\pm$ 9.67e-03 \\ 
5.25e-01 &  5.00e-02 &  5.05e-02  $\pm$ 8.46e-04 $\pm$ 5.05e-03 & 5.72e-02 $\pm$ 8.68e-04 $\pm$ 5.72e-03 \\ 
5.75e-01 &  5.00e-02 &  4.09e-02  $\pm$ 6.57e-04 $\pm$ 2.45e-03 & 4.42e-02 $\pm$ 7.46e-04 $\pm$ 2.65e-03 \\ 
6.25e-01 &  5.00e-02 &  3.36e-02  $\pm$ 5.81e-04 $\pm$ 2.36e-03 & 3.43e-02 $\pm$ 6.76e-04 $\pm$ 2.40e-03 \\ 
6.75e-01 &  5.00e-02 &  2.75e-02  $\pm$ 7.10e-04 $\pm$ 2.75e-03 & 2.89e-02 $\pm$ 7.72e-04 $\pm$ 2.89e-03 \\ 
7.25e-01 &  5.00e-02 &  2.47e-02  $\pm$ 1.03e-03 $\pm$ 7.41e-04 & 2.74e-02 $\pm$ 1.25e-03 $\pm$ 8.22e-04 \\ 
\hline  
\end{tabular}  
\end{center}  
\end{scriptsize}  
\end{table}  
\begin{table}[!h] 
\begin{scriptsize} 
\begin{center} 
\caption{Invariant yields for $K^{-}$ and $K^{+}$ at mid-rapidity for the multiplicity bin: $Nch_{0-2}$ in ${\bf pp}$ collisions at 200 GeV. \label{tab:Tables/ppKaonSpectrum02.tex}} 
\begin{tabular}{|c|c|c|c|} 
\hline   
 $p_{T}$ (GeV/c) &  $\Delta p_{T}$ (GeV/c) & $K^{-}$ & $K^{+}$ \\ \hline 
2.25e-01 &  5.00e-02 &  2.98e-02  $\pm$ 2.59e-03 $\pm$ 8.95e-04 & 2.28e-02 $\pm$ 1.64e-03 $\pm$ 6.85e-04 \\ 
2.75e-01 &  5.00e-02 &  2.80e-02  $\pm$ 1.71e-03 $\pm$ 2.80e-04 & 3.19e-02 $\pm$ 1.78e-03 $\pm$ 3.19e-04 \\ 
3.25e-01 &  5.00e-02 &  2.29e-02  $\pm$ 1.25e-03 $\pm$ 4.58e-04 & 2.77e-02 $\pm$ 1.40e-03 $\pm$ 5.54e-04 \\ 
3.75e-01 &  5.00e-02 &  1.98e-02  $\pm$ 1.00e-03 $\pm$ 5.95e-04 & 2.40e-02 $\pm$ 1.15e-03 $\pm$ 7.20e-04 \\ 
4.25e-01 &  5.00e-02 &  1.67e-02  $\pm$ 8.48e-04 $\pm$ 2.00e-03 & 1.94e-02 $\pm$ 9.41e-04 $\pm$ 2.32e-03 \\ 
4.75e-01 &  5.00e-02 &  1.36e-02  $\pm$ 6.48e-04 $\pm$ 1.76e-03 & 1.78e-02 $\pm$ 9.71e-04 $\pm$ 2.32e-03 \\ 
5.25e-01 &  5.00e-02 &  1.17e-02  $\pm$ 6.13e-04 $\pm$ 1.17e-03 & 1.40e-02 $\pm$ 6.75e-04 $\pm$ 1.40e-03 \\ 
5.75e-01 &  5.00e-02 &  9.25e-03  $\pm$ 5.57e-04 $\pm$ 5.55e-04 & 1.10e-02 $\pm$ 5.74e-04 $\pm$ 6.58e-04 \\ 
6.25e-01 &  5.00e-02 &  6.56e-03  $\pm$ 5.09e-04 $\pm$ 4.59e-04 & 8.13e-03 $\pm$ 5.12e-04 $\pm$ 5.69e-04 \\ 
6.75e-01 &  5.00e-02 &  6.73e-03  $\pm$ 6.99e-04 $\pm$ 6.73e-04 & 8.37e-03 $\pm$ 6.38e-04 $\pm$ 8.37e-04 \\ 
7.25e-01 &  5.00e-02 &  7.16e-03  $\pm$ 1.61e-03 $\pm$ 2.15e-04 & 5.42e-03 $\pm$ 8.36e-04 $\pm$ 1.63e-04 \\ 
\hline  
\end{tabular}  
\end{center}  
\end{scriptsize}  
\end{table}  
\begin{table}[!h] 
\begin{scriptsize} 
\begin{center} 
\caption{Invariant yields for $K^{-}$ and $K^{+}$ at mid-rapidity for the multiplicity bin: $Nch_{3-4}$ in ${\bf pp}$ collisions at 200 GeV. \label{tab:Tables/ppKaonSpectrum34.tex}} 
\begin{tabular}{|c|c|c|c|} 
\hline   
 $p_{T}$ (GeV/c) &  $\Delta p_{T}$ (GeV/c) & $K^{-}$ & $K^{+}$ \\ \hline 
2.25e-01 &  5.00e-02 &  1.38e-01  $\pm$ 5.56e-03 $\pm$ 4.13e-03 & 1.26e-01 $\pm$ 4.87e-03 $\pm$ 3.78e-03 \\ 
2.75e-01 &  5.00e-02 &  1.21e-01  $\pm$ 3.71e-03 $\pm$ 1.21e-03 & 1.16e-01 $\pm$ 3.52e-03 $\pm$ 1.16e-03 \\ 
3.25e-01 &  5.00e-02 &  1.02e-01  $\pm$ 2.76e-03 $\pm$ 2.05e-03 & 1.02e-01 $\pm$ 2.73e-03 $\pm$ 2.03e-03 \\ 
3.75e-01 &  5.00e-02 &  8.28e-02  $\pm$ 2.15e-03 $\pm$ 2.49e-03 & 8.81e-02 $\pm$ 2.22e-03 $\pm$ 2.64e-03 \\ 
4.25e-01 &  5.00e-02 &  6.60e-02  $\pm$ 1.87e-03 $\pm$ 7.92e-03 & 7.36e-02 $\pm$ 1.95e-03 $\pm$ 8.83e-03 \\ 
4.75e-01 &  5.00e-02 &  6.29e-02  $\pm$ 1.65e-03 $\pm$ 8.18e-03 & 6.12e-02 $\pm$ 1.64e-03 $\pm$ 7.96e-03 \\ 
5.25e-01 &  5.00e-02 &  4.42e-02  $\pm$ 1.35e-03 $\pm$ 4.42e-03 & 4.93e-02 $\pm$ 1.40e-03 $\pm$ 4.93e-03 \\ 
5.75e-01 &  5.00e-02 &  3.62e-02  $\pm$ 1.16e-03 $\pm$ 2.17e-03 & 3.80e-02 $\pm$ 1.20e-03 $\pm$ 2.28e-03 \\ 
6.25e-01 &  5.00e-02 &  2.83e-02  $\pm$ 1.08e-03 $\pm$ 1.98e-03 & 2.97e-02 $\pm$ 1.10e-03 $\pm$ 2.08e-03 \\ 
6.75e-01 &  5.00e-02 &  2.06e-02  $\pm$ 1.03e-03 $\pm$ 2.06e-03 & 2.48e-02 $\pm$ 1.21e-03 $\pm$ 2.48e-03 \\ 
7.25e-01 &  5.00e-02 &  2.19e-02  $\pm$ 1.91e-03 $\pm$ 6.56e-04 & 2.20e-02 $\pm$ 1.72e-03 $\pm$ 6.60e-04 \\ 
\hline  
\end{tabular}  
\end{center}  
\end{scriptsize}  
\end{table}  
\begin{table}[!h] 
\begin{scriptsize} 
\begin{center} 
\caption{Invariant yields for $K^{-}$ and $K^{+}$ at mid-rapidity for the multiplicity bin: $Nch_{5-6}$ in ${\bf pp}$ collisions at 200 GeV. \label{tab:Tables/ppKaonSpectrum56.tex}} 
\begin{tabular}{|c|c|c|c|} 
\hline   
 $p_{T}$ (GeV/c) &  $\Delta p_{T}$ (GeV/c) & $K^{-}$ & $K^{+}$ \\ \hline 
2.25e-01 &  5.00e-02 &  2.64e-01  $\pm$ 1.07e-02 $\pm$ 7.91e-03 & 2.13e-01 $\pm$ 1.09e-02 $\pm$ 6.40e-03 \\ 
2.75e-01 &  5.00e-02 &  2.15e-01  $\pm$ 6.88e-03 $\pm$ 2.15e-03 & 2.01e-01 $\pm$ 7.62e-03 $\pm$ 2.01e-03 \\ 
3.25e-01 &  5.00e-02 &  1.70e-01  $\pm$ 4.96e-03 $\pm$ 3.39e-03 & 1.73e-01 $\pm$ 5.40e-03 $\pm$ 3.46e-03 \\ 
3.75e-01 &  5.00e-02 &  1.36e-01  $\pm$ 3.86e-03 $\pm$ 4.08e-03 & 1.46e-01 $\pm$ 4.23e-03 $\pm$ 4.39e-03 \\ 
4.25e-01 &  5.00e-02 &  1.14e-01  $\pm$ 3.43e-03 $\pm$ 1.37e-02 & 1.23e-01 $\pm$ 3.59e-03 $\pm$ 1.48e-02 \\ 
4.75e-01 &  5.00e-02 &  1.06e-01  $\pm$ 3.04e-03 $\pm$ 1.38e-02 & 1.22e-01 $\pm$ 4.15e-03 $\pm$ 1.59e-02 \\ 
5.25e-01 &  5.00e-02 &  7.57e-02  $\pm$ 2.48e-03 $\pm$ 7.57e-03 & 8.99e-02 $\pm$ 2.63e-03 $\pm$ 8.99e-03 \\ 
5.75e-01 &  5.00e-02 &  5.69e-02  $\pm$ 2.10e-03 $\pm$ 3.41e-03 & 6.92e-02 $\pm$ 2.28e-03 $\pm$ 4.15e-03 \\ 
6.25e-01 &  5.00e-02 &  5.21e-02  $\pm$ 2.07e-03 $\pm$ 3.65e-03 & 5.21e-02 $\pm$ 2.03e-03 $\pm$ 3.65e-03 \\ 
6.75e-01 &  5.00e-02 &  3.96e-02  $\pm$ 2.21e-03 $\pm$ 3.96e-03 & 4.05e-02 $\pm$ 2.27e-03 $\pm$ 4.05e-03 \\ 
7.25e-01 &  5.00e-02 &  3.75e-02  $\pm$ 3.09e-03 $\pm$ 1.12e-03 & 4.33e-02 $\pm$ 4.09e-03 $\pm$ 1.30e-03 \\ 
\hline  
\end{tabular}  
\end{center}  
\end{scriptsize}  
\end{table}  
\begin{table}[!h] 
\begin{scriptsize} 
\begin{center} 
\caption{Invariant yields for $K^{-}$ and $K^{+}$ at mid-rapidity for the multiplicity bin: $Nch_{7-8}$ in ${\bf pp}$ collisions at 200 GeV. \label{tab:Tables/ppKaonSpectrum78.tex}} 
\begin{tabular}{|c|c|c|c|} 
\hline   
 $p_{T}$ (GeV/c) &  $\Delta p_{T}$ (GeV/c) & $K^{-}$ & $K^{+}$ \\ \hline 
2.25e-01 &  5.00e-02 &  3.17e-01  $\pm$ 1.75e-02 $\pm$ 9.51e-03 & 2.83e-01 $\pm$ 1.54e-02 $\pm$ 8.49e-03 \\ 
2.75e-01 &  5.00e-02 &  2.69e-01  $\pm$ 1.15e-02 $\pm$ 2.69e-03 & 2.66e-01 $\pm$ 1.11e-02 $\pm$ 2.66e-03 \\ 
3.25e-01 &  5.00e-02 &  2.20e-01  $\pm$ 8.40e-03 $\pm$ 4.41e-03 & 2.18e-01 $\pm$ 8.28e-03 $\pm$ 4.35e-03 \\ 
3.75e-01 &  5.00e-02 &  1.90e-01  $\pm$ 6.73e-03 $\pm$ 5.70e-03 & 2.07e-01 $\pm$ 7.05e-03 $\pm$ 6.20e-03 \\ 
4.25e-01 &  5.00e-02 &  1.60e-01  $\pm$ 6.09e-03 $\pm$ 1.92e-02 & 1.68e-01 $\pm$ 6.34e-03 $\pm$ 2.01e-02 \\ 
4.75e-01 &  5.00e-02 &  1.42e-01  $\pm$ 5.21e-03 $\pm$ 1.84e-02 & 1.40e-01 $\pm$ 5.34e-03 $\pm$ 1.81e-02 \\ 
5.25e-01 &  5.00e-02 &  1.07e-01  $\pm$ 4.42e-03 $\pm$ 1.07e-02 & 1.17e-01 $\pm$ 4.55e-03 $\pm$ 1.17e-02 \\ 
5.75e-01 &  5.00e-02 &  7.52e-02  $\pm$ 3.75e-03 $\pm$ 4.51e-03 & 8.90e-02 $\pm$ 3.95e-03 $\pm$ 5.34e-03 \\ 
6.25e-01 &  5.00e-02 &  7.40e-02  $\pm$ 3.59e-03 $\pm$ 5.18e-03 & 7.20e-02 $\pm$ 3.78e-03 $\pm$ 5.04e-03 \\ 
6.75e-01 &  5.00e-02 &  6.13e-02  $\pm$ 4.49e-03 $\pm$ 6.13e-03 & 6.14e-02 $\pm$ 4.15e-03 $\pm$ 6.14e-03 \\ 
7.25e-01 &  5.00e-02 &  7.40e-02  $\pm$ 8.44e-03 $\pm$ 2.22e-03 & 6.71e-02 $\pm$ 7.84e-03 $\pm$ 2.01e-03 \\ 
\hline  
\end{tabular}  
\end{center}  
\end{scriptsize}  
\end{table}  
\begin{table}[!h] 
\begin{scriptsize} 
\begin{center} 
\caption{Invariant yields for $K^{-}$ and $K^{+}$ at mid-rapidity for the multiplicity bin: $Nch_{9-...}$ in ${\bf pp}$ collisions at 200 GeV. \label{tab:Tables/ppKaonSpectrum9100.tex}} 
\begin{tabular}{|c|c|c|c|} 
\hline   
 $p_{T}$ (GeV/c) &  $\Delta p_{T}$ (GeV/c) & $K^{-}$ & $K^{+}$ \\ \hline 
2.25e-01 &  5.00e-02 &  4.27e-01  $\pm$ 2.42e-02 $\pm$ 1.28e-02 & 3.98e-01 $\pm$ 2.14e-02 $\pm$ 1.20e-02 \\ 
2.75e-01 &  5.00e-02 &  3.75e-01  $\pm$ 1.61e-02 $\pm$ 3.75e-03 & 3.99e-01 $\pm$ 1.62e-02 $\pm$ 3.99e-03 \\ 
3.25e-01 &  5.00e-02 &  3.02e-01  $\pm$ 1.18e-02 $\pm$ 6.05e-03 & 3.50e-01 $\pm$ 1.27e-02 $\pm$ 7.00e-03 \\ 
3.75e-01 &  5.00e-02 &  2.72e-01  $\pm$ 9.70e-03 $\pm$ 8.17e-03 & 2.79e-01 $\pm$ 1.02e-02 $\pm$ 8.37e-03 \\ 
4.25e-01 &  5.00e-02 &  2.17e-01  $\pm$ 8.00e-03 $\pm$ 2.60e-02 & 2.28e-01 $\pm$ 8.70e-03 $\pm$ 2.74e-02 \\ 
4.75e-01 &  5.00e-02 &  1.89e-01  $\pm$ 1.09e-02 $\pm$ 2.46e-02 & 2.38e-01 $\pm$ 1.35e-02 $\pm$ 3.09e-02 \\ 
5.25e-01 &  5.00e-02 &  1.76e-01  $\pm$ 6.54e-03 $\pm$ 1.76e-02 & 2.03e-01 $\pm$ 6.80e-03 $\pm$ 2.03e-02 \\ 
5.75e-01 &  5.00e-02 &  1.09e-01  $\pm$ 5.81e-03 $\pm$ 6.52e-03 & 1.34e-01 $\pm$ 5.69e-03 $\pm$ 8.05e-03 \\ 
6.25e-01 &  5.00e-02 &  7.36e-02  $\pm$ 5.28e-03 $\pm$ 5.15e-03 & 1.13e-01 $\pm$ 5.13e-03 $\pm$ 7.90e-03 \\ 
6.75e-01 &  5.00e-02 &  9.02e-02  $\pm$ 6.50e-03 $\pm$ 9.02e-03 & 9.52e-02 $\pm$ 5.83e-03 $\pm$ 9.52e-03 \\ 
7.25e-01 &  5.00e-02 &  7.43e-02  $\pm$ 9.57e-03 $\pm$ 2.23e-03 & 9.62e-02 $\pm$ 1.07e-02 $\pm$ 2.88e-03 \\ 
\hline  
\end{tabular}  
\end{center}  
\end{scriptsize}  
\end{table}  
%
\begin{table}[!h] 
\begin{scriptsize} 
\begin{center} 
\caption{Invariant yields for $\overline{p}$ and $p$ at mid-rapidity in {\bf Minimum Bias} ${\bf pp}$ collisions at 200 GeV.} \label{tab:Tables/ppPSpectrummb.tex} 
\begin{tabular}{|c|c|c|c|} 
\hline   
 $p_{T}$ (GeV/c) &  $\Delta p_{T}$ (GeV/c) & $\overline{p}$ & $p$ \\ \hline 
4.25e-01 &  5.00e-02 &  4.36e-02  $\pm$ 6.73e-04 $\pm$ 4.36e-04 & 4.70e-02 $\pm$ 3.65e-04 $\pm$ 4.70e-04 \\ 
4.75e-01 &  5.00e-02 &  4.15e-02  $\pm$ 6.19e-04 $\pm$ 4.15e-04 & 4.27e-02 $\pm$ 3.49e-04 $\pm$ 4.27e-04 \\ 
5.25e-01 &  5.00e-02 &  3.58e-02  $\pm$ 5.45e-04 $\pm$ 3.58e-04 & 3.89e-02 $\pm$ 3.38e-04 $\pm$ 3.89e-04 \\ 
5.75e-01 &  5.00e-02 &  3.10e-02  $\pm$ 4.84e-04 $\pm$ 3.10e-04 & 3.39e-02 $\pm$ 3.22e-04 $\pm$ 3.39e-04 \\ 
6.25e-01 &  5.00e-02 &  2.72e-02  $\pm$ 4.37e-04 $\pm$ 2.72e-04 & 2.85e-02 $\pm$ 3.01e-04 $\pm$ 2.85e-04 \\ 
6.75e-01 &  5.00e-02 &  2.27e-02  $\pm$ 3.84e-04 $\pm$ 4.54e-04 & 2.55e-02 $\pm$ 2.91e-04 $\pm$ 5.09e-04 \\ 
7.25e-01 &  5.00e-02 &  1.92e-02  $\pm$ 3.41e-04 $\pm$ 3.84e-04 & 2.21e-02 $\pm$ 2.77e-04 $\pm$ 4.42e-04 \\ 
7.75e-01 &  5.00e-02 &  1.59e-02  $\pm$ 3.01e-04 $\pm$ 3.19e-04 & 1.90e-02 $\pm$ 2.61e-04 $\pm$ 3.81e-04 \\ 
8.25e-01 &  5.00e-02 &  1.44e-02  $\pm$ 2.80e-04 $\pm$ 7.19e-04 & 1.68e-02 $\pm$ 2.48e-04 $\pm$ 8.38e-04 \\ 
8.75e-01 &  5.00e-02 &  1.20e-02  $\pm$ 2.45e-04 $\pm$ 6.02e-04 & 1.43e-02 $\pm$ 2.33e-04 $\pm$ 7.17e-04 \\ 
9.25e-01 &  5.00e-02 &  9.98e-03  $\pm$ 2.16e-04 $\pm$ 3.99e-04 & 1.27e-02 $\pm$ 2.18e-04 $\pm$ 5.09e-04 \\ 
9.75e-01 &  5.00e-02 &  8.19e-03  $\pm$ 1.91e-04 $\pm$ 2.46e-04 & 1.06e-02 $\pm$ 2.00e-04 $\pm$ 3.17e-04 \\ 
1.03e+00 &  5.00e-02 &  7.11e-03  $\pm$ 1.73e-04 $\pm$ 2.13e-04 & 9.27e-03 $\pm$ 3.47e-04 $\pm$ 2.78e-04 \\ 
1.08e+00 &  5.00e-02 &  5.48e-03  $\pm$ 1.54e-04 $\pm$ 1.64e-04 & 7.65e-03 $\pm$ 4.34e-04 $\pm$ 2.30e-04 \\ 
\hline  
\end{tabular}  
\end{center}  
\end{scriptsize}  
\end{table}  
\begin{table}[!h] 
\begin{scriptsize} 
\begin{center} 
\caption{Invariant yields for $\overline{p}$ and $p$ at mid-rapidity for the multiplicity bin: $Nch_{0-2}$ in ${\bf pp}$ collisions at 200 GeV.} \label{tab:Tables/ppPSpectrum02.tex} 
\begin{tabular}{|c|c|c|c|} 
\hline   
 $p_{T}$ (GeV/c) &  $\Delta p_{T}$ (GeV/c) & $\overline{p}$ & $p$ \\ \hline 
4.25e-01 &  5.00e-02 &  1.15e-02  $\pm$ 6.10e-04 $\pm$ 1.15e-04 & 1.70e-02 $\pm$ 2.29e-04 $\pm$ 1.70e-04 \\ 
4.75e-01 &  5.00e-02 &  1.21e-02  $\pm$ 5.83e-04 $\pm$ 1.21e-04 & 1.42e-02 $\pm$ 2.00e-04 $\pm$ 1.42e-04 \\ 
5.25e-01 &  5.00e-02 &  9.67e-03  $\pm$ 4.96e-04 $\pm$ 9.67e-05 & 1.12e-02 $\pm$ 1.73e-04 $\pm$ 1.12e-04 \\ 
5.75e-01 &  5.00e-02 &  7.48e-03  $\pm$ 4.14e-04 $\pm$ 7.48e-05 & 8.80e-03 $\pm$ 1.50e-04 $\pm$ 8.80e-05 \\ 
6.25e-01 &  5.00e-02 &  6.97e-03  $\pm$ 3.83e-04 $\pm$ 6.97e-05 & 6.54e-03 $\pm$ 1.29e-04 $\pm$ 6.54e-05 \\ 
6.75e-01 &  5.00e-02 &  5.20e-03  $\pm$ 3.18e-04 $\pm$ 1.04e-04 & 5.27e-03 $\pm$ 1.17e-04 $\pm$ 1.05e-04 \\ 
7.25e-01 &  5.00e-02 &  4.76e-03  $\pm$ 3.21e-04 $\pm$ 9.51e-05 & 4.20e-03 $\pm$ 1.08e-04 $\pm$ 8.39e-05 \\ 
7.75e-01 &  5.00e-02 &  3.39e-03  $\pm$ 2.36e-04 $\pm$ 6.77e-05 & 3.43e-03 $\pm$ 1.01e-04 $\pm$ 6.87e-05 \\ 
8.25e-01 &  5.00e-02 &  2.57e-03  $\pm$ 2.01e-04 $\pm$ 1.29e-04 & 2.60e-03 $\pm$ 9.12e-05 $\pm$ 1.30e-04 \\ 
8.75e-01 &  5.00e-02 &  2.14e-03  $\pm$ 1.88e-04 $\pm$ 1.07e-04 & 2.32e-03 $\pm$ 8.98e-05 $\pm$ 1.16e-04 \\ 
9.25e-01 &  5.00e-02 &  1.87e-03  $\pm$ 1.72e-04 $\pm$ 7.47e-05 & 1.89e-03 $\pm$ 8.53e-05 $\pm$ 7.55e-05 \\ 
9.75e-01 &  5.00e-02 &  1.42e-03  $\pm$ 1.47e-04 $\pm$ 4.27e-05 & 1.52e-03 $\pm$ 8.24e-05 $\pm$ 4.57e-05 \\ 
1.03e+00 &  5.00e-02 &  1.16e-03  $\pm$ 1.13e-04 $\pm$ 3.48e-05 & 1.36e-03 $\pm$ 8.00e-05 $\pm$ 4.07e-05 \\ 
1.08e+00 &  5.00e-02 &  1.07e-03  $\pm$ 1.22e-04 $\pm$ 3.20e-05 & 1.29e-03 $\pm$ 8.53e-05 $\pm$ 3.88e-05 \\ 
\hline  
\end{tabular}  
\end{center}  
\end{scriptsize}  
\end{table}  
\begin{table}[!h] 
\begin{scriptsize} 
\begin{center} 
\caption{Invariant yields for $\overline{p}$ and $p$ at mid-rapidity for the multiplicity bin: $Nch_{3-4}$ in ${\bf pp}$ collisions at 200 GeV.} \label{tab:Tables/ppPSpectrum34.tex} 
\begin{tabular}{|c|c|c|c|} 
\hline   
 $p_{T}$ (GeV/c) &  $\Delta p_{T}$ (GeV/c) & $\overline{p}$ & $p$ \\ \hline 
4.25e-01 &  5.00e-02 &  4.04e-02  $\pm$ 1.12e-03 $\pm$ 4.04e-04 & 4.30e-02 $\pm$ 5.89e-04 $\pm$ 4.30e-04 \\ 
4.75e-01 &  5.00e-02 &  3.91e-02  $\pm$ 1.04e-03 $\pm$ 3.91e-04 & 3.84e-02 $\pm$ 5.65e-04 $\pm$ 3.84e-04 \\ 
5.25e-01 &  5.00e-02 &  3.35e-02  $\pm$ 9.09e-04 $\pm$ 3.35e-04 & 3.60e-02 $\pm$ 5.59e-04 $\pm$ 3.60e-04 \\ 
5.75e-01 &  5.00e-02 &  2.90e-02  $\pm$ 8.03e-04 $\pm$ 2.90e-04 & 3.06e-02 $\pm$ 5.27e-04 $\pm$ 3.06e-04 \\ 
6.25e-01 &  5.00e-02 &  2.37e-02  $\pm$ 6.98e-04 $\pm$ 2.37e-04 & 2.57e-02 $\pm$ 4.95e-04 $\pm$ 2.57e-04 \\ 
6.75e-01 &  5.00e-02 &  1.96e-02  $\pm$ 6.08e-04 $\pm$ 3.91e-04 & 2.21e-02 $\pm$ 4.68e-04 $\pm$ 4.41e-04 \\ 
7.25e-01 &  5.00e-02 &  1.69e-02  $\pm$ 5.44e-04 $\pm$ 3.38e-04 & 1.91e-02 $\pm$ 4.45e-04 $\pm$ 3.82e-04 \\ 
7.75e-01 &  5.00e-02 &  1.39e-02  $\pm$ 4.72e-04 $\pm$ 2.78e-04 & 1.56e-02 $\pm$ 4.08e-04 $\pm$ 3.13e-04 \\ 
8.25e-01 &  5.00e-02 &  1.21e-02  $\pm$ 4.29e-04 $\pm$ 6.03e-04 & 1.40e-02 $\pm$ 3.92e-04 $\pm$ 6.99e-04 \\ 
8.75e-01 &  5.00e-02 &  9.23e-03  $\pm$ 3.70e-04 $\pm$ 4.62e-04 & 1.18e-02 $\pm$ 3.63e-04 $\pm$ 5.90e-04 \\ 
9.25e-01 &  5.00e-02 &  7.65e-03  $\pm$ 3.31e-04 $\pm$ 3.06e-04 & 1.05e-02 $\pm$ 3.41e-04 $\pm$ 4.20e-04 \\ 
9.75e-01 &  5.00e-02 &  6.42e-03  $\pm$ 2.93e-04 $\pm$ 1.93e-04 & 8.57e-03 $\pm$ 3.08e-04 $\pm$ 2.57e-04 \\ 
1.03e+00 &  5.00e-02 &  5.45e-03  $\pm$ 2.73e-04 $\pm$ 1.63e-04 & 7.31e-03 $\pm$ 2.96e-04 $\pm$ 2.19e-04 \\ 
1.08e+00 &  5.00e-02 &  3.66e-03  $\pm$ 2.26e-04 $\pm$ 1.10e-04 & 5.84e-03 $\pm$ 2.60e-04 $\pm$ 1.75e-04 \\ 
\hline  
\end{tabular}  
\end{center}  
\end{scriptsize}  
\end{table}  
\begin{table}[!h] 
\begin{scriptsize} 
\begin{center} 
\caption{Invariant yields for $\overline{p}$ and $p$ at mid-rapidity for the multiplicity bin: $Nch_{5-6}$ in ${\bf pp}$ collisions at 200 GeV.} \label{tab:Tables/ppPSpectrum56.tex} 
\begin{tabular}{|c|c|c|c|} 
\hline   
 $p_{T}$ (GeV/c) &  $\Delta p_{T}$ (GeV/c) & $\overline{p}$ & $p$ \\ \hline 
4.25e-01 &  5.00e-02 &  6.92e-02  $\pm$ 2.04e-03 $\pm$ 6.92e-04 & 6.46e-02 $\pm$ 1.23e-03 $\pm$ 6.46e-04 \\ 
4.75e-01 &  5.00e-02 &  6.38e-02  $\pm$ 1.85e-03 $\pm$ 6.38e-04 & 6.13e-02 $\pm$ 1.23e-03 $\pm$ 6.13e-04 \\ 
5.25e-01 &  5.00e-02 &  5.53e-02  $\pm$ 1.63e-03 $\pm$ 5.53e-04 & 5.79e-02 $\pm$ 1.21e-03 $\pm$ 5.79e-04 \\ 
5.75e-01 &  5.00e-02 &  4.57e-02  $\pm$ 1.42e-03 $\pm$ 4.57e-04 & 5.09e-02 $\pm$ 1.15e-03 $\pm$ 5.09e-04 \\ 
6.25e-01 &  5.00e-02 &  4.29e-02  $\pm$ 1.31e-03 $\pm$ 4.29e-04 & 4.25e-02 $\pm$ 1.05e-03 $\pm$ 4.25e-04 \\ 
6.75e-01 &  5.00e-02 &  3.40e-02  $\pm$ 1.12e-03 $\pm$ 6.80e-04 & 4.14e-02 $\pm$ 1.03e-03 $\pm$ 8.27e-04 \\ 
7.25e-01 &  5.00e-02 &  2.93e-02  $\pm$ 1.01e-03 $\pm$ 5.87e-04 & 3.40e-02 $\pm$ 9.33e-04 $\pm$ 6.79e-04 \\ 
7.75e-01 &  5.00e-02 &  2.46e-02  $\pm$ 8.86e-04 $\pm$ 4.91e-04 & 2.91e-02 $\pm$ 8.55e-04 $\pm$ 5.81e-04 \\ 
8.25e-01 &  5.00e-02 &  2.28e-02  $\pm$ 8.26e-04 $\pm$ 1.14e-03 & 2.64e-02 $\pm$ 8.10e-04 $\pm$ 1.32e-03 \\ 
8.75e-01 &  5.00e-02 &  1.82e-02  $\pm$ 7.31e-04 $\pm$ 9.10e-04 & 2.12e-02 $\pm$ 7.21e-04 $\pm$ 1.06e-03 \\ 
9.25e-01 &  5.00e-02 &  1.47e-02  $\pm$ 6.36e-04 $\pm$ 5.89e-04 & 1.85e-02 $\pm$ 6.62e-04 $\pm$ 7.38e-04 \\ 
9.75e-01 &  5.00e-02 &  1.26e-02  $\pm$ 5.61e-04 $\pm$ 3.79e-04 & 1.67e-02 $\pm$ 1.64e-03 $\pm$ 5.01e-04 \\ 
1.03e+00 &  5.00e-02 &  1.06e-02  $\pm$ 5.29e-04 $\pm$ 3.17e-04 & 1.26e-02 $\pm$ 5.46e-04 $\pm$ 3.78e-04 \\ 
1.08e+00 &  5.00e-02 &  7.92e-03  $\pm$ 4.39e-04 $\pm$ 2.38e-04 & 1.14e-02 $\pm$ 5.77e-04 $\pm$ 3.42e-04 \\ 
\hline  
\end{tabular}  
\end{center}  
\end{scriptsize}  
\end{table}  
\begin{table}[!h] 
\begin{scriptsize} 
\begin{center} 
\caption{Invariant yields for $\overline{p}$ and $p$ at mid-rapidity for the multiplicity bin: $Nch_{7-8}$ in ${\bf pp}$ collisions at 200 GeV.} \label{tab:Tables/ppPSpectrum78.tex} 
\begin{tabular}{|c|c|c|c|} 
\hline   
 $p_{T}$ (GeV/c) &  $\Delta p_{T}$ (GeV/c) & $\overline{p}$ & $p$ \\ \hline 
4.25e-01 &  5.00e-02 &  9.34e-02  $\pm$ 3.53e-03 $\pm$ 9.34e-04 & 9.05e-02 $\pm$ 2.27e-03 $\pm$ 9.05e-04 \\ 
4.75e-01 &  5.00e-02 &  8.31e-02  $\pm$ 3.12e-03 $\pm$ 8.31e-04 & 8.47e-02 $\pm$ 2.24e-03 $\pm$ 8.47e-04 \\ 
5.25e-01 &  5.00e-02 &  7.58e-02  $\pm$ 2.83e-03 $\pm$ 7.58e-04 & 8.12e-02 $\pm$ 2.22e-03 $\pm$ 8.12e-04 \\ 
5.75e-01 &  5.00e-02 &  6.52e-02  $\pm$ 2.51e-03 $\pm$ 6.52e-04 & 7.13e-02 $\pm$ 2.09e-03 $\pm$ 7.13e-04 \\ 
6.25e-01 &  5.00e-02 &  6.02e-02  $\pm$ 2.30e-03 $\pm$ 6.02e-04 & 6.46e-02 $\pm$ 1.98e-03 $\pm$ 6.46e-04 \\ 
6.75e-01 &  5.00e-02 &  5.14e-02  $\pm$ 2.04e-03 $\pm$ 1.03e-03 & 5.66e-02 $\pm$ 1.85e-03 $\pm$ 1.13e-03 \\ 
7.25e-01 &  5.00e-02 &  4.03e-02  $\pm$ 1.74e-03 $\pm$ 8.07e-04 & 4.87e-02 $\pm$ 1.70e-03 $\pm$ 9.73e-04 \\ 
7.75e-01 &  5.00e-02 &  3.36e-02  $\pm$ 1.53e-03 $\pm$ 6.72e-04 & 4.47e-02 $\pm$ 1.61e-03 $\pm$ 8.95e-04 \\ 
8.25e-01 &  5.00e-02 &  3.17e-02  $\pm$ 1.44e-03 $\pm$ 1.59e-03 & 3.78e-02 $\pm$ 1.47e-03 $\pm$ 1.89e-03 \\ 
8.75e-01 &  5.00e-02 &  2.76e-02  $\pm$ 1.34e-03 $\pm$ 1.38e-03 & 3.19e-02 $\pm$ 1.36e-03 $\pm$ 1.59e-03 \\ 
9.25e-01 &  5.00e-02 &  2.23e-02  $\pm$ 1.10e-03 $\pm$ 8.94e-04 & 2.83e-02 $\pm$ 1.23e-03 $\pm$ 1.13e-03 \\ 
9.75e-01 &  5.00e-02 &  1.87e-02  $\pm$ 1.05e-03 $\pm$ 5.60e-04 & 2.35e-02 $\pm$ 1.39e-03 $\pm$ 7.06e-04 \\ 
1.03e+00 &  5.00e-02 &  1.68e-02  $\pm$ 9.61e-04 $\pm$ 5.03e-04 & 2.19e-02 $\pm$ 1.10e-03 $\pm$ 6.57e-04 \\ 
1.08e+00 &  5.00e-02 &  1.28e-02  $\pm$ 8.30e-04 $\pm$ 3.85e-04 & 1.51e-02 $\pm$ 1.15e-03 $\pm$ 4.54e-04 \\ 
\hline  
\end{tabular}  
\end{center}  
\end{scriptsize}  
\end{table}  
\begin{table}[!h] 
\begin{scriptsize} 
\begin{center} 
\caption{Invariant yields for $\overline{p}$ and $p$ at mid-rapidity for the multiplicity bin: $Nch_{9-...}$ in ${\bf pp}$ collisions at 200 GeV.} \label{tab:Tables/ppPSpectrum9100.tex} 
\begin{tabular}{|c|c|c|c|} 
\hline   
 $p_{T}$ (GeV/c) &  $\Delta p_{T}$ (GeV/c) & $\overline{p}$ & $p$ \\ \hline 
4.25e-01 &  5.00e-02 &  1.18e-01  $\pm$ 4.74e-03 $\pm$ 1.18e-03 & 1.27e-01 $\pm$ 3.40e-03 $\pm$ 1.27e-03 \\ 
4.75e-01 &  5.00e-02 &  1.14e-01  $\pm$ 4.36e-03 $\pm$ 1.14e-03 & 1.17e-01 $\pm$ 3.24e-03 $\pm$ 1.17e-03 \\ 
5.25e-01 &  5.00e-02 &  9.54e-02  $\pm$ 3.79e-03 $\pm$ 9.54e-04 & 1.06e-01 $\pm$ 3.06e-03 $\pm$ 1.06e-03 \\ 
5.75e-01 &  5.00e-02 &  9.08e-02  $\pm$ 3.51e-03 $\pm$ 9.08e-04 & 9.85e-02 $\pm$ 2.93e-03 $\pm$ 9.85e-04 \\ 
6.25e-01 &  5.00e-02 &  7.60e-02  $\pm$ 3.08e-03 $\pm$ 7.60e-04 & 8.05e-02 $\pm$ 2.63e-03 $\pm$ 8.05e-04 \\ 
6.75e-01 &  5.00e-02 &  6.94e-02  $\pm$ 2.84e-03 $\pm$ 1.39e-03 & 7.03e-02 $\pm$ 2.43e-03 $\pm$ 1.41e-03 \\ 
7.25e-01 &  5.00e-02 &  6.14e-02  $\pm$ 2.55e-03 $\pm$ 1.23e-03 & 6.86e-02 $\pm$ 2.38e-03 $\pm$ 1.37e-03 \\ 
7.75e-01 &  5.00e-02 &  5.29e-02  $\pm$ 2.29e-03 $\pm$ 1.06e-03 & 5.90e-02 $\pm$ 2.19e-03 $\pm$ 1.18e-03 \\ 
8.25e-01 &  5.00e-02 &  5.03e-02  $\pm$ 2.17e-03 $\pm$ 2.51e-03 & 5.60e-02 $\pm$ 2.11e-03 $\pm$ 2.80e-03 \\ 
8.75e-01 &  5.00e-02 &  4.18e-02  $\pm$ 1.94e-03 $\pm$ 2.09e-03 & 4.79e-02 $\pm$ 4.40e-03 $\pm$ 2.39e-03 \\ 
9.25e-01 &  5.00e-02 &  3.45e-02  $\pm$ 1.75e-03 $\pm$ 1.38e-03 & 4.33e-02 $\pm$ 1.81e-03 $\pm$ 1.73e-03 \\ 
9.75e-01 &  5.00e-02 &  2.64e-02  $\pm$ 1.46e-03 $\pm$ 7.93e-04 & 3.59e-02 $\pm$ 1.56e-03 $\pm$ 1.08e-03 \\ 
1.03e+00 &  5.00e-02 &  2.35e-02  $\pm$ 1.35e-03 $\pm$ 7.05e-04 & 3.33e-02 $\pm$ 1.55e-03 $\pm$ 1.00e-03 \\ 
1.08e+00 &  5.00e-02 &  2.06e-02  $\pm$ 1.23e-03 $\pm$ 6.19e-04 & 2.64e-02 $\pm$ 1.38e-03 $\pm$ 7.91e-04 \\ 
\hline  
\end{tabular}  
\end{center}  
\end{scriptsize}  
\end{table}  
\clearpage
%
\begin{table}[!h] 
\begin{scriptsize} 
\begin{center} 
\caption{Invariant yields for $\pi^{-}$ and $\pi^{+}$ at mid-rapidity in {\bf Minimum Bias} ${\bf dAu}$ collisions at 200 GeV. \label{tab:Tables/dauPionSpectrummb.tex}} 
\begin{tabular}{|c|c|c|c|} 
\hline   
 $p_{T}$ (GeV/c) &  $\Delta p_{T}$ (GeV/c) & $\pi^{-}$  & $\pi^{+}$  \\ \hline 
2.25e-01 &  5.00e-02 &  6.52e+00  $\pm$ 1.18e-02 $\pm$ 6.52e-02 & 6.61e+00 $\pm$ 9.77e-03 $\pm$  6.61e-02 \\ 
2.75e-01 &  5.00e-02 &  4.85e+00  $\pm$ 6.05e-03 $\pm$ 4.85e-02 & 4.83e+00 $\pm$ 5.99e-03 $\pm$  4.83e-02 \\ 
3.25e-01 &  5.00e-02 &  3.62e+00  $\pm$ 4.69e-03 $\pm$ 3.62e-02 & 3.60e+00 $\pm$ 4.61e-03 $\pm$  3.60e-02 \\ 
3.75e-01 &  5.00e-02 &  2.70e+00  $\pm$ 3.74e-03 $\pm$ 2.70e-02 & 2.71e+00 $\pm$ 3.74e-03 $\pm$  2.71e-02 \\ 
4.25e-01 &  5.00e-02 &  2.06e+00  $\pm$ 3.06e-03 $\pm$ 8.23e-02 & 2.06e+00 $\pm$ 3.06e-03 $\pm$  8.24e-02 \\ 
4.75e-01 &  5.00e-02 &  1.57e+00  $\pm$ 2.51e-03 $\pm$ 6.27e-02 & 1.57e+00 $\pm$ 2.51e-03 $\pm$  6.30e-02 \\ 
5.25e-01 &  5.00e-02 &  1.20e+00  $\pm$ 2.09e-03 $\pm$ 4.81e-02 & 1.21e+00 $\pm$ 2.11e-03 $\pm$  4.84e-02 \\ 
5.75e-01 &  5.00e-02 &  9.30e-01  $\pm$ 1.76e-03 $\pm$ 3.72e-02 & 9.36e-01 $\pm$ 1.76e-03 $\pm$  3.74e-02 \\ 
6.25e-01 &  5.00e-02 &  7.25e-01  $\pm$ 1.50e-03 $\pm$ 4.35e-02 & 7.29e-01 $\pm$ 1.51e-03 $\pm$  4.37e-02 \\ 
6.75e-01 &  5.00e-02 &  5.70e-01  $\pm$ 1.34e-03 $\pm$ 1.71e-02 & 5.74e-01 $\pm$ 1.34e-03 $\pm$  1.72e-02 \\ 
\hline  
\end{tabular}  
\end{center}  
\end{scriptsize}  
\end{table}  
\begin{table}[!h] 
\begin{scriptsize} 
\begin{center} 
\caption{Invariant yields for $\pi^{-}$ and $\pi^{+}$ at mid-rapidity in {\bf 0 - 20 \%} ${\bf dAu}$ collisions at 200 GeV. \label{tab:Tables/dauPionSpectrumcent.tex}} 
\begin{tabular}{|c|c|c|c|} 
\hline   
 $p_{T}$ (GeV/c) &  $\Delta p_{T}$ (GeV/c) & $\pi^{-}$  & $\pi^{+}$  \\ \hline 
2.25e-01 &  5.00e-02 &  1.16e+01  $\pm$ 3.42e-02 $\pm$ 1.16e-01 & 1.19e+01 $\pm$ 2.98e-02 $\pm$  1.19e-01 \\ 
2.75e-01 &  5.00e-02 &  8.70e+00  $\pm$ 1.83e-02 $\pm$ 8.70e-02 & 8.69e+00 $\pm$ 1.81e-02 $\pm$  8.69e-02 \\ 
3.25e-01 &  5.00e-02 &  6.51e+00  $\pm$ 1.42e-02 $\pm$ 6.51e-02 & 6.49e+00 $\pm$ 1.39e-02 $\pm$  6.49e-02 \\ 
3.75e-01 &  5.00e-02 &  4.89e+00  $\pm$ 1.13e-02 $\pm$ 4.89e-02 & 4.94e+00 $\pm$ 1.14e-02 $\pm$  4.94e-02 \\ 
4.25e-01 &  5.00e-02 &  3.74e+00  $\pm$ 9.25e-03 $\pm$ 1.49e-01 & 3.77e+00 $\pm$ 9.32e-03 $\pm$  1.51e-01 \\ 
4.75e-01 &  5.00e-02 &  2.87e+00  $\pm$ 7.65e-03 $\pm$ 1.15e-01 & 2.90e+00 $\pm$ 7.70e-03 $\pm$  1.16e-01 \\ 
5.25e-01 &  5.00e-02 &  2.22e+00  $\pm$ 6.41e-03 $\pm$ 8.89e-02 & 2.23e+00 $\pm$ 6.46e-03 $\pm$  8.94e-02 \\ 
5.75e-01 &  5.00e-02 &  1.73e+00  $\pm$ 5.41e-03 $\pm$ 6.90e-02 & 1.74e+00 $\pm$ 5.45e-03 $\pm$  6.98e-02 \\ 
6.25e-01 &  5.00e-02 &  1.35e+00  $\pm$ 4.64e-03 $\pm$ 8.10e-02 & 1.37e+00 $\pm$ 4.68e-03 $\pm$  8.20e-02 \\ 
6.75e-01 &  5.00e-02 &  1.07e+00  $\pm$ 4.15e-03 $\pm$ 3.21e-02 & 1.09e+00 $\pm$ 4.18e-03 $\pm$  3.26e-02 \\ 
\hline  
\end{tabular}  
\end{center}  
\end{scriptsize}  
\end{table}  
\begin{table}[!h] 
\begin{scriptsize} 
\begin{center} 
\caption{Invariant yields for $\pi^{-}$ and $\pi^{+}$ at mid-rapidity in {\bf 20 - 40 \%} ${\bf dAu}$ collisions at 200 GeV. \label{tab:Tables/dauPionSpectrummid.tex}} 
\begin{tabular}{|c|c|c|c|} 
\hline   
 $p_{T}$ (GeV/c) &  $\Delta p_{T}$ (GeV/c) & $\pi^{-}$  & $\pi^{+}$  \\ \hline 
2.25e-01 &  5.00e-02 &  8.84e+00  $\pm$ 1.99e-02 $\pm$ 8.84e-02 & 8.53e+00 $\pm$ 2.38e-02 $\pm$  8.53e-02 \\ 
2.75e-01 &  5.00e-02 &  6.29e+00  $\pm$ 1.43e-02 $\pm$ 6.29e-02 & 6.23e+00 $\pm$ 1.43e-02 $\pm$  6.23e-02 \\ 
3.25e-01 &  5.00e-02 &  4.69e+00  $\pm$ 1.12e-02 $\pm$ 4.69e-02 & 4.65e+00 $\pm$ 1.12e-02 $\pm$  4.65e-02 \\ 
3.75e-01 &  5.00e-02 &  3.51e+00  $\pm$ 9.06e-03 $\pm$ 3.51e-02 & 3.48e+00 $\pm$ 9.01e-03 $\pm$  3.48e-02 \\ 
4.25e-01 &  5.00e-02 &  2.66e+00  $\pm$ 7.39e-03 $\pm$ 1.07e-01 & 2.66e+00 $\pm$ 7.38e-03 $\pm$  1.07e-01 \\ 
4.75e-01 &  5.00e-02 &  2.04e+00  $\pm$ 6.09e-03 $\pm$ 8.16e-02 & 2.04e+00 $\pm$ 6.09e-03 $\pm$  8.16e-02 \\ 
5.25e-01 &  5.00e-02 &  1.57e+00  $\pm$ 5.11e-03 $\pm$ 6.27e-02 & 1.58e+00 $\pm$ 5.13e-03 $\pm$  6.32e-02 \\ 
5.75e-01 &  5.00e-02 &  1.21e+00  $\pm$ 4.28e-03 $\pm$ 4.86e-02 & 1.22e+00 $\pm$ 4.28e-03 $\pm$  4.86e-02 \\ 
6.25e-01 &  5.00e-02 &  9.45e-01  $\pm$ 3.66e-03 $\pm$ 5.67e-02 & 9.54e-01 $\pm$ 3.65e-03 $\pm$  5.73e-02 \\ 
6.75e-01 &  5.00e-02 &  7.43e-01  $\pm$ 3.27e-03 $\pm$ 2.23e-02 & 7.49e-01 $\pm$ 3.28e-03 $\pm$  2.25e-02 \\ 
\hline  
\end{tabular}  
\end{center}  
\end{scriptsize}  
\end{table}  
\begin{table}[!h] 
\begin{scriptsize} 
\begin{center} 
\caption{Invariant yields for $\pi^{-}$ and $\pi^{+}$ at per-rapidity in {\bf 40 - 100 \%} ${\bf dAu}$ collisions at 200 GeV. \label{tab:Tables/dauPionSpectrumper.tex}} 
\begin{tabular}{|c|c|c|c|} 
\hline   
 $p_{T}$ (GeV/c) &  $\Delta p_{T}$ (GeV/c) & $\pi^{-}$  & $\pi^{+}$  \\ \hline 
2.25e-01 &  5.00e-02 &  4.16e+00  $\pm$ 1.04e-02 $\pm$ 4.16e-02 & 4.10e+00 $\pm$ 9.60e-02 $\pm$  4.10e-02 \\ 
2.75e-01 &  5.00e-02 &  3.07e+00  $\pm$ 6.16e-03 $\pm$ 3.07e-02 & 3.00e+00 $\pm$ 6.29e-02 $\pm$  3.00e-02 \\ 
3.25e-01 &  5.00e-02 &  2.28e+00  $\pm$ 4.93e-03 $\pm$ 2.28e-02 & 2.27e+00 $\pm$ 4.95e-02 $\pm$  2.27e-02 \\ 
3.75e-01 &  5.00e-02 &  1.69e+00  $\pm$ 3.92e-03 $\pm$ 1.69e-02 & 1.65e+00 $\pm$ 3.90e-02 $\pm$  1.65e-02 \\ 
4.25e-01 &  5.00e-02 &  1.28e+00  $\pm$ 3.20e-03 $\pm$ 5.14e-02 & 1.31e+00 $\pm$ 3.24e-02 $\pm$  5.22e-02 \\ 
4.75e-01 &  5.00e-02 &  9.66e-01  $\pm$ 2.61e-03 $\pm$ 3.86e-02 & 9.23e-01 $\pm$ 2.57e-02 $\pm$  3.69e-02 \\ 
5.25e-01 &  5.00e-02 &  7.31e-01  $\pm$ 2.16e-03 $\pm$ 2.93e-02 & 7.55e-01 $\pm$ 2.21e-02 $\pm$  3.02e-02 \\ 
5.75e-01 &  5.00e-02 &  5.64e-01  $\pm$ 1.81e-03 $\pm$ 2.26e-02 & 5.67e-01 $\pm$ 1.83e-02 $\pm$  2.27e-02 \\ 
6.25e-01 &  5.00e-02 &  4.38e-01  $\pm$ 1.55e-03 $\pm$ 2.63e-02 & 4.60e-01 $\pm$ 1.57e-02 $\pm$  2.76e-02 \\ 
6.75e-01 &  5.00e-02 &  3.40e-01  $\pm$ 1.37e-03 $\pm$ 1.02e-02 & 3.32e-01 $\pm$ 1.28e-02 $\pm$  9.96e-03 \\ 
\hline  
\end{tabular}  
\end{center}  
\end{scriptsize}  
\end{table}  
%
\begin{table}[!h] 
\begin{scriptsize} 
\begin{center} 
\caption{Invariant yields for $K^{-}$ and $K^{+}$ at mid-rapidity in {\bf Minimum Bias} ${\bf dAu}$ collisions at 200 GeV. \label{tab:Tables/dauKaonSpectrummb.tex}} 
\begin{tabular}{|c|c|c|c|} 
\hline   
 $p_{T}$ (GeV/c) &  $\Delta p_{T}$ (GeV/c) & $K^{-}$ & $K^{+}$ \\ \hline 
2.25e-01 &  5.00e-02 &  5.20e-01  $\pm$ 4.65e-03 $\pm$ 1.56e-02 & 4.92e-01 $\pm$ 4.29e-03 $\pm$  1.48e-02 \\ 
2.75e-01 &  5.00e-02 &  4.37e-01  $\pm$ 2.90e-03 $\pm$ 4.37e-03 & 4.31e-01 $\pm$ 2.77e-03 $\pm$  4.31e-03 \\ 
3.25e-01 &  5.00e-02 &  3.66e-01  $\pm$ 2.15e-03 $\pm$ 7.32e-03 & 3.71e-01 $\pm$ 2.10e-03 $\pm$  7.41e-03 \\ 
3.75e-01 &  5.00e-02 &  3.15e-01  $\pm$ 1.74e-03 $\pm$ 2.84e-02 & 3.26e-01 $\pm$ 1.73e-03 $\pm$  2.93e-02 \\ 
4.25e-01 &  5.00e-02 &  2.63e-01  $\pm$ 1.60e-03 $\pm$ 3.68e-02 & 2.75e-01 $\pm$ 1.60e-03 $\pm$  3.85e-02 \\ 
4.75e-01 &  5.00e-02 &  2.17e-01  $\pm$ 1.48e-03 $\pm$ 2.38e-02 & 2.26e-01 $\pm$ 1.50e-03 $\pm$  2.49e-02 \\ 
5.25e-01 &  5.00e-02 &  1.86e-01  $\pm$ 1.27e-03 $\pm$ 1.30e-02 & 2.07e-01 $\pm$ 1.27e-03 $\pm$  1.45e-02 \\ 
5.75e-01 &  5.00e-02 &  1.63e-01  $\pm$ 1.14e-03 $\pm$ 1.47e-02 & 1.77e-01 $\pm$ 1.10e-03 $\pm$  1.60e-02 \\ 
6.25e-01 &  5.00e-02 &  1.34e-01  $\pm$ 9.86e-04 $\pm$ 6.72e-03 & 1.44e-01 $\pm$ 9.94e-04 $\pm$  7.20e-03 \\ 
6.75e-01 &  5.00e-02 &  1.17e-01  $\pm$ 1.13e-03 $\pm$ 9.36e-03 & 1.28e-01 $\pm$ 1.12e-03 $\pm$  1.03e-02 \\ 
7.25e-01 &  5.00e-02 &  1.02e-01  $\pm$ 1.69e-03 $\pm$ 5.09e-03 & 1.05e-01 $\pm$ 1.63e-03 $\pm$  5.24e-03 \\ 
\hline  
\end{tabular}  
\end{center}  
\end{scriptsize}  
\end{table}  
\begin{table}[!h] 
\begin{scriptsize} 
\begin{center} 
\caption{Invariant yields for $K^{-}$ and $K^{+}$ at mid-rapidity in {\bf 0 - 20 \%} ${\bf dAu}$ collisions at 200 GeV. \label{tab:Tables/dauKaonSpectrumcent.tex}} 
\begin{tabular}{|c|c|c|c|} 
\hline   
 $p_{T}$ (GeV/c) &  $\Delta p_{T}$ (GeV/c) & $K^{-}$ & $K^{+}$ \\ \hline 
2.25e-01 &  5.00e-02 &  9.47e-01  $\pm$ 1.41e-02 $\pm$ 2.84e-02 & 8.93e-01 $\pm$ 1.29e-02 $\pm$  2.68e-02 \\ 
2.75e-01 &  5.00e-02 &  7.85e-01  $\pm$ 8.72e-03 $\pm$ 7.85e-03 & 7.78e-01 $\pm$ 8.34e-03 $\pm$  7.78e-03 \\ 
3.25e-01 &  5.00e-02 &  6.70e-01  $\pm$ 6.52e-03 $\pm$ 1.34e-02 & 6.68e-01 $\pm$ 6.33e-03 $\pm$  1.34e-02 \\ 
3.75e-01 &  5.00e-02 &  5.75e-01  $\pm$ 5.27e-03 $\pm$ 5.17e-02 & 5.94e-01 $\pm$ 5.25e-03 $\pm$  5.34e-02 \\ 
4.25e-01 &  5.00e-02 &  4.81e-01  $\pm$ 4.86e-03 $\pm$ 6.74e-02 & 5.04e-01 $\pm$ 4.86e-03 $\pm$  7.06e-02 \\ 
4.75e-01 &  5.00e-02 &  4.07e-01  $\pm$ 4.66e-03 $\pm$ 4.48e-02 & 4.30e-01 $\pm$ 4.62e-03 $\pm$  4.73e-02 \\ 
5.25e-01 &  5.00e-02 &  3.52e-01  $\pm$ 3.91e-03 $\pm$ 2.46e-02 & 3.72e-01 $\pm$ 3.93e-03 $\pm$  2.60e-02 \\ 
5.75e-01 &  5.00e-02 &  3.09e-01  $\pm$ 3.45e-03 $\pm$ 2.78e-02 & 3.26e-01 $\pm$ 3.54e-03 $\pm$  2.93e-02 \\ 
6.25e-01 &  5.00e-02 &  2.57e-01  $\pm$ 3.13e-03 $\pm$ 1.29e-02 & 2.70e-01 $\pm$ 3.08e-03 $\pm$  1.35e-02 \\ 
6.75e-01 &  5.00e-02 &  2.21e-01  $\pm$ 3.48e-03 $\pm$ 1.77e-02 & 2.45e-01 $\pm$ 3.47e-03 $\pm$  1.96e-02 \\ 
7.25e-01 &  5.00e-02 &  1.91e-01  $\pm$ 5.06e-03 $\pm$ 9.53e-03 & 2.02e-01 $\pm$ 5.07e-03 $\pm$  1.01e-02 \\ 
\hline  
\end{tabular}  
\end{center}  
\end{scriptsize}  
\end{table}  
\begin{table}[!h] 
\begin{scriptsize} 
\begin{center} 
\caption{Invariant yields for $K^{-}$ and $K^{+}$ at mid-rapidity in {\bf 20 - 40 \%} ${\bf dAu}$ collisions at 200 GeV. \label{tab:Tables/dauKaonSpectrummid.tex}} 
\begin{tabular}{|c|c|c|c|} 
\hline   
 $p_{T}$ (GeV/c) &  $\Delta p_{T}$ (GeV/c) & $K^{-}$ & $K^{+}$ \\ \hline 
2.25e-01 &  5.00e-02 &  6.65e-01  $\pm$ 1.12e-02 $\pm$ 2.00e-02 & 6.32e-01 $\pm$ 1.03e-02 $\pm$  1.90e-02 \\ 
2.75e-01 &  5.00e-02 &  5.63e-01  $\pm$ 6.98e-03 $\pm$ 5.63e-03 & 5.50e-01 $\pm$ 6.63e-03 $\pm$  5.50e-03 \\ 
3.25e-01 &  5.00e-02 &  4.75e-01  $\pm$ 5.18e-03 $\pm$ 9.49e-03 & 4.83e-01 $\pm$ 5.08e-03 $\pm$  9.66e-03 \\ 
3.75e-01 &  5.00e-02 &  4.06e-01  $\pm$ 4.18e-03 $\pm$ 3.65e-02 & 4.28e-01 $\pm$ 4.21e-03 $\pm$  3.85e-02 \\ 
4.25e-01 &  5.00e-02 &  3.43e-01  $\pm$ 3.88e-03 $\pm$ 4.81e-02 & 3.58e-01 $\pm$ 3.88e-03 $\pm$  5.01e-02 \\ 
4.75e-01 &  5.00e-02 &  2.76e-01  $\pm$ 3.67e-03 $\pm$ 3.04e-02 & 2.95e-01 $\pm$ 3.66e-03 $\pm$  3.25e-02 \\ 
5.25e-01 &  5.00e-02 &  2.44e-01  $\pm$ 3.11e-03 $\pm$ 1.71e-02 & 2.64e-01 $\pm$ 3.13e-03 $\pm$  1.85e-02 \\ 
5.75e-01 &  5.00e-02 &  2.15e-01  $\pm$ 2.77e-03 $\pm$ 1.93e-02 & 2.30e-01 $\pm$ 2.83e-03 $\pm$  2.07e-02 \\ 
6.25e-01 &  5.00e-02 &  1.72e-01  $\pm$ 2.44e-03 $\pm$ 8.61e-03 & 1.88e-01 $\pm$ 2.43e-03 $\pm$  9.42e-03 \\ 
6.75e-01 &  5.00e-02 &  1.57e-01  $\pm$ 2.73e-03 $\pm$ 1.26e-02 & 1.68e-01 $\pm$ 2.79e-03 $\pm$  1.34e-02 \\ 
7.25e-01 &  5.00e-02 &  1.31e-01  $\pm$ 4.00e-03 $\pm$ 6.53e-03 & 1.44e-01 $\pm$ 4.25e-03 $\pm$  7.20e-03 \\ 
\hline  
\end{tabular}  
\end{center}  
\end{scriptsize}  
\end{table}  
\begin{table}[!h] 
\begin{scriptsize} 
\begin{center} 
\caption{Invariant yields for $K^{-}$ and $K^{+}$ at per-rapidity in {\bf 40 - 100 \%} ${\bf dAu}$ collisions at 200 GeV. \label{tab:Tables/dauKaonSpectrumper.tex}} 
\begin{tabular}{|c|c|c|c|} 
\hline   
 $p_{T}$ (GeV/c) &  $\Delta p_{T}$ (GeV/c) & $K^{-}$ & $K^{+}$ \\ \hline 
2.25e-01 &  5.00e-02 &  3.21e-01  $\pm$ 4.88e-03 $\pm$ 9.63e-03 & 2.99e-01 $\pm$ 4.47e-03 $\pm$  8.97e-03 \\ 
2.75e-01 &  5.00e-02 &  2.73e-01  $\pm$ 3.06e-03 $\pm$ 2.73e-03 & 2.71e-01 $\pm$ 2.92e-03 $\pm$  2.71e-03 \\ 
3.25e-01 &  5.00e-02 &  2.25e-01  $\pm$ 2.24e-03 $\pm$ 4.51e-03 & 2.31e-01 $\pm$ 2.20e-03 $\pm$  4.61e-03 \\ 
3.75e-01 &  5.00e-02 &  1.96e-01  $\pm$ 1.82e-03 $\pm$ 1.76e-02 & 1.99e-01 $\pm$ 1.80e-03 $\pm$  1.80e-02 \\ 
4.25e-01 &  5.00e-02 &  1.61e-01  $\pm$ 1.66e-03 $\pm$ 2.25e-02 & 1.68e-01 $\pm$ 1.64e-03 $\pm$  2.36e-02 \\ 
4.75e-01 &  5.00e-02 &  1.24e-01  $\pm$ 1.52e-03 $\pm$ 1.36e-02 & 1.34e-01 $\pm$ 1.53e-03 $\pm$  1.47e-02 \\ 
5.25e-01 &  5.00e-02 &  1.09e-01  $\pm$ 1.29e-03 $\pm$ 7.64e-03 & 1.22e-01 $\pm$ 1.30e-03 $\pm$  8.55e-03 \\ 
5.75e-01 &  5.00e-02 &  9.63e-02  $\pm$ 1.16e-03 $\pm$ 8.67e-03 & 1.05e-01 $\pm$ 1.12e-03 $\pm$  9.41e-03 \\ 
6.25e-01 &  5.00e-02 &  7.67e-02  $\pm$ 9.97e-04 $\pm$ 3.84e-03 & 8.23e-02 $\pm$ 9.92e-04 $\pm$  4.11e-03 \\ 
6.75e-01 &  5.00e-02 &  6.76e-02  $\pm$ 1.15e-03 $\pm$ 5.41e-03 & 7.36e-02 $\pm$ 1.12e-03 $\pm$  5.89e-03 \\ 
7.25e-01 &  5.00e-02 &  6.16e-02  $\pm$ 1.80e-03 $\pm$ 3.08e-03 & 6.11e-02 $\pm$ 1.67e-03 $\pm$  3.06e-03 \\ 
\hline  
\end{tabular}  
\end{center}  
\end{scriptsize}  
\end{table}  
%
\begin{table}[!h] 
\begin{scriptsize} 
\begin{center} 
\caption{Invariant yields for $\overline{p}$ and $p$ at mid-rapidity in {\bf Minimum Bias} ${\bf dAu}$ collisions at 200 GeV. \label{tab:Tables/dauPSpectrummb.tex}} 
\begin{tabular}{|c|c|c|c|} 
\hline   
 $p_{T}$ (GeV/c) &  $\Delta p_{T}$ (GeV/c) & $\overline{p}$ & $p$ \\ \hline 
3.75e-01 &  5.00e-02 &  1.65e-01  $\pm$ 1.05e-03 $\pm$ 1.65e-03 & 1.53e-01 $\pm$ 2.13e-03 $\pm$  1.53e-03 \\ 
4.25e-01 &  5.00e-02 &  1.45e-01  $\pm$ 8.75e-04 $\pm$ 1.45e-03 & 1.53e-01 $\pm$ 2.10e-03 $\pm$  1.53e-03 \\ 
4.75e-01 &  5.00e-02 &  1.32e-01  $\pm$ 7.73e-04 $\pm$ 1.32e-03 & 1.41e-01 $\pm$ 2.04e-03 $\pm$  1.41e-03 \\ 
5.25e-01 &  5.00e-02 &  1.17e-01  $\pm$ 6.92e-04 $\pm$ 1.17e-03 & 1.32e-01 $\pm$ 2.07e-03 $\pm$  1.32e-03 \\ 
5.75e-01 &  5.00e-02 &  1.06e-01  $\pm$ 6.31e-04 $\pm$ 1.06e-03 & 1.19e-01 $\pm$ 2.11e-03 $\pm$  1.19e-03 \\ 
6.25e-01 &  5.00e-02 &  9.43e-02  $\pm$ 5.67e-04 $\pm$ 1.89e-03 & 1.06e-01 $\pm$ 1.96e-03 $\pm$  2.12e-03 \\ 
6.75e-01 &  5.00e-02 &  8.41e-02  $\pm$ 5.17e-04 $\pm$ 1.68e-03 & 9.91e-02 $\pm$ 1.94e-03 $\pm$  1.98e-03 \\ 
7.25e-01 &  5.00e-02 &  7.47e-02  $\pm$ 4.71e-04 $\pm$ 7.47e-04 & 8.85e-02 $\pm$ 1.90e-03 $\pm$  8.85e-04 \\ 
7.75e-01 &  5.00e-02 &  6.57e-02  $\pm$ 4.29e-04 $\pm$ 1.97e-03 & 7.64e-02 $\pm$ 1.77e-03 $\pm$  2.29e-03 \\ 
8.25e-01 &  5.00e-02 &  5.88e-02  $\pm$ 3.99e-04 $\pm$ 2.94e-03 & 6.80e-02 $\pm$ 1.55e-03 $\pm$  3.40e-03 \\ 
8.75e-01 &  5.00e-02 &  5.45e-02  $\pm$ 3.75e-04 $\pm$ 2.72e-03 & 6.01e-02 $\pm$ 1.12e-03 $\pm$  3.01e-03 \\ 
9.25e-01 &  5.00e-02 &  4.77e-02  $\pm$ 3.35e-04 $\pm$ 9.55e-04 & 5.38e-02 $\pm$ 1.05e-03 $\pm$  1.08e-03 \\ 
9.75e-01 &  5.00e-02 &  4.08e-02  $\pm$ 3.05e-04 $\pm$ 8.16e-04 & 4.74e-02 $\pm$ 9.78e-04 $\pm$  9.48e-04 \\ 
1.03e+00 &  5.00e-02 &  3.50e-02  $\pm$ 2.91e-04 $\pm$ 2.45e-03 & 3.86e-02 $\pm$ 8.70e-04 $\pm$  2.70e-03 \\ 
1.08e+00 &  5.00e-02 &  3.12e-02  $\pm$ 4.27e-04 $\pm$ 9.37e-04 & 3.38e-02 $\pm$ 8.57e-04 $\pm$  1.01e-03 \\ 
\hline  
\end{tabular}  
\end{center}  
\end{scriptsize}  
\end{table}  
\begin{table}[!h] 
\begin{scriptsize} 
\begin{center} 
\caption{Invariant yields for $\overline{p}$ and $p$ at mid-rapidity in {\bf 0 - 20 \%} ${\bf dAu}$ collisions at 200 GeV. \label{tab:Tables/dauPSpectrumcent.tex}} 
\begin{tabular}{|c|c|c|c|} 
\hline   
 $p_{T}$ (GeV/c) &  $\Delta p_{T}$ (GeV/c) & $\overline{p}$ & $p$ \\ \hline 
3.75e-01 &  5.00e-02 &  2.92e-01  $\pm$ 3.16e-03 $\pm$ 2.92e-03 & 2.78e-01 $\pm$ 2.18e-03 $\pm$  2.78e-03 \\ 
4.25e-01 &  5.00e-02 &  2.55e-01  $\pm$ 2.63e-03 $\pm$ 2.55e-03 & 2.50e-01 $\pm$ 1.95e-03 $\pm$  2.50e-03 \\ 
4.75e-01 &  5.00e-02 &  2.35e-01  $\pm$ 2.34e-03 $\pm$ 2.35e-03 & 2.32e-01 $\pm$ 1.84e-03 $\pm$  2.32e-03 \\ 
5.25e-01 &  5.00e-02 &  2.12e-01  $\pm$ 2.11e-03 $\pm$ 2.12e-03 & 2.22e-01 $\pm$ 1.79e-03 $\pm$  2.22e-03 \\ 
5.75e-01 &  5.00e-02 &  1.93e-01  $\pm$ 1.93e-03 $\pm$ 1.93e-03 & 2.04e-01 $\pm$ 1.69e-03 $\pm$  2.04e-03 \\ 
6.25e-01 &  5.00e-02 &  1.74e-01  $\pm$ 1.75e-03 $\pm$ 3.48e-03 & 1.92e-01 $\pm$ 1.62e-03 $\pm$  3.84e-03 \\ 
6.75e-01 &  5.00e-02 &  1.57e-01  $\pm$ 1.61e-03 $\pm$ 3.14e-03 & 1.75e-01 $\pm$ 1.53e-03 $\pm$  3.50e-03 \\ 
7.25e-01 &  5.00e-02 &  1.43e-01  $\pm$ 1.49e-03 $\pm$ 1.43e-03 & 1.58e-01 $\pm$ 1.42e-03 $\pm$  1.58e-03 \\ 
7.75e-01 &  5.00e-02 &  1.26e-01  $\pm$ 1.35e-03 $\pm$ 3.78e-03 & 1.44e-01 $\pm$ 1.34e-03 $\pm$  4.31e-03 \\ 
8.25e-01 &  5.00e-02 &  1.14e-01  $\pm$ 1.27e-03 $\pm$ 5.68e-03 & 1.31e-01 $\pm$ 1.26e-03 $\pm$  6.53e-03 \\ 
8.75e-01 &  5.00e-02 &  1.01e-01  $\pm$ 1.14e-03 $\pm$ 5.06e-03 & 1.23e-01 $\pm$ 1.18e-03 $\pm$  6.13e-03 \\ 
9.25e-01 &  5.00e-02 &  8.80e-02  $\pm$ 1.05e-03 $\pm$ 1.76e-03 & 1.08e-01 $\pm$ 1.08e-03 $\pm$  2.16e-03 \\ 
9.75e-01 &  5.00e-02 &  7.92e-02  $\pm$ 9.81e-04 $\pm$ 1.58e-03 & 9.54e-02 $\pm$ 1.02e-03 $\pm$  1.91e-03 \\ 
1.03e+00 &  5.00e-02 &  6.90e-02  $\pm$ 9.14e-04 $\pm$ 4.83e-03 & 8.09e-02 $\pm$ 9.63e-04 $\pm$  5.66e-03 \\ 
1.08e+00 &  5.00e-02 &  6.21e-02  $\pm$ 9.03e-04 $\pm$ 1.86e-03 & 7.29e-02 $\pm$ 8.57e-04 $\pm$  2.19e-03 \\ 
\hline  
\end{tabular}  
\end{center}  
\end{scriptsize}  
\end{table}  
\input{Tables/dauPSpectrummid}
\input{Tables/dauPSpectrumper}
%
\clearpage
\begin{table}[!h] 
\begin{scriptsize} 
\begin{center} 
\caption{Invariant yields for $\pi^{-}$ and $\pi^{+}$ at mid-rapidity in 0 - 5 \% ${\bf Au-Au}$ collisions at 62.4 GeV. \label{tab:Tables/auau62PionSpectrum05.tex}} 
\begin{tabular}{|c|c|c|c|} 
\hline   
 $p_{T}$ (GeV/c) &  $\Delta p_{T}$ (GeV/c) & $\pi^{-}$  & $\pi^{+}$  \\ \hline 
2.25e-01 &  5.00e-02 &  3.00e+02  $\pm$ 3.04e-01 $\pm$ 6.01e+00 & 2.95e+02 $\pm$ 3.02e-01 $\pm$  5.91e+00 \\ 
2.75e-01 &  5.00e-02 &  2.33e+02  $\pm$ 2.48e-01 $\pm$ 4.66e+00 & 2.27e+02 $\pm$ 2.40e-01 $\pm$  4.54e+00 \\ 
3.25e-01 &  5.00e-02 &  1.80e+02  $\pm$ 1.79e-01 $\pm$ 3.59e+00 & 1.75e+02 $\pm$ 1.78e-01 $\pm$  3.50e+00 \\ 
3.75e-01 &  5.00e-02 &  1.39e+02  $\pm$ 1.42e-01 $\pm$ 1.39e+00 & 1.36e+02 $\pm$ 1.41e-01 $\pm$  1.36e+00 \\ 
4.25e-01 &  5.00e-02 &  1.07e+02  $\pm$ 1.17e-01 $\pm$ 1.07e+00 & 1.05e+02 $\pm$ 1.17e-01 $\pm$  1.05e+00 \\ 
4.75e-01 &  5.00e-02 &  8.32e+01  $\pm$ 9.68e-02 $\pm$ 8.32e-01 & 8.21e+01 $\pm$ 9.64e-02 $\pm$  8.21e-01 \\ 
5.25e-01 &  5.00e-02 &  6.47e+01  $\pm$ 8.12e-02 $\pm$ 6.47e-01 & 6.38e+01 $\pm$ 8.06e-02 $\pm$  6.38e-01 \\ 
5.75e-01 &  5.00e-02 &  5.05e+01  $\pm$ 6.89e-02 $\pm$ 1.01e+00 & 4.99e+01 $\pm$ 6.86e-02 $\pm$  9.97e-01 \\ 
6.25e-01 &  5.00e-02 &  3.96e+01  $\pm$ 5.99e-02 $\pm$ 7.91e-01 & 3.90e+01 $\pm$ 5.99e-02 $\pm$  7.80e-01 \\ 
6.75e-01 &  5.00e-02 &  3.14e+01  $\pm$ 5.65e-02 $\pm$ 9.41e-01 & 3.09e+01 $\pm$ 5.65e-02 $\pm$  9.27e-01 \\ 
\hline  
\end{tabular}  
\end{center}  
\end{scriptsize}  
\end{table}  
\input{Tables/auau62PionSpectrum510}
\input{Tables/auau62PionSpectrum1020}
\input{Tables/auau62PionSpectrum2030}
\input{Tables/auau62PionSpectrum3040}
\input{Tables/auau62PionSpectrum4050}
\input{Tables/auau62PionSpectrum5060}
\input{Tables/auau62PionSpectrum6070}
\input{Tables/auau62PionSpectrum7080}
%
%
\input{Tables/auau62KaonSpectrum05}
\input{Tables/auau62KaonSpectrum510} 
\input{Tables/auau62KaonSpectrum1020}
\input{Tables/auau62KaonSpectrum2030}
\input{Tables/auau62KaonSpectrum3040}
\input{Tables/auau62KaonSpectrum4050}
\input{Tables/auau62KaonSpectrum5060}
\input{Tables/auau62KaonSpectrum6070}
\input{Tables/auau62KaonSpectrum7080}

\newpage 
%
%
\clearpage

\input{Tables/auau62PSpectrum05}
\input{Tables/auau62PSpectrum510}
\input{Tables/auau62PSpectrum1020}
\input{Tables/auau62PSpectrum2030}
\input{Tables/auau62PSpectrum3040}
\input{Tables/auau62PSpectrum4050}
\input{Tables/auau62PSpectrum5060}
\input{Tables/auau62PSpectrum6070}
\input{Tables/auau62PSpectrum7080}

\clearpage
\samepage

\chapter{Comparison of $dca$ and $N_{fit}$ distributions from real data and from embedding }\label{app:dataembed}

\begin{sidewaysfigure}[!h]
\begin{center}
\resizebox{.225\textwidth}{!}{\includegraphics{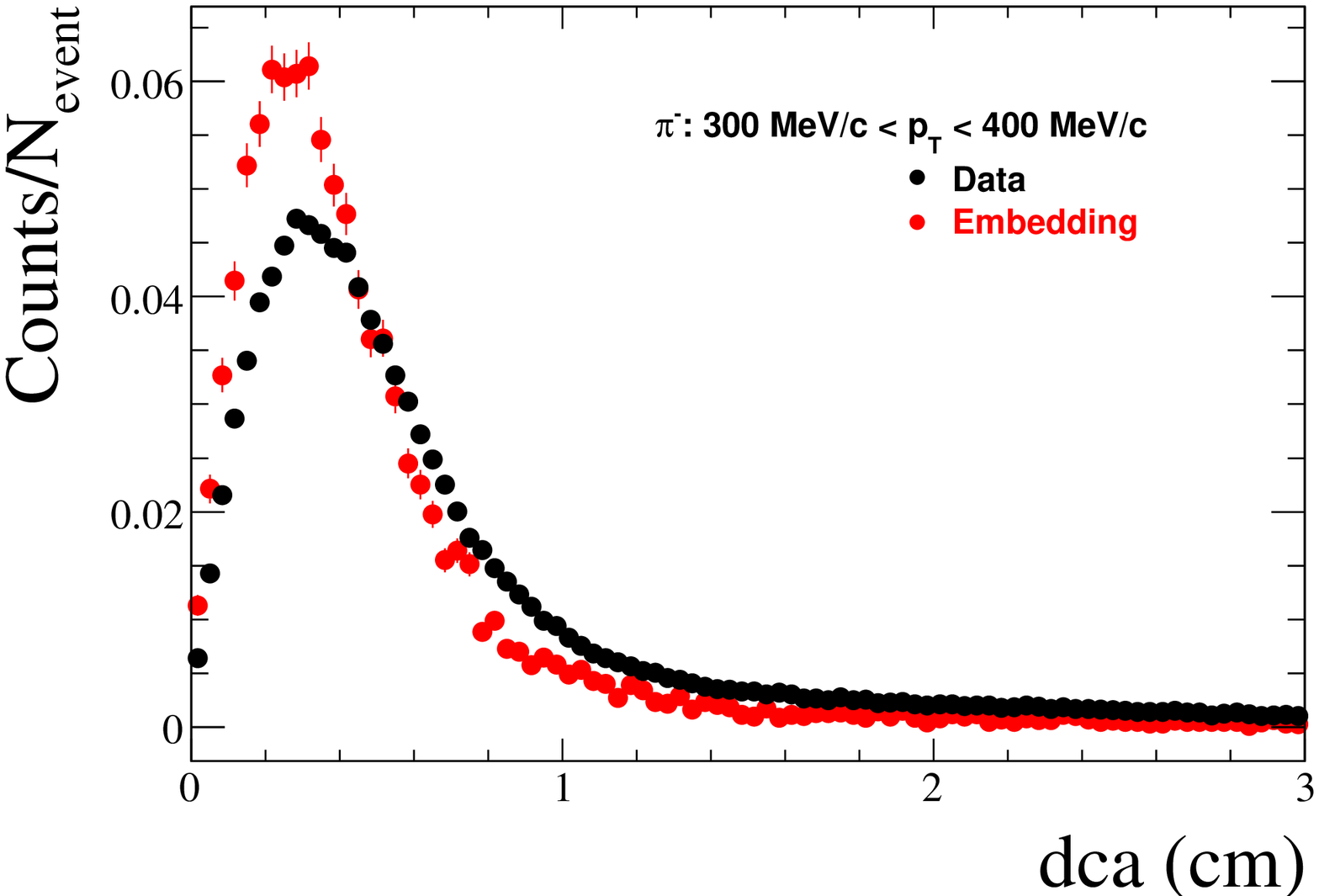}}
\resizebox{.225\textwidth}{!}{\includegraphics{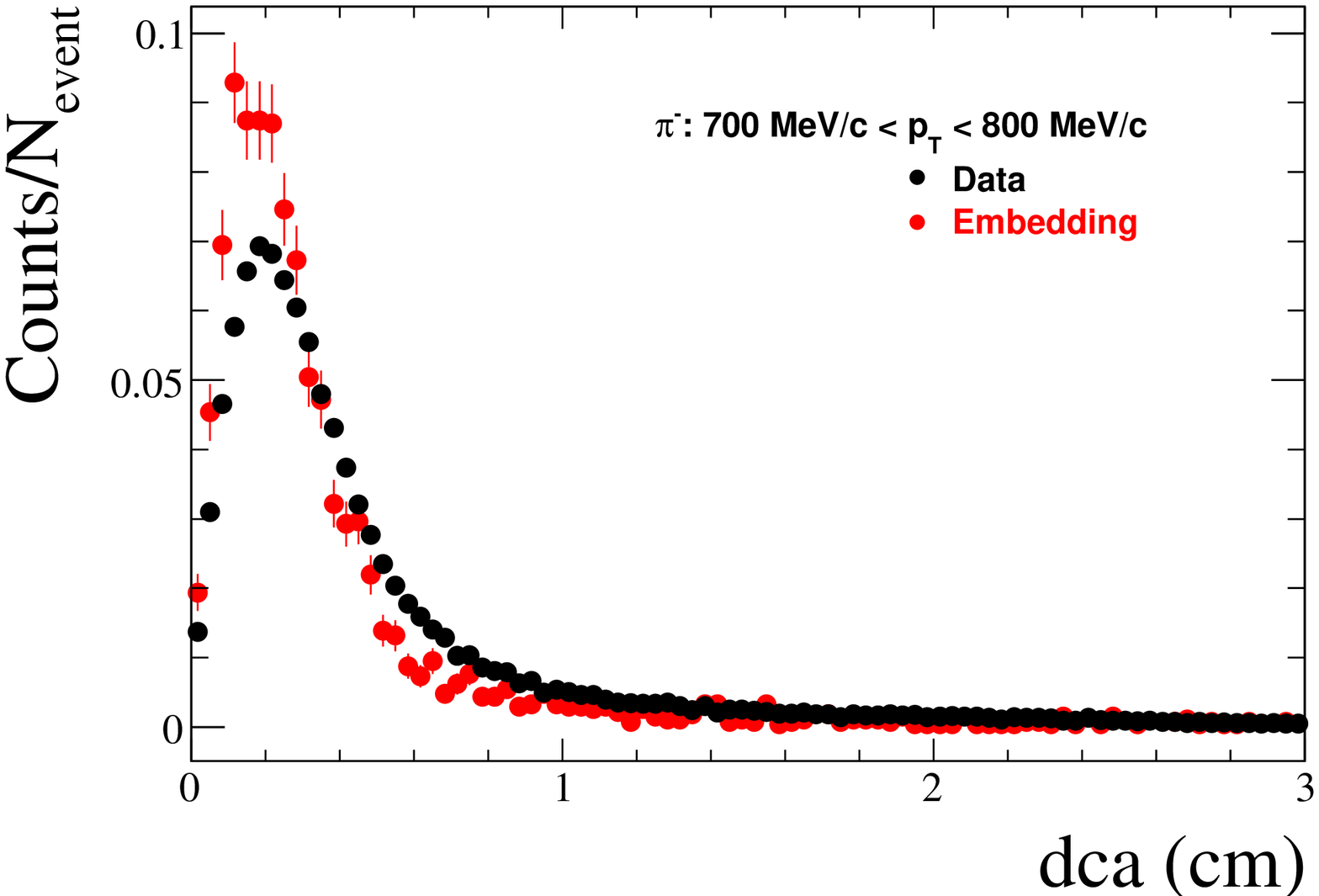}}
\resizebox{.225\textwidth}{!}{\includegraphics{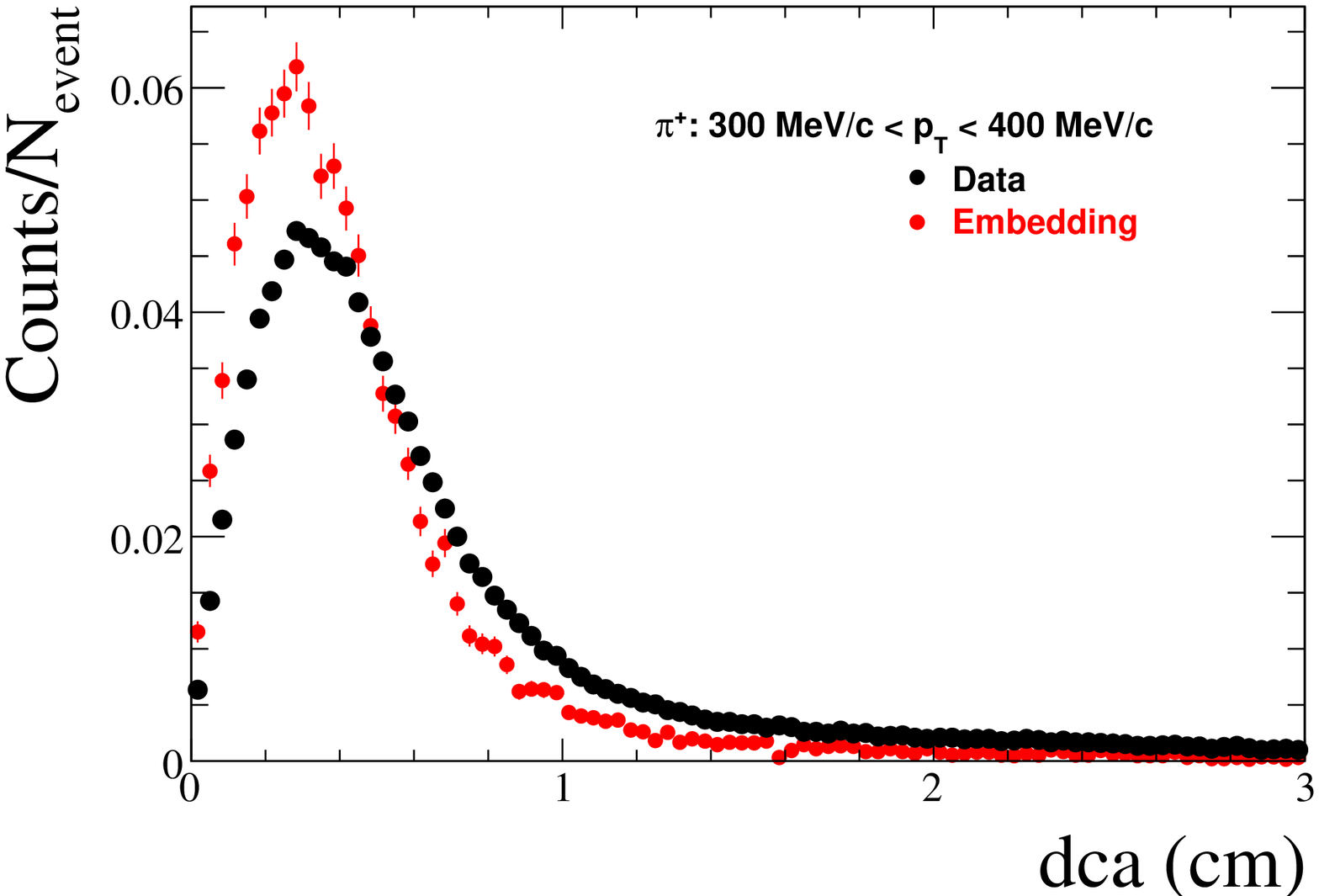}}
\resizebox{.225\textwidth}{!}{\includegraphics{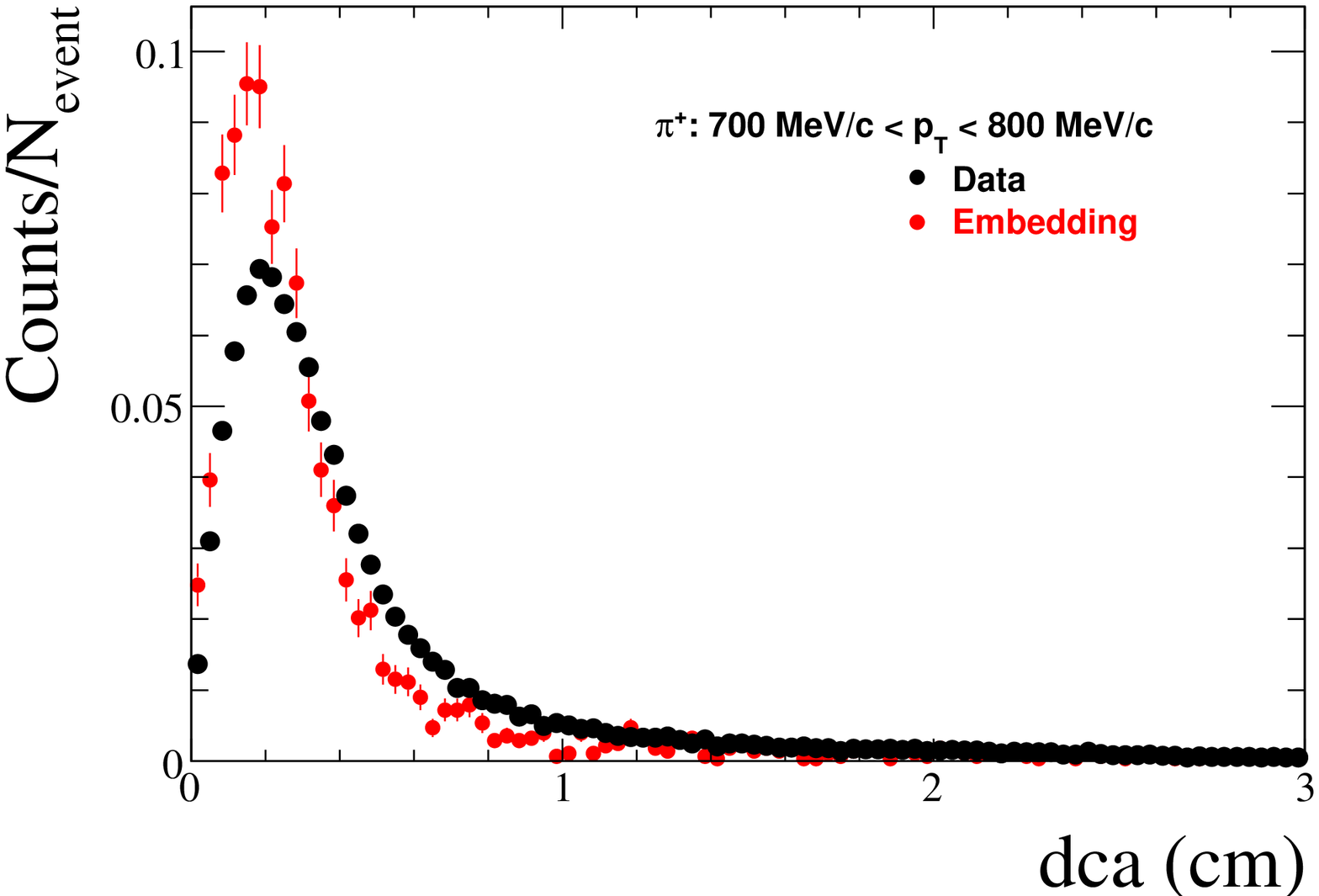}}\\
\resizebox{.225\textwidth}{!}{\includegraphics{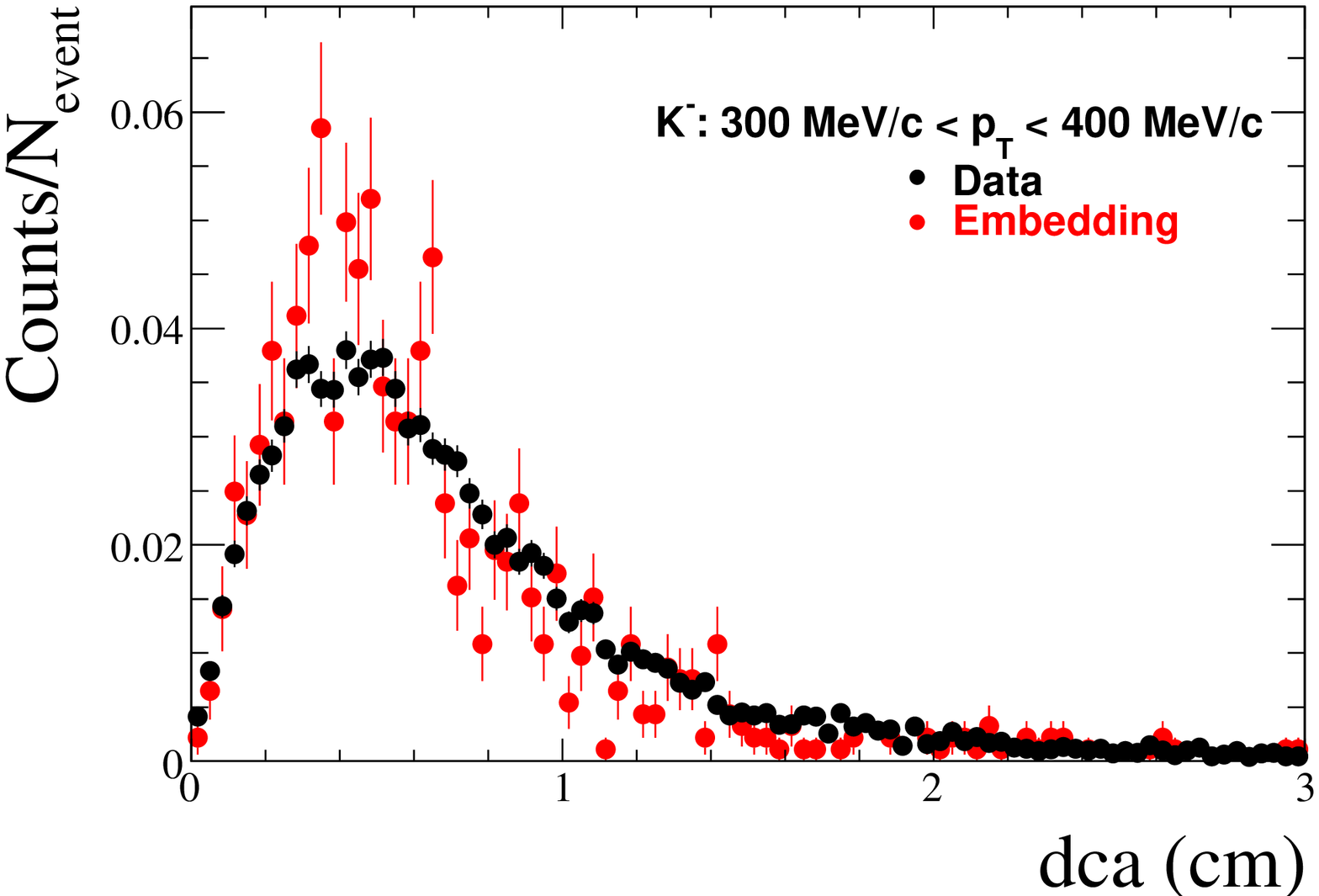}}
\resizebox{.225\textwidth}{!}{\includegraphics{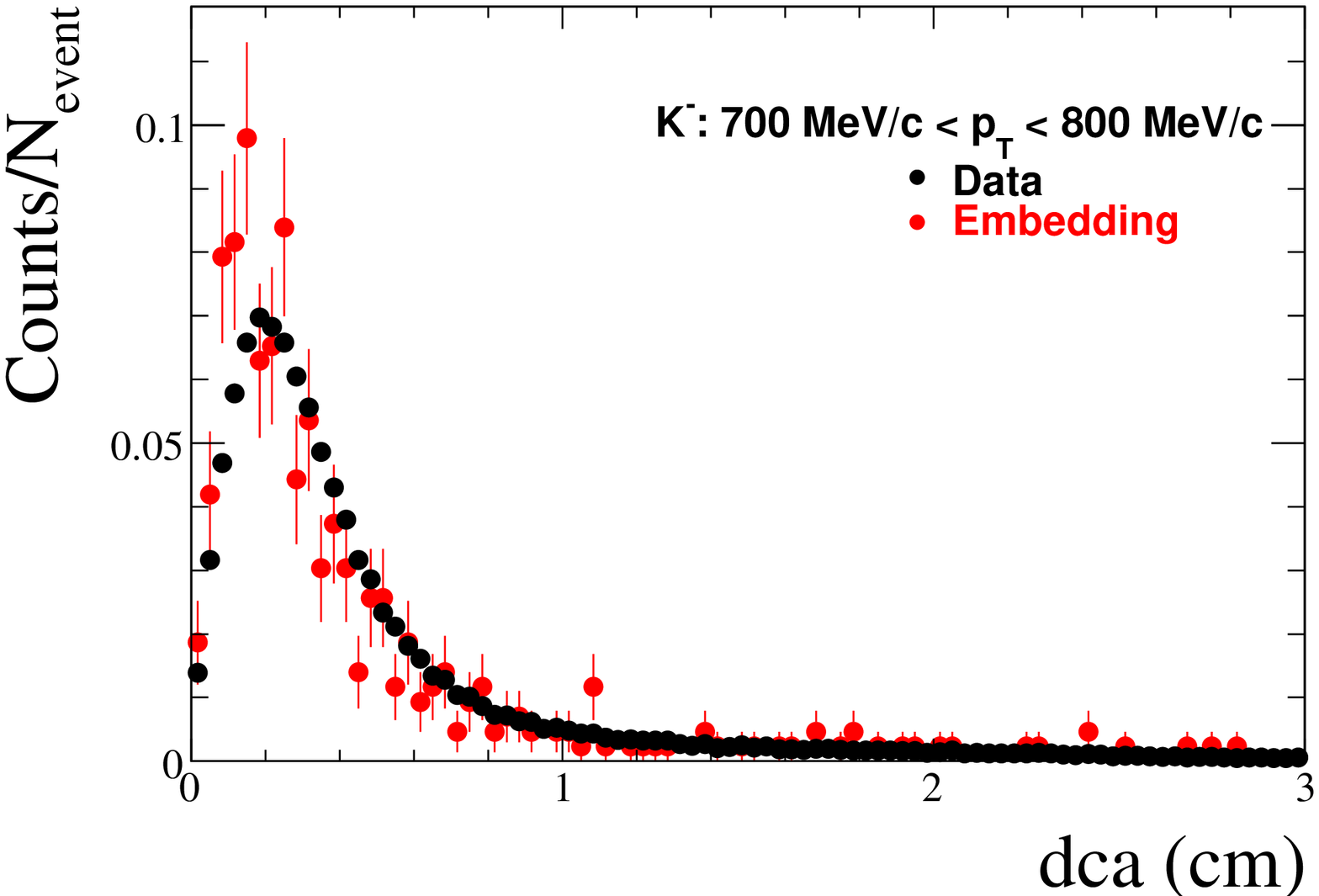}}
\resizebox{.225\textwidth}{!}{\includegraphics{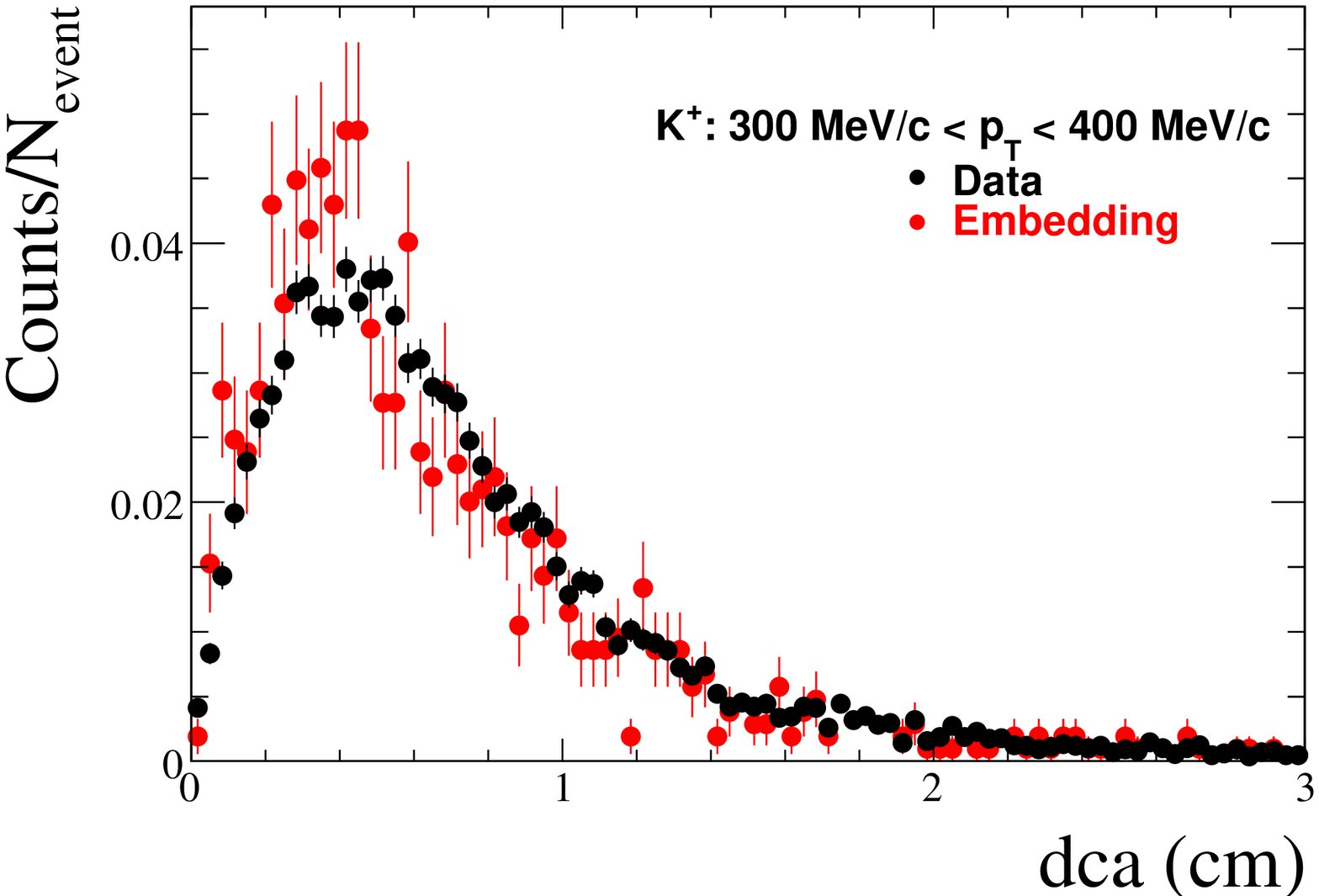}}
\resizebox{.225\textwidth}{!}{\includegraphics{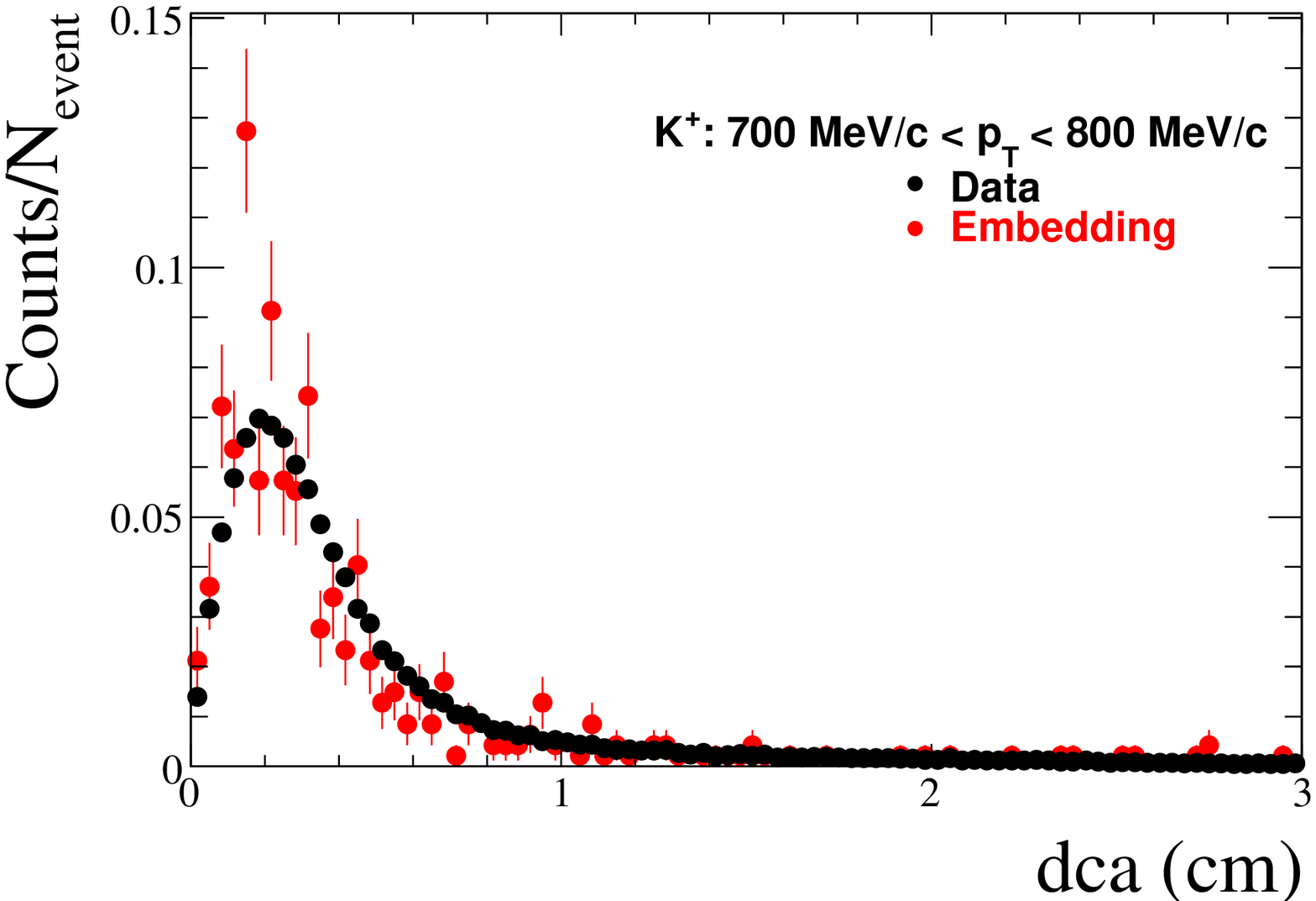}}\\
\resizebox{.225\textwidth}{!}{\includegraphics{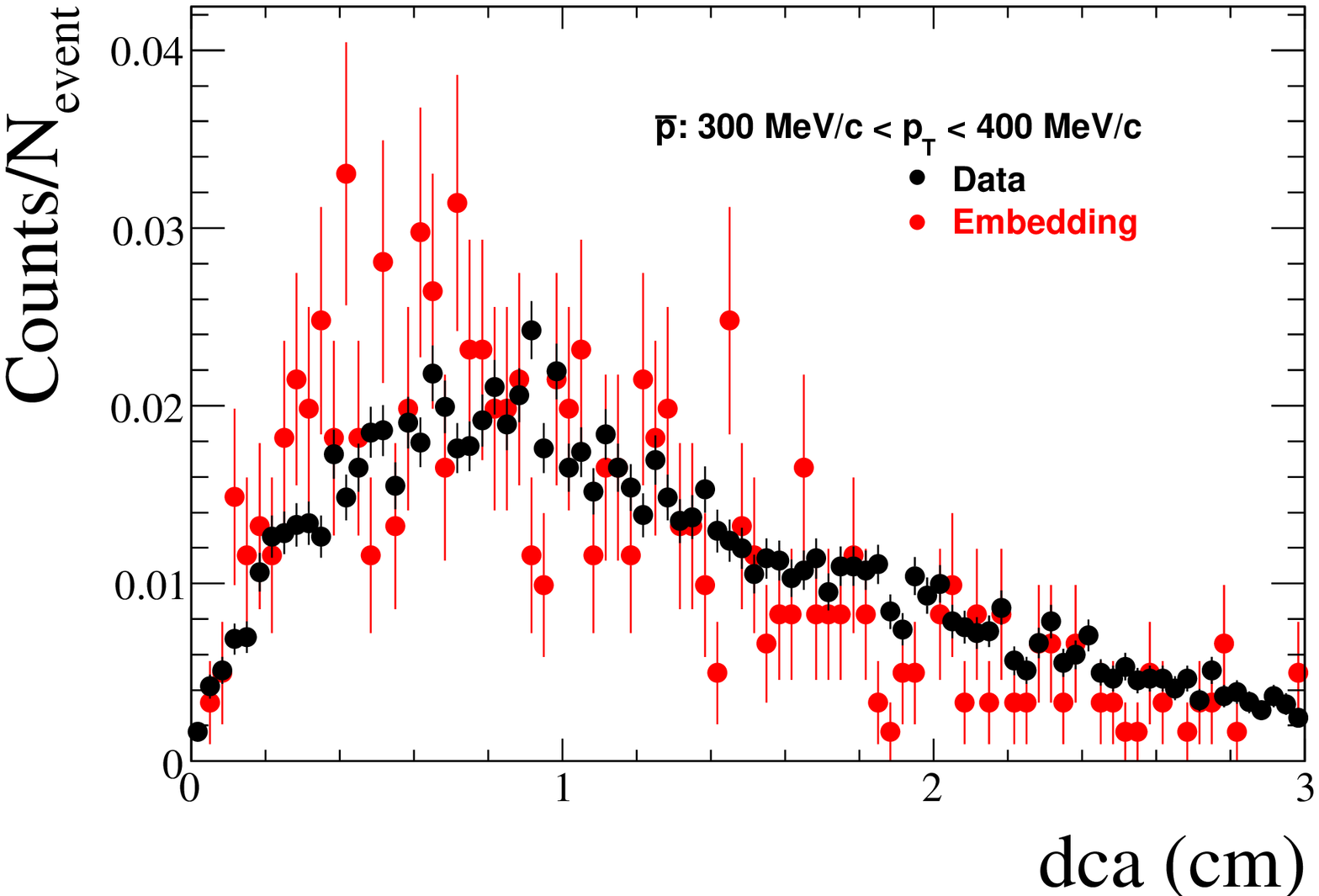}}
\resizebox{.225\textwidth}{!}{\includegraphics{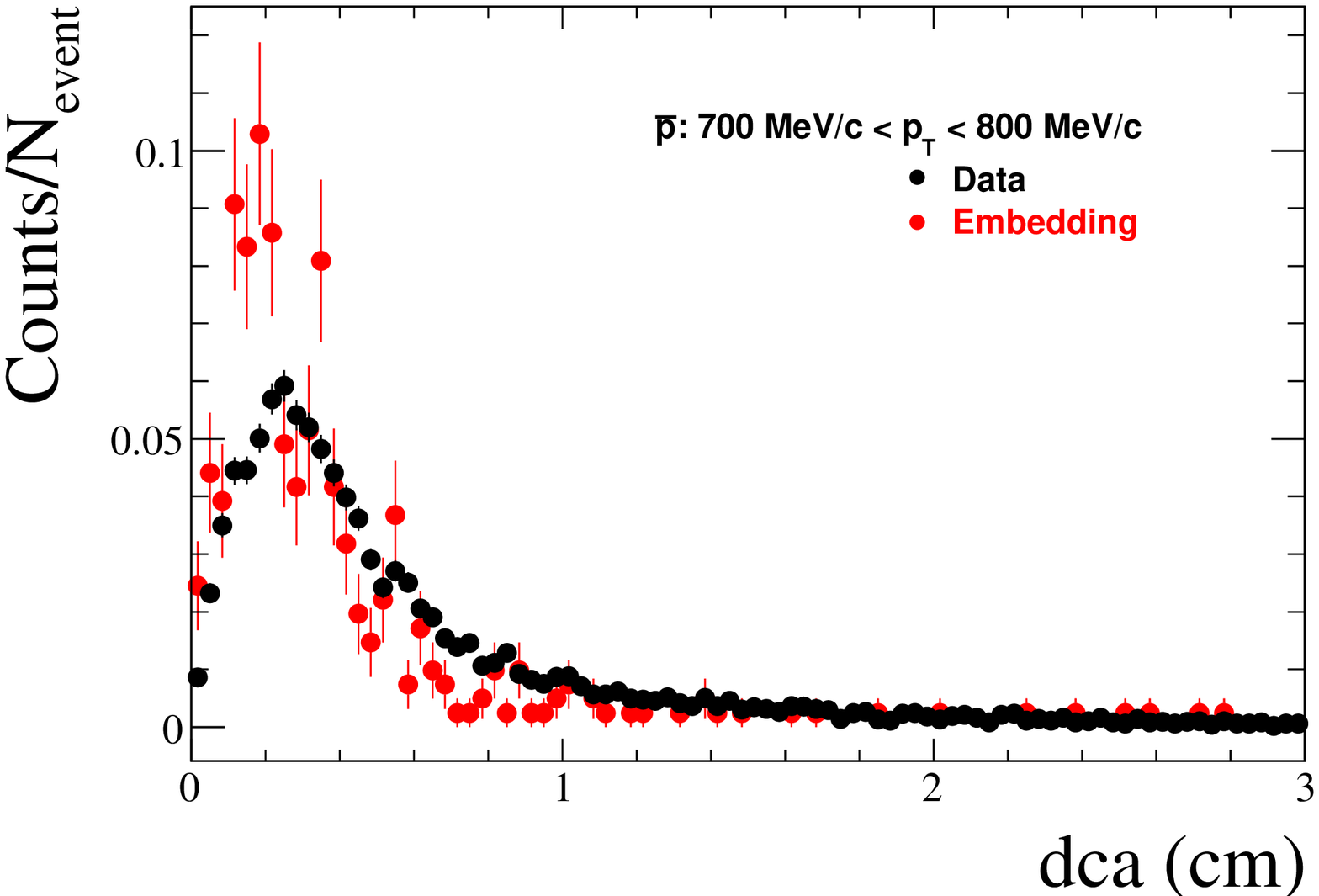}}
\resizebox{.225\textwidth}{!}{\includegraphics{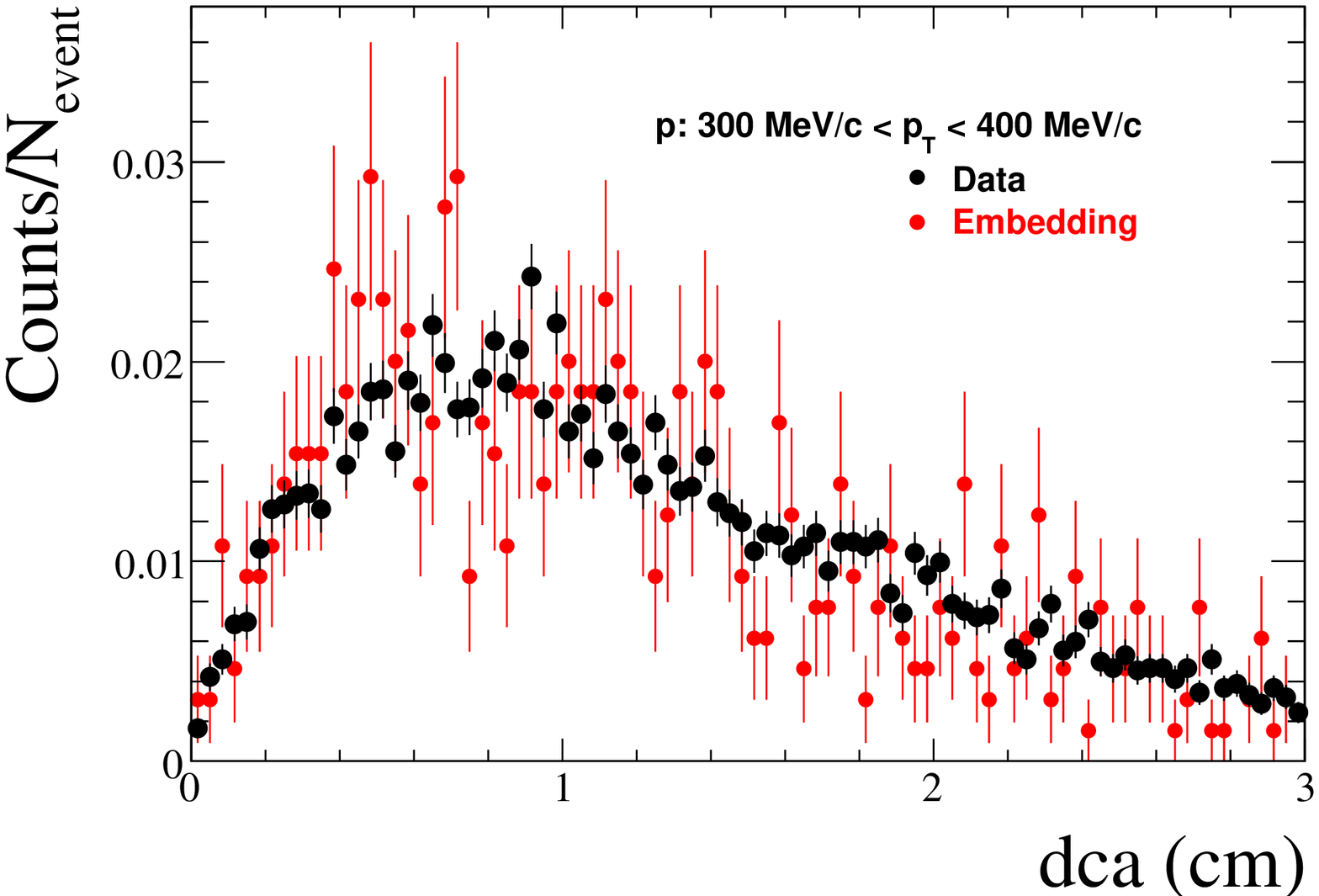}}
\resizebox{.225\textwidth}{!}{\includegraphics{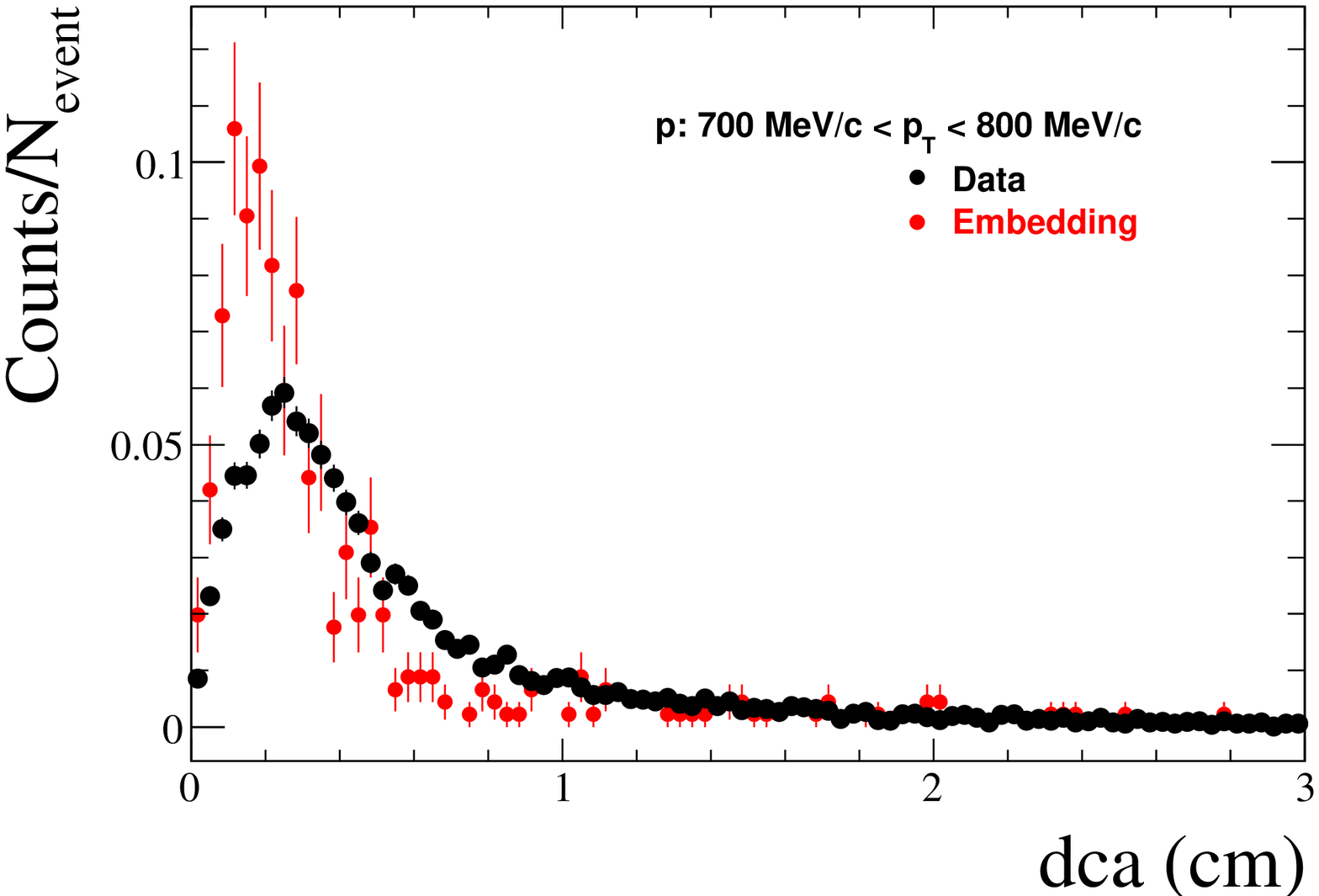}}\\
\caption{Comparison of $dca$ from real data and embedding in 200 GeV minimum bias pp collisions.}
\label{fig:pp_dca_data_embedding}
\end{center} 
\end{sidewaysfigure}
\begin{sidewaysfigure}[!h]
\begin{center}
\resizebox{.225\textwidth}{!}{\includegraphics{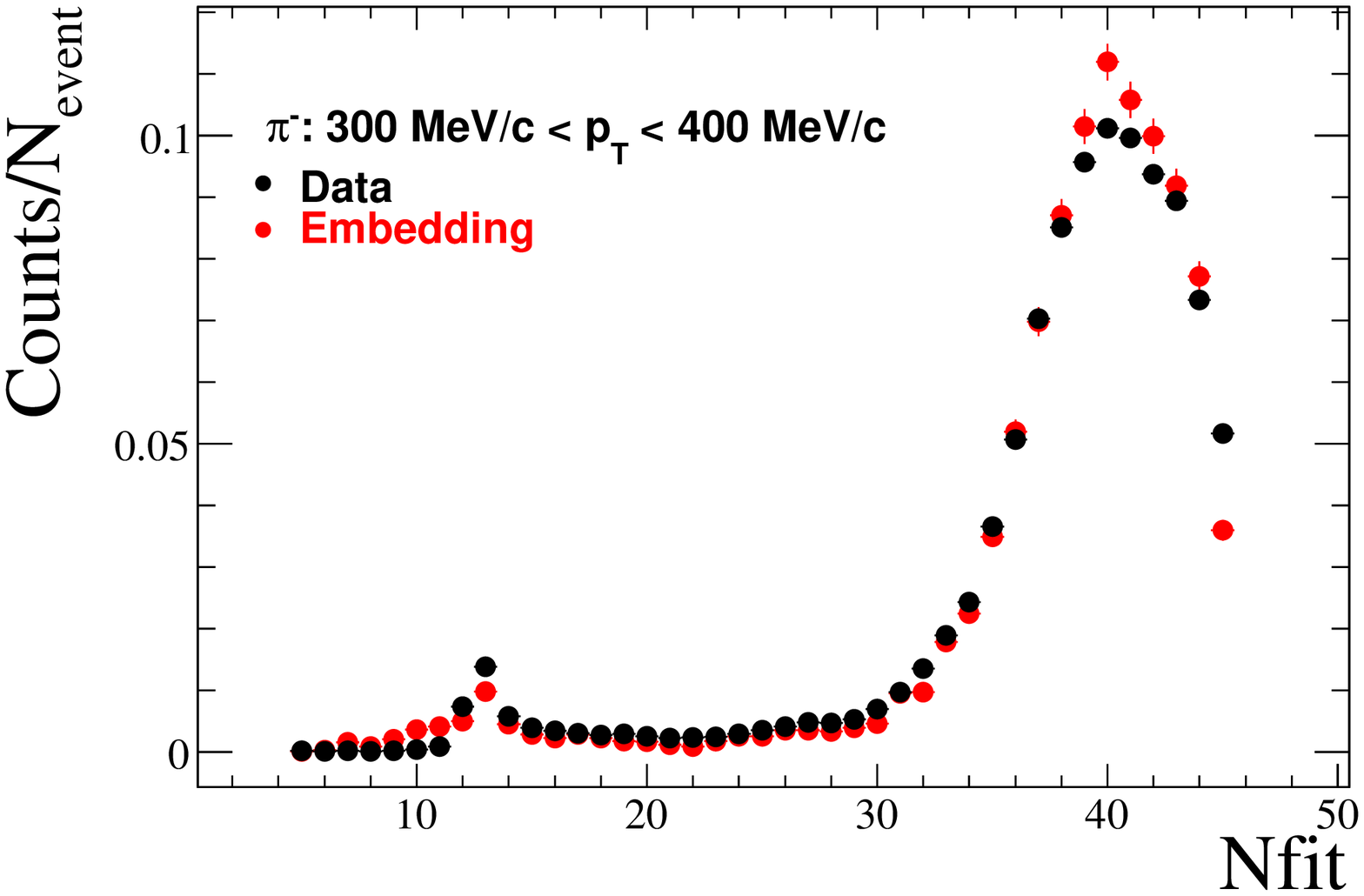}}
\resizebox{.225\textwidth}{!}{\includegraphics{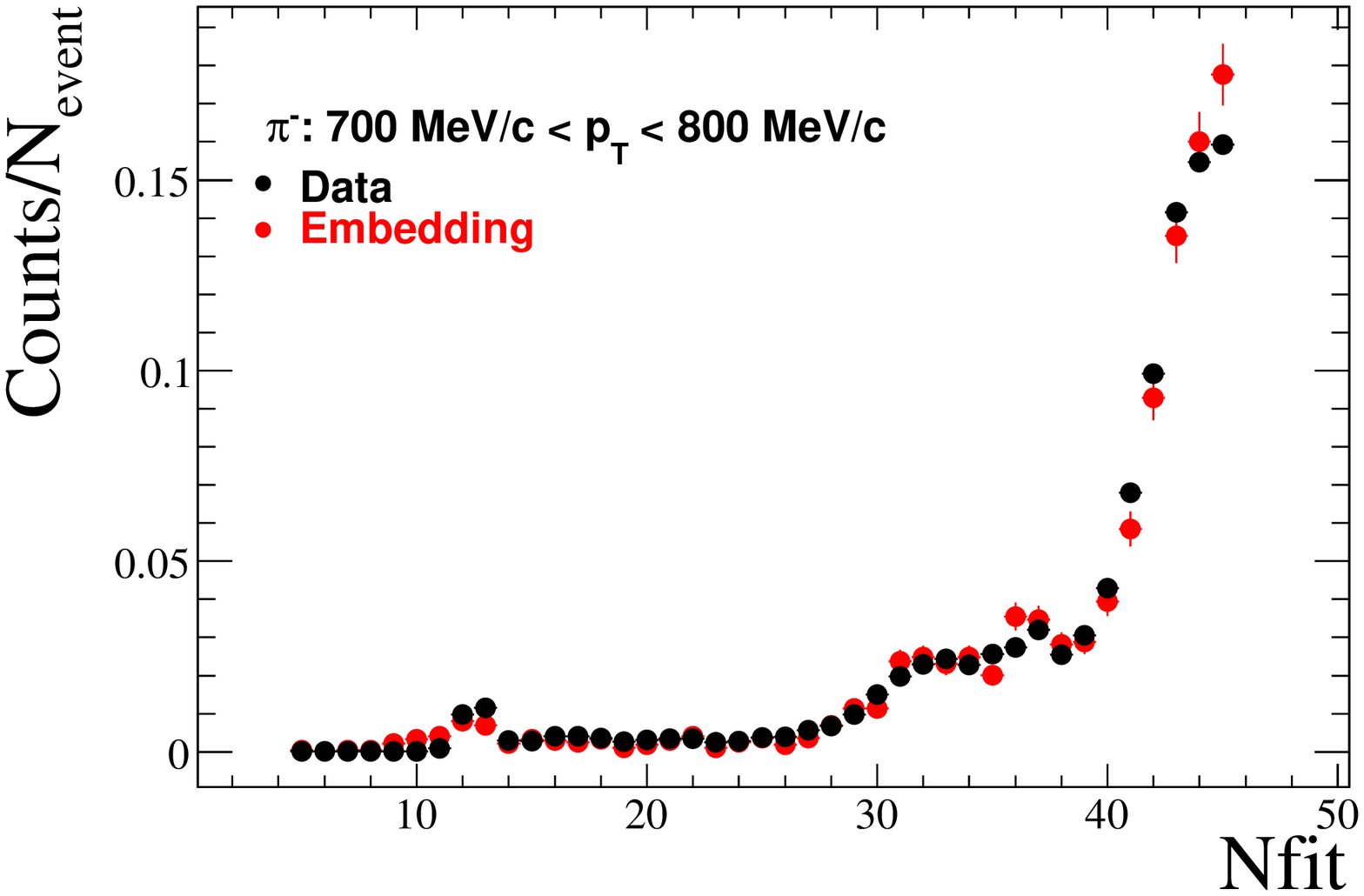}}
\resizebox{.225\textwidth}{!}{\includegraphics{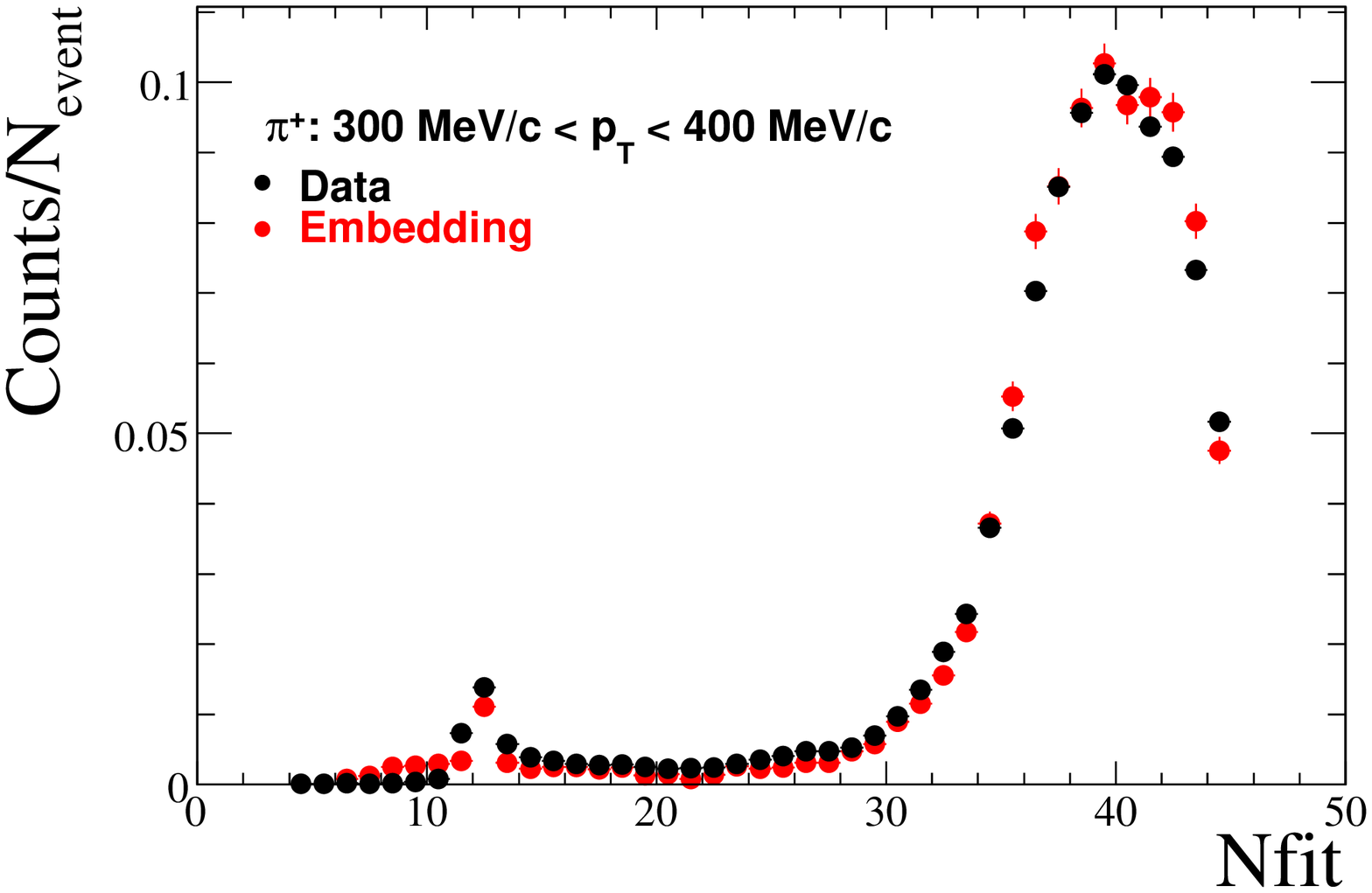}}
\resizebox{.225\textwidth}{!}{\includegraphics{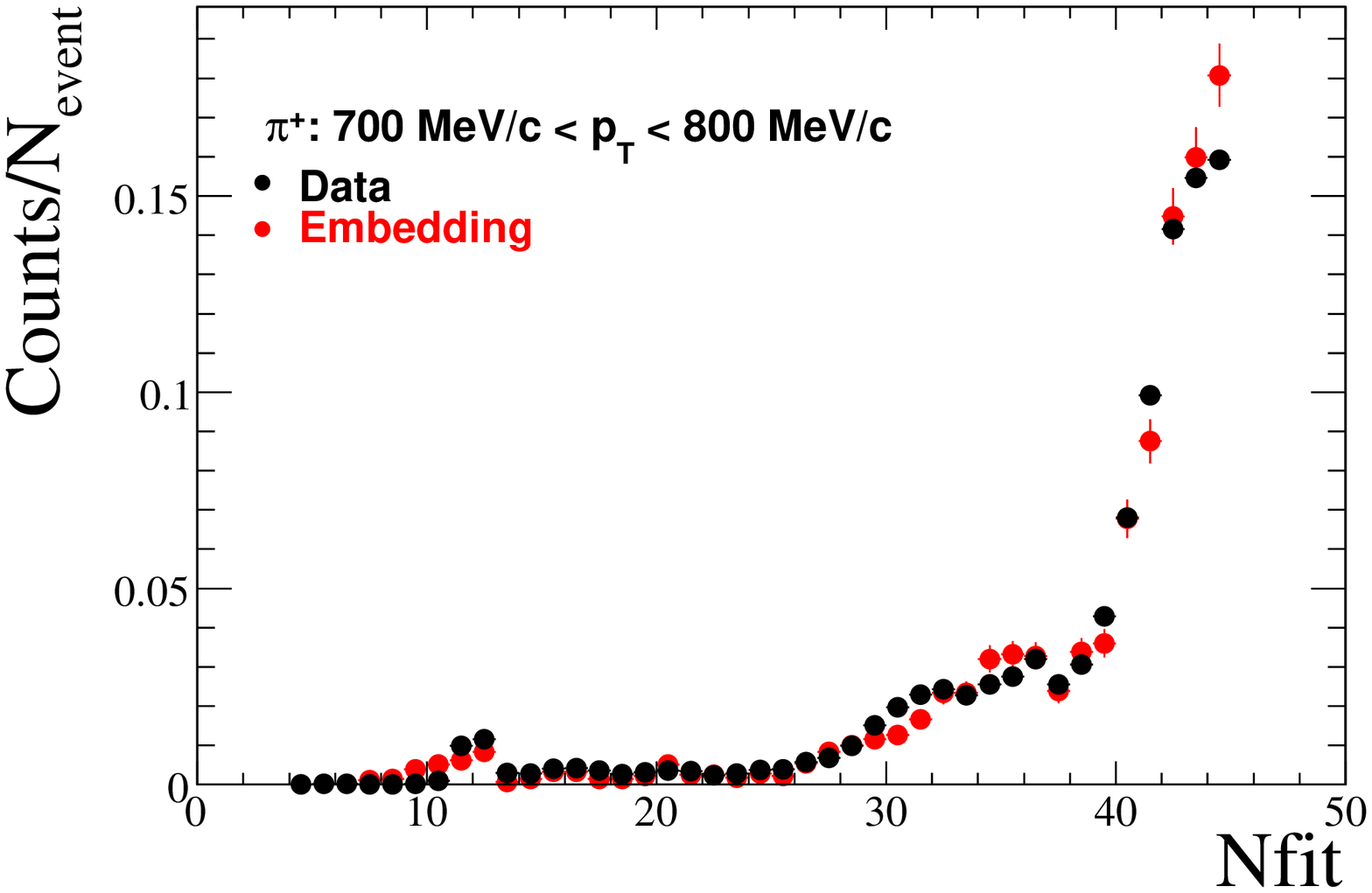}}\\
\resizebox{.225\textwidth}{!}{\includegraphics{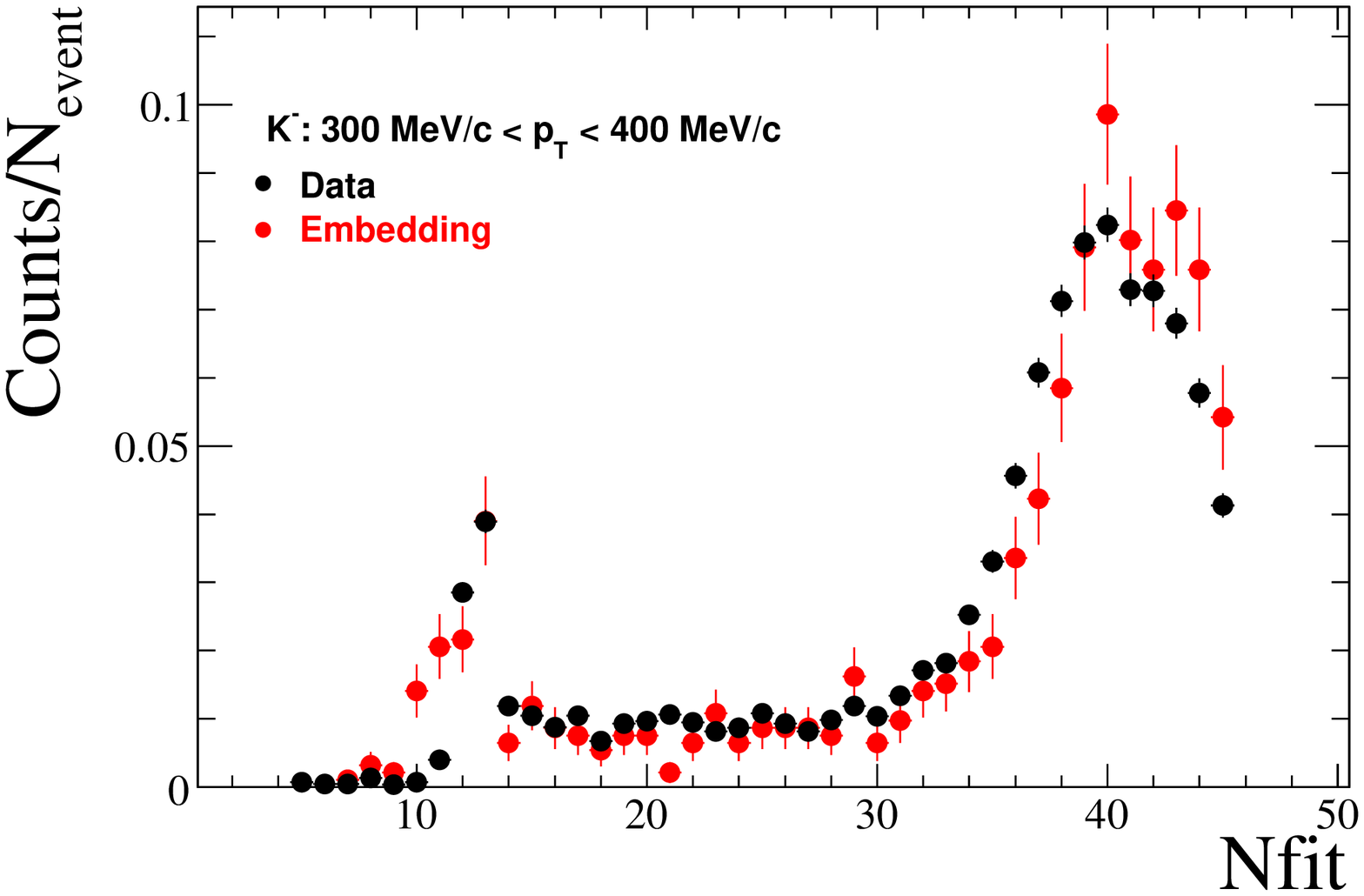}}
\resizebox{.225\textwidth}{!}{\includegraphics{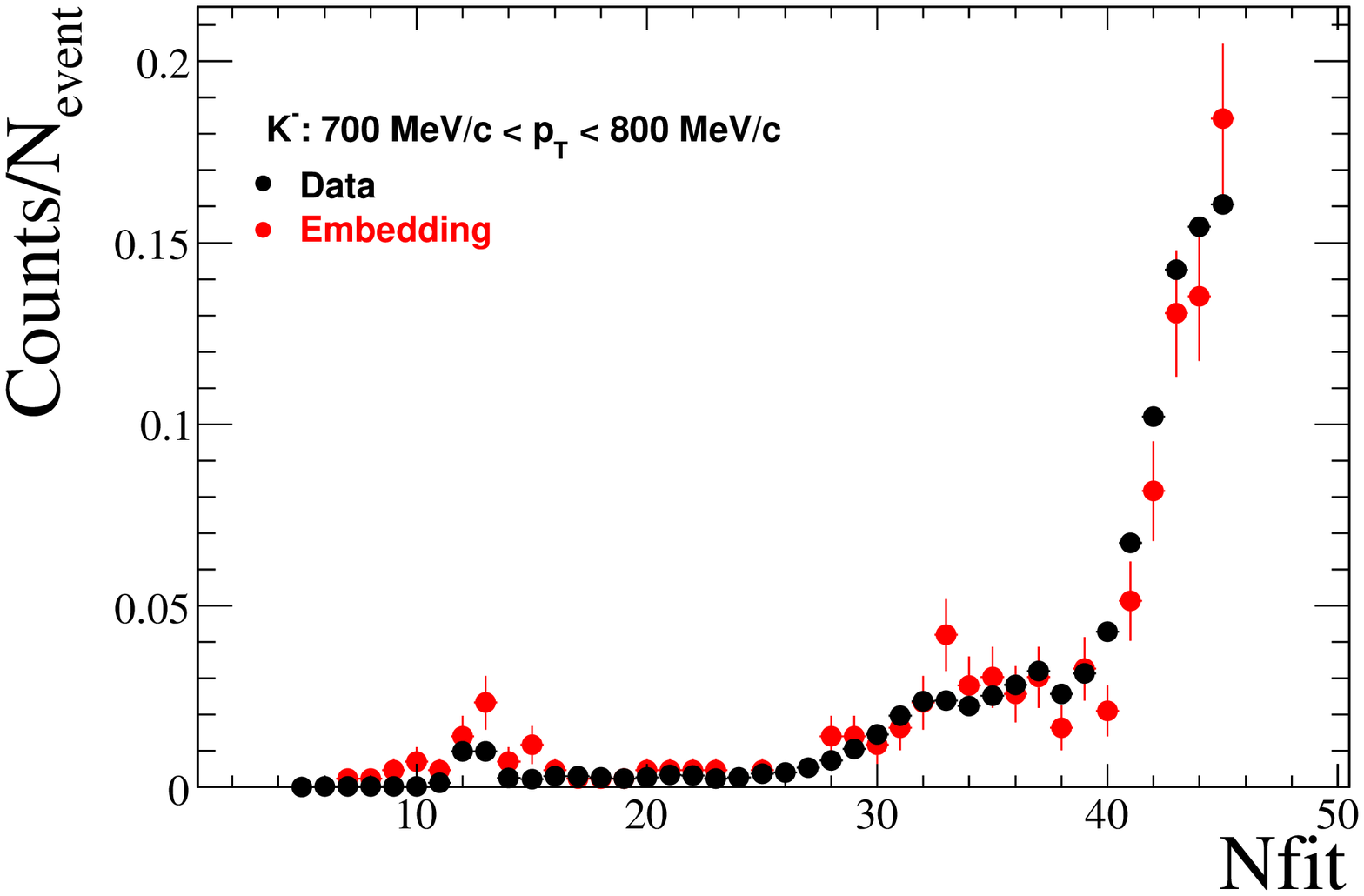}}
\resizebox{.225\textwidth}{!}{\includegraphics{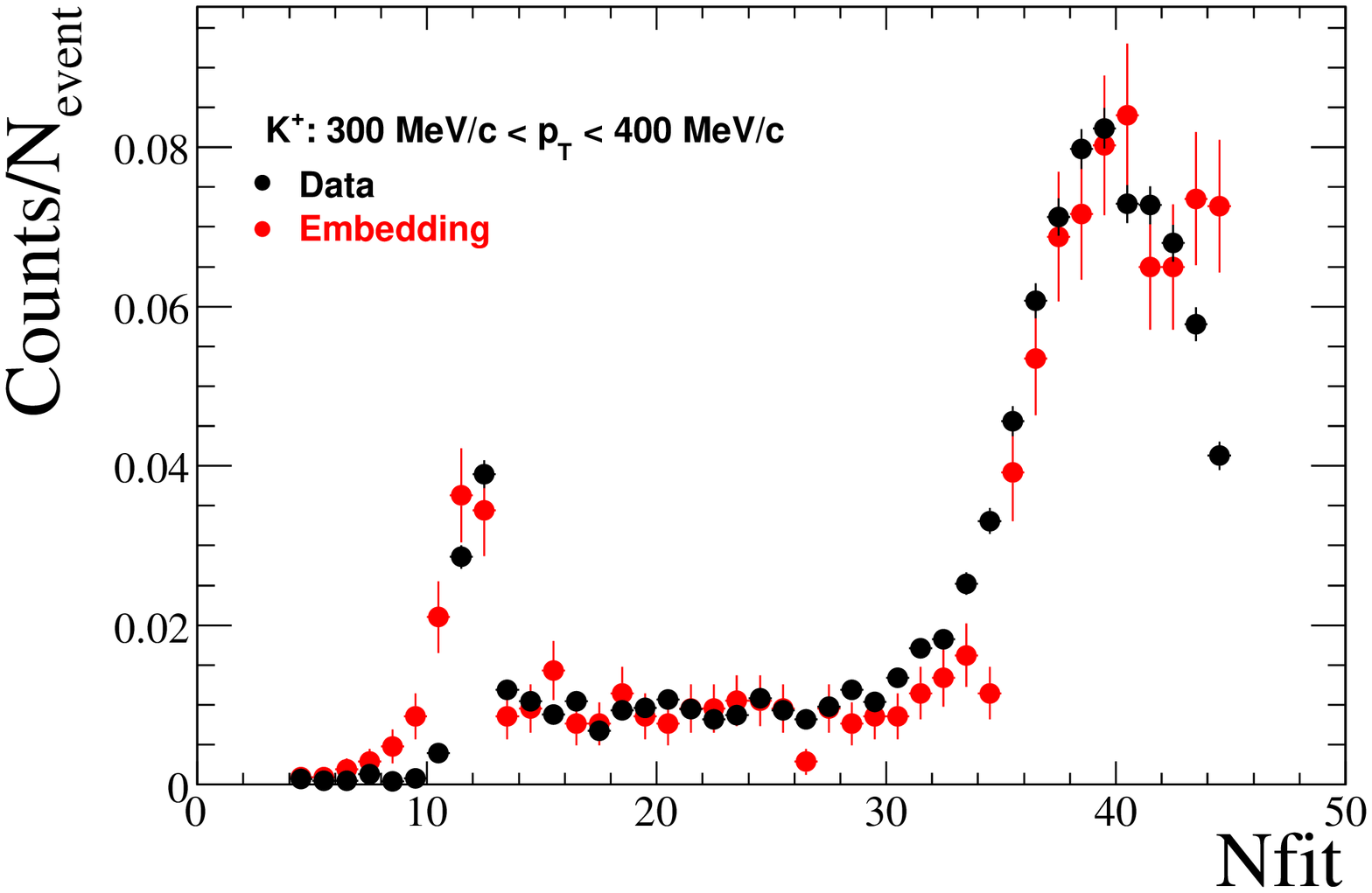}}
\resizebox{.225\textwidth}{!}{\includegraphics{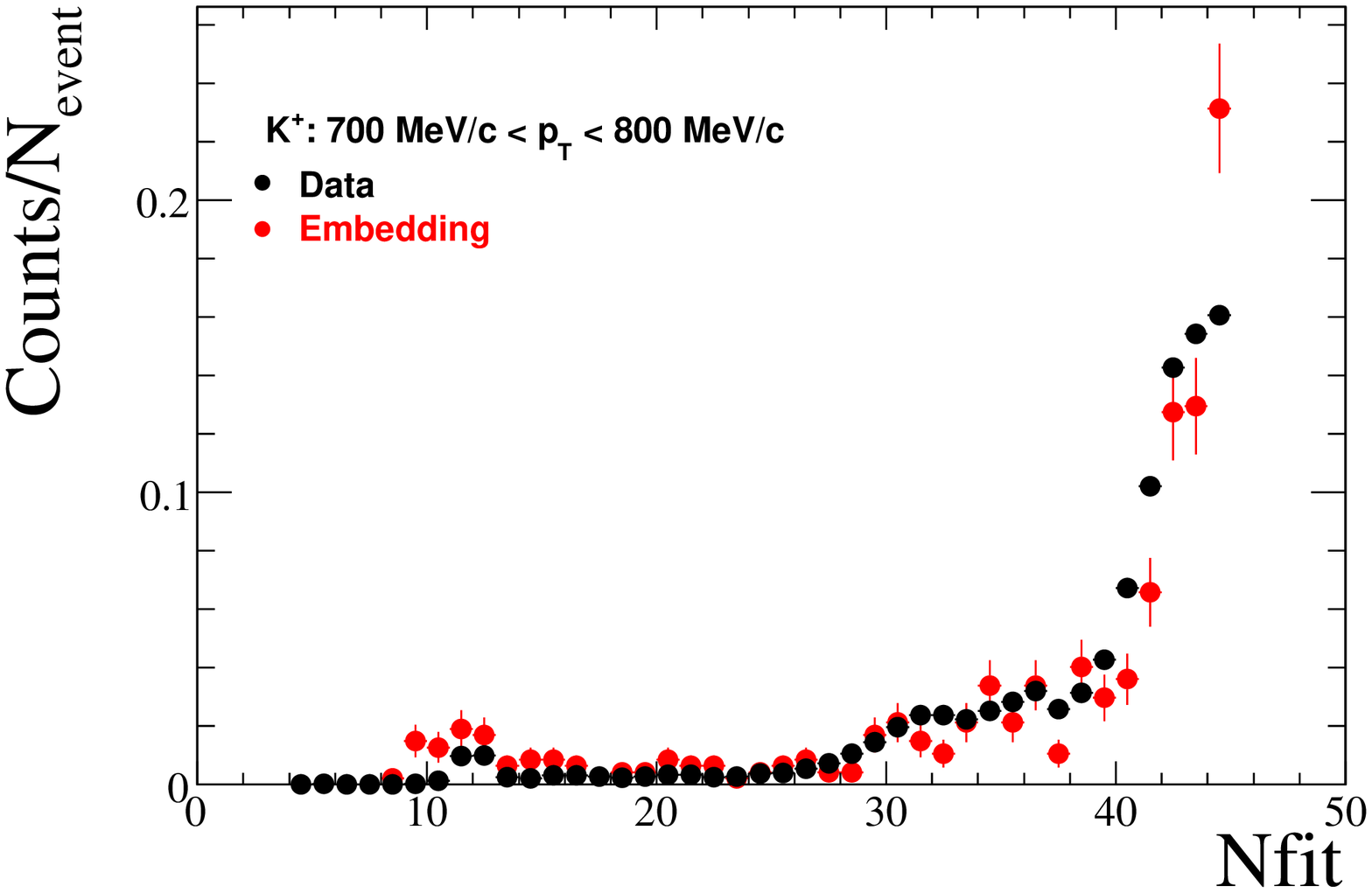}}\\
\resizebox{.225\textwidth}{!}{\includegraphics{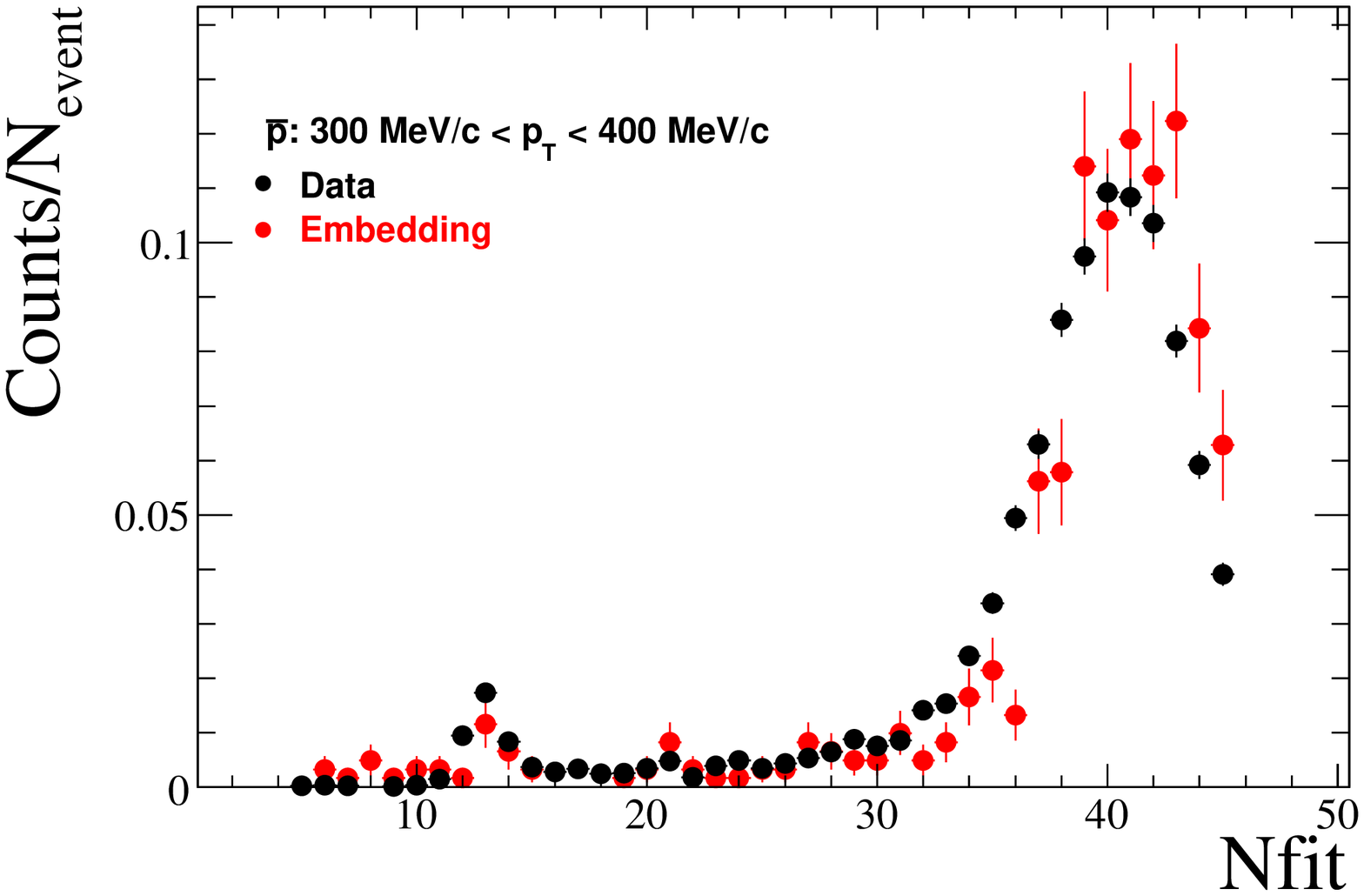}}
\resizebox{.225\textwidth}{!}{\includegraphics{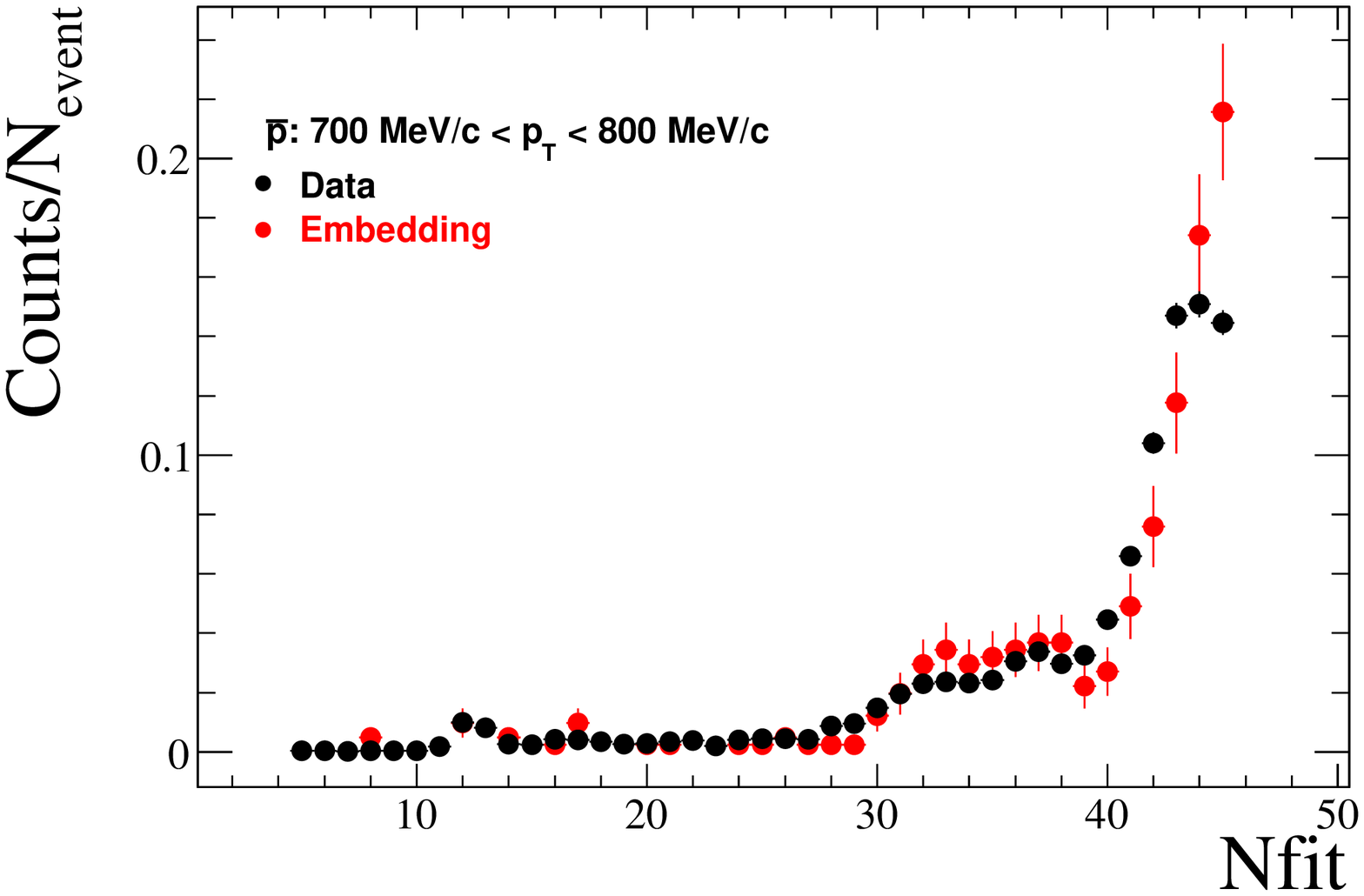}}
\resizebox{.225\textwidth}{!}{\includegraphics{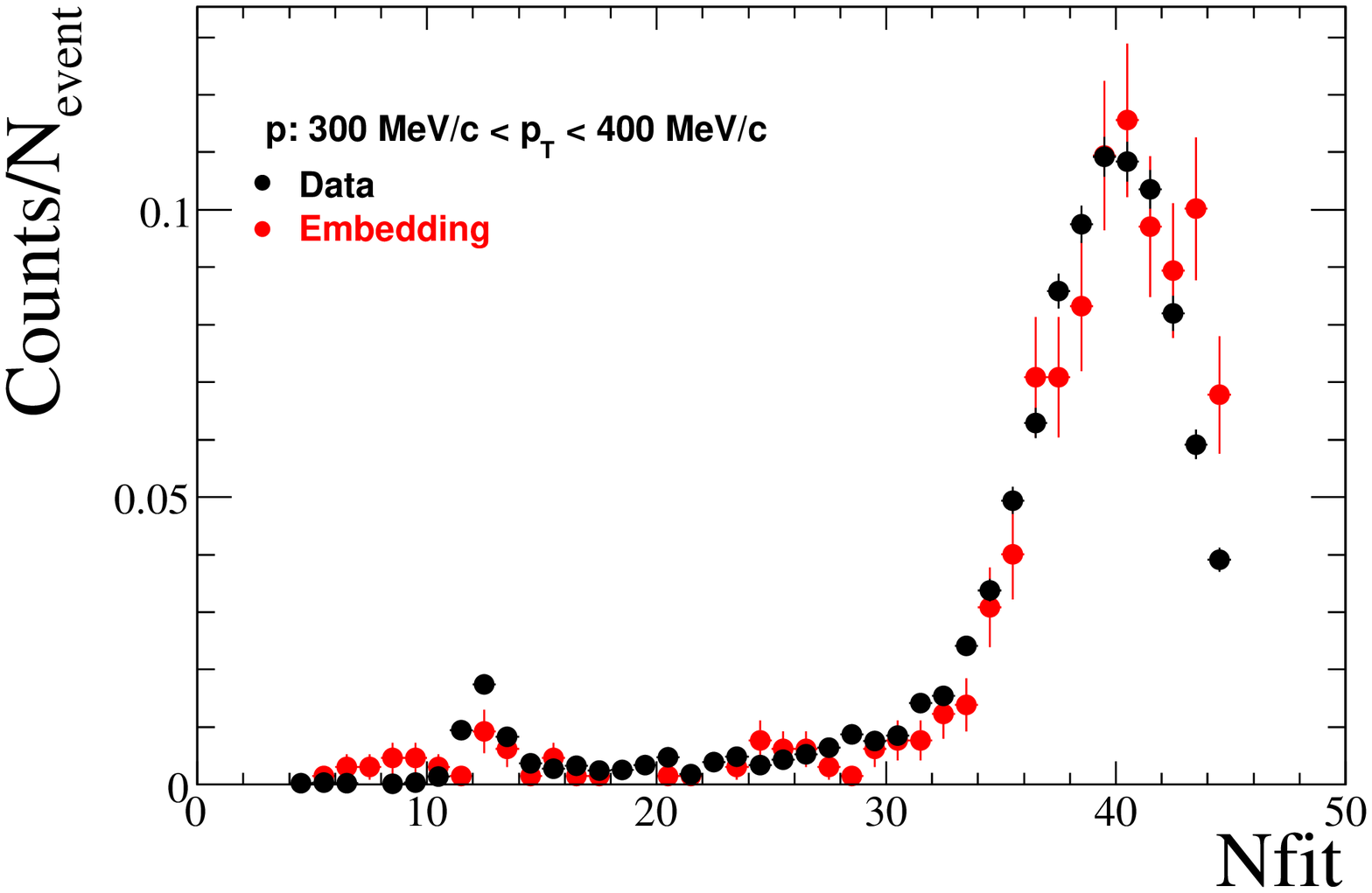}}
\resizebox{.225\textwidth}{!}{\includegraphics{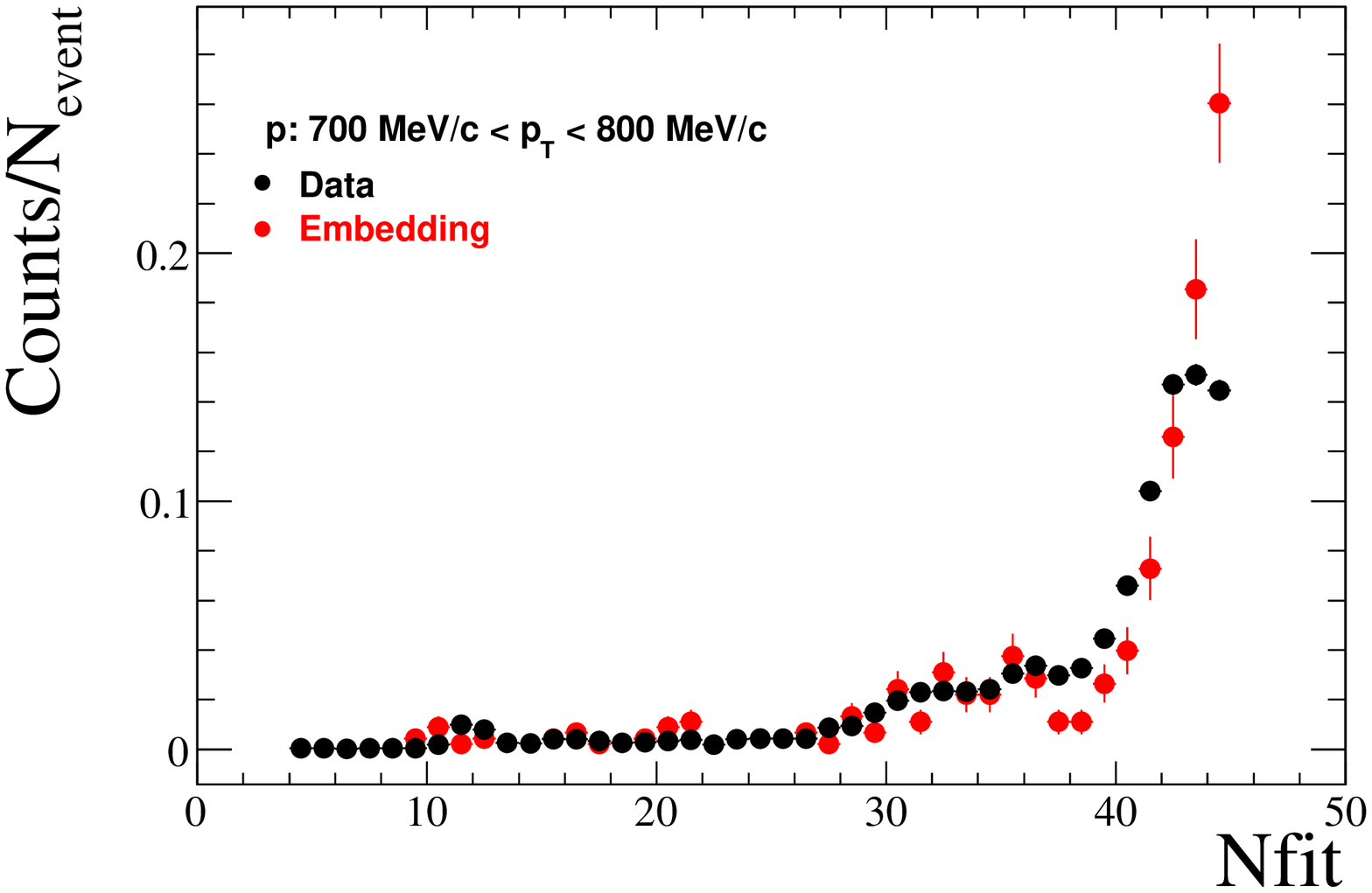}}\\
\caption{Comparison of $N_{fit}$ from real data and embedding in 200 GeV minimum bias pp collisions.}
\label{fig:pp_nfit_data_embedding}
\end{center} 
\end{sidewaysfigure}
\begin{sidewaysfigure}[!h]
\begin{center}
\resizebox{.225\textwidth}{!}{\includegraphics{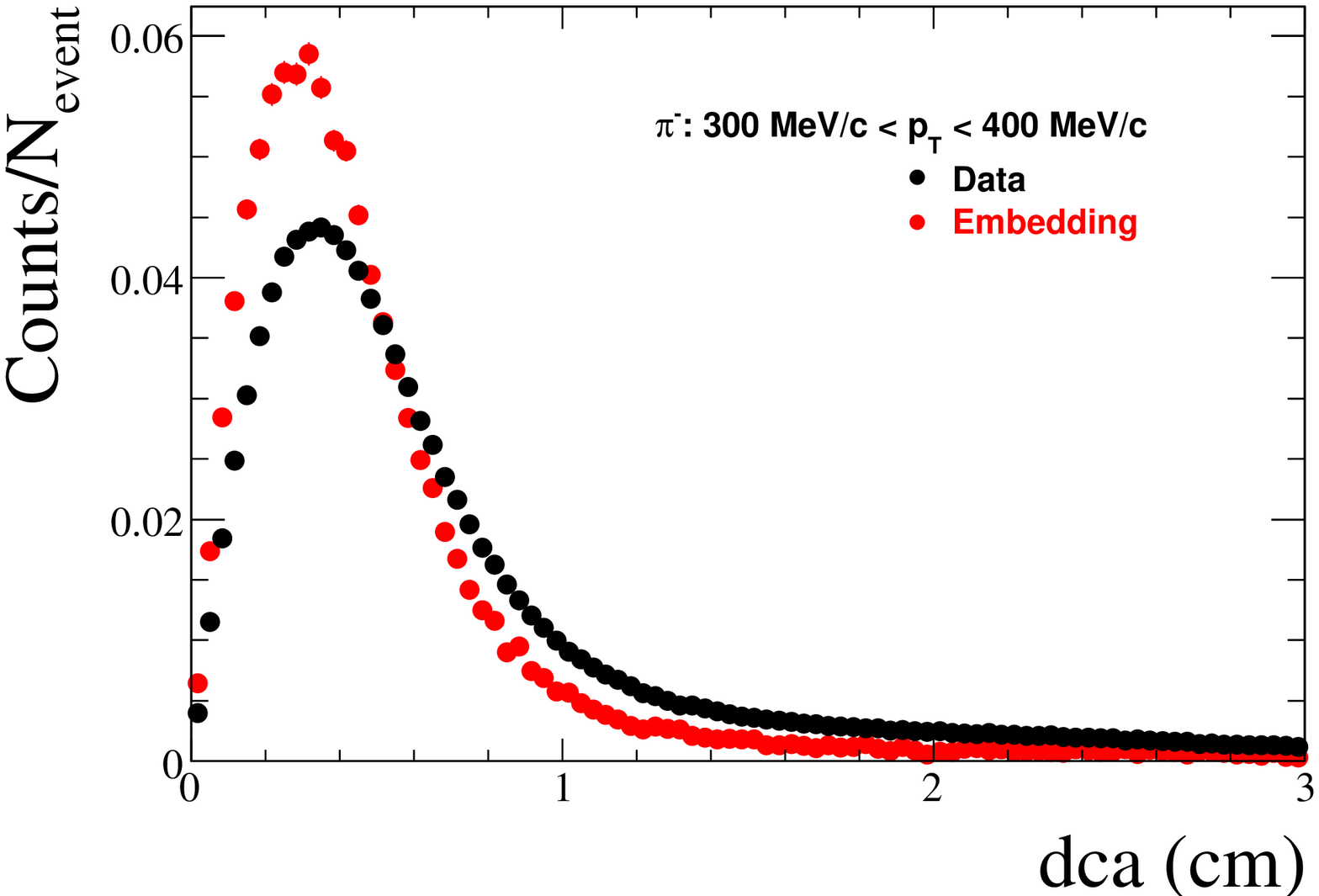}}
\resizebox{.225\textwidth}{!}{\includegraphics{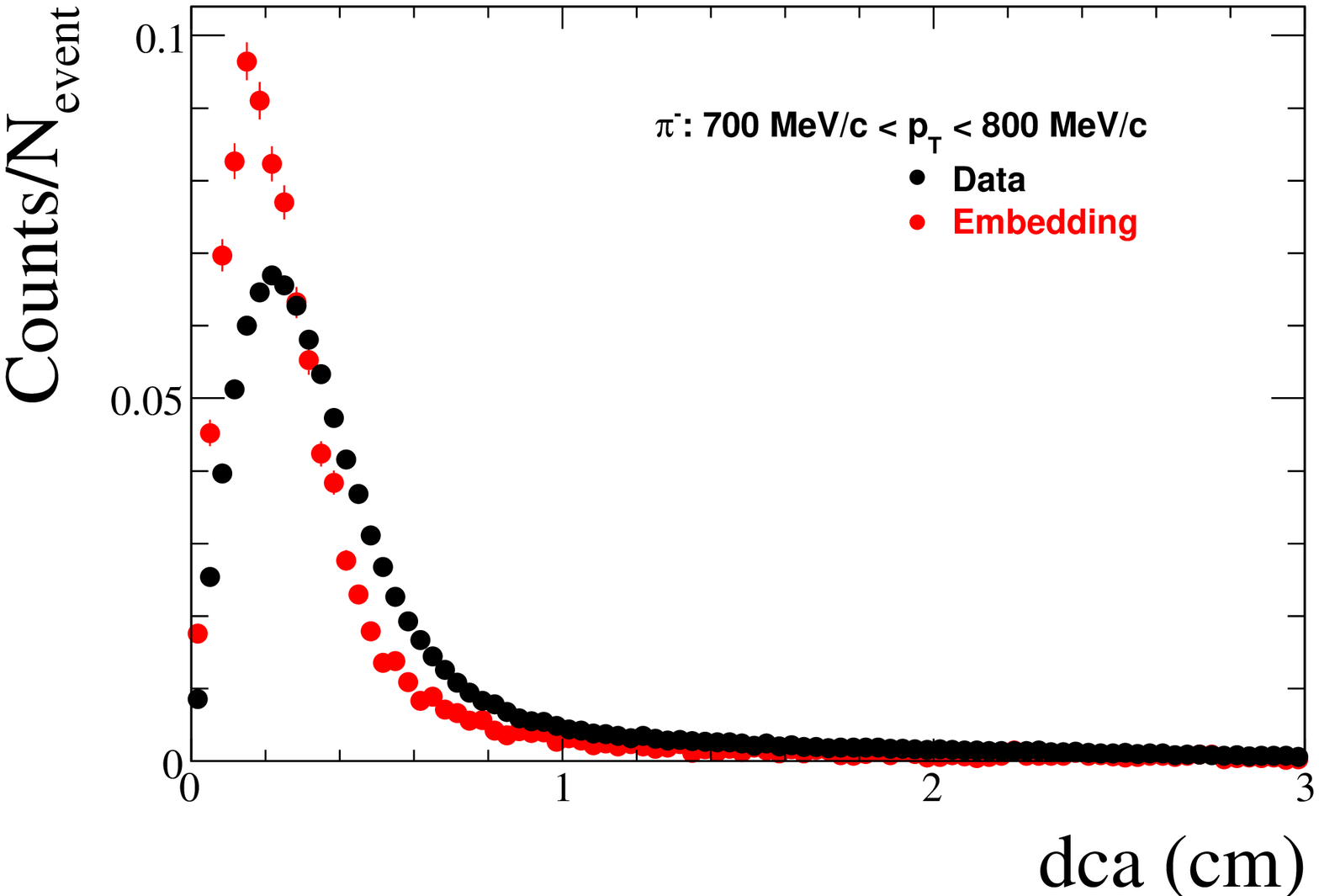}}
\resizebox{.225\textwidth}{!}{\includegraphics{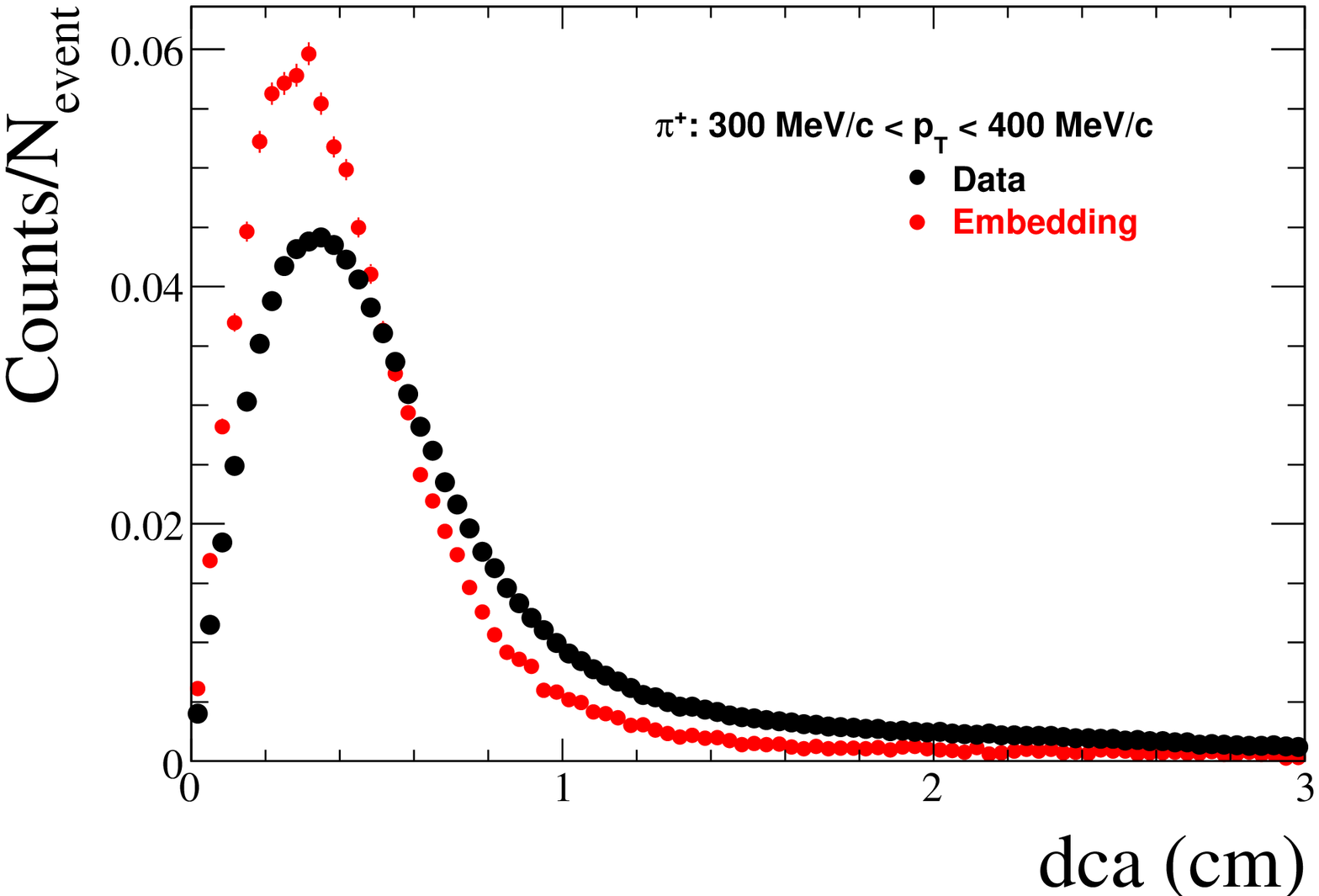}}
\resizebox{.225\textwidth}{!}{\includegraphics{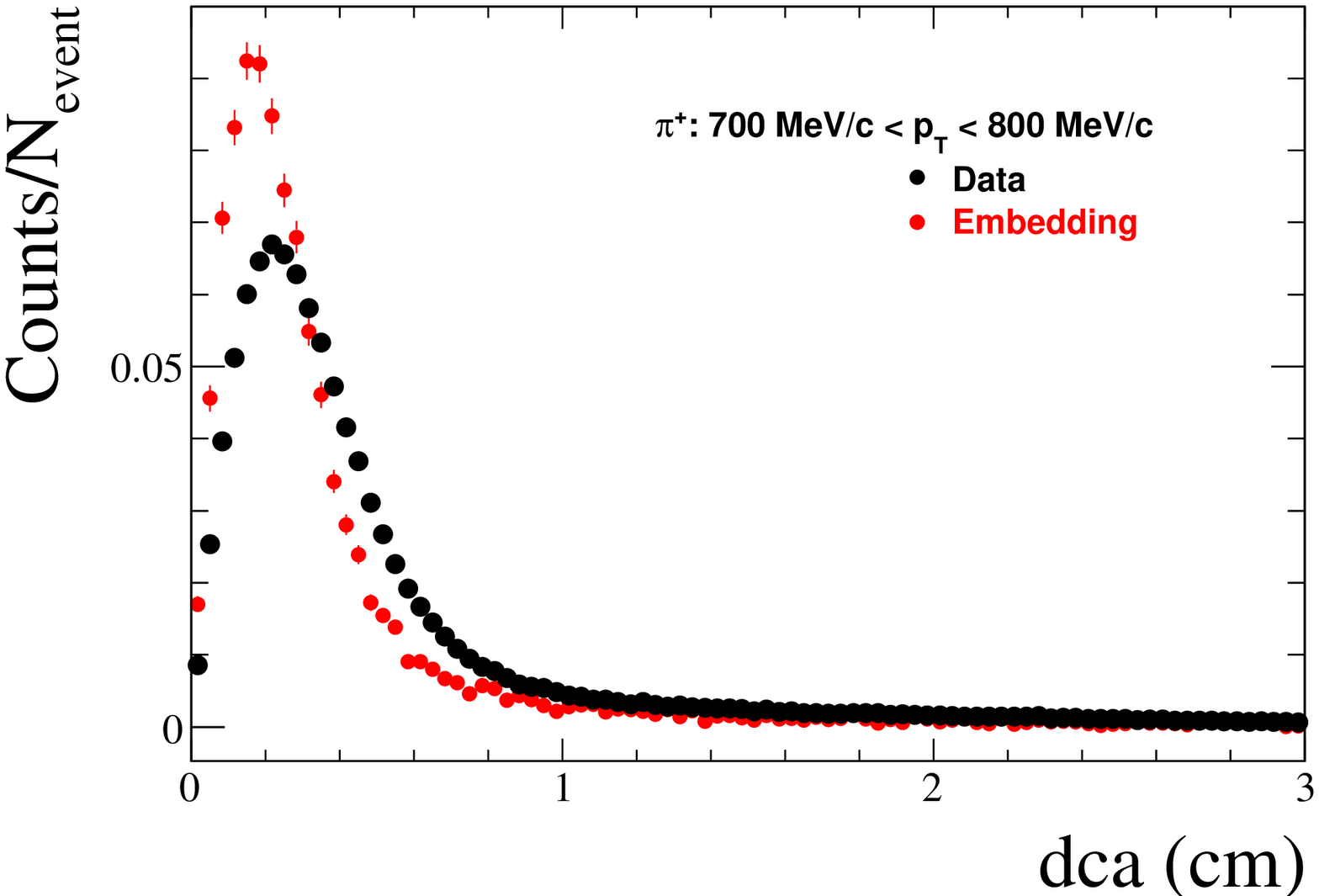}}\\
\resizebox{.225\textwidth}{!}{\includegraphics{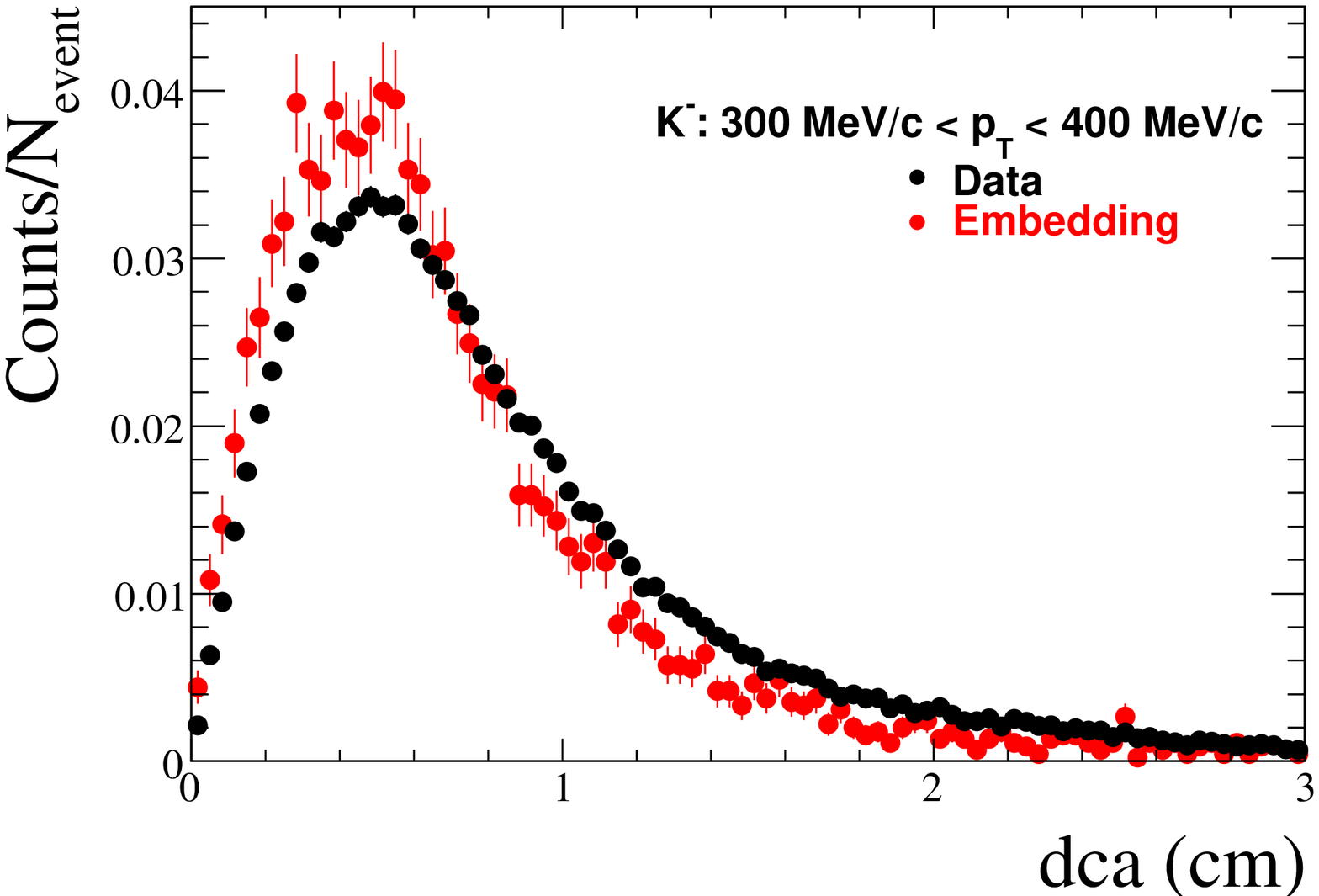}}
\resizebox{.225\textwidth}{!}{\includegraphics{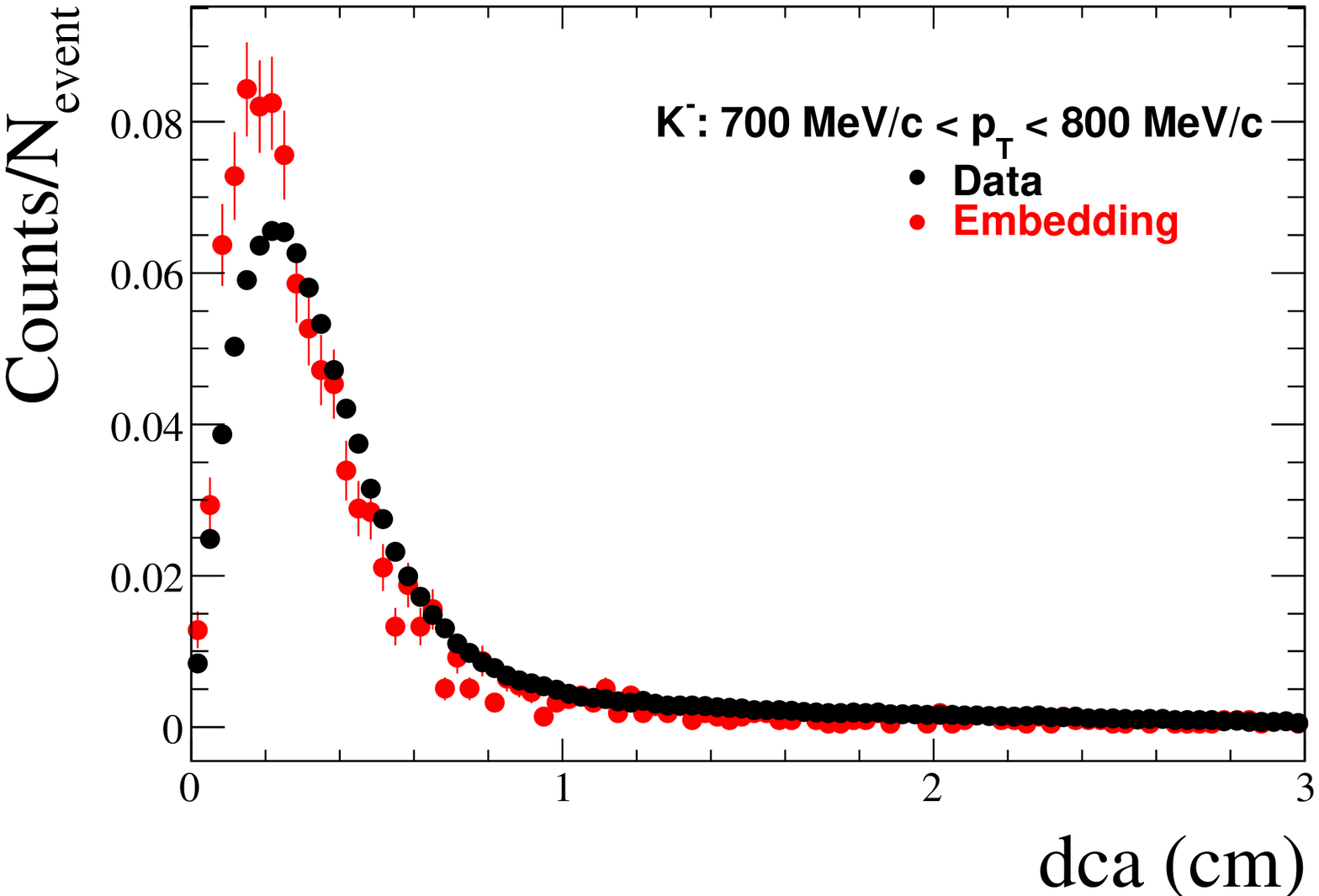}}
\resizebox{.225\textwidth}{!}{\includegraphics{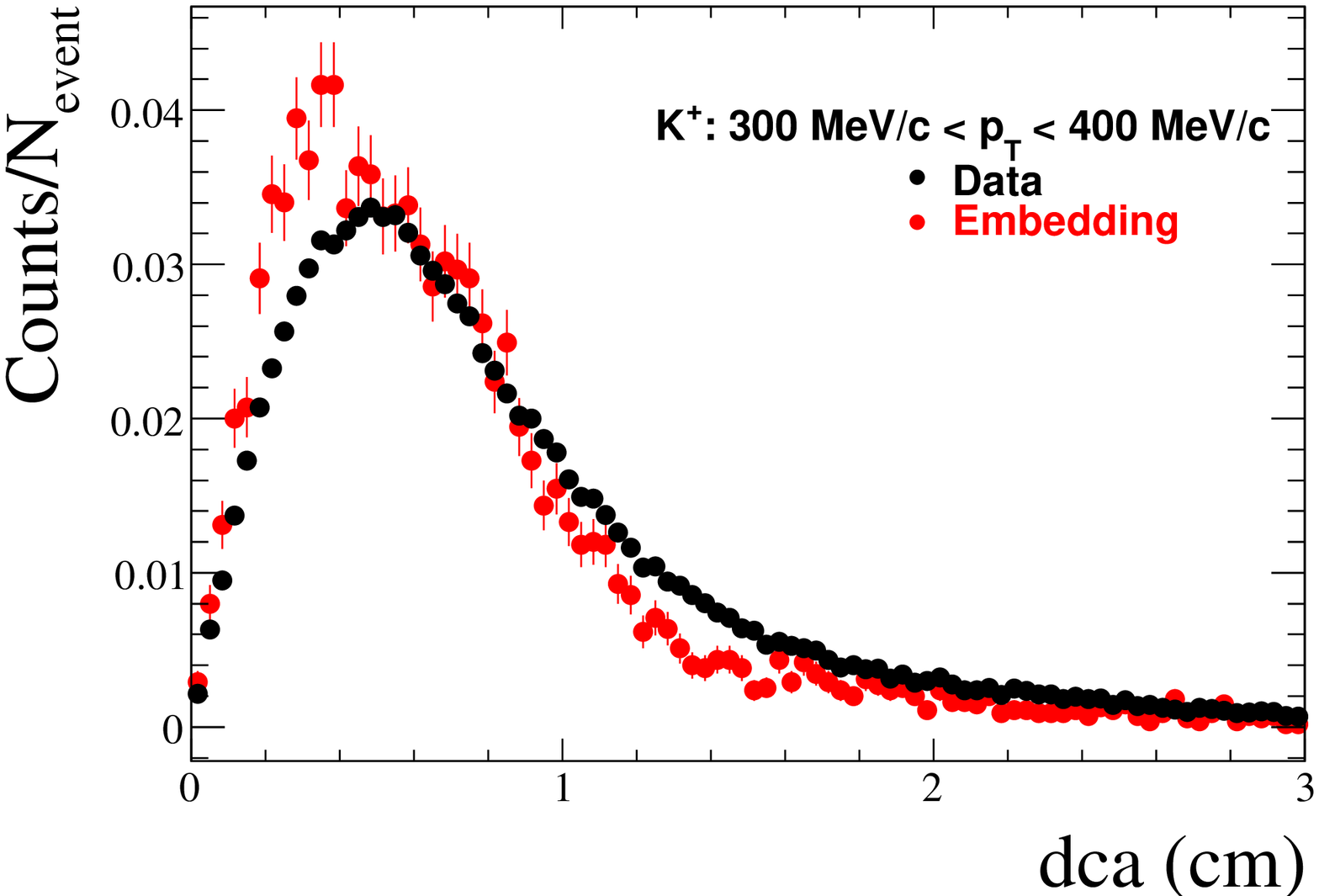}}
\resizebox{.225\textwidth}{!}{\includegraphics{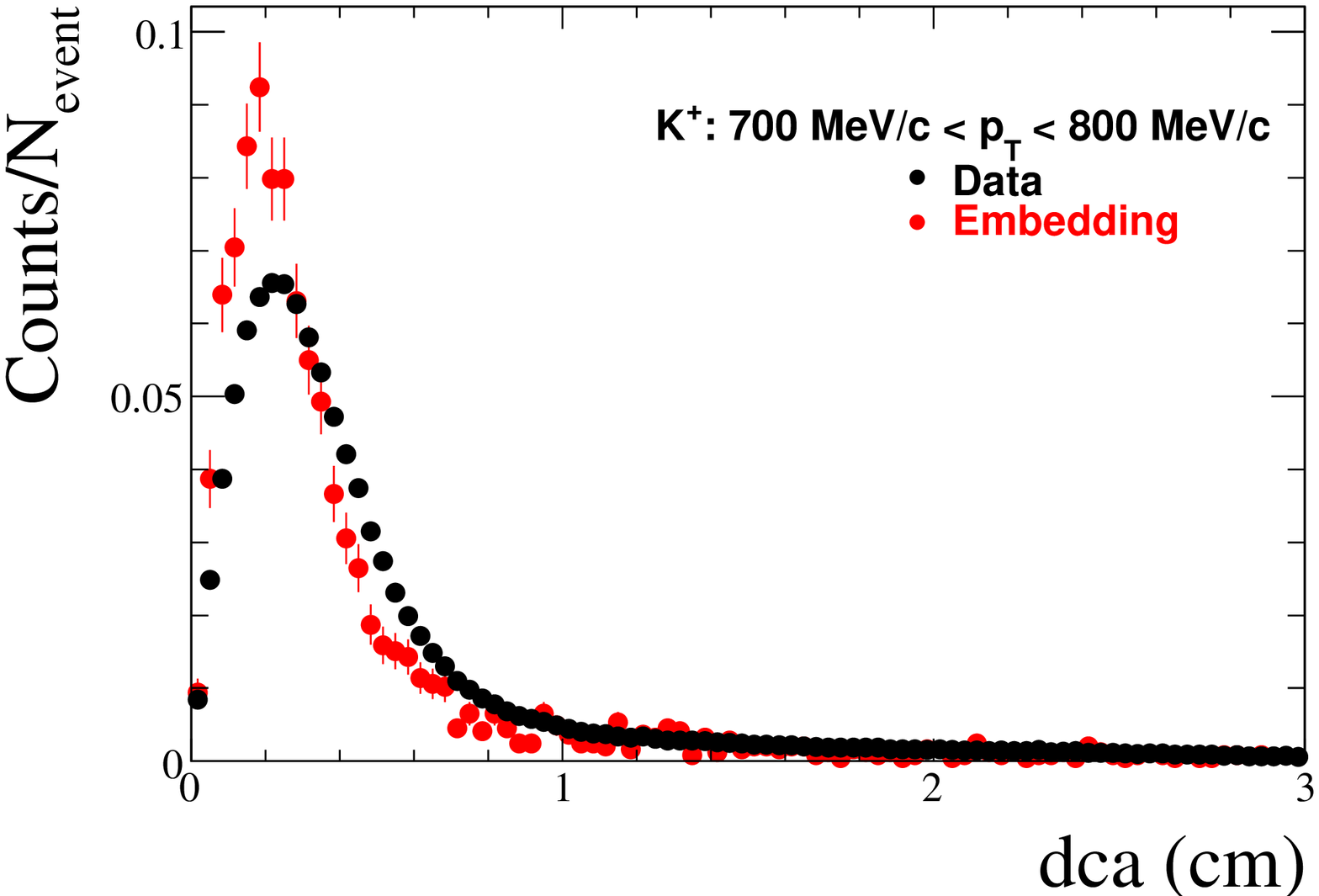}}\\
\resizebox{.225\textwidth}{!}{\includegraphics{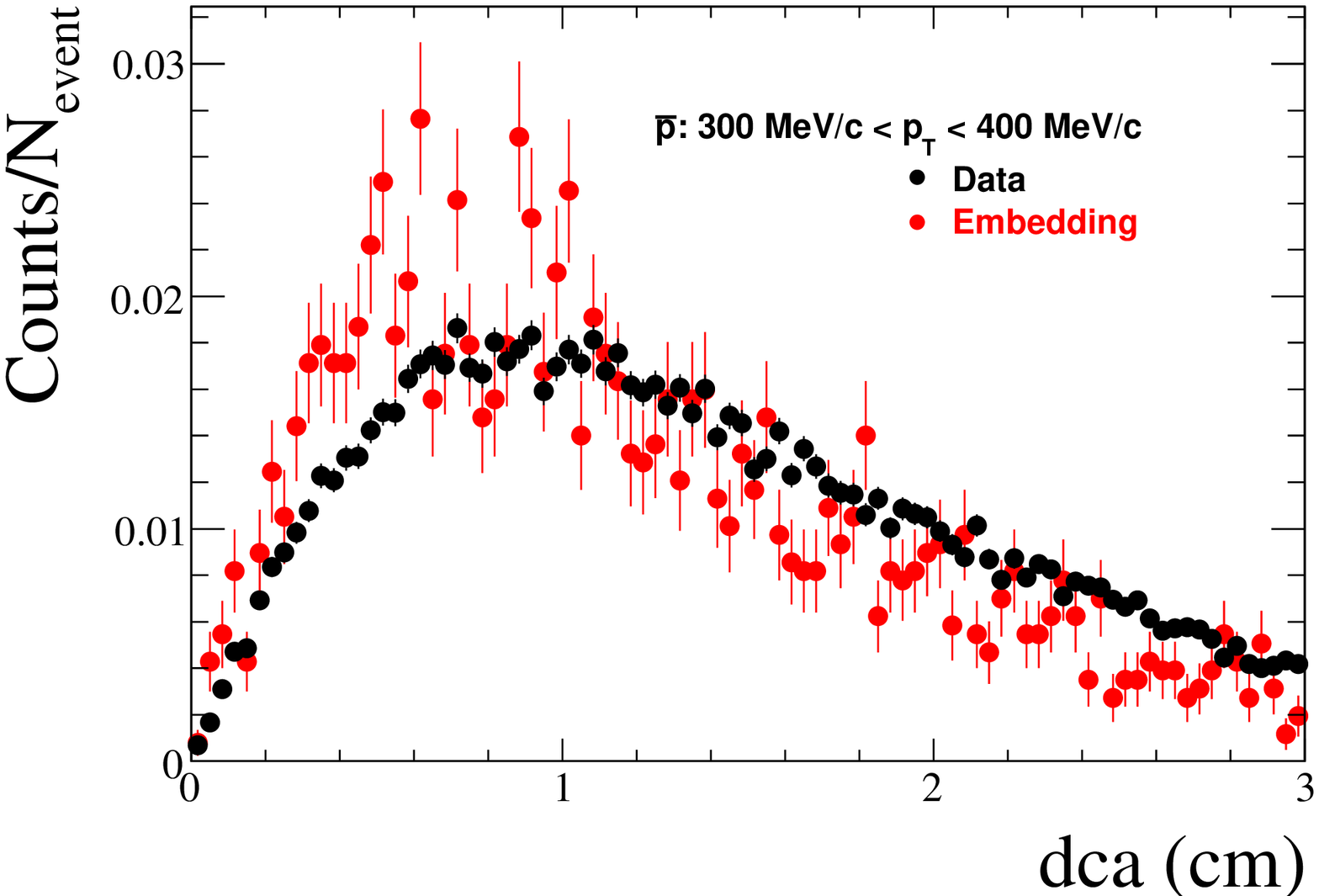}}
\resizebox{.225\textwidth}{!}{\includegraphics{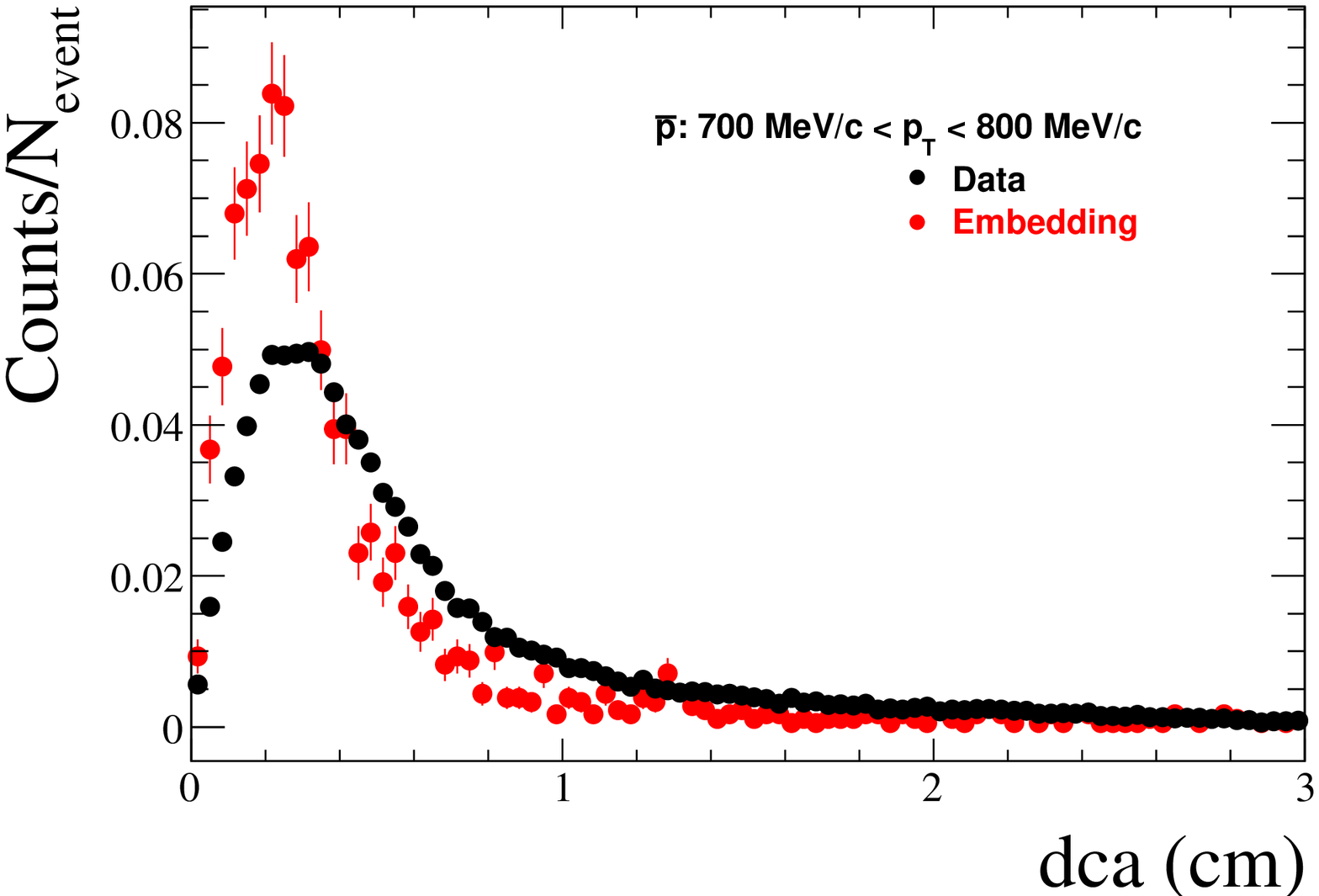}}
\resizebox{.225\textwidth}{!}{\includegraphics{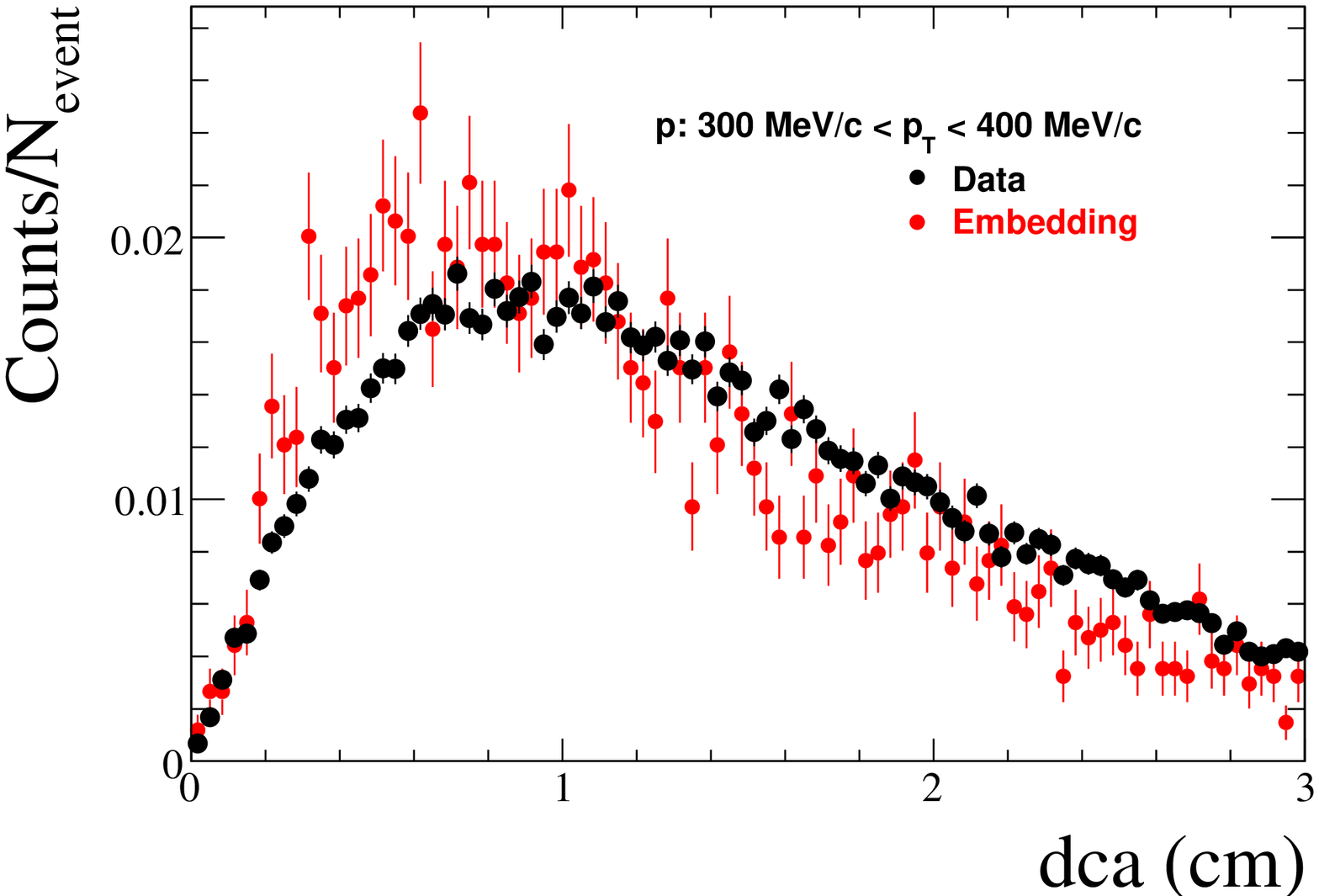}}
\resizebox{.225\textwidth}{!}{\includegraphics{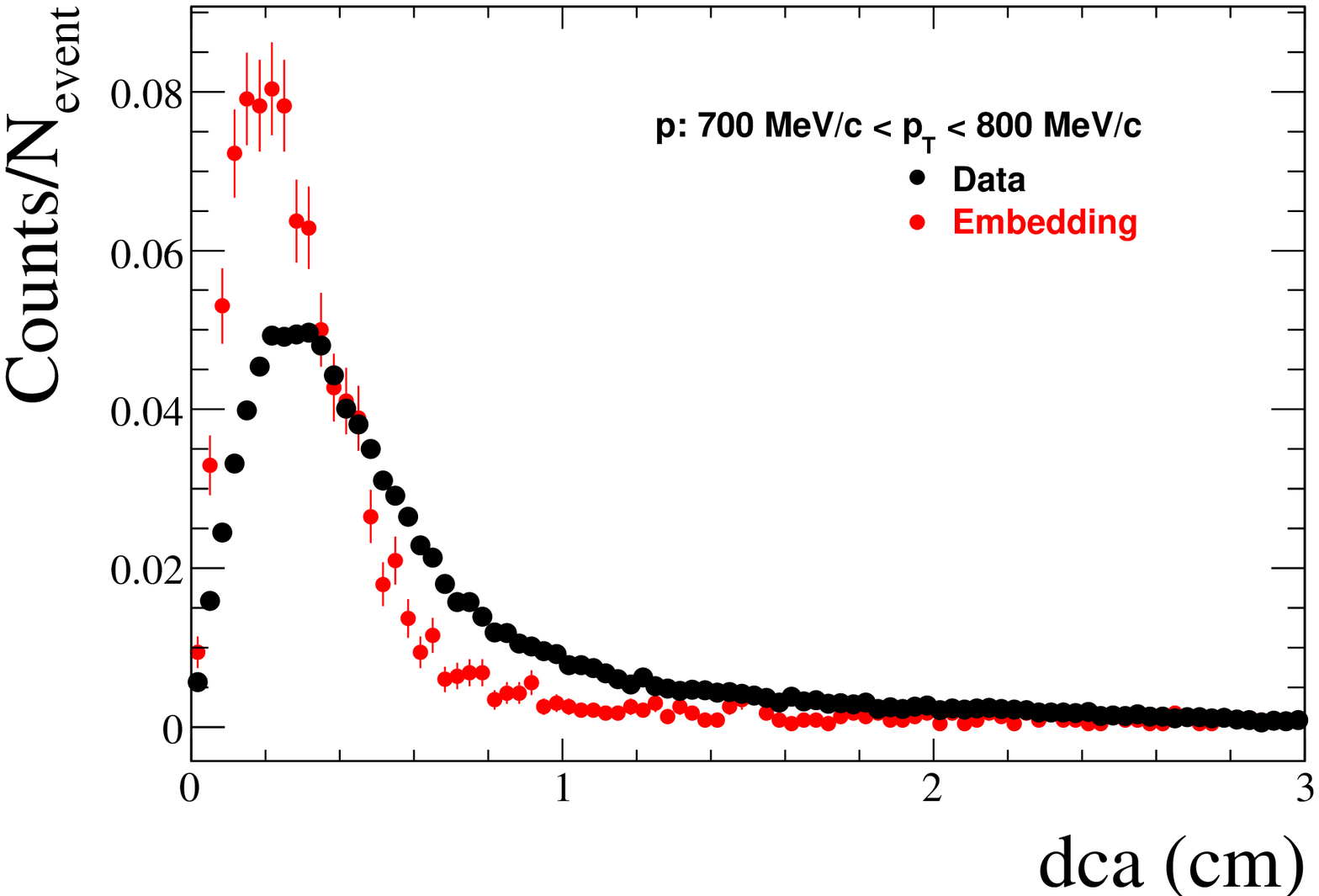}}\\
\caption{Comparison of $dca$ from real data and embedding in 200 GeV minimum bias dAu collisions.}
\label{fig:dau_dca_data_embedding}
\end{center} 
\end{sidewaysfigure}
\begin{sidewaysfigure}[!h]
\begin{center}
\resizebox{.225\textwidth}{!}{\includegraphics{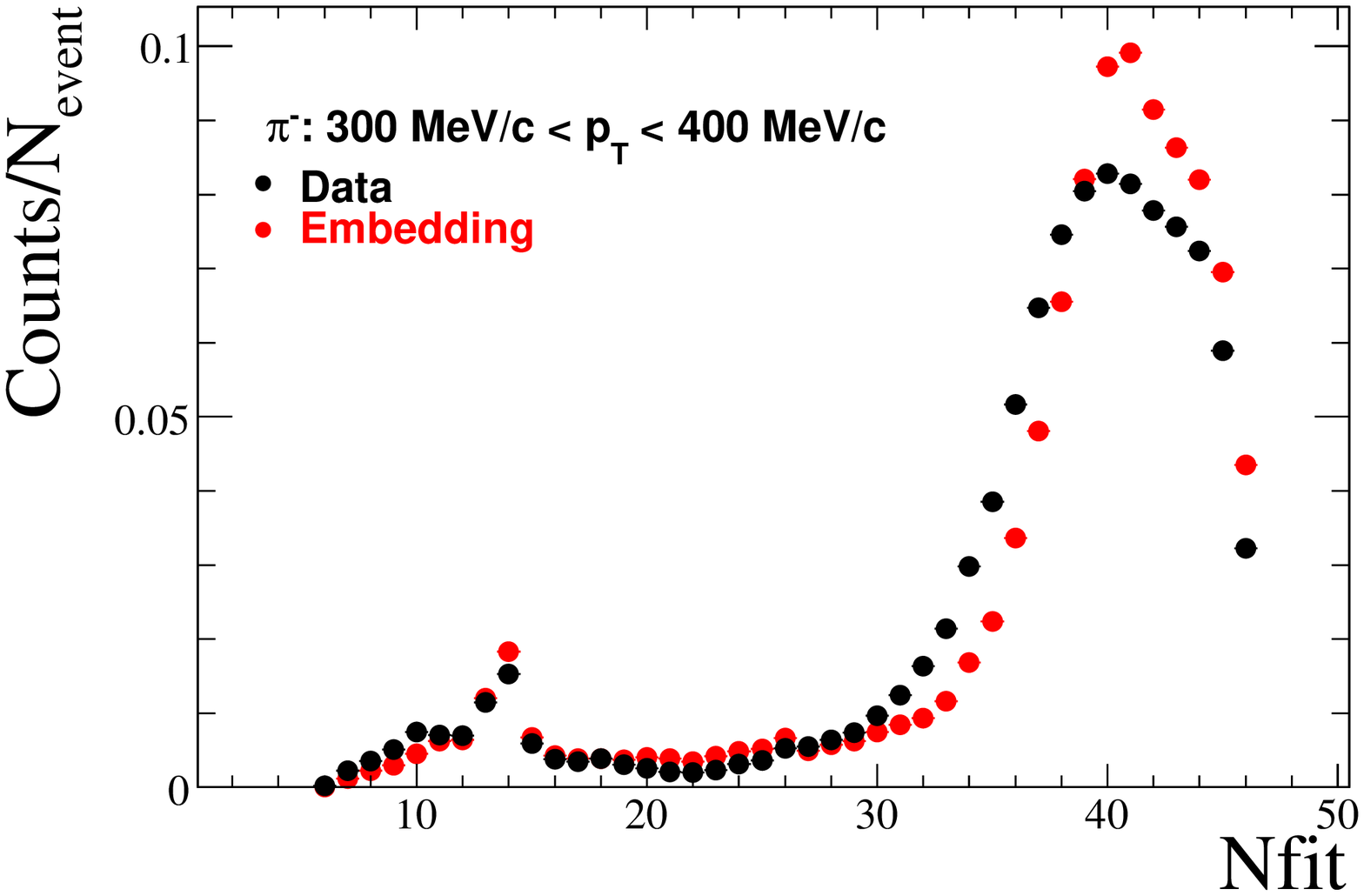}}
\resizebox{.225\textwidth}{!}{\includegraphics{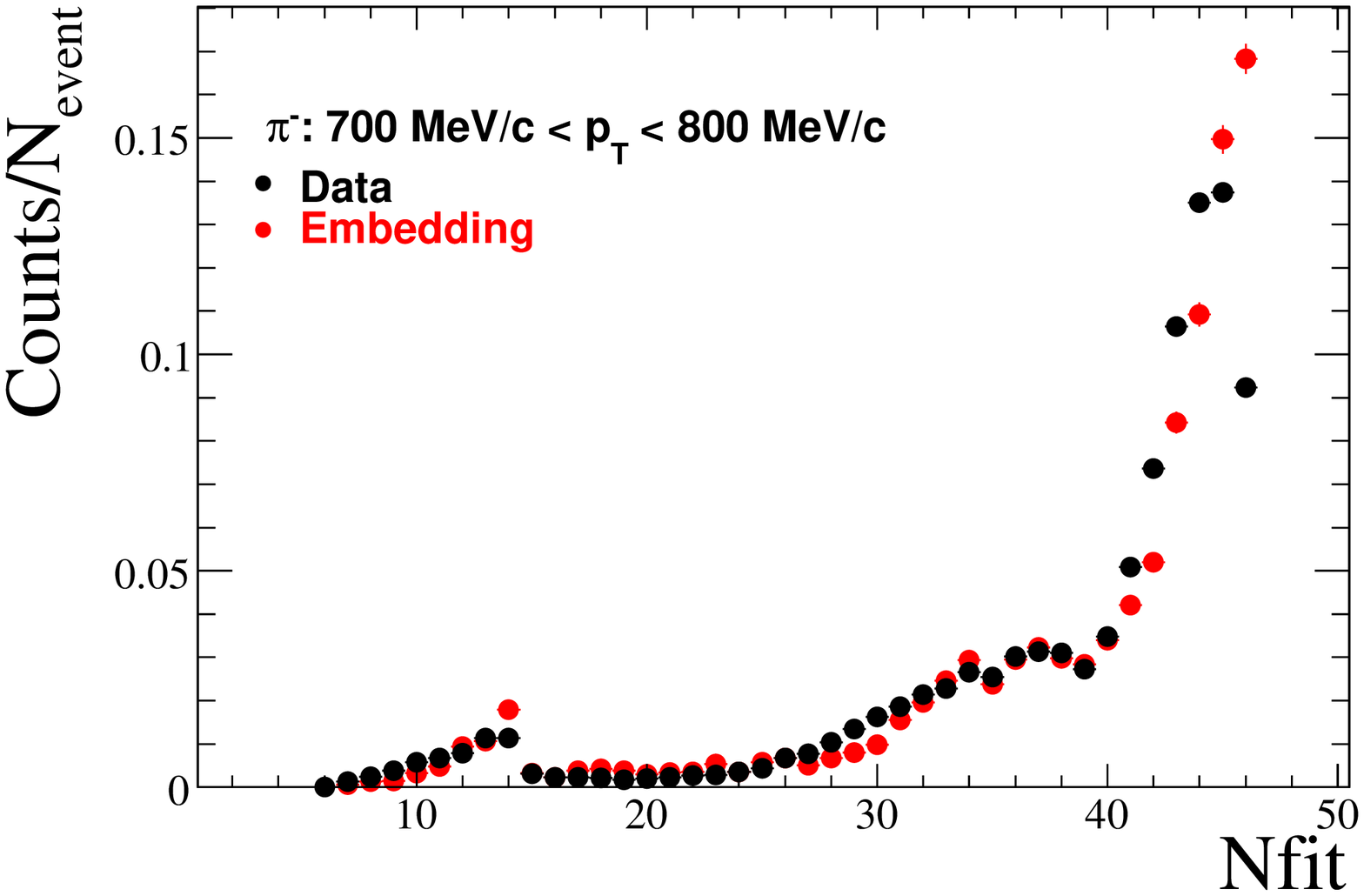}}
\resizebox{.225\textwidth}{!}{\includegraphics{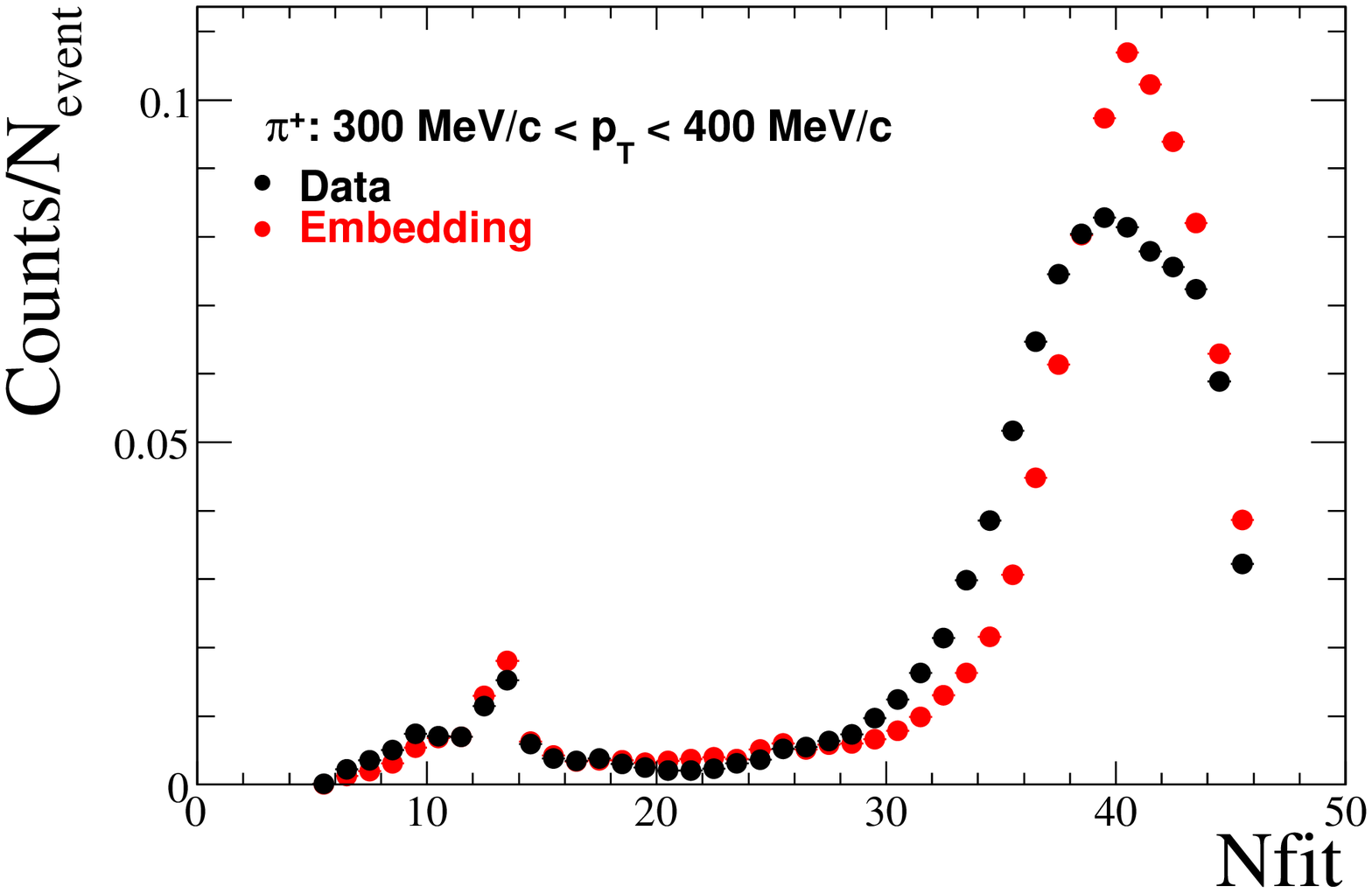}}
\resizebox{.225\textwidth}{!}{\includegraphics{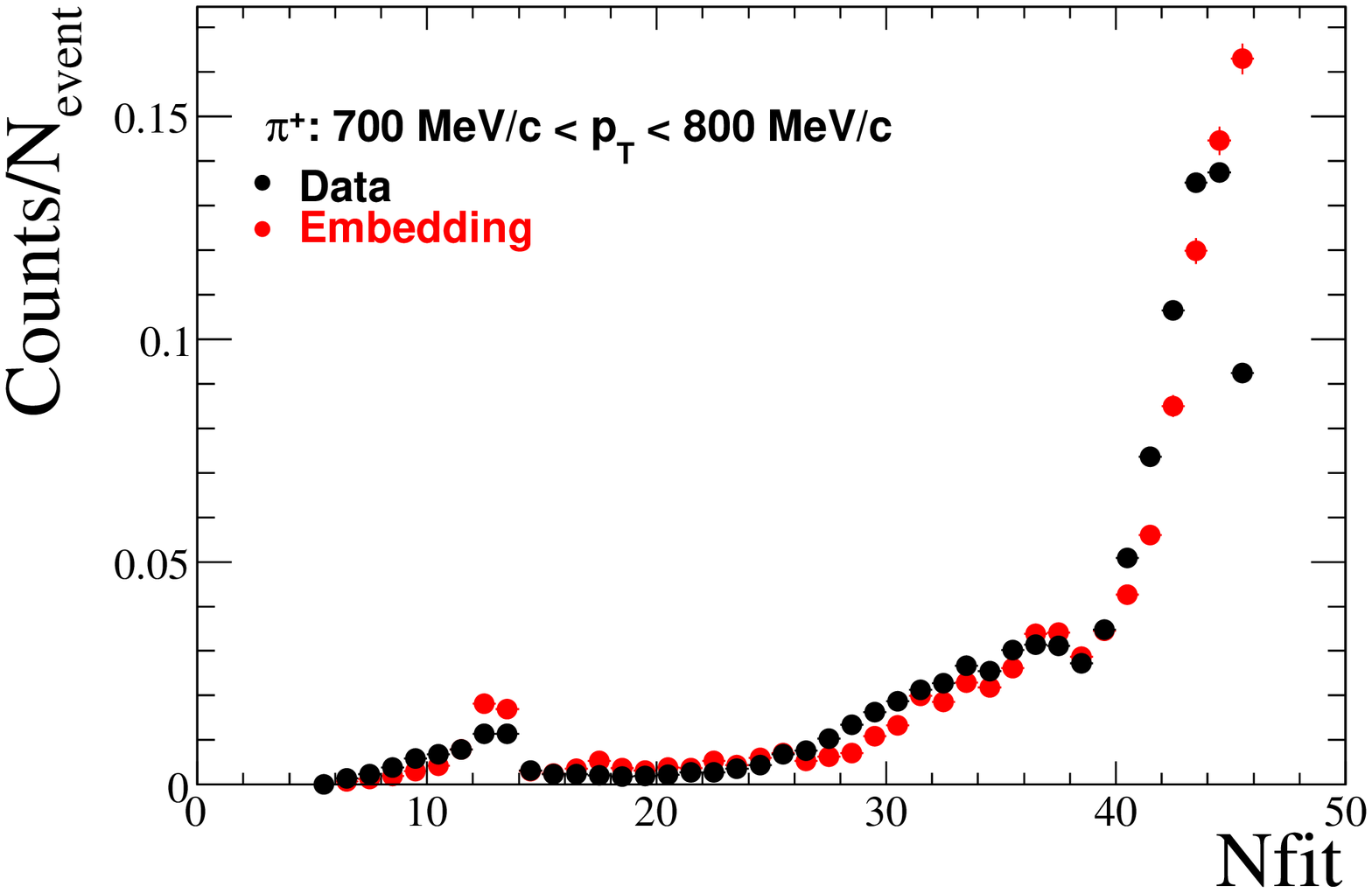}}\\
\resizebox{.225\textwidth}{!}{\includegraphics{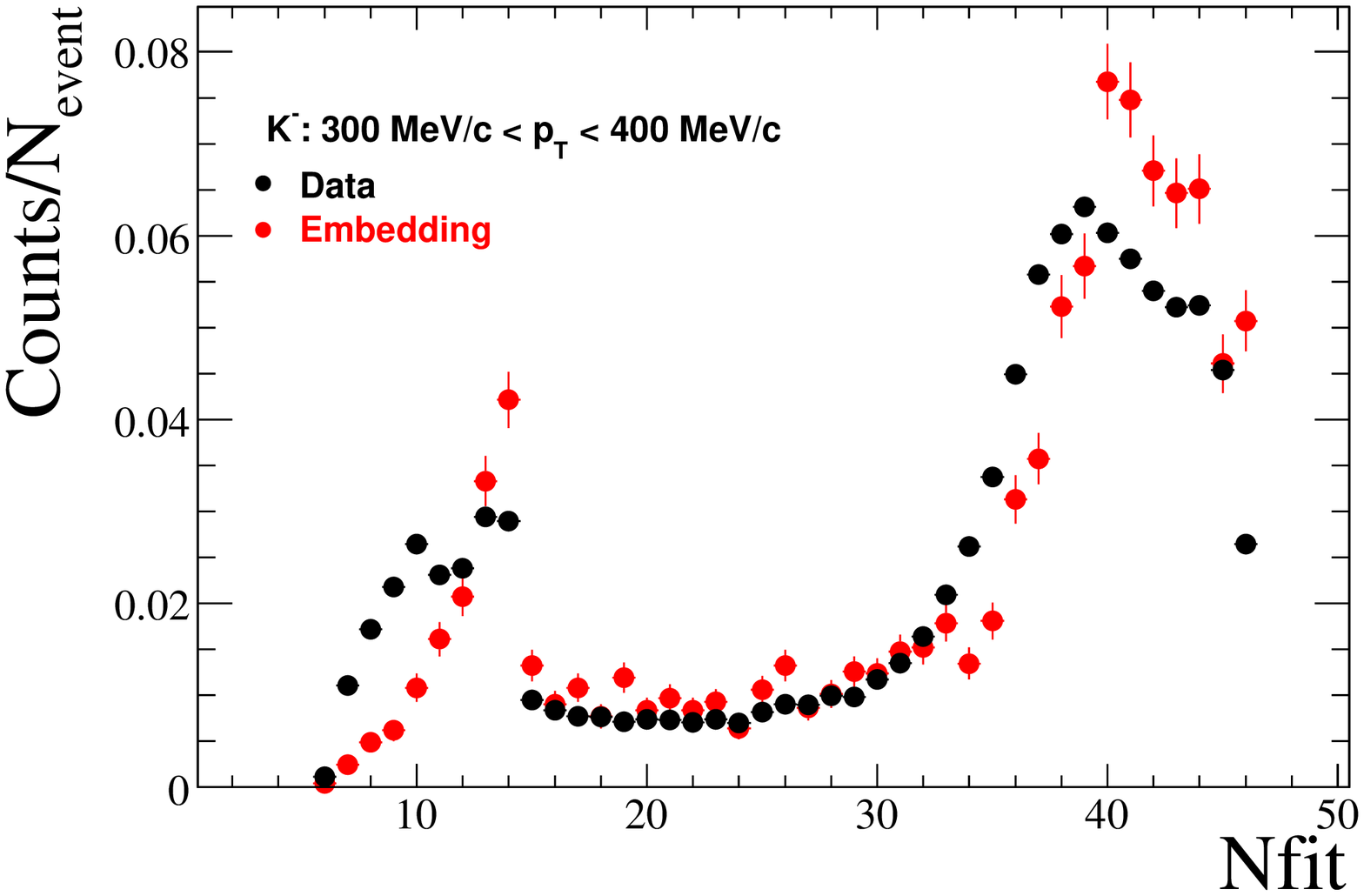}}
\resizebox{.225\textwidth}{!}{\includegraphics{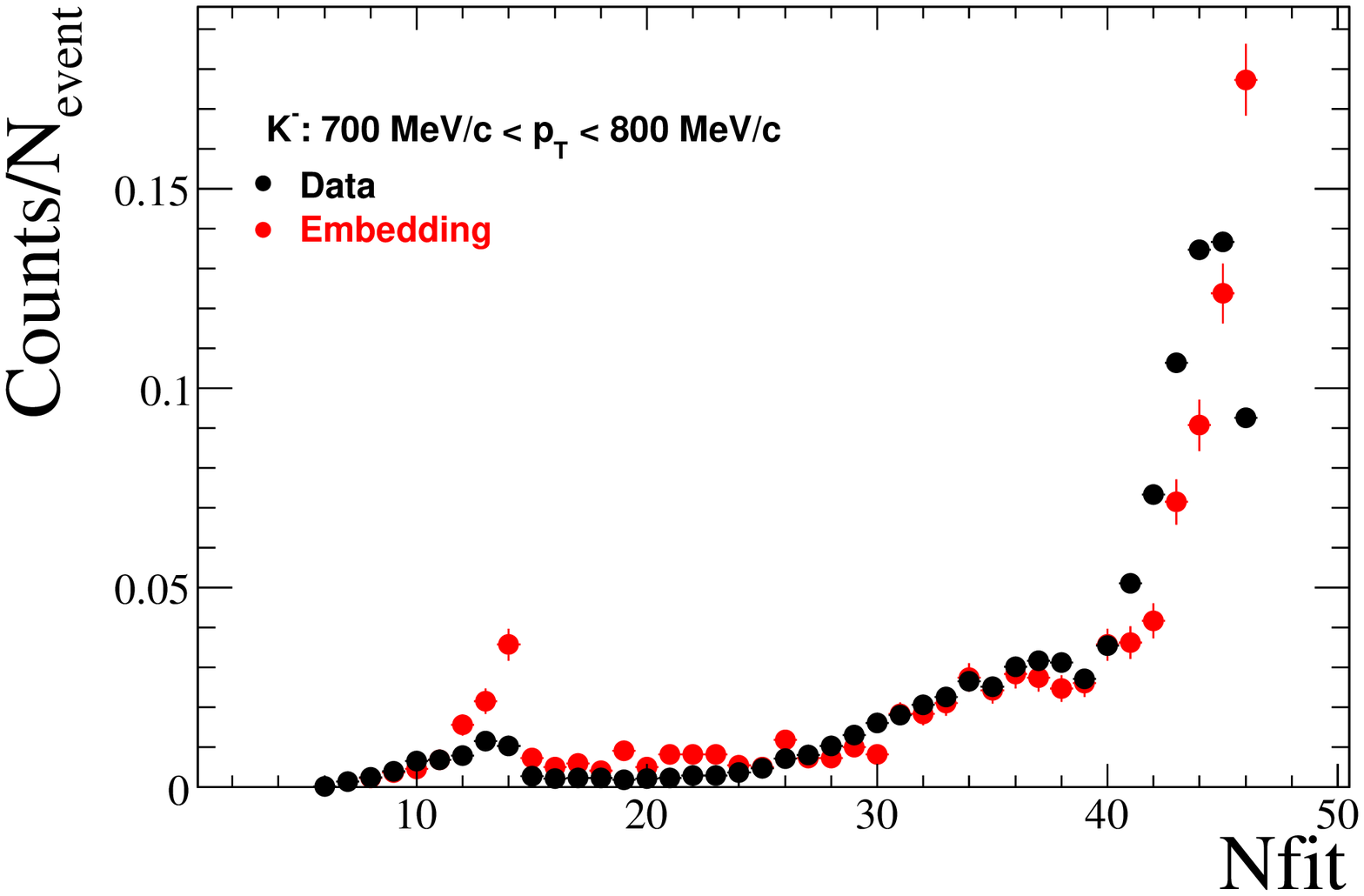}}
\resizebox{.225\textwidth}{!}{\includegraphics{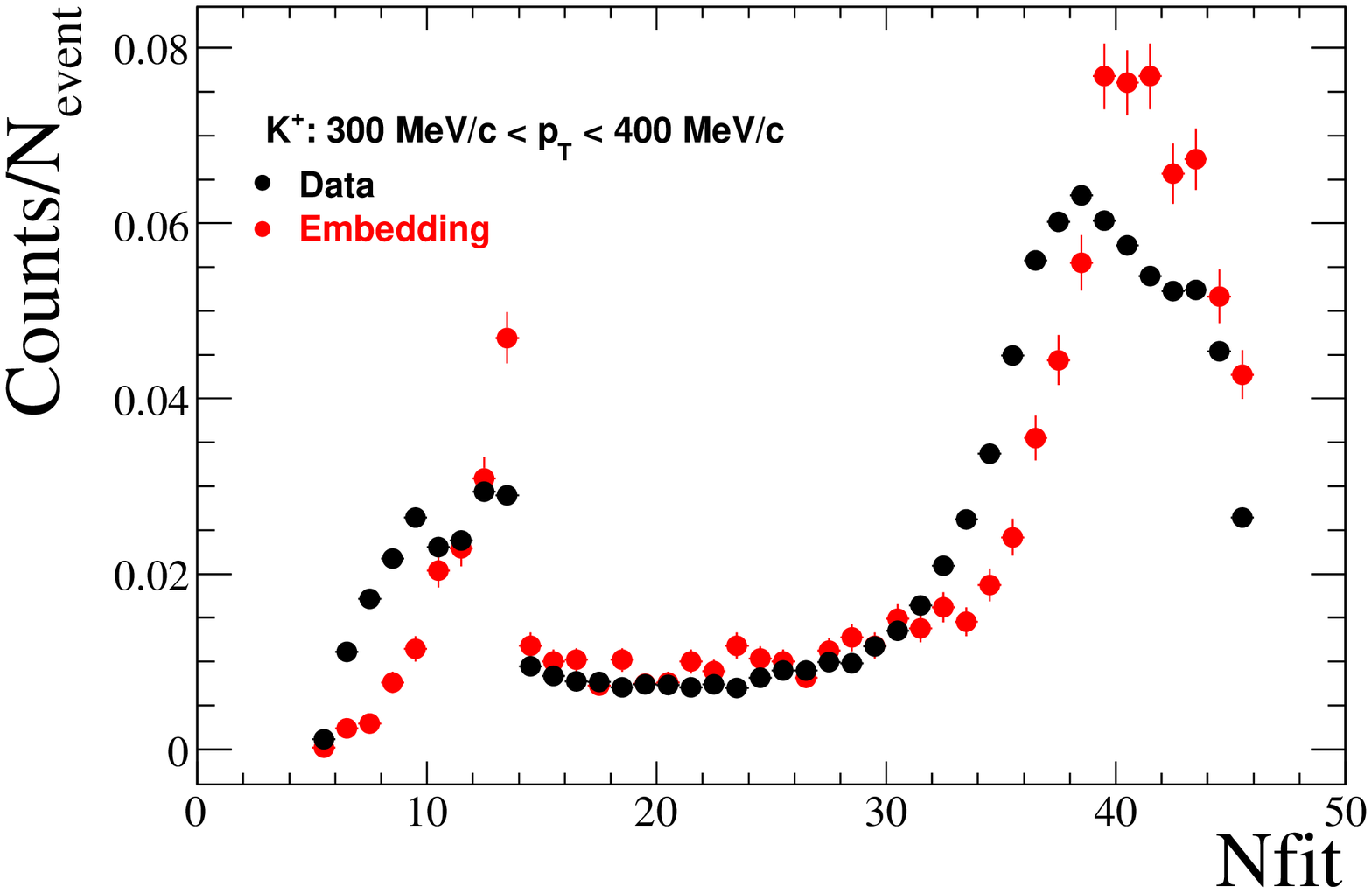}}
\resizebox{.225\textwidth}{!}{\includegraphics{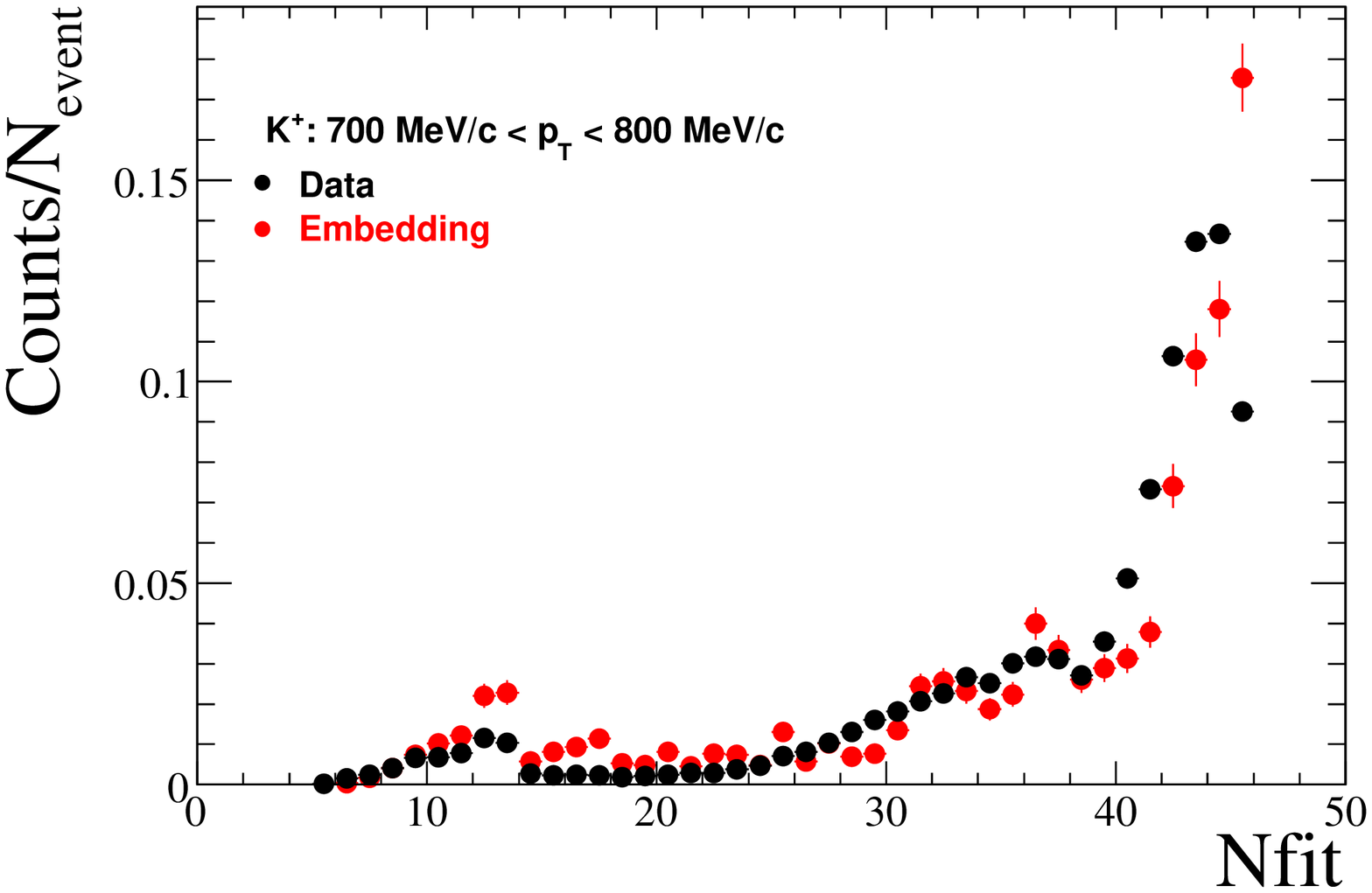}}\\
\resizebox{.225\textwidth}{!}{\includegraphics{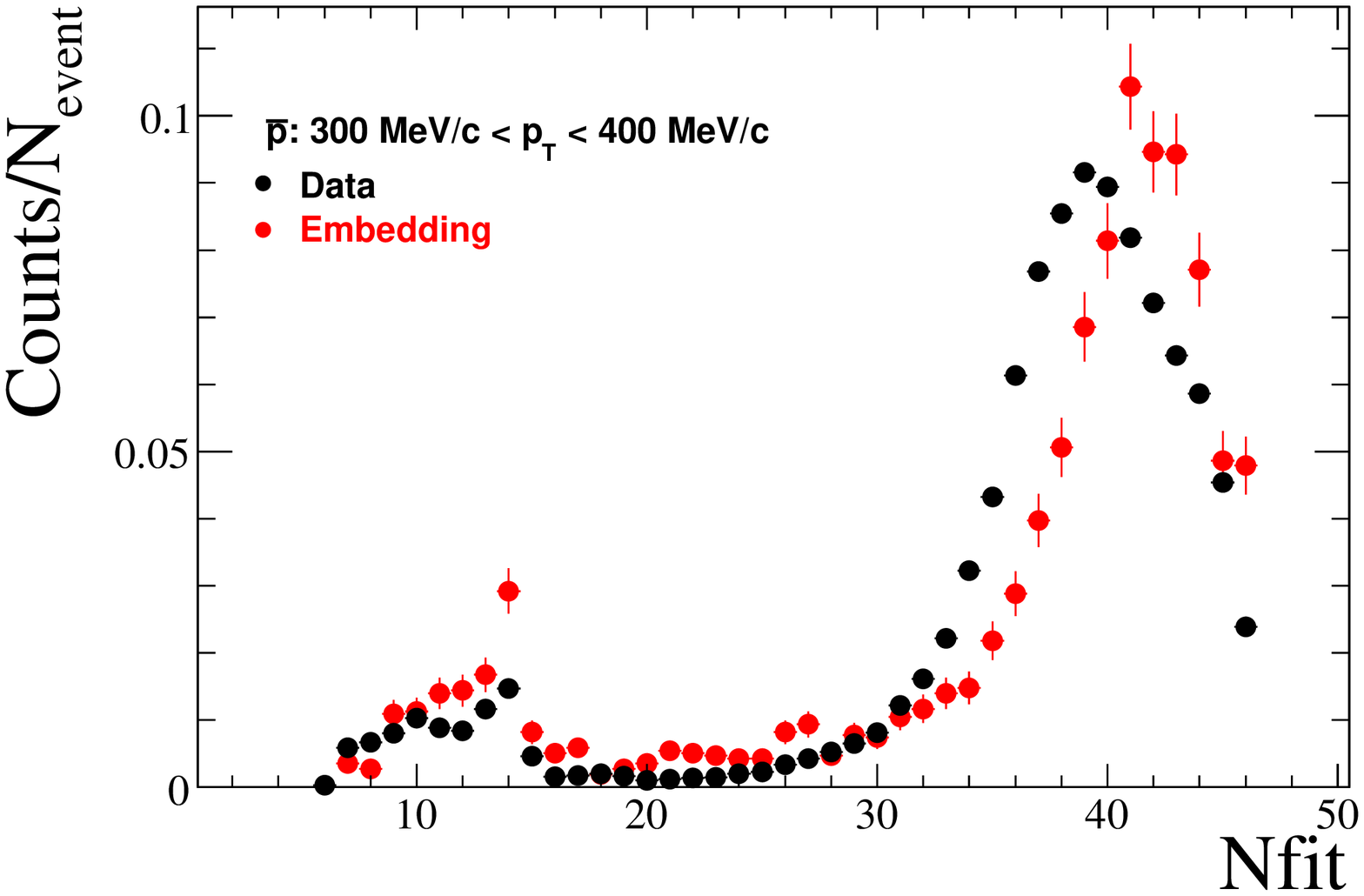}}
\resizebox{.225\textwidth}{!}{\includegraphics{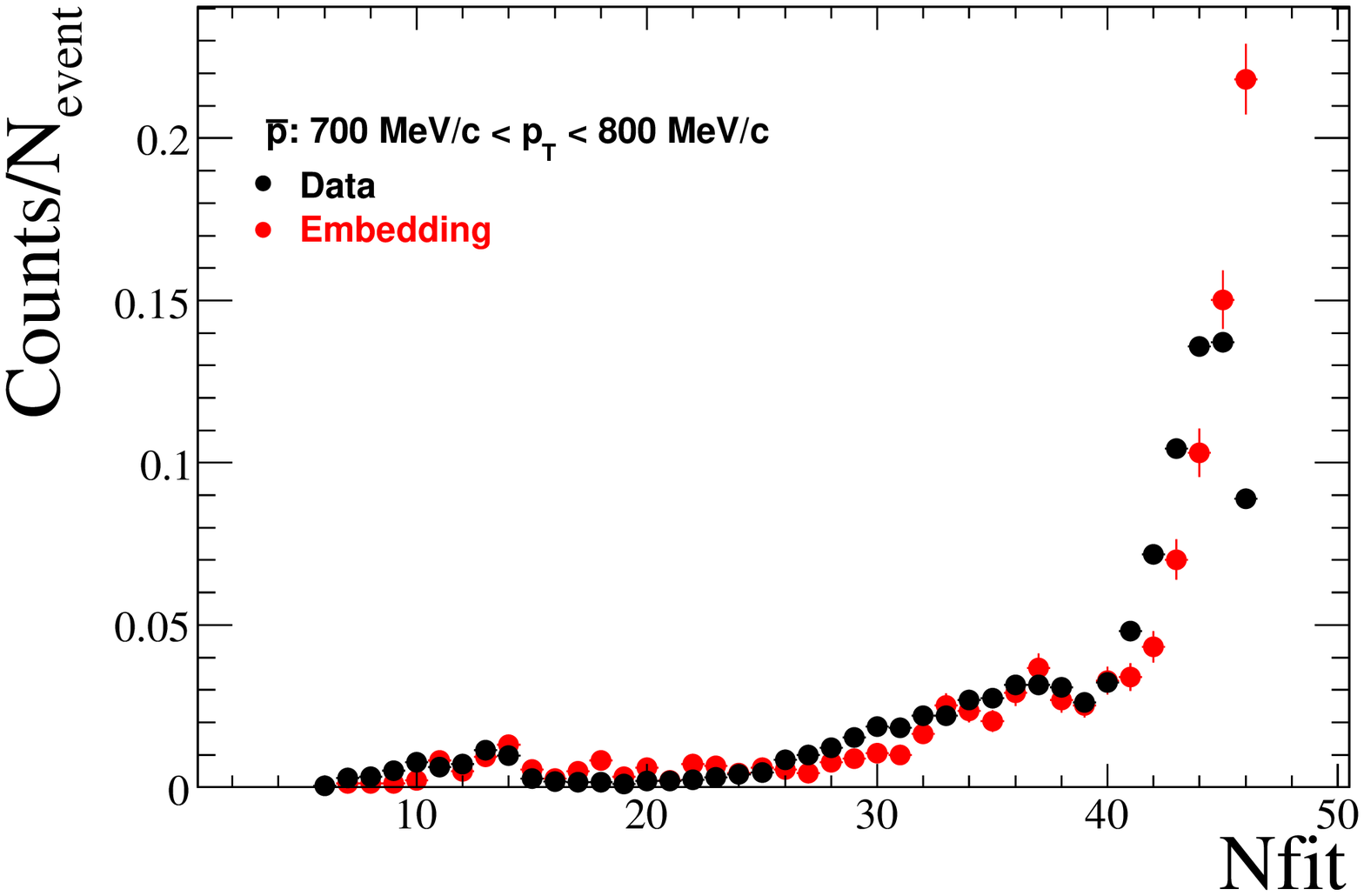}}
\resizebox{.225\textwidth}{!}{\includegraphics{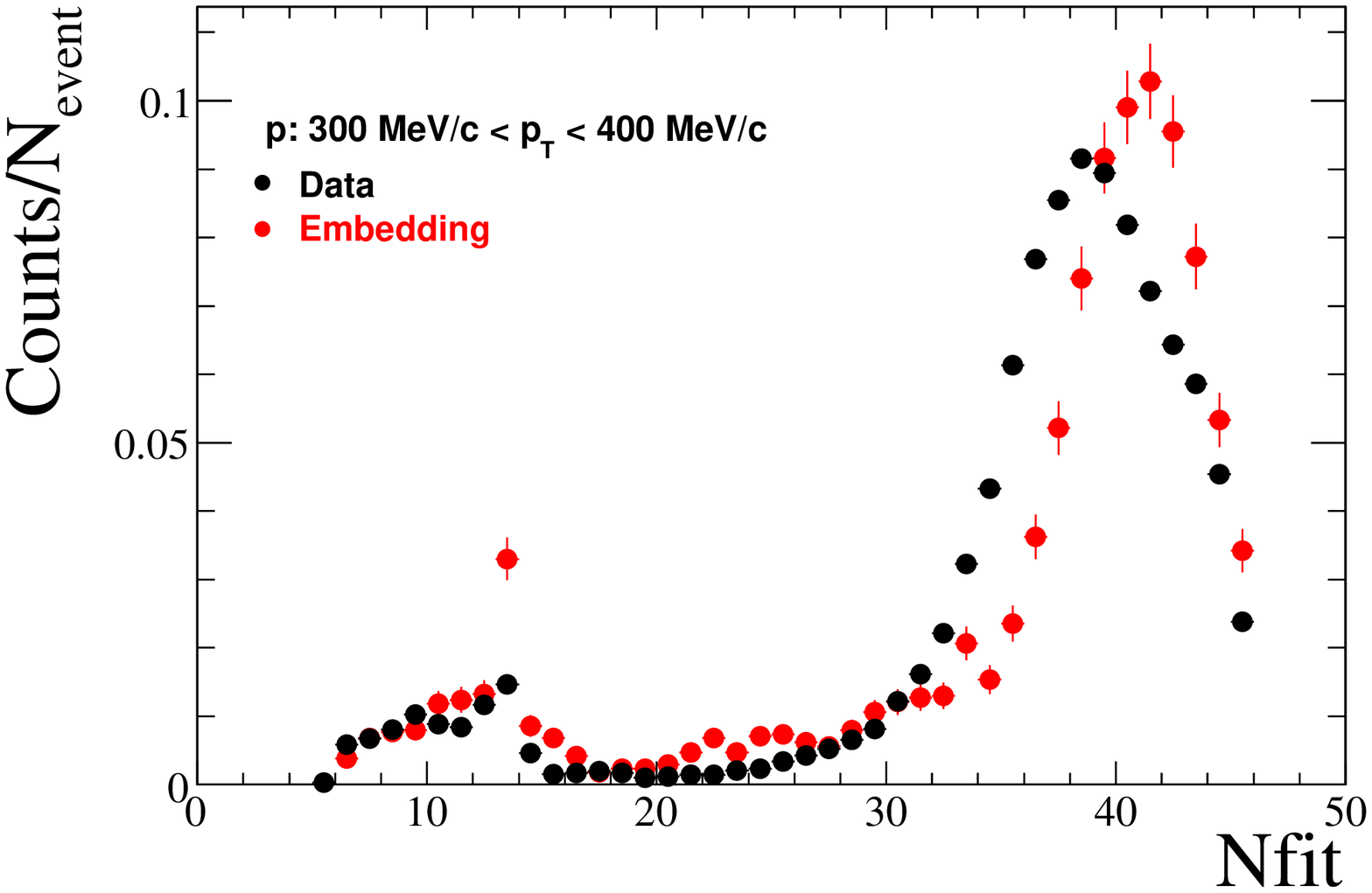}}
\resizebox{.225\textwidth}{!}{\includegraphics{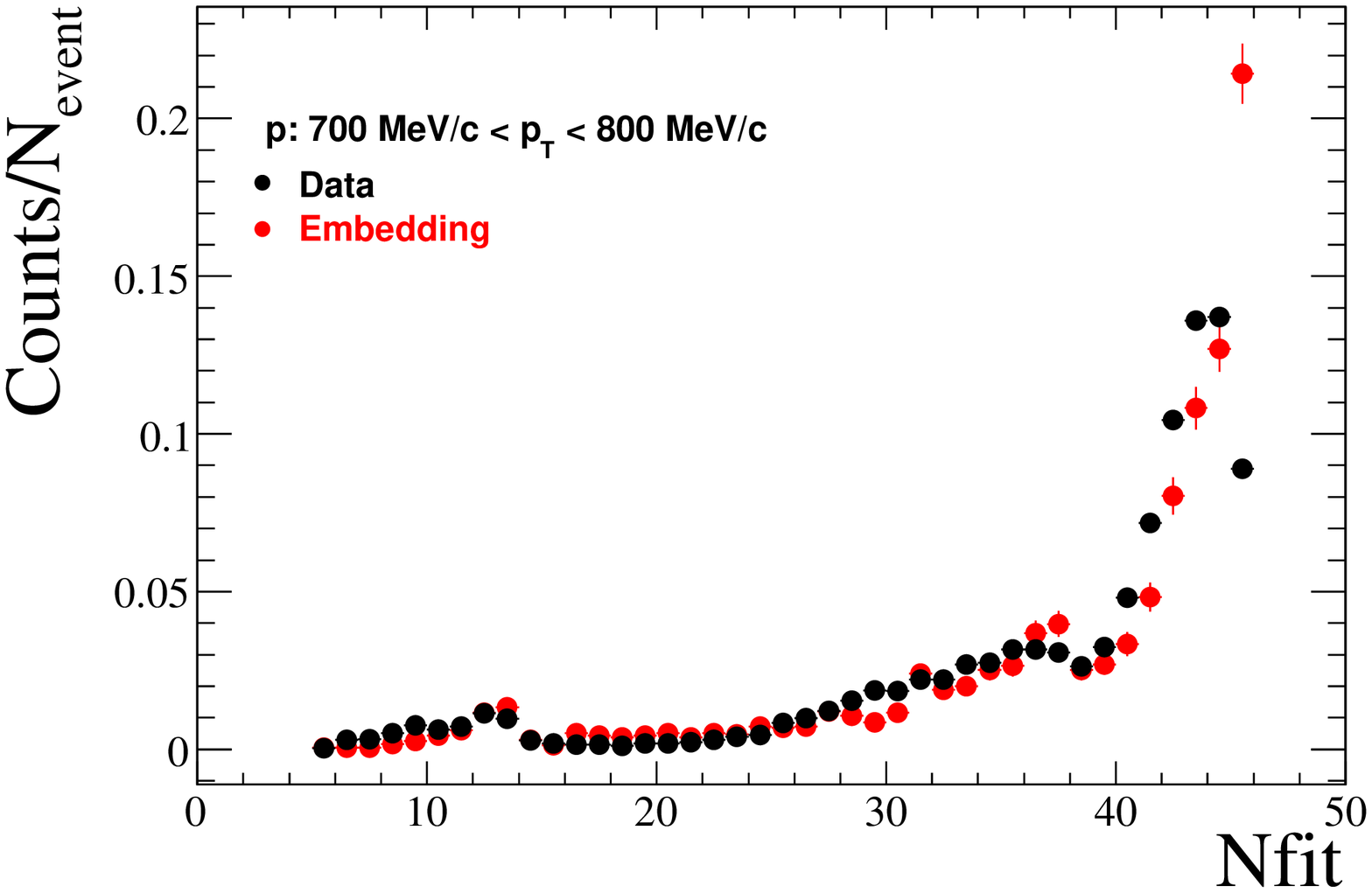}}\\
\caption{Comparison of $N_{fit}$ from real data and embedding in 200 GeV minimum bias dAu collisions.}
\label{fig:dau_nfit_data_embedding}
\end{center} 
\end{sidewaysfigure}
\begin{sidewaysfigure}[!h]
\begin{center}
\resizebox{.225\textwidth}{!}{\includegraphics{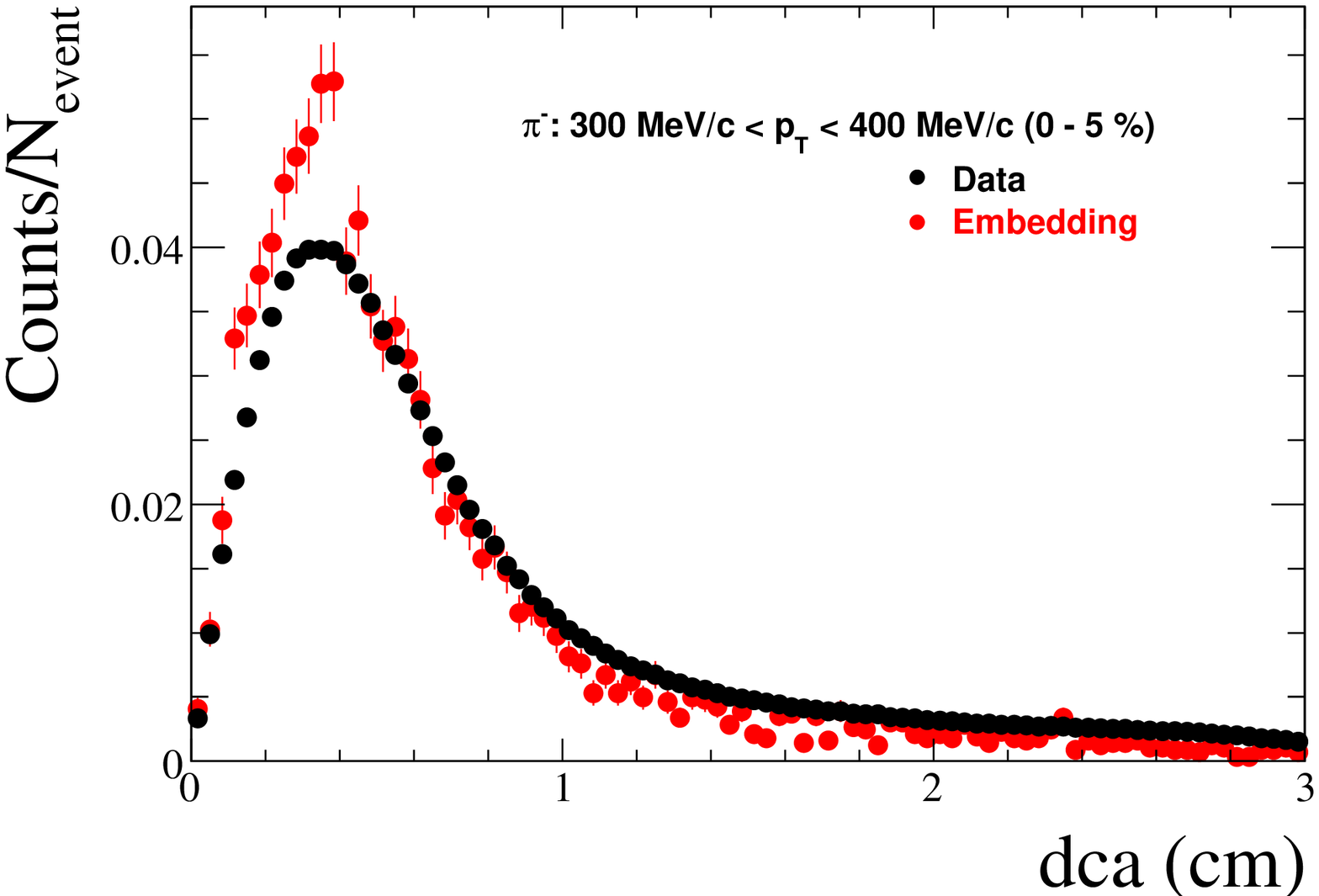}}
\resizebox{.225\textwidth}{!}{\includegraphics{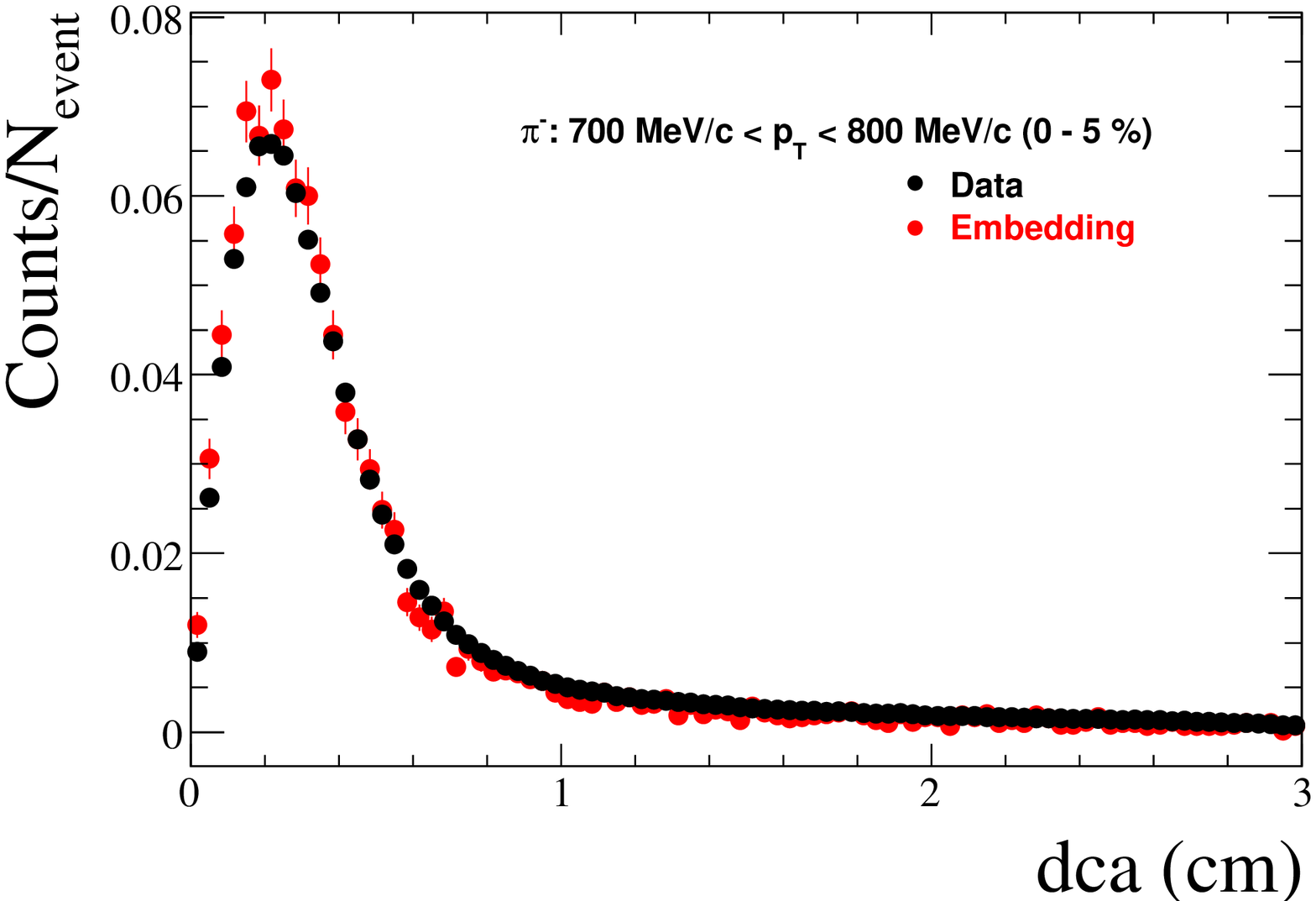}}
\resizebox{.225\textwidth}{!}{\includegraphics{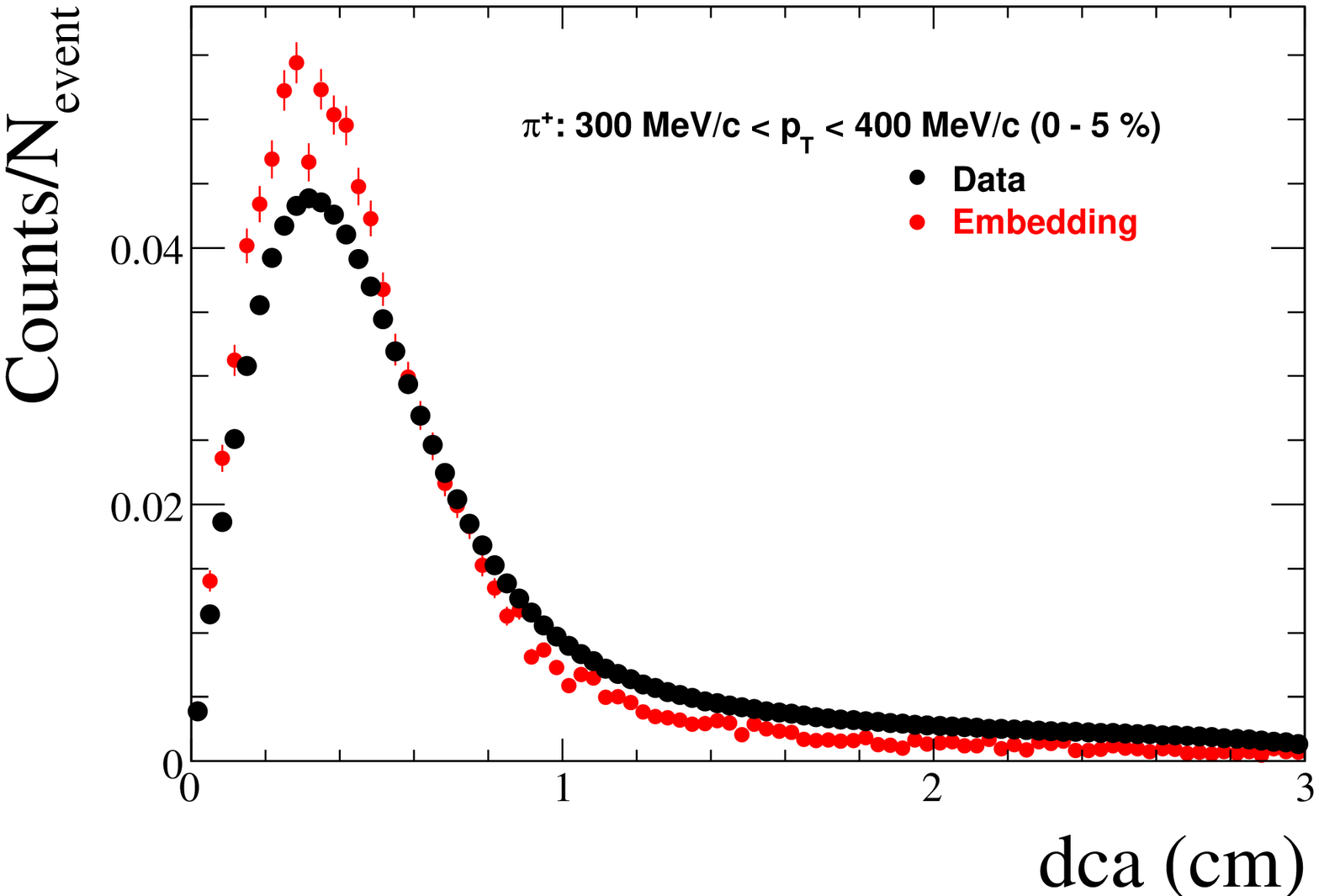}}
\resizebox{.225\textwidth}{!}{\includegraphics{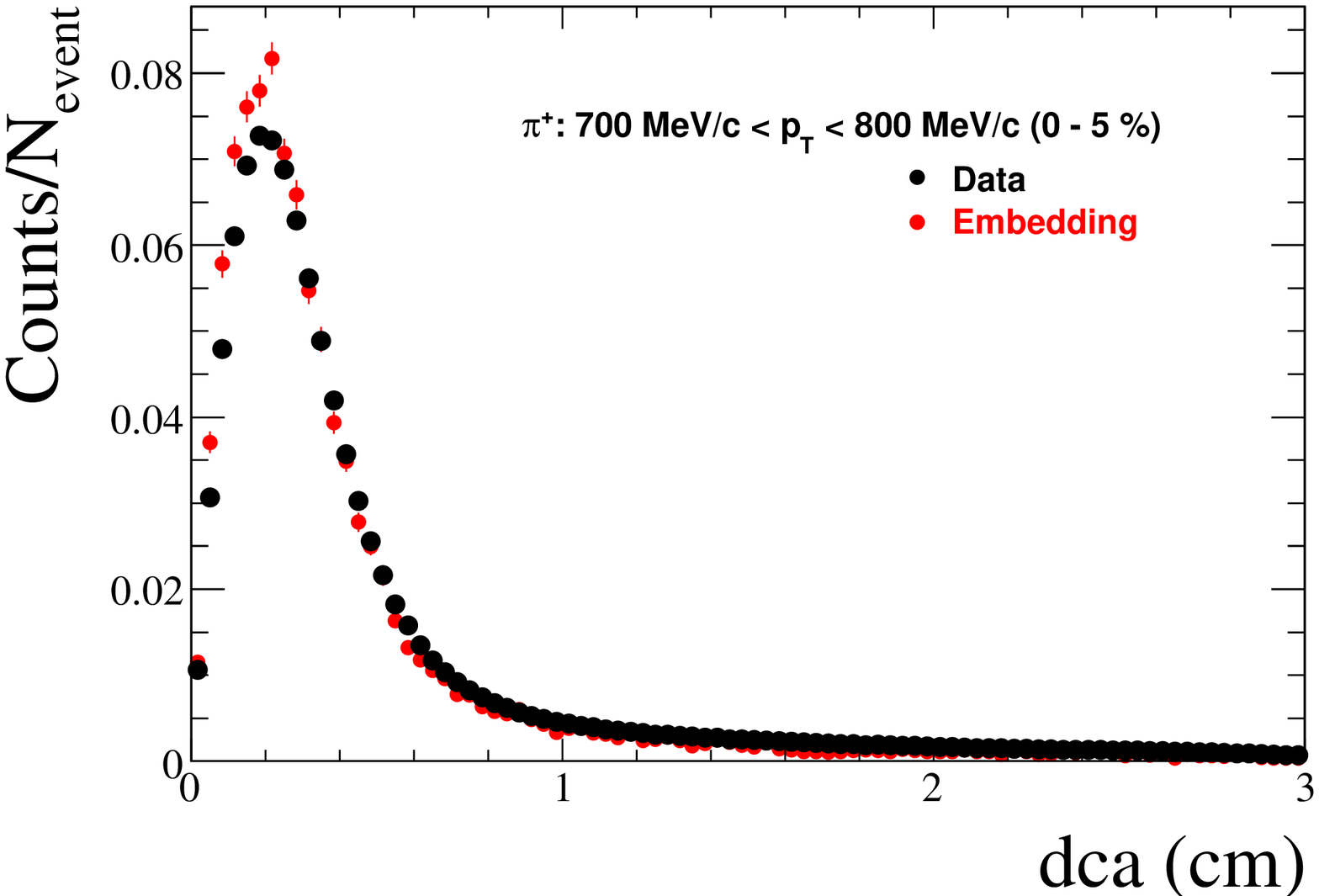}}\\
\resizebox{.225\textwidth}{!}{\includegraphics{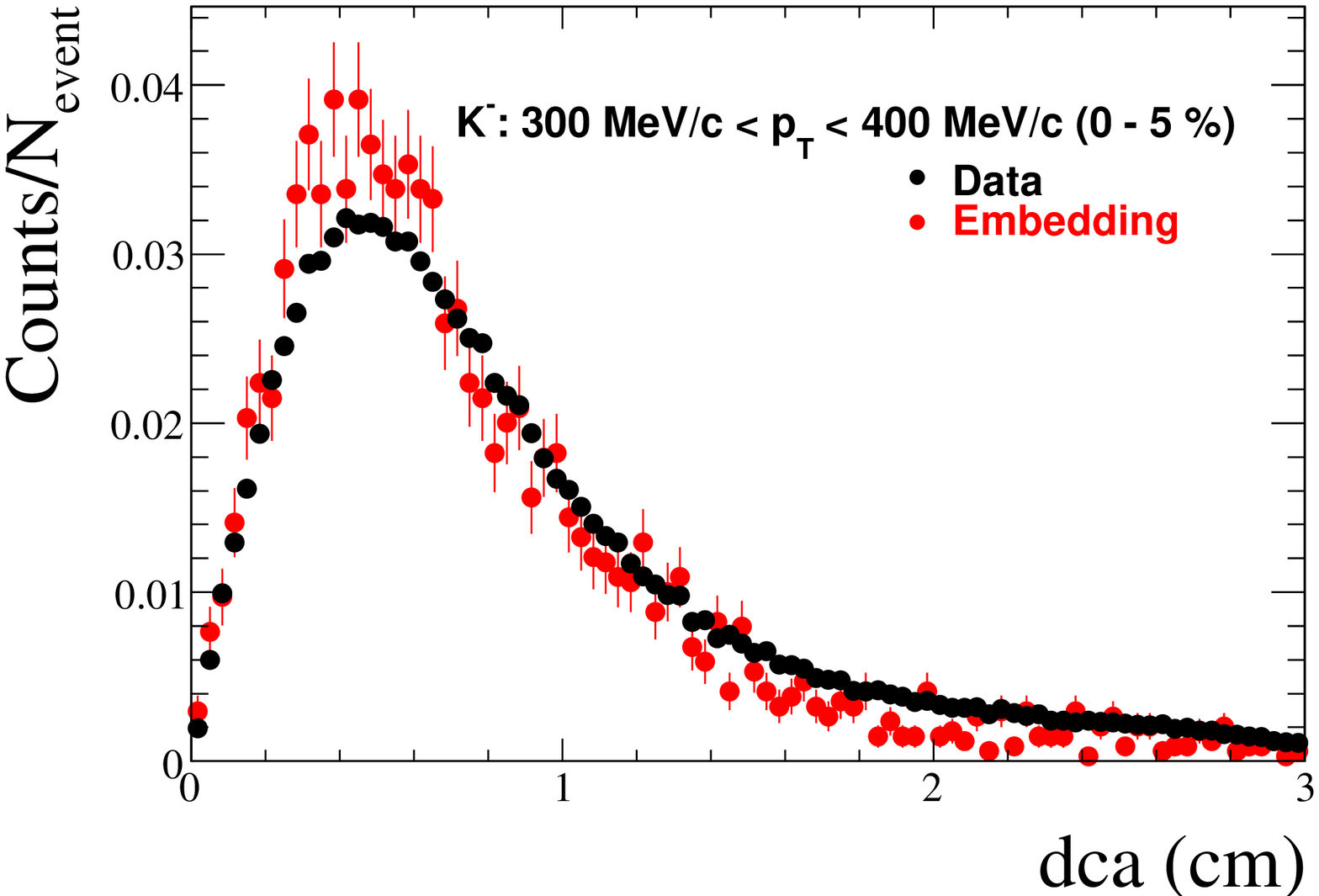}}
\resizebox{.225\textwidth}{!}{\includegraphics{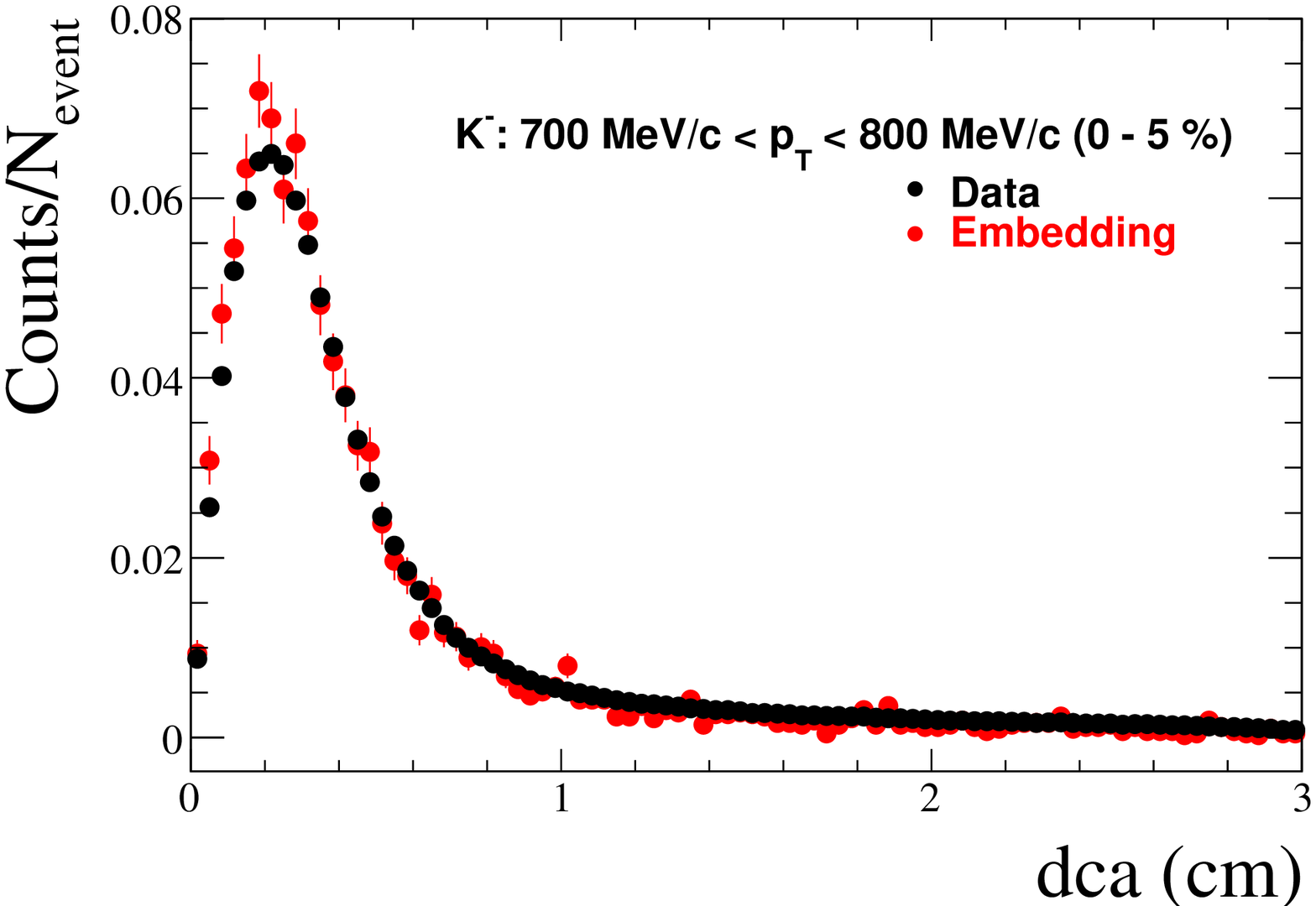}}
\resizebox{.225\textwidth}{!}{\includegraphics{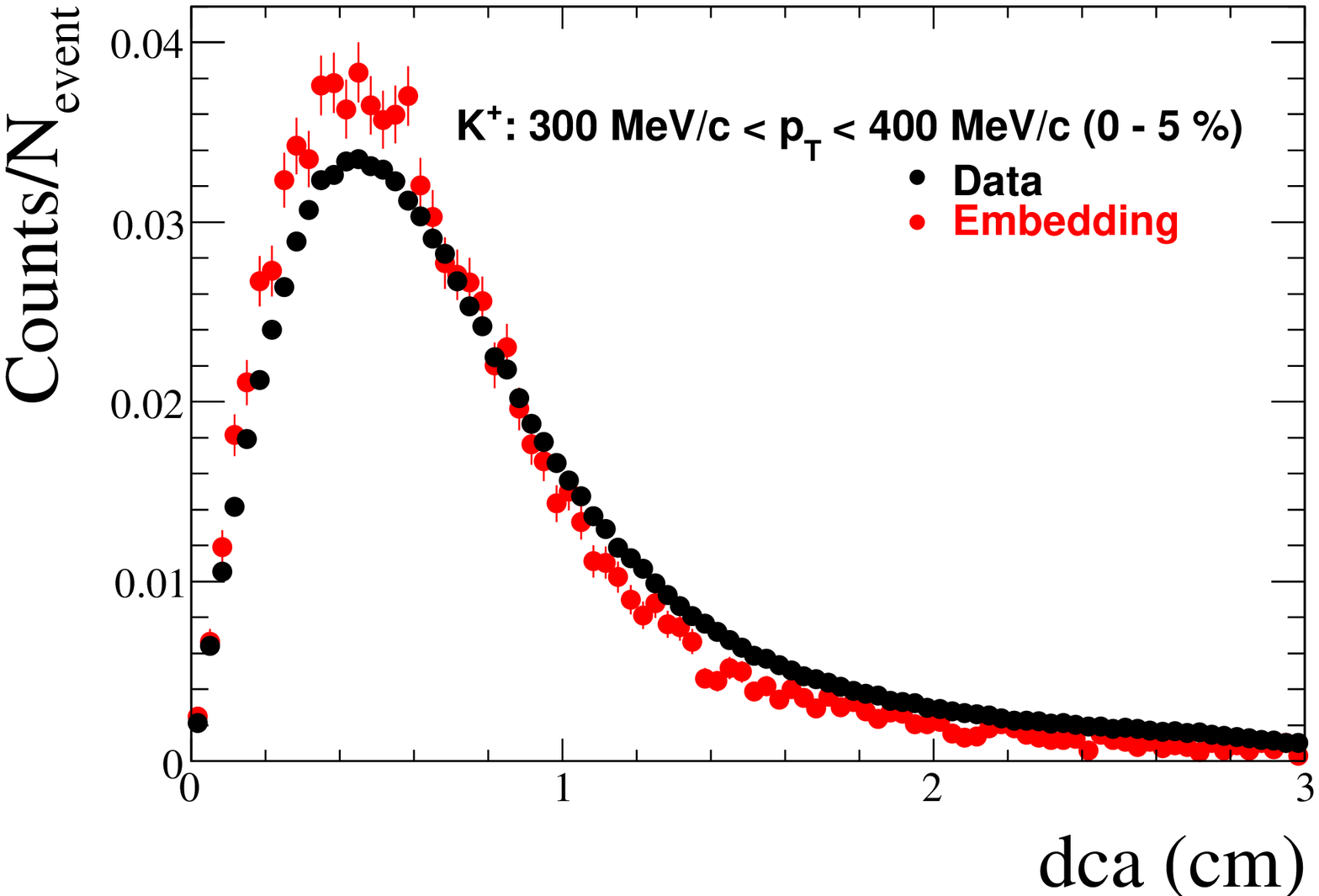}}
\resizebox{.225\textwidth}{!}{\includegraphics{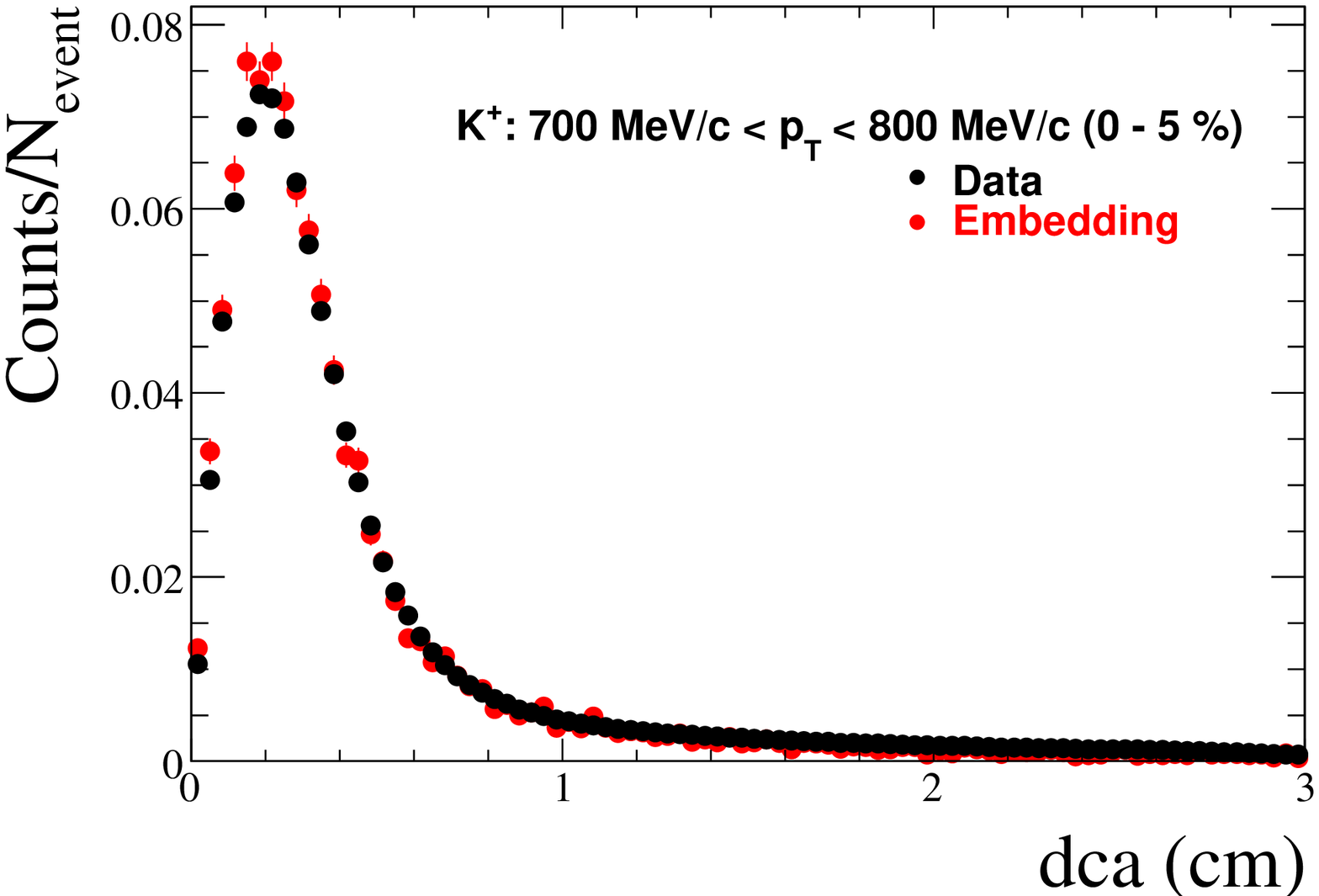}}\\
\resizebox{.225\textwidth}{!}{\includegraphics{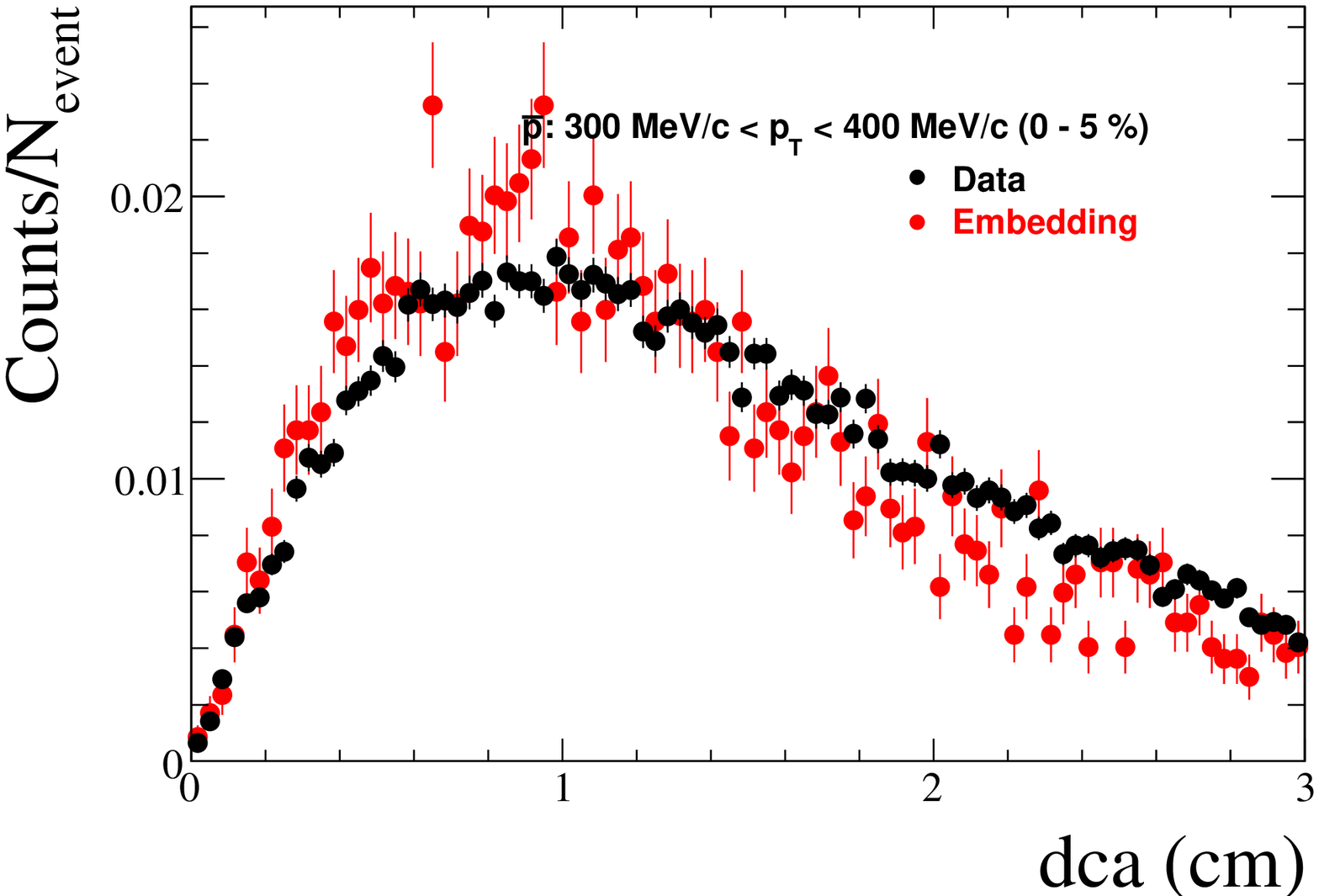}}
\resizebox{.225\textwidth}{!}{\includegraphics{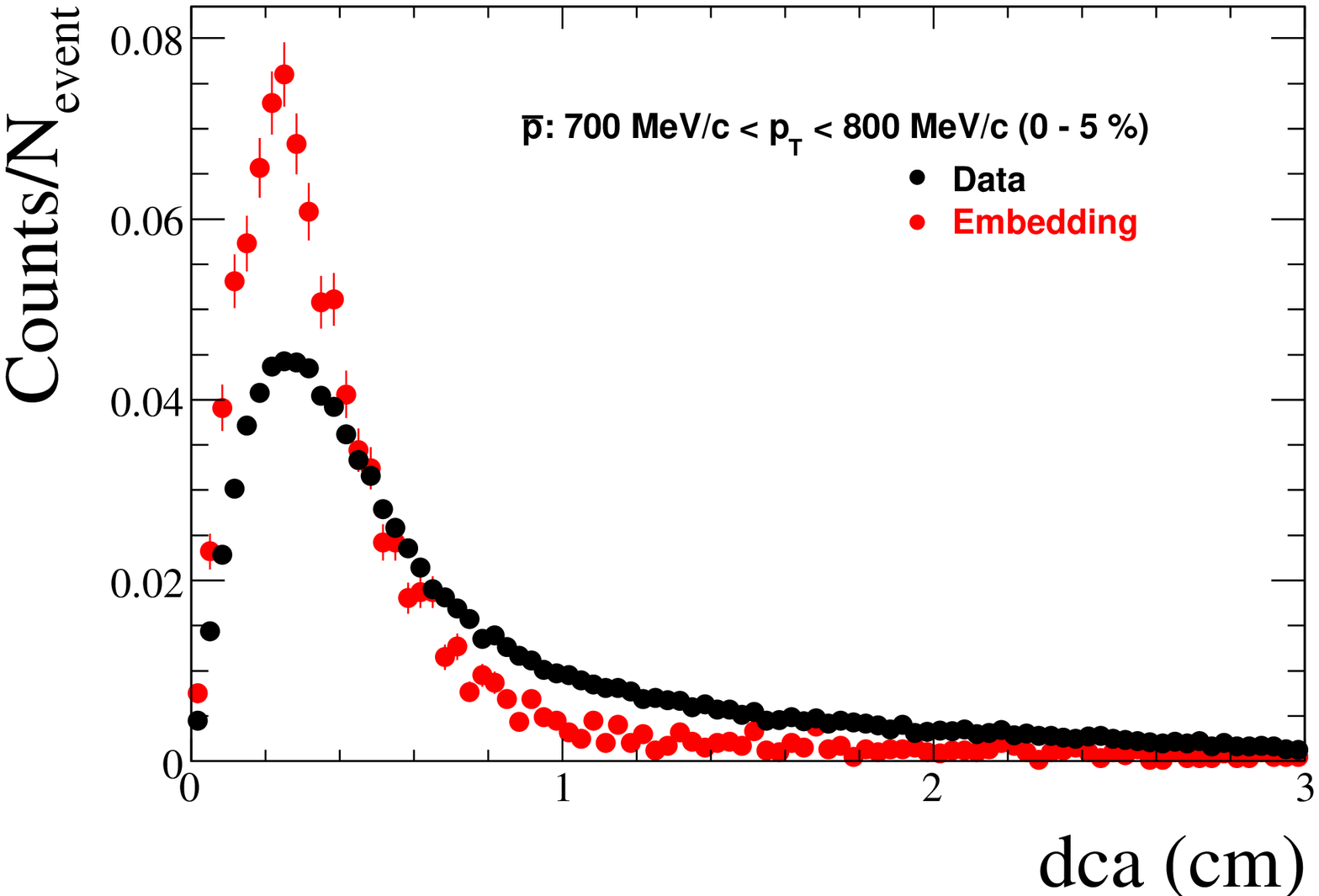}}
\resizebox{.225\textwidth}{!}{\includegraphics{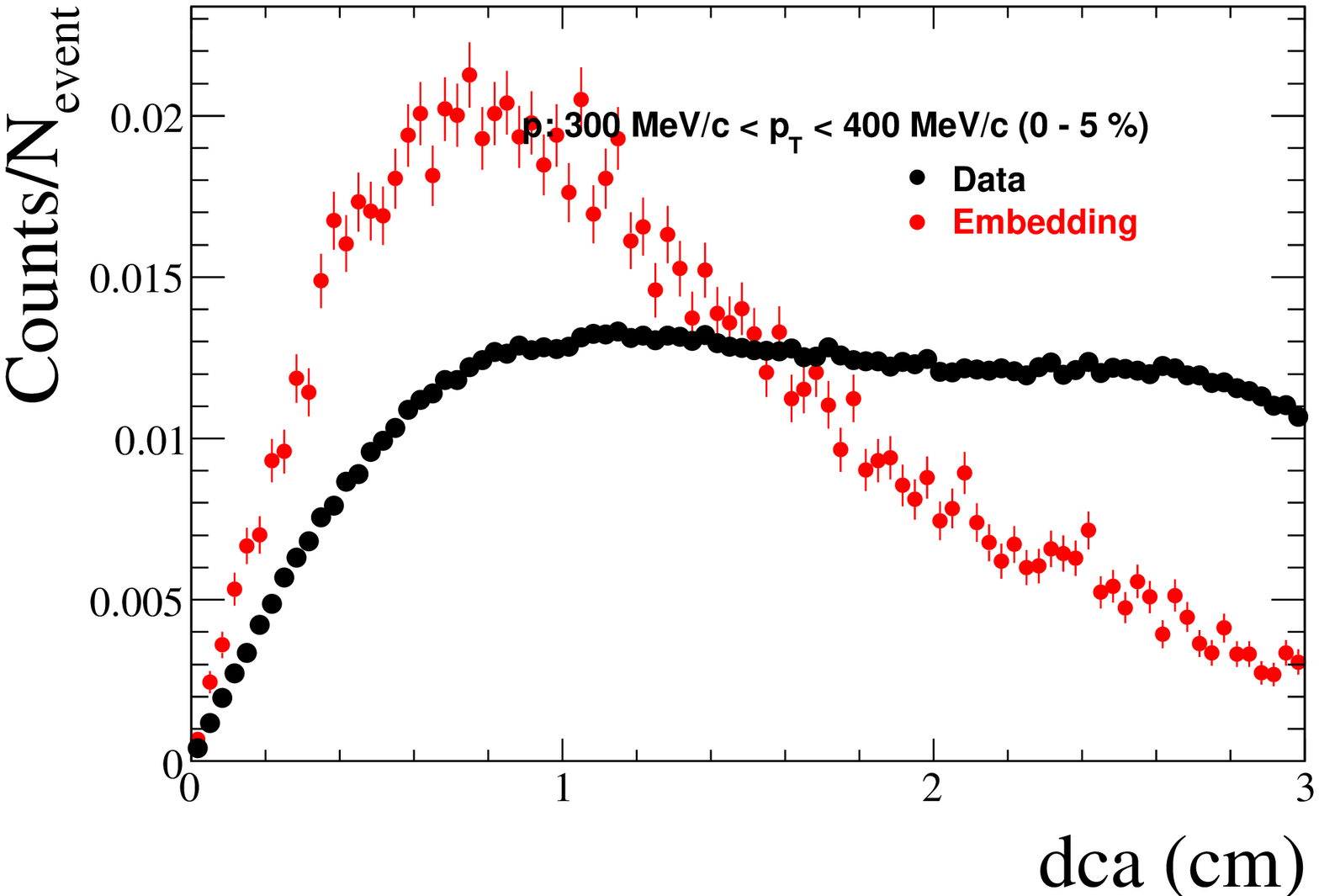}}
\resizebox{.225\textwidth}{!}{\includegraphics{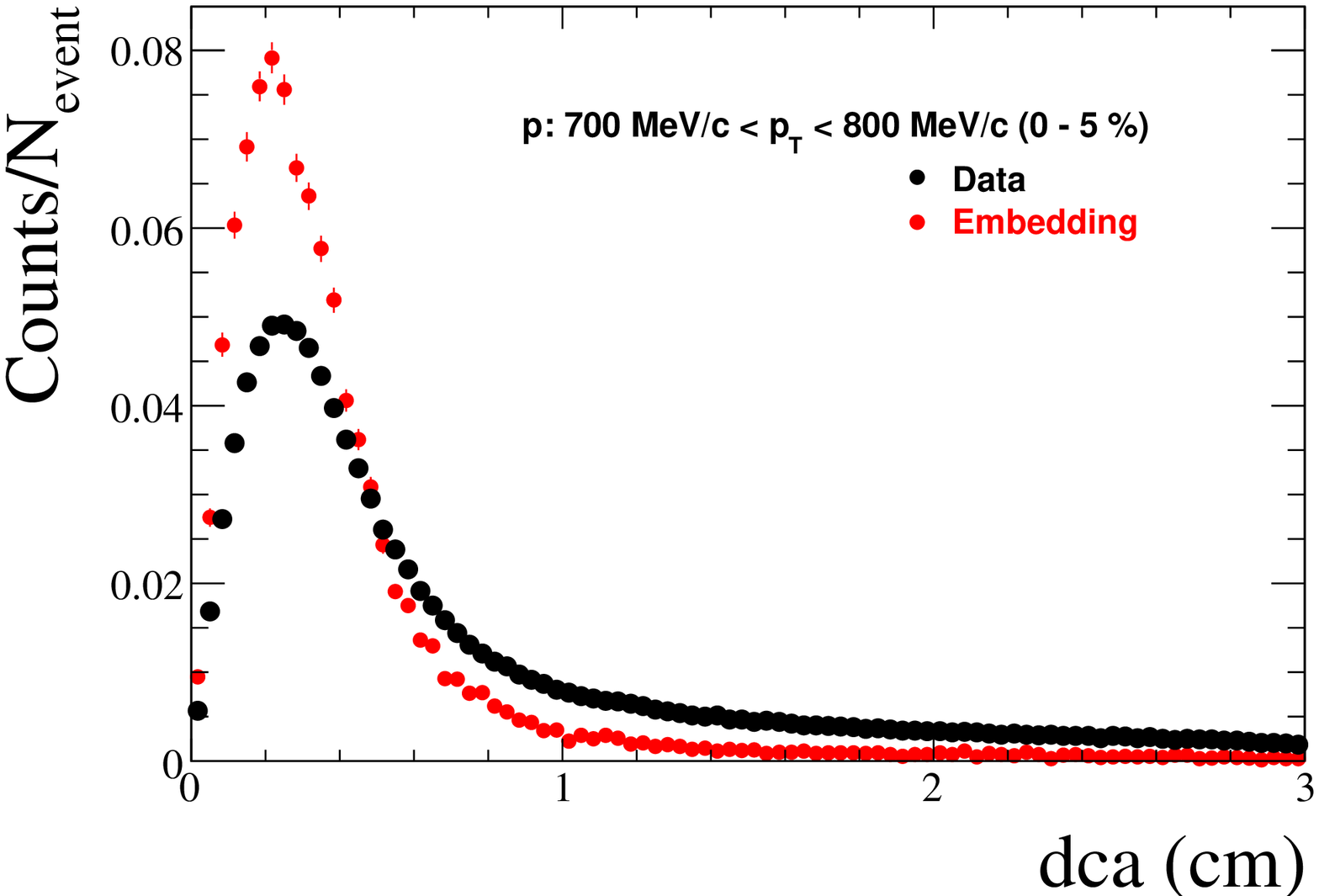}}\\
\caption{Comparison of $dca$ from real data and embedding in 62.4 GeV central (0-5\%) Au-Au collisions.}
\label{fig:auau62central_dca_data_embedding}
\end{center} 
\end{sidewaysfigure}
\begin{sidewaysfigure}[!h]
\begin{center}
\resizebox{.225\textwidth}{!}{\includegraphics{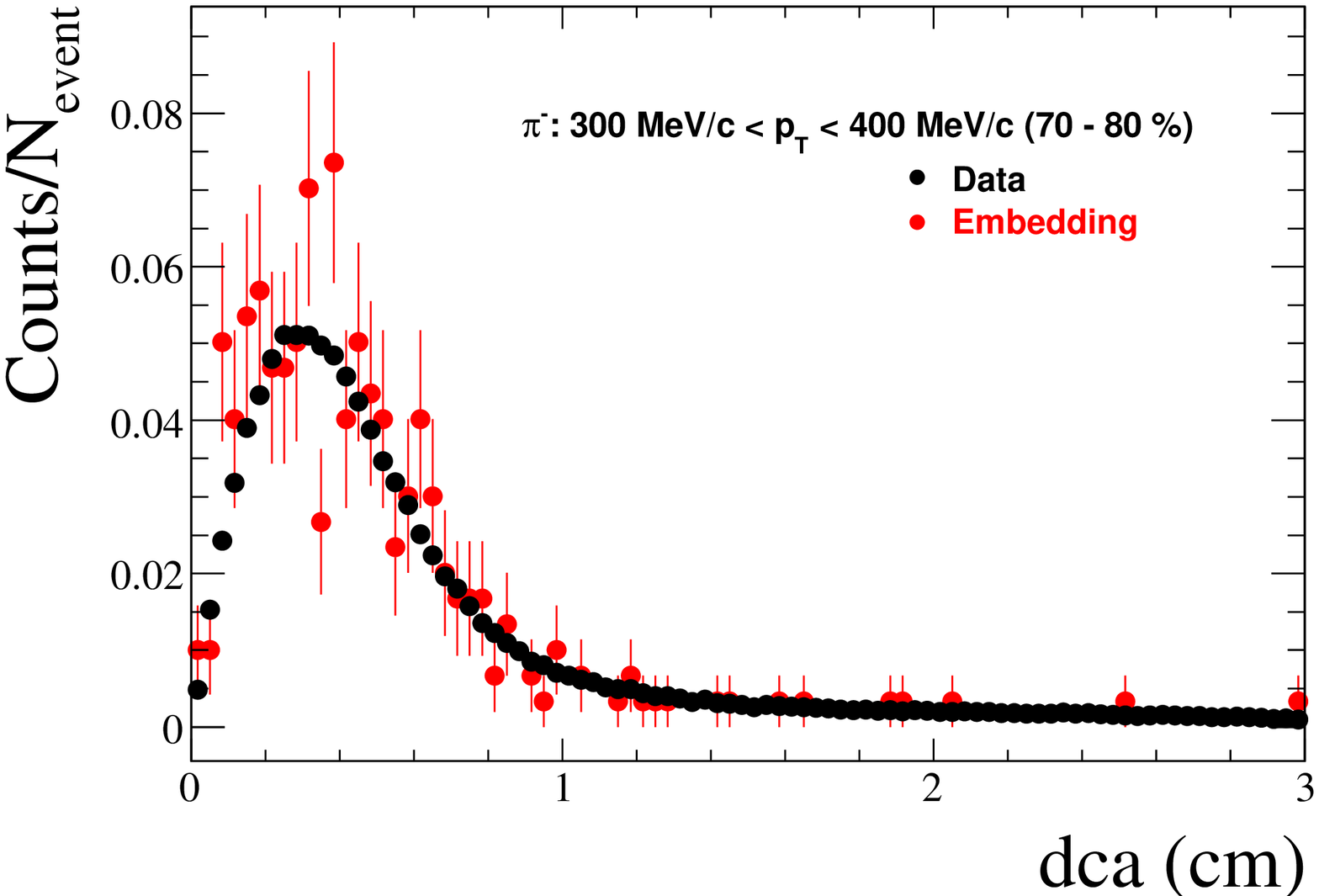}}
\resizebox{.225\textwidth}{!}{\includegraphics{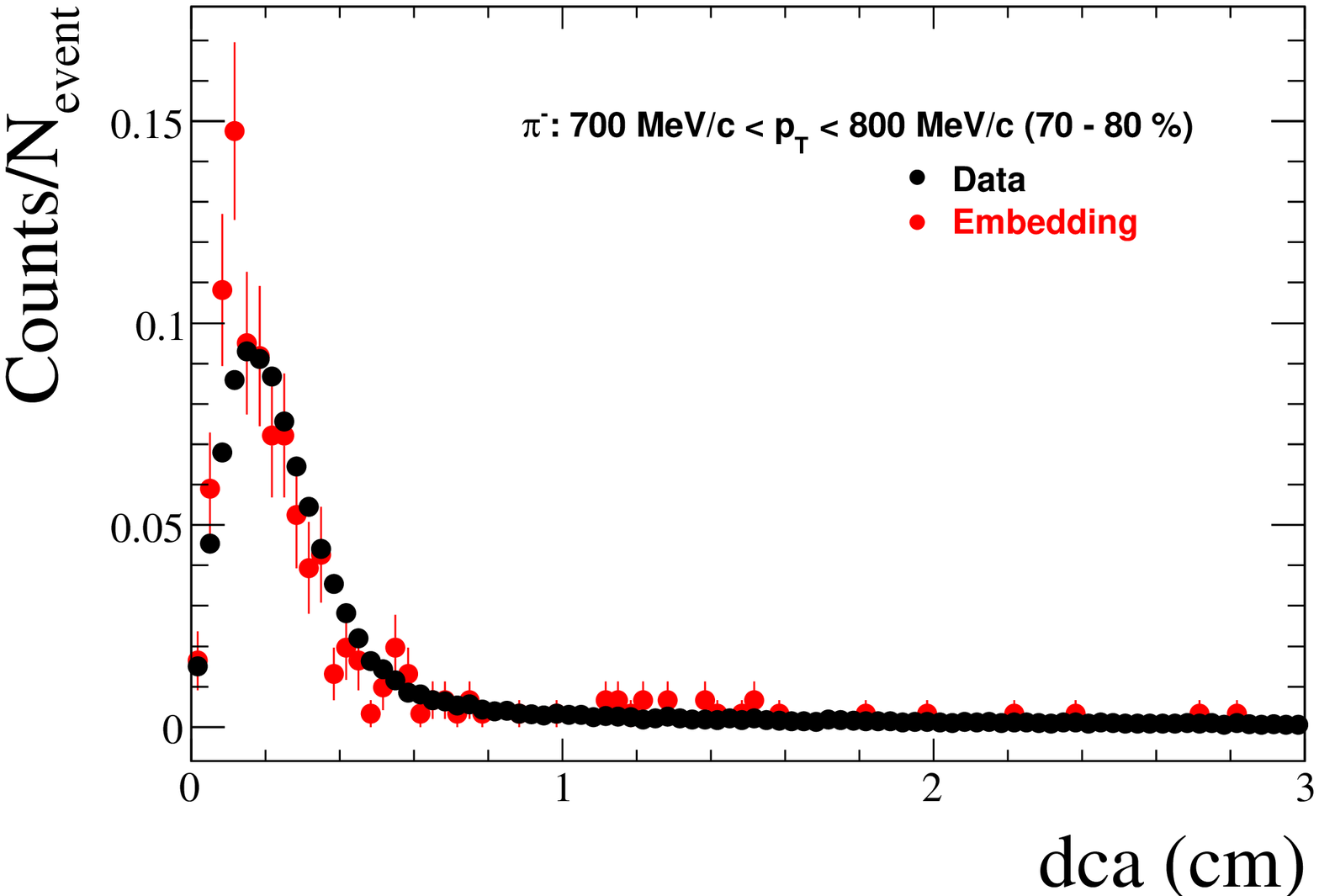}}
\resizebox{.225\textwidth}{!}{\includegraphics{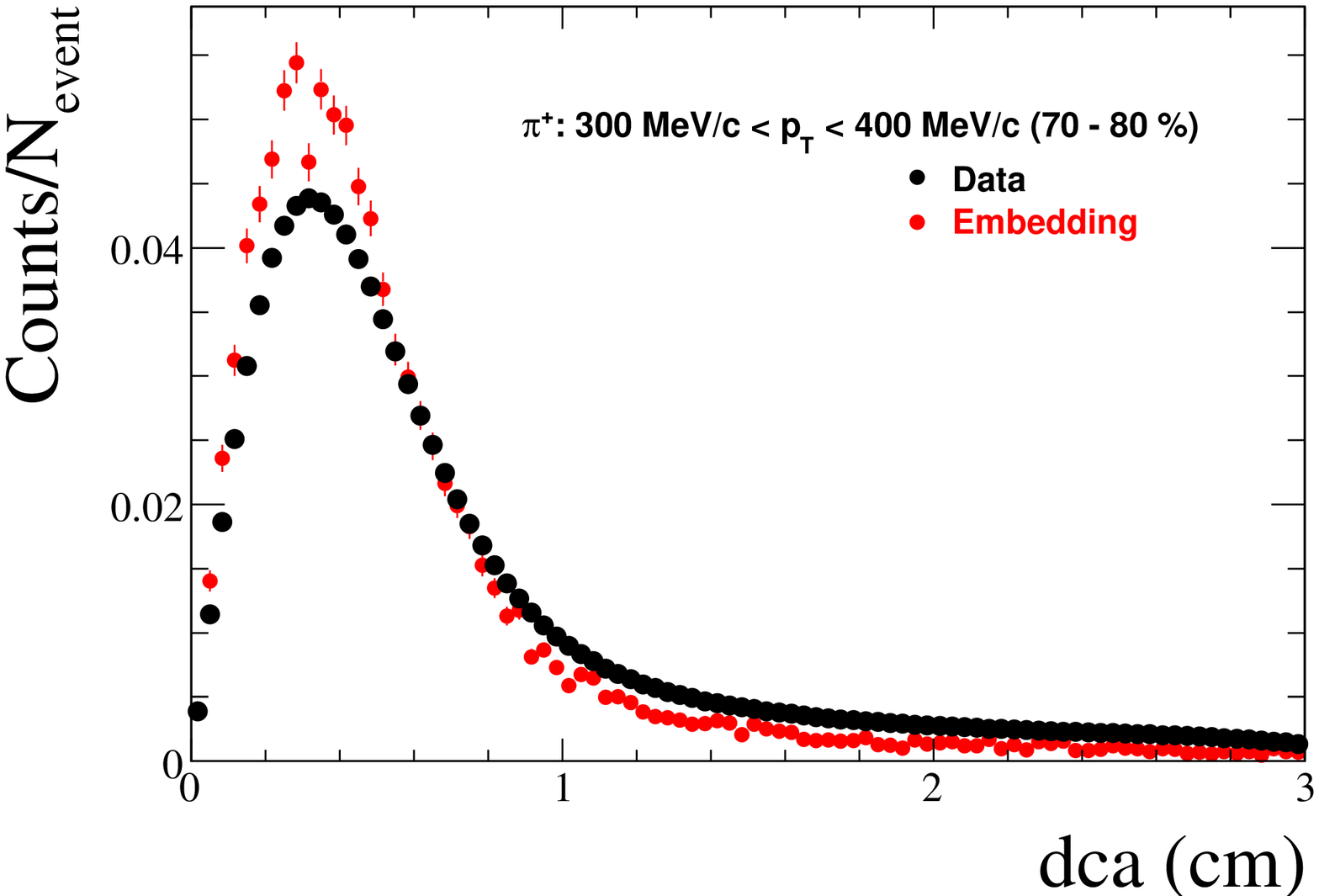}}
\resizebox{.225\textwidth}{!}{\includegraphics{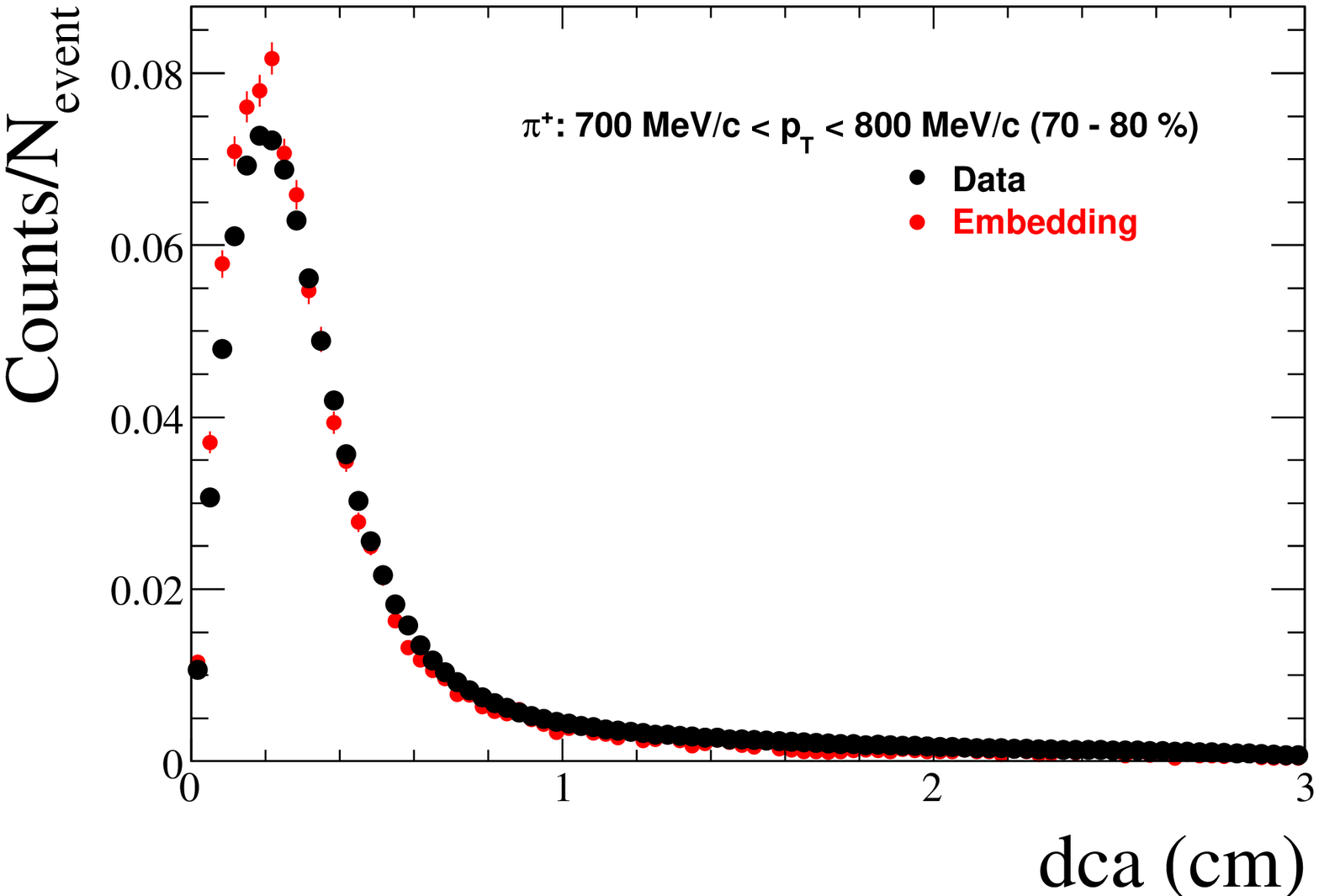}}\\
\resizebox{.225\textwidth}{!}{\includegraphics{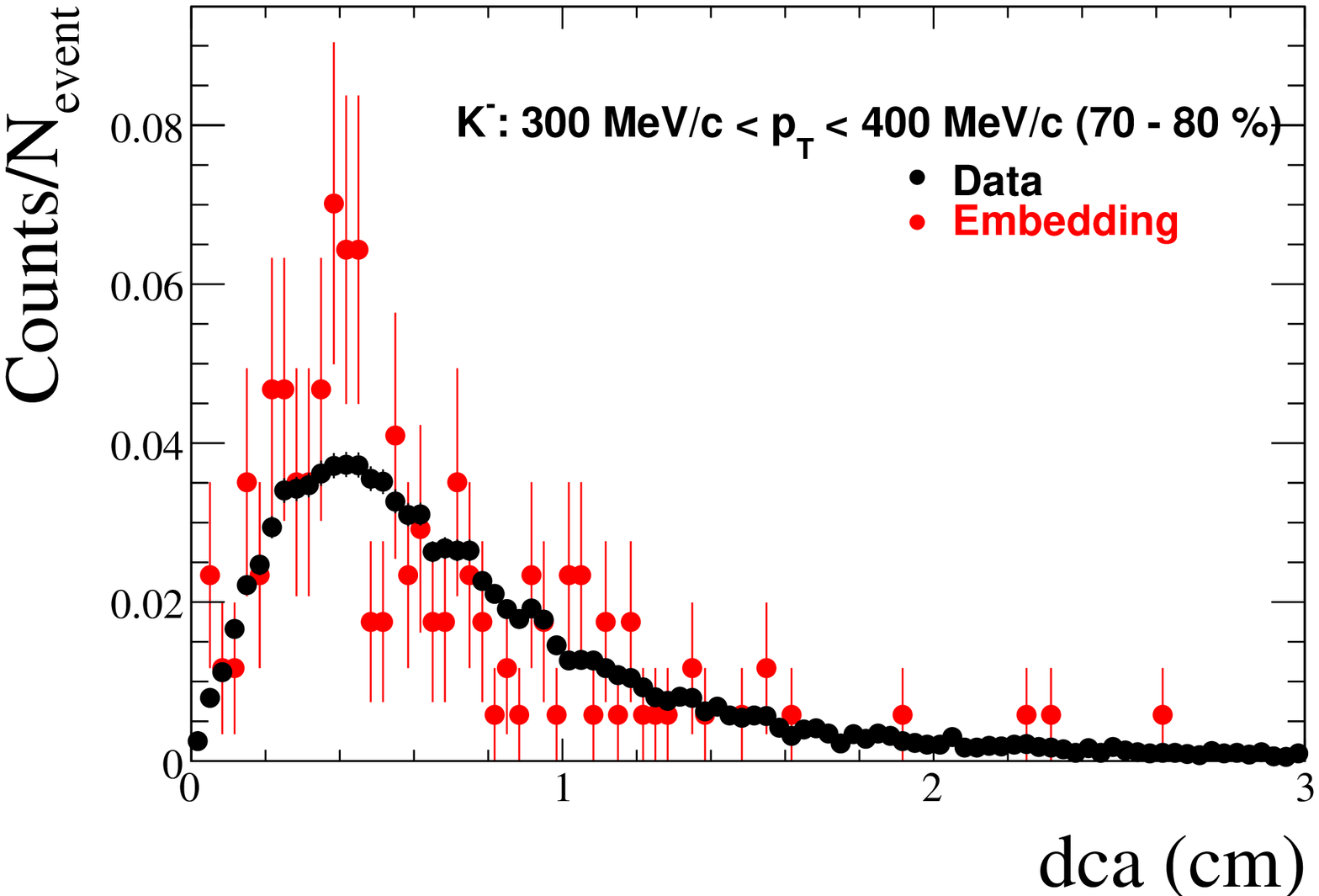}}
\resizebox{.225\textwidth}{!}{\includegraphics{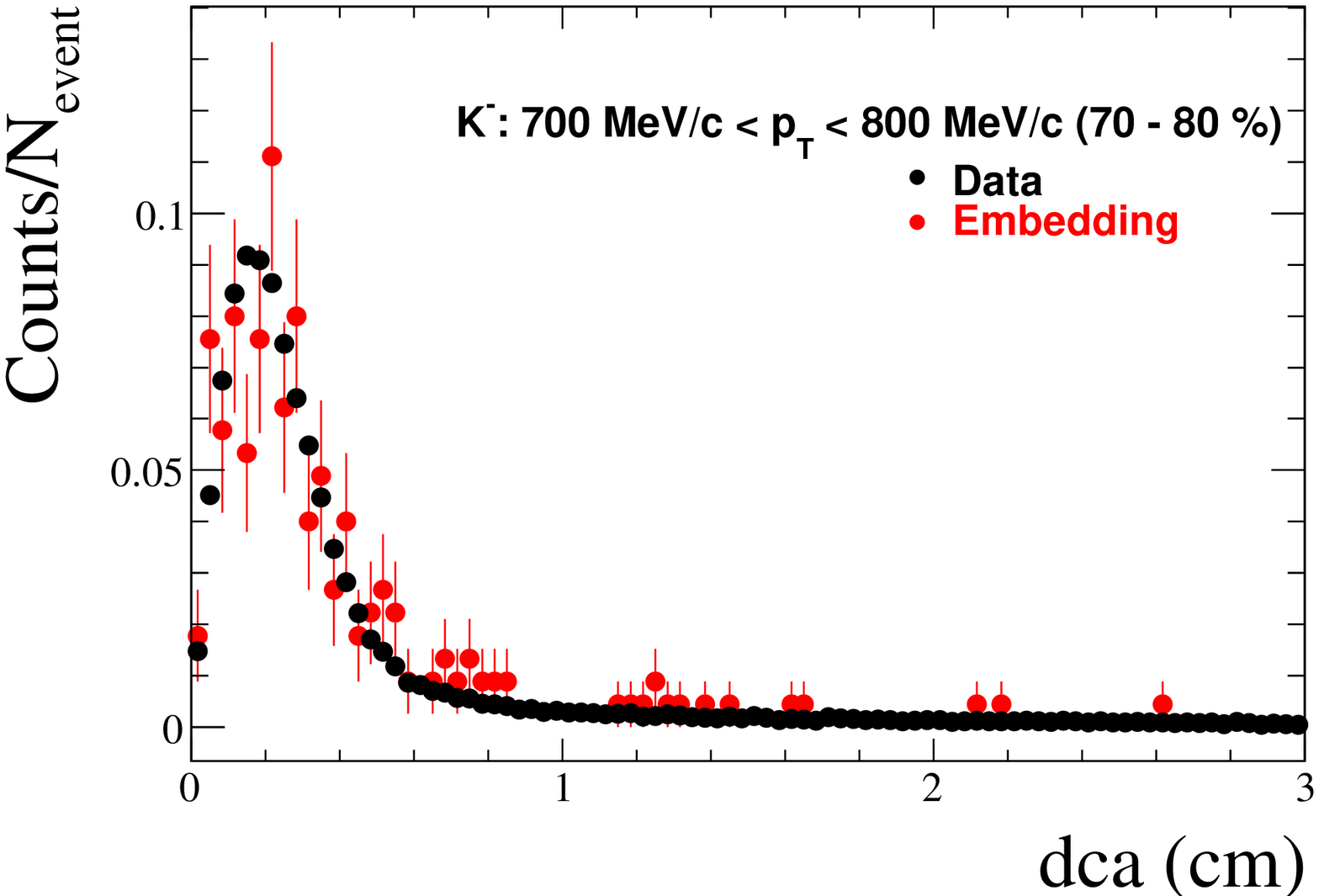}}
\resizebox{.225\textwidth}{!}{\includegraphics{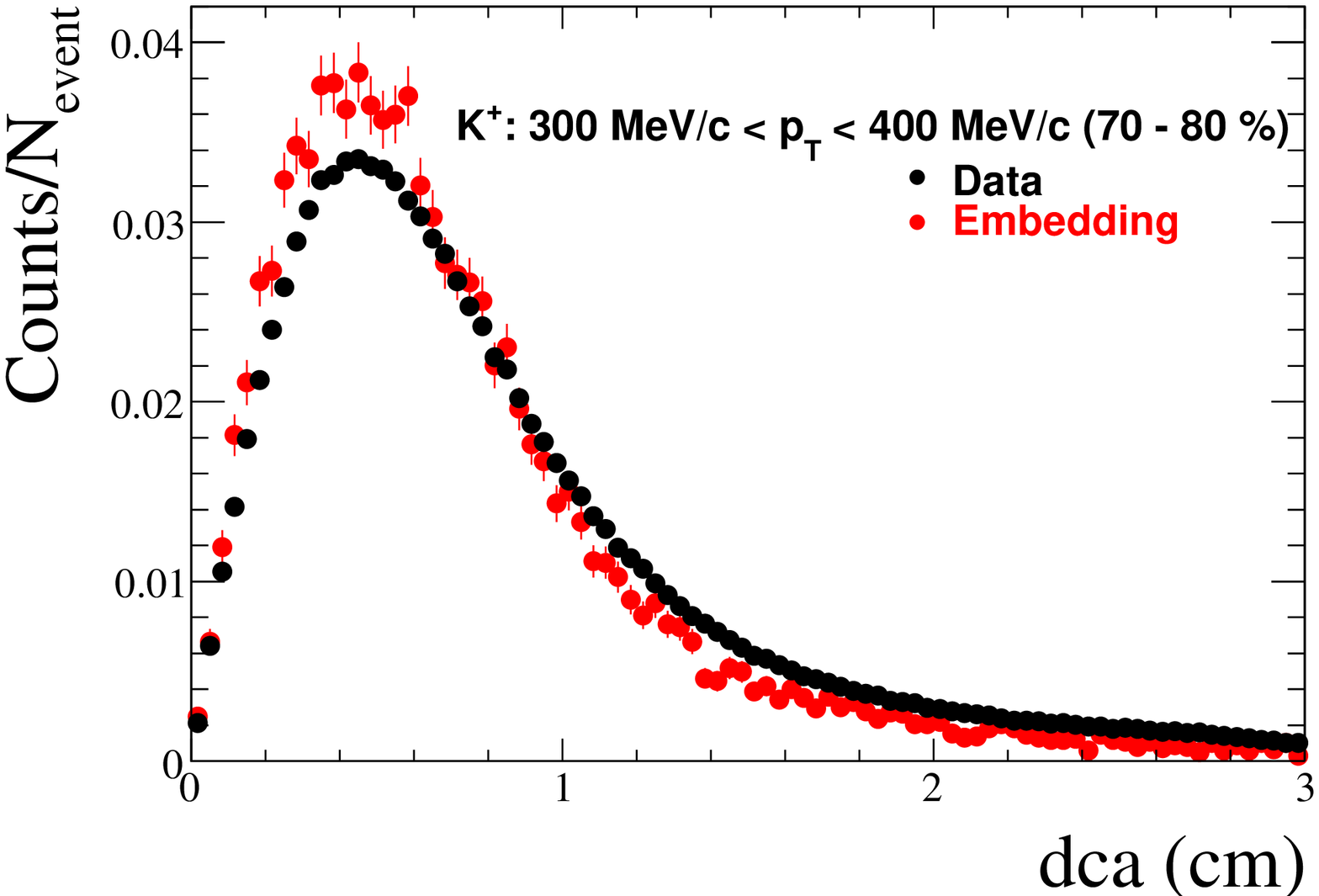}}
\resizebox{.225\textwidth}{!}{\includegraphics{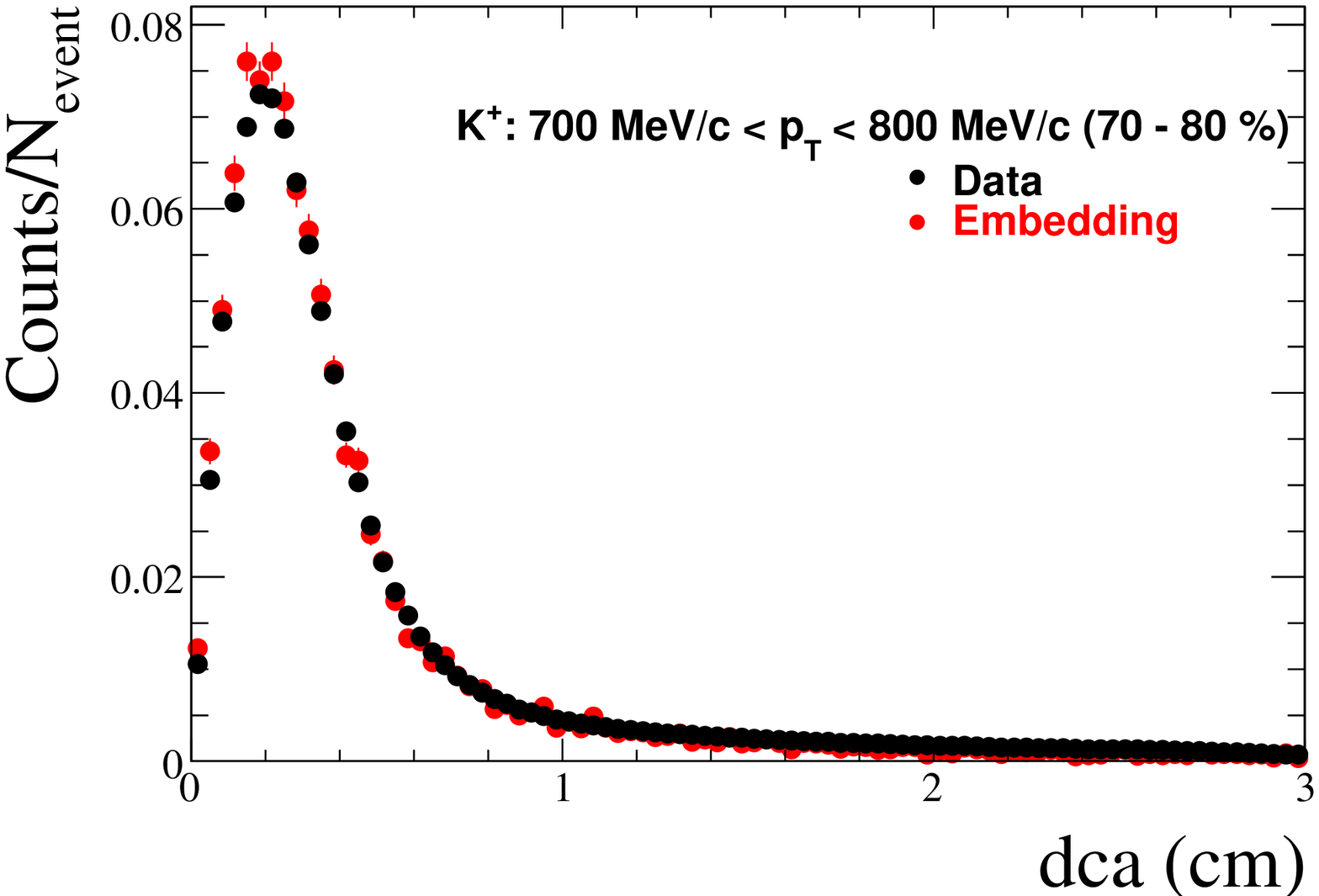}}\\
\resizebox{.225\textwidth}{!}{\includegraphics{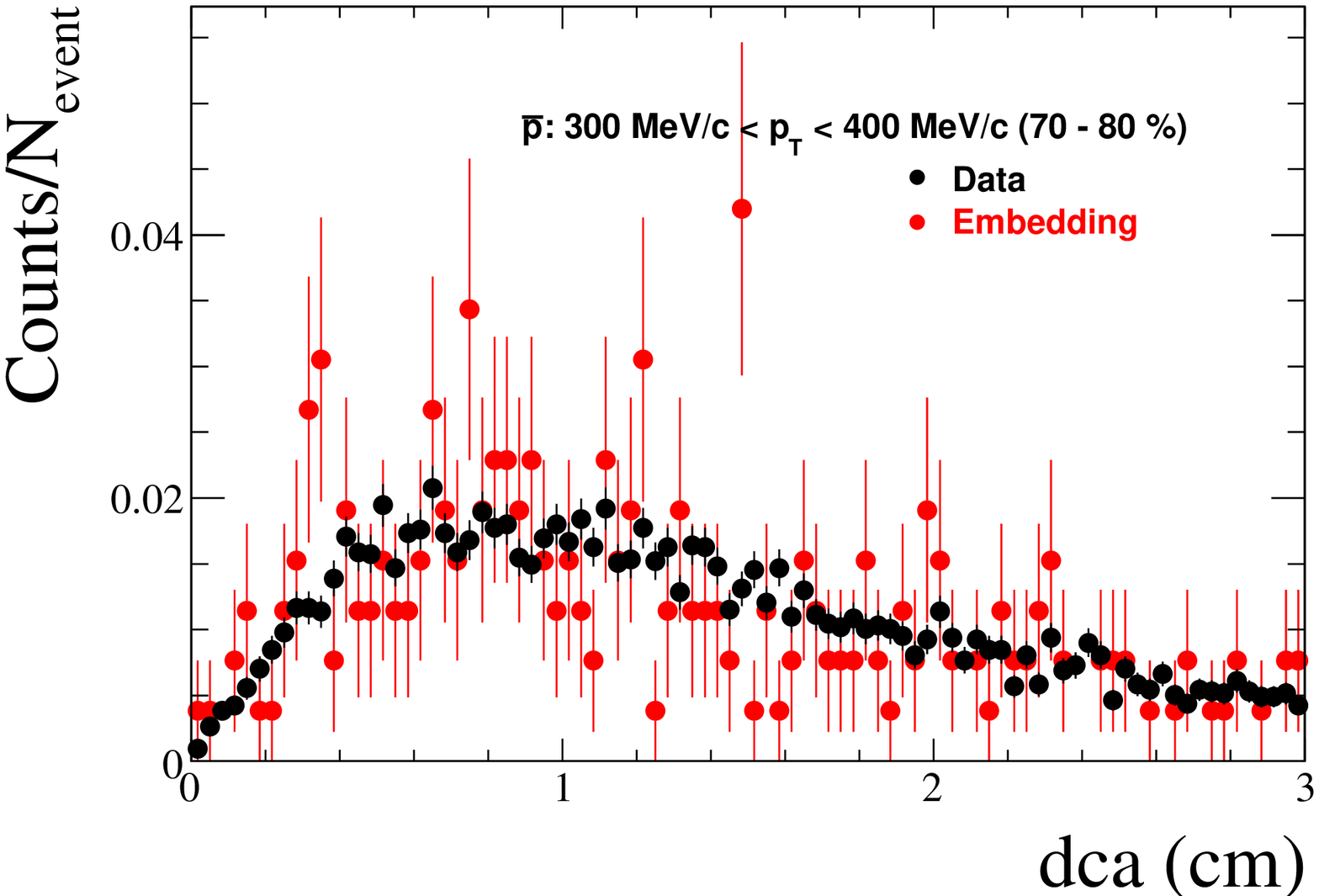}}
\resizebox{.225\textwidth}{!}{\includegraphics{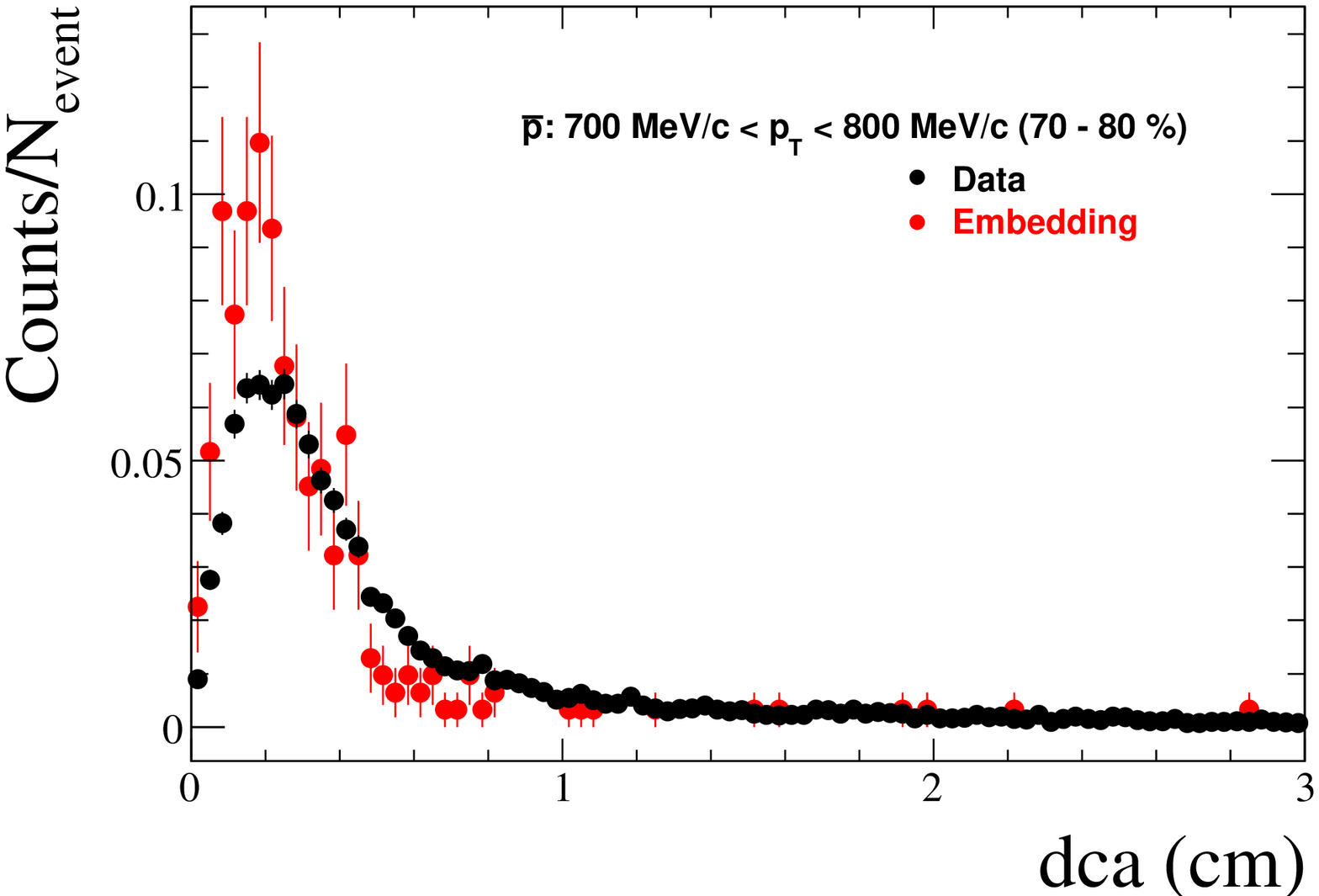}}
\resizebox{.225\textwidth}{!}{\includegraphics{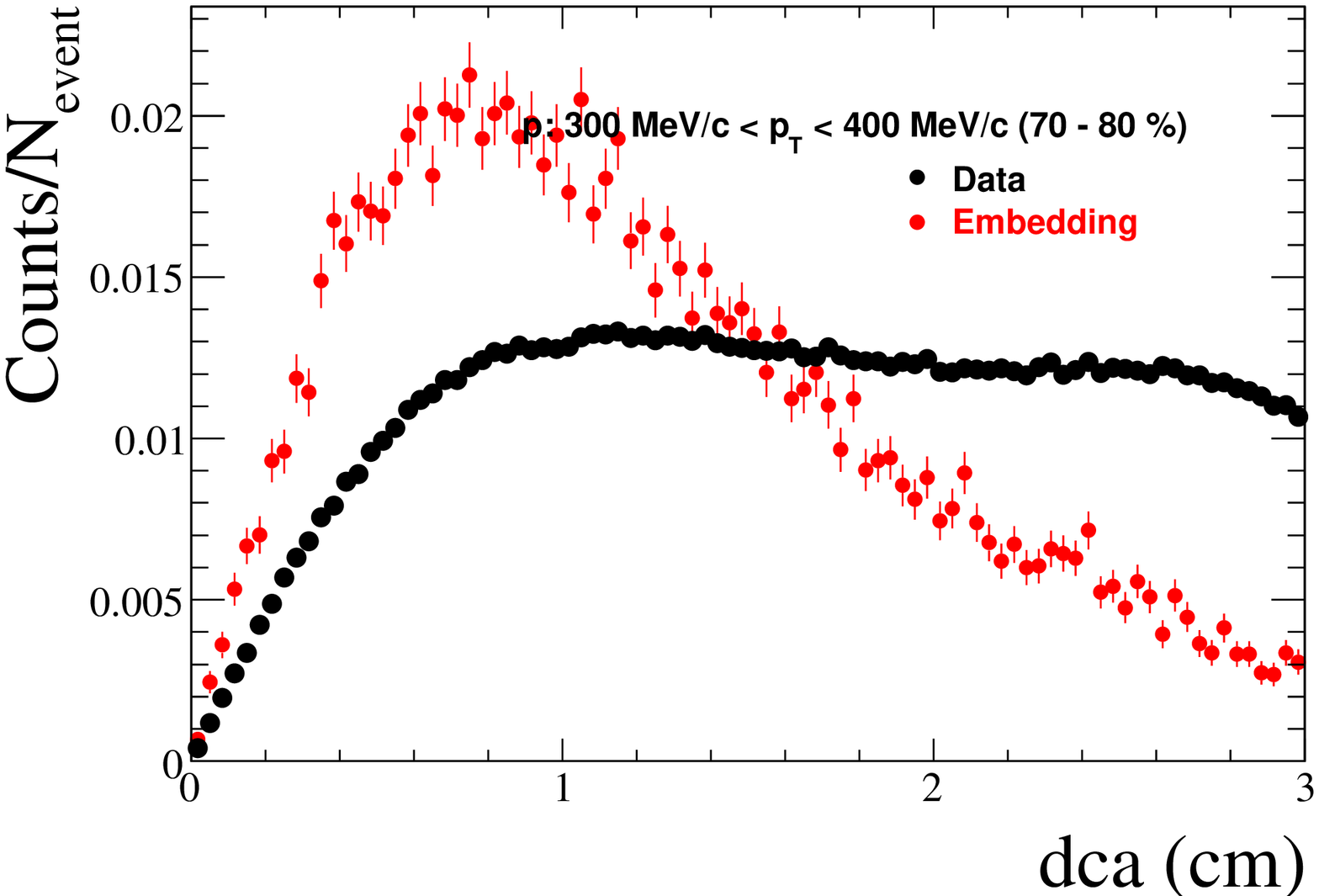}}
\resizebox{.225\textwidth}{!}{\includegraphics{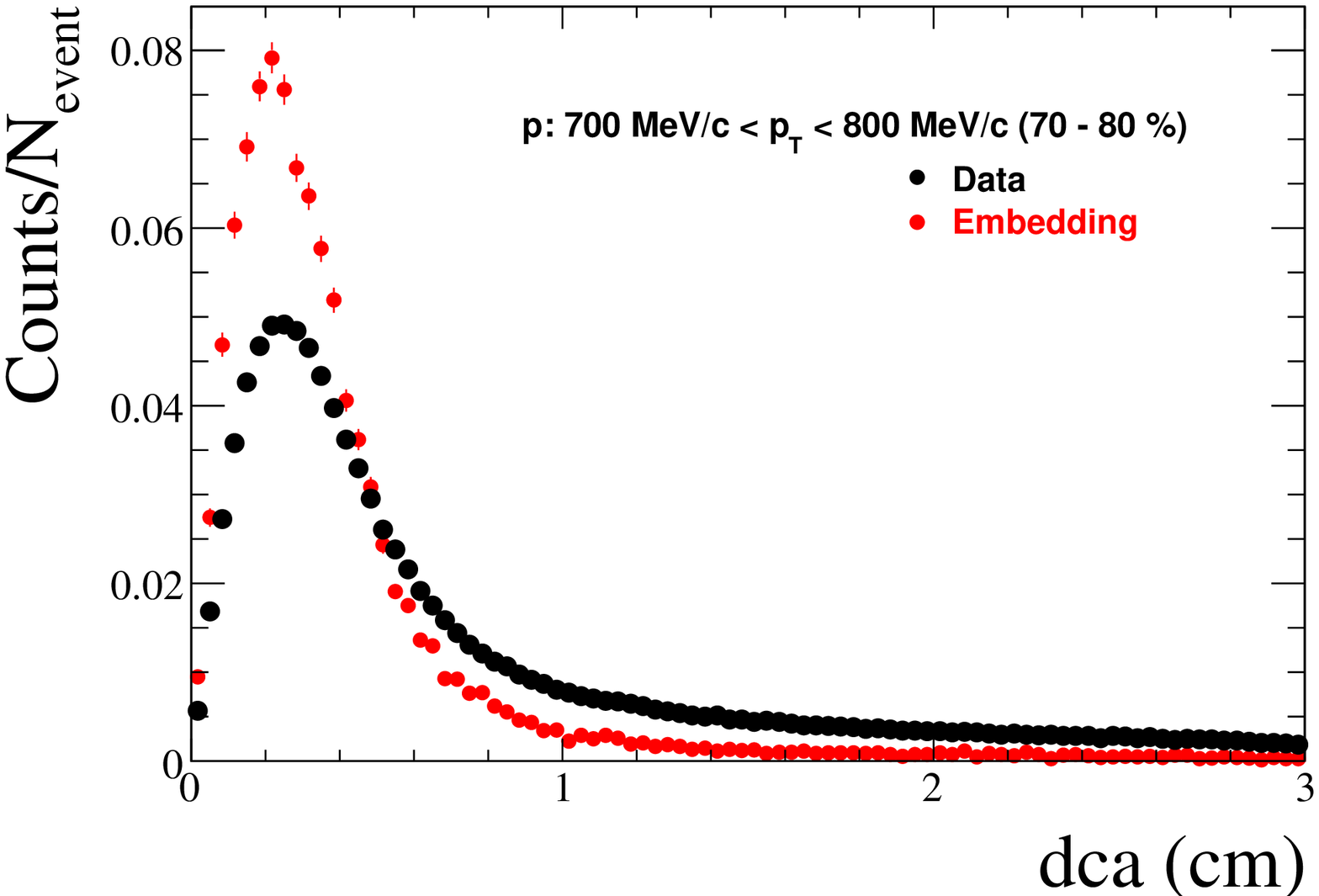}}\\
\caption{Comparison of $dca$ from real data and embedding in 62.4 GeV peripheral (70-80\%) Au-Au collisions.}
\label{fig:auau62peripheral_dca_data_embedding}

\end{center} 
\end{sidewaysfigure}
%
%
%
\begin{sidewaysfigure}[!h]
\begin{center}
\resizebox{.225\textwidth}{!}{\includegraphics{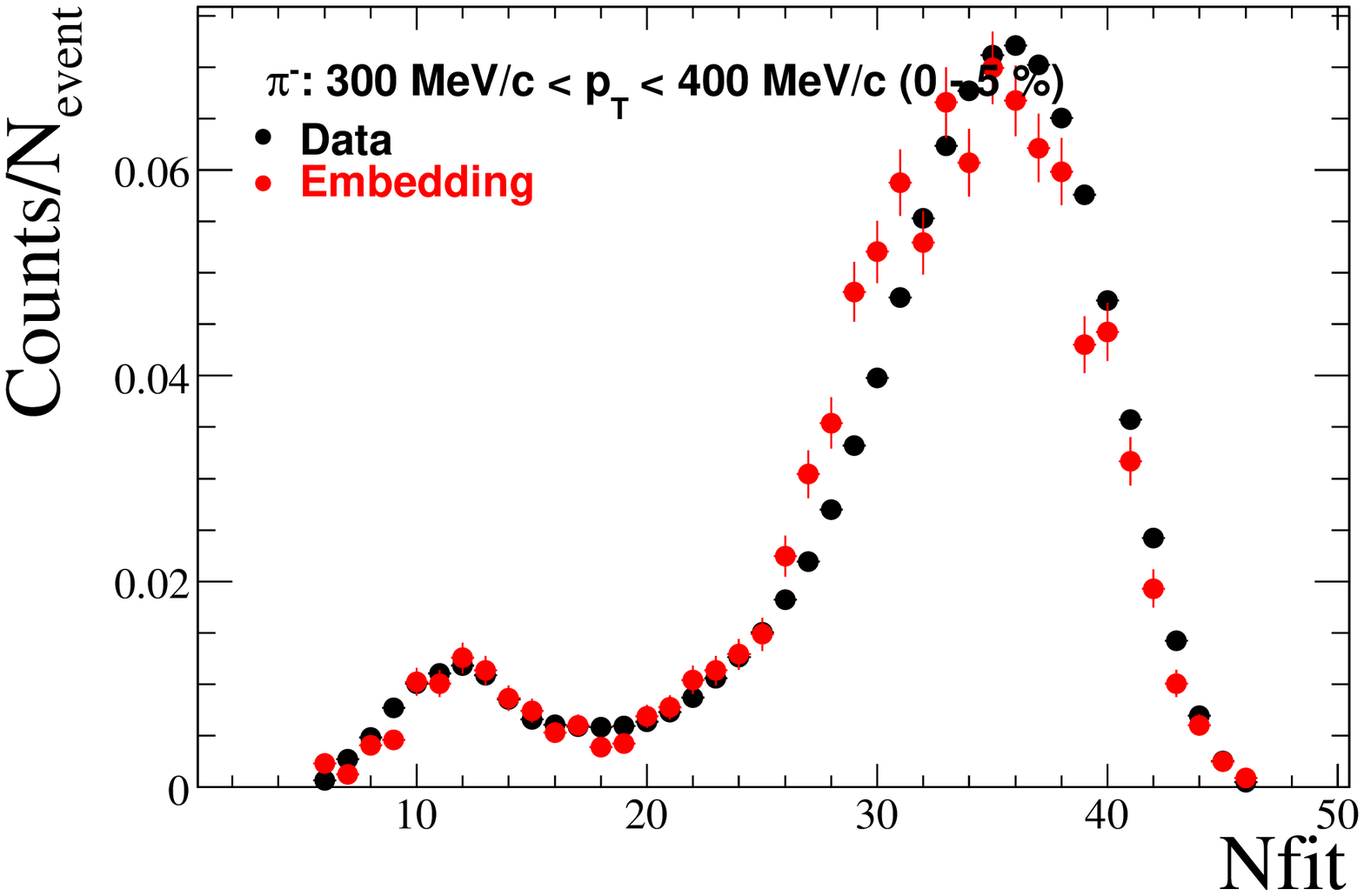}}
\resizebox{.225\textwidth}{!}{\includegraphics{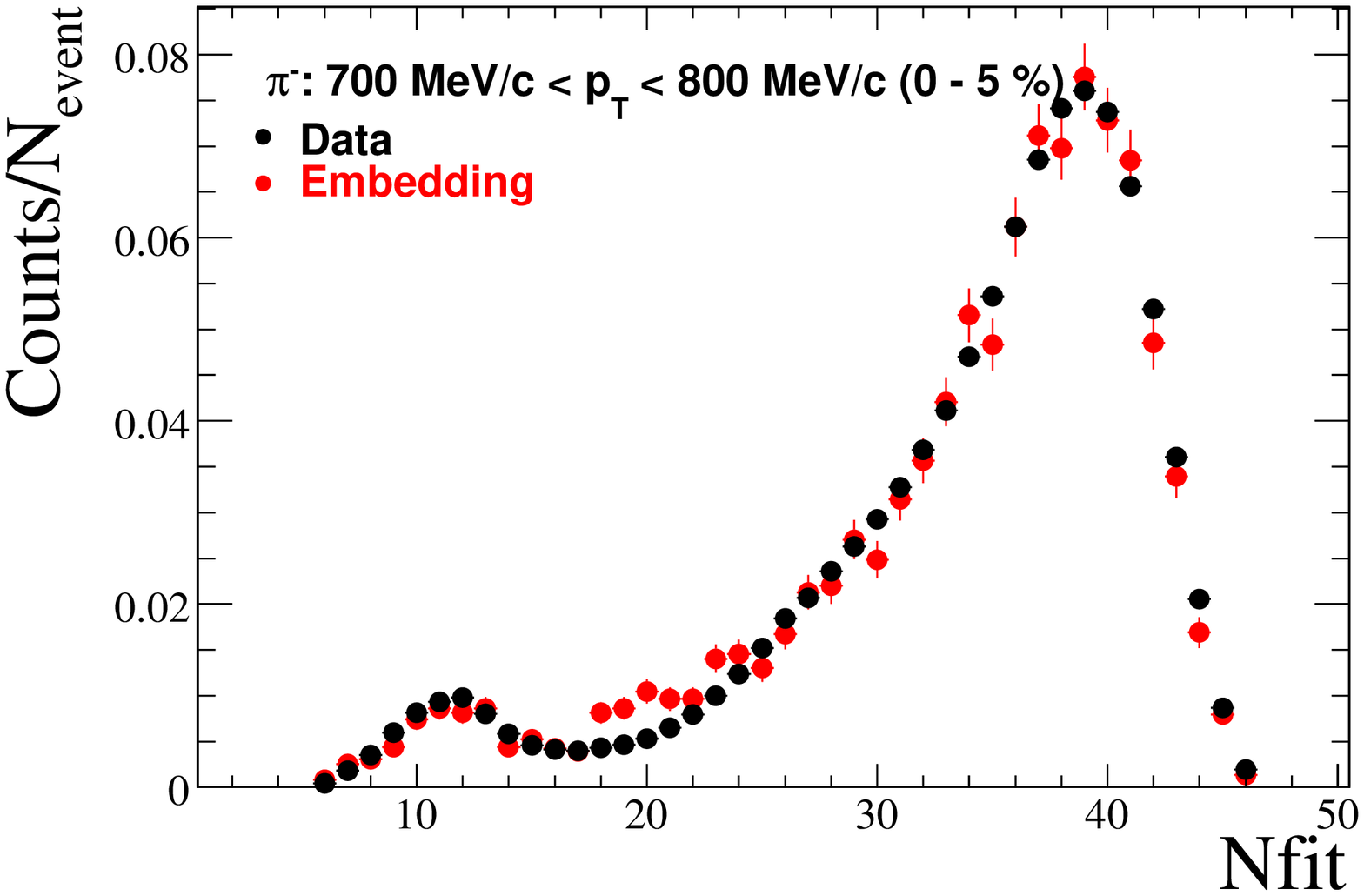}}
\resizebox{.225\textwidth}{!}{\includegraphics{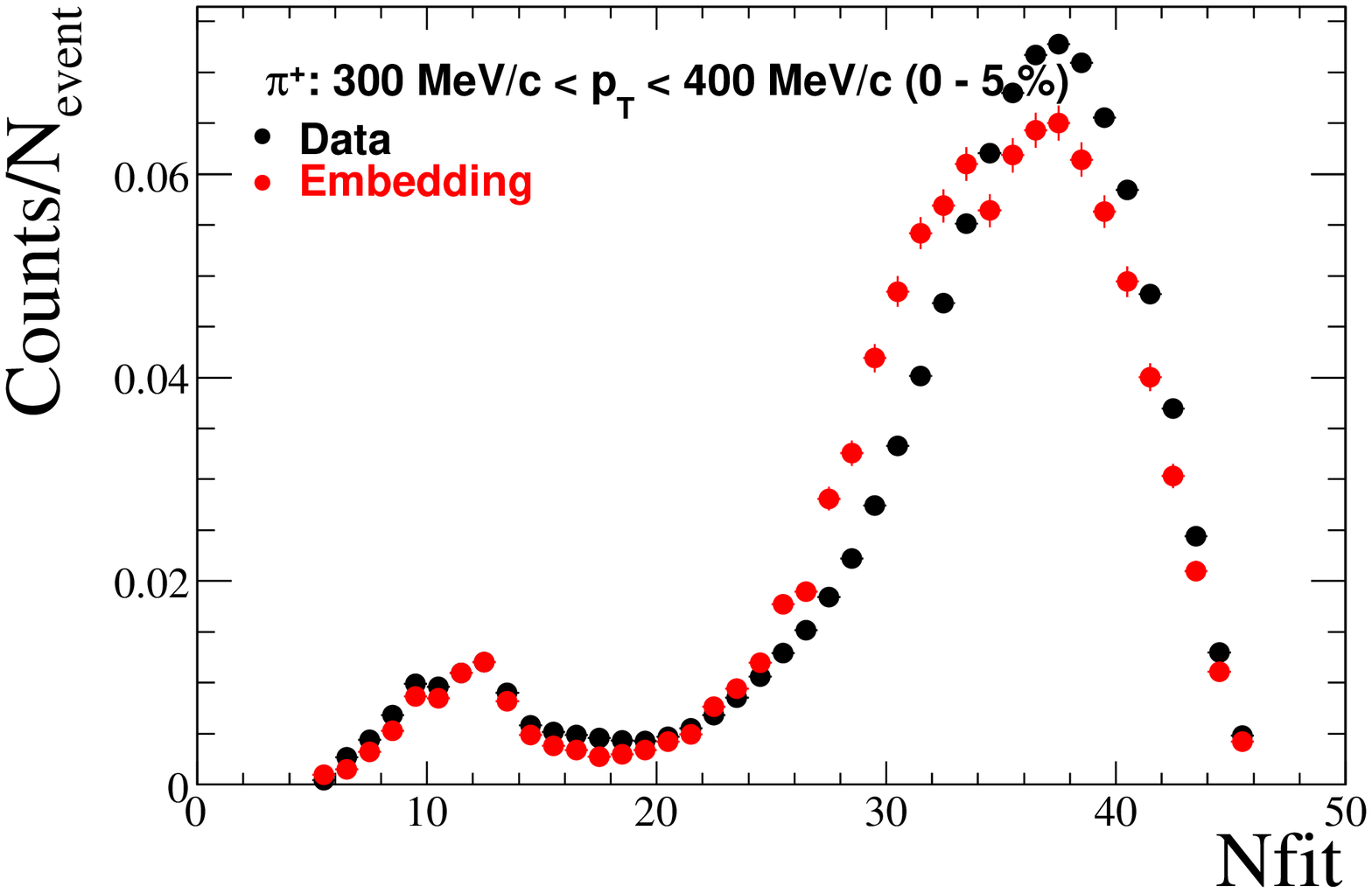}}
\resizebox{.225\textwidth}{!}{\includegraphics{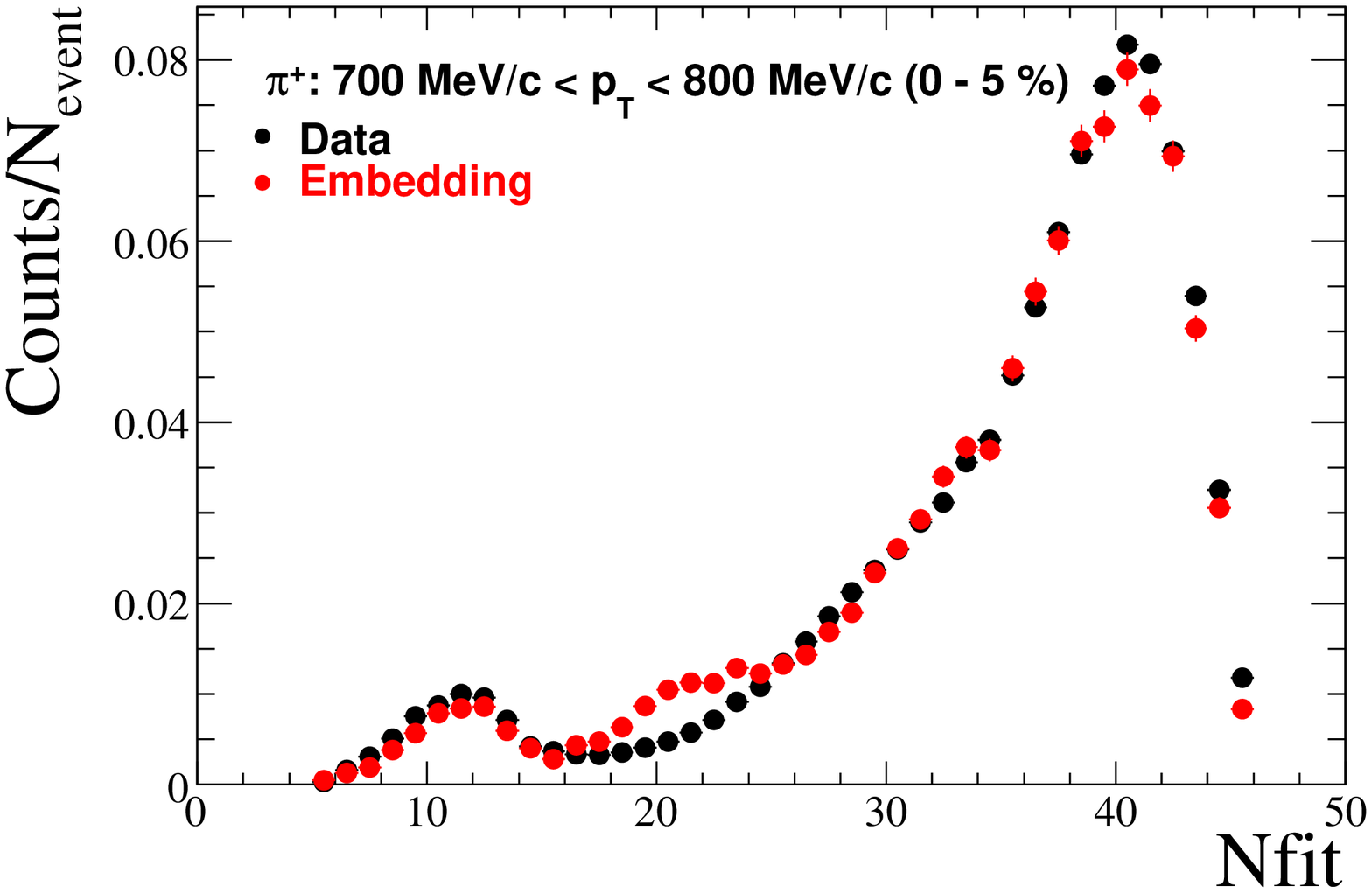}}\\
\resizebox{.225\textwidth}{!}{\includegraphics{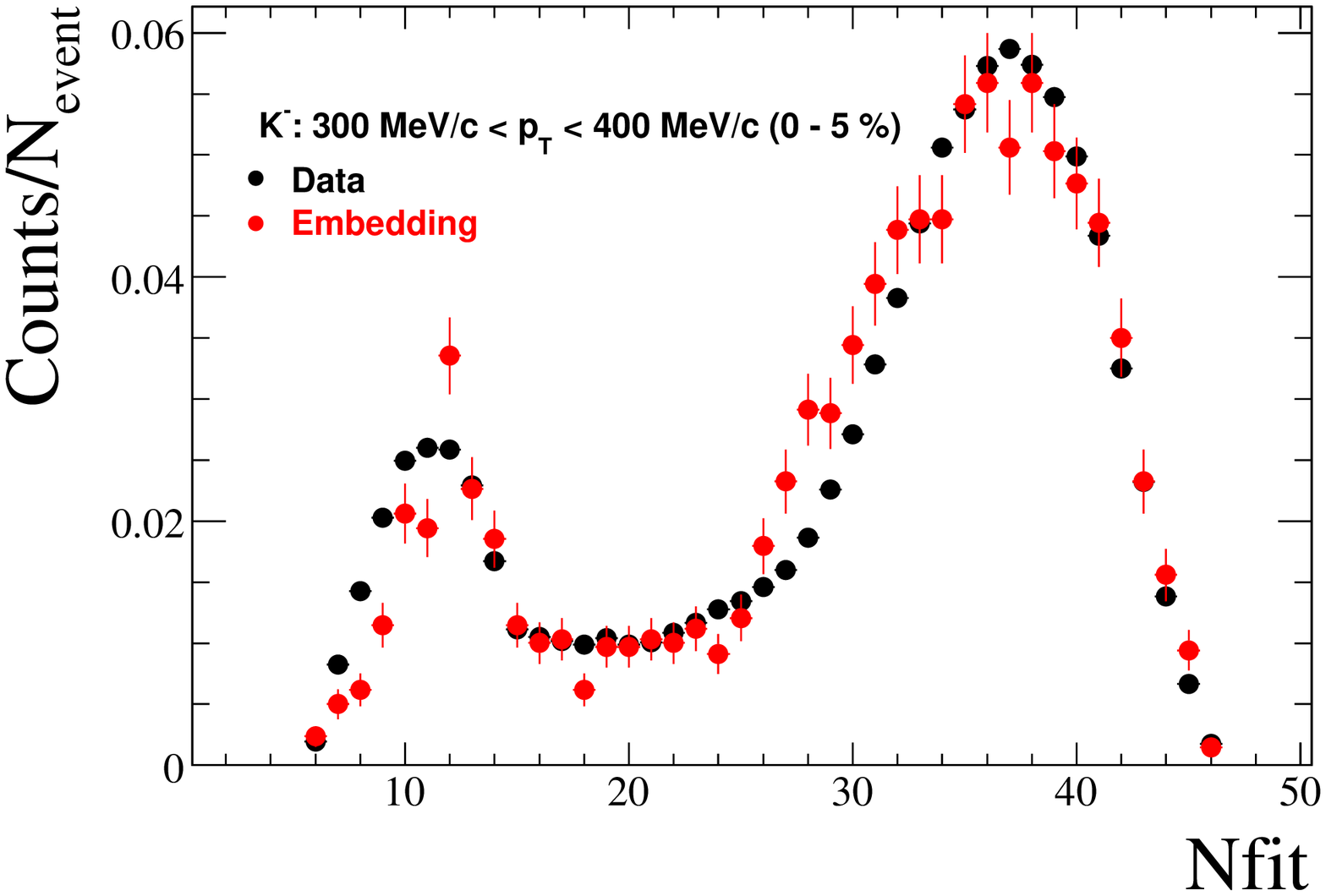}}
\resizebox{.225\textwidth}{!}{\includegraphics{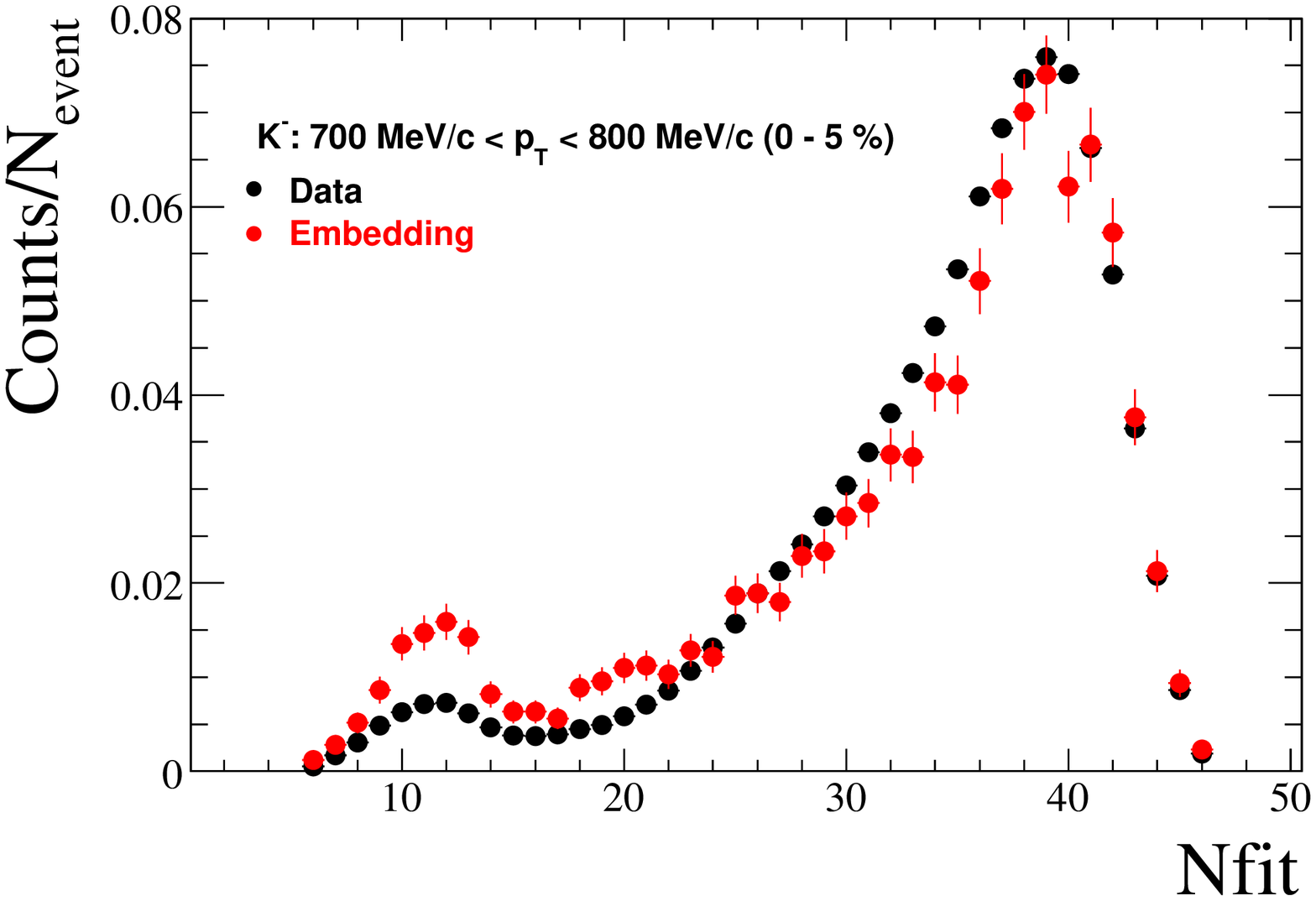}}
\resizebox{.225\textwidth}{!}{\includegraphics{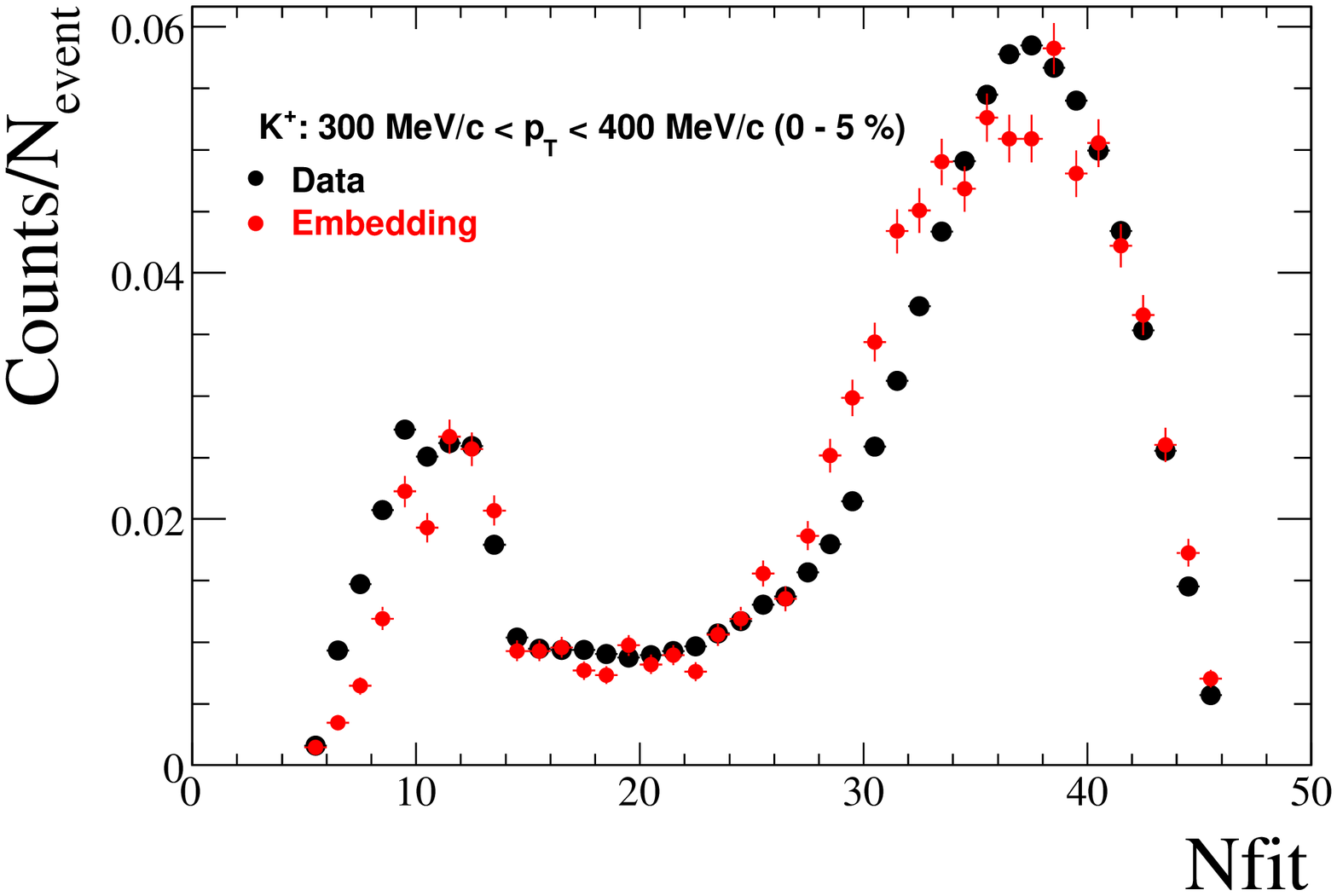}}
\resizebox{.225\textwidth}{!}{\includegraphics{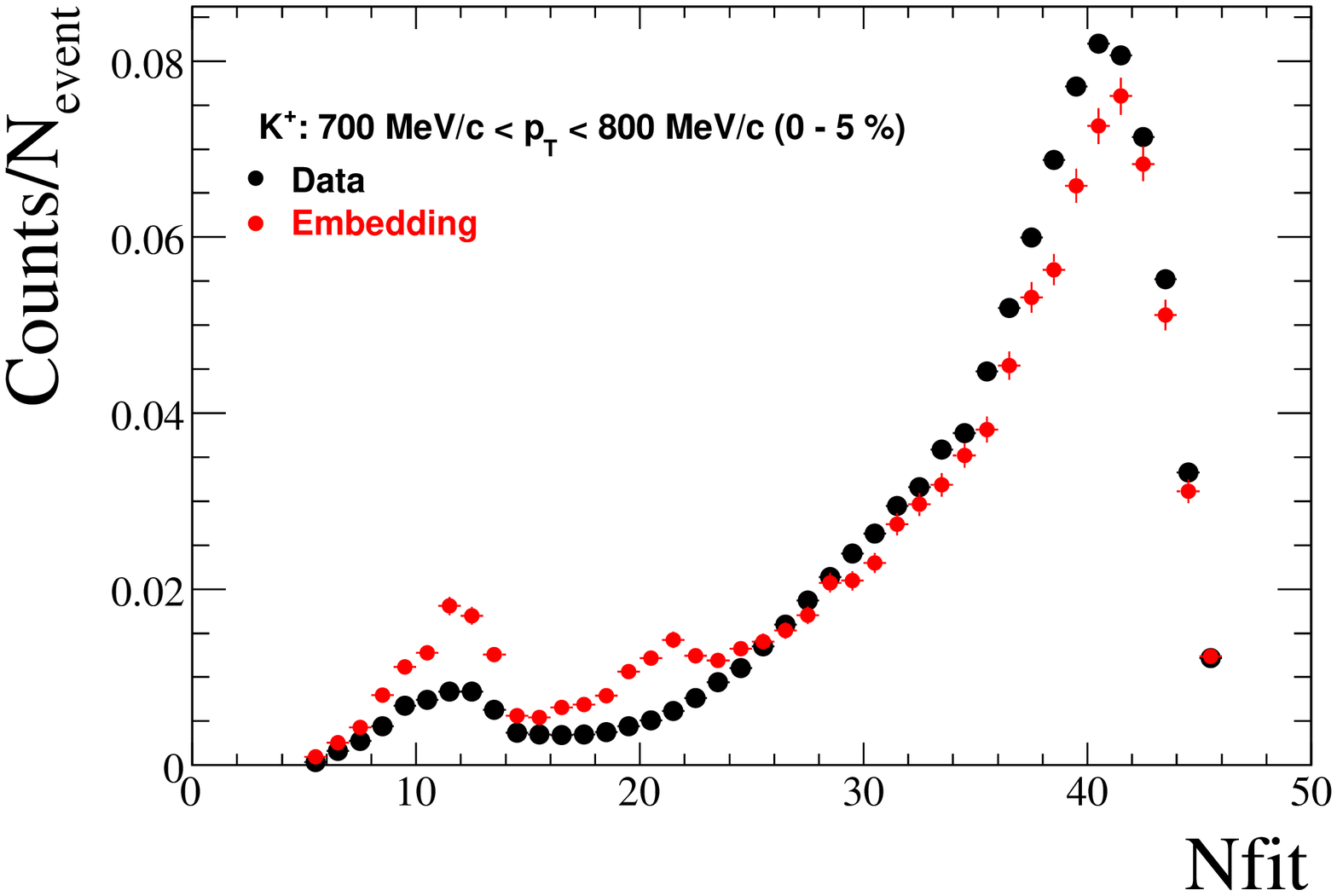}}\\
\resizebox{.225\textwidth}{!}{\includegraphics{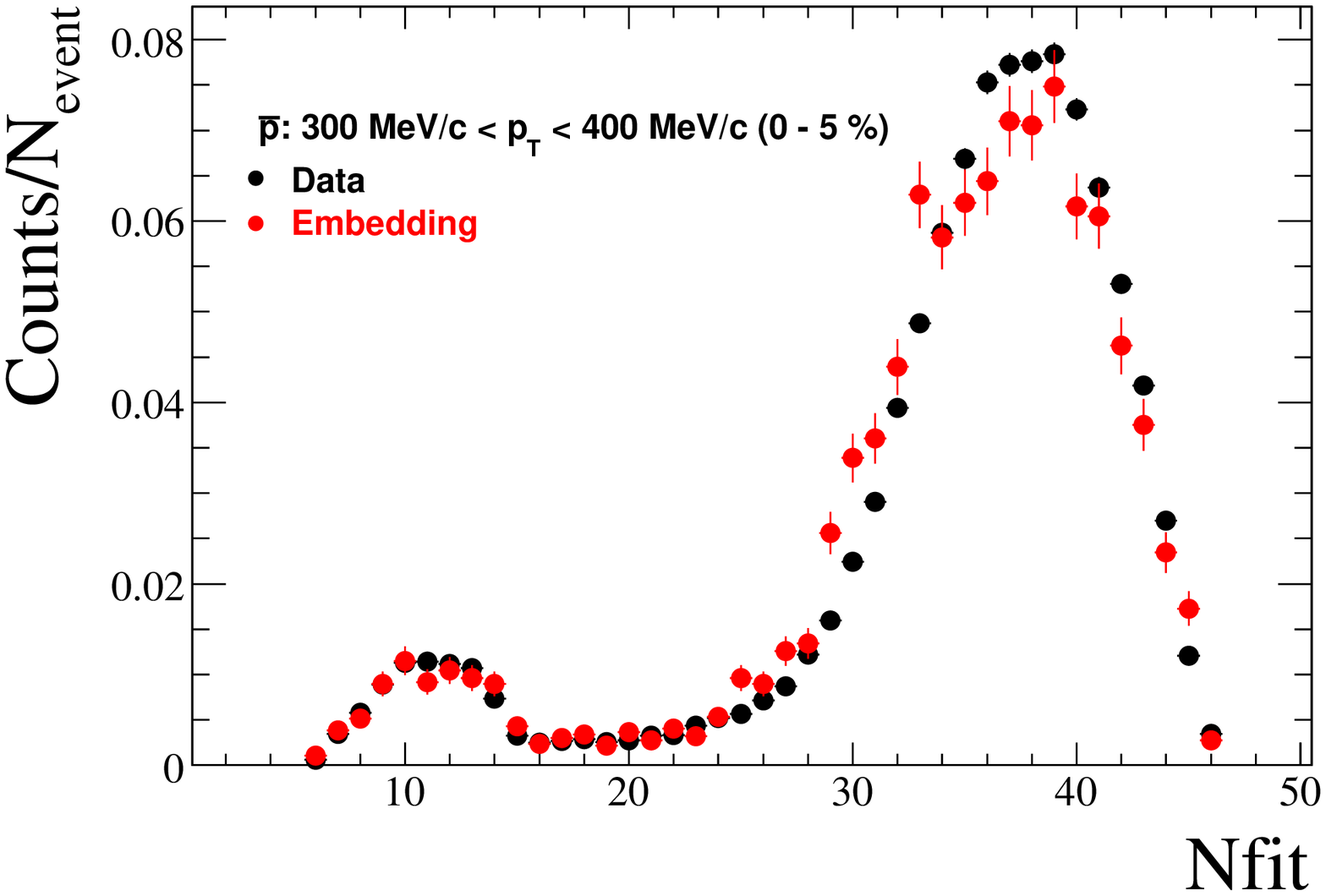}}
\resizebox{.225\textwidth}{!}{\includegraphics{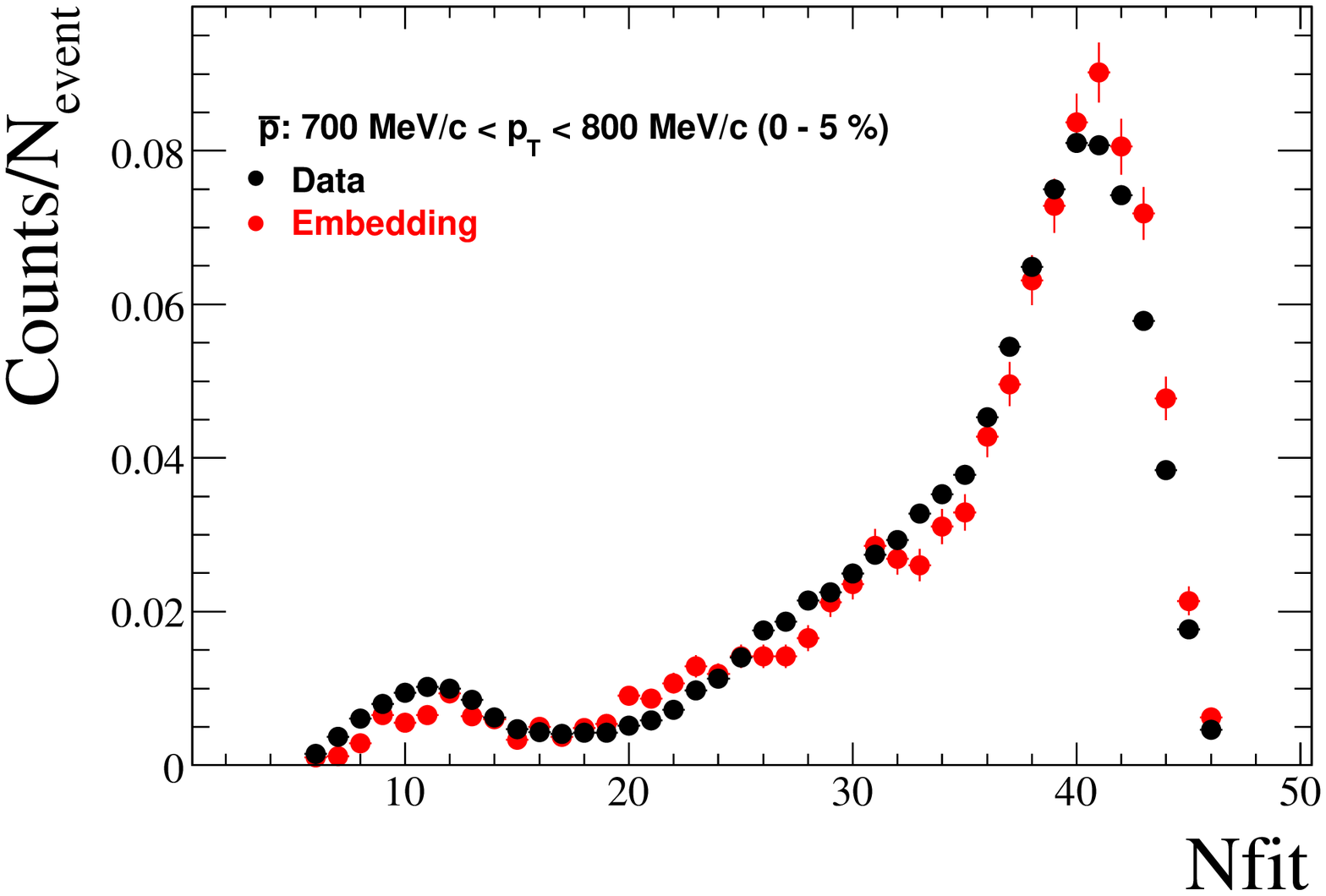}}
\resizebox{.225\textwidth}{!}{\includegraphics{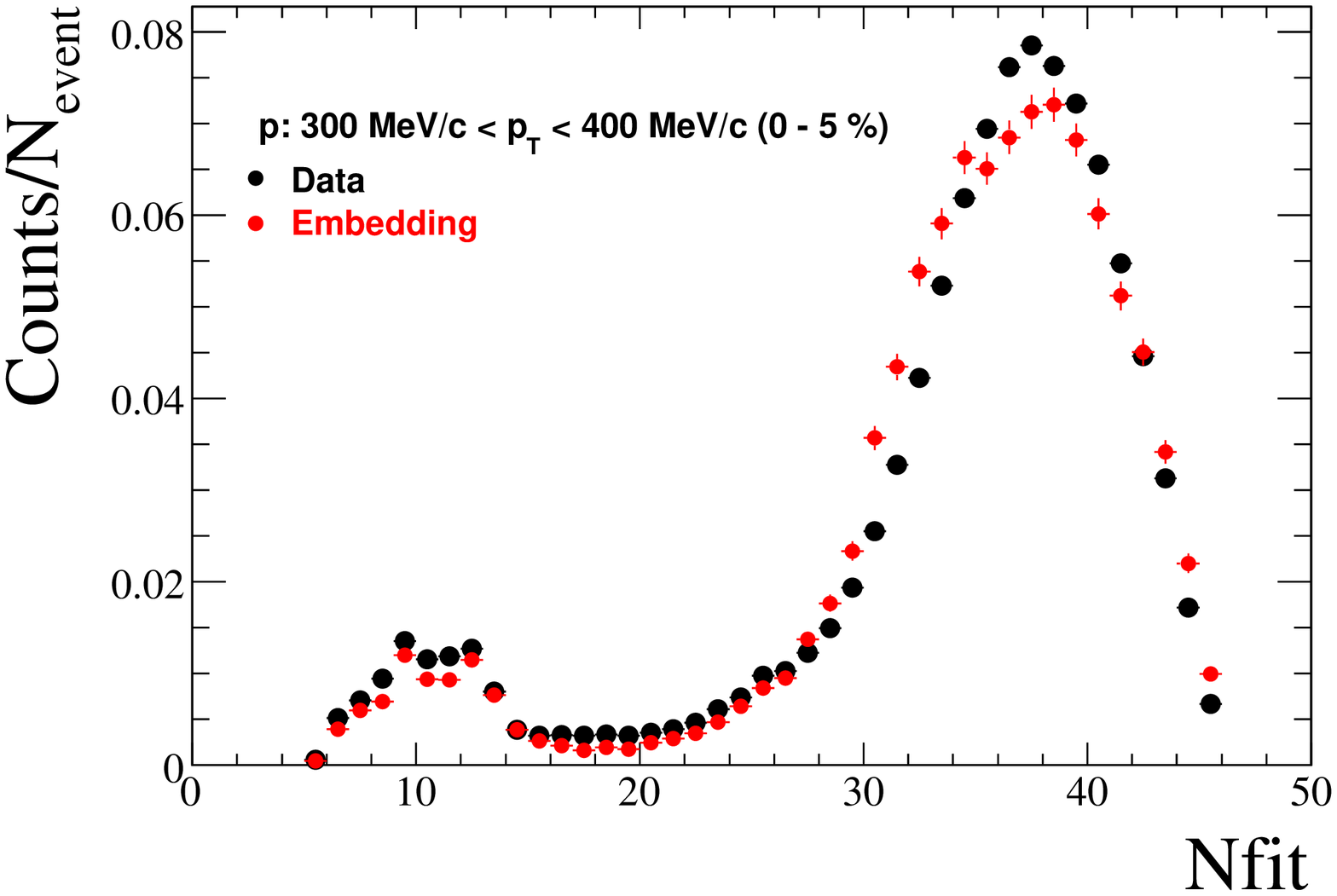}}
\resizebox{.225\textwidth}{!}{\includegraphics{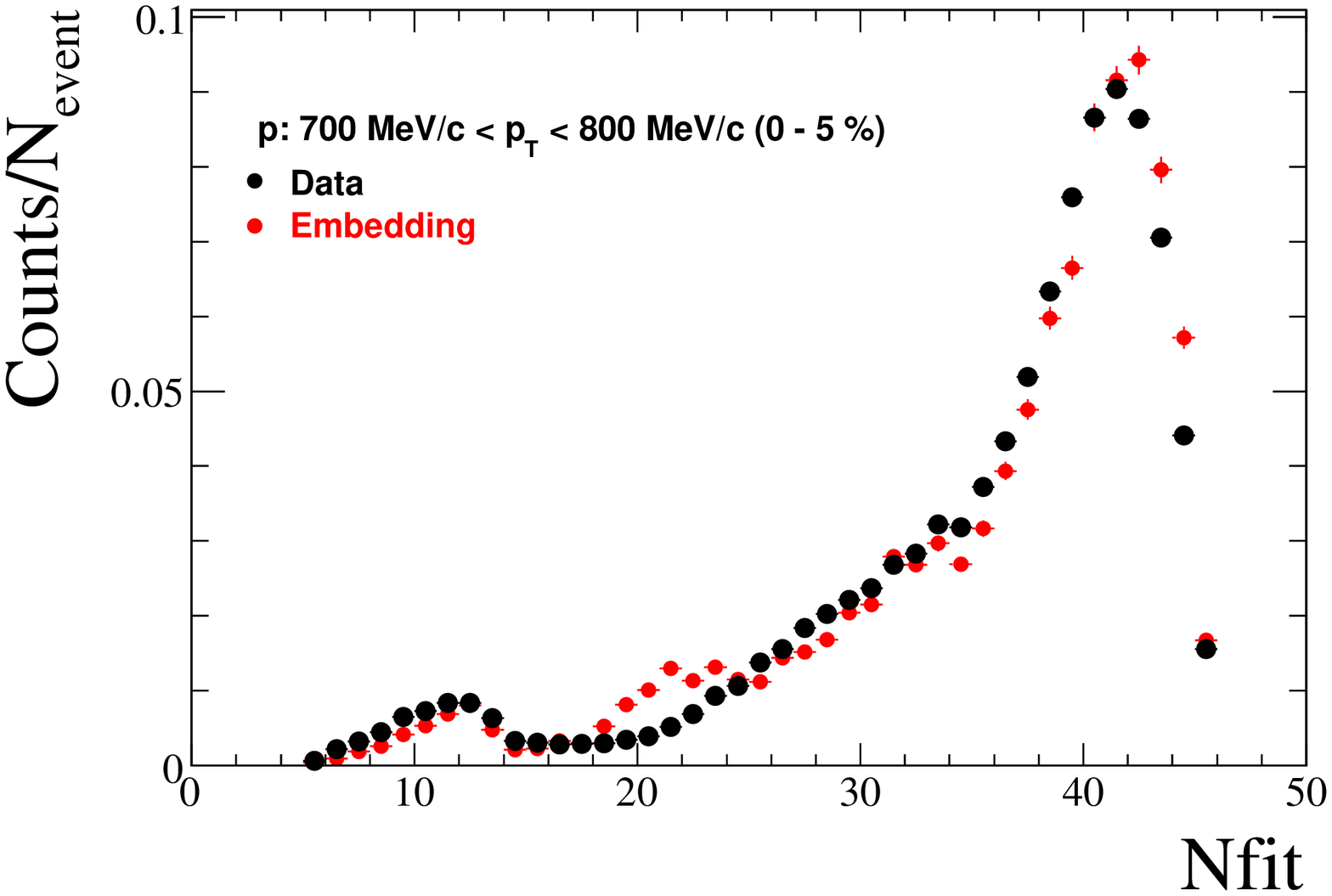}}\\
\caption{Comparison of $N_{fit}$ from real data and embedding in 62.4 GeV central (0-5\%) Au-Au collisions.}
\label{fig:auau62central_nfit_data_embedding}
\end{center} 
\end{sidewaysfigure}

\begin{sidewaysfigure}[!t]
\begin{center}
\resizebox{.22\textwidth}{!}{\includegraphics{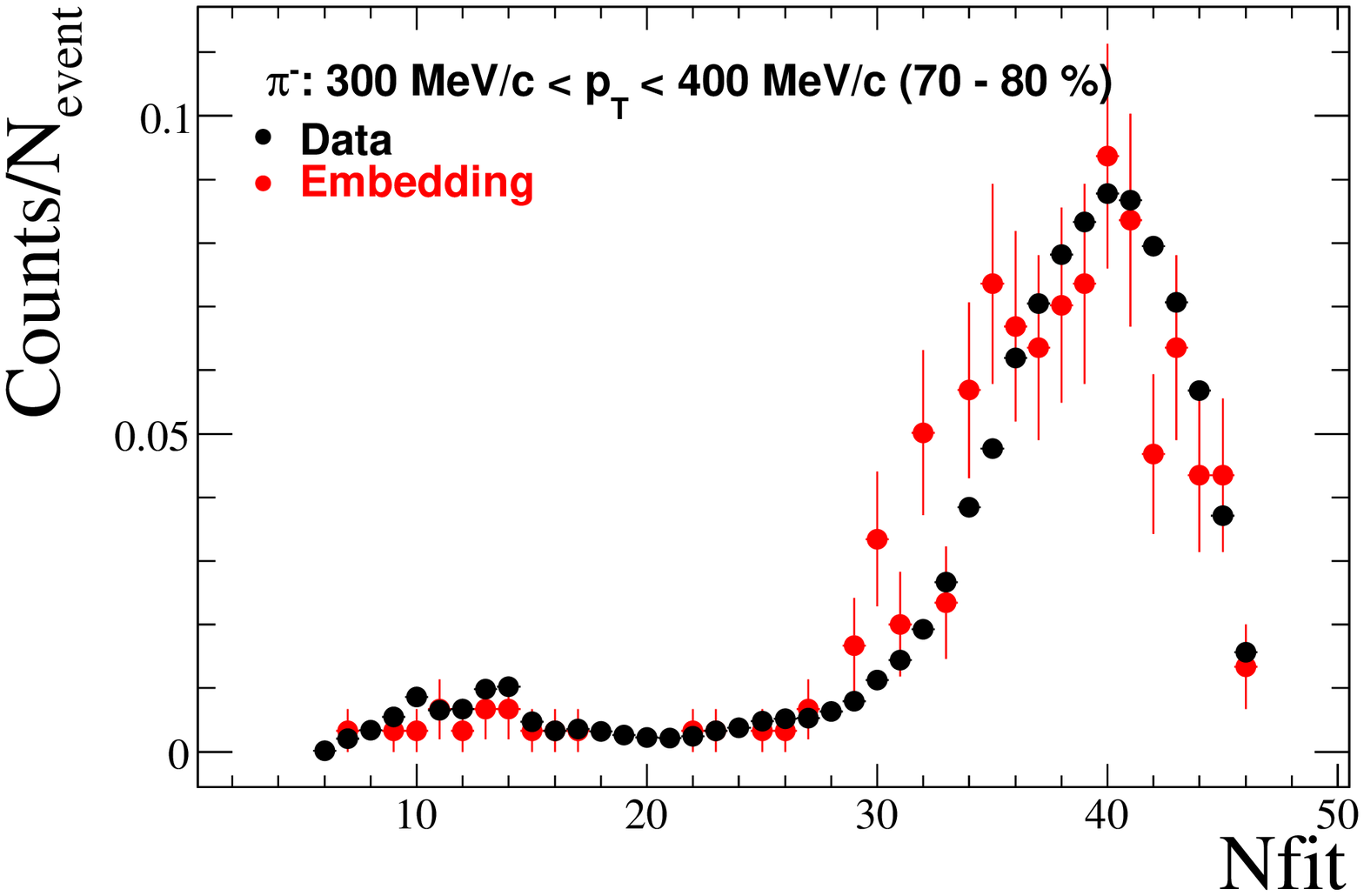}}
\resizebox{.22\textwidth}{!}{\includegraphics{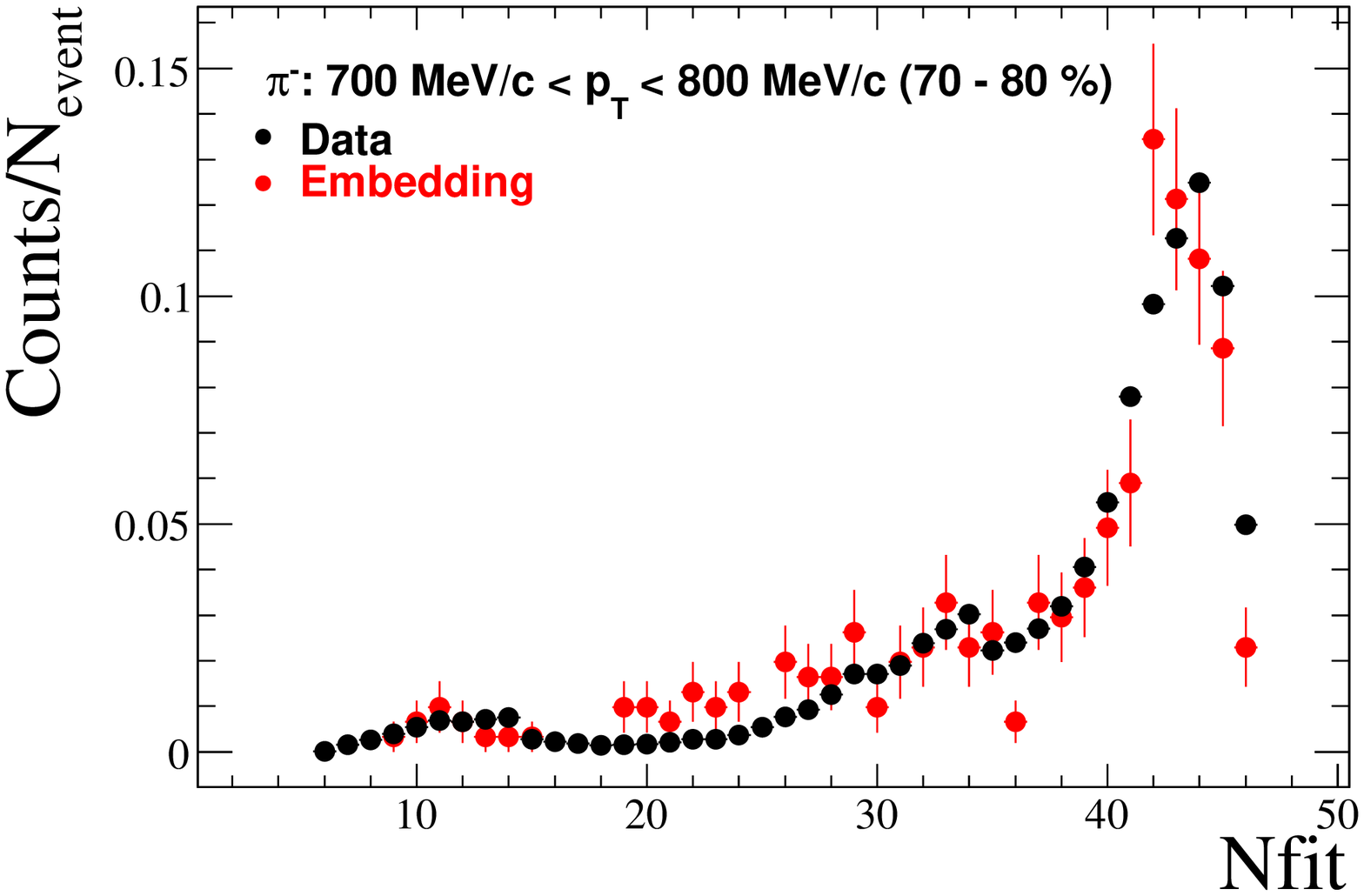}}
\resizebox{.22\textwidth}{!}{\includegraphics{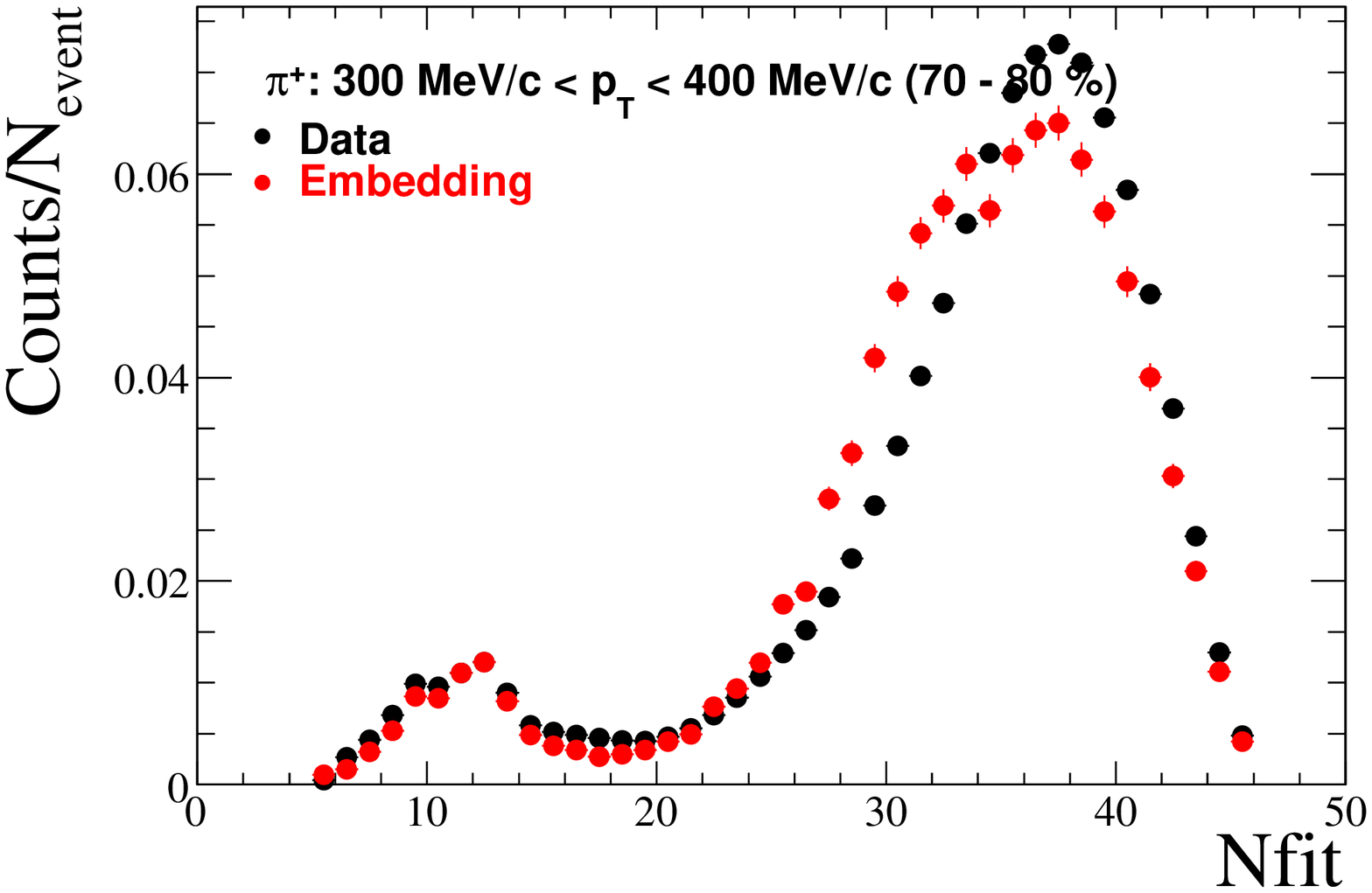}}
\resizebox{.22\textwidth}{!}{\includegraphics{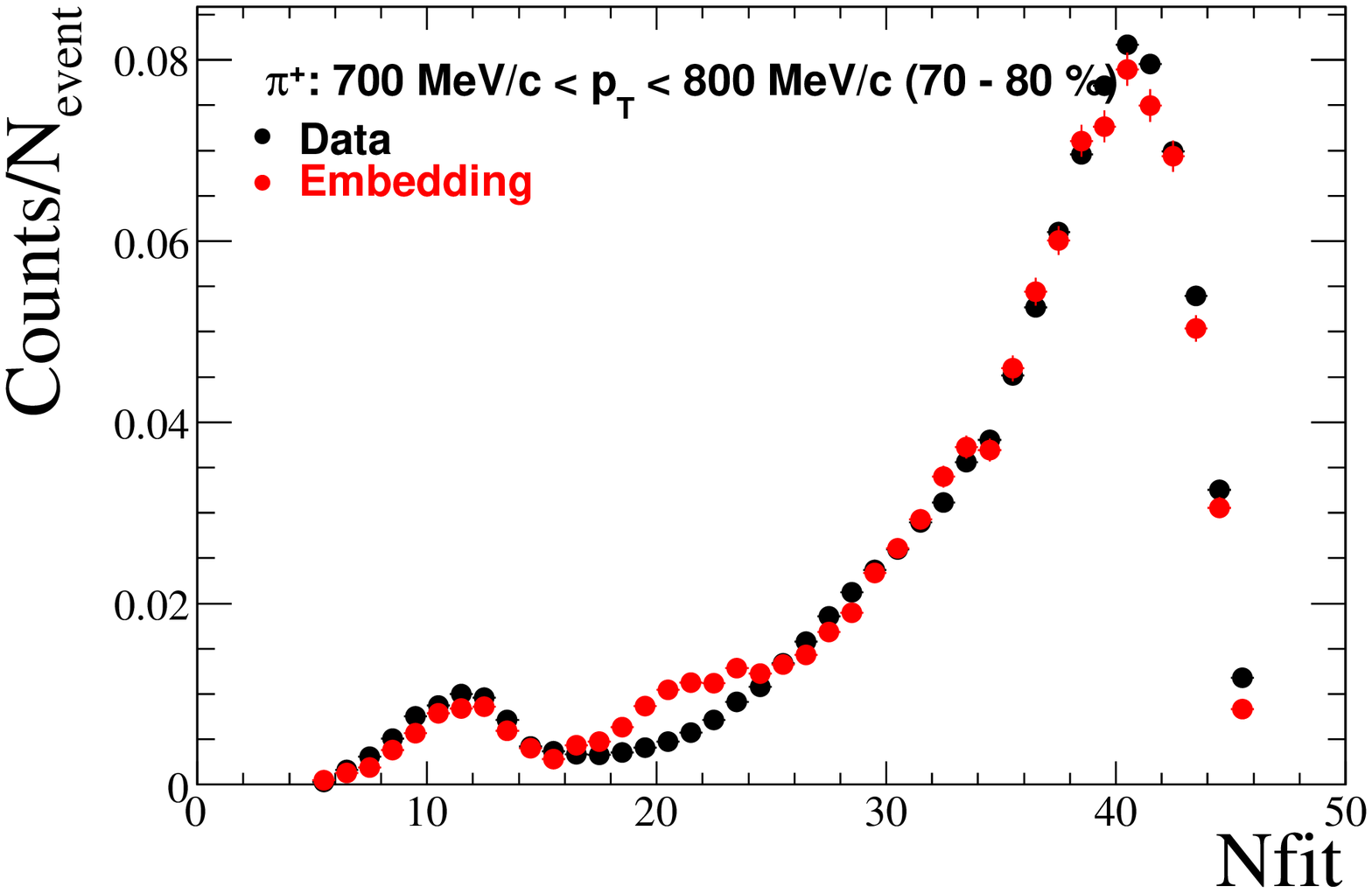}}\\
\resizebox{.22\textwidth}{!}{\includegraphics{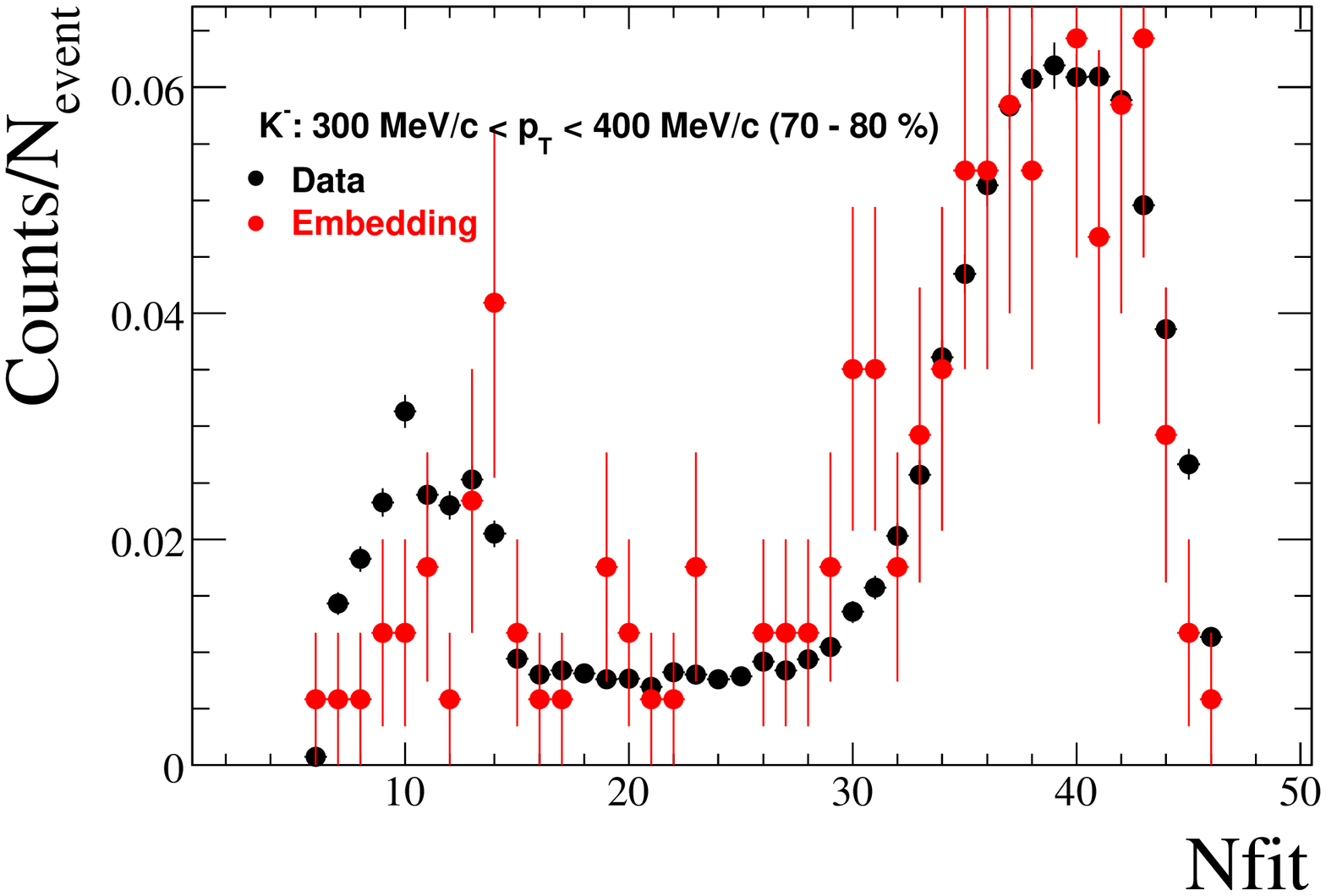}}
\resizebox{.22\textwidth}{!}{\includegraphics{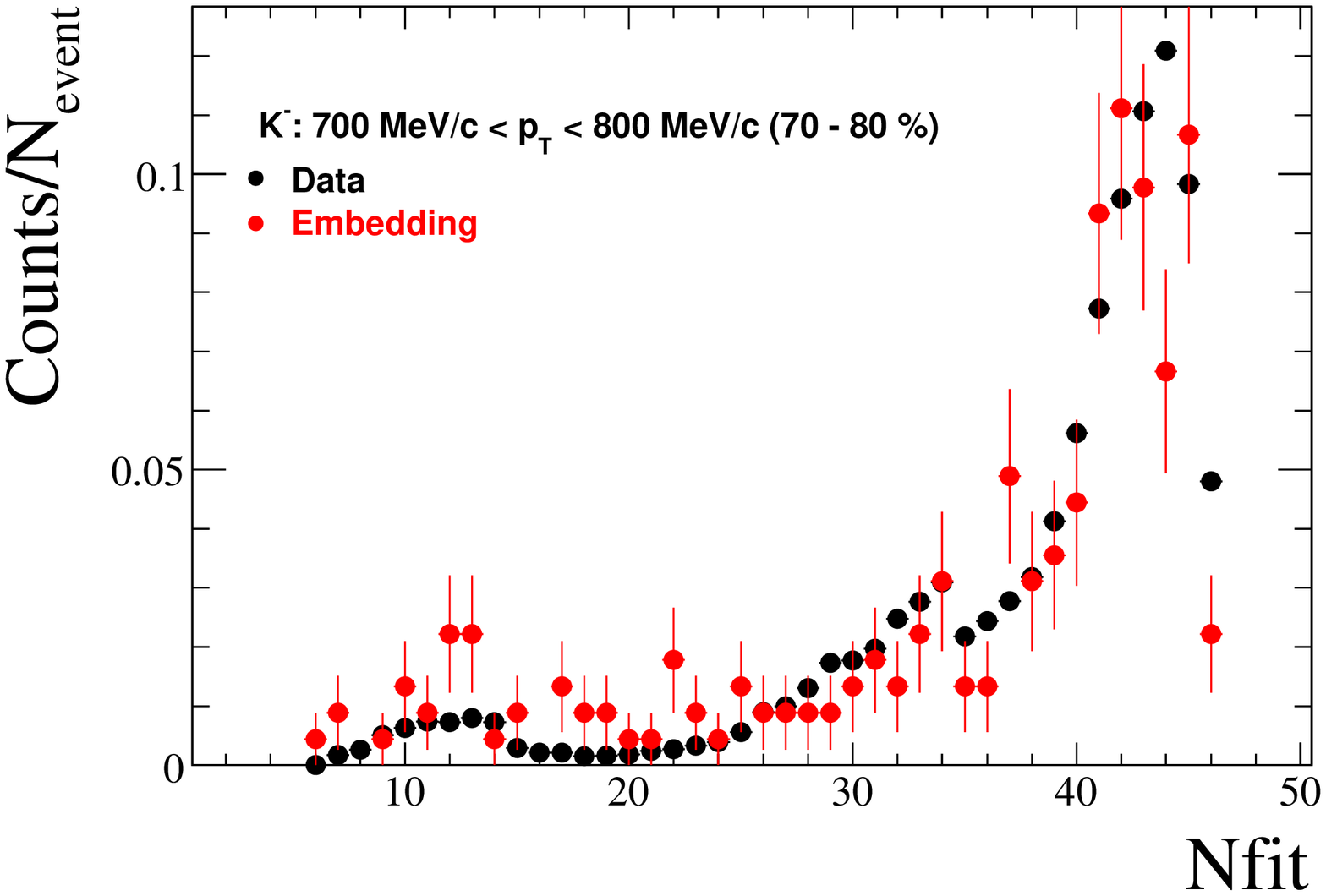}}
\resizebox{.22\textwidth}{!}{\includegraphics{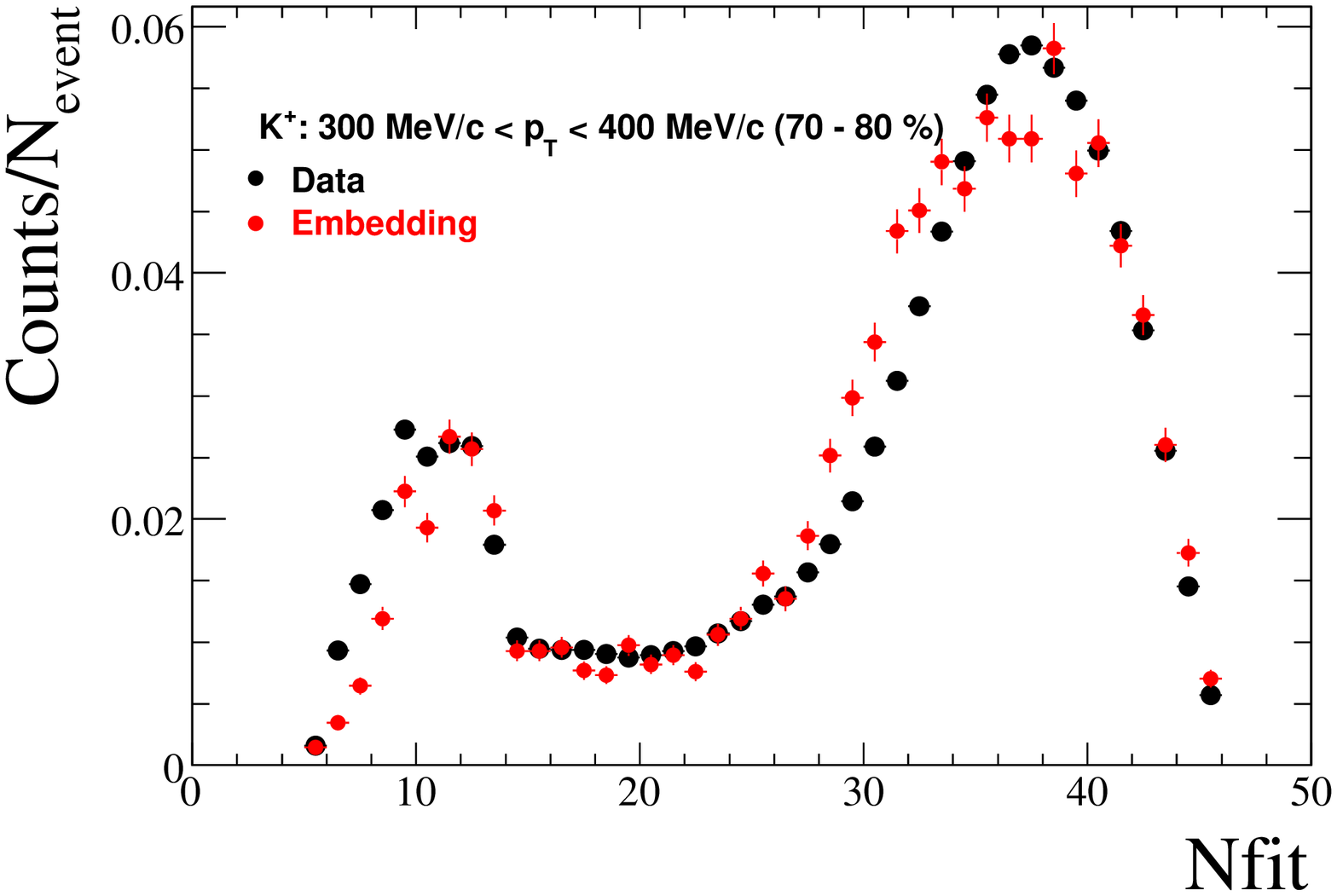}}
\resizebox{.22\textwidth}{!}{\includegraphics{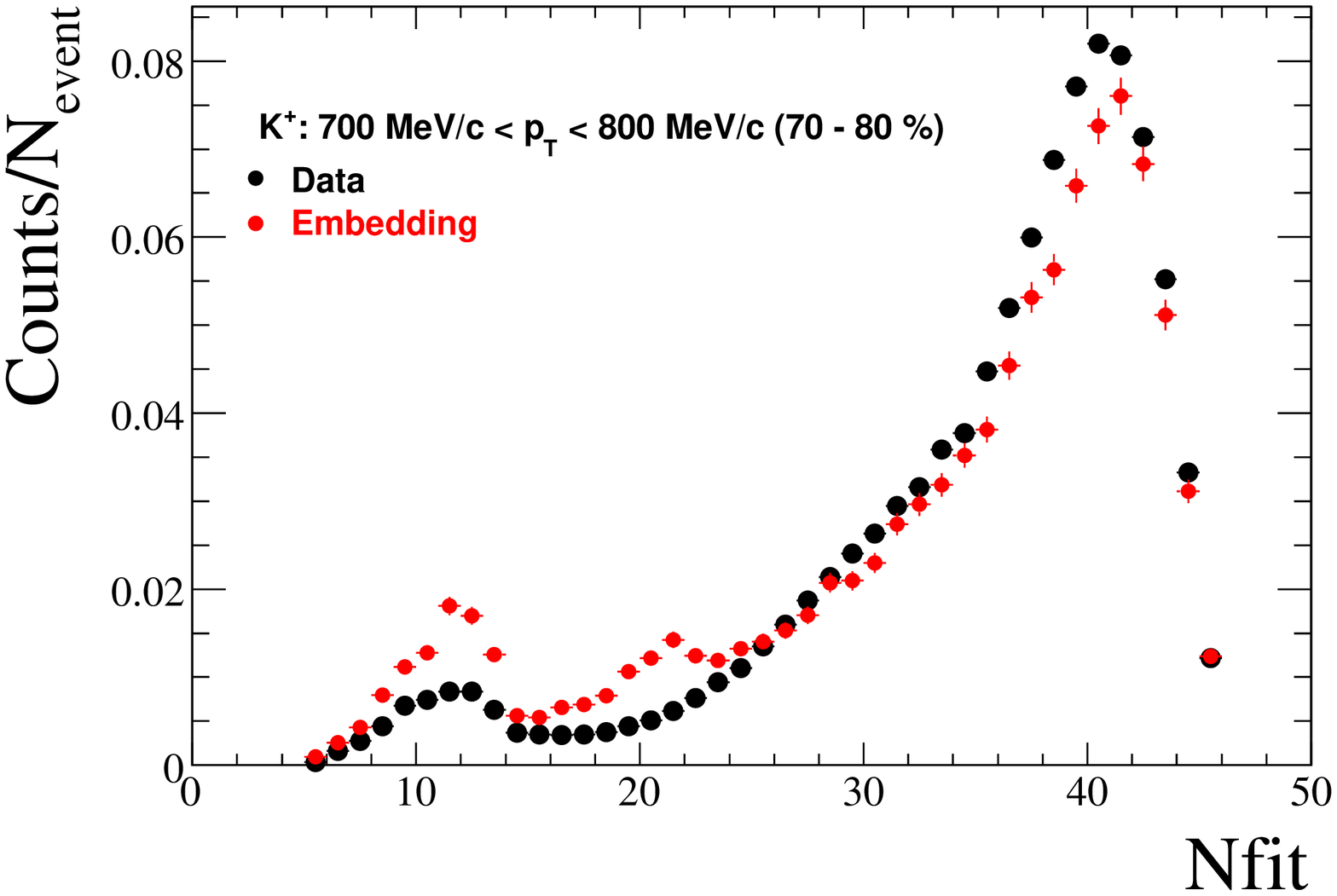}}\\
\resizebox{.22\textwidth}{!}{\includegraphics{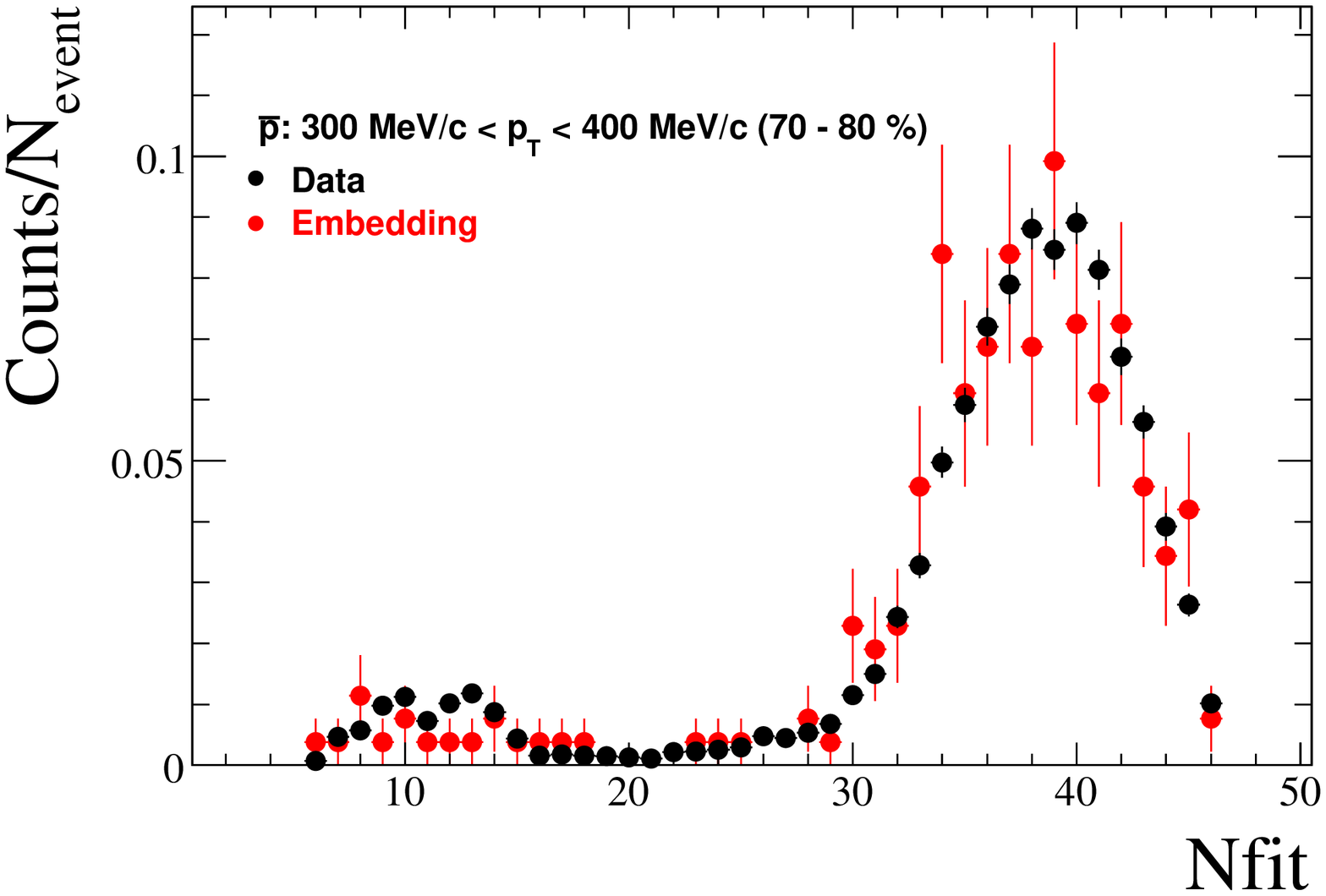}}
\resizebox{.22\textwidth}{!}{\includegraphics{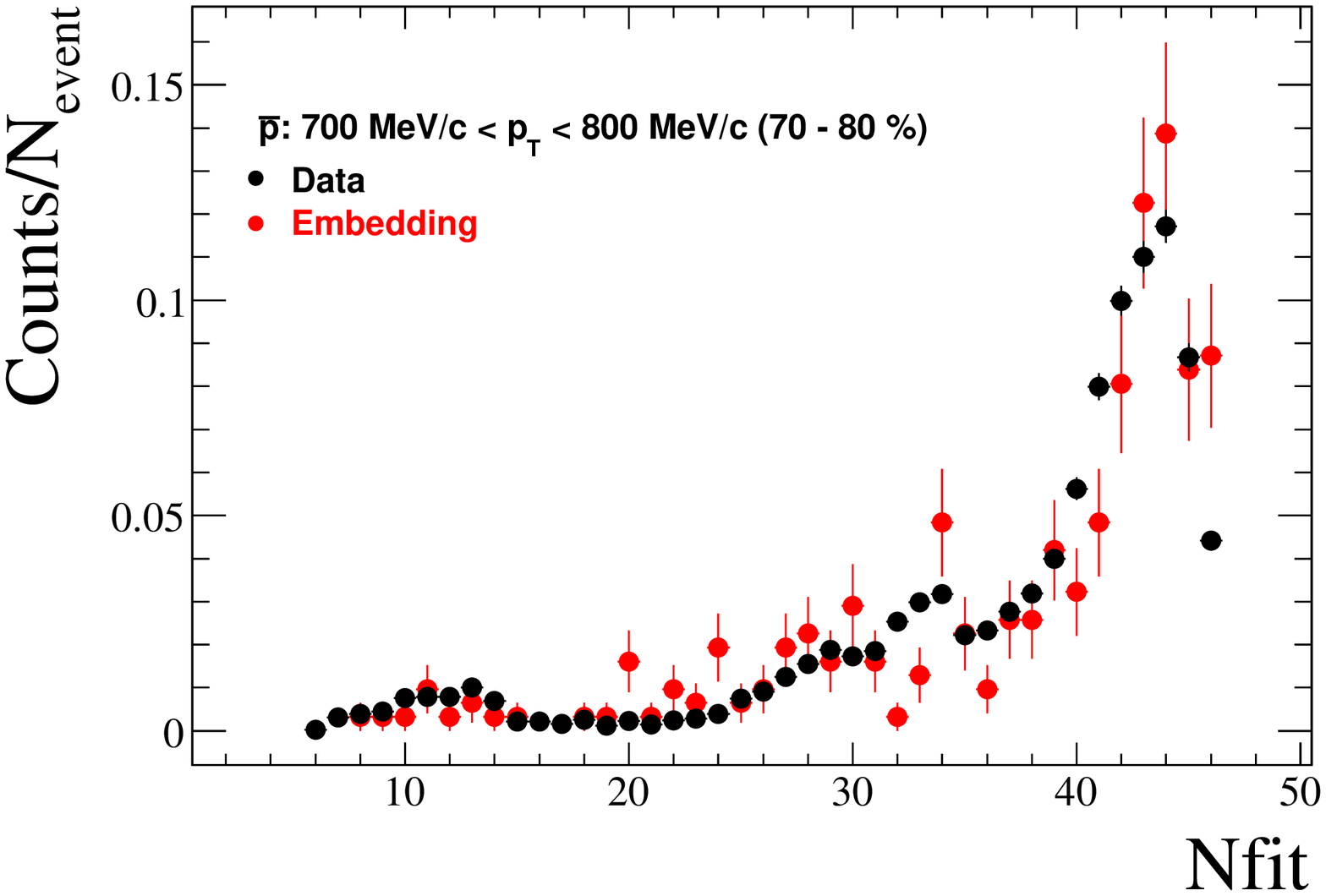}}
\resizebox{.22\textwidth}{!}{\includegraphics{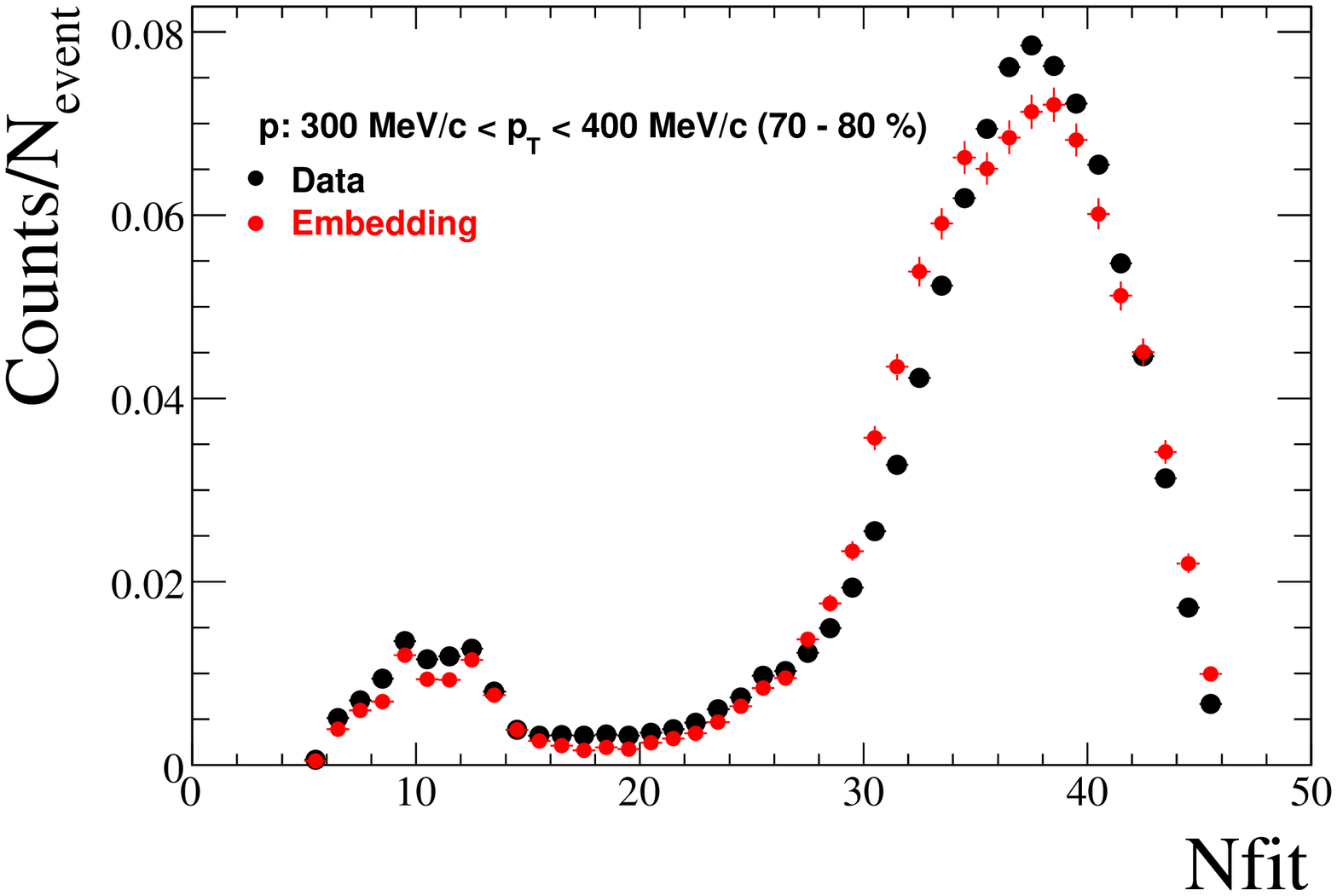}}
\resizebox{.22\textwidth}{!}{\includegraphics{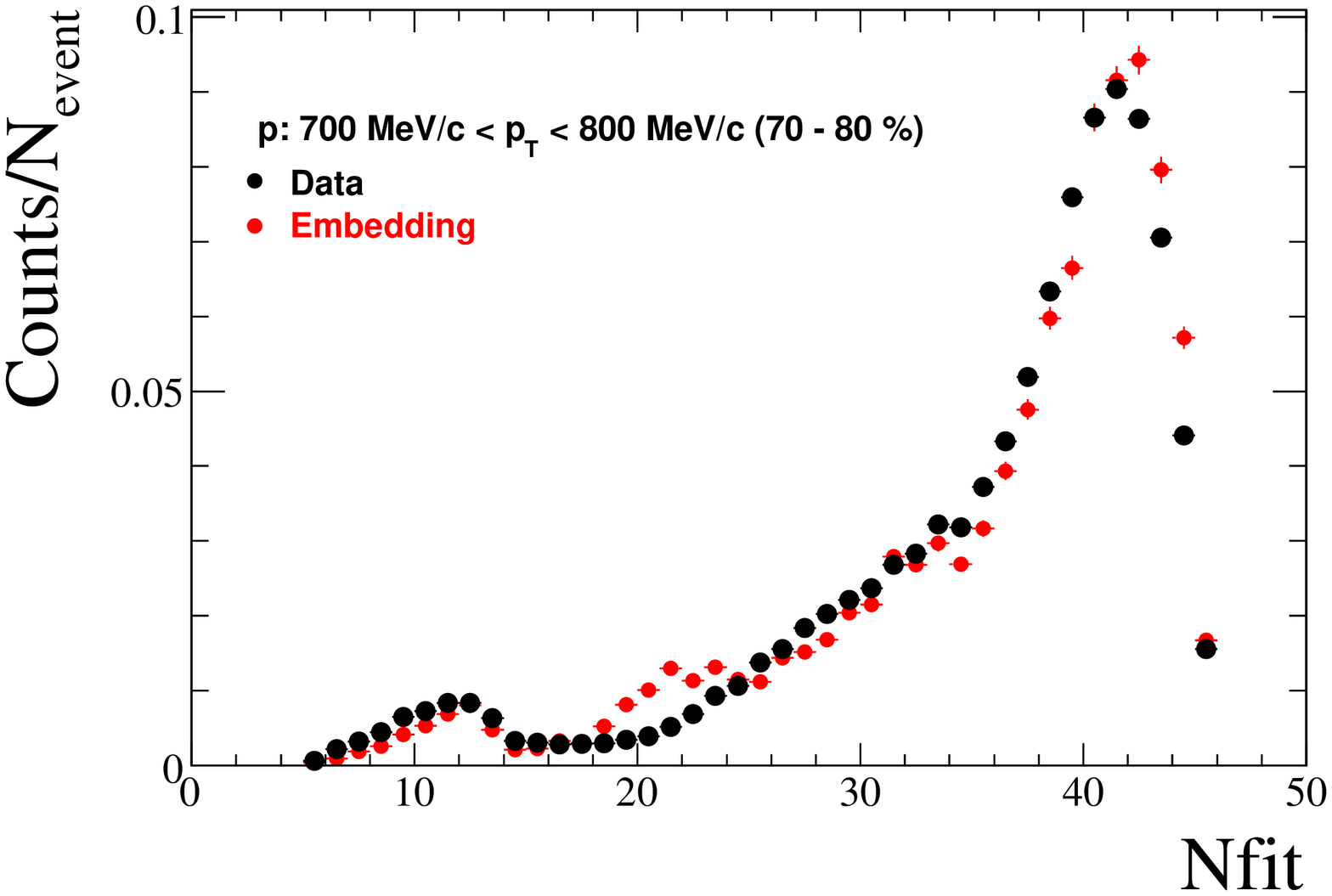}}
\caption{Comparison of $N_{fit}$ from real data and embedding in 62.4 GeV peripheral (70-80\%) Au-Au collisions.}
\label{fig:auau62peripheral_nfit_data_embedding}
\end{center} 
\end{sidewaysfigure}

%
%


%
%
%
%

\begin{vita}
Levente Moln\'{a}r was born June 15th 1977 in Hodmezovasarhely, Hungary.  He graduated from University of Debrecen in May 2000 where he obtained a Master of Science degree in Physics and a Special English-Hungarian, Hungarian-English Translator degree. He also worked as an undergraduate teaching assistant in the Physics Department at University of Debrecen.
He served as a teaching assistant at Purdue University from 2000 through 2002, and as a research assistant from 2002 through 2006. He received his Ph.D. in experimental high energy nuclear physics from Purdue University in May 2006. His dissertation was significant contribution to the understanding of particle production at low transverse momentum in high energy heavy ion collisions. It is titled ``Systematics of identified particle production in \lowercase{pp}, \lowercase{d}A\lowercase{u} and A\lowercase{u}-A\lowercase{u} collisions at RHIC energies''.
\end{vita}

\end{document}